\documentclass[12pt]{book}

\usepackage{hyperref}
\usepackage{amscd}
\usepackage{amsmath,amssymb,cite}
\usepackage{mathrsfs}
\usepackage{graphicx,color}
\usepackage{fancyhdr}

\newenvironment{abstracts} {
\thispagestyle{empty}
  \begin{center}
  \vspace*{1.5cm}
  {\Large {\bfseries Abstract}}
  \end{center}
  \vspace{0.5cm}
   \begin{quote}}
{\end{quote}}

\newenvironment{acknowledgements}{
    \thispagestyle{empty}
    \begin{center}
\vspace*{1.5cm}
{\Large {\bfseries Acknowledgements}}
\end{center}
\vspace{0.5cm}
\begin{quote}}
{\end{quote}}

\newenvironment{declaration}{
    \thispagestyle{empty}
    \begin{center}
\vspace*{1.5cm}
{\Large {\bfseries Declaration}}
\end{center}
\vspace{0.5cm}
\begin{quote}}
{\end{quote}}

\newcommand{\f}[2]{\frac{#1}{#2}}
\newcommand{\ko}[1]{\left( #1 \right)}
\newcommand{\kko}[1]{\left[ #1 \right]}
\newcommand{\kkko}[1]{\left\{ #1 \right\}}
\newcommand{\abs}[1]{\left| #1 \right|}
\newcommand{\ket}[1]{\left| #1 \right\rangle}
\newcommand{\kket}[1]{| #1 \rangle}
\newcommand{\bra}[1]{\left\langle #1 \right|}
\newcommand{\vev}[1]{\left\langle #1 \right\rangle}
\newcommand{\vvev}[1]{\langle #1 \rangle}

\newcommand{\bmt}[1]{{{\mbox{\boldmath$ #1 $}}}}
\newcommand{\komoji}[1]{\mbox{$#1$}}
\newcommand{\ord}[1]{{\mathcal O\mbox{\small $\left(#1\right)$}}}

\newcommand{\pint}{\makebox[0pt][l]{\hspace{3.4pt}$-$}\int}
\newcommand{\spint}{\makebox[0pt][l]{\hspace{1.6pt}$-$}\int}

\newcommand{\hf}[2]{\mbox{\large $\frac{#1}{#2}$}}

\DeclareMathOperator{\sn}{sn}
\DeclareMathOperator{\cn}{cn}
\DeclareMathOperator{\dn}{dn}
\DeclareMathOperator{\diag}{diag}
\DeclareMathOperator{\tr}{Tr}

\def\Gs{{G\hspace{-2.6mm}/\hspace{.6mm}}}
\def\ps{{p\hspace{-1.8mm}/}}

\def\check{{\mbox{\small $\surd$}}}
\def\batsu{{\mbox{\small $\times$}}}

\def\hs{\hat\sigma}
\def\ds{\displaystyle}
\def\pa{\partial}
\def\eq{\equiv}
\def\be{\beta}
\def\bsig{{\sigma}}
\def\bth{{\theta}}
\def\om{\omega}
\def\al{\alpha}
\def\as{\alpha_{\rm s}}
\def\ag{\alpha_{\rm g}}
\def\ga{\gamma}
\def\vp{\varphi}
\def\th{\theta}
\def\sig{\sigma}
\def\tsig{{\widetilde\sigma}}
\def\ttau{{\widetilde\tau}}
\def\Th{\Theta}
\def\bTh{{\bf \Theta}}
\def\no{\nonumber}
\def\lam{\lambda}
\def\Lam{\Lambda}
\def\ep{{\epsilon}}

\def\half{{\mbox{$\f{1}{2}$}}}
\def\cN{{\mathcal N}}
\def\cZ{{\mathcal Z}}
\def\cT{{\mathcal{T}}}
\def\fD{{\mathfrak{D}}}
\def\cW{{\mathcal W}}

\def\eK{\mathrm{\bf K}}
\def\eE{\mathrm{\bf E}}
\def\eZ{\mathrm{\bf Z}}
\def\cM{{\mathcal M}}

\def\ts{{\tau\leftrightarrow\sigma}}
\def\bmE{{\mbox{\boldmath$E$}}}
\def\bmK{{\mbox{\boldmath$K$}}}

\def\O{{\mathcal{O}}}
\def\cC{{\mathcal{C}}}
\def\bcC{{{\mbox{\boldmath$\mathcal C$}}}}
\def\bcP{{{\mbox{\boldmath$\mathcal P$}}}}
\def\bcK{{{\mbox{\boldmath$\mathcal K$}}}}

\def\cQ{{\mathcal{Q}}}
\def\cE{{\mathcal{E}}}

\def\cP{{\mathcal{P}}}
\def\cK{{\mathcal{K}}}
\def\cX{{\mathcal{X}}}
\def\cY{{\mathcal{Y}}}
\def\cO{{\mathcal{O}}}
\def\fC{{\mathfrak{C}}}
\def\fR{{\mathfrak{R}}}
\def\fL{{\mathfrak{L}}}
\def\fP{{\mathfrak{P}}}
\def\fK{{\mathfrak{K}}}
\def\fQ{{\mathfrak{Q}}}
\def\fS{{\mathfrak{S}}}
\def\fJ{{\mathfrak{J}}}

\def\half{{\mbox{$\f{1}{2}$}}}
\def\ichi{{\bf{1}}}
\def\c{{\bf{c}}}

\def\A{{\bf{A}}}
\def\B{{\bf{B}}}
\def\C{{\bf{C}}}
\def\D{{\bf{D}}}
\def\a{{\bf{a}}}
\def\b{{\bf{b}}}
\def\c{{\bf{c}}}
\def\d{{\bf{d}}}

\def\xp{x^{+}}
\def\xm{x^{-}}
\def\yp{y^{+}}
\def\ym{y^{-}}

\def\Im{\mathrm{Im}\, }
\def\Re{\mathrm{Re}\, }

\def\cS{{\mathcal{S}}}
\def\cE{{\mathcal{E}}}
\def\cJ{{\mathcal{J}}}
\def\cA{{\mathcal A}}
\def\cB{{\mathcal B}}
\def\cH{{\mathcal H}}
\def\cL{{\mathcal L}}
\def\const{{\rm const.}}

\def\tlambda{{\widetilde \lambda}}

\def\bb#1{\mathbb{#1}}
\def\pare#1{\left( #1\right)}
\def\bpare#1{\left\{ #1\right\}}
\def\cpare#1{\left[ #1\right]}
\def\defeq{\,\raisebox{0.4pt}{:}\hspace{-1.2mm}=}
\def\ds{\displaystyle}
\def\ssp{\hspace{0.3mm}}
\def\sn{\,\mathrm{sn}}
\def\cn{\,\mathrm{cn}}
\def\dn{\,\mathrm{dn}}
\def\iom{i\hspace{0.2mm}\omega}
\def\iomm#1{i\hspace{0.2mm}\omega_{#1}}
\def\tomm#1{i\hspace{0.2mm}\tilde\omega_{#1}}
\def\Re{\mathop{\rm Re}\nolimits\,}
\def\Im{\mathop{\rm Im}\nolimits\,}
\def\eK{\mathrm{\bf K}}
\def\eE{\mathrm{\bf E}}
\def\RS#1{{$\mathbb{R}\times S^{#1}$}}
\def\AdS#1{{$AdS_{#1}$}}
\def\AdSS#1{{$AdS_{#1} \times S^1$}}

\renewcommand{\eqref}[1]{$\pare{\rm \ref{#1}}$}

\def\bx{{\hspace{0.6mm}\unitlength 0.1in
\begin{picture}(1.00,1.00)(10.00,-13.00)
\special{pn 8}%
\special{pa 1000 1200}%
\special{pa 1100 1200}%
\special{pa 1100 1300}%
\special{pa 1000 1300}%
\special{pa 1000 1200}%
\special{fp}%
\end{picture}%
\hspace{0.7mm}}}
\def\sbx{{\hspace{0.6mm}\unitlength 0.1in
\begin{picture}(1.00,1.00)(10.00,-13.00)
\special{pn 8}%
\special{pa 1000 1200}%
\special{pa 1100 1200}%
\special{pa 1100 1300}%
\special{pa 1000 1300}%
\special{pa 1000 1200}%
\special{fp}%
\special{pn 8}%
\special{pa 1100 1200}%
\special{pa 1000 1300}%
\special{fp}%
\end{picture}%
\hspace{0.7mm}}}
\def\vtwosbx{{\hspace{1.0mm}\unitlength 0.1in
\begin{picture}(1.00,2.00)(16.00,-20.00)
\special{pn 8}%
\special{pa 1600 1900}%
\special{pa 1700 1900}%
\special{pa 1700 2000}%
\special{pa 1600 2000}%
\special{pa 1600 1900}%
\special{fp}%
\special{pn 8}%
\special{pa 1700 1900}%
\special{pa 1600 2000}%
\special{fp}%
\special{pn 8}%
\special{pa 1600 1800}%
\special{pa 1700 1800}%
\special{pa 1700 1900}%
\special{pa 1600 1900}%
\special{pa 1600 1800}%
\special{fp}%
\special{pn 8}%
\special{pa 1700 1800}%
\special{pa 1600 1900}%
\special{fp}%
\end{picture}
\hspace{1.0mm}}}
\def\Qbx{\,{\hspace{0mm}
\unitlength 0.1in
\begin{picture}(5.95,1.35)(19.55,-19.55)
\special{pn 8}%
\special{pa 1955 1855}%
\special{pa 2055 1855}%
\special{pa 2055 1955}%
\special{pa 1955 1955}%
\special{pa 1955 1855}%
\special{fp}%
\special{pn 8}%
\special{pa 2055 1855}%
\special{pa 2155 1855}%
\special{pa 2155 1955}%
\special{pa 2055 1955}%
\special{pa 2055 1855}%
\special{fp}%
\special{pn 8}%
\special{pa 2480 1855}%
\special{pa 2580 1855}%
\special{pa 2580 1955}%
\special{pa 2480 1955}%
\special{pa 2480 1855}%
\special{fp}%
\special{pn 8}%
\special{pa 2160 1855}%
\special{pa 2210 1855}%
\special{fp}%
\special{pn 8}%
\special{pa 2160 1955}%
\special{pa 2210 1955}%
\special{fp}%
\special{pn 8}%
\special{pa 2430 1855}%
\special{pa 2530 1855}%
\special{fp}%
\special{pn 8}%
\special{pa 2430 1955}%
\special{pa 2530 1955}%
\special{fp}%
\put(23.2800,-19.1300){\makebox(0,0){$\cdots$}}%
\end{picture}%
\hspace{0mm}}~}
\def\Qsbx{\,{\hspace{0mm}
\unitlength 0.1in
\begin{picture}(5.95,1.35)(19.55,-19.55)
\special{pn 8}%
\special{pa 2055 1855}%
\special{pa 1955 1955}%
\special{fp}%
\special{pn 8}%
\special{pa 1955 1855}%
\special{pa 2055 1855}%
\special{pa 2055 1955}%
\special{pa 1955 1955}%
\special{pa 1955 1855}%
\special{fp}%
\special{pn 8}%
\special{pa 2055 1855}%
\special{pa 2155 1855}%
\special{pa 2155 1955}%
\special{pa 2055 1955}%
\special{pa 2055 1855}%
\special{fp}%
\special{pn 8}%
\special{pa 2155 1855}%
\special{pa 2055 1955}%
\special{fp}%
\special{pn 8}%
\special{pa 2480 1855}%
\special{pa 2580 1855}%
\special{pa 2580 1955}%
\special{pa 2480 1955}%
\special{pa 2480 1855}%
\special{fp}%
\special{pn 8}%
\special{pa 2155 1955}%
\special{pa 2205 1905}%
\special{fp}%
\special{pn 8}%
\special{pa 2580 1855}%
\special{pa 2480 1955}%
\special{fp}%
\special{pn 8}%
\special{pa 2160 1855}%
\special{pa 2210 1855}%
\special{fp}%
\special{pn 8}%
\special{pa 2160 1955}%
\special{pa 2210 1955}%
\special{fp}%
\special{pn 8}%
\special{pa 2430 1855}%
\special{pa 2530 1855}%
\special{fp}%
\special{pn 8}%
\special{pa 2430 1955}%
\special{pa 2530 1955}%
\special{fp}%
\put(23.2800,-19.1300){\makebox(0,0){$\cdots$}}%
\special{pn 8}%
\special{pa 2430 1905}%
\special{pa 2480 1855}%
\special{fp}%
\end{picture}%
\hspace{0mm}}~}
\def\twobx{{\bx\hspace{-1.3mm}\bx}}
\def\twosbx{{\sbx\hspace{-1.3mm}\sbx}}

\makeatletter

\renewenvironment{thebibliography}[1]%
{\chapter*{\bibname\@mkboth{\bibname}{\bibname}}%
 \addcontentsline{toc}{chapter}{\bibname}%
   \list{\@biblabel{\@arabic\c@enumiv}}%
        {\settowidth\labelwidth{\@biblabel{#1}}%
         \leftmargin\labelwidth
         \advance\leftmargin\labelsep
         \@openbib@code
         \usecounter{enumiv}%
         \let\p@enumiv\@empty
         \renewcommand\theenumiv{\@arabic\c@enumiv}}%
   \sloppy
   \clubpenalty4000
   \@clubpenalty\clubpenalty
   \widowpenalty4000%
   \sfcode`\.\@m}
  {\def\@noitemerr
    {\@latex@warning{Empty `thebibliography' environment}}%
   \endlist}

\makeatother

\begin{document}

\renewcommand{\maketitle}{%
    \thispagestyle{empty}
\null\vfill
  \begin{center}
{ \LARGE \bf Aspects of Integrability in AdS/CFT Duality \par}
{\large \vspace*{35mm} {{\includegraphics[width=15mm]{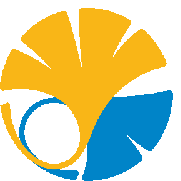}} \par} \vspace*{25mm}}
    {{\large \sc Keisuke Okamura} \par}
{\large \vspace*{4ex}
    {\em {Department of Physics, Faculty of Science,} \par}
\vspace*{1ex}
    {\em {University of Tokyo} \par}
\vspace*{25mm}
    {{A dissertation submitted for the degree of} \par}
\vspace*{1ex}
    {{Doctor of Philosophy} \par}
\vspace*{1ex}
    {{at the University of Tokyo} \par}
\vspace*{20mm}
    {December 2007}}
  \end{center}
  \null\vfill
}

\maketitle

\newpage
\quad 
\newpage

\begin{acknowledgements}

I am indebted to many people for my personal and scientific growth during these three years of study and work.\\[-5mm]

First of all, I would like to thank my advisor Yutaka Matsuo, for having chosen me as a Ph.D student, for all his help in my studies, and for all the opportunities he gave me that made my Ph.D days so joyful.
I still remember that it was he who suggested to me during my Master's course to try looking at a phenomena called ``AdS/CFT duality'' for my research area.\\[-6mm]

I would like to express my sincere gratitude to Nicholas Dorey for guiding me into research projects with enthusiasm and generosity, and for making me part of a lively and stimulating research environment.
I should give a lot of credit to him for this thesis.
He opened my eyes on many important aspects of the subject, and made the works contained in this thesis possible.
I certainly remember the many exciting and fruitful discussions we had, and I highly value his constant encouragement during my work.\\[-6mm]

I am especially grateful to Heng-Yu Chen, who has helped me in many ways.
I thank him not only for the scientific collaboration but also for the precious friendship.
I can never imagine my Ph.D days without his exclusive support.

I enjoyed collaboration with Yasuyuki Hatsuda, Hirotaka Hayashi, Ryo Suzuki, Yastoshi Takayama, Beno\^it Vicedo and Kentaroh Yoshida.
I have much benefited from discussions with them.\\[-6mm]

My gratitude also goes to all the former and present members of the University of Tokyo for the stimulating scientific environment they create, and the valuable source of knowledge they provide.
I would like to especially thank Tohru Eguchi, Kazuo Fujikawa, Kota Ideguchi, Yosuke Imamura, Teruhiko Kawano, Yu Nakayama, Yuji Sugawara, Yuji Tachikawa, Taizan Watari and Futoshi Yagi.
I am also grateful to our secretary Mami Hara and our former secretary
Michiko Ishiyama for helping me with administrative issues.\\[-6mm]

A number of distinguished physicists have helped me along the way.
It was a pleasure having discussions with Changrim Ahn, Diego Hofman, Juan Maldacena, Joseph Minahan, Soo-Jong Rey, Kazuhiro Sakai, Yuji Satoh, Matthias Staudacher, Arkady Tseytlin and Konstantin Zarembo.
The illuminating comments they gave me were very useful in completing this thesis.\\[-6mm]

It was an unforgettable experience for me to work in the Department of Applied Mathematics and Theoretical Physics (DAMTP), Cambridge University, as a visiting scholar.
I am so grateful to Nicholas Dorey for hosting me there.
I would like to thank many people in DAMTP\,: Joseph Conlon, Diego Correa, David Kagan, Rui F.\ Lima Matos, Keisuke Ohashi, Kerim Suruliz, David Tong, Beno\^it Vicedo, and many others for their hospitality and warm friendship.

I would also like to thank Heng-Yu Chen, Michael Green, Malcolm Perry, Amanda Stagg, John Turner, for their assistance in arranging my stay at DAMTP and Trinity College.
Special thanks to Yee-San Teoh and Pei-Jung Yang.\\[-6mm]

I would like to thank the organisers of {``Gauge Fields \& Strings''} held from 17th till 27th September 2007 at Isaac Newton Institute (INI), which is a part of the INI programme {``Strong Fields, Integrability and Strings''}.
The lecture series were useful in the completion of the introduction/review part of this thesis.\\[-6mm]

I would like to thank Nicholas Dorey, Yoichi Kazama, Yuji Satoh, Matthias Staudacher, and many others for valuable comments and feedback on the manuscript of this thesis.
I am especially grateful to Matthias Staudacher for kindly agreeing to take the role of an external referee for this thesis, for his very careful reading of it and for the many useful discussions, comments and suggestions.\\[-6mm]

This work was supported in part by Japan Society for the Promotion of Science (JSPS) Research Fellowships for Young Scientists.\\[-3mm]

Last but not least, I would like to thank my parents and my family, who gave me all their support and attention during my graduate student days. 

\end{acknowledgements}
\addcontentsline{toc}{chapter}{Acknowledgments}

\newpage

\begin{declaration}
The research described in this dissertation was carried out in Department of Physics, Faculty of Science, 
University of Tokyo, and also in DAMTP, Centre for Mathematical Sciences, Cambridge University, between April 2005 and December 2007.
The results are original except where reference is made to the work of others.
Following is the list of my original works mainly discussed in this thesis\,:\\[-3mm]

\begin{enumerate}
\item
{\sf \bfseries ``Dyonic giant magnons''}
  \\[2mm]
  {}H.~Y.~Chen, N.~Dorey and K.~Okamura
  \\{}{\em JHEP} {\bf 0609}, 024 (2006)
  \href{http://arXiv.org/abs/hep-th/0605155}{{\tt hep-th/0605155}}
  \\[-3mm]
%
\item
{\sf \bfseries ``On the scattering of magnon boundstates''}
  \\[2mm]
  {}H.~Y.~Chen, N.~Dorey and K.~Okamura
  \\{}{\em JHEP} {\bf 0611}, 035 (2006)
  \href{http://arXiv.org/abs/hep-th/0608047}{{\tt hep-th/0608047}}
  \\[-3mm]
%
\item
{\sf \bfseries ``The asymptotic spectrum of the \bmt{{\mathcal N} = 4} super Yang-Mills spin chain''}
  \\[2mm]
  {}H.~Y.~Chen, N.~Dorey and K.~Okamura
  \\{}{\em JHEP} {\bf 0703}, 005 (2007)
  \href{http://arXiv.org/abs/hep-th/0610295}{{\tt hep-th/0610295}}
  \\[-3mm]
%
\item
{\sf \bfseries ``Large winding sector of AdS/CFT''}
  \\[2mm]
  {}H.~Hayashi, K.~Okamura, R.~Suzuki and B.~Vicedo
  \\{}{\em JHEP} {\bf 0711}, 033 (2007)
  \href{http://arXiv.org/abs/arXiv:0709.4033 [hep-th]}{{\tt arXiv:0709.4033
  [hep-th]}}
\\[-3mm]
%
\item
{\sf \bfseries ``Singularities of the magnon boundstate S-matrix''}
  \\[2mm]
  {}N.~Dorey and K.~Okamura
  \\{}{\em JHEP} {\bf 0803}, 037 (2008)
  \href{http://arXiv.org/abs/arXiv:0712.4068 [hep-th]}{{\tt arXiv:0712.4068
  [hep-th]}}
\\[-3mm]
\end{enumerate}

\noindent
These papers are references \cite{Chen:2006ge, Chen:2006gq, Chen:2006gp,Hayashi:2007bq,Dorey:2007an} in the bibliography and are the main subject matter of Chapters \ref{chap:DGM}, \ref{chap:S-matrices in HM}, \ref{chap:Asymptotic}, \ref{chap:OS}, \ref{chap:Singularities}, respectively, of this thesis.
None of the original works contained in this dissertation has been submitted by me for any other degree, diploma or similar qualification.\\[-3mm]

Below is the list of my other papers with collaborators, some of which are also discussed in this dissertation\,:\\[-3mm]

\begin{enumerate}
\setcounter{enumi}{5}
\item
{\sf \bfseries ``Open spinning strings and AdS/dCFT duality''}
  \\[2mm]
  {}K.~Okamura, Y.~Takayama and K.~Yoshida
  \\{}{\em JHEP} {\bf 0601}, 112 (2006)
  \href{http://arXiv.org/abs/hep-th/0511139}{{\tt hep-th/0511139}}
  \\[-3mm]
%
\item
{\sf \bfseries ``The anatomy of gauge/string duality in Lunin-Maldacena background''}
  \\[2mm]
  {}H.~Y.~Chen and K.~Okamura
  \\{}{\em JHEP} {\bf 0602}, 054 (2006)
  \href{http://arXiv.org/abs/hep-th/0601109}{{\tt hep-th/0601109}}
  \\[-3mm]
%
\item
{\sf \bfseries ``Higher loop Bethe ansatz for open spin-chains in AdS/CFT''}
  \\[2mm]
  {}K.~Okamura and K.~Yoshida
  \\{}{\em JHEP} {\bf 0609}, 081 (2006)
  \href{http://arXiv.org/abs/hep-th/0604100}{{\tt hep-th/0604100}}
  \\[-3mm]
%
\item
{\sf \bfseries ``A perspective on classical strings from complex sine-Gordon solitons''}
  \\[2mm]
  {}K.~Okamura and R.~Suzuki
  \\{}{\em Phys.\ Rev.}  {\bf D75}, 046001 (2007)
  \href{http://arXiv.org/abs/hep-th/0609026}{{\tt hep-th/0609026}}
  \\[-3mm]
%
\item
{\sf \bfseries ``Emergent classical strings from matrix model''}
  \\[2mm]
  {}Y.~Hatsuda and K.~Okamura
  \\{}{\em JHEP} {\bf 0703}, 077 (2007)
  \href{http://arXiv.org/abs/hep-th/0612269}{{\tt hep-th/0612269}}\\
\end{enumerate}

\begin{flushright}
{Keisuke Okamura}\\
\textit{Tokyo, Japan}\\
\textit{20th December 2007}
\end{flushright}

\end{declaration}
\addcontentsline{toc}{chapter}{Declaration}

\newpage

\begin{abstracts}

In this dissertation, we discuss how our understanding of the large-$N$ spectrum of AdS/CFT has been deepened by integrability-based approaches.\\[-5mm]

We begin with a comprehensive review of the integrability of the gauge theory spin-chain and that of the string sigma model.
In the light of the AdS/CFT duality, they should be just two ways of describing the same underlying integrability, and it is believed that the unified integrability can be characterised by a set of Bethe ansatz equations which is valid for all values of the 't Hooft coupling.\\[-5mm]

The key objects we consider in testing the conjectured Bethe ansatz equations are multi-spin AdS/CFT solitons.
By studying the asymptotic spectrum of the AdS/CFT in the infinite spin/R-charge limit, we first identify the corresponding solitonic counterparts in the context of the AdS/CFT, which are the so-called dyonic giant magnons and the SYM magnon boundstates.
Then we show that the S-matrix computed directly from the string solitons scattering precisely reproduces the prediction from the conjecture.
We further perform an analyticity test by studying the singularities of the conjectured magnon boundstate S-matrix and checking the physicality conditions.
These tests give strong positive supports for the integrability of large-$N$ AdS/CFT as well as the specific form of the conjectured Bethe ansatz equations.\\[-5mm]

Concerning the string theory integrability, we also provide a detailed study of certain classical string solutions on $AdS_{5}\times S^{5}$\,.
These are called helical strings, which are constructed in such a way they correspond to generic soliton solutions of (Complex) sine/sinh-Gordon equations via the so-called Pohlmeyer reduction procedure.
Furthermore, we describe them in terms of algebro-geometric data as finite-gap solutions, giving a complete map of the elliptic string solutions.

\end{abstracts}
\addcontentsline{toc}{chapter}{Abstract}

\newpage
\quad 
\newpage

\pagestyle{plain}

\tableofcontents

\part[Introduction]
	{Introduction\label{part:0}}

\chapter[The Maldacena conjecture]
	{The Maldacena conjecture\label{chap:intro}}

It is an established notion that the Standard Model successfully describes three out of the four types of fundamental interactions in nature.
Those three already unified are the electromagnetic, strong and weak forces.
A quantum field theory for the remaining force, gravity, has not yet been found to date, as the fields of gravity are non-renormalisable.
However, there is one very attractive candidate that can unify all the four natural forces.
It is string theory, which offers a consistent quantum field theory of gravity.
Remarkably, it is conjectured that string theories in certain backgrounds are dual to particular gauge theories.

It has been ten years since the discovery of AdS/CFT correspondence \cite{Maldacena:1997re} in 1997, which provided the first concrete realisation of the gauge/string duality proposed by 't Hooft in 1974.
It also provided the first concrete example of the so-called the holographic principle, which had been proposed independently of gauge/string theory.

In this introduction, we will first give a heuristic derivation for the Maldacena's original argument for the AdS/CFT correspondence.\footnote{For a comprehensive review on the AdS/CFT correspondence and its applications, see \cite{Aharony:1999ti}.}
In particular, we discuss the correspondence between four-dimensional $SU(N)$ $\cN=4$ super Yang-Mills (SYM) theory and type IIB superstring theory on $AdS_{5}\times S^{5}$ background with R-R flux.
This AdS${}_{5}$/CFT${}_{4}$ correspondence is the best-studied example of AdS/CFT.
Then we briefly describe how integrability can be used to test the proposed duality, giving an overview of the progress and developments in the recent years.

\section[Large-$N$ gauge theory]
	{Large-\bmt{N} gauge theory}

Let us consider $SU(N)$ Yang-Mills theory with or without adjoint matter fields.
For the standard quantum chromodynamics (QCD), the rank of the gauge group is $N=3$\,.
The perturbative QCD picture in which quarks and gluons are the fundamental degrees of freedom is only valid in the weak coupling (high-energy) region, and they would not be fundamental elements in the strongly coupled (low-energy) dynamics.
To study the low-energy physics, we are going to define a special limit invented by 't Hooft.

The Yang-Mills coupling constant $g_{\rm YM}$ undergoes dynamical transmutation due to the renormalisation flow, which is governed by the {asymptotically free} $\beta$-function
\begin{equation}
\beta(g_{\rm YM})=\f{dg_{\rm YM}}{d\ln \mu}=-\f{g_{\rm YM}^{3}}{(4\pi^{2})}\cdot\f{11}{3}\,N+\cO(g_{\rm YM}^{5})\,.
\label{beta QCD}
\end{equation}
Ignoring $\cO(g_{\rm YM}^{5})$ terms, the differential equation (\ref{beta QCD}) can be solved easily, giving $g_{\rm YM}(\mu)^{2}N\propto\ln(\Lam_{\rm QCD}/\mu)$\,, where $\Lam_{\rm QCD}$ is the dynamical QCD scale.
The coupling can thus be determined dynamically, and there are no other free parameters in the QCD (with $N=3$).
To study the low-energy physics, the crucial idea by 't Hooft was to regard the rank of the gauge group $N$ as a free parameter, and send it to infinity and use $1/N$ as an expansion parameter of the theory.
Then it can be verified that the following limit called the {\em 't Hooft limit},
\begin{equation}
N\to \infty\,,\qquad \mbox{$\lam\eq g_{\rm YM}^{2}N\, :$~fixed}\,,
\label{'t Hooft limit}
\end{equation}
provides a well-defined limit, keeping $\Lam_{\rm QCD}$ fixed.

\begin{figure}[t]
\begin{center}
\vspace{.3cm}
\includegraphics[scale=1.0]{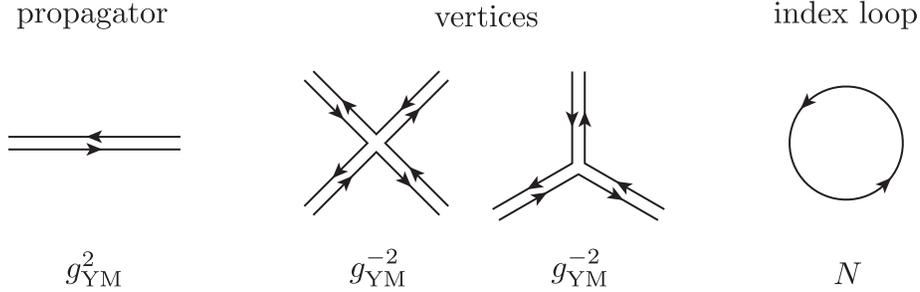}
\vspace{.3cm}
\caption{\small Feynman rules for Lagrangian (\ref{SU(N) Lag}) in terms of the double-line notation (only the dependence on $g_{\rm YM}$ and $N$ is stated).}
\label{fig:Feynman}
\end{center}
\end{figure}

Let us see the implication of the limit for the $SU(N)$ Yang-Mills theory.
The Lagrangian of the theory is given by, schematically,
\begin{equation}
\cL=\f{1}{g_{\rm YM}^{2}}\kko{\tr(\pa X_{i}\pa X_{j})+c_{(3)}^{ijk}\tr(X_{i}X_{j}X_{k})+c_{(4)}^{ijkl}\tr(X_{i}X_{j}X_{k}X_{l})}\,,
\label{SU(N) Lag}
\end{equation}
where $X_{i}$ stand for arbitrary fields such as gauge fields, adjoint matters.
They are all $N\times N$ Hermitian matrix-valued fields in the adjoint representation, $X^{a}_{b}=(X^{b}_{a})^{\dagger}$\,, and also $X^{a}_{a}=0$ due to the tracelessness.
The first term in the Lagrangian is essentially the kinetic term.
The Feynman rules can be read off from the Lagrangian (\ref{SU(N) Lag}), and they are displayed in Figure \ref{fig:Feynman}.
We employed the double-line notation introduced by 't Hooft\,;
since all the fields are in the adjoint representation, each field carries colour indices of fundamental and antifundamental representations.
These two representations are indicated by arrows opposite to each other.
The merit of this prescription is that in this way we can classify Feynman diagrams topologically.
As an illustration, let us consider the following vacuum-vacuum amplitudes\,:
\bigskip
\newsavebox{\boxPlanar}
\sbox{\boxPlanar}{\includegraphics{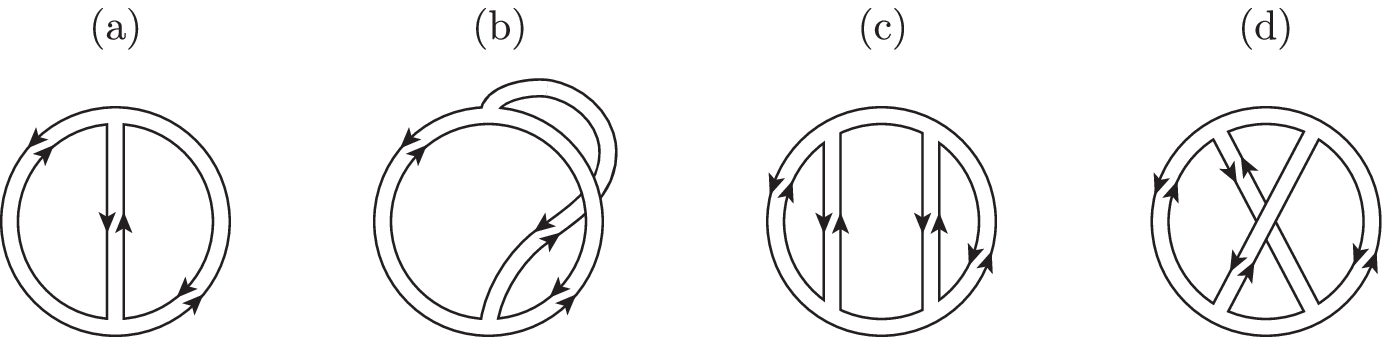}}
\newlength{\bwPlanar}
\settowidth{\bwPlanar}{\usebox{\boxPlanar}} 
\begin{center}
\parbox{\bwPlanar}{\usebox{\boxPlanar}}
\end{center}
\bigskip
Using the Feynman rules listed in Figure \ref{fig:Feynman}, the $g_{\rm YM}$ and $N$ dependence of each amplitude can be evaluated as
\begin{alignat}{5}
&\mbox{(a)} & ~ &\sim ~ (g_{\rm YM}^{2})^{3}\times (g_{\rm YM}^{-2})^{2}\times N^{3}
&&=(g_{\rm YM}^{2}N)N^{2}\,,\no\\
&\mbox{(b)} & ~ &\sim ~ (g_{\rm YM}^{2})^{3}\times (g_{\rm YM}^{-2})^{2}\times N
&&=g_{\rm YM}^{2}N\,,\no\\
&\mbox{(c)} & ~ &\sim ~ (g_{\rm YM}^{2})^{6}\times (g_{\rm YM}^{-2})^{4}\times N^{4}
&&=(g_{\rm YM}^{2}N)^{2}N^{2}\,,\no\\
&\mbox{(d)} & ~ &\sim ~ (g_{\rm YM}^{2})^{6}\times (g_{\rm YM}^{-2})^{2}\times N^{2}
&&=(g_{\rm YM}^{2}N)^{2}\,.\no
\end{alignat}
By noticing the dependence on $g_{\rm YM}^{2}N$ and $N$\,, one finds the diagram (a) and (c) dominate over diagram (b) and (d) in the large-$N$ limit\,.
Actually we can always organise perturbation theory in $g_{\rm YM}^{2}N$ and $1/N$\,, and the perturbation theory simplifies considerably in the 't Hooft limit (\ref{'t Hooft limit}).

Let us see what this limit means from a viewpoint of the topology of Feynman diagrams.
The diagrams (a) and (c) can be drawn on a two-dimensional surface with the topology of a sphere without self-crossing, while (b) and (d) cannot, and can only be embedded on a two-dimensional surface with the topology of a two-torus.
In this way, by employing the double-line notation, any Feynman diagram perturbatively expanded in powers of $1/N$ can be viewed as a polygonisation of a two-dimensional surface with $V$ vertices, $E$ edges and $F$ faces, which correspond to the vertices, propagators and index loops of Feynman diagrams.
In general, a diagram with $V$ vertices, $E$ propagators and $F$ index loops can be organised as
\begin{equation}
(g_{\rm YM}^{2})^{E}\times (g_{\rm YM}^{-2})^{V}\times N^{F}
=(g_{\rm YM}^{2}N)^{E-V}N^{F-E+V}=\lam^{E-V}N^{\chi}\,,
\end{equation}
where $\chi=F-E+V$ is a topological invariant called the Euler character of two-dimensional surface.
Writing it as $\chi=2-2g$\,, $g$ represents the genus of the surface.
The partition function $\mathscr Z$ can be doubly expanded in terms of $1/N$ and $\lam$ as
\begin{equation}
\mathscr Z\ko{\mbox{\large $\f{1}{N}$},\lam}=\sum_{g=0}^{\infty}N^{2-2g}{\mathscr F}_{g}(\lam)=\sum_{g=0}^{\infty}N^{2-2g}\sum_{n=0}^{\infty}\al_{g,n}\,\lam^{n}\,.
\end{equation}
The leading contribution with $g=0$ for fixed $\lam$\,, which dominates in the large-$N$ limit, is the contributions of {\em planar} diagrams, since it can be drawn on a plane without self-crossing.
All other diagrams are called non-planar diagrams.
The above argument can be generalised to correlation functions as $\langle \cO_{1}\dots\cO_{n} \rangle\sim\sum_{g=0}^{\infty}N^{2-2g-n}{\mathscr F}_{g}(\lam)$\,, where $\cO_{i}$ are single-trace local operators.
Another remark is that, in the context of the AdS/CFT duality, the two parameters $1/N$ and $\lam$ actually roughly correspond to the string coupling constant $g_{\rm s}$ and the string inverse tension $\al'$\,, as we will see below.

\section[The AdS$_{5}$/CFT$_{4}$ Correspondence]
	{The AdS\bmt{_{5}}/CFT\bmt{_{4}} Correspondence}

In physics it often happens that studying a system from several different points of view reveals a profound fact about the system.
It is just what happened to Maldacena who arrived at the celebrated AdS/CFT duality conjecture by looking at the same brane system in two distinct ways.

\subsection{Description I\,: D-branes interact with closed strings}

Let us consider the type IIB string theory in ten-dimensional flat spacetime $\mathbb R^{1,9}$ in the background of $N$ D3-branes, which are sitting on top of each other at $x^{4}=\dots =x^{9}=0$\,.
We use $x^{\mu}=(t,x_{1},x_{2},x_{3})$ to denote the coordinates on the coincident D3-branes\,; the first coordinate is the time variable and the rest three are the spatial variables.
In this setup, there are two kinds of the excitations\,: one is closed strings which live in ten dimensions, including gravitons $g_{\mu\nu}$\,, NS-NS two-form flux $B_{\mu\nu}$\,, a dilaton $\phi$\,, an axion $\chi$\,, R-R two-form potential $C_{\mu\nu}$ and a four-form potential $C_{\mu\nu\rho\sig}$\,.
The other is open strings whose ends are confined to D-branes, and the excitation modes on the branes include gauge fields $A_{\mu}$\,, scalars $\Phi_{i}$ and fermions $\Psi$\,.
Hence the action for this system is given by the sum of the actions for the brane (four-dimensional field theory on the branes plus higher derivative terms $\cO(l_{\rm s}^{2})$\,, where $l_{\rm s}$ is the string scale), the closed string (ten-dimensional type IIB supergravity plus higher derivative terms $\cO(l_{\rm s}^{2})$), and the interaction between brane modes and bulk modes.
Schematically it is represented as
\begin{equation}
S=S_{\rm brane}+S_{\rm bulk}+S_{\rm int}\,.
\label{S=S+S+S}
\end{equation}
If we wish to decouple the low-energy physics on the branes (which is described by $SU(N)$\,, $\cN=4$ SYM theory in four dimensions) and the closed strings in the bulk, we just need to switch off the Newton's constant and take the low-energy limit ($S_{\rm int}\to 0$).
This amounts to taking the following {\em decoupling limit}
\begin{equation}
g_{\rm YM}\,:\,\mbox{fixed}\quad \mbox{and}\quad l_{\rm s}\to 0\,.
\label{decoupling}
\end{equation}
Then the bulk theory governed by $S_{\rm bulk}$ becomes free, and $S_{\rm brane}$ becomes equivalent to the action of the four-dimensional $\cN=4$ SYM.

\begin{figure}[t]
\begin{center}
\vspace{.3cm}
\includegraphics[scale=1.0]{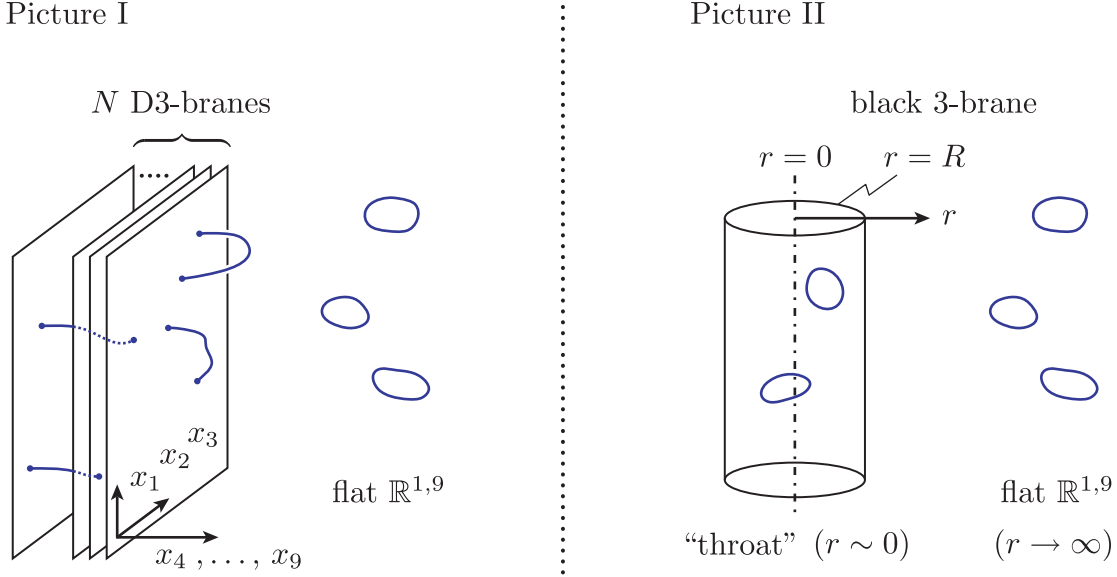}
\vspace{.3cm}
\caption{\small Two ways to look at the same system\,: Description I (open string point of view) is shown in the picture I and Description II (closed string point of view) is shown in the picture II.}
\label{fig:ads-cft}
\end{center}
\end{figure}

\subsection{Description II\,: D-branes turn into black brane geometry}

We are going to replace the above picture with completely equivalent but different, purely gravitational point of view.
First notice when the number of the branes $N$ is very large, we can allow the branes to back-react on the geometry of the bulk spacetime.
One can write down a complicated geometry in the ten-dimensional supergravity which has the same (macroscopic) quantum number as the $N$ D3-branes, which is known as a black 3-brane geometry.
The metric and the self-dual five-form flux are given by 
\begin{align}
&ds^{2}_{(10)}=H(r)^{-1/2}\ko{-dt^{2}+{(dx^{\mu})}^{2}}+H(r)^{1/2}\ko{dr^{2}+r^{2}d\Omega_{5}^{2}}\,,
\label{black p-brane}\\ 
&F_{(5)}=\ko{1+*}dt\wedge dx^{1}\wedge dx^{2}\wedge dx^{3}\wedge dH^{-1}\,,
\end{align}
where the harmonic function in six-dimension is defined as
\begin{equation}
H(r)=1+\f{R^{4}}{r^{4}}\,,\qquad R^{4}=4\pi g_{\rm s}\al'{}^{2}N=\al'{}^{2}\lam\,.
\label{H(r)}
\end{equation}
Here $r$ is the radial distance away from the branes, and $\Omega_{5}$ is a five-sphere transverse to the branes.
The metric (\ref{black p-brane}) is a solution of equations of motion for the type IIB supergravity, in which the branes are the source of the gravity $g_{\mu\nu}$ and the four-form potential $C_{\mu\nu\rho\sig}$\,.
The second equation in (\ref{H(r)}) defines the relation between the radius parameter $R$ and the magnitude of the flux.
On one hand, in the region $r\gg R$\,, the harmonic function tends to $H(r)\to 1$\,, and the black 3-brane geometry asymptotes a flat $\mathbb R^{1,9}$ spacetime, where closed strings propagating freely.
On the other hand, in the region $r\ll R$\,, it asymptotically becomes the metric of an $AdS_{5}\times S^{5}$ as will be shown momentarily.

\paragraph{}
As in the case of the description I, we can split the system into two regions.
To see this, let us consider an energy flux of closed string excitations, say a dilaton wave, scattering off the brane.
When scattered by the branes, some open strings are excited, and the dilaton is reflected back.
Let us solve the massless scalar wave equation for the dilaton, $\square\phi=0$\,, where $\square$ is the d'Alembertian operator defined for the metric (\ref{black p-brane}).
Plugging the ansatz in the form $\phi(x_{\mu},r,\Omega_{5})=e^{ip^{\mu}x_{\mu}}Y_{\ell}(\Omega_{5})\varphi(r)$ into the wave equation, one obtains a one-dimensional differential equation for $\varphi(r)$\,, which can be solved exactly.
However, for our purpose, to see the decoupling of the two regions, it suffices to consider the $r\gg R$ (asymptotically flat region) and $r\ll R$ ($AdS_{5}\times S^{5}$) behaviors of the solution.
One then finds the absorption cross-section as $\sig\sim \om^{3}R^{8}$ with the energy of the wave $\om\eq p^{0}$ \cite{Klebanov:1997kc}.
The result tells us that as the incident energy gets lower, it becomes harder for the dilaton to be absorbed by the branes.
In other words, there is an effective barrier between the two regions $r\gg R$ and $r\ll R$\,.

Let us see more precisely how we can decouple the $AdS_{5}\times S^{5}$ region and the $\mathbb R^{1,9}$ region in the low-energy limit.
The prescription for obtaining a string theory on the $AdS_{5}\times S^{5}$ decoupled from the asymptotically flat region is to send the string scale $l_{\rm s}$ (and therefore the Planck length $l_{\rm P}$ and the Newton's constant in ten-dimension $G_{10}=g_{\rm s}^{2}l_{\rm s}^{8}=l_{\rm P}^{8}$) to zero while keeping $g_{\rm s}$ fixed.
This is the same limit as the decoupling limit (\ref{decoupling}) we already saw in the description I.
Note we also keep $\lam=4\pi g_{\rm s}N$ fixed.
At the same time, we also want to keep the energy finite down the throat, $\sqrt{\al'}E_{r}$ fixed, where $E_{r}$ is the typical energy scale at some point $r\ll R$\,.
The energy $E_{\infty}$ measured at the asymptotically flat region $r\gg R$ undergoes a redshift due to the $H^{-1/2}$ factor in front of $dt^{2}$\,, so we have $E_{\infty}=H^{-1/4}E_{r}$\,.
For $r\ll R$\,, it becomes $E_{\infty}\sim r\al'{}^{-1/2}\lam^{-1/4}E_{r}$\,.
In particular, $E_{\infty}\to 0$ as $r\to 0$\,.
In view of $E_{\infty}\sim (r/\al')\lam^{-1/4}(\al'{}^{1/2}E_{r})$\,, for our purpose to decouple the two asymptotic regions while retaining finite energy on both sides, we should also keep $r/\al'$ fixed as $r\to 0$\,, along with taking (\ref{decoupling}).
By taking this {\em ``near-horizon'' limit}, the metric of the black brane (\ref{black p-brane}) becomes
\begin{equation}
ds^{2}_{10}~\sim~ ds^{2}_{(AdS_{5}\times S^{5})}=\kko{\f{r^{2}}{R^{2}}\ko{-dt^{2}+{(dx^{\mu})}^{2}}+\f{R^{2}}{r^{2}}dr^{2}}+R^{2}d\Omega_{5}^{2}\,.
\label{AdSS metric-1}
\end{equation}
Obviously the last term represents the metric of a five-sphere with radius $R$\,, and the remaining part in the square parentheses represents five-dimensional anti-de-Sitter (AdS) space $AdS_{5}$\,.
The five-form flux integrated over the sphere gives $\int_{S^{5}}F_{(5)}=N$\,, which is the same as the charge of the $N$ D3-branes in the first description.
Thus we have shown the original system has decoupled to the free supergravity in ten-dimensional flat region and the superstring on the near-horizon region, that is on $AdS_{5}\times S^{5}$\,.

\paragraph{}
It is convenient to introduce the so-called Poincar\'e coordinate, $z\eq R^{2}/r$\,, and rewrite the metric as
\begin{equation}
ds^{2}_{(AdS_{5}\times S^{5})}= R^{2}\kko{\f{-dt^{2}+{(dx^{\mu})}^{2}+dz^{2}}{z^{2}}+d\Omega_{5}^{2}}\,.
\label{AdSS metric-2}
\end{equation}
In the new variable, the horizon at $r=0$ and the boundary at $r=\infty$ (which is actually $r\sim R$) of the AdS space correspond to $z=\infty$ and $z=0$\,, respectively.

\subsection{The AdS$\bmt{{}_{5}}$/CFT$\bmt{{}_{4}}$ Correspondence}

\subsubsection*{The conjecture}

We have so far described the same system from two different points of view.
In the description I, we observed that in the decoupling limit (\ref{decoupling}),
\begin{align}
&\mbox{$N$ D3-branes interacting with closed strings in flat $\mathbb R^{1,9}$}\label{picture I-1}\\
&\quad \longrightarrow\quad 
\mbox{Four-dimensional $SU(N)$ $\cN=4$ SYM}
~ \oplus ~
\mbox{Free IIB SUGRA in flat $\mathbb R^{1,9}$}\,.\label{picture I-2}
\end{align}
On the other hand, in the description II, we found, by taking the same limit (\ref{decoupling}),
\begin{align}
&\mbox{$N$ D3-branes as black 3-brane geometry in ten-dimensional SUGRA}\label{picture II-1}\\
&\quad \longrightarrow\quad 
\mbox{Type IIB superstring in $AdS_{5}\times S^{5}$}
~ \oplus ~
\mbox{Free IIB SUGRA in flat $\mathbb R^{1,9}$}\,.\label{picture II-2}
\end{align}
By equating (\ref{picture I-1}) and (\ref{picture II-1}) and noticing that we have the same free IIB supergravity in the second terms in (\ref{picture I-2}) and (\ref{picture II-2}), we arrive at the relation,
\bigskip
\begin{equation}
\mbox{\it Four-dimensional $\it SU(N)$ $\it \cN=4$ SYM}
 ~~ = ~~ 
\mbox{\it Type IIB superstring in $\it AdS_{5}\times S^{5}$}\,.
\label{AdS-CFT}
\bigskip
\end{equation}
This is the celebrated conjecture by Maldacena, and is pursued in the current thesis.
As we will see in detail later in Section \ref{sec:N=4 SCFT}, in the four-dimensional $\cN=4$ SYM, the $\beta$-function for the coupling constant vanishes to all orders in perturbation theory, namely it is a conformal field theory (CFT).
Therefore the conjecture is usually referred to as the {\em AdS/CFT correspondence (duality)}.
In particular, we have discussed the AdS${}_{5}$/CFT${}_{4}$ case.
It states the bulk gravity theory defined in the AdS space is equivalent to the boundary field theory, and in this sense the AdS/CFT duality can be regarded as a manifestation of the holographic principle, which states all of the information contained in some region of space can be represented as a theory which lives on the boundary of that region.

\subsubsection*{The AdS/CFT dictionary}

Let us see how the parameters on both sides are related.
The type IIB string theory has a (non-perturbative) $SL(2,\mathbb Z)$ invariance, and has a complex coupling constant $\tau_{\rm s}=i/g_{\rm s}+\chi/2\pi$ which is formed of the expectation value of the dilaton $g_{\rm s}=e^{\vev{\phi}}$ and the axion $\chi$\,.
It can be viewed as a moduli parameter arising from the compactification of M-theory on a two-torus.
The $\cN=4$ Yang-Mills theory also has a complex coupling $\tau_{\rm g}=4\pi i/g_{\rm YM}^{2}+\theta/2\pi$\,, which exhibits an $SL(2,\mathbb Z)$ symmetry acting on a doublet of electric and magnetic charges.
Hence the correspondence $\tau_{\rm s}=\tau_{\rm g}$ of the coupling constants tells us $4\pi g_{\rm s}=g_{\rm YM}^{2}$\,, which is equal to the 't Hooft coupling divided by $N$\,, and also $\chi=\theta$\,.
We end up with the following relation between the parameters\,:
\begin{equation}
4\pi g_{\rm s}N=\f{R^{4}}{\al'{}^{2}}=\lam=g_{\rm YM}^{2}N\,.
\label{key relation}
\end{equation}
This is the central relation in the AdS${}_{5}$/CFT${}_{4}$ duality.

\subsubsection*{Correspondence of the global symmetry}

Both the four-dimensional $\cN=4$ SYM and the superstring theory on $AdS_{5}\times S^{5}$ have the same bosonic global symmetry, $SO(2,4)\times SO(6)$\,.
From the gauge theory perspective, $SO(2,4)$ and $SO(6)$ correspond to the four-dimensional conformal symmetry and the $\cN=4$ R-symmetry, respectively, while from the string theory perspective, they correspond to the isometries of $AdS_{5}$ and $S^{5}$\,, respectively.
Furthermore, the string background possesses thirty-two supercharges\,;
in the presence of parallel (coincident) D3-branes, the supersymmetry is broken and only half (sixteen) remain, however, in the near-horizon limit, the number of supersymmetries is doubled due to extra Killing spinors, thus recovering thirty-two supercharges.
They correspond to the sixteen supersymmetry and the sixteen superconformal symmetry charges of the gauge theory (see Section \ref{sec:N=4 SCFT}).
Thus both the gauge and string theory have the same number of supersymmetries and the full global symmetry of AdS${}_{5}$/CFT${}_{4}$ is $PSU(2,2|4)$\,.

\subsubsection*{The strong/weak nature}

Let us briefly discuss the valid regions for each description I and II.
The D3-brane tension of the system is given by $N$ times the one of a single brane, $T_{\rm brane}=Ng_{\rm s}^{-1}l_{\rm s}^{-4}$\,.
The Schwarzschild radius $a$ is determined by the condition that the brane tension and the gravity self-energy balance out, $G_{10}/a^{4}\times T_{\rm brane}\sim 1$\,, which yields $a\sim \lam^{1/4}l_{\rm s}$\,.
For the description I to be valid, the Schwarzschild radius must be much smaller than the string scale $l_{\rm s}$\,, which can be rephrased as $\lam\ll 1$\,.
On the other hand, for the description II where the system is described by classical supergravity to be valid, the relation $l_{\rm P}\, ( < l_{\rm s}) \ll a$ must be satisfied, {\em i.e.}, the curvature of the background must be much larger than the string scale.
This condition yields $\lam\gg 1$ (notice we fixed $g_{\rm s}$ small)\,.
Therefore the perturbative regime of gauge theory corresponds to non-perturbative regime of the string theory, and vice versa, namely, the duality is of a {\em strong/weak} type.

This strong/weak nature is a blessing and a misfortune at the same instance.
It can be seen as a blessing since once the duality is established, it enables us to study the non-perturbative regime of one of the theories by using the perturbative result of the other theory, and vice versa.
It is a misfortune since it is hard to prove the duality itself.
Definitely, we wish to turn this misfortune into a fullfledged blessing.

\subsection{The AdS/CFT spectrum}

\begin{figure}[t]
\begin{center}
\vspace{.3cm}
\includegraphics[scale=1.0]{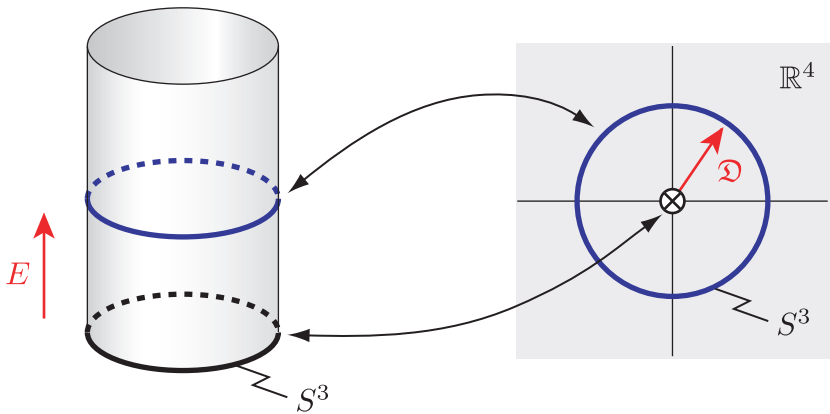}
\vspace{.3cm}
\caption{\small Conformal mapping of a state on $\mathbb R\times S^{3}$ to an operator on $\mathbb R^{4}$\,.}
\label{fig:conf-map}
\end{center}
\end{figure}

As we will see later in Section \ref{sec:string sigma} (see (\ref{metric-AdS})), in the global coordinate system, the metric on $AdS_{5}$ can be written as
\begin{equation}
ds^2 _{\left( {AdS_5 } \right)}  =  R^{2}\ko{d\rho ^2  - \cosh ^2 \rho \,dt^2  + \sinh ^2 \rho \, d\Omega_{3}}
\end{equation}
with $0\leq \rho<\infty$ and $-\infty<t<\infty$\,.
In the global coordinates, the boundary of $AdS_{5}$ where the gauge theory is supposed to live is considered to be located at $\rho\to \infty$\,.
By further rewriting it in terms of $\tan\al=\sinh\rho$\,, $0\leq \al<\pi/2$\,, the metric can be cast into 
\begin{equation}
ds^2 _{\left( {AdS_5 } \right)}=\f{R^{2}}{\cos^{2}\al}\ko{d\al^{2}-dt^{2}+\sin^{2}\al\, d\Omega_{3}}\,,
\label{metric AdS-3}
\end{equation}
from which we can read off the boundary ($\al\to\pi/2$) metric as $\mathbb R\times S^{3}$\,.\footnote{A conformal rescaling on the metric does not change the causal structure of the spacetime, so we are free to multiply the metric by $\cos^{2}\al/R^{2}$\,.}
Notice that $\cN=4$ SYM on the boundary $\mathbb R\times S^{3}$ can be mapped to $\cN=4$ SYM on $\mathbb R^{4}$ by using the conformal symmetry.
Via the state-operator mapping, a state on $\mathbb R\times S^{3}$ is mapped to a local operator on $\mathbb R^{4}$\,, see Figure \ref{fig:conf-map}.\footnote{This mapping requires a Wick rotation $iY_{0}\mapsto -iY_{0}$ in the string embedding coordinates (\ref{AdS-coord}).
Notice also the correspondence (\ref{charge correspondence}).}
The translation in the $t$\,-direction generating the energy $E(\f{R^{4}}{\al'{}^{2}}, g_{\rm s})$ of a string corresponds to, in view of the AdS/CFT dictionary, a dilatation in the $\mathbb R^{4}$ on the field theory side, which generates the conformal dimension $\Delta(\lam, \f{\lam}{N})$ of a local operator.
Thus we arrive at one of the key proposals of the AdS/CFT,
\begin{equation}
E=\Delta\,,
\label{E=D}
\end{equation}
with the identification of the parameters (\ref{key relation}).\footnote{We set the radius of $AdS_{5}\times S^{5}$ to be unity hereafter.}
Among many ways to test the duality, it is the strict $N\to \infty$ form of the relation (\ref{E=D}) that is in our main scope in this thesis.

\paragraph{}
Unfortunately, apart from the rather trivial BPS sector, it is hard to check the relation (\ref{E=D}) in general, because of the strong/weak nature of the duality as we emphasised.
In perturbation theory, string energies are obtained as expansions in large-$\lam$ as\footnote{Actually recent studies have revealed that the string theory has an essential singularity at strict $\lam\to \infty$ limit.
Hence the expansion (\ref{E=...}) should be understood as performed at sufficiently large but finite (and real) value of $\lam$\,.}
\begin{equation}
E(\lam)=
\sqrt{\lambda}\,\cE_{0}+\cE_{1}+\f{\cE_{2}}{\sqrt{\lambda}}+\f{\cE_{3}}{\sqrt{\lambda}^{2}}+\dots\,,
\label{E=...}
\end{equation}
while the conformal dimension of gauge theory operators are obtained as expansions in small-$\lam$\,,
\begin{equation}
\Delta(\lam)=\Delta_{0}+\lam \delta_{1}+\lam^{2}\delta_{2}+\lam^{3}\delta_{3}+\dots\,.
\label{D=...}
\end{equation}
Therefore one cannot compare $E(\lam)$ and $\Delta(\lam)$ directly before summing up all order contributions for each side, as long as one does the perturbation theory from opposite ends.
However, there is actually a nice sector where we can have an {\em almost} overlapping perturbative region, which we will discuss in the next chapter.

\chapter[Large spin/R-charge sector of AdS/CFT]
	{Large spin/R-charge sector of AdS/CFT\label{chap:outline}}

After the proposal of the AdS/CFT correspondence \cite{Maldacena:1997re,Gubser:1998bc,Witten:1998qj}, a wealth of tests has been done and has provided numerous positive supports for the conjecture.
They are, however, mainly restricted to the supergravity region $\al'\ll 1$\,, since the superstring theory in the curved background $AdS_{5}\times S^{5}$ has not been solved to date.
To go beyond the supergravity regime, and overcome the substantial strong/weak difficulty of the AdS/CFT, some clever idea is needed, which we discuss in this chapter.

\section{The BMN and Frolov-Tseytlin sectors}

\subsubsection*{``Near-BPS'' sector\,: plane-wave/SYM correspondence}

The situation was enormously improved after the discovery of the plane-wave/SYM correspondence by Berenstein, Maldacena and Nastase \cite{Berenstein:2002jq} in 2002.
It provided a tractable playground beyond the supergravity regime for the first time.
In the light-cone gauge, the string theory on the plane-wave background reduces to a free, massive two dimensional model \cite{Metsaev:1998it}, so that it can be quantised despite the presence of nonzero R-R flux, enabling us to obtain the exact free spectrum.\footnote{We will very briefly review the plane-wave geometry in Section \ref{sec:pp-wave}.}

The key idea is to take a $U(1)$\,-charge $J$ of $SO(6)$ very large, and define a new effective coupling constant $\tlambda\eq \lam/J^{2}$\,.
Indeed the perturbative regime is opposite between gauge and string theories (for the former being $\lam\ll 1$ while for the latter $\lam\gg 1$), but if we take the quantum number $J$ much larger than $\sqrt{\lam}$\,, the effective coupling $\tlambda$ can be very small even on the string theory side.
Therefore there is a chance that both theories can have an overlapping perturbative regime, where we can perturbatively access from both sides of the correspondence.

On the string theory side, the $U(1)$\,-charge represents an angular momentum that a string carries on $S^{5}$\,, while on the gauge theory side, it is a $U(1)$ R-charge of a local operator.
In this large-spin limit, of course the energies $E$ of string states and the conformal dimensions $\Delta$ of SYM operators are also very large (larger than $J$).
To be precise, the famous {\em BMN limit} is defined as
\begin{equation}
J\,,\,\, N\to \infty\,,\qquad 
\f{J}{\sqrt{N}} : {\rm fixed}\,,\qquad 
\tlambda\eq\f{\lambda}{J^{2}} : {\rm fixed}\,,\qquad 
E-J\,,\,\, \Delta-J : {\rm fixed}\,. 
\label{BMN limit}
\end{equation}
In this way one can establish a concrete AdS/CFT dictionary relating string states and SYM operators, namely string states with large angular momentum and SYM operators with large R-charge.
Moreover, the correspondence at the level of interacting-string/non-planar-diagrams has been also explored to some extent.

\paragraph{}
Soon after the work by BMN, a remarkable shortcut for some parts of their argument was provided by Gubser, Klebanov and Polyakov (GKP) \cite{Gubser:2002tv}, where the general framework for a worldsheet sigma model approach to this large-spin/R-charge sector of AdS/CFT was established.
They discovered that one can actually reach the BMN result by considering semiclassical solitonic solutions of $AdS_{5}\times S^{5}$ string worldsheet sigma model as the string theory duals of ``long'' SYM operators.
Starting with classical string theory in $AdS_{5}\times S^{5}$ and by computing the one-loop ($\al'$) correction to the string sigma model, one can reach the same free string spectrum as the one obtained by quantising the string theory on the plane-wave background (the Penrose limit of $AdS_{5}\times S^{5}$).

\subsubsection*{``Far-from-BPS'' sector\,: spinning-string/spin-chain correspondence}

The BMN sector is almost BPS in that it is a sector where a point-like string is circulating around a great circle of $S^{5}$ with large spin $J$\,.
The dual SYM operator contains a large number $J$ of one species out of the three complex scalar fields, say $\cZ$\,, plus a few ``impurity'' fields.
For example, a BMN operator with two scalar impurities looks like $\cO\sim \tr(\cZ^{J-2}\Phi_{i}\Phi_{j})+\dots$ (for more precise definition, see (\ref{BMN op})), where $\Phi_{i}$ and $\Phi_{j}$ are the impurities  with $SO(6)$ indices while $\cZ$ is the background field.

The further development was driven by Frolov and Tseytlin, who went farther from the BPS sector than BMN, and proposed more general sectors.
Their approach was based on the string sigma model perspective of GKP, and they found a more general mapping between macroscopic string soliton solutions and SYM operators with large quantum numbers.
The limit they considered was to send $\lam\to \infty$ while keeping $\lam'\eq \lam/J^{2}$ fixed, where $J=J_{1}+J_{2}+J_{3}$\,, now with macroscopic number $J_{2}+J_{3}$ of impurities, $(J_{2}+J_{3})/J\sim\cO(1)$\,.
Here $J_{1,2,3}$ are the three Cartan charges of the $SO(6)$\,, which are three independent angular momenta on the string theory side, and the numbers of three complex scalar fields on the gauge theory side.
These quantum numbers are often referred to as {\em ``spins''}.
A remarkable feature about the Frolov-Tseytlin strings is that as long as at least one of the angular momenta is sufficiently large, the classical computation of the string energy becomes exact, dropping out all the $\al'$\,-corrections \cite{Frolov:2002av, Frolov:2003qc, Frolov:2003tu}.
This feature simplifies the problem on the string theory side considerably, since it enables us to consider only classical string solutions.

The gauge theory dual operators of the Frolov-Tseytlin strings take the form $\cO\sim \tr(\cZ^{J_{1}}\cW^{J_{2}}\cY^{J_{3}})+\mbox{permutations}$\,.
For the test of the AdS/CFT proposal (\ref{E=D}) at the classical level, one needs to find the scaling dimensions of such ``long'' operators with a macroscopic number of impurities, $J_{2}+J_{3}\sim \cO(J)\to \infty$\,.
Computing the dimensions of such an operator, however, is not an easy task since the operator mixing problem is complicated, and so we need some trick to achieve it.
In the next section, we will see that the diagonalisation of the mixing matrix can be done by the Bethe ansatz method, after mapping the (subsectors of) $\cN=4$ SYM to some {\em integrable spin-chain} systems.

\paragraph{}
Before that, we shall give some general argument on the expansion of string energy and SYM scaling dimension.
In the large spin limit, the classical string energy $E_{0}\eq \sqrt{\lam}\,\cE_{0}$\,, which is the leading term in the expansion (\ref{E=...}), scales as $E_{0}\sim J$\,.\footnote{It can be compared to the flat space result $E_{0}\sim\sqrt{J}$\,.}
This is consistent with the fact that on the gauge theory side the scaling dimension scales as $\Delta \sim J$ with the bare dimension $J$\,.
Let us assume this expansion goes on in higher orders as
\begin{equation}
E_{0}=J\ko{1+\f{\ep_{1}\lambda}{J^{2}}+\f{\ep_{2}\lambda^{2}}{J^{4}}+\dots}\,,
\qquad \mbox{\em i.e.,}\quad \cE_{0}=\cJ\ko{1+\f{\ep_{1}}{\cJ^{2}}+\f{\ep_{2}}{\cJ^{4}}+\dots}
\label{assumption 1}
\end{equation}
with $\cJ\eq J/\sqrt{\lam}$\,.
Under this assumption of {\em BMN scaling} in the Frolov-Tseytlin sector, we have a chance to compare it to the scaling dimensions of SYM operators directly, since the scaling dimensions are computed perturbatively in powers of $\lam$ and might also have expansion form that respects the BMN scaling.

If we also assume the worldsheet quantum ($\al'$) corrections enter as $1/J$\,-correction for each perturbative order in (\ref{assumption 1}), the total energy can be organised into the following double-expanded form in powers of $\lambda/J^{2}$ and $1/J$\,,
\begin{align}
E_{\rm tot}\ko{J;\lambda}&=E_{0}+\sum\limits_{n=1}^{\infty}E_{n}
=\overbrace{J+J\sum\limits_{k=1}^{\infty}\f{\ep_{k}\lambda^{k}}{J^{2k}}}^{\text{\footnotesize classical energy}}
+\overbrace{J\sum\limits_{n=1}^{\infty}\sum\limits_{k=1}^{\infty}\f{\ep_{nk}\lambda^{k}}{J^{n+2k}}}^{\text{\footnotesize quantum correction}}\no\\[2mm]
&=J\left[ 1+ \sum\limits_{k=1}^{\infty}\ko{\ep_{k}+\sum\limits_{n=1}^{\infty}\f{\ep_{nk}}{J^{n}}}\ko{\f{\lambda}{J^{2}}}^{k}\right]\,.\label{double expansion}
\end{align}
Then if we take the BMN limit (\ref{BMN limit}) in (\ref{double expansion}), all the quantum $\al'\sim 1/J$\,-corrections drop off to reduce $E_{\rm tot}$ to its classical part $E_{0}$\,.
This means the classical string energies provide the leading contribution to the true quantum spectrum in the limit.

On the gauge theory side, suppose the scaling dimensions of SYM operators can be also doubly-expanded in powers of $\lam/J^{2}$ and inverse-``length'' $1/J$ in the same way as in (\ref{double expansion}),
\begin{align}
\Delta\ko{J;\lambda}&=J\left[ 1+\sum\limits_{k=1}^{\infty}\ko{\delta_{k}+\sum\limits_{n=1}^{\infty}\f{\delta_{kn}}{J^{n}}}\ko{\f{\lambda}{J^{2}}}^{k}\right]\,.
\label{double expansion-Delta}
\end{align}
If it is correct, then again, if we take the limit (\ref{BMN limit}), the finite-size correction parts drop off.
Hence in order to check the AdS/CFT central relation (\ref{E=D}) in the planar and far from BPS regime, checking the relations
\begin{equation}\label{c=a?}
\ep_{k}\, \stackrel{?}{=}\, \delta_{k}\qquad \ko{k=1, 2, \dots}
\end{equation}
serves as non-trivial and, importantly, quantitative tests \cite{Beisert:2003xu, Beisert:2003ea,Serban:2004jf}.
We need to compute $\ep_{k}$ and $\delta_{k}$ on each side.
In doing so, perhaps the greatest difficulty is found on the gauge theory side.
For simple operators like Konishi operators, the loop ($\lam$) corrections to the scaling dimensions can be readily achieved by the usual diagrammatical manner (it still gets very hard as we go to higher orders though).
However, as the number of fields contained in the local operator grows, which are actually the class of operators that should be compared with (semi)classical string states carrying large spins, such diagrammatic techniques are no longer useful since the operator mixing problem becomes more and more involved.

\section{Integrability in AdS/CFT}

At the end of the last section, we mentioned the technical problem concerning the operator mixing in the gauge theory, for ``long'' operators.
A special symmetry called {\em integrability}, which the gauge theory is proved to possess at least to the first few perturbative orders, can be a clue to overcome the difficulty.\footnote{For recent review articles on various aspects of integrability in AdS/CFT (and the spinning-string/spin-chain correspondence), we refer to the articles by Tseytlin \cite{Tseytlin:2003ii,Tseytlin:2004xa}, Zarembo \cite{Zarembo:2004hp}, Beisert \cite{Beisert:2004ry}, Swanson \cite{Swanson:2005wz,Swanson:2007dh}, Plefka \cite{Plefka:2005bk} and Minahan \cite{Minahan:2006sk}.}

\subsubsection*{Integrability in gauge theory}

Integrability arises on the gauge theory side as a quantum symmetry of local operator mixing.
It was discovered in the seminal work by Minahan and Zarembo \cite{Minahan:2002ve} that the planar dilatation operator of the $SO(6)$ scalar sector of $\mathcal N = 4$ SYM at one-loop can be identified with the Hamiltonian of an integrable $SO(6)$ spin-chain.
In particular, restricting the sector to $SU(2)$\,, their dilatation operator gives the Hamiltonian of a Heisenberg XXX${}_{1/2}$ spin-chain, which is a very well-known integrable model.
At the one-loop level, this way of mapping field theory dilatation operators to integrable spin-chain Hamiltonians was generalised to the full $PSU(2,2|4)$ sector by Beisert and Staudacher \cite{Beisert:2003yb} which is based on the complete one-loop dilatation operator of \cite{Beisert:2003jj}.
Thus on the gauge theory side, in the planar limit, the problem of computing the spectrum of local operators, which requires diagonalising the dilatation operators, was reformulated as diagonalising integrable spin-chain Hamiltonians.
By virtue of the integrability, the so-called {\em Bethe ansatz} method \cite{Bethe:1931hc,Faddeev:1996iy} can be used to achieve it, yielding the eigenvalues of the dilatation operator/spin-chain Hamiltonian.\footnote{The integrable spin-chain approach to the gauge theory spectrum was already applied earlier in the study of high-energy QCD.
See, {\em e.g.}, \cite{Lipatov:1993yb,Faddeev:1994zg,Braun:1998id}.}

As such, the reformulation was successful at the one-loop level, and one naturally seeks for higher-loop integrability.
In the closed $SU(2)$ subsector, it was shown that the dilatation operator is consistent with integrability at the two-loop level \cite{Beisert:2003tq}.
In the paper, based on the integrability assumption as well as on additional assumption of the BMN scaling, the planar dilatation operator of the closed $SU(2)$ sector of $\cN = 4$ SYM was constructed to three-loops.
The three-loop integrability conjecture made in \cite{Beisert:2003tq} was indeed confirmed by an explicit calculation \cite{Kotikov:2004er}, and it was further shown that integrability is consistent with three-loops for the larger closed $SU(2|3)$ subsector \cite{Beisert:2003ys}.
An all-order asymptotic Bethe ansatz equation for gauge theory was also proposed by assuming all-order BMN scaling and perturbative integrability \cite{Beisert:2004hm}.

\subsubsection*{Integrability in string theory}

The AdS/CFT correspondence then leads one to expect integrability on the string theory side as well, since the scaling dimensions of the gauge theory operators are identified with the energies of dual string states as we discussed.
Indeed, on the string theory side, the classical ($\lam\gg 1$) string sigma-model on $AdS_{5}\times S^{5}$ is also integrable, admitting a Lax representation \cite{Bena:2003wd} (see also \cite{Dolan:2003uh, Mandal:2002fs, Alday:2003zb, Vallilo:2003nx, Dolan:2004ps}).
The integrability therefore arises in the string theory side as a classical symmetry of string worldsheet theory.
It was shown in \cite{Arutyunov:2003uj, Arutyunov:2003za} that the $SO(2,4)\times SO(6)$ string sigma model evaluated on a particular type of rotating string ansatz falls into the class of so-called Neumann-Rosochatius integrable systems.
In \cite{Kazakov:2004qf} (see also \cite{Dorey:2006mx,Dorey:2006zj}), the classical sigma model on $\mathbb R\times S^{3}$ was solved in terms of spectral data ({\em i.e.}, hyperelliptic curves endowed with meromorphic differentials), and the classical string solutions were described as finite-gap solutions.
Their formalism was generalised to the sigma models on $AdS_{3}\times S^{1}$ \cite{Kazakov:2004nh}, $\mathbb R\times S^{5}$ \cite{Beisert:2004ag} and to the full $AdS_{5}\times S^{5}$ \cite{Beisert:2005bm} sectors.

Using the integrability techniques/properties, the AdS/CFT proposal (\ref{E=D}) was tested in the far-from-BPS sector in the form of (\ref{c=a?}).
In Sections \ref{sec:discrepancy}, we will explicitly see the results for $k=1,2,3$ cases for particular solutions, where we will find, surprisingly, the equality $a_{k}=c_{k}$ is true for the first two levels $k=1,2$\,, while the third order, this matching breaks down, $a_{3}\neq c_{3}$\,.
This mismatch has been (infamously) known as the {\em ``three-loop discrepancy''} \cite{Serban:2004jf}.
It was argued some non-trivial interpolation of string-energy/SYM-dimension may occur when going from weak ($\lam\ll 1$) to strong ($\lam\gg 1$) coupling.
To account for the interpolation, the so-called the {\em dressing factor} was introduced \cite{Arutyunov:2004vx}.

\subsubsection*{Unifying the gauge and string theory integrability}

As we have seen, integrability was observed in rather different ways in gauge and string theory.
With the aim of comparing the integrable structures themselves directly, the programme of constructing the underlying AdS/CFT Bethe ansatz equations has been intensively pursued.
The approach taken was to use perturbative results on both gauge and string theory sides as well as other established or expected symmetries/properties, and also some sophisticated guesses that are waiting to be justified, as guides to yet-to-be uncovered exact answers.
In this way, an all-order asymptotic Bethe ansatz equation for the AdS/CFT system was proposed based on the S-matrix approach of Staudacher \cite{Staudacher:2004tk}, and subsequently refined \cite{Beisert:2005fw}.
Remarkably, it was found by Beisert \cite{Beisert:2005tm} that the structure of the S-matrix for the full $\cN=4$ model can be completely fixed by symmetry argument only, up to an overall scalar phase.
Interestingly, the phase turned out essentially the dressing phase which is expected to cure the three-loop discrepancy.

\subsubsection*{Testing the conjectured AdS/CFT S-matrix}

In \cite{Hofman:2006xt}, a different large-spin limit was
proposed by Hofman and Maldacena to serve as a new playground in testing the conjectured AdS/CFT S-matrix, especially the aforementioned scalar-phase/dressing-phase.
In this limit, both $J_{1}$ and $E$ go to infinity while the
difference $E-J_{1}$ and the coupling $\lam$ are kept finite.
The worldsheet quantum corrections drop out in this limit, which
simplifies the comparison of the two spectra considerably. 
So-called {\em giant magnons} are string solutions living in this sector, which is the string theory dual of an isolated magnon propagating on an asymptotic SYM spin-chain considered by Beisert \cite{Beisert:2005tm}.
The correspondence between such a solitonic string state and a SYM magnon excitation was generalised to multi-spin/magnon-boundstate case by Dorey in \cite{Dorey:2006dq} and extended in \cite{Chen:2006ge,Chen:2006gq,Chen:2006gp}, providing further and stronger tests for the conjectured S-matrix.

After years of strenuous efforts for constructing the best asymptotic Bethe ansatz equations, it has now reached a considerably refined form, accounting for the quantum corrections to the classical string worldsheet as well.
There has been increasing evidence and positive support for the conjectured Bethe ansatz equations \cite{Beisert:2005tm,Janik:2006dc,Eden:2006rx,Arutyunov:2006iu,Beisert:2006ib,Beisert:2006ez,Dorey:2007xn} that is supposed to be valid for all values of $\lam$\,.
The conjectured S-matrix is inconsistent with the BMN scaling hypothesis in the weak coupling region, breaking it from the fourth loop order.
So the true scenario would be, assuming the conjectured S-matrix is correct, that the assumption for the expansion form (\ref{double expansion-Delta}) for the gauge theory was not quite right to begin with.

\section{Outline of the thesis\label{sec:outline}}

Here we sketch how the original works \cite{Chen:2006ge,Chen:2006gq,Chen:2006gp,Hayashi:2007bq,Dorey:2007an} are distributed along the thesis.

In Part \ref{part:1}, we discuss the integrability in gauge theory.
In Chapter \ref{chap:N=4 spin-chain}, after reviewing some relevant aspects of $\cN=4$ SYM, we discuss the one-loop renormalisation problem for the $SO(6)$ sector, and demonstrate how the resulting dilatation operator is identified with an integrable spin-chain Hamiltonian by following \cite{Minahan:2002ve}.
Chapter \ref{chap:MZ} includes some introductory material to the Bethe ansatz method.
In particular, we discuss the $SU(2)$ case in detail, for which we also review the higher-loop integrability.

\paragraph{}
In Part \ref{part:string}, we move on to the $AdS_{5}\times S^{5}$ string sigma model and study its classical integrability.
We first review the Frolov-Tseytlin strings on $\mathbb R\times S^{3}$ in Chapter \ref{chap:FT}, then compare the energies with the energies of SYM spin-chain states.
For a particular set of solutions, we explicitly see the mismatch of the energy coefficients of gauge and string theory at the third loop order, and explain the need of the dressing phase.
We also review the finite-gap problem approach to the classical string spectrum, by reviewing the work of KMMZ \cite{Kazakov:2004qf}.
Furthermore, we present the most general elliptic string solutions on $\mathbb R\times S^{3}$ in Chapter \ref{chap:OS}, which is based on the original works \cite{Okamura:2006zv,Hayashi:2007bq}.
They are also interpreted as finite-gap solutions.

\paragraph{}
Having discussed the integrability observed on each side of the AdS/CFT correspondence, in Part \ref{part:5}, we try to unify them in the form of Bethe ansatz equations.
Chapter \ref{chap:Asymptotic} is based on the original work \cite{Chen:2006gp}, in which we discuss the asymptotic spectrum of the $\cN=4$ SYM spin-chain.
We show that the boundstates of $Q$ magnons form a certain short representation of dimension $16Q^{2}$\,.
We also derive the exact dispersion relation for the magnon boundstates by purely group theoretic means.
In Chapter \ref{chap:dressing}, we summarise the current knowledge about the ``AdS/CFT S-matrix'' that is supposed to interpolate between gauge and string theory S-matrix.
Chapters \ref{chap:DGM} and \ref{chap:S-matrices in HM} are based on the original works \cite{Chen:2006ge} and \cite{Chen:2006gq}, respectively.
On the string theory side, we generalise the giant magnon solutions of Hofman and Maldacena \cite{Hofman:2006xt} to the two-spin cases, which we call dyonic giant magnons.
The energy-spin relation for our solution is shown to precisely agree with the dispersion relation for the SYM magnon boundstates \cite{Dorey:2006dq,Chen:2006ge}.
The scattering phase-shift computed directly from the dyonic giant magnon scattering also agrees with the one obtained using the conjectured S-matrix \cite{Chen:2006gq}, thus giving a positive support for the conjecture.
In Chapter \ref{chap:Singularities}, following the original work \cite{Dorey:2007an}, we examine the singular structures of the conjectured S-matrix.
By considering physical processes involving one or more on-shell intermediate particles belonging to the known BPS spectrum of \cite{Dorey:2006dq,Chen:2006ge}, we perform further analyticity tests for the conjectured S-matrix.

Finally, Part \ref{part:6} is devoted to the conclusion.

\newpage

\quad 

\vfill 

\subsubsection*{Map of the thesis\,:}

\vfill 

\begin{center}
\unitlength 0.1in
\begin{picture}(70.00,75.80)(5.50,-75.80)
\put(38.0000,-2.0000){\makebox(0,0){Part \ref{part:0}\,: Introduction}}%
%
\special{pn 8}%
\special{pa 3800 1800}%
\special{pa 3800 3600}%
\special{dt 0.045}%
\special{pa 3800 3600}%
\special{pa 3800 3599}%
\special{dt 0.045}%
\put(20.0000,-22.0000){\makebox(0,0){Part \ref{part:1}\,: Integrability in Gauge Theory}}%
\put(56.0000,-22.0000){\makebox(0,0){Part \ref{part:string}\,: Integrability in String Theory}}%
%
\special{pn 8}%
\special{pa 3400 1400}%
\special{pa 2600 1800}%
\special{fp}%
\special{sh 1}%
\special{pa 2600 1800}%
\special{pa 2669 1788}%
\special{pa 2648 1776}%
\special{pa 2651 1752}%
\special{pa 2600 1800}%
\special{fp}%
%
\special{pn 8}%
\special{pa 4200 1400}%
\special{pa 5000 1800}%
\special{fp}%
\special{sh 1}%
\special{pa 5000 1800}%
\special{pa 4949 1752}%
\special{pa 4952 1776}%
\special{pa 4931 1788}%
\special{pa 5000 1800}%
\special{fp}%
%
\special{pn 8}%
\special{pa 2800 0}%
\special{pa 4800 0}%
\special{pa 4800 400}%
\special{pa 2800 400}%
\special{pa 2800 0}%
\special{fp}%
%
\special{pn 8}%
\special{pa 400 2000}%
\special{pa 3600 2000}%
\special{pa 3600 2400}%
\special{pa 400 2400}%
\special{pa 400 2000}%
\special{fp}%
%
\special{pn 8}%
\special{pa 4000 2000}%
\special{pa 7200 2000}%
\special{pa 7200 2400}%
\special{pa 4000 2400}%
\special{pa 4000 2000}%
\special{fp}%
\put(42.0000,-26.0000){\makebox(0,0)[lt]{$\bullet$ Spinning/rotating strings}}%
\put(42.0000,-30.0000){\makebox(0,0)[lt]{$\bullet$ Helical strings \cite{Okamura:2006zv,Hayashi:2007bq}}}%
\put(42.0000,-34.0000){\makebox(0,0)[lt]{$\bullet$ Classical string Bethe equations}}%
\put(6.0000,-26.0000){\makebox(0,0)[lt]{$\bullet$ Mapping to integrable spin-chains}}%
\put(6.0000,-30.0000){\makebox(0,0)[lt]{$\bullet$ Bethe ansatz for SYM spin-chains}}%
\put(6.0000,-34.0000){\makebox(0,0)[lt]{$\bullet$ Thermodynamic limit}}%
\put(38.0000,-48.0000){\makebox(0,0){Part \ref{part:5}\,: Unifying AdS- \& CFT- integrabilities}}%
%
\special{pn 8}%
\special{pa 1800 4600}%
\special{pa 5800 4600}%
\special{pa 5800 5000}%
\special{pa 1800 5000}%
\special{pa 1800 4600}%
\special{fp}%
%
\special{pn 8}%
\special{pa 2600 4000}%
\special{pa 3400 4400}%
\special{fp}%
\special{sh 1}%
\special{pa 3400 4400}%
\special{pa 3349 4352}%
\special{pa 3352 4376}%
\special{pa 3331 4388}%
\special{pa 3400 4400}%
\special{fp}%
%
\special{pn 8}%
\special{pa 5000 4000}%
\special{pa 4200 4400}%
\special{fp}%
\special{sh 1}%
\special{pa 4200 4400}%
\special{pa 4269 4388}%
\special{pa 4248 4376}%
\special{pa 4251 4352}%
\special{pa 4200 4400}%
\special{fp}%
\put(38.0000,-38.0000){\makebox(0,0){{\small ``Three-loop discrepancy''}}}%
%
\special{pn 8}%
\special{pa 3600 3500}%
\special{pa 4000 3500}%
\special{fp}%
\special{sh 1}%
\special{pa 4000 3500}%
\special{pa 3933 3480}%
\special{pa 3947 3500}%
\special{pa 3933 3520}%
\special{pa 4000 3500}%
\special{fp}%
\special{pa 4000 3500}%
\special{pa 3600 3500}%
\special{fp}%
\special{sh 1}%
\special{pa 3600 3500}%
\special{pa 3667 3520}%
\special{pa 3653 3500}%
\special{pa 3667 3480}%
\special{pa 3600 3500}%
\special{fp}%
%
\special{pn 8}%
\special{pa 3800 4000}%
\special{pa 3800 4400}%
\special{dt 0.045}%
\special{pa 3800 4400}%
\special{pa 3800 4399}%
\special{dt 0.045}%
%
\special{pn 8}%
\special{pa 3800 5890}%
\special{pa 3800 6290}%
\special{dt 0.045}%
\special{pa 3800 6290}%
\special{pa 3800 6289}%
\special{dt 0.045}%
%
\special{pn 8}%
\special{pa 3600 6090}%
\special{pa 4000 6090}%
\special{fp}%
\special{sh 1}%
\special{pa 4000 6090}%
\special{pa 3933 6070}%
\special{pa 3947 6090}%
\special{pa 3933 6110}%
\special{pa 4000 6090}%
\special{fp}%
\special{pa 4000 6090}%
\special{pa 3600 6090}%
\special{fp}%
\special{sh 1}%
\special{pa 3600 6090}%
\special{pa 3667 6110}%
\special{pa 3653 6090}%
\special{pa 3667 6070}%
\special{pa 3600 6090}%
\special{fp}%
\put(18.0000,-60.0000){\makebox(0,0)[lt]{$\bullet$ Magnon boundstate \cite{Chen:2006gp}}}%
\put(42.0000,-60.0000){\makebox(0,0)[lt]{$\bullet$ Dyonic giant magnons \cite{Chen:2006ge}}}%
\put(18.0000,-56.0000){\makebox(0,0)[lt]{$\bullet$ The AdS/CFT Bethe ansatz equations}}%
\put(18.0000,-64.0000){\makebox(0,0)[lt]{$\bullet$ Magnon boundstate S-matrix \cite{Chen:2006gq}, analyticity tests \cite{Dorey:2007an}}}%
\put(26.0000,-6.0000){\makebox(0,0)[lt]{$\bmt{\check}$\,  What is AdS/CFT correspondence}}%
\put(26.0000,-10.0000){\makebox(0,0)[lt]{$\bmt{\check}$\,  How to compare the AdS/CFT spectra}}%
\put(18.0000,-52.0000){\makebox(0,0)[lt]{$\bullet$ The dressing phase factor}}%
\put(38.0000,-73.8000){\makebox(0,0){Part \ref{part:6}\,: Conclusion}}%
%
\special{pn 8}%
\special{pa 2800 7180}%
\special{pa 4800 7180}%
\special{pa 4800 7580}%
\special{pa 2800 7580}%
\special{pa 2800 7180}%
\special{fp}%
%
\special{pn 8}%
\special{pa 3800 6740}%
\special{pa 3800 7040}%
\special{fp}%
\special{sh 1}%
\special{pa 3800 7040}%
\special{pa 3820 6973}%
\special{pa 3800 6987}%
\special{pa 3780 6973}%
\special{pa 3800 7040}%
\special{fp}%
\end{picture}%
\end{center}

\vfill

\part[Integrability in Super Yang-Mills Spin-Chain]
	{Integrability in Super Yang-Mills Spin-Chain\label{part:1}}

\chapter[Integrable Spin-Chains from $\mathcal{N}=4$ Super Yang-Mills]
	{Integrable Spin-Chains from $\boldsymbol{\mathcal{N}=4}$ Super Yang-Mills\label{chap:N=4 spin-chain}}

\section[The ${\cN=4}$ super Yang-Mills theory]
	{The \bmt{\cN=4} super Yang-Mills theory}

As we saw in Chapter \ref{chap:intro}, the dual description of the $AdS_{5}\times S^{5}$ superstring is believed to be the $\cN=4$ SYM theory, which is the maximally supersymmetric Yang-Mills theory in four dimensional spacetime.
The field content of the theory comprises a vector (gluon) $A_{\mu}$ ($\mu=0,1,2,3$), six real scalars $\Phi_{i}$ ($i=1,\dots,6$) and four Weyl spinors (gluinos), which can be written as a sixteen component ten-dimensional Majorana-Weyl spinor $\Psi$\,.
They have bare dimensions $\Delta_{0}[A_{\mu}]=1$\,, $\Delta_{0}[\Phi_{i}]=1$ and $\Delta_{0}[\Psi]=3/2$ respectively.
All fields are in the adjoint representation of the gauge group $SU(N)$ and are represented by $X(x)=\sum_{a=0}^{N^{2}-1}X^{a}(x)T^{a}$\,, where $X$ stands for either of the fields $\Phi_{i}$\,, $A_{\mu}$ and $\Psi$\,, and $T^{a}$ ($a=0,\dots,N^{2}-1$) are the generators of $U(N)$ ($N\times N$ Hermitian matrices)\,.\footnote{The $a=0$ component, $(T^{a=0})_{mn}=\delta_{mn}/\sqrt{2N}$ corresponds to the $U(1)$ generator.}
The $\cN=4$ SYM is defined by the action
\begin{align}
S=\f{1}{2g_{\rm YM}^{2}}\int d^{4}x\,\cL_{\cN=4}(A_{\mu},\Phi_{I},\Psi)
\end{align}
where the Lagrangian density is given by
\begin{align}
\cL_{\cN=4}=\tr\bigg\{
-\mbox{\large $\f{1}{2}$}(F_{\mu\nu})^{2}+\ko{D_{\mu}\Phi_{I}}^{2}-\sum_{i<j}[\Phi_{i},\Phi_{j}]^{2}
+i\bar \Psi \Gamma^{\mu}D_{\mu}\Psi-\bar \Psi\Gamma^{i}[\Phi_{i},\Psi]
\bigg\}\,.
\end{align}
Here the covariant derivatives $D_{\mu}$ and the field strength $F_{\mu\nu}$ are defined as
\begin{align}
D_{\mu}(\mbox{\small $\star$})=\pa_{\mu}-i[A_{\mu},\mbox{\small $\star$}\,]\,,\qquad 
F_{\mu\nu}=\pa_{\mu}A_{\nu}-\pa_{\nu}A_{\mu}+[A_{\mu},A_{\nu}]\,,
\end{align}
and $\Gamma^{A}=(\Gamma^{\mu},\Gamma^{I})$ is the $16\times 16$ Dirac matrices in ten-dimensional spacetime which satisfy $\tr(\Gamma^{A}\Gamma^{B})=16\delta^{AB}$\,.
``Letters'' (components of the fundamental multiplet) consists of $\cX_{A}\in \{\Phi_{i},\Psi,\overline{\Psi},F_{\mu\nu},D_{\mu}(\star)\}$\,, from which ``words'' (generic single-trace local operator) $\cO_{A_{1}\dots A_{L}}=\tr(\cX_{A_{1}}\dots \cX_{A_{L}})$ and ``sentences'' $\sum_{L}C^{A_{1}\dots A_{L}}\cO_{A_{1}\dots A_{L}}$ are generated.

In the $\cN=4$ SYM theory, there is no scale parameter (hence all fields are massless), so that the theory is scale-invariant at least at the classical level.
In fact, the scale-invariance survives even at the quantum level, since the $\be$-function for the gauge coupling constant $g_{\rm YM}$ vanishes to all-orders in perturbation theory.
Hence the theory is conformally invariant, namely the $\cN=4$ SYM is a {\em conformal field theory (CFT)}.

\section[The $\cN=4$ superconformal symmetry]
	{The \bmt{\cN=4} superconformal symmetry\label{sec:N=4 SCFT}}

\subsection{The four-dimensional conformal algebra}

In four dimensions, there are 15 generators in a conformal algebra\,: four generators of space-time translations $P_{\mu}$\,, six generators of Lorentz transformations $L_{\mu\nu}$\,, four special conformal transformations $K_{\mu}$\,, and a generator of scaling transformation, or a dilatation, $\fD$\,.

An infinitesimal spacetime transformation $\delta_\ep x^\mu=\ep\xi^{\mu}(x)$ is generated by the energy-momentum tensor, and the associated current is $T^{\mu}_{\nu}\xi^{\nu}$\,.
For a local operator located at $x$\,, by integrating the current over the sphere of radius $R\to 0$ centred at $x$\,, the variation of the operator under the infinitesimal transformation becomes
\begin{equation}
\delta_{\ep}\cO(x)=-\ep\lim_{R\to 0}\oint_{S_{R}}dy^{3}\,n^{\mu}T_{\mu\nu}(y)\xi^{\nu}(y)\cO(x)\,.
\label{conformal variation}
\end{equation}
Here $n^{\mu}$ is a unit vector normal to the sphere (pointing outwards).

If a spacetime transformation is a symmetry, then the corresponding current is conserved.
In particular let us focus on a scale transformation.
If $\cO(x)$ is invariant under an infinitesimal dilatation $\delta_{\ep}x^{\mu}=\ep x^{\mu}$\,, then the dilatation current $\fD^{\mu}=T^{\mu}_{\nu}x^{\nu}$ is conserved, $\pa_{\mu}\fD^{\mu}=0$\,.
On the other hand, by differentiating the terms separately, we have $\pa_{\mu}\fD^{\mu}=\pa_{\mu} T^{\mu}_{\nu}x^{\nu}+T^{\mu}_{\nu}\delta^{\nu}_{\mu}=T^{\mu}_{\mu}$\,, where we have used the fact the energy-momentum tensor is conserved.
The above argument results in that the scale invariance is equivalent to the tracelessness of the energy-momentum tensor $T^{\mu}_{\nu}$\,.\footnote{In quantum theory, the trace of the energy-momentum tensor has an anomaly which is proportional to the $\beta$-function.
For $\cN=4$ SYM, the $\beta$-function vanishes to all orders, so the theory remains scale-invariant even at the quantum level.}

Actually, the tracelessness of the energy-momentum tensor implies more symmetry and the associated conserved currents.
Taking the derivative of the current $T^{\mu}_{\nu}\xi^{\nu}$ associated with the vector field $\xi^{\nu}$ yields
\begin{equation}
\pa_{\mu}(T^{\mu}_{\nu}\xi^{\nu})=T^{\mu}_{\nu}\pa_{\mu}\xi^{\nu}
=\mbox{\large $\f{1}{2}$}\,T^{\mu\nu}(\pa_{\mu}\xi_{\nu}+\pa_{\nu}\xi_{\mu})\,,
\end{equation}
where we used the symmetry of the energy-momentum tensor to symmetrise the indices of $\pa\xi$\,.
It then follows that the conformal current $T^{\mu}_{\nu}\xi^{\nu}$ is conserved if $\xi^{\mu}$ is a so-called conformal Killing vector satisfying the following conformal Killing equation
\begin{equation}
\pa_{\mu}\xi_{\nu}+\pa_{\nu}\xi_{\mu}=\mbox{\large $\f{1}{2}$}\,\pa_{\rho}\xi^{\rho}\eta_{\mu\nu}\,.
\label{Killing}
\end{equation}
Its solutions form a Lie algebra with respect to the Lie bracket $[\eta,\xi]^{\mu}=\eta^{\nu}\pa_{\nu}\xi^{\mu}-\xi^{\mu}\pa_{\nu}\eta^{\mu}$\,.
The complete set of solutions for the Killing equation is given by\,:
\begin{alignat}{5}
&\mbox{$\circ$\, Space-time translations $P_{\mu}$}&\quad\mbox{F}&\quad &\xi_\mu&=c_{\mu}\quad (\const),\label{conformal P}\\
&\mbox{$\circ$\, Lorentz transformations $L_{\mu\nu}$}&\quad\mbox{F}&\quad &\xi_\mu&=\om_{\mu\nu}x^\nu\quad (\om_{\mu\nu}=-\om_{\nu\mu})\,,\label{conformal M}\\
&\mbox{$\circ$\, Dilatation $\fD$}&\quad\mbox{F}&\quad &\xi_\mu&=\Lam x_\mu\,,\label{conformal D}\\
&\mbox{$\circ$\, Special conformal transformations $K_{\mu}$}&\quad\mbox{F}&\quad &\xi_\mu&=2c_\nu x^\nu x_\mu -x_\nu x^\nu c_\mu\,.\label{conformal K}
\end{alignat}
These generators satisfy the conformal algebra,
\begin{align}
&[\fD,P_{\mu}]=-iP_{\mu}\,,\\
&[\fD,K_{\mu}]=iK_{\mu}\,,\\
&[\fD,L_{\mu\nu}]=0\,,\\
&[P_{\mu},K_{\nu}]=2i(L_{\mu\nu}-\eta_{\mu\nu}\fD)\,,\\
&[L_{\mu\nu},P_{\lam}]=-i(\eta_{\mu\lam}P_{\nu}-\eta_{\lam\nu}P_{\mu})\,,\\
&[L_{\mu\nu},K_{\lam}]=-i(\eta_{\mu\lam}K_{\nu}-\eta_{\lam\nu}K_{\mu})\,,\\
&[L_{\mu\nu},L_{\rho\sig}]=i(\eta_{\nu\rho}L_{\mu\sig}+\eta_{\mu\sig}L_{\nu\rho}-\eta_{\mu\rho}L_{\nu\sig}-\eta_{\nu\sig}L_{\mu\rho})\,.
\end{align}
These commutation relations can be arranged into a single compact form as
\begin{equation}
[M_{IJ},M_{KL}]=i(\eta_{JK}M_{IL}-\eta_{JL}M_{IK}-\eta_{IK}M_{JL}+\eta_{IL}M_{JK})\,,
\end{equation}
where we re-labeled the generators as
\begin{equation}
M_{\mu 5}=\f{1}{2}\ko{P_{\mu}+K_{\mu}}\,,\qquad 
M_{4 \mu}=\f{1}{2}\ko{P_{\mu}-K_{\mu}}\,,\qquad 
M_{54}=\fD\,,
\end{equation}
with the generalised metric $\eta_{IJ}=\diag(\eta_{\mu\nu};+1,-1)=\diag(-1,+1,+1,+1;+1,-1)$ ($I,J=0,1,\dots,5$)\,.
Hence the elements of the conformal group can be regarded as rotation generators in a space with metric $\eta_{IJ}$\,, which is $SO(2,4)$\,.
This is the same symmetry as the isometry of $AdS_{5}$ part of the $AdS_{5}\times S^{5}$ background of the string theory.

\paragraph{}
Under a scaling transformation $x\mapsto \Lam x$\,, a local operator in the field theory scales as $\cO(x)\mapsto \Lam^{-\Delta_{\cO}}\cO(\Lam x)$\,, where $\Delta_{\cO}$ is the scaling dimension of $\cO(x)$\,.
The dilatation generator $\fD$ acts on the operator as $[\fD,\cO(x)]=-i(\Delta-x\pa_{x})\cO(x)$\,, and so when $x=0$\,, it just counts the scaling dimension (multiplied by $-i$).
Acting with $\fD$ on the commutator $[K_{\mu},\cO(0)]$ and using the Jacobi identity, one finds
\begin{equation}
[\fD,[K_{\mu},\cO(0)]]=[[\fD,K_{\mu}],\cO(0)]+[K_{\mu},[\fD,\cO(0)]]
=-i(\Delta-1)[K_{\mu},\cO(0)]\,,
\end{equation}
which means that the operator $[K_{\mu},\cO(0)]$ has a scaling dimension lower than that of $\cO(0)$ by one.
In other words, $K_{\mu}$ lowers the dimension by one.
Notice that, operators with negative dimension are impossible in a unitary quantum field theory.
Therefore, successive application of $K_{\mu}$ to any operator with definite dimension must at some point yield zero.
One can then define a {\em conformal primary operator} $\cO\, (\neq 0)$ as an operator satisfying
\begin{equation}
[K_{\mu},\cO(0)]=0\,.
\end{equation}
Any descendant operators can be constructed by acting with generators of the conformal algebra on the conformal primary operator.
Note that $P_{\mu}$ plays the role of a raising operator, and $\fD$ is the Cartan component of the algebra.

\paragraph{}
For a conformal primary operator $\cO(x)$\,, the action of conformal generators (\ref{conformal variation}) goes as
\begin{equation}
\delta_{\ep}\cO(x)=\ep\kko{\xi^{\mu}\pa_{\mu}+\f{\Delta}{4}\pa_{\mu}\xi^{\mu}+\f{1}{2}\pa_{\mu}\xi_{\nu}\Sigma^{\mu\nu}}\cO(x)\,.
\end{equation}
Here $\Sigma^{\mu\nu}$ is a Lorentz generator acting on the indices of $\cO$\,.
For $\cO(0)$\,, it then follows
\begin{eqnarray}
&P_{\mu}\cdot \cO(0)=\pa_{\mu}\cO(0)\,,\qquad
L_{\mu\nu}\cdot \cO(0)=\Sigma_{\mu\nu}^{(S_{1},S_{2})}\cO(0)\,,&\no\\
&\fD\cdot \cO(0)=\Delta\cO(0)\,,\qquad
K_{\mu}\cdot \cO(0)=0\,,&\no
\label{Delta, S1, S2}
\end{eqnarray}
where the action of any generator $\mathfrak G\in\{ P_{\mu}, L_{\mu\nu}, \fD, K_{\mu} \}$ on $\cO$ is defined by $\mathfrak G\cdot \cO\eq i[\mathfrak G,\cO]$\,.
In the second relation, $\Sigma_{\mu\nu}^{(S_{1},S_{2})}$ indicates $\cO$ is an $(S_{1},S_{2})$-tensor in the Lorentz indices.
The action (\ref{Delta, S1, S2}) thus defines the highest-weight state of $SO(2,4)$ characterised by three Cartan charges $(\Delta; S_{1},S_{2})$\,, which are the charges of $SO(1,1)\times SO(1,3)\subset SO(2,4)$\,.
Note that this is an infinite-dimensional representation due to the non-compactness of the algebra.

\subsection[The $\cN=4$ superconformal algebra]
	{The \bmt{\cN=4} superconformal algebra}

The $\cN=4$ SYM theory also possesses maximal supersymmetry (SUSY) in four-dimensional spacetime, and is invariant under supersymmetry transformations.
There are eight supercharges $Q_{\alpha}^{a}$\,, eight superconformal charges $S_{\alpha a}$\,, and their conjugate charges $\bar Q_{\dot \alpha a}$\,, $\bar S_{\dot \alpha}^{a}$ $(\al=1,\,2$\,; $\dot\al=1,\,2$\,; $a=1,\dots,4)$\,.
The (anti-)commutation relations are given by\,:
\begin{alignat}{3}
&[K^{\mu},Q_{\alpha}^{a}]=(\sigma^{\mu})_{\alpha\dot\beta}\ep^{\dot\be\dot\ga}\bar S_{\dot \ga}^{a}\,,&\qquad &
	[K^{\mu},\bar Q_{\dot \alpha a}]=(\sigma^{\mu})_{\beta\dot\alpha}\ep^{\be\ga}S_{\ga a}\,,\\
&[P^{\mu},S_{\alpha a}]=(\sigma^{\mu})_{\alpha\dot\beta}\ep^{\dot\be\dot\ga}\bar Q_{\dot \ga a}\,,&\qquad &
	[P^{\mu},\bar S_{\dot \alpha}^{a}]=(\sigma^{\mu})_{\beta\dot\alpha}\ep^{\be\ga}Q_{\ga}^{a}\,,\\
&[L^{\mu\nu},Q_{\alpha}^{a}]=-i\ko{\sigma^{\mu\nu}}_{\alpha\beta}\ep^{\be\ga}Q_{\ga}^{a}\,,&\qquad &
	[L^{\mu\nu},\bar Q_{\dot \alpha a}]=-i\ko{\bar \sigma^{\mu\nu}}_{\dot \alpha\dot \beta}\ep^{\dot\be\dot \ga}\bar Q_{\dot \ga a}\,,\\
&[L^{\mu\nu},S_{\alpha a}]=-i\ko{\sigma^{\mu\nu}}_{\alpha\beta}\ep^{\be\ga}S_{\ga a}\,,&\qquad &
	[L^{\mu\nu},\bar S_{\dot \alpha}^{a}]=-i\ko{\bar \sigma^{\mu\nu}}_{\dot \alpha\dot \beta}\ep^{\dot\be\dot \ga}\bar S_{\dot \ga}^{a}\,,\\
&[\fD,Q_{\alpha}^{a}]=-\mbox{\large $\f{i}{2}$}\,Q_{\alpha}^{a}\,,&\qquad &
	[\fD,\bar Q_{\dot \alpha a}]=-\mbox{\large $\f{i}{2}$}\,\bar Q_{\dot \alpha a}\,,\\
&[\fD,S_{\alpha a}]=+\mbox{\large $\f{i}{2}$}\,S_{\alpha a}\,,&\qquad &
	[\fD,\bar S_{\dot \alpha}^{a}]=+\mbox{\large $\f{i}{2}$}\,\bar S_{\dot \alpha}^{a}\,,\\
&\{ Q_{\alpha}^{a}, \bar Q_{{\dot \beta} b}\}=(\sigma^{\mu})_{\alpha\dot\beta}\delta^{a}{}_{b}P_{\mu}\,,&\qquad &
\{ S_{\alpha a}, \bar S_{{\dot \beta}}^{b}\}=(\sigma^{\mu})_{\alpha\dot\beta}\delta_{a}{}^{b}K_{\mu}\,,
\end{alignat}
and all the other (anti-)commutators vanish.
Here we defined $\sig^{\mu\nu}=\f{1}{4}\sig^{[\mu}\bar\sig^{\nu]}$ and $\bar \sig^{\mu\nu}=\f{1}{4}\bar \sig^{[\mu}\sig^{\nu]}$ with $\sig^{\mu}=(-{\bf 1},\sig^{i})$ and $\bar \sig^{\mu}=(-{\bf 1},-\sig^{i})$\,, where $\sig^{i=1,2,3}$ are Pauli matrices
\begin{equation}
\sigma^1 = \mbox{\small $\Bigg (\begin{array}{cc}
        0 & 1 \\
        1 & 0
\end{array}\Bigg )$}
\, ,\quad 
\sigma^2 = \mbox{\small $\Bigg (\begin{array}{cr}
        0 & -i \\
        i &  0
\end{array}\Bigg )$}
\, ,\quad 
\sigma^3 = \mbox{\small $\Bigg (\begin{array}{cr}
        1 &  0 \\
        0 & -1
\end{array}\Bigg )$}\,;\qquad 
[ \sigma^i , \sigma^j ] = i\epsilon_{ijk}\sigma^k\,.
\end{equation}
A new set of generators $R_{ij}$ appears when we compute the commutation relation between $Q_{\alpha}^{a}$ and $S_{\beta b}$ (or $\bar Q_{\dot \alpha a}$ and $\bar S^{b}_{\dot\beta}$),
\begin{align}
&\{ Q_{\alpha}^{a}, S_{\beta b} \}=+(\sig^{ij})^{a}{}_{b}\ep_{\al\be}R_{ij}+i(\sig^{\mu\nu})_{\al\be}\delta^{a}{}_{b}L_{\mu\nu}-i\ep_{\al\be}\delta^{a}{}_{b}\fD\,,\\
&\{ \bar Q_{\dot \alpha a}, \bar S^{b}_{\dot\beta} \}=-(\sig^{ij})_{a}{}^{b}\ep_{\dot\al\dot\be}R_{ij}+i(\bar\sig^{\mu\nu})_{\dot\al\dot\be}\delta_{a}{}^{b}L_{\mu\nu}-i\ep_{\dot\al\dot\be}\delta_{a}{}^{b}\fD\,.
\end{align}
The generators $R_{ij}$ $(i,j=1,\dots, 6)$ are the $SU(4)$ R-symmetry generators.
We denote three Cartan generators of the $SU(4)$ as $(R_{12},R_{34},R_{56})$\,, and the corresponding charges as $(J_{1},J_{2},J_{3})$\,.
For later purpose, we combine the six real scalar fields $\Phi_{i}$ into three complex scalar fields,\footnote{In the $\cN=1$ language, they and their complex conjugate are the lowest components of the holomorphic and antiholomorphic superfields.
Three of the four Weyl fermions are the spinors of the chiralmultiplet, and the fourth spinor together with the vector $A_{\mu}$ form the vectormultiplet.}
\begin{gather}
\begin{split} 
\cZ=\Phi_{1}+i\Phi_{2}\,,\qquad 
\cW=\Phi_{3}+i\Phi_{4}\,,\qquad 
\cY=\Phi_{5}+i\Phi_{6}\,,\\
\overline\cZ=\Phi_{1}-i\Phi_{2}\,,\qquad 
\overline\cW=\Phi_{3}-i\Phi_{4}\,,\qquad 
\overline\cY=\Phi_{5}-i\Phi_{6}\,,
\end{split} 
\label{WYZ}
\end{gather}
such that the action of the R-symmetry generators become
\begin{alignat}{5}
&[R_{12},\cZ]=\cZ\,,&\qquad& [R_{34},\cW]=\cW\,,&\qquad& [R_{56},\cY]=\cY\,,\no\\
&[R_{12},\overline \cZ]=-\overline\cZ\,,&\qquad& [R_{34},\overline\cW]=-\overline\cW\,,&\qquad& [R_{56},\overline\cY]=-\overline\cY\,,\no
\end{alignat}
and all other commutators between $R_{ij}$ and $\cZ,\cW,\cY$ are zeros.
Those R-charges together with (\ref{Delta, S1, S2}) tells us that the gauge theory spectrum is characterised by
\begin{equation}
\{ \Delta\,;\, S_{1}\,,\, S_{2}\, ; J_{1}\,,\, J_{2}\,,\, J_{3} \}\,.
\label{CFT spectrum}
\end{equation}
The conformal symmetry is $SO(2,4)\cong SU(2,2)$ and the R-symmetry is $SO(6)\cong SU(4)$\,, so that the superconformal symmetry becomes $SU(2,2|4)$\,.
Actually the full global symmetry of the $\cN=4$ SYM theory is $PSU(2,2|4)$\,.
The generators can be schematically expressed in terms of $(4+4)\times (4+4)$ supermatrix as
\begin{equation}
\left(\begin{array}{c|c}
P_{\mu}, L_{\mu\nu}, \fD, K_{\mu} & Q_{\al}^{a},\bar S_{\dot \al}^{a} \\
\hline
\bar Q_{\dot \al a}, S_{\al a} & R_{ij}\end{array}\right)\,.
\end{equation}
In this representation, the conformal symmetry and the R-symmetry are represented as $SU(2,2)$ and $SU(4)$ respectively, and the supercharges and superconformal charges form bi-spinors of $SU(2,2)\times SU(4)$\,.

\paragraph{}
Now we can introduce the notion of {\em superconformal primary operators}, or {\em chiral primary (BPS) operators}.
From the superconformal algebra, we have
\begin{align}
[\{ Q_{\alpha}^{a}, S_{\beta b} \}, \cO(0)]
=[(\sig^{ij})^{a}{}_{b}\ep_{\al\be}R_{ij}+i(\sig^{\mu\nu})_{\al\be}\delta^{a}{}_{b}L_{\mu\nu}-i\ep_{\al\be}\delta^{a}{}_{b}\fD, \cO(0)]\,.
\label{QSO}
\end{align}
Chiral primary operators are defined such that they satisfy $[Q_{\alpha}^{a},\cO(0)]=0$ for some $\al$ and $a$\,, and $[S_{\beta b},\cO(0)]=0$ for all $\be$ and $b$\,.
For chiral primaries satisfying $[Q_{\alpha}^{a}, \cO(0)]=0$\,, the LHS of (\ref{QSO}) becomes zero by definition, because it can be rewritten as $\{ Q_{\alpha}^{a}, [S_{\beta b}, \cO(0)]\}+\{ S_{\beta b}, [Q_{\alpha}^{a}, \cO(0)]\}$ using a graded Jacobi identity.
If $\cO(0)$ is a scalar, then $[L_{\mu\nu}, \cO(0)]=0$\,, so that the scaling dimension of $\cO(0)$ is in direct correspondence with its R-charge.

Operators $\cO_{1}=\tr(\cZ^{J_{1}})$\,, $\cO_{2}=\tr(\cW^{J_{2}})$\,, and $\cO_{3}=\tr(\cY^{J_{3}})$ with R-charges assignment $(J_{1},0,0)$\,, $(0,J_{2},0)$\,, and $(0,0,J_{3})$ respectively, are chiral primary operators.
Indeed, by evaluating the RHS of (\ref{QSO}) explicitly, one finds $[\{ Q_{\alpha}^{a}, S_{\beta b} \}, \cO_{i}]=(-iJ_{i}+0-(-iJ_{i}))\cO_{i}=0$\,.
Actually these three chiral primaries are annihilated by eight of the sixteen super(conformal) charges, 
so they are half-BPS operators.
Their scaling dimensions do not receive quantum corrections and remain at their classical (bare) values $\Delta_{0}[\cO_{i}]=J_{i}$ at all orders in $\lam$\,.
For non-BPS operators, however, the scaling dimensions $\Delta$ receive non-trivial $\lam$\,-correction, and the problem of computing $\Delta(\lam)$ in perturbation theory generally gets harder as the operator becomes longer and more complicated.

\subsection{Anomalous dimension}

In the $\cN=4$ SYM theory, even though the gauge coupling constant $g_{\rm YM}$ is not renormalised, gauge invariant local composite operators are renormalised in general.
Classically the scaling dimension of the gauge invariant operator is simply the sum of the individual dimensions of the constituent fields, but at the quantum level it acquires a so-called {\em anomalous dimension}.

Suppose we manage to find a conformal primary operator, then we can use the symmetry to restrict the correlation functions.
In particular, it fixes the two-point function completely.
To see this, recall the conformal Ward identity for a three-point function with energy-momentum operator $T_{\mu\nu}$ and two conformal primary operator $\hat\cO$ (we consider scalar operators for simplicity),
\begin{equation}
\lim\limits_{R\to \infty}\oint_{S_{R}} d^{3}y\, n^{\mu}y^{\nu}\vev{T_{\mu\nu}(y)\hat\cO(x)\hat\cO(0)}=0\,.
\label{conformal Ward identity}
\end{equation}
Here $y^{\nu}$ is a conformal Killing vector, $S_{R}$ is a three-sphere of radius $R$ containing both $x=x$ and $x=0$ inside, and $n^{\mu}$ is a unit vector normal to $S_{R}$\,.
The LHS (\ref{conformal Ward identity}) can be integrated by parts as
\begin{equation}
\int d^{4}y\vev{\pa_{\mu}T_{\nu}^{\mu}(y)\hat\cO(x)\hat\cO(0)}
-\lim\limits_{\ep\,\to\,0}\ko{\oint_{S_{\ep}(x)}+\oint_{S_{\ep}(0)}}d^{3}y\, n^{\mu}y^{\nu}\vev{T_{\mu\nu}\hat\cO(x)\hat\cO(0)}\,.
\end{equation}
The first integral vanishes because of the conservation of the conformal current, and the remaining two integrals give variations of the operators under the conformal transformation, $\vvev{(\delta\hat\cO(x))\cO(0)}$ and $\vvev{\hat\cO(x)(\delta\hat\cO(0))}$\,.
It then leads to the following differential equation for the correlation function,
\begin{equation}
\ko{x^{\mu}\pa_{\mu}+2\Delta_{\cO}}\vev{\hat\cO(x)\hat\cO(0)}=0\,,
\end{equation}
which can be easily solved to give
\begin{equation}
\vev{\hat\cO(x)\hat\cO(0)}=\f{C_{\cO}(\Lam)}{|x|^{2\Delta_{\cO}}}\,.
\label{<OO>}
\end{equation}
Here $C_{\cO}(\Lam)$ is a constant depending on the cut-off scale $\Lam$\,.
We see that the conformal dimension $\Delta_{\cO}$ of $\hat\cO$ appears in the exponent.
By decomposing $\Delta_{\cO}$ into its bare (classical) part and anomalous (quantum) part, which depends on the coupling constant $\lam$\,, as
\begin{equation}
\Delta_{\cO}(\lam)=\Delta_{0}+\gamma(\lam)\,,
\end{equation}
the RHS of (\ref{<OO>}) can be perturbatively expanded in powers of $\gamma$ (which is supposed to be very small in perturbation theory) as
\begin{equation}
\vev{\hat\cO(x)\hat\cO(0)}
=\f{C_{\cO}(\Lam)\cdot \Lam^{2\gamma}}{|x|^{2\Delta_{0}}}\kko{1-2\ga\ln\ko{\Lam |x|}+2\ga^{2}\ln^{2}\ko{\Lam |x|}+\dots}\,.
\label{<OO>2}
\end{equation}
From this expression, the conformal invariance appears to be broken at each order of $\gamma$\,, but it is an artifact of the perturbative expansion and of course the theory itself is conformally invariant, as is obvious in the original expression (\ref{<OO>}).

\section[One-loop dilatation operator for $SO(6)$ sector]
	{One-loop dilatation operator for \bmt{SO(6)} sector}

In this section, by reviewing \cite{Minahan:2002ve}, we explicitly perform the one-loop renormalisation of operators in the $SO(6)$ sector in the planar limit.
They are single-trace operators made up of $L$ scalar fields $\Phi_{i}$\,,
\begin{equation}
\cO_{i_{1},\dots,i_{L}}(x)=\tr\ko{\Phi_{i_{1}}(x)\dots\Phi_{i_{L}}(x)}\,,\qquad i_{k}\in\{ 1,\dots,6 \}\,,\quad L\gg 1\,,
\label{SO(6) op}
\end{equation}
where all the scalars are located at the same spacetime point $x$\,.
In general, an operator of the form (\ref{SO(6) op}) is not diagonal with respect to the renormalisation flow in that $\langle \cO_{i_{1},\dots, i_{L}}(0) \cO^{j_{1},\dots, j_{L}}(x) \rangle$ is not proportional to $\delta_{i_{1},\dots, i_{L}}^{j_{1},\dots, j_{L}}$\,.
Therefore, there occurs an operator mixing among operators with the same bare dimension and R-charge.
Let us see how the resulting mixing matrix can be identified with the Hamiltonian of an integrable $SO(6)$ spin-chain as advertised in the introduction.

To begin with, let us consider the following correlation function,
\newsavebox{\boxMZ}
\sbox{\boxMZ}{\includegraphics{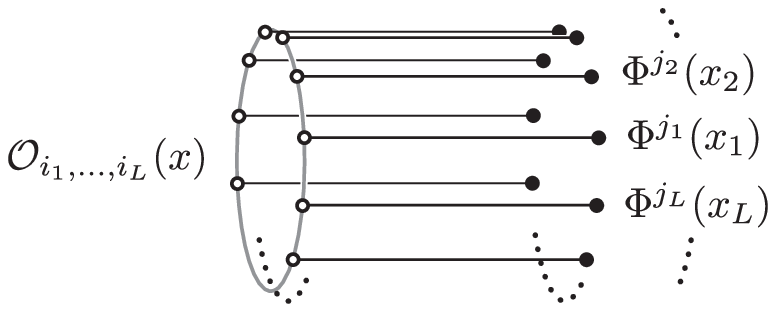}}
\newlength{\bwMZ}
\settowidth{\bwMZ}{\usebox{\boxMZ}} 
\begin{equation}
\vev{\cO_{i_{1},\dots,i_{L}}(x)\Phi^{j_{1}}(x_{1})\dots \Phi^{j_{L}}(x_{L})}
= ~ \parbox{\bwMZ}{\usebox{\boxMZ}} ~ \,.
\end{equation}
This is divergent, so we need to renormalise it by introducing a UV cut-off $\Lam$\,.
As usual, we define the renormalised fields $\Phi^{j_{\ell}}_{\rm ren}$ and $\cO_{j_{1},\dots,j_{L}}^{\rm ren}$ as
\begin{align}
&\Phi^{j_{\ell}}(x_{\ell})=Z_{\Phi}(\Lam)^{1/2}\Phi^{j_{\ell}}_{\rm ren}(x_{\ell})\,,\quad (\ell=1,\dots,L)\,,\\
&\cO_{i_{1},\dots,i_{L}}(x)=Z_{\cO}(\Lam)^{j_{1},\dots,j_{L}}_{i_{1},\dots,i_{L}}\cO_{j_{1},\dots,j_{L}}^{\rm ren}(x)
\end{align}
with the wavefunction renormalisation factors $Z_{\Phi}(\Lam)$ and $Z_{\cO}(\Lam)$\,.
They can be determined such that the correlation function
\begin{align}
&\vev{\cO_{i_{1},\dots,i_{L}}^{\rm ren}(x)\Phi^{j_{1}}_{\rm ren}(x_{1})\dots \Phi^{j_{L}}_{\rm ren}(x_{L})}\no\\
&\qquad {}=Z_{\Phi}(\Lam)^{-L/2}\kko{Z_{\cO}(\Lam)^{-1}}{}_{i_{1},\dots,i_{L}}^{k_{1},\dots,k_{L}}
\vev{\cO_{k_{1},\dots,k_{L}}(x)\Phi^{j_{1}}(x_{1})\dots \Phi^{j_{L}}(x_{L})}\,,
\end{align}
becomes finite after the renormalisation.
The anomalous dimension matrix (the dilatation operator) $\mathfrak D$ for the composite operator $\cO_{i_{1},\dots,i_{L}}$ is then defined as
\begin{equation}
(\mathfrak D_{\cO})_{i_{1},\dots,i_{L}}^{j_{1},\dots,j_{L}}=
\kko{Z_{\cO}(\Lam)^{-1}}{}_{i_{1},\dots,i_{L}}^{k_{1},\dots,k_{L}}
\f{d\kko{Z_{\cO}(\Lam)}_{k_{1},\dots,k_{L}}^{j_{1},\dots,j_{L}}}{d\ln \Lam}\,.
\end{equation}
In this way, the dilatation operator can be perturbatively obtained by computing Feynman diagrams.
Let us write
\begin{equation}
\mathfrak D_{\cO}=\sum_{n=0}^{\infty}\mathfrak D_{\cO}^{(n)}\,,
\end{equation}
where $\mathfrak D_{\cO}^{(n)}$ is the dilatation operator at the $n$\,-loop order, $\mathfrak D_{\cO}^{(n)}\sim \cO(\lam^{n})$\,.
Schematically, the operation is represented as
\newsavebox{\boxDkloop}
\sbox{\boxDkloop}{\includegraphics{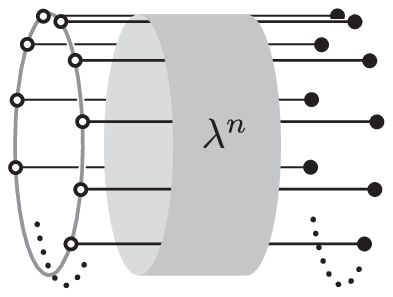}}
\newlength{\bwDkloop}
\settowidth{\bwDkloop}{\usebox{\boxDkloop}} 
\newsavebox{\boxDtree}
\sbox{\boxDtree}{\includegraphics{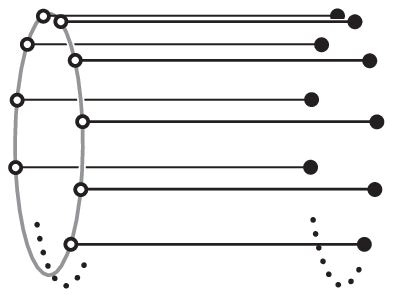}}
\newlength{\bwDtree}
\settowidth{\bwDtree}{\usebox{\boxDtree}} 
\begin{equation}
\mathfrak D_{\cO} \cdot \parbox{\bwDtree}{\usebox{\boxDtree}} 
~ = ~ \sum_{n=0}^{\infty} \, \parbox{\bwDkloop}{\usebox{\boxDkloop}} ~ \,.
\end{equation}
The $n=0$ piece, $\mathfrak D_{\cO}^{(0)}$\,, just counts the number of the fields in the trace, giving $\Delta_{0}[\cO]=L$\,.
The eigenstates $\{ \hat \cO_{A} \}$ of $\mathfrak D_{\cO}$ are the conformal operators, whose two-point functions take the form of (\ref{<OO>}), {\em i.e.}, $\langle\hat \cO_{A}(x)\hat \cO_{B}(0)\rangle=C_{A}\delta_{AB}|x|^{-2\Delta_{A}}$ with $\Delta_{A}=L+\gamma_{A}(\lam)$\,.
The anomalous dimensions $\gamma_{A}(\lam)$ are computed in perturbation theory\,: $\gamma_{A}(\lam)=\sum_{n=1}^{\infty}\ga_{n}\lam^{n}=\ga_{1}\lam+\ga_{2}\lam^{2}+\cdots$\,.
Note that only two of the $L$ legs are relevant in the current one-loop computation, and only the nearest-neighbour interactions are relevant since non-planar interactions are suppressed in the large-$N$ limit, as we saw in the introduction.

In computing the diagrams, we take the Feynman gauge, in which the propagators for scalars and vectors take the same form
\begin{equation}
\vev{\Phi_{i}^{a}(x)\Phi_{j}^{b}(0)}=\f{g_{\rm YM}^{2}\delta_{ij}\delta^{ab}}{8\pi^{2}|x|^{2}}\,,\qquad 
\vev{A_{\mu}^{a}(x)A_{\nu}^{b}(0)}=\f{g_{\rm YM}^{2}\delta_{\mu\nu}\delta^{ab}}{8\pi^{2}|x|^{2}}\,.
\end{equation}
Let us set $x_{1}=\dots =x_{L}=0$\,, then the tree-loop diagram gives
\newsavebox{\boxKtree}
\sbox{\boxKtree}{\includegraphics{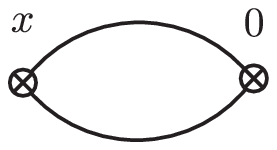}}
\newlength{\bwKtree}
\settowidth{\bwKtree}{\usebox{\boxKtree}} 
\begin{equation}
\vev{\cO_{i_{1},\dots,i_{L}}(x)\Phi^{j_{1}}(0)\dots \Phi^{j_{L}}(0)}_{\rm tree}
= ~ \parbox{\bwKtree}{\usebox{\boxKtree}} ~ 
= \ko{\f{g_{\rm YM}^{2}}{8\pi^{2}|x|^{2}}}^{L}\eq \cM_{\rm tree}
\end{equation}
At the one-loop, there are three diagrams that contribute to the two-point function\,: (a) scalar four-vertex, (b) gluon exchange, and (c) self-energy diagrams.
Diagramatically,
\newsavebox{\boxDvertex}
\sbox{\boxDvertex}{\includegraphics{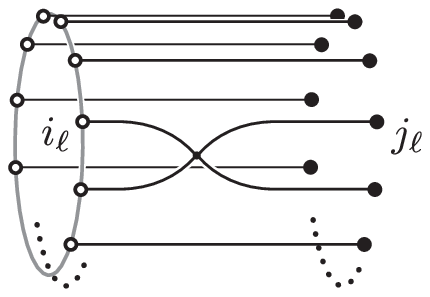}}
\newlength{\bwDvertex}
\settowidth{\bwDvertex}{\usebox{\boxDvertex}} 
\newsavebox{\boxDgluon}
\sbox{\boxDgluon}{\includegraphics{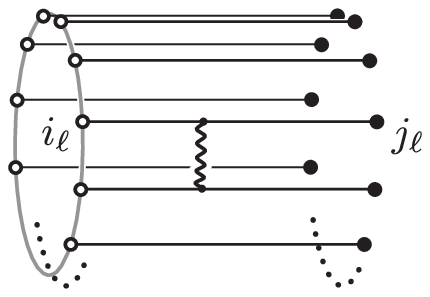}}
\newlength{\bwDgluon}
\settowidth{\bwDgluon}{\usebox{\boxDgluon}} 
\newsavebox{\boxDself}
\sbox{\boxDself}{\includegraphics{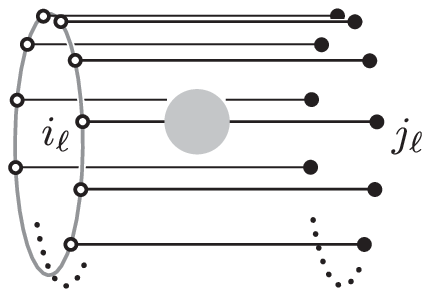}}
\newlength{\bwDself}
\settowidth{\bwDself}{\usebox{\boxDself}} 
\begin{align}
\mathfrak D_{\cO}^{(1)} \cdot \parbox{\bwDtree}{\usebox{\boxDtree}} 
\quad &= \quad \sum_{\ell=1}^{L} \, \underbrace{\parbox{\bwDvertex}{\usebox{\boxDvertex}} ~}_{\mbox{$\vphantom{\int}$\footnotesize (a)}} +{}\no\\[2mm]
&\hspace{-2.6cm} {}+ ~ \sum_{\ell=1}^{L} \, \underbrace{\parbox{\bwDgluon}{\usebox{\boxDgluon}} ~}_{\mbox{$\vphantom{\int}$\footnotesize (b)}} 
+ ~ \sum_{\ell=1}^{L} \, \underbrace{\parbox{\bwDself}{\usebox{\boxDself}} ~}_{\mbox{$\vphantom{\int}$\footnotesize (c) }} 
\,.
\end{align}
The scalar four-vertex diagrams can be evaluated easily.
The Feynman rules are given by
\newsavebox{\boxVertex}
\sbox{\boxVertex}{\includegraphics{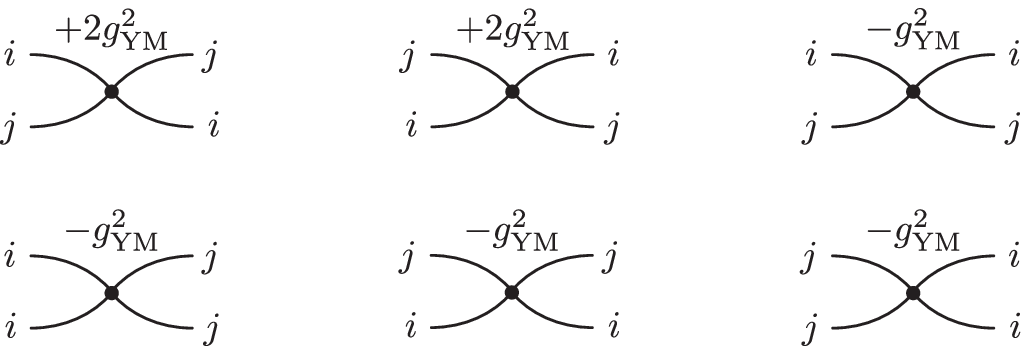}}
\newlength{\bwVertex}
\settowidth{\bwVertex}{\usebox{\boxVertex}} 
\begin{center}
\parbox{\bwVertex}{\usebox{\boxVertex}} ~ ,
\end{center}
\newsavebox{\boxKvertex}
\sbox{\boxKvertex}{\includegraphics{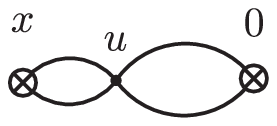}}
\newlength{\bwKvertex}
\settowidth{\bwKvertex}{\usebox{\boxKvertex}} 
\newsavebox{\boxKgluon}
\sbox{\boxKgluon}{\includegraphics{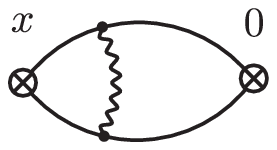}}
\newlength{\bwKgluon}
\settowidth{\bwKgluon}{\usebox{\boxKgluon}} 
\newsavebox{\boxKself}
\sbox{\boxKself}{\includegraphics{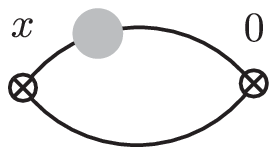}}
\newlength{\bwKself}
\settowidth{\bwKself}{\usebox{\boxKself}} 
\vspace{.3cm}
from which we can read off the $SO(6)$ flavour structure as $2\delta_{i_{\ell}}^{j_{\ell+1}}\delta_{i_{\ell+1}}^{j_{\ell}}-\delta_{i_{\ell}}^{j_{\ell}}\delta_{i_{\ell+1}}^{j_{\ell+1}}-\delta_{i_{\ell}i_{\ell+1}}\delta^{j_{\ell}j_{\ell+1}}$\,.
Then the four-vertex diagram is computed as
\begin{align}
\mbox{(a)}_{\ell,\,\ell+1}~&= ~ \parbox{\bwKvertex}{\usebox{\boxKvertex}} ~\no\\[2mm]
&\hspace{-.5cm}{}= ~ \ko{\f{g_{\rm YM}^{2}}{8\pi^{2}|x|^{2}}}^{L-2}\times 
\kko{\int d^{4}u\, \ko{\f{g_{\rm YM}^{2}}{8\pi^{2}|x-u|^{2}}}^{2}
\ko{\f{g_{\rm YM}^{2}}{8\pi^{2}|u|^{2}}}^{2}}\times {}\no\\
&{}\times g_{\rm YM}^{-2}\ko{2\delta_{i_{\ell}}^{j_{\ell+1}}\delta_{i_{\ell+1}}^{j_{\ell}}-\delta_{i_{\ell}}^{j_{\ell}}\delta_{i_{\ell+1}}^{j_{\ell+1}}-\delta_{i_{\ell}i_{\ell+1}}\delta^{j_{\ell}j_{\ell+1}}}\times N\,.
\label{(a) tmp}
\end{align}
The integral over $u$ is divergent as $u\to x$ and $u\to 0$\,, so we must regularise it.
By introducing a cutoff $\Lam$ and using the integral formula
\begin{equation}
\int_{|x-u|,|u|>1/\Lam}\f{d^{4}u}{|x-u|^{4}|u|^{4}}
\sim\f{8\pi^{2}}{|x|^{4}}\ln(\Lam|x|)\,,
\end{equation}
the expression (\ref{(a) tmp}) can be evaluated as
\begin{equation}
\mbox{(a)}_{\ell,\,\ell+1}=
\cM_{\rm tree}\times 
\kko{\f{\lam}{8\pi^{2}}\ln(\Lam |x|)} \ko{2\delta_{i_{\ell}}^{j_{\ell+1}}\delta_{i_{\ell+1}}^{j_{\ell}}-\delta_{i_{\ell}}^{j_{\ell}}\delta_{i_{\ell+1}}^{j_{\ell+1}}-\delta_{i_{\ell}i_{\ell+1}}\delta^{j_{\ell}j_{\ell+1}}}\,.
\end{equation}
The remaining two diagrams (b) and (c), both having the same $SO(6)$ flavour structure $\delta_{i_{\ell}}^{j_{\ell}}\delta_{i_{\ell+1}}^{j_{\ell+1}}$\,, also give contributions proportional to $\ln (\Lam |x|)$\,,
\begin{align}
\ko{\mbox{(b)}+\mbox{(c)}}_{\ell,\,\ell+1}
~&= ~ \parbox{\bwKgluon}{\usebox{\boxKgluon}} ~ + ~ \parbox{\bwKself}{\usebox{\boxKself}}\no\\[2mm]
&=\cM_{\rm tree}\times 
\kko{-C\cdot \f{\lam}{8\pi^{2}}\ln(\Lam |x|)}\delta_{i_{\ell}}^{j_{\ell}}\delta_{i_{\ell+1}}^{j_{\ell+1}}\,.
\end{align}
The proportionality constant $C$ can be also determined by computing the Feynman diagrams.\footnote{Actually the value of $C$ does not affect the integrability of the resulting anomalous dimension matrix.}
The one-loop contributions to the self-energy diagram (c) come from gluon exchange, fermion loop, scalar four-vertex and gluon-scalar four-vertex, and they can be evaluated in the usual manner.
But actually we can do better, since we know that for BPS operators the total contribution $\mbox{(a)}+\mbox{(b)}+\mbox{(c)}$ should be finite.
We will use this observation soon.
Collecting all the contributions, we have
\begin{align}
&\vev{\cO_{i_{1},\dots,i_{L}}(x)\Phi^{j_{1}}(0)\dots \Phi^{j_{L}}(0)}_{\mbox{\scriptsize 1-loop}}
= \cM_{\rm tree}\times \kko{-\f{\lam}{8\pi^{2}}\ln(\Lam |x|)}\sum_{\ell=1}^{L}\cH_{\ell,\ell+1}\,,\no\\
&\qquad \mbox{where}\quad \cH_{\ell,\ell+1}= (1+C)I_{\ell,\ell+1}+K_{\ell,\ell+1}-2P_{\ell,\ell+1}\,.
\end{align}
We have introduced three operators acting on $(\mathbb R^{6})_{\ell}\otimes (\mathbb R^{6})_{\ell+1}$\,, which are the identity, trace and permutation operators, defined as, respectively,
\begin{align}
I_{\ell,\ell+1}\eq \delta_{i_{\ell}}^{j_{\ell}}\delta_{i_{\ell+1}}^{j_{\ell+1}}\,,\qquad 
K_{\ell,\ell+1}\eq \delta_{i_{\ell}i_{\ell+1}}\delta^{j_{\ell}j_{\ell+1}}\,,\qquad 
P_{\ell,\ell+1}\eq \delta_{i_{\ell}}^{j_{\ell+1}}\delta_{i_{\ell+1}}^{j_{\ell}}\,.
\end{align}
They act on the tensor product $\vec a\otimes \vec b\in \mathbb R^{6}\otimes \mathbb R^{6}$ as
\begin{align}
I_{\ell,\ell+1}(\vec a\otimes \vec b)=\vec a\otimes \vec b\,,\quad 
K_{\ell,\ell+1}(\vec a\otimes \vec b)=(\vec a\cdot \vec b)\sum_{i=1}^{6}{\rm e}^{i}\otimes {\rm e}^{i}\,,\quad 
P_{\ell,\ell+1}(\vec a\otimes \vec b)=\vec b\otimes \vec a\,,
\end{align}
where $\vec a$ and $\vec b$ are six-vectors and $\{ {\rm e}^{i} \}$ is a set of orthogonal unit vectors in $\mathbb R^{6}$\,.
For a BPS state $\ket{0}\eq\tr(\cZ^{L})$\,, these operators act as $I_{\ell,\ell+1}\ket{0}=\ket{0}$\,, $K_{\ell,\ell+1}\ket{0}=0$ and $P_{\ell,\ell+1}\ket{0}=\ket{0}$\,.
Recall the total operator $\cH_{\ell,\ell+1}$ should annihilate the BPS state, $\cH_{\ell,\ell+1}\ket{0}=0$\,.
This fixes the constant $C$ as $C=1$\,.
Thus we obtain the one-loop dilatation operator as
\begin{equation}
\mathfrak D^{(1)} = \f{\lam}{16\pi^{2}}\sum_{\ell=1}^{L}\cH_{\ell,\ell+1}\,,\qquad 
\cH_{\ell,\ell+1}=2I_{\ell,\ell+1}+K_{\ell,\ell+1}-2P_{\ell,\ell+1}\,.
\label{SO(6) ham}
\end{equation}
A crucial observation made in \cite{Minahan:2002ve} was that the dilatation operator (\ref{SO(6) ham}) is identical to the Hamiltonian of an $SO(6)$ integrable spin-chain as we see in the next section.

\section[Mapping to $SO(6)$ integrable spin-chain]
	{Mapping to \bmt{SO(6)} integrable spin-chain}

For an $SO(n)$ integrable spin-chain, the Hilbert space is the product of the spaces at each site, $\mathscr H=(\mathbb R^{n})_{\ell=1}\otimes \dots \otimes (\mathbb R^{n})_{\ell=L}$\,.
Observables at each site are $n^{L}\times n^{L}$ matrices constructed from the identity, trace and permutation matrices.
The R-matrix acting on $(\mathbb R^{n})_{a}\otimes (\mathbb R^{n})_{b}$ with spectral parameter $u$ is defined as
\begin{equation}
R_{ab}(u)=\f{2}{n-2}\kko{u\ko{u+1-\f{n}{2}}I_{ab}+uK_{ab}-\ko{u+1-\f{n}{2}}P_{ab}}\,.
\end{equation}
It satisfies the Yang-Baxter equation for $(\mathbb R^{n})_{a}\otimes (\mathbb R^{n})_{b}\otimes (\mathbb R^{n})_{c}$\,,
\begin{equation}
R_{ab}(u)R_{ac}(u+v)R_{bc}(v)=R_{bc}(v)R_{ac}(u+v)R_{ab}(u)\,.
\label{YB RRR}
\end{equation}
Using the R-matrices we can construct the monodromy matrix $\Omega_{0}(u)$ as
\begin{equation}
\Omega_{0}(u)=R_{01}(u)R_{02}(u) \dots R_{0L}(u)\,,
\label{monodromy}
\end{equation}
where the index $0$ refers to an auxiliary space $(\mathbb R^{n})_{\ell=0}$\,.
The transfer matrix $t(u)$ is constructed by taking the trace of the monodromy matrix over the auxiliary space,
\begin{equation}
t(u)=\tr_{0}\Omega_{0}(u)\,.
\label{transfer}
\end{equation}
Using the Yang-Baxter equation (\ref{YB RRR}) for the R-matrices, it can be verified that an R-matrix and two monodromy matrices also satisfy the Yang-Baxter equation,
\begin{equation}
R_{ab}(u-v)\Omega_{a}(u)\Omega_{b}(v)
=\Omega_{b}(v)\Omega_{a}(u)R_{ab}(u-v)\,.
\label{YB ROmOm}
\end{equation}
It is understood as $R_{ab}(u-v){}^{k_{a}k_{b}}_{i_{a}i_{b}}\Omega_{a}(u){}^{i_{a}}_{j_{a}}\Omega_{b}(v)^{i_{b}}_{j_{b}}=\Omega_{b}(v){}^{k_{b}}_{l_{b}}\Omega_{a}(u){}^{k_{a}}_{l_{a}}R_{ab}(u-v){}^{l_{a}l_{b}}_{j_{a}j_{b}}$ when written in terms of the components explicitly.\footnote{The $({}^{i\, k}_{j\, l})$ component of the direct product $A\otimes B$ of two matrices $A$ and $B$ are defined as $\ko{A\otimes B}{}^{i\, k}_{j\, l}\eq A^{i}_{j}B^{k}_{l}$\,, where $A^{i}_{j}$ is the $(i,j)$ component of the matrix $A$ and the similar for $B^{k}_{l}$\,.
For example, in the case of two $2\times 2$ matrices $A$ and $B$\,,
\begin{equation}
\left( {A \otimes B} \right) = \left( {\begin{array}{*{20}c}
   {\left( {A \otimes B} \right){}_{11}^{11} } & {\left( {A \otimes B} \right){}_{12}^{11} } & {\left( {A \otimes B} \right){}_{21}^{11} } & {\left( {A \otimes B} \right){}_{22}^{11} }  \\
   {\left( {A \otimes B} \right){}_{11}^{12} } & {\left( {A \otimes B} \right){}_{12}^{12} } & {\left( {A \otimes B} \right){}_{21}^{12} } & {\left( {A \otimes B} \right){}_{22}^{12} }  \\
   {\left( {A \otimes B} \right){}_{11}^{21} } & {\left( {A \otimes B} \right){}_{12}^{21} } & {\left( {A \otimes B} \right){}_{21}^{21} } & {\left( {A \otimes B} \right){}_{22}^{21} }  \\
   {\left( {A \otimes B} \right){}_{11}^{22} } & {\left( {A \otimes B} \right){}_{12}^{22} } & {\left( {A \otimes B} \right){}_{21}^{22} } & {\left( {A \otimes B} \right){}_{22}^{22} }  \\
\end{array}} \right) = \left( {\begin{array}{*{20}c}
   {A_1^1 B_1^1 } & {A_1^1 B_2^1 } & {A_2^1 B_1^1 } & {A_2^1 B_2^1 }  \\
   {A_1^1 B_1^2 } & {A_1^1 B_2^2 } & {A_2^1 B_1^2 } & {A_2^1 B_2^2 }  \\
   {A_1^2 B_1^1 } & {A_1^2 B_2^1 } & {A_2^2 B_1^1 } & {A_2^2 B_2^1 }  \\
   {A_1^2 B_1^2 } & {A_1^2 B_2^2 } & {A_2^2 B_1^2 } & {A_2^2 B_2^2 }  \\
\end{array}} \right)\,.\no
\end{equation}
}
Rewriting it as  $\Omega_{a}(u){}^{i_{a}}_{j_{a}}\Omega_{b}(v)^{i_{b}}_{j_{b}}=R_{ab}^{-1}(u-v){}^{i_{a}i_{b}}_{k_{a}k_{b}}\Omega_{b}(v){}^{k_{b}}_{l_{b}}\Omega_{a}(u){}^{k_{a}}_{l_{a}}R_{ab}(u-v){}^{l_{a}l_{b}}_{j_{a}j_{b}}$ and taking the trace over both indices $a$ and $b$\,, we find
\begin{equation}
\tr_{a}\ko{\Omega_{a}(u)}\tr_{b}\ko{\Omega_{b}(v)}
=\tr_{b}\ko{\Omega_{b}(v)}\tr_{a}\ko{\Omega_{a}(u)}\,,
\label{[t,t]=0}
\end{equation}
{\em i.e.}, the transfer matrices commute, $[t(u),t(v)]=0$\,.
If the R-matrix is at least linear in $u$\,, from the definition (\ref{monodromy}) and (\ref{transfer}), the transfer matrix $t(u)=\sum_{n}t_{n}u^{n}$ is at least of order $u^{L}$\,.
Then it follows from (\ref{[t,t]=0}) that there are at least $L$ independent commuting charges.
Their number becomes infinite in the scaling limit $L\to \infty$\,.
The transfer matrix can be expanded in powers of $u$ as
\begin{equation}
t(u)=t_{0}-\f{2u}{n-2}\,t_{0}\sum_{\ell=1}^{L}\kko{I_{\ell,\ell+1}-K_{\ell,\ell+1}+\f{n-2}{2}P_{\ell,\ell+1}}+\dots\,,
\end{equation}
from which we can read off the Hamiltonian (essentially the $t_{1}$ part),
\begin{equation}
H_{SO(n)}=\left.\f{n-2}{2}\f{d}{du}\,\ln t(u)\right|_{u=0}
=\sum_{\ell=1}^{L}\kko{-I_{\ell,\ell+1}+K_{\ell,\ell+1}-\f{n-2}{2}P_{\ell,\ell+1}}\,.
\end{equation}
When setting $n=6$\,, this Hamiltonian is essentially the same as the dilatation operator for the $SO(6)$ sector of $\cN=4$ SYM we computed in the previous subsection, see (\ref{SO(6) ham}).\footnote{As we already mentioned, for the integrability, only the ratio of the coefficients of $K_{\ell,\ell+1}$ and $P_{\ell,\ell+1}$ is important and we are free to add any multiple of the identity operator to the Hamiltonian.}

\paragraph{}
In summary, the dilatation operator for the $SO(6)$ sector of $\cN=4$ SYM is given by, at the one-loop level,
\begin{align}
\mathfrak D_{\cO}&=\mathfrak D_{\cO}^{(0)}+\mathfrak D_{\cO}^{(1)}+\cO(\lam^{2})
=L+\f{\lam}{16\pi^{2}}\sum_{\ell=1}^{L}\cH_{\ell,\ell+1}+\cO(\lam^{2})\,,\no\\
&\mbox{where}\quad \cH_{\ell,\ell+1}=2I_{\ell,\ell+1}+K_{\ell,\ell+1}-2P_{\ell,\ell+1}\,,
\label{1-loop SO(6)}
\end{align}
and $\mathfrak D_{\cO}^{(1)}$ is essentially identical to the Hamiltonian of an integrable $SO(6)$ spin-chain.
The eigenstate $\hat\cO$ of $\mathfrak D_{\cO}^{(1)}$\,, that is the SYM operators with definite scaling dimensions, are mapped to the eigenvectors of the $SO(6)$ Hamiltonian.
If we decompose the eigenvector into single trace operators of (\ref{SO(6) op}), 
\begin{equation}
\hat\cO[\psi]=\psi^{i_{1}\dots i_{L}}\cO_{i_{1}\dots i_{L}}\,,
\label{SO(6) op 2}
\end{equation}
the coefficient $\psi^{i_{1}\dots i_{L}}$ can be regarded as a ``wavefunction'' living in the Hilbert space $\mathscr H=(\mathbb R^{6})^{\otimes L}$\,.
The number of possible operators is roughly given by $6^{L}$\,, divided by a cyclic permutation reflecting the trace condition of the SYM operators.
The diagonalisation of the anomalous dimension matrix (the dilatation operator) is equivalent to the diagonalisation of the integrable spin-chain Hamiltonian, so that the Bethe ansatz method can be implemented. We will see how this method works in the next chapter.
Before doing so, let us check our one-loop result (\ref{1-loop SO(6)}) through some simple examples.

\subsubsection*{Examples\,: chiral primary and Konishi operators}

The simplest example would be chiral primary operators, whose scaling dimensions are protected and so the anomalous dimensions are zero at all coupling regions. 
For the chiral primaries, the wavefunction $\psi^{i_{1}\dots i_{L}}$ in (\ref{SO(6) op 2}) is traceless and symmetric with respect to the $SO(6)$ indices $i_{1},\dots, i_{L}$\,.
Hence we have $K_{1,2}\ket{\rm CPO}=0$ and $P_{1,2}\ket{\rm CPO}=\ket{\rm CPO}$\,.
Plugging them into the energy formula (\ref{1-loop SO(6)}), indeed we get $\mathfrak D_{\cO}\cdot \cO_{\rm CPO}=\Delta_{\cO_{\rm CPO}}\cO_{\rm CPO}$ with scaling dimension
\begin{equation}
\Delta_{\cO_{\rm CPO}}=L+\f{\lam}{16\pi^{2}}\sum_{\ell=1}^{L}\ko{2+0-2}+\cO(\lam^{2})=L+0\cdot\lam+\cO(\lam^{2})\,,
\end{equation}
leading to the vanishing anomalous dimension at the one-loop.
Of course the result $\Delta_{\cO_{\rm CPO}}-L=0$ is true to all orders.

The second example is the Konishi operator,
\begin{equation}
\cK=\sum_{I=1}^{6}\tr\ko{\Phi_{I}\Phi_{I}}\,,
\label{Konishi}
\end{equation}
which corresponds to a spin-chain state with only two sites ($L=2$).
Since the wavefunction is simply given by $\psi^{i_{1}i_{2}}=\delta^{i_{1}i_{2}}$\,, the action of the trace and the permutation operators becomes $K_{1,2}\ket{\cK}=6\ket{\cK}$ and $P_{1,2}\ket{\cK}=\ket{\cK}$\,.
The bare conformal dimension is $\Delta_{0}[\cK]=1+1=2$\,.
Then we get $\mathfrak D_{\cO}\cdot \cK=\Delta_{\cK}\cK$\,, where the scaling dimension is given by
\begin{equation}
\Delta_{\cK}=2+\f{\lam}{16\pi^{2}}\sum_{\ell=1}^{2}\ko{2+6-2}+\cO(\lam^{2})=2+\f{3\lam}{4\pi^{2}}+\cO(\lam^{2})\,.
\end{equation}
Comparing it with the perturbative result computed from the two-point function $\vev{\cK(0)\cK(x)}$ known in the literature \cite{Anselmi:1998ms},
indeed we see the matching up to one-loop.

For such Konishi operators, there are no other operators which can mix with them, so the diagonalisation of the dilatation operator is trivial in this regard.
However, when we consider more complicated and ``longer'' operators, the mixing problem gets more involved since the number of operators of the same length grows rapidly with the length.
Hence we need some clever idea to deal with it.
This is where the integrability of the dilatation operator plays a crucial role.
We will demonstrate the power of the integrability in the next chapter.

\chapter[The Bethe Ansatz for $\mathcal N=4$ SYM Spin-Chains]
	{The Bethe Ansatz for $\boldsymbol{\mathcal N=4}$ SYM Spin-Chains\label{chap:MZ}}

In this chapter, we will see how the dilatation operator of the integrable planar $\cN=4$ SYM spin-chain can be diagonalised by the Bethe ansatz method.
We first discuss the $SU(2)$ subsector rather in detail including higher-loop orders, then briefly discuss the $SO(6)$ and the full $PSU(2,2|4)$ sectors as well, both at the one-loop level.
We postpone the discussion of all-loop conjecture for the full $PSU(2,2|4)$ model (based on integrability assumption as well as other inputs) in Chapter \ref{chap:Asymptotic}.

\section[The Bethe ansatz for $SU(2)$ sector at one-loop]
	{The Bethe ansatz for \bmt{SU(2)} sector at one-loop}

\begin{figure}[t]
\begin{center}
\vspace{.3cm}
\includegraphics[scale=0.9]{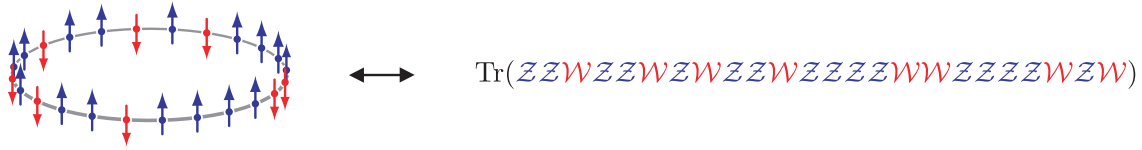}
\vspace{.3cm}
\caption{\small Heisenberg spin-chain state $\leftrightarrow$ SYM operator in the $SU(2)$ sector.}
\label{fig:spin-chain2}
\end{center}
\end{figure}

The full ${\cal N}=4$ SYM spin-chain has a couple of closed subsectors\,; they are the $SU(2)$\,, $SL(2)$\,, $SU(2|3)$\,, and the full $PSU(2,2|4)$ sectors.
For the moment we restrict our attention to the $SU(2)$ sector, which is the simplest and most studied closed subsector of $\cN=4$ SYM.
Operators in this sector are built of two of the three complex scalar fields, say $\cZ$ and $\cW$\,.
Taking the half-BPS operator $\tr(\cZ^{L})$ as the vacuum state of the spin-chain, and regard $\cW$ as the only impurity (magnon) field, any states in this sector can be represented as
\begin{equation}
\cO=\tr\ko{\cZ^{L-M}\cW^{M}}+\mbox{permutations}\,,
\label{SU(2) op}
\end{equation}
where $M$ is the number of impurities.
There are roughly $2^{L}$ possible states, and the Hilbert space is given by $\mathscr H=(\mathbb C^{2})^{\otimes L}$ up to a cyclic permutation.
Only operators of the same bare dimension $L$ and the same R-charge $(J_{1},J_{2},J_{3})=(L-M,M,0)$ mix under renormalisation, and we need to diagonalise them in order to acquire well-defined scaling dimensions.
The $SU(2)$ sector is holomorphic ({\em i.e.}, $K_{\ell,\ell+1}=0$) so the $SO(6)$ Hamiltonian (\ref{SO(6) ham}) reduces to\footnote{In this section we use $x=1,2,\dots$ instead of $\ell=1,2,\dots$ to label the sites.}
\begin{equation}
H_{SU(2)}=\f{\lam}{8\pi^{2}}\sum_{x=1}^{L}\ko{I_{x,x+1}-P_{x,x+1}}
=\f{\lam}{16\pi^{2}}\sum_{x=1}^{L}\ko{I_{x}\cdot I_{x+1}-\vec\sig_{x}\cdot \vec\sig_{x+1}}\,,
\label{1-loop Ham}
\end{equation}
where we used $P_{a,b}=\f{1}{2}\ko{I_{a,b}+\vec\sig_{a}\cdot\vec\sig_{b}}$\,.
Remarkably, this is exactly the same as the Hamiltonian of ferromagnetic Heisenberg (XXX${}_{1/2}$) spin-chain.
Identifying the complex scalar fields of SYM and the up- and down- spins of Heisenberg spin-chain as
\begin{equation}
\uparrow~\eq~\cZ\quad \mbox{and}\quad 
\downarrow~\eq~\cW\,,
\end{equation}
any SYM operators in the $SU(2)$ sector can be mapped to corresponding Heisenberg spin-chain states, see Figure \ref{fig:spin-chain2}.
In particular, the half-BPS operator $\tr(\cZ^{L})$ corresponds to the ferromagnetic vacuum state of the Heisenberg spin-chain, 
\begin{equation}
\kket{\underbrace{\uparrow\dots\uparrow\vphantom{\half}}_{L}}~\eq~
\tr(\underbrace{\cZ\dots\cZ\vphantom{\half}}_{L})\,,
\label{spin-chain vacuum}
\end{equation}
which has a vanishing eigenvalue of the Hamiltonian (\ref{1-loop Ham}).

The mixing problem involves diagonalising a $2^{L}\times 2^{L}$ matrix, which, for small $L$\,, can be done directly.
For large $L$\,, this is almost hopeless.
However, by virtue of the integrability of the Heisenberg spin-chain Hamiltonian, we can achieve the diagonalisation by using a very powerful method known in solid state physics, namely the {\em Bethe ansatz method}.
The vacuum state (\ref{spin-chain vacuum}) plays the role of the reference state in the Bethe ansatz setup, and we will denote it as $\ket{0}$ as before.

We first drop the cyclicity condition that reflects the trace structures of the SYM operators, 
and denote a spin-chain state with magnons located at sites $x_{1},\dots,x_{M}$ with $x_{i}<x_{j}$ $(i<j)$ as $\ket{x_{1},x_{2},\dots,x_{M}}$\,.
Alternatively, we can write it as $\ket{x_{1},x_{2},\dots,x_{M}}\eq \sig^{-}_{x_{1}}\sig^{-}_{x_{2}}\dots \sig^{-}_{x_{M}}\ket{0}$\,, where $\sig^{-}_{x}$ is the standard creation operator that turns ``$\uparrow$'' ($\cZ$) at site $x$ into ``$\downarrow$'' ($\cW$) at the same site.
For example, a spin-chain state $\ket{\uparrow \downarrow \uparrow \uparrow \uparrow \downarrow \uparrow \uparrow \dots  \uparrow \downarrow \uparrow}$ is denoted as $\ket{2,6,\dots,L-1}=\sig^{-}_{2}\sig^{-}_{6}\dots\sig^{-}_{L-1}\ket{0}$\,.

Let us construct an eigenstate of the Hamiltonian (\ref{1-loop Ham}) for an arbitrary number $M$ of impurities.
The problem is to find out the correct coefficients $\psi(x_{1},\dots, x_{M})$ of $\ket{M}=\sum_{x_{1},\dots, x_{M}}\psi(x_{1},\dots, x_{M})\ket{x_{1},x_{2},\dots,x_{M}}$ such that $\ket{M}$ becomes an eigenstate, then to compute the eigenvalue, or the spin-chain energy, $\ep^{(M)}$ where $H_{SU(2)}\ket{M}=\ep^{(M)}\ket{M}$\,.
Let us start with the one-magnon problem.

\paragraph{$\bullet$ One-magnon case.}
The Fourier-transformed wavefunction,
\begin{align}
\ket{p}&=\sum_{1\leq x\leq L}\psi(x)\sig^{-}_{x}\ket{0}=\sum_{1\leq x\leq L}\psi(x)\ket{x}
\quad \mbox{with}\quad \psi(x)=\f{1}{\sqrt{L}}\,e^{ipx}
\end{align}
is trivially an eigenfunction of the Hamiltonian (\ref{1-loop Ham})\,:
\begin{align}
H\ket{p}&=\f{\lam}{8\pi^{2}}\sum_{1\leq x\leq L}\ko{2\psi(x)\ket{x}-\psi(x)\ket{x-1}-\psi(x)\ket{x+1}}\cr
&=\f{\lam}{8\pi^{2}}\sum_{1\leq x\leq L}\ko{2\psi(x)\ket{x}-\psi(x+1)\ket{x}-\psi(x-1)\ket{x}}\cr
&=\f{\lam}{8\pi^{2}}\ko{2-e^{-ip}-e^{ip}}\sum_{1\leq x\leq L}\psi(x)\ket{x}
=\f{\lam}{2\pi^{2}}\sin^{2}\ko{\f{p}{2}}\ket{p}\,.
\end{align}
The energy of the one-magnon state is thus given by
\begin{equation}
\ep^{(1)}(p)=\f{\lam}{2\pi^{2}}\sin^{2}\ko{\f{p}{2}}\,.
\label{ene M=1}
\end{equation}
The periodic boundary condition requires $p=2\pi n/L$ with $n$ integers.
Note, however, there is no corresponding SYM operator for this one-magnon state in general, because of the trace (cyclicity) condition imposed on the SYM operators which implies $p=0$\,.\footnote{Actually operators with a single magnon $\cO_{i}=\tr(\Phi_{i}\cZ^{J-1})$ are chiral primary (BPS) operators.}

\begin{figure}[t]
\begin{center}
\vspace{.3cm}
\includegraphics[scale=0.9]{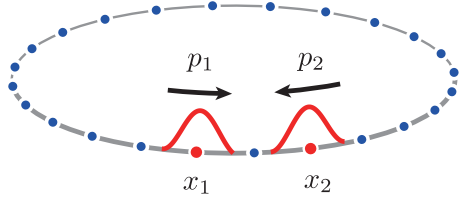}
\vspace{.3cm}
\caption{\small Two-magnon state in a spin-chain.}
\label{fig:2-magnon}
\end{center}
\end{figure}

\paragraph{$\bullet$ Two-magnon case.}
Let us move on to a two-magnon state (Figure \ref{fig:2-magnon}).
This is the simplest case that is relevant to the SYM theory.
We define the wavefunction by
\begin{align}
\ket{p_{1},p_{2}}&=\sum_{1\leq x_{1}<x_{2}\leq L}\psi(x_{1},x_{2})\sig^{-}_{x_{1}}\sig^{-}_{x_{2}}\ket{0}=\sum_{1\leq x_{1}<x_{2}\leq L}\psi(x_{1},x_{2})\ket{x_{1},x_{2}}\,,
\label{Bethe wf-2}
\end{align}
where $x_{1}$ and $x_{2}$ are the positions of the two magnons.
Imaginably, the coefficient $\psi(x_{1},x_{2})$ is given by almost the sum of two plane waves $e^{ip_{1}x_{1}+ip_{2}x_{2}}$ and $e^{ip_{2}x_{1}+ip_{1}x_{2}}$\,.
The coefficients when taking the linear combination of them turn out to be important, as their ratio describes the scattering phase-shift.
In position space, the Schr\"odinger equation $H\ket{p_{1},p_{2}}=\ep^{(2)}\ket{p_{1},p_{2}}$ with the two-body wavefunction (\ref{Bethe wf-2}) becomes
\begin{alignat}{3}
\mbox{for\quad $x_{1}+1<x_{2}$\,,}&\qquad &
\ep^{(2)}\psi (x_{1},x_{2})&=4\psi (x_{1},x_{2})-\psi (x_{1}-1,x_{2})-\psi (x_{1}+1,x_{2})-{}\no\\
&{}&{}&\qquad{}-\psi (x_{1},x_{2}-1)-\psi (x_{1},x_{2}+1)\,,\label{2-mag 1}\\[2mm]
\mbox{for\quad $x_{1}+1=x_{2}$\,,}&\qquad &
\ep^{(2)}\psi (x_{1},x_{2})&=2\psi (x_{1},x_{2})-\psi (x_{1}-1,x_{2})-\psi (x_{1},x_{2}+1)\,.\label{2-mag 2}
\end{alignat}
From (\ref{2-mag 1}) and (\ref{2-mag 2}), the following consistency condition follows
\begin{equation}
0=2\psi (x_{1},x_{1}+1)-\psi (x_{1},x_{1})-\psi (x_{1}-1,x_{1})\,.
\label{meeting}
\end{equation}
These difference equations (\ref{2-mag 1}), (\ref{2-mag 2}) and (\ref{meeting}) can be solved by an ansatz of the following form,
\begin{equation}
\psi(x_{1},x_{2})=A_{0}(p_{1},p_{2})\,e^{ip_{1}x_{1}+ip_{2}x_{2}}+A_{0}(p_{2},p_{1})\,e^{ip_{2}x_{1}+ip_{1}x_{2}}\,.
\label{Bethe wf}
\end{equation}
Then one finds that the wavefunction (\ref{Bethe wf-2}) with (\ref{Bethe wf}) indeed solves (\ref{2-mag 1}) and (\ref{2-mag 2}) if the dispersion relation is given by
\begin{equation}
\ep^{(2)}(p_{1},p_{2})
=\ep^{(1)}(p_{1})+\ep^{(1)}(p_{2})
=\f{\lam}{2\pi^{2}}\kko{\sin^{2}\ko{\f{p_{1}}{2}}+\sin^{2}\ko{\f{p_{2}}{2}}}
\label{1-loop DR M=2}
\end{equation}
{\em and} the S-matrix is given by
\begin{equation}
S(p_{1},p_{2})\eq \f{A_{0}(p_{2},p_{1})}{A_{0}(p_{1},p_{2})}=-\f{e^{ip_{1}+ip_{2}}-2e^{ip_{1}}+1}{e^{ip_{1}+ip_{2}}-2e^{ip_{2}}+1}\,.
\label{1-loop S-mat}
\end{equation}
The momenta $p_{j}$ are fixed once periodic boundary conditions $\psi(x_{1},x_{2})=\psi(x_{2},x_{1}+L)$ are imposed on the Bethe wavefunction (\ref{Bethe wf}). 
This requirement of the periodicity leads to the following {\em Bethe ansatz equations},
\begin{equation}
e^{ip_{1}L}=S(p_{1},p_{2})\quad \mbox{and}\quad e^{ip_{2}L}=S(p_{2},p_{1})\,.
\label{1-loop BAE M=2}
\end{equation}
There is an extra condition for the momenta which should be taken into account, that is the cyclicity condition resulted from the trace structure of the original SYM operators.
It is given by
\begin{equation}
p_{1}+p_{2}=0\,.
\label{mom cond M=2}
\end{equation}
We will also refer to this constraint as the zero-momentum condition.
\begin{figure}[t]
\begin{center}
\vspace{.3cm}
\includegraphics[scale=1.0]{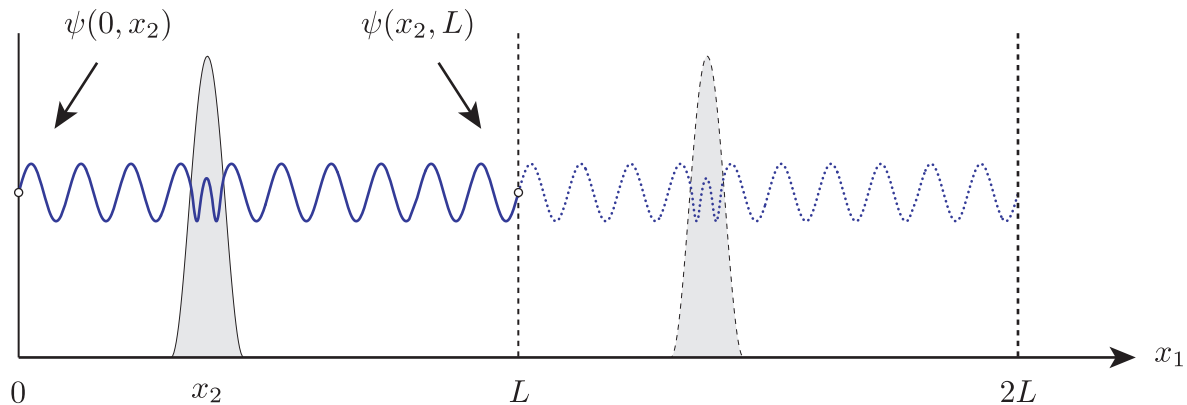}
\vspace{.3cm}
\caption{\small Periodicity condition for the two-magnon wavefunction leads to the Bethe ansatz equations (\ref{1-loop BAE M=2}).}
\label{fig:2-mag-wave}
\end{center}
\end{figure}

Note that, while a state with $p_{j}=2\pi n_{j}/L$ (which amounts to set the S-matrix unity) gives the energy $\ep$ to the accuracy of $\cO(L^{-1})$\,, a solution of the Bethe ansatz equation $p_{j}L=2\pi n_{j}+\delta(\{p_{j}\})$ (where $\delta$ is the scattering phase-shift, $S=e^{i\delta}$) gives $\ep$ to $\cO(e^{-R/L})$ where $R$ is the interaction range.

\paragraph{$\bullet$ \bmt{M}-magnon case.}

Since we know the Heisenberg spin-chain (the $SU(2)$ sector of the SYM spin-chain at the one-loop) is integrable, we can immediately generalise the result of the two-magnon case to the $M$-magnon case.
Actually, one definition of quantum integrability is that all scattering processes are factorised to two-body scatterings, in which the momenta are exchanged but not changed in magnitude.
Due to the factorisability of the S-matrix, the Bethe ansatz equations become
\begin{equation}
e^{ip_{k}L}=\prod_{\mbox{$j=1\atop j\neq k$}}^{M}S(p_{k},p_{j})\,,\qquad k=1,\dots,M
\label{1-loop BAE}
\end{equation}
with the same S-matrix as given in (\ref{1-loop S-mat}).
What this means is that the total phase-shift acquired by a magnon labeled by $k$ propagating along the spin-chain (LHS) must be equal to the sum of the two-body phase-shifts due to individual scattering with all other $M-1$ magnons (RHS) for the consistency.
The zero-momentum condition is given by
\begin{equation}
\sum_{k=1}^{M}p_{k}=0\,,
\label{momentum cond}
\end{equation}
and the dispersion relation is also readily generalised as
\begin{equation}
\ep^{(M)}(p_{1},\dots,p_{M})=\sum_{k=1}^{M}\ep^{(1)}(p_{k})
=\f{\lam}{2\pi^{2}}\sum_{k=1}^{M}\sin^{2}\ko{\f{p_{k}}{2}}\,.
\label{1-loop DR}
\end{equation}
It is convenient to rewrite all those equations and constraints in terms of so-called {\em rapidity} parameters defined by
\begin{equation}
u_{k}\eq \f{1}{2}\cot\ko{\f{p_{k}}{2}}\,,\quad k=1,\dots,M\,.
\label{def : rap}
\end{equation}
In terms of the rapidities, (\ref{1-loop BAE}), (\ref{momentum cond}) and (\ref{1-loop DR}) are cast into the following set of algebraic constraints and a relation on the rapidities,
\begin{alignat}{3}
&\mbox{\sl Bethe ansatz equation:}&\qquad 
	&\ko{\f{u_{k}+i/2}{u_{k}-i/2}}^{L}=\prod_{\mbox{$j=1\atop j\neq k$}}^{M}\f{u_{k}-u_{j}+i}{u_{k}-u_{j}-i}\,,\quad k=1,\dots,M\,,
	\label{1-loop BAE : u}\\
&\mbox{\sl Momentum condition:}&\qquad 
	&\ko{e^{i\sum_{k=1}^{M}p_{k}}={}}\prod_{k=1}^{M}\f{u_{k}+i/2}{u_{k}-i/2}=1\,,
	\label{momentum cond : u}\\
&\mbox{\sl Energy formula:}&\qquad 
	&\ep^{(M)}=\f{\lam}{8\pi^{2}}\sum_{k=1}^{M}\f{1}{u_{k}^{2}+1/4}\,.
	\label{1-loop DR : u}
\end{alignat}
This set of the equations completely determines the spectrum of the Hamiltonian (\ref{1-loop Ham}).
Note that since the momenta can be complex in general, so are the rapidities.

For our purpose to check the AdS/CFT in far-from-BPS sector (at the one-loop level), what we are going to do is to find a set of rapidities $\{ u_{k} \}$ that satisfies both (\ref{1-loop BAE : u}) and (\ref{momentum cond : u})\,, then plug it into the energy formula (\ref{1-loop DR : u}) to obtain the coefficient $\delta_{1}\eq \ep^{(M)}$ (recall $\Delta=L+\ga=L\big[1+\delta_{1}\tlambda+\mathcal O(\tlambda^{2})\big]$ with $\tlambda\eq \lam/L^{2}$\,, see (\ref{double expansion-Delta})).
The obtained $\delta_{1}$ is to be compared with the corresponding string theory result $\ep_{1}$\,. 

\paragraph{}
Here we make some remark on the hidden set of higher conserved charges $\cQ_{r}$ of the spin-chain, which mutually commute, $[\cQ_{r},\cQ_{s}]=0$\,.
Once we could diagonalise the Hamiltonian by solving the Bethe ansatz equation (\ref{1-loop BAE : u}), it means we could at the same time diagonalise all the other higher conserved charges as well.
It is because an eigenstate of the transfer matrix $t(u)$ is also determined by a set of Bethe roots (solution of the Bethe ansatz equation) $\{ u_{k} \}$\,, with eigenvalue
\begin{equation}
t(u)=\ko{u+\f{i}{2}}^{L}\prod_{k=1}^{M}\f{u-u_{k}-i}{u-u_{k}}+\ko{u-\f{i}{2}}^{L}\prod_{k=1}^{M}\f{u-u_{k}+i}{u-u_{k}}\,.
\end{equation}
The higher conserved charges are generated by the formula
\begin{equation}
\cQ_{r+1}=\left.-\f{i}{2}\f{1}{r!}\f{d^{r}}{du^{r}}\,\ln t(u)\right|_{u=i/2}\,,
\label{higher charge}
\end{equation}
and in particular the energy (\ref{1-loop DR : u}) agrees with the second charge $\cQ_{2}$\,, up to irrelevant constant and a multiplicative factor (the coupling constant).

\begin{figure}[t]
\begin{center}
\vspace{.3cm}
\includegraphics[scale=1.0]{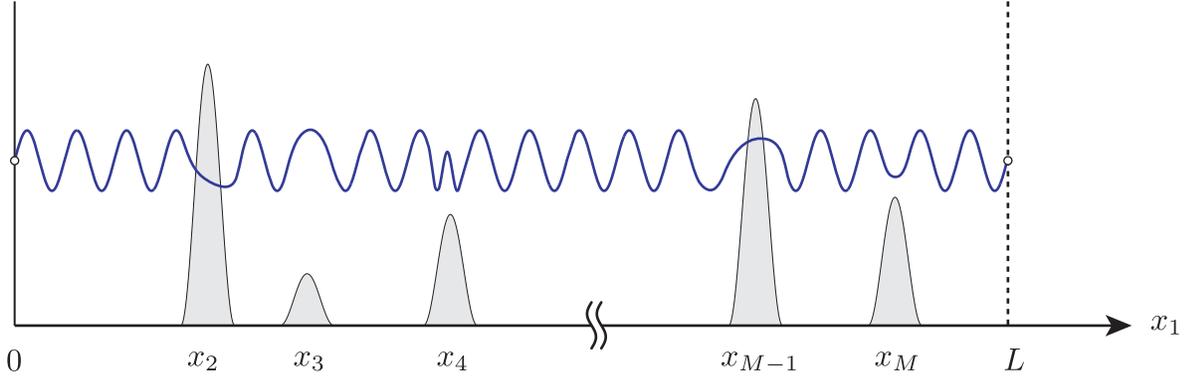}
\vspace{.3cm}
\caption{\small Periodicity condition for the $M$-magnon wavefunction leads to the Bethe ansatz equations (\ref{1-loop BAE}).}
\label{fig:M-mag-wave}
\end{center}
\end{figure}

\subsubsection*{BMN formula from Bethe Ansatz approach}

As a check of the applicability of the Bethe ansatz approach, let us see how it reproduces the known result in the BMN (near-BPS) sector.
We go back to the simplest two-magnon case, {\em i.e.}, we consider a BMN operator with two impurities\,:
\begin{equation}
\cO_{ij}^{(n)}=\sum_{k=1}^{J}e^{2\pi i n k/J}\tr\ko{\Phi_{i}\cZ^{k-1}\Phi_{j}\cZ^{J-k-1}}\,.
\label{BMN op}
\end{equation}
Because of the momentum condition (\ref{mom cond M=2}), we can set $p\eq p_{1}=-p_{2}$\,, or $u\eq u_{1}=-u_{2}$\,.
Plugging it into the Bethe ansatz equations (\ref{1-loop BAE M=2}), we are left with a single equation
\begin{equation}
\ko{\f{u+i/2}{u-i/2}}^{L}=\f{2u+i}{2u-i}\,,
\qquad \mbox{\em i.e.}, \qquad 
\ko{\f{u+i/2}{u-i/2}}^{L-1}=1\,,
\end{equation}
which immediately gives $p=2\pi n/(L-1)$\,.
Substituting it into (\ref{1-loop DR M=2}) and taking the BMN limit (\ref{BMN limit}), we obtain
\begin{equation}
\ep^{(2)}(p,-p)
=2\times \f{\lam}{2\pi^{2}}\sin^{2}\ko{\f{p}{2}}
=\f{\lam}{\pi^{2}}\sin^{2}\ko{\f{\pi n}{L-1}}
\quad \xrightarrow[]{\mbox{\footnotesize BMN limit}} \quad \f{n^{2}\lam}{L^{2}}\,,
\end{equation}
which reproduces the known result \cite{Berenstein:2002jq},
\begin{equation}
\Delta_{\cO_{ij}^{(n)}}=L+\gamma\,,\qquad 
\gamma=\f{n^{2}\lam}{L^{2}}+\cO(g_{\rm YM}^{2})\,.
\label{BMN op ene}
\end{equation}
On the string theory side, as we will see in the end of Section \ref{sec:rot-string-ansatz}, this result matches with the energy of a string state on plane-wave background.
The zero-momentum condition (\ref{momentum cond : u}) matches with the level matching condition.

The matching of the gauge and string theory spectra in the BMN limit was extended to two-loops in \cite{Gross:2002su}.
For a single magnon case, under the assumption of a ``dilute gas'' approximation, it was even argued that the spectra match to all orders in perturbation theory (in the form of the full square-root structure) \cite{Santambrogio:2002sb}.
However, recent study has revealed that we cannot actually apply the ``dilute gas'' approximation, due to the existence of the so-called dressing phase factor in the conjectured all-order S-matrix.
We will come back to this point later in Section \ref{sec:dressing phase}.

\paragraph{}
The BMN case we have seen corresponds to a situation where a small number of magnons propagate on the chain almost freely with a very simple scattering matrix.
When the number of magnons becomes macroscopically large, however, the interactions among magnons cannot longer be neglected and we need to solve the algebraic equation (\ref{1-loop BAE : u}) with the constraints (\ref{momentum cond : u}).
We will see how it is achieved in the thermodynamic limit in Section \ref{sec:thermodynamic}.

\section[Higher loops in $SU(2)$ sector]
	{Higher loops in \bmt{SU(2)} sector\label{sec:higher SU(2)}}

As we go higher beyond one-loop, the corresponding spin-chain becomes more and more long-ranged.
For convenience let us define a new gauge coupling constant which is related to the 't Hooft coupling as
\begin{equation}
g\eq \f{\sqrt{\lam}}{4\pi}\,.
\end{equation}
We can expand the dilatation operator in powers of the coupling,
\begin{equation}
\mathfrak D(g)=\sum_{\ell=1}^{L}\sum_{n=0}^{\infty}g^{2n}\cH_{n}=\sum_{\ell=1}^{L}\ko{1+g^{2}\cH_{1}+g^{4}\cH_{2}+g^{6}\cH_{3}+\dots}\,,
\label{Hamiltonian SU(2)-3}
\end{equation}
where the perturbative Hamiltonians are found to be, up to three-loop order,\footnote{In the $SU(2)$ sector, assuming the higher-order integrability, the dilatation operator is conjectured to five-loop order (with some undetermined constants which do not affect the integrability and the BMN scaling) \cite{Beisert:2004hm}.
Recently the four-loop piece was obtained in \cite{Beisert:2007hz}.
The analysis is based on certain inputs, which are free from integrability assumption.} \cite{Beisert:2003tq,Beisert:2003ys,Eden:2004ua}
\begin{align}
\cH_{0}&=1\,,\\[2mm]
\cH_{1}&=\mbox{\large $\f{1}{2}$}\ko{1-\vec\sig_{\ell}\cdot\vec\sig_{\ell+1}}\,,\label{SU(2) 1-loop}\\[2mm]
\cH_{2}&=-\ko{1-\vec\sig_{\ell}\cdot\vec\sig_{\ell+1}}+\mbox{\large $\f{1}{4}$}\ko{1-\vec\sig_{\ell}\cdot\vec\sig_{\ell+2}}\,,\label{SU(2) 2-loop}\\[2mm]
\cH_{3}&=\mbox{\large $\f{15}{4}$}\ko{1-\vec\sig_{\ell}\cdot\vec\sig_{\ell+1}}-\mbox{\large $\f{3}{2}$}\ko{1-\vec\sig_{\ell}\cdot\vec\sig_{\ell+2}}+\mbox{\large $\f{1}{4}$}\ko{1-\vec\sig_{\ell}\cdot\vec\sig_{\ell+3}}-{}\no\\
&\qquad {}-\mbox{\large $\f{1}{8}$}\ko{1-\vec\sig_{\ell}\cdot\vec\sig_{\ell+3}}\ko{1-\vec\sig_{\ell+1}\cdot\vec\sig_{\ell+2}}+\mbox{\large $\f{1}{8}$}\ko{1-\vec\sig_{\ell}\cdot\vec\sig_{\ell+2}}\ko{1-\vec\sig_{\ell+1}\cdot\vec\sig_{\ell+3}}\,.\label{SU(2) 3-loop}
\end{align}
The above result was first proposed in \cite{Beisert:2003tq} by assuming the integrability beyond the two-loops, and was proved by combining algebraic \cite{Beisert:2003ys} and field theoretic \cite{Eden:2004ua} computation.
Note that $\cH_{3}$ contains not only $\vec\sig_{\ell}\cdot \vec\sig_{\ell+k}$ terms but also $(\vec\sig_{\ell}\cdot \vec\sig_{\ell+k})(\vec\sig_{\ell+k'}\cdot \vec\sig_{\ell+k''})$ terms.
In general, $n$-loop piece contains up to $(n+1)$ nearest neighbour interactions.

In \cite{Serban:2004jf}, Serban and Staudacher succeeded in emulating the dilatation operator for the $SU(2)$ sector of SYM to a long-range integrable spin-chain called Inozemtsev spin-chain \cite{JStatistPhys.59.1143} up to the three-loops,\footnote{A redefinition of the coupling constant and the charges is needed.} thus proving the three-loop integrability.
Then they proposed a three-loop Bethe ansatz, with which they could compute the anomalous dimensions of long operators dual to circular and folded strings of \cite{Frolov:2003xy}.
We will review the comparison later in Section \ref{sec:discrepancy}.
Remarkably, the embedding of $SU(2)$ spin-chain to the three-loop order to the Inozemtsev model turns out inconsistent with the perturbative BMN scaling.
It breaks the scaling at four-loop.
However, see the next section, where we see there actually exists a novel long-range integrable spin-chain that respects the BMN scaling to all orders.

\paragraph{}
As briefly mentioned, the Bethe ansatz method, with some modification, can be also applied to higher order (therefore it is long-range) interactions of SYM when computing anomalous dimensions.
Such a generalised Bethe ansatz method is called {\em perturbative asymptotic Bethe ansatz (PABA)}, invented by Staudacher to analyse gauge theory (and its string dual) \cite{Serban:2004jf,Staudacher:2004tk}.
The idea is to modify the form of the two-magnon Bethe wavefunction as
\begin{align}\label{PABA-closed}
\begin{split}
\psi (x_{1},x_{2})&=\big(1+F\ko{|x_{2}-x_{1}|,p_{1},p_{2};g}\big) A\ko{p_{1},p_{2};g}\, e^{i\ko{p_{1}x_{1}+p_{2}x_{2}}}+{}\\
&\qquad {}+\big(1+{\widetilde F}\ko{|x_{2}-x_{1}|,p_{2},p_{1};g}\big) {\widetilde A}\ko{p_{2},p_{1};g}\, e^{i\ko{p_{2}x_{1}+p_{1}x_{2}}}
\end{split}
\end{align}
with the correction factors
\begin{equation}\label{f-closed}
A(p_{1},p_{2};g)=\sum_{k=0}^{\infty} g^{2k} A_{k}(p_{1},p_{2})\,,\qquad 
F\ko{|d|,p_{1},p_{2};g}=\sum_{k=1}^{\infty} g^{|d|+2k} F_{k}(|d|,p_{1},p_{2})\,,
\end{equation}
and the same for ${\widetilde A}$ and ${\widetilde F}$\,.
As $|d|$ becomes large, the fudge functions exponentially tend to zero, so that in the leading order in $g$\,, it reproduces the original asymptotic Bethe wavefunction (\ref{Bethe wf}).
Applying the three-loop Hamiltonian (\ref{Hamiltonian SU(2)-3}) to the modified Bethe wavefunction (\ref{PABA-closed}) leads to a set of difference equations for $|x_{2}-x_{1}|=1,2,3$\,.
They can be solved only when the coefficients $F_{k}$\,, $\widetilde F_{k}$ and $A_{k}$\,, $\widetilde A_{k}$ are properly fine-tuned.
Then the S-matrix is computed as the ratio of $A$ and ${\widetilde A}$\,, and admits a $g$-expansion as
\begin{equation}
S\ko{p_{1},p_{2};g}=\f{{\widetilde A}\ko{p_{2},p_{1};g}}{A\ko{p_{1},p_{2};g}}
=\sum_{k=0}^{\infty}g^{2k}S^{(k)}\ko{p_{1},p_{2}}\,.
\label{paba S}
\end{equation}
The fudge factors $F$ and $\widetilde F$ capture the long-range interactions whereas $S$ is responsible for the nearest-neighbour scattering. 
The perturbatively corrected S-matrix (\ref{paba S}) does not change the form of the Bethe ansatz equations (\ref{1-loop BAE}), but does correct the charges perturbatively as $\cQ_{r}(g)=\cQ_{r}^{(0)}+\sum_{k=1}^{\infty}g^{k}\cQ_{r}^{(k)}$\,.

The PABA technique was first adapted in \cite{Serban:2004jf} to extract the bulk S-matrix for ${SU}(1|1)$ sector \cite{Staudacher:2004tk} from the $SU(2|3)$ results of \cite{Beisert:2003ys}, then used to analyse the S-matrices for the ${SU}(1|2)$ and ${SU}(1,1|2)$ sectors \cite{Beisert:2005fw}, and also for plane-wave matrix model \cite{Fischbacher:2004iu,Klose:2005cv}.
It was also applied to open spin-chain systems with boundaries, such as a giant graviton system with an open string excitation \cite{Agarwal:2006gc,Okamura:2006zr}, and a defect CFT system \cite{Okamura:2006zr}.

\section{The BDS model\label{sec:BDS}}

We have argued that the integrable structure of the $SU(2)$ sector of $\cN=4$ SYM up to three-loop can be embedded to that of the Inozemtsev spin-chain, but at the same time the embedding forces us to give up the BMN scaling beyond three-loops.
In \cite{Beisert:2004hm}, Beisert, Dippel and Staudacher proposed a novel type of long-range integrable spin-chain, together with an all-loop asymptotic Bethe ansatz for the spin-chain.
We will call it the {\em BDS model}.
\begin{figure}[t]
\begin{center}
\vspace{.3cm}
\includegraphics[scale=0.9]{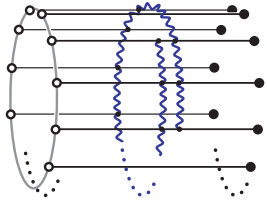}
\vspace{.3cm}
\caption{\small Wrapping interaction.}
\label{fig:wrapping}
\end{center}
\end{figure}

The three key assumptions for the BDS model are\,: $(1)$ integrability, $(2)$ field theoretic considerations (structural consistency with general features of SYM perturbation theory), and $(3)$ qualitative BMN scaling behavior.
Requirements of the latter two properties uniquely define perturbative scheme to construct dilatation operator up to the wrapping order, $\cO(\lam^{L-1})$\,.\footnote{For the wrapping issue, see below (\ref{higher charges BDS}).}
As compared to the Inozemtsev model, which violates the BMN scaling at the four-loop order, the BDS model respects the BMN scaling to all-order by construction.

The conjectured all-loop Bethe ansatz equation is give by
\begin{equation}
e^{i p_{j}L}=\prod_{\mbox{$k=1\atop k\neq j$}}^{M} S_{\rm BDS}(p_{j},p_{k})
\qquad \mbox{for}\quad 
j=1,\dots, M\,,
\label{BDS bethe eq}
\end{equation}
where $M$ is the number of magnons, and the S-matrix is given by
\begin{equation}
S_{\rm BDS}(p_{j},p_{k})=\f{u(p_{j})-u(p_{k})-(i/g)}{u(p_{j})-u(p_{k})+(i/g)}\,.
\label{S-matrix}
\end{equation}
We call it as the BDS S-matrix.
The rapidity variable $u$ for the BDS S-matrix is given in terms of the momentum by
\begin{equation}
u(p)=\f{1}{2g}\cot\ko{\f{p}{2}}\sqrt{1+16g^{2}\sin^{2}\ko{\f{p}{2}}}\,.
\label{rap}
\end{equation}
It should be emphasised that the S-matrix cannot have any non-trivial overall phase for it to possess the correct BMN scaling.
Notice also, when we take $g\ll 1$ limit, the BDS S-matrix reduces to the one-loop (Heisenberg) S-matrix of (\ref{1-loop BAE : u}).

It is convenient to introduce complex variables $x^{\pm}$ called
{\em spectral parameters} introduced in \cite{Beisert:2004jw}, which are related to the rapidity (\ref{rap}) via the formulae\footnote{The spectral parameters $x^{\pm}$ introduced here are ``rescaled'' ones which will turn out to be useful when we investigate the strong coupling region.
In many (earlier) literatures, ``un-rescaled'' spectral parameters ${\rm x}^{\pm}$ are employed, which are related to the ones here as ${\rm x}^{\pm}=gx^{\pm}$\,.\label{foot:x}}
\begin{equation}
x^{\pm}(u)=x\ko{u\pm\mbox{\large $\f{i}{2g}$}}
\quad \mbox{where}\quad 
x(u)=\f{1}{2}\ko{u+\sqrt{u^{2}-4}}\quad \ko{~\mbox{or}~~u=x+\f{1}{x}~}\,.
\label{x(u)}
\end{equation}
In terms of the spectral parameters, the rapidity can be expressed as
\begin{align}
u(x^{\pm}) &=\f{1}{2}\kko{\ko{x^{+}+\f{1}{x^{+}}}+\ko{x^{-}+\f{1}{x^{-}}}}\\[2mm]
&=x^{+}+\f{1}{x^{+}}-\f{i}{2g}
=x^{-}+\f{1}{x^{-}}+\f{i}{2g}\,,
\label{constraint xpm}
\end{align}
and the BDS S-matrix is cast into the form
\begin{equation}
S_{\rm BDS}(x_{j}^{\pm}, x_{k}^{\pm})=\f{x_{j}^{+}-x_{k}^{-}}{x_{j}^{-}-x_{k}^{+}}\cdot
\f{1-1/(x_{j}^{+}x_{k}^{-})}{1-1/(x_{j}^{-}x_{k}^{+})}\,.
\label{S in x}
\end{equation}

As usual, for each solution of the Bethe ansatz equations (\ref{BDS bethe eq}), the energy $E$ of the corresponding state is simply the sum of the energies $\ep_{j}=\ep(p_{j})$ of the individual magnons, $E=\sum_{j=1}^{M}\ep_{j}$\,.
The energy of each magnon is determined by the relation,\footnote{As we will see in Chapter \ref{chap:Asymptotic}, the dispersion relation (\ref{disp}) is not mere a conjecture but can be actually derived though symmetry argument (with additional field theory input).
To be precise, it is a BPS relation under a centrally-extended $SU(2|2)$ algebra \cite{Beisert:2005tm}.}
\begin{equation}
\epsilon_{j}= \sqrt{1+16g^{2}\sin^{2}\ko{\f{p_{j}}{2}}}\,.
\label{disp}
\end{equation}
The dispersion relation (\ref{disp}) is equivalent to the constraint
\begin{equation}
\bigg(x_{j}^{+}+\frac{1}{x_{j}^{+}}\bigg) - \bigg(x_{j}^{-}+\frac{1}{x_{j}^{-}}\bigg)=\frac{i}{g}\,.
\end{equation}
The asymptotically exact formulae for the magnon momenta and energies are written as, in terms of the spectral parameters, 
\begin{align}
p_{j}&=p(x_{j}^{\pm})=\f{1}{i}\ln\bigg(\f{x_{j}^{+}}{x_{j}^{-}}\bigg)\,,\label{p}\\[2mm]
\ep_{j}&=\ep(x_{j}^{\pm})=\f{g}{i}\kko{\bigg(x_{j}^{+}-\f{1}{x_{j}^{+}}\bigg)-\bigg(x_{j}^{-}-\f{1}{x_{j}^{-}}\bigg)}=1+2ig\bigg(\f{1}{x_{j}^{+}}-\f{1}{x_{j}^{-}}\bigg)\,,
\label{Delta}
\end{align}
and all the local commuting charges are given by
\begin{align}
q_{r+1}(u)&=\f{2\sin\ko{rp/2}}{r}\ko{\f{\sqrt{1+16g^{2}\sin^{2}\ko{p/2}}}{4\sin\ko{p/2}}}^{r}\,,
\qquad r=1,2,\dots\,,
\label{higher charges2}
\end{align}
or more compactly, in terms the spectral parameters as
\begin{align}
q_{r+1}(x^{\pm})=\f{i}{r}\kko{\f{1}{(x^{+})^{r}}-\f{1}{(x^{-})^{r}}}\,,
\qquad r=1,2,\dots\,.
\label{higher charges}
\end{align}
At each order $r$\,, the total charge of the $M$ excitations is given by the sum of individual charges,
\begin{equation}
\cQ_{r}=\sum_{k=1}^{M}q_{r}(x^{\pm}_{k})\,,
\label{higher charges BDS}
\end{equation}
and in particular, $\cQ_{2}$ is related to the magnon energy as $\ep=1+2g\cQ_{2}$\,.
The higher charges (\ref{higher charges2}) or (\ref{higher charges}) will play an important role in the construction of the AdS/CFT S-matrix later in Chapter \ref{chap:dressing}.

Some remarks on the BDS model are in order.
First, the Bethe ansatz equation (\ref{BDS bethe eq}) is applicable in ``asymptotic'' region, which means, it is only valid up to $g^{2(L-1)}\sim \lam^{L-1}$ order, when the length of the spin-chain is $L$\,.\footnote{The finite-size correction is exponentially small in the $L\to \infty$ limit.}
Beyond that order, we should include into the gauge theory computations the so-called {\em wrapping interactions} \cite{Beisert:2004hm,Ambjorn:2005wa,Janik:2007wt}, which arise when
the interaction stretches all around the spin-chain states (SYM single-trace operators), see Figure \ref{fig:wrapping}.
Currently we do not know how to incorporate those wrapping interactions into gauge theory Bethe ansatz equations.\footnote{However, see \cite{Janik:2007wt} for recent progress on the string theory side.}

Second, in an attempt at the non-asymptotic, all-loop Bethe ansatz for the $SU(2)$ sector of $\cN=4$ SYM, it is argued that the long-range BDS spin-chain model is identical to the strong-coupling approximation of half-filled one-dimensional Hubbard model (Hubbard chain) up to the wrapping order \cite{Rej:2005qt}.
The Hubbard model is a short-range model of itinerant fermions, whose Hamiltonian is given by
\begin{equation}
H_{\rm Hubbard}=-t\sum_{\ell=1}^{L}\kko{\,\sum_{\sig=\uparrow,\downarrow}\ko{c_{\ell,\sig}^{\dagger}c_{\ell+1,\sig}+c_{\ell+1,\sig}^{\dagger}c_{\ell,\sig}}-U\,c_{\ell,\uparrow}^{\dagger}c_{\ell,\uparrow}c_{\ell,\downarrow}^{\dagger}c_{\ell,\downarrow}}\,,
\end{equation}
where $c_{\ell,\sig}^{\dagger}$ and $c_{\ell,\sig}$ are fermionic creation and annihilation operators respectively, satisfying $\{ c_{\ell,\sig},c_{\ell',\sig'} \}=\{ c_{\ell,\sig}^{\dagger},c_{\ell',\sig'}^{\dagger} \}=0$ and $\{ c_{\ell,\sig},c_{\ell',\sig'} \}=\delta_{\ell,\ell'}\delta_{\sig,\sig'}$\,.
It has two coupling constants\,: $t$ is the coupling of the kinetic nearest-neighbour hopping term, and $U$ is the coupling of the density potential.
By comparing the ground state energy of the half-filled band of the Hubbard model with the energy of the antiferromagnetic state ({\em i.e.}, with highest possible anomalous dimension) of the BDS model, it was found that they coincide under identification of the couplings $t=-1/2g$ and $U=1/g$\,.
They were also able to derive the asymptotic Bethe ansatz equations (\ref{BDS bethe eq}) from the Lieb-Wu equations \cite{PhysRevLett.20.1445,PhysicaA.321.1} of the Hubbard model.

Third, the BDS model is an old conjecture for the all-loop gauge theory, and now that we have better knowledge about both the perturbative gauge theory and the AdS/CFT, we know the BDS conjecture can no longer be a correct candidate capturing the all-loop gauge theory.
However, the significance of their model never fades in the study of the integrable structure of AdS/CFT.
Actually it is believed to describe all but the so-called dressing factor which we will discuss later in Chapter \ref{chap:dressing}.

\section[Nested Bethe ansatz for $\cN=4$ SYM at one-loop]
	{Nested Bethe ansatz for \bmt{\cN=4} SYM at one-loop}

The Bethe ansatz method can be formulated for any Lie algebra, and so any symmetry sector of the $\cN=4$ SYM.
Compared with the simplest $SU(2)$ sector at one-loop (Heisenberg spin-chain) case, in general, the Bethe ansatz equations have to be extended to the so-called nested Bethe ansatz.
It reads
\begin{equation}
\ko{\f{u_{\al,k}+iq_{\al}/2}{u_{\al,k}-iq_{\al}/2}}^{L}
=\prod_{\mbox{$j=1,\dots,K_{\be}\atop \be=1,\dots,7$}}^{(\al,k)\neq (\be,j)}
\f{u_{\al,k}-u_{\be,j}+iM_{\al\be}/2}{u_{\al,k}-u_{\be,j}-iM_{\al\be}/2}\,,
\label{BAE general}
\end{equation}
where $M_{\al\be}$ is the Cartan matrix and $q_{\al}$ is the Dynkin labels of the highest weight representation.
By solving the Bethe ansatz equations (\ref{BAE general}), the magnon momentum $p$ and the energy $E$ are obtained from the formulae
\begin{equation}
e^{ip}=\prod_{(\al,k)}\f{u_{\al,k}+iq_{\al}/2}{u_{\al,k}-iq_{\al}/2}\,,\qquad 
E=\sum_{(\al,k)}\f{q_{\al}}{u_{\al,k}^{2}+q_{\al}^{2}/4}\,.
\label{p and E gen}
\end{equation}
For the simplest Heisenberg case, there was no nesting and we only had $M_{\al\be}=M_{11}=2$ and $q_{\al}=q_{1}=1$\,, so that (\ref{BAE general}) and (\ref{p and E gen}) reduce to (\ref{1-loop BAE : u}) and (\ref{momentum cond : u})\,-\,(\ref{1-loop DR : u}), respectively.
In this section, we collect some results for more complicated cases\,: the nested Bethe ansatz for the $SO(6)$ and $PSU(2,2|4)$ sector of $\cN=4$ SYM.

\subsection[The $SO(6)$ sector at one-loop]
	{The \bmt{SO(6)} sector at one-loop\label{sec:SO(6) one-loop}}
\newsavebox{\boxDynkinSOvi}
\sbox{\boxDynkinSOvi}{\includegraphics{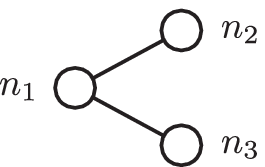}}
\newlength{\bwDynkinSOvi}
\settowidth{\bwDynkinSOvi}{\usebox{\boxDynkinSOvi}} 
The rank of the Cartan algebra of $SO(6)$ is three, and the three simple roots are represented as $\vec\al_{1}=(1,-1,0)$\,, $\vec\al_{2}=(0,1,-1)$\,, $\vec\al_{3}=(0,1,1)$\,.
The vector representation of the highest-weight state is given by $\vec q=(1,0,0)$\,, satisfying $\vec q\cdot \vec\al_{1}=1$ and $\vec q\cdot \vec\al_{2}=\vec q\cdot \vec\al_{3}=0$\,.
The elements of the Cartan matrix are obtained through the definition $M_{\al\be}=\vec\al_{\al}\cdot\vec\al_{\be}$\,, as
\begin{equation}
M_{\al\be}=
\mbox{\footnotesize $\left(
\begin{array}{ccc}
2 & -1 & -1   \\
-1 & 2  & 0   \\
-1& 0& 2   
\end{array}
\right)$}\,,\qquad 
q_{\al}=
\mbox{\footnotesize $\left(\begin{array}{c}
1 \\
0 \\
0 \end{array}\right)$}\,.
\label{Cartan SO(6)}
\end{equation}
Denoting the number of Bethe roots associated with each simple root $\al_{1}$\,, $\al_{2}$ and $\al_{3}$ of $SO(6)$ as $n_{1}$\,, $n_{2}$ and $n_{3}$\,, the Dynkin diagram is given by
\begin{center}
\vspace{.5cm}
\parbox{\bwDynkinSOvi}{\usebox{\boxDynkinSOvi}}\,.
\vspace{.5cm}
\end{center}
The total representation becomes $\vec w=L\vec q-n_{1}\vec\al_{1}-n_{2}\vec\al_{2}-n_{3}\vec\al_{3}=(L-n_{1},n_{1}-n_{2}-n_{3},n_{2}-n_{3})$\,.
The Dynkin indices for the highest weight state $\vec w=(J_{1}, J_{2}, J_{3})$ with the three Cartan charges $J_{1,2,3}$ are $[\vec w\cdot \vec\al_{2}, \vec w\cdot \vec\al_{1}, \vec w\cdot \vec\al_{3}]=[n_{1}-2n_{2},L-2n_{1}+n_{2}+n_{3},n_{1}-2n_{3}]=[J_{2}+J_{3}, J_{1}-J_{2}, J_{2}-J_{3}]$\,.
In particular, for the complex scalars $\cZ$\,, $\cW$\,, $\cY$\,, and their complex conjugates,
\begin{alignat}{5}
&\cZ~&:~(1,0,0)&=\vec q\,,&\qquad 
&\overline{\cZ}~&:~(-1,0,0)&=\vec w-2\vec\al_{1}-\vec\al_{2}-\vec\al_{3}\,,\\
&\cW~&:~(0,1,0)&=\vec q-\vec\al_{1}\,,&\qquad 
&\overline{\cW}~&:~(0,-1,0)&=\vec q-\vec\al_{1}-\vec\al_{2}-\vec\al_{3}\,,\\
&\cY~&:~(0,0,1)&=\vec w-\vec\al_{1}-\vec\al_{2}\,,&\qquad 
&\overline{\cY}~&:~(0,0,-1)&=\vec q-\vec\al_{1}-\vec\al_{3}\,,
\end{alignat}
the Dynkin indices are obtained in the following way\,:
\begin{equation}
\begin{CD}
\cZ~[0,1,0]@>{\mbox{$\vphantom{\f{}{}}$ \footnotesize $-\vec\al_{1}$}}>>
\cW~[1,-1,1]@>{\mbox{$\vphantom{\f{}{}}$ \footnotesize $-\vec\al_{2}$}}>> 
\cY~[-1,0,1]@.\\
	@.
	@VV{\mbox{$\vphantom{\f{}{}}$ \footnotesize $-\vec\al_{3}$}}V
	@VV{\mbox{$\vphantom{\f{}{}}$ \footnotesize $-\vec\al_{3}$}}V 
	@.\\
{}@.
\overline{\cY}~[1,0,-1]@>{\mbox{$\vphantom{\f{}{}}$ \footnotesize $-\vec\al_{2}$}}>>
\overline{\cW}~[-1,1,-1]@>{\mbox{$\vphantom{\f{}{}}$ \footnotesize $-\vec\al_{1}$}}>>
\overline{\cZ}~[0,-1,0]\,.
\end{CD}
\label{SO(6) fields}
\end{equation}
Notice that $\overline \cZ$ is not a fundamental excitation, but a composite field containing two fundamental excitations.

For any $SO(6)$ operator, the R-charge is bounded by the bare dimension from above,
\begin{equation}
\Delta_{0}\geq L=J_{1}+J_{2}+J_{3}\,.
\label{saturation}
\end{equation}
In particular, for holomorphic operators without containing $\overline{\cZ}$\,, $\overline{\cW}$ nor $\overline{\cY}$ requires $n_{3}=0$\,, in which case the relation (\ref{saturation}) is saturated.

Let us see a couple of more examples.
For a highest-weight state with two $u_{1,j}$ roots $\vec w=(L-2,2,0)$ in a length-$L$ chain, the Dynkin indices become $[2,L-4,2]$\,.
For a Konishi operator (\ref{Konishi}) corresponding to $(n_{1},n_{2},n_{3})=(2,1,1)$\,, it is given by $[0,0,0]$\,, {\em i.e.}, the Konishi operator is an $SO(6)$ singlet.

Under one-loop renormalisation, there is no operator mixing between $SO(6)$ scalar operators (\ref{SO(6) op}) and operators containing gluons or fermions.\footnote{For example, in the $SU(3)$ sector, three scalars with different flavours can mix into two fermions, preserving the bare dimension, spin and R-charge \cite{Beisert:2003ys}.
In the $SO(6)$ sector, there can be mixing into the full $SU(2,2|4)$ sector.
However, see \cite{Minahan:2004ds}, where it was argued that, even at higher loop orders, the operator mixing outside the $SU(3)$ or $SO(6)$ are suppressed in the thermodynamic limit.}
By plugging (\ref{Cartan SO(6)}) into the general formula (\ref{BAE general}), a set of nested Bethe ansatz equations for the $SO(6)$ sector is obtained\,:
\begin{align}
\ko{\f{u_{1,k}+i/2}{u_{1,k}-i/2}}^{L}
&=\prod_{\mbox{$j=1\atop j\neq k$}}^{n_{1}}\f{u_{1,k}-u_{1,j}+i}{u_{1,k}-u_{1,j}-i}
	\prod_{j=1}^{n_{2}}\f{u_{1,k}-u_{2,j}-i/2}{u_{1,k}-u_{2,j}+i/2}
	\prod_{j=1}^{n_{3}}\f{u_{1,k}-u_{3,j}-i/2}{u_{1,k}-u_{3,j}+i/2}\,,\\
1&=\prod_{\mbox{$j=1\atop j\neq k$}}^{n_{2}}\f{u_{2,k}-u_{2,j}+i}{u_{2,k}-u_{2,j}-i}
	\prod_{j=1}^{n_{1}}\f{u_{2,k}-u_{1,j}-i/2}{u_{2,k}-u_{1,j}+i/2}\,,\\
1&=\prod_{\mbox{$j=1\atop j\neq k$}}^{n_{3}}\f{u_{3,k}-u_{3,j}+i}{u_{3,k}-u_{3,j}-i}
	\prod_{j=1}^{n_{1}}\f{u_{3,k}-u_{1,j}-i/2}{u_{3,k}-u_{1,j}+i/2}\,.
\end{align}
Special cases $[0,J,0]$\,, $[0,J-J',0]$ ($J\geq J'$) and $[J'-J,0,J'+J]$ ($J\leq J'$), corresponding to $(J,0,0)$\,, $(J,J',J')$ and $(J',J',J)$ respectively, are studied in \cite{Engquist:2003rn}.

\subsection[The $PSU(2,2|4)$ sector at one-loop]
	{The \bmt{PSU(2,2|4)} sector at one-loop\label{sec:full one-loop}}

As we saw in Section \ref{sec:N=4 SCFT}, the full global symmetry of the $\cN=4$ SYM theory is $PSU(2,2|4)$\,.
In \cite{Beisert:2003yb}, Beisert and Staudacher constructed the complete one-loop Bethe ansatz equations for the full $PSU(2,2|4)$ sector,\footnote{The conjectured all-order asymptotic Bethe ansatz for $PSU(2,2|4)$ super spin-chain will be discussed later in Section \ref{sec:full S-matrix}.} by combining the $SO(6)$ result of Minahan and Zarembo \cite{Minahan:2002ve} and earlier results in QCD \cite{Lipatov:1993yb,Faddeev:1994zg,Braun:1998id,Braun:1999te,Belitsky:1999bf}.
They restrict the complete one-loop dilatation operator of $\cN=4$ SYM worked out by Beisert \cite{Beisert:2003jj} to the planar sector, and obtained the corresponding $SU(2,2|4)$ super spin-chain.

There are seven types of roots in the complete super spin-chain.
In contrast to a classical semi-simple algebra, in such a superalgebra case, their choice is not unique.
For a particular choice of Dynkin diagram,
\newsavebox{\boxSdynkin}
\sbox{\boxSdynkin}{\includegraphics{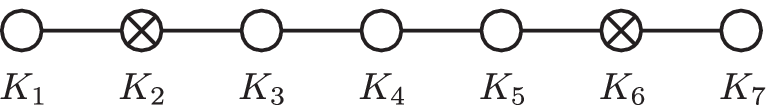}}
\newlength{\bwSdynkin}
\settowidth{\bwSdynkin}{\usebox{\boxSdynkin}} 
\newsavebox{\boxDynkinSO}
\sbox{\boxDynkinSO}{\includegraphics{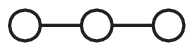}}
\newlength{\bwDynkinSO}
\settowidth{\bwDynkinSO}{\usebox{\boxDynkinSO}} 
\newsavebox{\boxDynkinF}
\sbox{\boxDynkinF}{\includegraphics{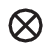}}
\newlength{\bwDynkinF}
\settowidth{\bwDynkinF}{\usebox{\boxDynkinF}} 
\newsavebox{\boxDynkinB}
\sbox{\boxDynkinB}{\includegraphics{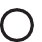}}
\newlength{\bwDynkinB}
\settowidth{\bwDynkinB}{\usebox{\boxDynkinB}} 
\begin{center}
\vspace{.5cm}
\parbox{\bwSdynkin}{\usebox{\boxSdynkin}}\,,
\vspace{.5cm}
\end{center}
where $0\leq K_{1}\leq K_{2}\leq K_{3}\leq K_{4}\geq K_{5}\geq K_{6}\geq K_{7}\geq 0$\,,
\label{1-loop K}
the Cartan matrix and the highest weight vector are given by \cite{Beisert:2003yb}
\begin{equation}
M_{\al\be}=
\mbox{\footnotesize $\left(
\begin{array}{c|c|ccc|c|c}
-2 & 1 &  &  &  &  &  \\
\hline
1 & 0 & -1 &  &  &  &  \\
\hline
& -1 & 2 & -1 &  &  &  \\
&  & -1 & 2 & -1 &  &  \\
&  &  & -1 & 2 & -1 &  \\
\hline
&  &  &  & -1 & 0 & 1 \\
\hline
&  &  &  &  & 1 & -2
\end{array}
\right)$}\,,\qquad 
q_{\al}=
\mbox{\footnotesize $\left(\begin{array}{c}
0 \\
\hline
0 \\
\hline
0 \\
1 \\
0 \\
\hline
0 \\
\hline
0\end{array}\right)$}\,.
\end{equation}
The bosonic roots $k=1,3,4,5,7$ (\parbox{\bwDynkinB}{\usebox{\boxDynkinB}}) of the same flavour repulse each other, and the fermionic roots $k=2,6$ (\parbox{\bwDynkinF}{\usebox{\boxDynkinF}}) do not feel each other.
Any roots of adjacent flavours attract each other, and form boundstates called stacks\,: $u_{\al,k}-u_{\al,k+1}\sim \cO(L^{0})$\,.
Stacks containing $u_{\al,4}$ form states that carry momentum and energy, while other states are ``auxiliary'' (they merely change flavours of excitations).

The $SU(2)$ sector is represented by only exciting the central node, and the $M$\,-magnon state just corresponds to $K_{k}=\delta_{k4}M$\,.
In addition to the bosonic $SU(2)$ sector (\parbox{\bwDynkinB}{\usebox{\boxDynkinB}}), there are two other rank-one sectors in the $\cN=4$ SYM theory\,: the fermionic $SU(1|1)$ sector
which consists of operators of the form $\tr(\cZ^{L-M}\Psi^{M})+\mbox{permutations}$\,, and the derivative $SL(2)$ sector
which consists of $\tr(D^{M}\cZ^{L})+\mbox{permutations}$.
Here $D$ is the adjoint light-cone covariant derivative.
Actually it can be shown \cite{Staudacher:2004tk} that the $SU(1|1)$ and $SL(2)$ representation can be further reduced to \parbox{\bwDynkinF}{\usebox{\boxDynkinF}} with $q=0$ and \parbox{\bwDynkinB}{\usebox{\boxDynkinB}} with $q=-1$\,, which means, the corresponding one-loop Bethe ansatz equations for each sector are given by,
\begin{alignat}{5}
&\ko{\f{u_{k}+i/2}{u_{k}-i/2}}^{L}=-1\,,&\quad& k=1,\dots,M\,,&\qquad &\mbox{for~~$SU(1|1)$}\,,\label{1-loop BAE : u SU(1|1)}\\[2mm]
&\ko{\f{u_{k}+i/2}{u_{k}-i/2}}^{L}=\prod_{\mbox{$j=1\atop j\neq k$}}^{M}\f{u_{k}-u_{j}-i}{u_{k}-u_{j}+i}\,,&\quad& k=1,\dots,M\,,&\qquad &\mbox{for~~$SL(2)$}\,.\label{1-loop BAE : u SL(2)}
\end{alignat}
Notice the $SU(1|1)$ S-matrix describes free motion of fermions with $p_{k}=2\pi n_{k}/L$\,, and the $SL(2)$ S-matirx is just the reciprocal of the one for the $SU(2)$ (Heisenberg spin-chain), (\ref{1-loop BAE : u}).\footnote{For the all-loop conjecture for these three rank-one sectors, see (\ref{all-loop rank-1}) in Chapter \ref{chap:dressing}.}
Furthermore, the $SO(6)$ sector can be read off as \parbox{\bwDynkinSO}{\usebox{\boxDynkinSO}}\,, which corresponds to the $3\times 3$ matrix in the centre of the Cartan matrix.\footnote{Notice the numbers $n_{1}$\,, $n_{2}$ and $n_{3}$ of the simple roots $\al_{1}$\,, $\al_{2}$ and $\al_{3}$ of the $SO(6)$ representation we saw in Section \ref{sec:SO(6) one-loop} correspond to $K_{4}$\,, $K_{3}$ and $K_{5}$\,, respectively in the $SU(4)$ representation used here.}

The quantum numbers (\ref{CFT spectrum}) are related to the excitation mode numbers $K_{1},\dots,K_{7}$ as
\begin{eqnarray}
& \ds \Delta=L+\half K_{2}+\half K_{6}+2g^{2}\sum_{k=1}^{K_{4}}\f{1}{u_{4,k}^{2}+1/4}\,,&{}\no\\[2mm]
&S_{1}=\half K_{2} - K_{1}\,,\qquad 
S_{2}=\half K_{6} - K_{7}\,;\qquad 
J_{1}=L-K_{4}+\half K_{2}+\half K_{6}\,,&{}\no\\
&J_{2}=K_{4}-K_{3}-K_{5}+\half K_{2}+\half K_{6}\,,\qquad 
J_{3}=K_{3}-K_{5}-\half K_{2}+\half K_{6}\,.&{}\no
\end{eqnarray}
In particular, the state with all vanishing mode numbers $(K_{1},\dots, K_{7})=(0,\dots,0)$ represents the BPS vacuum $\tr(\cZ^{L})$ with $\Delta=L$ \,.

\section{Thermodynamic Limit of SYM Spin-Chain\label{sec:thermodynamic}}

When the rapidities acquire imaginary parts, they can form boundstates corresponding to classical spin-waves.
Those boundstates are often referred to as {\em Bethe strings}.
There is a finite-size correction in the real direction towards the imaginary axis, and the further away the root locates from the real line, the correction becomes larger.
Hence the string is bending outward about the origin, see the left diagram of Figure \ref{fig:scaling-limit}.
We are going to solve the Bethe equation (\ref{1-loop BAE : u}) in the {\em ``thermodynamic'' (``scaling'') limit}, where
\begin{equation}
L\to \infty\,,\quad M\to \infty\,,\quad \mbox{while}\quad \al_{\rm g}\eq\f{M}{L}\,:\, \mbox{fixed.}\quad \ko{0\leq \al_{\rm g}< \mbox{\large $\f{1}{2}$}}
\label{thermodynamic}
\end{equation}
Here $\al_{\rm g}$ is a parameter which measures the proportion of impurities $\cW$ to the background fields $\cZ$\,, and called filling fraction.
The thermodynamic limit simplifies the problem of solving the Bethe ansatz equations technically, reducing the discrete form to the integral form.
Also, this limit is needed to compare with the semiclassical string results of \cite{Beisert:2003xu,Beisert:2003ea}.

\subsection{Thermodynamic limit of one-loop BAE}

The thermodynamic limit (\ref{thermodynamic}) is reached by first taking the logarithm of (\ref{1-loop BAE : u}) and (\ref{momentum cond : u}),
\begin{align}
L\ln \ko{\f{u_{k}+i/2}{u_{k}-i/2}}-2\pi i n_{k}
=\sum_{\mbox{$j=1\atop j\neq k$}}^{M}\ln \ko{\f{u_{k}-u_{j}+i}{u_{k}-u_{j}-i}}\,,\quad k=1,\dots,M\,.
\label{log BAE}
\end{align}
Here $n_{k}$ are integers corresponding to the branches of the log.
Since $p\approx 2\pi n/L\sim 1/L$ (for fixed $n$), the relation (\ref{def : rap}) implies $u\sim 1/p\sim L$ in the thermodynamic limit.
This motivates us to define a rescaled variable $x\eq u/L$\,, with which (\ref{log BAE}) is rewritten as
%
\begin{figure}[t]
\begin{center}
\vspace{.3cm}
\includegraphics[scale=1.0]{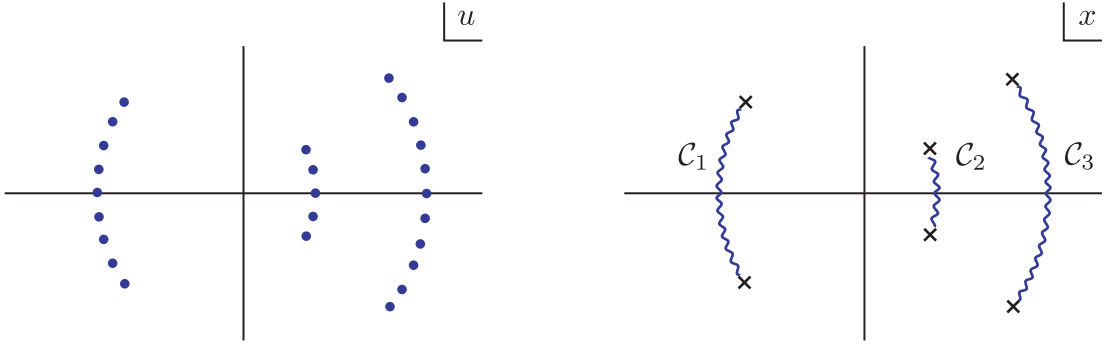}
\vspace{.3cm}
\caption{\small Before (left) and after (right) taking the scaling limit $L\to \infty$, where $u=Lx$\,.}
\label{fig:scaling-limit}
\end{center}
\end{figure}
%
\begin{equation}
\f{1}{x_{k}}-2\pi n_{k}=\f{2}{L}\sum_{\mbox{$j=1\atop j\neq k$}}^{M}\f{1}{x_{k}-x_{j}}\,,\quad k=1,\dots,M\,.
\label{BAE delta}
\end{equation}
Introducing the Bethe root density function $\rho_{\rm g}(x)$ whose support is given by $\cC\eq \cC_{1}\cup\dots\cup\cC_{K}$\,,
\begin{equation}
\rho_{\rm g}(x)\eq \f{1}{L}\sum_{k=1}^{M}\delta(x-x_{k})\,,\qquad 
\int_{\cC}dx\, \rho_{\rm g}(x)=\sum_{j=1}^{K}\int_{\cC_{j}}dx\, \rho_{\rm g}(x)=\f{M}{L}=\al_{\rm g}\,,
\end{equation}
the Bethe ansatz equations can be further translated into the following integral form,
\begin{equation}
\f{1}{x}-2\pi n_{k}=2\pint_{\cC}dy\,\f{\rho_{\rm g}(y)}{x-y}\,,\quad x\in \cC_{k}\,.
\label{int BAE 1}
\end{equation}
See the right diagram of Figure \ref{fig:scaling-limit}.
The symbol $\spint$ in (\ref{int BAE 1}) is to be understood in the principal part sense, reflecting that we needed to omit the
$j= k$ piece in (\ref{1-loop BAE : u}).
Those roots belonging to the same contour $\cC_{j}$ have the same mode number $n_{j}$\,.

In this thermodynamic limit, the total momentum $P\eq \sum_{k=1}^{M}p_{k}$ and the spin-chain energy are also expressed in integral forms\,; (\ref{1-loop BAE}), (\ref{momentum cond}) and (\ref{1-loop DR}) become, respectively,
\begin{align}
&\f{1}{x}-2\pi n_{j}=2\pint_{\cC}dy\,\f{\rho_{\rm g}(y)}{x-y}\,,\quad x\in \cC_{j}\,,
	\label{1-loop BAE : x}\\[2mm]
&P=\int_{\cC}dx\,\f{\rho_{\rm g}(x)}{x}=2\pi m\,,\quad m\in\mathbb Z\,,
	\label{momentum cond : x}\\[2mm]
&\ga=\f{\lam}{8\pi^{2}L}\int_{\cC}dx\,\f{\rho_{\rm g}(x)}{x^{2}}\,.
	\label{1-loop DR : x}
\end{align}
It is also convenient to introduce the resolvent defined by
\begin{equation}
G_{\rm g}(x)\eq \f{1}{L}\sum_{j=1}^{M}\f{1}{x-x_{j}}
=\int_{\cC}dy\,\f{\rho_{\rm g}(y)}{x-y}\,.
\end{equation}
In terms of the resolvent, (\ref{1-loop BAE : x})\,-\,(\ref{1-loop DR : x}) can be expressed as
\begin{align}
&\f{1}{x}-2\pi n_{j}=2\Gs{}_{\rm \hspace{-.1mm}g}(x)\,,\quad x\in \cC_{j}\,,
	\label{1-loop BAE : x 2}\\[2mm]
&P=-\oint_{\cC}\f{dx}{2\pi i}\,\f{G_{\rm g}(x)}{x}
	=-G_{\rm g}(0)=2\pi m\,,\quad m\in\mathbb Z\,,
	\label{momentum cond : x 2}\\[2mm]
&\ga=- \f{\lam}{8\pi^{2}L}\oint_{\cC}\f{dx}{2\pi i}\,\f{G_{\rm g}(x)}{x^{2}}
	=\left.- \f{\lam}{8\pi^{2}L}\f{dG_{\rm g}(x)}{dx}\right|_{x=0}\,.
	\label{1-loop DR : x 2}
\end{align}
Here $\Gs{}_{\rm \hspace{-.1mm}g}(x)\eq G_{\rm g}(x\pm i0)\pm i\pi \delta(x)$\,, where $\pm i0$ refer to before and after the cut.
By Taylor-expanding the resolvent around $x=0$\,, independent $L$ conserved charges are generated.
In general, the $n$-th charge is expressed by the resolvent as
\begin{equation}
\mathcal Q_{n}\eq -\f{1}{L^{n}}\left.\f{d^{n}G_{\rm g}(x)}{dx^{n}}\right|_{x=0}\,.
\end{equation}
In particular, $P\eq \mathcal Q_{0}$ and $\ga\eq \mbox{\large $\f{\lam}{8\pi^{2}}$}\mathcal Q_{1}$\,.

\paragraph{}
In \cite{Beisert:2003xu,Beisert:2003ea}, the one- and two- cut solutions (rational and elliptic solutions, respectively) at the one-loop level are studied and compared with semiclassical folded/circular strings of \cite{Frolov:2003qc,Frolov:2003xy}.
For the rational solutions, the comparison was performed to the two-loops in \cite{Kazakov:2004qf}, and for the elliptic solutions, it was performed to the three-loops in \cite{Serban:2004jf}.
We will demonstrate the comparison up to the one-loop for the rational strings, and to the three-loop for the elliptic strings, later in Section \ref{sec:discrepancy}.
For that purpose, below we will explicitly solve the corresponding set of the Bethe ansatz equations.
We will obtain the energy expression of the form
\begin{equation}
\Delta(\ag;\tlambda)=L+\ga(\ag;\tlambda)\quad \mbox{with} \quad 
\ga(\ag;\tlambda)=L\kko{\tlambda\,\delta_{1}\ko{\ag}+\tlambda^{2}\,\delta_{2}\ko{\ag}+\dots} \,.
\label{Delta expansion}
\end{equation}
The energy coefficients $\delta_{k}$ are to be compared with the energy coefficients $\ep_{k}$ of the corresponding string solutions, by identifying $\ag$ with $\as$\,, where $\as$ is the ``spin-fraction'' of the spinning string.

\subsection{Rational solutions at one-loop order\label{sec:rational 1-loop}}

Let us start compute the anomalous dimension for the rational (or the one-cut\,:\,$K=1$) case, following the trick used in \cite{Lubcke:2004dg}.
By setting $n_{k}=n$ in (\ref{BAE delta}), the rescaled Bethe ansatz equation becomes
\begin{equation}
\f{1}{x_{k}}-2\pi n=\f{2}{L}\sum_{\mbox{$j=1\atop j\neq k$}}^{M}\f{1}{x_{k}-x_{j}}\,,\quad k=1,\dots,M\,.
\label{BAE delta 1-cut}
\end{equation}
By multiplying both sides of (\ref{BAE delta 1-cut}) by $1/(x-x_{k})$ and summing over $k$\,, one obtains
\begin{equation}
G_{\rm g}(x)^{2}+\f{1}{L}\,G_{\rm g}'(x)=\ko{\f{1}{x}-2\pi n}G_{\rm g}(x)+2\pi m\,,
\label{G(x)-LZ}
\end{equation}
where the momentum condition
\begin{equation}
P=-G_{\rm g}(0)=\f{1}{L}\sum_{j=1}^{M}\f{1}{x_{j}}=2\pi m
\label{P2}
\end{equation}
was taken into account. 
This equation can be solved order by order in $1/L$\,, using the following expansion of the resolvent,
\begin{equation}
G_{\rm g}(x)=G_{\rm g}^{(0)}(x)+\hf{1}{L}\,G_{\rm g}^{(1)}(x)+\cO(\hf{1}{L^{2}})\,.
\end{equation}
At the leading order, (\ref{G(x)-LZ}) reduces to a quadratic equation, $xG_{\rm g}^{(0)}(x)^{2}-\ko{1-2\pi n x}G_{\rm g}^{(0)}(x)-2\pi m=0$\,.
Of the two roots of this equation, we should choose
\begin{equation}
G_{\rm g}^{(0)}(x)=\f{1}{2x}\ko{1-2\pi nx+\sqrt{(1-2\pi nx)^{2}+8\pi n\ag x}}
\label{G^{(0)}}
\end{equation}
so that the resolvent has the asymptotic behavior $G_{\rm g}(x)\sim \hf{1}{L}\cdot \hf{M}{x}=\hf{\ag}{x}$ in $x\to \infty$\,.
Finally, by plugging (\ref{G^{(0)}}) into the energy formula (\ref{1-loop DR : x 2}), one obtains the anomalous dimension for this one-cut solutions in the leading order in $1/L$ as
\begin{equation}
\ga=\left.- \f{\lam}{8\pi^{2}L}\f{dG_{\rm g}^{(0)}(x)}{dx}\right|_{x=0}
=\f{n^{2}\ag (1-\ag)\lam}{2L}\,,
\label{rational 1-loop}
\end{equation}
that is, in light of (\ref{Delta expansion}), the leading energy coefficient $\delta_{1}$ is given by
\begin{equation}
\delta_{1}=\f{1}{2}\,n^{2}\ag (1-\ag)=\f{1}{2}\,m(n-m)\,.
\label{delta1 rational}
\end{equation}
We have used $\ag=m/n$\,, since `$\sum_{k=1}^{M}(\ref{BAE delta 1-cut})$' together with (\ref{P2}) implies $mL=nM$\,.

It is straightforward to include the contribution of the first finite-size correction $\hf{1}{L}\,G_{\rm g}^{(1)}(x)$\,.
The result is given by just multiplying (\ref{rational 1-loop}) by a factor of $\big(1+\hf{1}{L}\big)$\,, that is, $\delta_{1}=\delta_{11}=\hf{1}{2}\,n^{2}\ag (1-\ag)$ for this one-cut solution (the notation $\delta_{kn}$ is defined in (\ref{double expansion-Delta})).

It is also straightforward to generalise the analysis to higher loops, albeit a bit tedious.
The two- and the three-loop pieces $\delta_{2}$ and $\delta_{3}$ were obtained in \cite{Kazakov:2004qf} and \cite{Minahan:2004ds}, respectively.

\subsection{Elliptic solutions at three-loop order\label{sec:elliptic 3-loop}}

Next let us solve the three-loop extended version of the set of equations (\ref{1-loop BAE : x 2}\,-\,\ref{1-loop DR : x 2}) for the elliptic (or the two-cut\,:\,$K=2$) cases.
We saw in Section \ref{sec:higher SU(2)} that the higher-loop dilatation operator (spin-chain Hamiltonian) is given by (\ref{SU(2) 2-loop}), (\ref{SU(2) 3-loop}).
The BDS model is consistent with them up to the three-loop order by construction.
Let us expand the BDS rapidity (\ref{rap}) in powers of $g^{2}$ and take the thermodynamic limit $L\to \infty$ with the BMN coupling $\sim g^{2}/L^{2}$ kept fixed.
This leads the LHS of the Bethe equation (\ref{BDS bethe eq}) to the following form,
\begin{equation}
e^{i p_{j}}\sim \f{1}{x_{j}}+\f{2\tilde g^{2}}{x_{j}^{3}}+\f{6\tilde g^{4}}{x_{j}^{5}}+\cO(\tilde g^{6})\,,
\end{equation}
where we defined $\mbox{\large $\f{1}{2}$}\cot\mbox{\large $\ko{\f{p}{2}}$}\eq L x$ and $16\pi^{2}\tilde g^{2}\eq \lam/L^{2}$\,.
In the thermodynamic limit, the integral Bethe ansatz equations and the energy formula (\ref{1-loop DR : x}) are corrected to, up to the three-loop order,
\begin{align}
&\f{1}{x}+\f{2\tilde g^{2}}{x^{3}}+\f{6\tilde g^{4}}{x^{5}}-2\pi n_{j}=2\Gs{}_{\rm \hspace{-.1mm}g}(x)\,,\quad x\in \cC_{j}\,,\\[2mm]
&\f{\gamma}{L}=-\oint_{\cC}\f{d  x}{2\pi i }\,G_{\rm g}\ko{x}\left(
\f{2\tilde g^{2}}{x^{2}}+
 \f{6\tilde g^{4}}{x^{4}}+
  \f{20\tilde g^{6}}{x^{6}}+
  \cO(\tilde g^{8})
   \right)\,.\label{anomalous_dimension}
\end{align}
As we saw for the one-loop case, in order to compute the energy density $\ga/L$\,, we have only to find out the form of $G_{\rm g}(x)$ expanded around $x=0$\,.
It is convenient to introduce a so-called {\em quasi-momentum},\footnote{Not to be confused with the momentum of a magnon.}
\begin{equation}\label{quasi-momentum}
p_{\rm g}\ko{x}=G_{\rm g}\ko{x}-\f{1}{2x}-\f{\tilde g^{2}}{x^{3}}-\f{3\tilde g^{4}}{x^{5}}+\cO(\tilde g^{6})\,.
\end{equation}
It is an Abelian integral for the meromorphic differential $dp_{\rm g}$ which has integer periods on a hyperelliptic curve $\Sigma$\,.

\subsubsection*{General \bmt{K}-cut case}

First let us consider the most general $K$-cut case, in which the elliptic curve is given by
\begin{align}\label{Riemann_surface K}
\Sigma\,:~y^{2}&=x^{2K}+c_{1}x^{2K-1}+\dots+c_{2K-1}x+c_{2K}=\prod_{j=1}^{2K}(x-x_{j})\,.
\end{align}
In order for the charges computed from the curve to be real, the branch-points $x_{1},\dots,x_{2K}$ in (\ref{Riemann_surface K}) must satisfy a so-called reality condition, that is, the set $\{ x_{j} \}_{j=1,\dots,2K}$ and its complex conjugate $\{ x_{j}^{*} \}_{j=1,\dots,2K}$ must be the same.
However, for the moment we relax the reality condition, and consider the most general case.
Let us expand the differential as
\begin{equation}
d p_{\rm g}=\f{d  x}{y}\sum_{k=-5}^{K-1}\f{a_{k}}{x^{-k+1}}\,.
\label{coeff:a}
\end{equation}

\begin{figure}[t]
\begin{center}
\vspace{.3cm}
\includegraphics[scale=0.9]{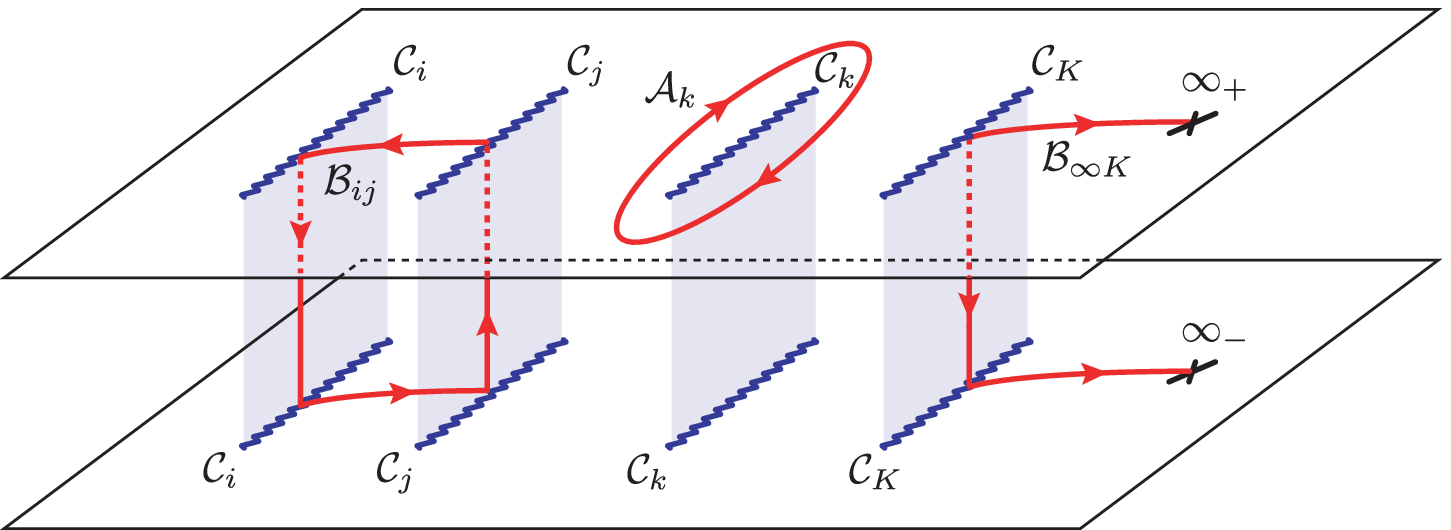}
\vspace{.3cm}
\caption{\small Definitions of $\cA$\,- and $\cB$\,-cycles.}
\label{fig:cycles}
\end{center}
\end{figure}

There are two kinds of cycles defined for the hyperelliptic curve\,; they are the $\cB$\,-cycle and the $\cA$\,-cycle.
See Figure \ref{fig:cycles} for the definition of the cycles relative to the two cuts.
Their periods give integer moduli to the solutions.
The $\cB$\,-cycle conditions are given by
\begin{gather}
\begin{split} 
&\oint_{\cB_{kK}}dp_{\rm g}=2\pi(n_{k}-n_{K})\,,\qquad (k=1,\dots,K-1)\,,\\[2mm]
&\oint_{\infty_{-}}^{\infty_{+}}dp_{\rm g}=p_{\rm g}(\infty_{+})-p_{\rm g}(\infty_{-})=2\pi n_{K}\,,
\end{split} 
\label{B cycle}
\end{gather}
where $\infty_{+}$ and $\infty_{-}$ mean points $x=\infty$ on the upper (physical) and the lower (unphysical) sheet, respectively.
Here $\cB_{ij}$ is a $\cB$\,-cycle that goes across the cuts $\cC_{i}$ and $\cC_{j}$\,.
In addition to the two square root branch cuts, if there are condensate cuts as well, the value of $p_{\rm g}(x)$ jumps by $2\pi m_{k}$ ($m_{k}\in \mathbb Z$) when it goes across the $\cA_{k}$\,-cycle.
Thus we have the conditions
\begin{equation}
\oint_{\cA_{k}}dp_{\rm g}=2\pi m_{k}\,.\qquad (k=1,\dots K-1)\,.
\label{A cycle}
\end{equation}

Let us count the numbers of free parameters and see how they can be fixed.
In general $K$-cut case, the elliptic Riemann surface (\ref{Riemann_surface}) is characterised by $2K$ coefficients in which $2K$ branch-points are encoded.
In addition, the differential $dp_{\rm g}$ has $K+5$ free parameters $a_{k}$\,.
Hence, there are $3K+5$ free parameters to be fixed in total.
In our three-loop computation, the first six coefficients $a_{0},\dots, a_{-5}$ are fixed by the conditions that the integration of (\ref{coeff:a}) matches up to (\ref{quasi-momentum}).
The $\cB$\,-cycle conditions (\ref{B cycle}) and the $\cA$\,-cycle conditions (\ref{A cycle}) kill $K+(K-1)$ degrees of freedom.
The rest $K$ degrees of freedom are fixed by the $K$ inputs for the partial filling fractions for each cut, $\al_{j}$ ($j=1,\dots,K$), defined by
\begin{gather}
\begin{split} 
\al_{k}&=\f{1}{2\pi i}\oint_{\cA_{k}}dx \, p_{\rm g}(x)=\int_{\cC_{k}}dx \, \rho_{\rm g}(x)\,,\qquad (k=1,\dots,K-1)\,,\\[2mm]
\al_{K}&=\f{1}{2\pi i}\oint_{\cC_{K}}dx \, p_{\rm g}(x)=\int_{\cC_{K}}dx \, \rho_{\rm g}(x)\qquad \mbox{with}\quad \al_{\rm g}=\sum_{j=1}^{K}\al_{j}\,.
\end{split}
\end{gather}
From the normalisation condition, the degree of freedom of the total filling fraction $\al_{\rm g}$ is directly related to $a_{1}$ as
\begin{equation}
a_{1}=\f{1}{2}-\ag\,.
\label{a1}
\end{equation}
In this setup, the Bethe ansatz equation is rewritten in the form of a Riemann-Hilbert problem defining the quasi-momentum $p_{\rm g}(x)$ from its discontinuity 
\begin{equation}
\ps{}_{\rm \hspace{-.1mm}g}(x)=2\pi n_{j}\,,\quad x\in \cC_{j}
\label{RH 3-loop}
\end{equation}
on every cut $\cC_{j}$\,.
So what one should do in order to compute the anomalous dimension at given $\ag$ is to solve the Riemann-Hilbert problem (\ref{RH 3-loop}) to find out the set of coefficients $\{c_{i}\}$ that satisfy all the cycle conditions, then plug it into the energy formula (\ref{ene 3-loop}).

\paragraph{}
Let us illustrate how the above procedure works by taking the one-cut ($K=1$) solution at the one-loop as a simple example.
In this case, the differential can be just set as
\begin{equation}
dp_{\rm g}=\f{dx}{\sqrt{x^{2}+c_{1}x+c_{2}}}\ko{\f{a_{-1}}{x^{2}}+\f{a_{0}}{x}}\,.
\label{quasi-momentum 1C1L}
\end{equation}
The condition that the integration of (\ref{quasi-momentum 1C1L}) matches up to (\ref{quasi-momentum}) (at the one-loop level) fixes $a_{-1}$ and $a_{0}$ in terms of $c_{1}$ and $c_{2}$ as
\begin{equation}
a_{-1}=\f{\sqrt{c_{2}}}{2}\,,\qquad 
a_{0}=\f{c_{1}}{4\sqrt{c_{2}}}\,.
\end{equation}
Then the resolvent is obtained as
\begin{equation}
G_{\rm g}(x)=p_{\rm g}(x)+\f{1}{2x}
=-\f{c_{1}}{4c_{2}}+\ko{-\f{1}{4c_{2}}+\f{c_{1}^{2}}{16c_{2}^{2}}}x\,.
\end{equation}
By plugging it into (\ref{anomalous_dimension}), the general one-loop anomalous dimension formula is found to be
\begin{equation}
\f{\gamma}{L}=-2\tilde g^{2}\ko{\f{c_{1}^{2}}{16c_{2}^{2}}-\f{1}{4c_{2}}}\,.
\label{ene 1C1L}
\end{equation}
Let us apply the formula to derive the one-loop energy of the one-cut solution investigated in Section \ref{sec:rational 1-loop}.
From (\ref{G^{(0)}}), one finds the curve $\Sigma$ for the one-cut solution is given by
\begin{equation}
y^{2}=\f{1}{(2\pi n)^{2}}\kko{ (1-2\pi nx)^{2}+8\pi n\ag x}
=x^{2}-\f{1-2\ag}{\pi n}\,x+\f{1}{4\pi^{2}n^{2}}\,,
\end{equation}
from which one can read off the coefficients in (\ref{Riemann_surface}) as 
\begin{equation}
c_{1}=-\f{1-2\ag}{\pi n}\,,\qquad 
c_{2}=\f{1}{4\pi^{2}n^{2}}\,.
\end{equation}
By plugging these coefficients into (\ref{ene 1C1L}), one can reproduce the one-loop anomalous dimension (\ref{rational 1-loop}) as expected.

\subsubsection*{General two-cut (elliptic) case}

Now let us particularly concentrate on the elliptic case, $K=2$\,.
The hyperelliptic curve reduces to an elliptic curve with two cuts, 
\begin{align}\label{Riemann_surface}
\begin{split}
\Sigma\,:~y^{2}&=x^{4}+c_{1}x^{3}+c_{2}x^{2}+c_{3}x+c_{4}\\
&=(x-x_{1})(x-x_{2})(x-x_{3})(x-x_{4})\,.
\end{split}
\end{align}
As explained, the coefficients $a_{0},\dots, a_{-5}$ are fixed by imposing the condition that the integration of (\ref{coeff:a}) matches up to (\ref{quasi-momentum}).
The results are a bit lengthy and we collect it in Appendix \ref{app:a_{-5}-a_{0}}.
Putting all into (\ref{anomalous_dimension}), we obtain the anomalous dimension formula at the three-loop order, 
\begin{align}
\frac{\gamma}{L} &=  -2{\tilde g}^{2}\left( \frac{{{c_3^2}}}{16{{c_4^2}}} - 
     \frac{{c_2}}{4{c_4}} + 
     \frac{{a_1}}{{\sqrt{{c_4}}}} \right)  \no\\[2mm]
   &\hspace{0.5cm}{}- 2{\tilde g}^{4}\left( \frac{45 {{c_3^4}}}
      {256 {{c_4^4}}} - 
     \frac{27 {c_2} {{c_3^2}}}
      {32 {{c_4^3}}} + 
     \frac{3 {a_1} {{c_3^2}}}
      {8 {{c_4}}^{{5/2}}} + 
     \frac{9 {{c_2^2}}}{16 {{c_4^2}}} + 
     \frac{{c_1} {c_3}}{{{c_4^2}}} - 
     \frac{{a_1} {c_2}}
      {2 {{c_4}}^{{3/2}}} - 
     \frac{7}{4 {c_4}} \right) \no\\[2mm]
   &\hspace{0.5cm}{}- 4{\tilde g}^{6}\left( \frac{285 {{c_3}}^6}
      {1024 {{c_4}}^6} - 
     \frac{467 {c_2} {{c_3^4}}}
      {256 {{c_4}}^5} + 
     \frac{35 {a_1} {{c_3^4}}}
      {128 {{c_4}}^{{9/2}}} + 
     \frac{195 {{c_2^2}} {{c_3^2}}}
      {64 {{c_4^4}}} + 
     \frac{2{c_1} {{c_3^3}}}{ {{c_4^4}}}- 
     \frac{15 {a_1} {c_2} {{c_3^2}}}
      {16 {{c_4}}^{{7/2}}} - 
     \frac{13 {{c_2^3}}}{16 {{c_4^3}}} \right.\no\\[2mm]
     &\hspace{1.0cm}\left.{}- 
     \frac{9 {c_1} {c_2} {c_3}}
      {2 {{c_4^3}}}  - 
     \frac{39 {{c_3^2}}}{16 {{c_4^3}}} + 
     \frac{3 {a_1} {{c_2^2}}}
      {8 {{c_4}}^{{5/2}}} + 
     \frac{3 {a_1} {c_1} {c_3}}
      {4 {{c_4}}^{{5/2}}} + 
     \frac{21 {{c_1^2}}}{16 {{c_4^2}}} + 
     \frac{3 {c_2}}{{{c_4^2}}} - 
     \frac{{a_1}}{2 {{c_4}}^{{3/2}}}
     \right) +\O({\tilde g}^{8})\,.
\label{ene 3-loop}
\end{align}

\begin{figure}[t]
\begin{center}
\vspace{.3cm}
\includegraphics[scale=1.0]{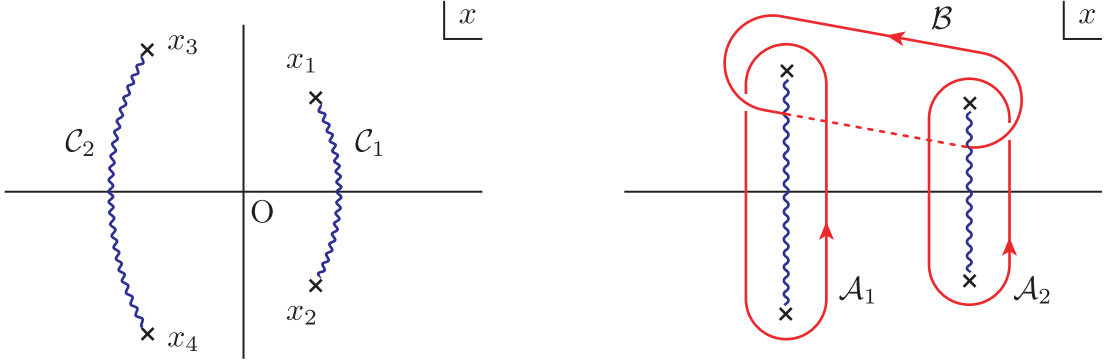}
\vspace{.3cm}
\caption{\small General two-cut configuration and the definitions of the cycles.}
\label{fig:2cuts}
\end{center}
\end{figure}

\subsubsection*{Symmetric two-cut case}

Let us further restrict our attention to symmetric two-cut cases.
Without loss of generality we can set $(x_{1},x_{2},x_{3},x_{4})$ as $(a,-a,b,-b)$ with $a=b^{*}$\,, so that they satisfy the reality condition.
The Riemann surface (\ref{Riemann_surface}) is then given by
\begin{equation}\label{symmetric_Riemann_surface}
y^{2}=\ko{x^{2}-a^{2}}\ko{x^{2}-b^{2}}\,,
\end{equation}
with two cuts $[a,b]$ and $[-b,-a]$ on a double cover of the complex plane.
The differential (\ref{coeff:a}) now reduces to
\begin{align}
d p_{\rm g}&=-\f{d  x}{\sqrt{\ko{b^{2}-x^{2}}\ko{x^{2}-a^{2}}}}\left[ \f{1}{2}-\ag-\f{ab}{2x^{2}}
+6{\tilde g}^{2}\ko{-\f{ab}{x^{4}}+\f{1}{2x^{2}}\f{a^{2}+b^{2}}{ab}}\right.\no\\[2mm]
&\hspace{5.0cm}\left.{}+ 15{\tilde g}^{4}\ko{-\f{ab}{x^{6}}+\f{1}{2x^{4}}\f{a^{2}+b^{2}}{ab}+\f{1}{8x^{2}}\f{\ko{a^{2}-b^{2}}^{2}}{a^{3}b^{3}} } \right]\,.
\end{align}
The periods of $d p_{\rm g}$ can be expressed through the complete integrals of the first and the second kind.\footnote{For the definition of the complete elliptic integrals, see Appendix \ref{app:Elliptic}.}
We assign the mode number $n$ and $-n$ to the two cuts, normalising the $\cB$-period of $d p_{\rm g}$ to $4\pi n$\,.  We also put a condensate of density $m$ between the cuts, giving the $\cA$-period $2\pi m$\,, {\em i.e.},
\begin{equation}
\oint_{\cA} d p_{\rm g}=2\pi m\,,\quad \qquad \oint_{\cB} d p_{\rm g}=4\pi n\,.
\label{peropds mn}
\end{equation}
Converting them into the standard Legendre form with the help of the elliptic
integral formulae listed in Appendix \ref{app:Elliptic Functions}, we obtain
\begin{align}
2\pi i m&=\f{1}{br}\left[ \ko{1-2\ag}r\eK(1-r)-\eE(1-r)
\right]+\f{{\tilde g}^{2}}{b^{3}r^{3}}\left[
2r^{2}\eK(1-r) -(1+r^{2})\eE(1-r)\right]\no\\
&\hspace{1.0cm}{}+\f{3{\tilde g}^{4}}{4b^{5}r^{5}}
\left[4r^{2}(1+r^{2})\eK(1-r)
-\ko{3+2r^{2}+3r^{4}}\eE(1-r)\right]\,,\label{A-cycles}
\end{align}
for the $\cA$\,-cycle, and 
\begin{align}
2\pi n&=\f{1}{br}\left[ \ko{1-2\ag}r\eK(r)-\eE(r)-\eK(r)
\right]+\f{{\tilde g}^{2}}{b^{3}r^{3}}\left[
(1+r^{2})\eE(r) -(1-r^{2})\eK(r)\right]\no\\
&\hspace{1.0cm}{}+\f{3{\tilde g}^{4}}{4b^{5}r^{5}}
\left[\ko{3+2r^{2}+3r^{4}}\eE(r)
-\ko{3-2r^{2}-r^{4}}\eK(r)\right]\,, \label{B-cycles}
\end{align}
for the $\cB$\,-cycle.
Here the moduli $r^{2}$ is defined by
\begin{equation}
r^{2}=\f{a^{2}}{b^{2}}\,. 
\end{equation}
Expanding both the moduli $r^{2}$ and one of the endpoints $b$ 
in powers of $\tilde g^{2}$\,:
\begin{align}
&r^{2}(\ag;\tilde g)=r^{2}_{(0)}\ko{\ag}+\tilde g^{2}r^{2}_{(1)}\ko{\ag}+\tilde g^{4}r^{2}_{(2)}\ko{\ag}+\cO(\tilde g^{6})\,,
\label{r expansion}\\
&b(\ag;\tilde g)=b_{0}\ko{\ag}+\tilde g^{2}b_{1}\ko{\ag}+\tilde g^{4}b_{2}\ko{\ag}+\cO(\tilde g^{6})\,,
\label{b expansion}
\end{align}
and eliminating the unwanted parameters by using (\ref{A-cycles}) and (\ref{B-cycles}), one can obtain a set of equations relating $\al_{\rm g}$ and the energy coefficients with the moduli $r^{2}_{(k)}$ at each order.
From the $\O(\tilde g^{2})$ relation, we obtain
\begin{equation}
\label{fraction}
1-2\ag=\f{1}{r^{2}_{(0)}}\,\f{n\eE(1-r_{(0)})
+i m\left[ \eE(r_{(0)})-\eK(r_{(0)}) \right]}{n\eK(1-r_{(0)})-i  m\eK(r_{(0)})}\,.
\end{equation}
This relation can be viewed as defining the leading order moduli $r^{2}_{(0)}$ by the periods $(n,m)$ and $\ag$\,.
The energy density can be written down in terms of the moduli $r^{2}_{(0)}$ and the periods $m$ and $n$\,, and we display the most general result in Appendix \ref{sec:(m,n)}.

\begin{figure}[t]
\begin{center}
\vspace{.3cm}
\includegraphics[scale=1.0]{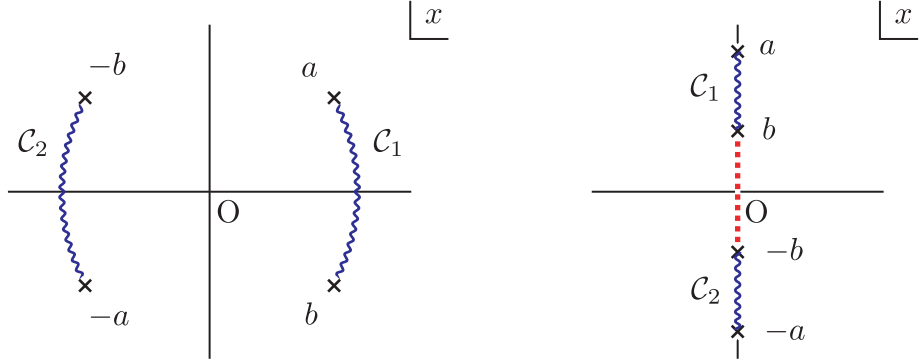}
\vspace{.3cm}
\caption{\small Symmetric two-cut solutions\,: the ``double contour'' solution (left) and the ``imaginary root'' solution (right).}
\label{fig:DC-IM}
\end{center}
\end{figure}

\subsubsection{``Double contour'' solution}
If set $m=0$\,, the so-called double-contour solution \cite{Beisert:2003xu,Beisert:2003ea} is obtained at the three-loop order.
See the left diagram in Figure \ref{fig:DC-IM}.
Before taking the thermodynamic limit, half of the roots are distributed along one of the Bethe strings $\cC_{1}$ in the region $\Re u<0$\,, while the other half $\cC_{2}$ locate at the mirror image of $\cC_{1}$\,.
The string centres of the both strings are on the real axis, and the distribution is totally symmetric with respect to the origin.
There are no condensate cuts (so that $m=0$) and the only non-zero period is of the $\cB$\,-cycle.
The energy coefficients are found by setting $m=0$ in the general result (\ref{formula}) to be
\begin{align}
\delta_1&=-\f{n^{2}}{2\pi^{2}}\,\eK\left(\eK-2\eE+\eK\,r^{2}_{(0)}\right)\,,\label{dc-1}\\[2mm]
\delta_2&=-\f{n^{4}}{8\pi^{4}}\,\eK^{3}\left[\eK-4\eE+2\ko{3\eK-2\eE}r^{2}_{(0)}+\eK\,r^{4}_{(0)}\right]\,,\label{dc-2}\\[2mm]
\delta_3&=-\f{n^{6}}{4\pi^{6}}\f{\eK^{5}}{\ko{\eE-\eK}(\eE-\eK\,r^{2}_{(0)})}\Big[
\eE^{2}\ko{\eK-3\eE}-\eE\ko{3\eK^{2}-11\eE\eK+2\eE^{2}}r^{2}_{(0)}\no\\[2mm]
&\hspace{1.0cm} {}+ \ko{4\eK^{3}-18\eE\eK^{2}+11\eE^{2}\eK-3\eE^{2}}r^{4}_{(0)}
+ \eK\ko{4\eK^{2}-3\eE\eK+\eE^{2}}r^{6}_{(0)}\Big]\,.\label{dc-3}
\end{align}
Here we introduced the shorthand notations $\eK\equiv \eK(1-r_{(0)})$ and $\eE\equiv \eE(1-r_{(0)})$\,.\footnote{This three-loop result is first obtained in \cite{Serban:2004jf}.
The moduli $r^{2}_{(0)}$ used here is related to the one $q_{0}$ used in \cite{Serban:2004jf} as $r^{2}_{(0)}=1-q_{0}$\,.}

\subsubsection{``Imaginary-root'' solution}
Another well-known two-cut solution is the so-called imaginary root solution \cite{Beisert:2003xu,Beisert:2003ea}.
Before taking the thermodynamic limit, all the roots are distributed on the imaginary axis (so the mode number $n$ is zero), symmetrically with respect to the origin.
The roots close to the real axis are separated by $i/2m$ with $m$ integer, which actually corresponds to the $\cA$\,-cycle of the Riemann-Hilbert problem after the thermodynamic limit, while those farther away from the real axis has non-trivial density functions.
Denoting the root densities for the contours $\cC_{1}$ and $\cC_{2}$ as $\sig_{1}(x)$ and $\sig_{2}(x)$\,, where $x$ is the rescaled rapidity variable $x=u/L$ as before, the density function for the total distribution is given by
\begin{equation}
\rho(x)=\rho(i\xi)=
\left\{\begin{array}{cc}
2m & \mbox{$-s<\xi<s$} \\
\sig_{1}(x) & \mbox{$s<\xi<t$}\\
\sig_{2}(x) & \mbox{$-t<\xi<-s$}\\
0 & \mbox{$|\xi|>t$} \end{array}\right.\qquad 
\mbox{with\quad $\sig_{1}(x)=\sig_{2}(-x)$}\,,
\label{IR solution}
\end{equation}
where $a=it$ and $b=is$\,.
Actually, in the limit $r^{2}=a^{2}/b^{2}=t^{2}/s^{2}\to 1$\,, when the two cuts collapse to two symmetric points on the imaginary axis, the imaginary root configuration becomes equivalent to the double contour configuration in the same limit.

Setting $n=0$ in (\ref{formula}), we obtain the three-loop anomalous dimension for the imaginary-root solution as\footnote{The moduli $r^{2}_{(0)}$ used here is related to the one $r_{0}$ used in \cite{Serban:2004jf} as $r^{2}_{(0)}=r_{0}$\,.}
\begin{align}
\hat\delta_1&=-\f{2m^{2}}{\pi^{2}}\,\hat\eK(\hat\eK-2\hat\eE-\hat\eK r^{2}_{(0)})\,,\label{im-1}\\[2mm]
\hat\delta_2&=-\f{2m^{4}}{\pi^{4}}\,\hat\eK^{3}\left[ -3\hat\eK+4\hat\eE+2(\hat\eK+2\hat\eE)r^{2}_{(0)}+\hat\eK r^{4}_{(0)} \right]\,,\label{im-2}\\[2mm]
\hat\delta_3&=-\f{16m^{6}}{\pi^{6}}\f{\hat\eK^{5}}{\hat\eE(-\hat\eK+\hat\eE+\hat\eK r^{2}_{(0)})}\Big[
(3\hat\eE-2\hat\eK)(\hat\eE-\hat\eK)^{2} +(2\hat\eE^{3}+5\hat\eE^{2}\hat\eK-13\hat\eE\hat\eK^{2}+6\hat\eK^{3}) r^{2}_{(0)} \no\\[2mm]
&\hspace{2.5cm} {}+(3\hat\eE^{3}+2\hat\eE^{2}\hat\eK+5\hat\eE\hat\eK^{2}-6\hat\eK^{3}) r^{4}_{(0)}
+ \hat\eK(\hat\eE^{2}+\hat\eE\hat\eK+2\hat\eK^{2}) r^{6}_{(0)}\Big]\,,\label{im-3}
\end{align}
where we introduced the shorthand notations $\hat\eK\equiv \eK(r_{(0)})$\,, $\hat\eE\equiv \eE(r_{(0)})$\,.

\paragraph{}
So far we have worked out the three-loop anomalous dimensions for particular SYM operators (Bethe root distributions) in the thermodynamic limit.
The inputs were the three-loop dilatation operator of (\ref{SU(2) 1-loop}\,-\,\ref{SU(2) 3-loop}), and there is no assumption about that.

The thermodynamic limit of the BDS spin-chain can be also worked out in the similar manner, which will be discussed later in Section \ref{sec:SBAE}, see (\ref{scaling limit BDS 2}) and (\ref{BA sig}).
It will be compared to the corresponding integral equations of string theory.
In \cite{Beisert:2004hm}, all-loop analyses of the folded and circular strings based on the BDS ansatz are also discussed.

\section[Appendices for Chapter \ref{chap:MZ}]{Appendices for Chapter \ref{chap:MZ}\,: Some computational formulae and results}

\subsection{Useful integral formulae\label{note:integral formula}}

Elliptic integral formulae listed below are useful in the intermediate calculation in Section \ref{sec:elliptic 3-loop}.  The moduli parameter $r^{2}$ is defined as $r^{2}=a^{2}/b^{2}$\,.  
We used the following integral formulae in the computation concerning the ${\cA}$-cycle of the $(n,m)$-elliptic solution\,:
\begin{align}
&\int_{a}^{b}\f{d  x}{\sqrt{\ko{b^{2}-x^{2}}\ko{x^{2}-a^{2}}}}=\f{1}{b}\,\eK(1-r)\,,\\[2mm]
&\int_{a}^{b}\f{d  x}{x^{2}\sqrt{\ko{b^{2}-x^{2}}\ko{x^{2}-a^{2}}}}=\f{1}{a^{2}b}\,\eE(1-r)\,,\\[2mm]
&\int_{a}^{b}\f{d  x}{x^{4}\sqrt{\ko{b^{2}-x^{2}}\ko{x^{2}-a^{2}}}}=\f{1}{3a^{4}b^{3}}\left[2\ko{a^{2}+b^{2}}\eE(1-r)-a^{2}\eK(1-r)\right]\,,\\[2mm]
&\int_{a}^{b}\f{d  x}{x^{6}\sqrt{\ko{b^{2}-x^{2}}\ko{x^{2}-a^{2}}}}=\f{1}{15a^{6}b^{5}}\left[\ko{8a^{4}+7a^{2}b^{2}+8b^{4}}\eE(1-r)\right.\no\\
&\hspace{8.0cm}\left.{}-4a^{2}\ko{a^{2}+b^{2}}\eK(1-r)\right]\,,
\end{align}
and the same for the ${\cB}$-cycle:
\begin{align}
&\int_{b}^{\infty}\f{d  x}{\sqrt{\ko{x^{2}-a^{2}}\ko{x^{2}-b^{2}}}}=\f{1}{b}\,\eK(r)\,,\\[2mm]
&\int_{b}^{\infty}\f{d  x}{x^{2}\sqrt{\ko{x^{2}-a^{2}}\ko{x^{2}-b^{2}}}}=\f{1}{a^{2}b}\left[\eK(r)-\eE(r)\right]\,,\\[2mm]
&\int_{b}^{\infty}\f{d  x}{x^{4}\sqrt{\ko{x^{2}-a^{2}}\ko{x^{2}-b^{2}}}}=\f{1}{3a^{4}b^{3}}\,\left[\ko{a^{2}+2b^{2}}\eK(r)-2\ko{a^{2}+b^{2}}\eE(r)\right]\,,\\[2mm]
&\int_{b}^{\infty}\f{d  x}{x^{6}\sqrt{\ko{x^{2}-a^{2}}\ko{x^{2}-b^{2}}}}=\f{1}{15a^{6}b^{5}}\,\left[\ko{4a^{4}+3a^{2}b^{2}+8b^{4}}\eK(r)\right.\no\\
&\hspace{8.0cm}\left.{}-\ko{8a^{4}+7a^{2}b^{2}+8b^{4}}\eE(r)\right]\,.
\end{align}

\newpage
\subsection{The coefficients in (\ref{coeff:a})\label{app:a_{-5}-a_{0}}}

We find the coefficients to be
\begin{align}
a_{0}&=\frac{3 }{32768 {\pi }^4 
      {{c_4}}^{{9/2}}}
      \left( -5 {\lambda }^2 {{c_3 ^2}} + 
        20 {\lambda }^2 {c_2} {c_4} + 
        128 {\pi }^2 \lambda  {{c_4 ^2}} \right)  
      \left( 5 {{c_3 ^3}} - 12 {c_2} {c_3} {c_4} + 
        8 {c_1} {{c_4 ^2}} \right)  \no\\[4mm]
&\hspace{1.0cm}{}+ 
   \frac{3  }{32768 {\pi }^4 {{c_4}}^{{9/2}}}
   \left( -3 {{c_3 ^2}} + 4 {c_2} {c_4} \right)
         \left( 5 {\lambda }^2 {{c_3 ^3}} - 
        20 {\lambda }^2 {c_2} {c_3} {c_4} + 
        40 {\lambda }^2 {c_1} {{c_4 ^2}} + 
        128 {\pi }^2 \lambda  {c_3} {{c_4 ^2}} \right) 
     \no\\[4mm]
&\hspace{1.0cm}{} - 
   \frac{15 {\lambda }^2 {c_3}  }{65536 {\pi }^4 
      {{c_4}}^{{9/2}}}
      \left( 35 {{c_3 ^4}} - 
        120 {c_2} {{c_3 ^2}} {c_4} + 
        96 {c_1} {c_3} {{c_4 ^2}} + 
        16 \left( 3 {{c_2 ^2}} - 4 {c_4} \right)  
         {{c_4 ^2}} \right) \no\\[4mm]
&\hspace{1.0cm}{}- 
   \frac{15 {\lambda }^2 }{65536 {\pi }^4 {{c_4}}^{{9/2}}}
      \left( -63 {{c_3}}^5 + 
        280 {c_2} {{c_3 ^3}} {c_4} - 
        240 {c_1} {{c_3 ^2}} {{c_4 ^2}} - 
        48 {c_3} \left( 5 {{c_2 ^2}} - 4 {c_4} \right)
            {{c_4 ^2}} + 192 {c_1} {c_2} {{c_4 ^3}}
        \right) \no\\[4mm]
&\hspace{1.0cm}{}
    - \frac{{c_3} }{65536 
      {\pi }^4 {{c_4}}^{{9/2}}} \left( 75 {\lambda }^2 {{c_3 ^4}} - 
        360 {\lambda }^2 {c_2} {{c_3 ^2}} {c_4} + 
        240 {\lambda }^2 {{c_2 ^2}} {{c_4 ^2}} + 
        480 {\lambda }^2 {c_1} {c_3} {{c_4 ^2}} \right.\no\\[4mm]
&\hspace{4.0cm}{}\left.{}+ 
        768 {\pi }^2 \lambda  {{c_3 ^2}} {{c_4 ^2}} - 
        960 {\lambda }^2 {{c_4 ^3}} - 
        3072 {\pi }^2 \lambda  {c_2} {{c_4 ^3}} - 
        16384 {\pi }^4 {{c_4 ^4}} \right)\,,\no\\[4mm]
a_{-1}&=  -\frac{1}{32768 
     {\pi }^4 {{c_4}}^{{7/2}}}\left( 75 {\lambda }^2 {{c_3 ^4}} - 
       360 {\lambda }^2 {c_2} {{c_3 ^2}} {c_4} + 
       240 {\lambda }^2 {{c_2 ^2}} {{c_4 ^2}} + 
       480 {\lambda }^2 {c_1} {c_3} {{c_4 ^2}} \right.\no\\[4mm]
&\hspace{4.0cm}{}\left.{} + 
       768 {\pi }^2 \lambda  {{c_3 ^2}} {{c_4 ^2}} - 
       960 {\lambda }^2 {{c_4 ^3}} - 
       3072 {\pi }^2 \lambda  {c_2} {{c_4 ^3}} - 
       16384 {\pi }^4 {{c_4 ^4}} \right) \,,\no\\[4mm]
a_{-2}&=  \frac{3}
     {4096 {\pi }^4 {{c_4}}^{{5/2}}} \left( 5 {\lambda }^2 {{c_3 ^3}} - 
       20 {\lambda }^2 {c_2} {c_3} {c_4} + 
       40 {\lambda }^2 {c_1} {{c_4 ^2}} + 
       128 {\pi }^2 \lambda  {c_3} {{c_4 ^2}} \right) \,,\no\\[4mm]
a_{-3}&=  \frac{3 }{2048 {\pi }^4 {{c_4}}^{{3/2}}}\left( -5 {\lambda }^2 {{c_3 ^2}} + 20 {\lambda }^2 {c_2} {c_4} + 128 {\pi }^2 \lambda  {{c_4 ^2}} \right) \,,\no\\[4mm]
a_{-4}&=  \frac{15 {\lambda }^2 {c_3}}{512 {\pi }^4 {\sqrt{{c_4}}}}\,,\no\\[4mm]
 a_{-5}&= \frac{15 {\lambda }^2 {\sqrt{{c_4}}}}{256 {\pi }^4}\,.\nonumber 
\end{align}

\newpage
\subsection{Anomalous dimension formula for generic \bmt{(m,n)}-solution\label{sec:(m,n)}}

Below we display the general formula for the elliptic solutions with periods $(n,m)$ (normalised as (\ref{peropds mn})) at the three-loop\,:
\begin{align}\label{formula}
\f{\gamma}{L}&=-8{\tilde g}^{2}( n\widetilde{\rm \bf K}  -i  m {\rm \bf K})\big[ -2n \widetilde{\rm \bf E} -2i  m {\rm \bf E} +n(1+r^{2}_{(0)}) \widetilde{\rm \bf K} +i  m (1-r^{2}_{(0)}){\rm \bf K}\big]\no\\[3mm]
&\quad {}-32{\tilde g}^{4}(n \widetilde{\rm \bf K} -i  m {\rm \bf K})^3\big[ -4n(1+r^{2}_{(0)})\widetilde{\rm \bf E}-4i  m(1+r^{2}_{(0)}){\rm \bf E}\no\\[2mm]
&\quad \hspace{3.8cm}{}+n(1+6r^{2}_{(0)}+r^{4}_{(0)})\widetilde{\rm \bf K}+i  m(3-2r^{2}_{(0)}-r^{4}_{(0)}){\rm \bf K}\big]\no\\[3mm]
&\quad {}-1024{\tilde g}^{6}{( n {\widetilde {\rm \bf K}} - i m {\rm \bf K}) }^5\Big\{ -{\big[ n {\widetilde {\rm \bf E}} + i    m ( {\rm \bf E} - {\rm \bf K} ) \big] }^2 ( 3 n {\widetilde {\rm \bf E}} +  3 i m {\rm \bf E} - n {\widetilde {\rm \bf K}} -  2 i m {\rm \bf K} )\no\\[2mm]
&\quad \hspace{.6cm}{}-\big[ n {\widetilde {\rm \bf E}} +  i m ( {\rm \bf E} - {\rm \bf K} )  \big] \big[ 2 n^2 {{\widetilde {\rm \bf E}}}^2 - 2 m^2 {{\rm \bf E}}^2 + 3 n^2 {{\widetilde {\rm \bf K}}}^2 +  5 i m n {\widetilde {\rm \bf K}} {\rm \bf K} + 6 m^2 {{\rm \bf K}}^2 \no\\[2mm]
&\quad \hspace{4.5cm}{}- m {\rm \bf E} ( 11 i n  {\widetilde {\rm \bf K}} + 7 m {\rm \bf K} ) +  i n {\widetilde {\rm \bf E}} ( 4 m {\rm \bf E} + 11 i n  {\widetilde {\rm \bf K}} + 7 m {\rm \bf K} ) \big] {r^{2}_{(0)}} \no\\[2mm]
&\quad \hspace{.3cm}{}+ \big[ -3 n^3 {{\widetilde {\rm \bf E}}}^3 +  3 i     m^3  {{\rm \bf E}}^3 +  4 n^3 {{\widetilde {\rm \bf K}}}^3 + 6 i  m n^2  {{\widetilde {\rm \bf K}}}^2  {\rm \bf K}+ 13 m^2 n {\widetilde {\rm \bf K}}  {{\rm \bf K}}^2 -  6 i    m^3 {{\rm \bf K}}^3\no\\[2mm]
&\quad \hspace{1.0cm}{}+   n^2 {{\widetilde {\rm \bf E}}}^2 (  -9 i m {\rm \bf E} + 11 n {\widetilde {\rm \bf K}} - 2 i m {\rm \bf K})\no\\[2mm]
&\quad \hspace{1.0cm}{}+ m^2 {{\rm \bf E}}^2 ( -11 n {\widetilde {\rm \bf K}} + 2 i m {\rm \bf K})  + i m {\rm \bf E}( -18 n^2 {{\widetilde {\rm \bf K}}}^2 + 14 i  m n {\widetilde {\rm \bf K}}  {\rm \bf K} + 5 m^2 {{\rm \bf K}}^2 ) \no\\[2mm]
&\quad \hspace{1.0cm}{}+ n {\widetilde {\rm \bf E}} ( 9 m^2 {{\rm \bf E}}^2 - 18 n^2 {{\widetilde {\rm \bf K}}}^2 + 14 i m n {\widetilde {\rm \bf K}} {\rm \bf K} + 5 m^2 {{\rm \bf K}}^2 + 2 m {\rm \bf E} (  11 i  n {\widetilde {\rm \bf K}} + 2 m {\rm \bf K} ) )\big]  r^{4}_{(0)}\no\\[2mm]
&\quad \hspace{.3cm}{}+ ( n {\widetilde {\rm \bf K}} -  i    m {\rm \bf K} ) \big[ n^2 {{\widetilde {\rm \bf E}}}^2 - m^2 {{\rm \bf E}}^2 + 4 n^2 {{\widetilde {\rm \bf K}}}^2  - 5 i   m n {\widetilde {\rm \bf K}} {\rm \bf K} - 2 m^2 {{\rm \bf K}}^2  \no\\[2mm]
&\quad \hspace{3.0cm}{} - m {\rm \bf E} ( 3 i n {\widetilde {\rm \bf K}} +  m {\rm \bf K} )+ i n {\widetilde {\rm \bf E}} ( 2 m {\rm \bf E} + 3 i   n {\widetilde {\rm \bf K}} + m {\rm \bf K}) \big]  r^{6}_{(0)} 
\Big\}\no\\[2mm]
&\quad \hspace{.3cm}{}\times \Big\{(n {\widetilde {\rm \bf E}} + i m {\rm \bf E} - n {\widetilde {\rm \bf K}} ) [ n {\widetilde {\rm \bf E}} + i m ( {\rm \bf E} - {\rm \bf K} )  + ( - n {\widetilde {\rm \bf K}} +  i m {\rm \bf K} )  r^{2}_{(0)}]\Big\}^{-1}
+\O(\tilde g^{8})\,.
\end{align}
Here for simplicity we denoted $\eK(r_{(0)})$\,, $\eE(r_{(0)})$\,, $\eK(1-r_{(0)})$ and $\eE(1-r_{(0)})$ as ${\rm \bf K}$\,, ${\rm \bf E}$\,, $\widetilde{\rm \bf K}$ and $\widetilde{\rm \bf E}$\,, respectively.

\part[Integrability in Classical String Theory]
	{Integrability in Classical String Theory\label{part:string}}

\chapter[Integrable Sigma Models from $AdS_{5}\times S^{5}$ Strings]
	{Integrable Sigma Models from \bmt{AdS_{5}\times S^{5}} Strings\label{chap:FT}}

\section[String sigma model on $AdS_{5}\times S^{5}$]
	{String sigma model on \bmt{AdS_{5}\times S^{5}}\label{sec:string sigma}}

\subsubsection*{Action and equations of motion}

The type IIB string theory on $AdS_{5}\times S^{5}$ is described as a non-linear sigma model on the supercoset $\mbox{\large $\f{PSU(2,2|4)}{{SO(1,4)}\times {SO(5)}}$}$\,.
In this thesis, we will mainly consider its bosonic part,
\begin{equation}
\f{{SU}(2,2)\times {SU}(4)}{{SO(1,4)}\times {SO(5)}}\cong
\f{{SO(2,4)}}{{SO(1,4)}}\times \f{{SO(6)}}{{SO(5)}}\,.
\end{equation}
Note that the group $SO(2,4)\times SO(6)$ is the isometry group of $AdS_{5}\times S^{5}$\,.
The type IIB superstring action with the full $PSU(2,2|4)$ symmetry is formulated by Metsaev and Tseytlin in \cite{Metsaev:1998it}.
The bosonic part of it is given by
\begin{equation}\label{action}
{\mathcal I}_{\rm bosonic} = \f{\sqrt \lambda}{2\pi}  \int d\tau \int_{0}^{2\pi} d\sigma \left( {\cL_{{AdS}_5 }  + \cL_{S^5 } } \right)\,,\\
\end{equation}
with
\begin{subequations}
\begin{align}
 \cL_{{AdS}_5 } &=  - \frac{1}{2}\,\ga^{ab} \eta ^{PQ} \partial _a Y_P \partial _b Y_Q   + \frac{1}{2}\,\widetilde \Lambda \left( {\eta ^{PQ} Y_P Y_Q   + 1} \right)\,,\label{AdS-Lagrangian}\\[2mm]
  \cL_{S^5 } &=  - \frac{1}{2}\,\ga^{ab} \delta^{MN}\partial _a X_M \partial _b X_N   + \frac{1}{2}\,\Lambda \left( {\delta^{MN}X_M X_N   - 1} \right)\,,
  \label{S-Lagrangian}
\end{align}
\end{subequations}
where $Y_{0,\dots,5}\left( {\tau ,\sigma } \right)$ and $X_{1,\dots,6}\left( {\tau ,\sigma } \right)$ are the embedding coordinates for $AdS_{5}$ and $S^{5}$ coordinates, respectively.
The worldsheet metric is taken as $\ga^{ab}=\left( { - , + } \right)$\,, and the target space metrics are $\eta ^{PQ} = \left( { - , + , + , + , + , - } \right)$ for the $AdS_{5}$ and $\delta ^{MN} = \left( { + , + , + , + , + , + } \right)$ for the $S^{5}$\,.
Two auxiliary fields $\widetilde \Lambda\left( {\tau ,\sigma } \right)$ and $\Lambda\left( {\tau ,\sigma } \right)$ are Lagrange multipliers to impose the sigma model constraints $\eta^{PQ}Y_PY_Q =-1$ and $X_M X_M =1$\,.
We are interested in closed string states, so we should impose periodic boundary conditions
\begin{equation}
Y_P\ko{\tau, \sigma+2\pi}=Y_P\ko{\tau, \sigma},\qquad 
X_M\ko{\tau, \sigma+2\pi}=X_M\ko{\tau, \sigma}.\label{PBC}
\end{equation}
The equations of motion for $Y_P$ and $X_M$ follow from the action (\ref{action}) as
\begin{equation}
(\partial^a\partial_a +\widetilde \Lambda)Y_P=0\,,
\qquad (\partial^a\partial_a +\Lambda)X_M=0\,.
\label{eom}
\end{equation}
It also follows that $\widetilde \Lambda  =  - \eta ^{PQ} \partial ^a Y_P \partial _a Y_Q $ and $\Lambda  = \partial ^a X_M \partial _a X_M $ for the auxiliary fields.
The classical energy-momentum tensor vanishes, $0 = \cT_{ab}\eq \delta {\mathcal I}_{{\rm bosonic}} / \delta \ga^{ab}$\,, which leads to the following set of Virasoro constraints\,:
\begin{align}
  0&= \cT_{\tau\tau}+\cT_{\sig\sig}=\eta ^{PQ} \left( {\partial _\tau  Y_P \partial _\tau  Y_Q   + \partial _\sigma  Y_P \partial _\sigma  Y_Q  } \right) + \partial _\tau  X_M \partial _\tau  X_M   + \partial _\sigma  X_M \partial _\sigma  X_M \,, \label{conf.g.c.1}\\
  0&= \cT_{\tau\sig}+\cT_{\sig\tau}=\eta ^{PQ} \partial _\tau  Y_P \partial _\sigma  Y_Q   + \partial _\tau  X_M \partial _\sigma  X_M \,,\label{conf.g.c.2}
\end{align}
with $\eta ^{PQ} Y_P Y_Q  =-1$ and $X_M X_M =1$\,.
Thus the $AdS_{5}$ and $S^{5}$ parts of the action are coupled at the classical level.

\subsubsection*{Global coordinates and global charges}

The action (\ref{action}) has ${SO}(2,4)\times {SO}(6)$ global symmetry.
In the global coordinates, $AdS_{5}\times S^{5}$ metric is written as $ds^2 _{\left( {AdS_5 }\times {\rm S}_{5} \right)}=ds^2 _{\left( {AdS_5 } \right)}+ds^2 _{\left(S_{5} \right)}$\,, where
\begin{align}
  ds^2 _{\left( {AdS_5 } \right)}  &=  {d\rho ^2  - \cosh ^2 \rho \,dt^2  + \sinh ^2 \rho \left( {d\gamma  ^2  + \cos ^2 \gamma  \,d\phi _1 ^2  + \sin ^2 \gamma  \,d\phi _2 ^2 } \right)} \,, \label{metric-AdS}\\ 
  ds^2 _{\left( {S^5 } \right)}  &={d\psi   ^2  + \cos ^2 \psi   \,d\varphi _3 ^2  + \sin ^2 \psi   \left( {d\theta^2  + \cos ^2 \theta~d\varphi _1 ^2  + \sin ^2 \psi \,d\varphi _2 ^2 } \right)} \,. \label{metric-S}
\end{align}
Throughout this thesis, we take the following global coordinates\,:\footnote{We must use the covering space of $AdS_{5}$ to avoid the AdS time $t$ becoming periodic.}
\begin{alignat}{3}
  \eta _1  &= Y_1  + i Y_2  \quad &~~
  \eta _2  &= Y_3  + i Y_4  \quad &~~
  \eta _0  &= Y_5  + i Y_0   \quad \cr
  {} \quad  &= \sinh \rho \,\cos \gamma  \,e ^{i \phi _1 } \,,\quad &~~
  {} \quad  &= \sinh \rho \,\sin \gamma  \,e ^{i \phi _2 } \,,\quad &~~
  {} \quad  &= \cosh \rho \,e ^{i t} \,;\label{AdS-coord}\quad \\[2mm]
  \xi _1  &= X_1  + i X_2  \quad &~~
  \xi _2  &= X_3  + i X_4  \quad &~~
  \xi _3  &= X_5  + i X_6  \quad \cr
  {}  \quad &= \sin \psi   \,\cos \theta\,e ^{i \varphi _1 } \,,\quad &~~
  {}  \quad &= \sin \psi   \,\sin \theta\,e^{i \varphi _2 } \,,\quad &~~
  {}  \quad &= \cos \psi   \,e^{i \varphi _3 } \,.\label{S-coord}
\end{alignat}
All the global charges are defined as the N\"other charges associated with shifts of the angular variables.
The associated Noether currents are given by
\begin{equation}
j_{PQ}^a= \sqrt{\lambda } (Y_P \partial^a Y_Q  - Y_Q  \partial^a Y_P)\,,\qquad 
j_{MN}^a= \sqrt{\lambda } (X_M \partial^a X_N  - X_N  \partial^a X_M)
\end{equation}
and are conserved, $\partial_a j_{PQ}^a=\partial_a j_{MN}^a=0$\,.
Correspondingly, the Noether charges are given by
\begin{align}
S_{PQ}&= \int_0^{2\pi} \frac{ d\sigma}{2\pi}\, j_{PQ}^{a=0}=\sqrt{\lambda } \int_0^{2\pi} \frac{ d\sigma}{2\pi}\ko{Y_P \partial_\tau Y_Q   -  Y_Q  \partial_\tau Y_P}\,,\label{AdS-charges}\\[2mm]
J_{MN}&= \int_0^{2\pi} \frac{ d\sigma}{2\pi}\, j_{MN}^{a=0}=\sqrt{\lambda } \int_0^{2\pi} \frac{ d\sigma}{2\pi}\ko{X_M \partial_\tau X_N   -  X_N  \partial_\tau X_M}\,.\label{S-charges}
\end{align}
Conventionally, let us define
\begin{alignat}{3}
 E &\equiv S_{50} ,\qquad &\quad S_1  &\equiv S_{12} ,\qquad &\quad S_2  &\equiv S_{34} \, ;\label{AdS-nonzero_charges}\\
 J_1 &\equiv J_{12} ,\qquad &\quad J_2  &\equiv J_{34} ,\qquad &\quad J_3  &\equiv J_{56}\, .\label{S-nonzero_charges}
\end{alignat}
For later purpose, we also introduce rescaled global charges which do not depend on $\lam$\,, and write them in calligraphic style\,: $E=\sqrt \lambda  \, \cE$\,, $S_{1,2} = \sqrt \lambda  \, \cS_{1,2}$ and $J_{1,2,3} = \sqrt \lambda  \, \cJ_{1,2,3}$\,.

In the global coordinates (\ref{AdS-coord}, \ref{S-coord}), the charges are interpreted as follows\,: $E$ is the target space energy which generates the shift of the AdS-time $t$\,; $S_{1,2}$ and $J_{1,2,3}$ are the $AdS_{5}$ and the $S^{5}$ spins which generate the shifts of angular variables $\phi_{1,2}$ and $\varphi_{1,2,3}$\,, respectively.
They are the $3+3$ Cartan generators of ${SO}(2,4)\times {SO}(6)$\,, which is the isometry of $AdS_{5}\times S^{5}$\,.
Actually they are related to the $SO(2,4)$ conformal charges of SYM we saw in (\ref{conformal P}\,-\,\ref{conformal K}) as
\begin{equation}
M_{\mu\nu}=S_{\mu\nu}\,,\quad 
K_{\mu}=S_{\mu 4}+S_{\mu 5}\,,\quad 
P_{\mu}=S_{\mu 4}-S_{\mu 5}\,,\quad 
\fD=S_{45}\,.
\label{charge correspondence}
\end{equation}
In particular, string energy corresponds to $E=\f{1}{2}(K_{0}+P_{0})$\,.\footnote{The combination $\f{1}{2}(K_{0}+P_{0})$ turns to $\fD$ after Euclideanisation of the AdS time coordinate in conformal mapping from $\mathbb R\times S^{3}$ to $\mathbb R^{4}$\,, see below (\ref{metric AdS-3}) in Introduction.}
In summary, the spectrum is characterised by
\begin{equation}
\{ E\,;\, S_{1}\,,\, S_{2}\, ; J_{1}\,,\, J_{2}\,,\, J_{3} \}\,.
\label{AdS spectrum}
\end{equation}
The AdS/CFT conjecture states the exact matching of the spectrum (\ref{AdS spectrum}) with the one for gauge theory (\ref{CFT spectrum}).

\section{Spinning/rotating strings\label{sec:rot-string-ansatz}}

\subsection{Rotating string ansatz\label{sec:rot-string-ansatz2}}

We are interested in obtaining (semi)classical string solutions which carry large spins on $AdS_{5}\times S^{5}$ so that all quantum sigma model corrections ($\al'$) are suppressed as we saw below (\ref{double expansion}).
For that purpose, we impose a so-called {\em rotating string ansatz} on the sigma model solution.
The simplest rotating string is a point-particle circulating around one of the great circle of $S^{5}$ follows from the ansatz,
\begin{alignat}{6}
\mbox{$AdS_{5}$\,:}&\qquad &
\rho&=0\,,\qquad &
\gamma &=0\,,\qquad &
\phi_1&=0\,,\qquad &
\phi_2&=0\,,\qquad &
t&=\kappa\tau\,;\label{SU(2) ansatz AdS}\\
\mbox{$S^{5}$\,:}&\qquad &
\psi  &=\f{\pi}{2}\,,\qquad &
\theta&=0\,,\qquad &
\varphi_1&=0\,,\qquad &
\varphi_2&=0\,,\qquad &
\varphi_3&=w\tau\,.\label{BPS ansatz}
\end{alignat}
The global charges are trivially computed as $E=\sqrt{\lam}\,\kappa$\,, $J_{1}=J_{2}=0$ and $J_{3}=\sqrt{\lam}\,\kappa$\,, and saturate the BPS relation $E=J$\,.
In order to find more non-trivial rotating solutions, we make the following ansatz 
\begin{alignat}{5}
\xi _i\ko{\tau,\sigma} &= r_i\ko{\sigma}\,\exp\kkko{i \varphi_{i}(\tau,\sig)}&&=r_i\ko{\sigma}\,\exp\kkko{i \kko{ w_i\tau+\alpha_i\ko{\sigma} }}\,,&\quad &\ko{i=1, \, 2, \, 3}\,,\label{rotating_string_ansatz}\\
\eta _r\ko{\tau,\sigma} &= s_r\ko{\sigma}\,\exp\kkko{i \phi_{r}(\tau,\sig)}&&=s_r\ko{\sigma}\,\exp\kkko{i \kko{ \om_r\tau+\beta_r\ko{\sigma} }}\,,&\quad &\ko{r=0, \, 1, \, 2}\label{rotating_string_ansatz AdS}
\end{alignat}
with $r_i,\, s_{r}\in \mathbb R$\,.
In the new coordinates, the sigma model constraints become $\delta^{ij} r_{i}r_{j}  = 1$ and $\eta^{rs}s_{r}s_{s}  = -1$ with $\delta^{ij}=(+,+,+)$ and $\eta^{rs}=(-,+,+)$\,.

This ansatz implies $(i)$ the string is rigid ($|\xi_{i}|$ and $|\eta_{r}|$ depend only on $\sig$ and not on $\tau$), and $(ii)$ the phases $\varphi_{1,2,3}$ and $\phi_{1,2}$ depend on $\ko{\tau,\sigma}$ only through the form $\varphi _i\ko{\tau,\sigma}=w_i\tau+\alpha_i\ko{\sigma}$\,, $\phi _r\ko{\tau,\sigma}=\om_{r}\tau+\beta_r\ko{\sigma}$\,.
The two constants $w_i$ and $\om_{r}$ play the role of constant angular velocities.
The periodic boundary conditions (\ref{PBC}) are rewritten as
\begin{alignat}{5}
&r_i\ko{\sigma+2\pi}&=r_i\ko{\sigma}\,,
&\qquad &\alpha_i\ko{\sigma+2\pi}&=\alpha_i\ko{\sigma}+2\pi m_i&\quad &\ko{m_{i}\in\mathbb Z}\,,\label{PBC for_alpha}\\
&s_{r}\ko{\sigma+2\pi}&=s_r\ko{\sigma}\,,
&\qquad &\beta_{r}\ko{\sigma+2\pi}&=\beta_{r}\ko{\sigma}+2\pi k_{r}&\quad &\ko{k_{0}=0\,; ~ k_{1,2}\in\mathbb Z}\,.\label{PBC for_alpha AdS}
\end{alignat}
Here $m_i$ and $k_{r}$ are the winding numbers along $\varphi _i$ and $\beta_{r}$ directions respectively.
Note that, however, the winding number in the time direction must be zero, $k_{0}=0$\,.

Under the rotating string ansatz (\ref{rotating_string_ansatz}), the AdS energy $E\eq S_{0}$\,, the $AdS_{5}$ spins (\ref{AdS-nonzero_charges}) and the $S^{5}$ spins (\ref{S-nonzero_charges}) become
\begin{alignat}{2}
  S_{r}&=\sqrt{\lambda}~\cS_{r},\qquad &\cS_{r}  &= \om_{r} \int_0^{2\pi } {\frac{{{d}\sigma }}{{2\pi }}\, s_{r}^2 \left( \sigma  \right)}\,,\qquad r=1,\,2\,,
  \label{S1}\\[2mm]
  J_i&=\sqrt{\lambda}~\cJ_i,\qquad &\cJ_i  &= w_i \int_0^{2\pi } {\frac{{{d}\sigma }}{{2\pi }}\, r_i^2 \left( \sigma  \right)}\,,\qquad i=1,\,2,\,3\,.
  \label{J1}
\end{alignat}
The sigma model constraints $\delta^{ij} r_{i}r_{j}  = 1$ and $\eta^{rs}s_{r}s_{s}  = -1$ can be rewritten in terms of the charges and angular velocities as
\begin{equation}
\frac{\cJ_1}{w_1}+\frac{\cJ_2}{w_2}+\frac{\cJ_3}{w_3}=1\,,\qquad 
\f{\cE}{\kappa}-\f{\cS_{1}}{\om_{1}}-\f{\cS_{2}}{\om_{2}}=1\,,
\label{rot.str.ans sigma model constraint}
\end{equation}
where we renamed $\om_{0}$ as $\kappa$\,.

\subsection{Mapping to Neumann-Rosochatius model}

As shown by Bena, Polchinski and Roiban, the $AdS_{5}\times S^{5}$ string sigma model admits an infinite number of local and non-local conserved currents \cite{Bena:2003wd}.
This can be seen in various ways, see for example the review by Tseytlin \cite{Tseytlin:2003ii} and the references therein.
In \cite{Arutyunov:2003uj,Arutyunov:2003za}, the authors were able to provide an explicit way of constructing generic solutions with required properties (periodic, finite-energy etc.).
It was shown that under the rotating string ansatze (\ref{rotating_string_ansatz}) and (\ref{rotating_string_ansatz AdS}), one can reduce the $AdS_{5}\times S^{5}$ string sigma model to a well-known one-dimensional integrable model called Neumann-Rosochatius model, thus giving an explicit way to understand the integrability.

The Lagrangian under the rotating string ansatz is cast into
\begin{align}
\cL_{(AdS_{5}\times S^{5})}\Big|_{\mbox{\scriptsize (\ref{rotating_string_ansatz})+(\ref{rotating_string_ansatz AdS})}}&=\f{1}{2}\kko{\delta^{ij}\ko{r_{i}'r_{j}'-w_{i}^{2}r_{i}r_{j}-\f{v_{i}v_{j}}{r_{i}r_{j}}}-\Lam\komoji{\ko{\delta^{ij}r_{i}r_{j}-1}}}+{}\no\\[2mm]
&\quad {}+\f{1}{2}\kko{\eta^{rs}\ko{s_{r}'s_{s}'-\om_{r}^{2}s_{r}s_{s}-\f{u_{r}u_{s}}{s_{r}s_{s}}}-\widetilde\Lam\komoji{\ko{\eta^{rs}s_{r}s_{s}+1}}}\,,
\label{NR}
\end{align}
where $v_{i}\eq r_{i}^{2}\al_{i}'$ ($i=1,2,3$) and $u_{r}\eq s_{r}^{2}\be_{r}'$ ($r=0,1,2$) are integrals of motion (however, notice $u_{0}=0$ due to $k_{0}=0$ as previously pointed out).
In the special case $v_{i}=0$ or $u_{r}=0$\,, the corresponding (generalised) Neumann-Rosochatius system reduces to the so-called $n=3$ Neumann system, which describes the system of a harmonic oscillator on a two-sphere.
Actually, each ($O(6)$ or $O(2,4)$) Neumann-Rosochatius system can be viewed as a special case of $n=6$ Neuman system describing an oscillator on a five-sphere, from which the integrability of the Lagrangian (\ref{NR}) directly follows.

The Neumann-Rosochatius system has only a few commuting integrals of motion, which merely specify the topology of the solution.
By contrast, there is infinite number of hidden higher commuting charges in the original two-dimensional string sigma model.
In fact, these infinite number of charges can be constructed from the integrals of motion of Neumann-Rosochatius system \cite{Arutyunov:2003rg}.

Although the Lagrangian (\ref{NR}) is decoupled to two Neumann-Rosochatius systems, $r_{i}$ and $s_{r}$ are coupled at the classical level through the Virasoro constraints,
\begin{align}
0&=\eta^{rs}\ko{s_{r}'s_{s}'+\om_{r}^{2}s_{r}s_{s}+\f{u_{r}u_{s}}{s_{r}s_{s}}}+
\delta^{ij}\ko{r_{i}'r_{j}'+w_{i}^{2}r_{i}r_{j}+\f{v_{i}v_{j}}{r_{i}r_{j}}}\,,\label{NR Virasoro}\\
0&=\eta^{rs}\om_{r}u_{r}+\delta^{ij}w_{i}v_{i}\,.
\end{align}
Let us focus on the sphere part and consider the case $u_{r}=0$ and $r_{i}=r_{i}(\sig)$\,, which will be actually our main focus in Section \ref{sec:elliptic},
then the first Virasoro constraint (\ref{NR Virasoro}) becomes a sine-Gordon equation.
The folded and circular strings we will discuss in Section \ref{sec:elliptic} are related to the solutions of the sine-Gordon equation, so they can be also viewed as special periodic solutions of the Neumann model.
For more general solutions of Neumann-Rosochatius model, the string energy can be computed but becomes a complicated implicit function of spins and topological numbers, see \cite{Arutyunov:2003za} for details.

\subsection{Strings on plane-wave\label{sec:pp-wave}}

Let us give a lightning review of a so-called BMN string.
It is almost a point-like, BPS string as in (\ref{SU(2) ansatz AdS}) and (\ref{BPS ansatz}), but includes small quantum fluctuations around the BPS string.
The geometry seen by the BMN strings is the plane-wave geometry, which emerges as the Penrose limit of the $AdS_{5}\times S^{5}$ background.
Let us define the null coordinates,
\begin{equation}
\tilde x^{+}\eq \f{t+\varphi_{3}}{2}={\mathrm{const.}}\,,\qquad 
\tilde x^{-}\eq \f{t-\varphi_{3}}{2}\,.
\label{light-like}
\end{equation}
The null geodesic corresponds to $t=\varphi_{3}$\,.
To study the geometry near the null trajectory (\ref{light-like}), let us introduce the new coordinates
\begin{equation}
x^{+}\eq \f{\tilde x^{+}}{\mu}\,,\qquad 
x^{-}\eq \mu R^{2}\tilde x^{-}\,,\qquad 
r\eq R\rho\,,\qquad 
y\eq R\theta\,,
\end{equation}
and take the Penrose limit $R\to \infty$\,.
Then the metric becomes, neglecting $\cO(R^{-2})$ terms,
\begin{align}
ds^{2}_{(AdS_{5}\times S^{5})}~\to~
&-2dx^{+}dx^{-}-\mu^{2}\ko{r^{2}+y^{2}}(dx^{+})^{2}+dr^{2}+r^{2}(d\Omega_{3})^{2}+dy^{2}+y^{2}(d\Omega'_{3})^{2}\no\\[2mm]
=&-2dx^{+}dx^{-}-\mu^{2}\sum_{i=1}^{8}(x^{i})^{2}(dx^{+})^{2}+\sum_{i=1}^{8}(dx^{i})^{2}\eq ds^{2}_{\rm (pw)}\,,
\label{pp-wave}
\end{align}
where $d\Omega_{3}$ and $d\Omega'_{3}$ are three-spheres in $S^{5}$ and $AdS_{5}$ respectively, and we also re-defined the spacetime coordinates as $x^{P}=Y_{P}$ and $x^{4+M}=X_{M}$ ($P=1,\dots,4$\,; $M=1,\dots,4$).

This is the metric of the plane-wave geometry.
The non-vanishing self-dual R-R five-form flux is $F_{(5)}=\mu dx^{+}\wedge (dx^{1}\wedge\dots\wedge dx^{4}+dx^{5}\wedge\dots\wedge dx^{8})$\,.
Notice that in the limit $\mu\to 0$\,, the plane-wave metric (\ref{pp-wave}) reduces to that of flat Minkowski spacetime.

The momenta $p^{\pm}$ conjugate to the light-cone coordinates $x^{\pm}$ are given by
\begin{align}
2p^{-}&=i\pa_{x^{+}}=i\mu\ko{\pa_{t}+\pa_{\varphi_{3}}}=\mu (E-J)\,,\label{E-J}\\[2mm]
2p^{+}&=i\pa_{x^{-}}=\f{i}{\mu R^{2}}\ko{\pa_{t}-\pa_{\varphi_{3}}}=\f{E+J}{\mu R^{2}}\label{E+J}\,.
\end{align}
As can be seen from (\ref{E-J}), the excitation energy above the BPS state, $E-J$\,, is given by the light-cone energy $H_{\rm LC}\eq 2p^{-}$ divided by $\mu$\,.
In order to describe string theory on the plane-wave, we need to keep $p^{\pm}$ finite while taking the Penrose limit of the $AdS_{5}\times S^{5}$\,.
This means we need to consider string-states/SYM-operators with $E\simeq J\sim R^{2}\to \infty$\,, and also if we keep $g_{\rm s}$ fixed, the AdS/CFT relation $R^{2}/\al'=\sqrt{\lam}=\sqrt{g_{\rm s}N}$ implies that $J\sim \sqrt{N}$\,.
Thus we arrive at the BMN limit (\ref{BMN limit}).

\paragraph{}
The string theory can be quantised in the light-cone gauge $x^{+}(\tau,\sig)=\al' p^{+}\tau$\,, giving rise to a free, massive two dimensional theory for the transverse degrees of freedom $x^{i}$\,,
\begin{equation}
{\mathcal I}=\f{1}{2\pi \al'}\int_{-\infty}^{\infty} d\tau \int_{0}^{2\pi \al' p^{+}}d\sig
\kko{\f{1}{2}(\pa_{a}x^{i})^{2}-\f{(\mu\al' p^{+})^{2}}{2}(x^{i})^{2}+\mbox{fermions}}\,.
\end{equation}
It is exactly solvable, leading to the following famous BMN formula \cite{Berenstein:2002jq},
\begin{equation}
\f{H_{\rm LC}}{\mu}=\sum_{n=-\infty}^{\infty}N_{n}\sqrt{1+\f{n^{2}}{(\mu\al'p^{+})^{2}}}
=\sum_{n=-\infty}^{\infty}N_{n}\sqrt{1+\f{n^{2}\lam}{J^{2}}}\,.
\label{BMN sqrt formula 2}
\end{equation}
The level matching condition reads $P=\sum_{n=-\infty}^{\infty}nN_{n}=0$\,, where $N_{n}=\al_{n}^{\dagger}{}^{i}\al_{n}^{i}$ ($[\al_{n}^{i},\al_{m}^{\dagger}{}^{j}]=\delta_{nm}\delta^{ij}$) is the number of excitations with mode $n$\,.
Operators with $N_{n}=0$ for all $n\neq 0$ are chiral, while the rest are non-chiral.
As the simplest non-trivial stringy excitation above the light-cone Fock-vacuum $\kket{0,p^{+}}$\,, let us consider $\al_{n}^{\dagger}{}^{i}\al_{-n}^{\dagger}{}^{j}\kket{0,p^{+}}$ with $1\leq i,j\leq 4$\,, {\em i.e.}, two excitation modes on the sphere side.
In this case, the energy formula (\ref{BMN sqrt formula 2}) becomes
\begin{equation}
\f{H_{\rm LC}}{\mu}=2\sqrt{1+\f{n^{2}\lam}{J^{2}}}\,.
\label{BMN sqrt formula}
\end{equation}
This can be compared to the gauge theory result for the BMN operator (\ref{BMN op}) with two impurities (magnons), (\ref{BMN op ene}).
We can see exact matching (at least) at one-loop.

\subsection[String sigma model on $\mathbb R\times S^{3}$]
	{String sigma model on \bmt{\mathbb R\times S^{3}}}

In the rest of this chapter, we will mainly focus on string states on $\mathbb R\times S^{3}$\,, which is the $SU(2)$ sector of the string theory.
In this sector, the rotating string ansatz becomes (\ref{SU(2) ansatz AdS}) for the AdS side, and
\begin{alignat}{6}
&\psi =\f{\pi}{2}\,,\quad &
&\theta=\theta\ko{\tau, \sigma}\,,\quad &
&\varphi_1=\varphi_1(\tau,\sig)\,,\quad &
&\varphi_2=\varphi_2(\tau,\sig)\,,\quad &
&\varphi_3=0\,,\label{SU(2) ansatz S}
\end{alignat}
for the sphere side.
Hence the string state has only two independent spins $J_{1}$ and $J_{2}$ on $S^{5}$\,.
We find it notationally convenient to use the complex coordinates introduced in (\ref{AdS-coord}) and (\ref{S-coord}).
In terms of them, the metric on $\mathbb R\times S^{3}$ can be written as
\begin{equation}
ds^{2}_{(\mathbb R\times S^3)} = -d\eta_{0}^{2}+\abs{d\xi_{1}}^{2}+\abs{d\xi_{2}}^{2}\,,
\end{equation}
and the Polyakov action becomes
\begin{equation}
S_{\mathbb R\times S^3} = - \f{\sqrt{\lambda}}{2}\int d\tau\int\f{d\sigma}{2\pi}\kkko{
\gamma^{ab}\kko{- \, \pa_{a}\eta_{\ssp 0} \, \pa_{b}\eta_{\ssp 0}+\pa_{a}\vec\xi\cdot\pa_{b}\vec\xi^{*}} + \Lambda(|\vec\xi|^{2}-1)
}\,.   \label{RtS3_action}
\end{equation}
We take the standard conformal gauge as before.
Then denoting the energy-momentum tensor following from the action (\ref{RtS3_action}) as $\cT_{ab}$\,, the Virasoro constraints are imposed as
\begin{equation}\label{string_Virasoro}
\begin{matrix}
0&= \ds \cT_{\sig\sig}=\cT_{\tau\tau} = - \hf{1}{2} \, (\pa_{\tau}\eta_{0})^{2} - \hf{1}{2} \, (\pa_{\sigma}\eta_{0})^{2} + \hf{1}{2} \, |\pa_{\tau}\vec\xi|^{2} + \hf{1}{2} \, |\pa_{\sig}\vec\xi|^{2}\,, \\[3mm]
0&=\cT_{\tau\sig}=\cT_{\sig\tau}= \Re \pare{ \pa_{\tau}\vec\xi\cdot \pa_{\sig}\vec\xi^{*} }\,. \hfill
\end{matrix}
\end{equation}
The equations of motion that follow from \eqref{RtS3_action} are
\begin{equation}
\pa_{a}\pa^{a}\eta_{0}=0\qquad
\mbox{and}\qquad
\pa_{a}\pa^{\ssp a}\vec\xi+(\pa_{a}\vec\xi\cdot\pa^{\ssp a}\vec\xi^{*})\vec\xi = \vec 0\,.
\label{string_eom}
\end{equation}
Note that the temporal gauge we took in the ansatz (\ref{SU(2) ansatz AdS}),
\begin{equation}
t=\kappa\tau\label{t=kappatau}\,,
\quad \mbox{\em i.e.},\quad 
\eta _0= e ^{it}=e ^{i\kappa\tau}\,,
\end{equation}
already solves the first equation of motion in (\ref{string_eom}).

\subsection{Rational circular (``constant radii'') solutions\label{sec:Rational circular}}

The simplest rotating string solution in a far-from-BPS sector follows from the ansatz $r_{i}(\sig)=a_{i}$ (constant), when the lagrange multiplier $\Lambda$ in (\ref{S-Lagrangian}) is constant.
One can check the following set of variables satisfies the equations of motion (\ref{eom}) and the sigma model constraint\,:
\begin{align}
&\xi_{1}\ko{\tau,\sigma}=r_{1}\ko{\sigma}\,e^{i\varphi_{1}\ko{\tau,\sigma}}
=a_{1}\,e^{i\ko{w_{1}\tau+m_{1}\sigma}}\,,\\
&\xi_{2}\ko{\tau,\sigma}=r_{2}\ko{\sigma}\,e^{i\varphi_{2}\ko{\tau,\sigma}}
=a_{2}\,e^{i\ko{w_{2}\tau+m_{2}\sigma}}\,,\\
&\Lambda=m_{1}^{2}-w_{1}^{2}=m_{2}^{2}-w_{2}^{2}\,,\qquad 
a_{1}^{2}+a_{2}^{2}=1\,.
\end{align}
The Virasoro conditions (\ref{conf.g.c.1}) and (\ref{conf.g.c.2}) become $\kappa^{2}=2\sum\nolimits_{i = 1}^2 a_i^2w_i^2+\Lambda$ and $\sum\nolimits_{i = 1}^2 {a_i^2w_i m_i }  = 0$\,.
These relations and the sigma model constraint (\ref{rot.str.ans sigma model constraint}) are rewritten in terms of  the energy $\cE=\kappa$ and spins $\cJ_i=a_i^2w_i$ of the constant radii solution as
\begin{equation}
\cE^2=\kappa^2= 
2\sum\limits_{i = 1}^2\sqrt{m_i^2-\Lambda}\,\cJ_i+\Lambda\,,\qquad 
\sum\limits_{i = 1}^2m_i\cJ_i=0,\qquad
\sum\limits_{i = 1}^2\frac{\cJ_i}{\sqrt{m_i^2-\Lambda}}=1\,.\label{33}
\end{equation}
Solving the last equation for $\Lam =\Lam\ko{\cJ_i, m_i}$\,, then plugging it into the first equation leads to the energy expression $\cE=\cE\mbox{$\ko{\cJ_i ; m_i}$}$ as function of $S^{5}$ spins $\cJ_i$ and the winding numbers $m_i$\,.
In the large spin limit $\cJ\equiv \sum_{i=1}^2\cJ_i\gg 1$\,, this can be expanded as
\begin{align}
\cE\ko{\cJ_i ; m_i}&=\cJ+\frac{1}{2\cJ}\sum\limits_{i=1}^2 m_i^2\frac{\cJ_i}{\cJ}+\dots\,,\no\\ 
&\mbox{\em i.e.},\quad 
E\ko{J_i ; m_i}=J\ko{1+\frac{\lambda}{2J^2}\sum\limits_{i=1}^2 m_i^2\frac{J_i}{J}+\dots}\label{E(J)}\,,
\end{align}
where $J\equiv \sum_{i=1}^2J_i$\,.
Recall that for given $\cJ_i$ and $m_i$\,, only such solutions that satisfy the second equation in (\ref{33}) make sense.
Taking this into account, we finally arrive at the following large-spin expansion of the energy (recall our notation of the energy coefficients $\ep_{k}$ from (\ref{double expansion}))
\begin{equation}
E=J\ko{1-\f{\lambda}{2J^{2}}\,m_{1}m_{2}+\dots}\,,\qquad \mbox{\em i.e.},\qquad 
\ep_{1}=-\f{1}{2}\,m_{1}m_{2}=\f{1}{2}\,|m_{1}m_{2}|\,.
\label{rational circular energy}
\end{equation}
Notice that since we take $w_{i}$ (and $J_{i}$) to be positive, in light of the second relation in (\ref{33}), the winding numbers must satisfy $m_{1}m_{2}<0$\,.
The spin-fraction, which is defined by the ratio of the second spin to the total spin
\begin{equation}
\as\equiv \f{\cJ_{2}}{\cJ}=1-\f{\cJ_{1}}{\cJ}\,,
\label{spin fraction}
\end{equation}
for this rational circular solution can be computed through the second relation in (\ref{33}) as
\begin{equation}
\as=\f{m_{1}}{m_{1}-m_{2}}\,.
\label{spin fraction-rational}
\end{equation}
The results (\ref{rational circular energy}) and (\ref{spin fraction-rational}) will be compared to the gauge theory counterpart later in Section \ref{sec:1,2 agreement}.

\subsection{Elliptic folded/circular strings\label{sec:elliptic}}

We have seen that an ansatz $r_{i}=\mbox{const.}$ results in a rational string solution.
Here we consider another type of rotating/spinning strings in the $SU(2)$ sector, with ``inhomogeneous'' profile given in terms of elliptic functions.
The rotating string ansatz we make is, (\ref{SU(2) ansatz S}) for the AdS side, and
\begin{alignat}{6}
&\psi =\f{\pi}{2}\,,\qquad &
&\theta=\theta\ko{\sigma}\,,\qquad &
&\varphi_1=w_1\tau\,,\qquad &
&\varphi_2=w_2\tau\,,\qquad &
&\varphi_3=0\label{elliptic ansatz}
\end{alignat}
for the sphere side.
This ansatz describes a string that stays at the centre of the $AdS_{5}$\,, while on $S^{5}$ it is extended along a great circle in $\psi$\,-direction with a {\em rigid} profile.
The centre of mass of the string rotates along the $\varphi_1$\,-direction with angular velocity $w_1$ (spin $J_{1}$) and it also spins about the centre of mass in the $\varphi_2$\,-direction with angular velocity $w_2$ (spin $J_{2}$).
As we will see, there are two classes of such elliptic strings\,; one is of folded type and the other is of circular type.

Under the ansatz (\ref{elliptic ansatz}), the non-zero charges are computed as
\begin{equation}
\cE = \kappa\,,\qquad 
\cJ_1 =w_1\int_0^{2\pi} \f{d\sigma}{2\pi}\cos^2\theta \,,\qquad 
\cJ_2 =w_2\int_0^{2\pi} \f{d\sigma}{2\pi}\sin^2\theta\,.
\end{equation}
The Virasoro conditions for this solution become
\begin{equation}
-\kappa^2+(\theta'{}^2 +w_1^2 \cos^2\theta + w_2^2 \sin^2\theta )=0\,.
\label{Virasoro elliptic}
\end{equation}
By differentiating (\ref{Virasoro elliptic}) with respect to $\sig$\,, (or equivalently, from the equation of motion for $\theta$), we reach the sine-Gordon equation
\begin{equation}
2\theta''+\ko{w_2^2-w_1^2}\sin\ko{2\theta}=0
\label{eom sG}
\end{equation}
as advertised before, which is easily integrated to give
\begin{equation}
\theta'{}^2=\ko{w_2^2-w_1^2}\ko{C-\sin^2{\theta}}\,,
\label{def:C}
\end{equation}
where we assumed $|w_2|\geq |w_1|$\,.
Here $C$ is the integration constant, and it plays the role of the elliptic moduli that controls the topology of the rigid string\,: we will have folded string for $0\leq C\leq 1$\,, while circular strings for $1\leq C$\,.
Notice that if we think of $\sig$ as playing the role of time, the equation of motion (\ref{eom sG}) or (\ref{def:C}) describes the motion of a planar pendulum in a gravitational field.
Under this identification, the folded and the circular strings correspond to, respectively, the oscillating solutions and the circulating solutions of a pendulum, which are special periodic solutions of the Neumann model.\footnote{Another example of such correspondence between integrable pendulum systems and classical strings is found in \cite{Chen:2006bh}, where particular solutions on the Lunin-Maldacena background \cite{Lunin:2005jy} were described by the motion of a spherical pendulum.
They are special periodic solutions of the Neumann-Rosochatius system.}

\subsubsection*{Two-spin folded strings}

\begin{figure}[htbp]
\begin{center}
\vspace{0.5cm}
\includegraphics[scale=0.9]{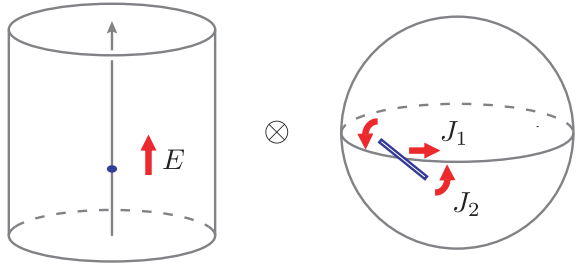}
\vspace{0.5cm}
\caption{\small Elliptic folded $(J_{1},J_{2})$-string.}
\label{fig:J1J2-folded}
\end{center}
\end{figure}

For $0\leq C\leq 1$\,, the string is folded, see Figure \ref{fig:J1J2-folded}.
It stretches over a great circle in the $\theta$-direction and spinning around its centre of mass in the $X^{3}$\,-\,$X^{4}$ plane with angular velocity $w_{2}$ (spin $J_{2}$).
The centre of mass itself moves along another orthogonal great circle in the $X^{1}$\,-\,$X^{2}$ plane with velocity $w_{1}$ (spin $J_{1}$).
Setting $C\equiv \sin^2\theta_0$ in (\ref{def:C}), we have
\begin{equation}
\theta'=\pm\sqrt{\ko{w_2^2-w_1^2}\ko{\sin^2\theta_0-\sin^2{\theta}}}\,,
\end{equation}
which describes a folded string folded in the region $[-\theta_0,\theta_0]$ in the $\theta$\,-direction.
We shall consider the case of a single fold\,; the number of folds $n$ is easy to restore at any stage.

Let us first compute $\cJ_1$ of the folded string and see its dependence on the moduli parameter $C=\sin^2\theta_0$\,.
For this purpose, we integrate the ``$+$''-branch of $\theta'$ over $0\leq \sig\leq \pi/2$ (equivalently, $0\leq \theta\leq \theta_{0}$) and multiply by four\,:
\begin{equation}
\cJ_1  =w_1\int_0^{2\pi} \f{d\sigma}{2\pi}\cos^2\theta 
= \f{2}{\pi}\f{w_1}{\sqrt{w_2^2-w_1^2}}\int_0^{\theta_0} \f{\cos^2\theta\, d\theta}{\sqrt{\sin^2\theta_0 -\sin^2\theta }}\,.
\end{equation}
By changing integration variables from $\theta$ to $\Psi$ such that $\sin\Psi\equiv \sin\theta/\sin\theta_0$\,, the integration can be performed as
\begin{equation}
\int_0^{\theta_0} \f{\cos^2\theta\, d\theta}{\sqrt{\sin^2\theta_0 -\sin^2\theta }}
=\int_0^{\pi/2} d\Psi\, \f{1}{\sqrt{1-\sin^2\theta_0\sin^2\Psi }}=\eE(\sin^2\theta_0)\,,
\end{equation}
where we used the definition of the complete elliptic integral of the second kind, $\eE$\,.\footnote{\,See Appendix \ref{app:Elliptic} for the definitions and properties of the complete elliptic integrals.}
Hence the first spin is given by
\begin{equation}
\cJ_1  =\f{2}{\pi}\f{w_1}{\sqrt{w_2^2-w_1^2}}\, \eE(\sin^2\theta_0)\,.
\end{equation}
In a similar way, for the second spin, we have
\begin{align}
\cJ_2  &=w_2\int_0^{2\pi} \f{d\sigma}{2\pi}\sin^2\theta
= \f{2}{\pi}\f{w_2}{\sqrt{w_2^2-w_1^2}}\int_0^{\theta_0} \f{\sin^2\theta\, d\theta}{\sqrt{\sin^2\theta_0 -\sin^2\theta }}\no\\[2mm]
&= \f{2}{\pi}\f{w_2}{\sqrt{w_2^2-w_1^2}}\int_0^{\pi/2} d\Psi\, \f{\sin^2\theta_0\sin^2\Psi }{\sqrt{1-\sin^2\theta_0\sin^2\Psi }}\no\\[2mm]
&= \f{2}{\pi}\f{w_2}{\sqrt{w_2^2-w_1^2}}\left[ \eK(\sin^2\theta_0)-\eE(\sin^2\theta_0) \right]\,.
\end{align}
Defining the elliptic moduli as $x^{2}\equiv \sin^2\theta_0\, (\geq 0)$\,, the results are summarised as
\begin{align}
\cJ_1=\f{2}{\pi}\f{w_1}{\sqrt{w_2^2-w_1^2}}\, \eE(x)\, ,\qquad 
\cJ_2= \f{2}{\pi}\f{w_2}{\sqrt{w_2^2-w_1^2}}\left[ \eK(x)-\eE(x) \right]\,.
\label{spins folded}
\end{align}
From the Virasoro condition (\ref{Virasoro elliptic}), the energy is evaluated as
\begin{equation}
\cE=\kappa=\sqrt{w_1^2+x^{2}\ko{w_2^2-w_1^2}}\,.
\label{energy folded}
\end{equation}
This set of equations yields the following pair of relations\,:
\begin{alignat}{2}
\ko{\f{\cE}{\eK(x)}}^2&&-\ko{\f{\cJ_1}{\eE(x)}}^2&=\f{4}{\pi^2}\,x^{2}\,,\label{1st-JJ}\\[2mm]
\ko{\f{\cJ_2}{\eK(x)-\eE(x)}}^2&&-\ko{\f{\cJ_1}{\eE(x)}}^2&=\f{4}{\pi^2}\, .\label{2nd-JJ}
\end{alignat}
Using these relations, one can express the energy as a function of spins, which will be demonstrated in Section \ref{sec:Expansion of energy} to the three-loop. 
One can also express the string profile in terms of the moduli, as
\begin{align}
r_1\ko{\sigma}=\cos\theta\ko{\sigma}={\mathrm{dn}}{\left.\ko{\hf{2}{\pi}\, \eK(x)\sigma \right| x}}\, ,\qquad 
r_2\ko{\sigma}=\sin\theta\ko{\sigma}=x\,{\mathrm {sn}}{\ko{\left.\hf{2}{\pi}\,\eK(x)\sigma \right| x}}\,.
\label{profile folded}
\end{align}
Later in Section \ref{sec:helical large-spin}, we will derive those expressions (\ref{profile folded}) (as well as their circular counterparts (\ref{profile circular})) as special cases of the so-called type $(i)$ helical strings constructed in \cite{Okamura:2006zv}.\footnote{The definitions of $\xi_{1}$ and $\xi_{2}$ are swapped between here and Section \ref{sec:helical large-spin}.}

\subsubsection*{Two-spin circular strings}

\begin{figure}[htbp]
\begin{center}
\vspace{0.5cm}
\includegraphics[scale=0.9]{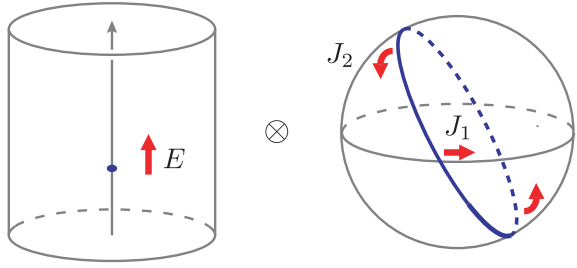}
\vspace{0.5cm}
\caption{\small Elliptic circular $(J_{1},J_{2})$-string.}
\label{fig:J1J2-circular}
\end{center}
\end{figure}

Next let us turn to the other elliptic solution which follows from the ansatz (\ref{elliptic ansatz})\,, namely the elliptic circular string.
As in the folded case, it rotates in the $\varphi_{1}$ and $\varphi_{2}$ planes with two large spins $J_{1}$ and $J_{2}$\,, but this time it wraps around a great circle in the $\theta$\,-direction.
The argument is the same as the folded case up to (\ref{def:C}).
Let us define a new moduli parameter by
\begin{equation}
y^{2}\eq \f{1}{C}\,,\qquad 0\leq y^{2}\leq 1\,,
\end{equation}
which is just the inverse of the moduli $x^{2}\eq C$ for the elliptic folded strings.
When $C\geq 1$\,, there is no such $\sigma$ that gives $\theta'\ko{\sigma}=0$\,, that is, there are no turning points in the $\theta$ direction, thus making the string circular rather than folded.
See Figure \ref{fig:J1J2-circular} for a diagram.
If we write the periodic boundary condition for $\theta(\sig)$ as
\begin{equation}
\theta\ko{\sigma+2\pi}=\theta\ko{\sigma}+2\pi m\,,
\end{equation}
the integer $m$ counts the number of winding into $\theta$ direction while $\sig$ goes from $0$ to $2\pi$\,.\footnote{Of course this is not a topological winding (there is no non-trivial cycle in $S^{5}$).}
For the moment we shall set $m=1$\,, and restore it only when we compute the energy coefficients.

In this circular case also, the two spins are expressed through the complete elliptic integrals as
\begin{align}
\cJ_1=\f{w_1}{y^{2}}\ko{\f{\eE(y)}{\eK(y)}+y^{2}-1}\, ,\qquad 
\cJ_2=-\f{w_2}{y^{2}}\ko{\f{\eE(y)}{\eK(y)}-1}
\label{spins circular}
\end{align}
and the energy is given by
\begin{equation}
\cE=\kappa=\sqrt{w_1^2+\f{1}{y^{2}}\ko{w_2^2-w_1^2}}\,,
\label{energy circular}
\end{equation}
so that in the circular case, we have
\begin{alignat}{2}
\ko{\f{\cE}{\eK(y)}}^2&&-\ko{\f{y^{2}\cJ_1}{(1-y^{2})\eK(y)-\eE(y)}}^2&=\f{4}{\pi^2}\, ,\label{1st-JJc}\\[2mm]
\ko{\f{\cJ_2}{\eK(y)-\eE(y)}}^2&&-\ko{\f{\cJ_1}{(1-y^{2})\eK(y)-\eE(y)}}^2&=\f{4}{\pi^2}\f{1}{y^{2}}\label{2nd-JJc}\, .
\end{alignat}
The profile of the radial coordinates become ({\em c.f.}, (\ref{profile folded}))
\begin{align}
r_1\ko{\sigma}=\cos\theta\ko{\sigma}={\mathrm{sn}}{\ko{\left.\hf{2}{\pi}\, \eK(y)\sigma \right| y}}\, ,\qquad 
r_2\ko{\sigma}=\sin\theta\ko{\sigma}={\mathrm {cn}}{\ko{\left.\hf{2}{\pi}\,\eK(y)\sigma \right| y}}\,.
\label{profile circular}
\end{align}
This elliptic circular string can be obtained as a special case of the so-called type $(ii)$ helical strings, as we will see in Section \ref{sec:helical large-spin}.

\subsection{Perturbative expansion of the energy\label{sec:Expansion of energy}}

We are now in the stage of performing the large-spin expansion of the energy of the long strings, in order to obtain the energy coefficients $\ep_{k}$\,.
The procedure is the same for both folded and circular strings, so we will first see the folded case in detail, then collect the results for the circular case.

Let us rewrite (\ref{1st-JJ}, \ref{2nd-JJ}) in terms of the total spin $\cJ=\cJ_{1}+\cJ_{2}$ and the spin-fraction (\ref{spin fraction}).
This leads (\ref{1st-JJ}, \ref{2nd-JJ}) to
\begin{alignat}{2}
\ko{\f{\cE/\cJ}{\eK(x)}}^2&{}&-\ko{\f{1-\as}{\eE(x)}}^2&=\f{4}{\pi^2}\f{x^{2}}{\cJ^{2}}\,,\label{1st-JJ2}\\[2mm]
\ko{\f{\as}{\eK(x)-\eE(x)}}^2&{}&-\ko{\f{1-\as}{\eE(x)}}^2&=\f{4}{\pi^2}\f{1}{\cJ^{2}}\,.\label{2nd-JJ2}
\end{alignat}
The large-spin limit we take is the following {\em Frolov-Tseytlin limit} ({\em c.f.}, the thermodynamic limit (\ref{thermodynamic}) considered on the gauge theory side),
\begin{equation}
\cJ=\cJ_{1}+\cJ_{2}\to \infty\,,\quad \cJ_{2}\to \infty\,,\quad \mbox{while}\quad \al_{\rm s}=\f{\cJ_{2}}{\cJ}\,:\, \mbox{fixed.}\quad \ko{0\leq \al_{\rm s}\leq 1}\,.
\label{Frolov-Tseytlin limit}
\end{equation}
In this limit, the energy and the moduli parameters are expanded in powers of the effective coupling $\tlambda\equiv \lambda/J^{2}=1/\cJ^{2}$ as
\begin{align}
\cE(\as;\tlambda)&=\cJ\left[1+\tlambda\,\epsilon_1\ko{\as}+\tlambda^{2}\,\epsilon_2\ko{\as}+\dots \right]
=\cJ+\cJ\sum\limits_{k=1}^{\infty}\epsilon_k\ko{\as}\tlambda^k\,,\label{E expansion}\\
x^{2}(\as;\tlambda)&=x^{2}_{(0)}\ko{\as}+\tlambda\, x^{2}_{(1)}\ko{\as}+\tlambda^{2}\, x^{2}_{(2)}\ko{\as}+\ldots
=\sum\limits_{k=0}^{\infty}x^{2}_{(k)}\ko{\as}\tlambda^k\,.\label{x expansion}
\end{align}
By plugging these expansions into (\ref{1st-JJ2}, \ref{2nd-JJ2}), we can obtain the expansion coefficients $x^{2}_{(k)}$ and $\epsilon_k$ order by order.
For example, by plugging (\ref{x expansion}) into (\ref{2nd-JJ2}) and comparing the $\cO(\tlambda^{0})$ order on both sides of the equality, we obtain
\begin{equation}
\as=1-\f{\eE(x_{(0)})}{\eK(x_{(0)})}\,,
\label{def:al_{s}}
\end{equation}
which determines the moduli $x^{2}_{(0)}$ for given spin-fraction $\al_{\rm s}$\,.
By using the relation (\ref{def:al_{s}}) at each order $\cO(\tlambda^{1})$\,, $\cO(\tlambda^{2})$\,, ${}\dots{}$\,, we can express $x^{2}_{(k)}$ in terms of (complicated function of) $x^{2}_{(0)}$\,.
By plugging so-obtained $x^{2}_{(k)}=x^{2}_{(k)}(x^{2}_{(0)})$ into (\ref{1st-JJ2}) together with (\ref{E expansion}), we reach the expression of $\epsilon_k=\epsilon_k(x^{2}_{(0)})$\,.
The results up to $\cO(\tlambda^{3})$ are displayed below (here we restore the folding number $n$)\,:\footnote{The moduli $x^{2}_{(0)}$ used here is related to the one $t_{0}$ used in \cite{Serban:2004jf} as $x^{2}_{(0)}=t_{0}$\,.}
\begin{align}
\epsilon_1&=-\f{2n^{2}}{\pi^{2}}\,\bmK\left(\bmK-\bmE-\bmK\,x^{2}_{(0)}\right)\,,\label{fold-1}\\[2mm]
\epsilon_2&=-\f{2n^{4}}{\pi^{4}}\,\bmK^{3}\left[\bmK-\bmE-2\ko{\bmK-\bmE}x^{2}_{(0)}+\bmK\,x^{4}_{(0)}\right]\,,\label{fold-2}\\[2mm]
\epsilon_3&=-\f{4n^{6}}{\pi^{6}}\f{\bmK^{5}}{\ko{\bmK-\bmE}^{2}-\bmK\ko{\bmK-2\,\bmE}x^{2}_{(0)}}\Big[\ko{\bmK-\bmE}^{3}+{}\no\\
&\hspace{3.5cm} {}+\ko{-4\,\bmK^{3}+14\,\bmK^{2}\bmE-17\,\bmK\bmE^{2}+7\,\bmE^{3}}x^{2}_{(0)}+{}\no\\
&\hspace{3.5cm} {}+\ko{6\,\bmK^{3}-20\,\bmK^{2}\bmE+21\,\bmK\bmE^{2}-7\,\bmE^{3}}x^{4}_{(0)}+{} \no\\
&\hspace{3.5cm} {}+\bmK\ko{-4\,\bmK^{2}+11\,\bmK\bmE-7\,\bmE^{2}}x^{6}_{(0)}+\bmK^{2}\ko{\bmK-2\,\bmE}x^{8}_{(0)}\Big]\,,\label{fold-3}
\end{align}
where we introduced the shorthand notations $\bmK\equiv \eK(x_{(0)})$ and $\bmE\equiv \eE(x_{(0)})$\,.

\paragraph{}
For the elliptic circular strings also, we can obtain the energy coefficients in the same way.
All we have to do is to first rewrite the relations (\ref{1st-JJc}, \ref{2nd-JJc}) as
\begin{alignat}{2}
\ko{\f{\cE/\cJ}{\eK(y)}}^2&&-\ko{\f{y^{2}\ko{1-\as}}{(1-y^{2})\eK(y)-\eE(y)}}^2&=\f{4}{\pi^2}\f{1}{\cJ^{2}}\, ,\\[2mm]
\ko{\f{\as}{\eK(y)-\eE(y)}}^2&&-\ko{\f{1-\as}{(1-y^{2})\eK(y)-\eE(y)}}^2&=\f{4}{\pi^2}\f{1}{y^{2}\cJ^{2}}\,,
\end{alignat}
then use the same algorithm as the folded case.
For given $\as$\,, the moduli $y^{2}_{(0)}$ is determined through the relation
\begin{equation}
\as = \f{1}{y^{2}_{(0)}}\ko{1-\f{\eE(y_{(0)})}{\eK(y_{(0)})}}\,,\quad \mbox{where}\quad 
y^{2}(\as;\tlambda)=\cJ+\sum\limits_{k=1}^{\infty}y^{2}_{(k)}\ko{\as}\tlambda^k\,.
\end{equation}
The energy coefficients are obtained as,\footnote{The moduli $y^{2}_{(0)}$ used here is related to the one $t_{0}$ used in \cite{Serban:2004jf} as $y^{2}_{(0)}=t_{0}$\,.}
restoring the winding number $m$\,,
\begin{align}
\hat\epsilon_1&=\f{2m^{2}}{\pi^{2}}\,\hat\bmK\hat\bmE\,,\label{circ-1}\\[2mm]
\hat\epsilon_2&=-\f{2m^{4}}{\pi^{4}}\,\hat\bmK^{3}\left[2\,\hat\bmE-\hat\bmK+(\hat\bmK-\hat\bmE)y^{2}_{(0)}\right]\,,\label{circ-2}\\[2mm]
\hat\epsilon_3&=-\f{4m^{6}}{\pi^{6}}\f{\hat\bmK^{5}}{\hat\bmK^{2}-\hat\bmE^{2}-\hat\bmK^{2}y^{2}_{(0)}}\Big[ -2\,\hat\bmK^{3}+9\,\hat\bmK^{2}\hat\bmE-14\,\hat\bmK\hat\bmE^{2}+7\,\hat\bmE^{3}+{} \no\\
&\hspace{3.5cm} {}+(5\,\hat\bmK^{3}-18\,\hat\bmK^{2}\bmE+21\,\hat\bmK\hat\bmE^{2}-7\,\hat\bmE^{3})y^{2}_{(0)}+{} \no\\
&\hspace{3.5cm} {}+(-4\,\hat\bmK^{3}+10\,\hat\bmK^{2}\hat\bmE-7\,\hat\bmK\hat\bmE^{2}+\hat\bmE^{3})y^{4}_{(0)}+\hat\bmK^{2}(\hat\bmK-\hat\bmE)y^{6}_{(0)}\Big]\,.\label{circ-3}
\end{align}
We used a hat to distinguish the circular variables from the folded ones, and we also introduced the shorthand notations $\hat\bmK\equiv \eK(y_{(0)})$ and $\hat\bmE\equiv \eE(y_{(0)})$ as in the folded case.

\begin{figure}[t]
\begin{center}
\vspace{.3cm}
\includegraphics[scale=0.9]{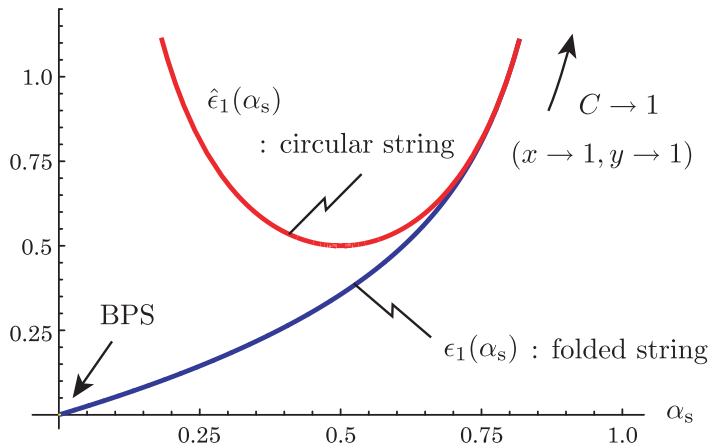}
\vspace{.3cm}
\caption{\small One-loop energies $\ep_{1}$ and $\hat \ep_{1}$ of the elliptic circular and folded strings, respectively, as functions of the spin fraction $\al_{\rm s}$\,.}
\label{fig:string-energy}
\end{center}
\end{figure}

\paragraph{}
The one-loop pieces $\ep_{1}$ and $\hat \ep_{1}$ of the elliptic circular and folded strings are shown in Figure \ref{fig:string-energy} as functions of their spin fraction $\al_{\rm s}$\,.
We see while the folded string has a BPS limit and its energy monotonically grows as $\al_{\rm s}$ increases, the circular string has a symmetric graph with respect to the half-filling $\al_{\rm s}=1/2$\,.
In the limit $\al_{\rm s}\to 1$\,, the energies of both folded and circular strings approach the same divergent value, since it corresponds to the $C\to 1$ limit which makes the string essentially identical.

\section{``Three-loop discrepancy''\label{sec:discrepancy}}

\subsection{The one-, two-loop agreement\,....\label{sec:1,2 agreement}}

Let us recall our strategy (\ref{c=a?}) to test the AdS/CFT conjecture, that is, let us check whether the coefficients of string energy $\ep_{k}$ and those of the scaling dimensions $\delta_{k}$ of SYM operators, both expanded in the BMN coupling, agree.\footnote{Comparison of the ``finite-size'' correction parts, $\ep_{nk}\stackrel{?}{=}\delta_{kn}$\,, is also challenging problem and have been explored to some extent.
See, {\em e.g.}, \cite{Callan:2003xr, Callan:2004ev, Beisert:2005mq, Park:2005ji, Minahan:2005mx, Hernandez:2005nf, Beisert:2005bv, SchaferNameki:2006ey}.
The results have been used to refine the form of the dressing phase factor of the conjectured S-matrix, see Section \ref{sec:dressing phase}.}
It can be done by identifying the corresponding solutions on each side, and relating the moduli parameters in a proper manner.

\subsubsection*{Rational circular strings vs.\ One-cut solutions of SYM spin-chain}

For the rational case, the rational circular string solution (Section \ref{sec:Rational circular}) and the one-cut configuration of the SYM Bethe roots (Section \ref{sec:rational 1-loop}) are the counterparts of each other.
Indeed, we see precise matching between $\ep_{1}$ of the former, (\ref{delta1 rational}), and $\delta_{1}$ of the latter, (\ref{rational circular energy}), under the identification of $(m_{1},m_{2})$ with $(m, m-n)$\,, which ensures $\as=\ag$\,.
Furthermore, in \cite{Kazakov:2004qf}, two-loop matching was explicitly shown, the result being 
\begin{equation}
\ep_{2}=\delta_{2}=-\f{1}{8}\,m(n-m)\kko{n^{2}-3m(n-m)}\,.
\end{equation}

\subsubsection*{Elliptic folded/circular strings vs.\ Two-cut solutions of SYM spin-chain}

For the elliptic case, it is known that the folded and circular string solutions of string theory (Section \ref{sec:elliptic}) correspond to the double-contour and the imaginary-root solutions of gauge theory (Section \ref{sec:elliptic 3-loop}).
We can apply the Gauss-Landen transformation,
\begin{equation}
x^{2}_{(0)}=\f{1}{y^{2}_{(0)}}=-\f{(1-r_{(0)})^{2}}{4r_{(0)}}\,,
\label{Gauss-Landen}
\end{equation}
to relate the moduli parameters of gauge and string theory.
Then we see precise matching the spectra, $\delta_{k}=\ep_{k}$\,, for both the folded/double-contour and circular/imaginary-root correspondence, {up to and including the two-loop order $\it k=2$}\,.
Concretely speaking, by using the map (\ref{Gauss-Landen}), the energy coefficients (\ref{fold-1}, \ref{fold-2}) and (\ref{circ-1}, \ref{circ-2}) of string theory are shown to precisely agree with (\ref{dc-1}, \ref{dc-2}) and (\ref{im-1}, \ref{im-2}) of gauge theory, respectively\,:
\begin{equation}
\delta_{1}=\ep_{1}\,,\quad \delta_{2}=\ep_{2}\,;\qquad 
\hat\delta_{1}=\hat\ep_{1}\,,\quad \hat\delta_{2}=\hat\ep_{2}\,.
\end{equation}
in quite a non-trivial manner.
The folding number of the folded string and the winding number of the circular string correspond to, respectively, the $\cB$- and $\cA$-cycles of the symmetric two-cut solutions (see (\ref{peropds mn})).

These one-, two-loop agreements can be also directly checked by other means, such as the effective sigma model approach \cite{Kruczenski:2003gt} which will be discussed in Section \ref{sec:Kruczenski}, or the algebraic approach (the Bethe ansatz), which will be demonstrated in the next section.

\subsection{....\,and the three-loop discrepancy}

In the elliptic sector, at three-loop order, however, the coefficients cease to agree, and we are faced with a mismatch\,:
\begin{equation}
\delta_{3}\neq\ep_{3}\,;\qquad
\hat\delta_{3}\neq\hat\ep_{3}\,.
\end{equation}
This mismatch is infamously known as the {\em three-loop discrepancy} \cite{Serban:2004jf}.

Actually, a similar mismatch is also found in the so-called near-BMN sector (which includes the first order $1/J$\,-correction) \cite{Callan:2003xr}.
The scaling dimension of BMN operator (\ref{BMN op}) in the near-BMN limit is known to be given by
\begin{equation}
\Delta-J=2+\f{n^{2}\lam}{J^{2}}\ko{1-\f{2}{J}}-\f{n^{4}\lam^{2}}{J^{4}}\ko{\f{1}{4}+\f{0}{J}}-\f{n^{6}\lam^{3}}{J^{6}}\ko{\f{1}{8}+\f{1/2}{J}}+\dots\,,
\end{equation}
while on the other hand, the energy of near-plane-wave string is computed as
\begin{equation}
E-J=2+\f{n^{2}\lam}{J^{2}}\ko{1-\f{2}{J}}-\f{n^{4}\lam^{2}}{J^{4}}\ko{\f{1}{4}+\f{0}{J}}-\f{n^{6}\lam^{3}}{J^{6}}\ko{\f{1}{8}+\f{0}{J}}+\dots\,.
\end{equation}
This kind of three-loop discrepancy is also known for three-impurity cases \cite{Callan:2004ev}.

\subsection{``Order-of-limits'' issue}

Nevertheless, those discrepancies do not immediately disprove the AdS/CFT conjecture.
It is a conjecture for a strong/weak duality, so there is a chance that some non-trivial interpolation occurs between $\lam\ll 1$ and $\lam \gg 1$\,.
There may exist some unknown contributions that do not appear in perturbative gauge theory but do contribute to the classical string, or vice versa.
Here we present some possible logic that could solve this puzzle.

As long as we stick to the BMN scaling hypothesis (\ref{double expansion-Delta}) to arbitrary orders and for arbitrary number of magnons,\footnote{In the $SU(2)$ sector, the number of magnons must of course be sufficiently smaller than the total number of sites of the spin-chain for the applicability of the Bethe ansatz.}
it seems unavoidable that one ends up with the three-loop discrepancy.
However, suppose in the true scenario, there actually exist terms that violate the BMN scaling.
Then it is possible to explain the mismatch by the ``order-of-limits'' mechanism, as first noticed in \cite{Serban:2004jf}.
The point is that the gauge and string theory actually employ different scaling procedures.
On the gauge theory side, one firstly compute the dilatation operator around $\lam=0$\,, then secondary take $J\to \infty$ in the thermodynamic limit.
On the string theory side, the order of limits are the opposite way round\,; the large-spin $J\to \infty$ is assumed in the first setup (so that the quantum correction drops off), and the BMN coupling $\lam/J^{2}$ expansion is performed afterward.
It can make difference in the final coefficients $\delta_{k}$ and $\ep_{k}$ even though we start with identical  function of $\lam$ and $J$\,.

To illustrate this, suppose the true interpolating function $F(\lam;J)$\,, which is to become $\Delta(\lam)$ when $\lam\ll 1$ while $E(\lam)$ when $\lam\gg 1$\,, contains the following toy-term (the last piece) that violates the BMN scaling manifestly,
\begin{equation}
F(\lam;J)=J\kko{1+a_{1}\,\f{\lam}{J^{2}}+a_{2}\,\ko{\f{\lam}{J^{2}}}^{2}+a_{3}\ko{\f{\lam}{J^{2}}}^{3}
+\dots}+c\ko{\f{\lam}{J^{2}}}^{3}\f{1+c'\lam^{J-3}}{1+c\lam^{J-3}}\,.
\label{interpolating toy model}
\end{equation}
in addition to those respects the BMN scaling (in the square bracket).
Then one finds that if one takes the string theory path, that is to first take the limit $J\to \infty$ then $\lam/J^{2}$ small, the third order coefficient yields $(a_{3}+c')(\lam/J^{2})^{3}$\,, while if one takes the gauge theory order, it results in $(a_{3}+c)(\lam/J^{2})^{3}$\,.
Indeed we see the mismatch $\ep_{3}\,(=a_{3}+c')\neq \delta_{3}\,(=a_{3}+c)$\,.\footnote{Furthermore, in the small-$\lam$ expansion for the gauge theory side, the extra term in (\ref{interpolating toy model}) gives rise to an $\cO(\lam^{J})$ term in the next leading order.
It may be seen as indicating the wrapping interaction which indeed begins to contribute from the $\lam^{J}$ order, although the current case is no more than an arbitrary toy model.
Note also that such kind of term diverges in the BMN limit (\ref{BMN limit}).}

Those terms aside from the BMN scaling result in some extra factor for the BDS S-matrix before the thermodynamic limit (\ref{S-matrix}) \cite{Arutyunov:2004vx,Staudacher:2004tk} (recall the BDS S-matrix is so constructed that it respects the BMN scaling to all orders).
This was the first hint for the need of the so-called {\rm dressing phase factor}.
We will investigate the dressing factor in more detail later in Chapter \ref{chap:dressing}.

\section{Classical String Bethe ansatz equations\label{sec:SBAE}}

The one- and two-loop agreements and the three-loop discrepancy between the spinning strings and the SYM Bethe strings we have seen are in fact more transparent at the level of integral equations.
The integral equations arise as the thermodynamic limit of Bethe equations on gauge theory side, and as a so-called {\em finite-gap problem} for the two-dimensional sigma model on string theory side.
This line of approach stems from the work by Kazakov, Marshakov, Minahan and Zarembo (KMMZ) \cite{Kazakov:2004qf}, where they showed how to describe classical string solutions on $\mathbb R\times S^{3}$ as finite-gap solutions, and how to compare them with the gauge theory counterparts.
In this section, we are going to give a brief review of the KMMZ formalism.

\subsubsection*{The \bmt{SU(2)} chiral principal field}

The sigma model on $\mathbb R\times S^{3}$ has an ${SU(2)}_{\rm L}\times {SU(2)}_{\rm R}\times \mathbb R$ global symmetry.
Let us define $SU(2)$ chiral principal field as
\begin{equation}
{\mathcal G}(\tau,\sig)\equiv 
\mbox{\small $\Bigg( 
\begin{array}{cc}
{X_1+i X_2}	&{X_3+i X_4}	\\
{-X_3+i X_4}	&{X_1-i X_2}	\\
\end{array} 
\Bigg)$}
=\mbox{\small $\Bigg( 
\begin{array}{rc}
\xi_1	&\xi_2	\cr
-\bar \xi_2	&\bar \xi_1	\cr
\end{array} 
\Bigg)$}\in {SU(2)}\,.
\end{equation}
The ${SU(2)}_{\rm L}$ and ${SU(2)}_{\rm R}$ correspond to the left action of the group element ${\mathcal G}\mapsto {\mathcal G}'{\mathcal G}$\,, and  the right action of the group element ${\mathcal G}\mapsto {\mathcal G}{\mathcal G}'$\,, respectively, and the associated left and right currents and charges are defined as
\begin{alignat}{5}
l_a&\equiv \ko{\pa_a{\mathcal G}}{\mathcal G}^{-1}\equiv \f{1}{2i}\, l_a^A\sigma^A, &\quad &
l_a^A=\tr\ko{i l_a\sigma^A}\,,&\quad &
Q_{\rm L}^A=\f{\sqrt{\lambda}}{4\pi}\int_{0}^{2\pi}d\sig\, l_{a=0}^A\,;\\[2mm]
j_a&\equiv {\mathcal G}^{-1}\ko{\pa_a{\mathcal G}}\equiv \f{1}{2i}\, j_a^A\sigma^A, &\quad &
j_a^A=\tr\ko{i j_a\sigma^A}\,,&\quad &
Q_{\rm R}^A=\f{\sqrt{\lambda}}{{4\pi}}\int_{0}^{2\pi}d\sig\, j_{a=0}^A\,,
\end{alignat}
where $\sig^{A}$ are Pauli matrices ($A=1,2,3$).
The rest $U(1)$ corresponds to the shift of the AdS time $t\mapsto t+t'$\,, and the associated charge,
\begin{equation}
E=\f{\sqrt{\lambda}}{2\pi}\int_{0}^{2\pi}d\sig\, \pa_{a=0}\,t=\sqrt{\lambda}\kappa\,,
\label{E=4 pi g kappa}
\end{equation}
is identified with the string energy as before ($t=\kappa\tau$).
Using the currents defined above, the sigma model action on $\mathbb R\times S^{3}$ is written as, in conformal gauge,
\begin{equation}
S=-\f{\sqrt{\lam}}{4\pi}\int d\tau \int_{0}^{2\pi}d\sig\kko{-(\pa_{a}t)^{2}+\f{1}{2}\tr (j_{a}^{2})}\,.
\end{equation}
The equations of motion that follow from the action are
\begin{align}
0&=\pa_{+}j_{-}+\pa_{-}j_{+}\,,\label{eom-SU(2)-1}\\
0&=\pa_{+}j_{-}-\pa_{-}j_{+}+[j_{+},j_{-}]\,,\label{eom-SU(2)-2}\\
0&=\pa_{+}\pa_{-}t\,.\label{eom-SU(2)-3}
\end{align}
The third equation for the AdS-time is already solved by the conformal gauge choice $t=\kappa\tau$\,.
The angular momenta $J_{1}$ and $J_{2}$ of the string are related to the third components of the left and right charges as
\begin{equation}
Q_{\rm L}^{A=3}=J_{1}+J_{2}\,,\qquad 
Q_{\rm R}^{A=3}=J_{2}-J_{1}\,.
\end{equation}
The Virasoro constraints are given by
\begin{equation}
\hf{1}{2}\tr(j_{+}^{2})=\hf{1}{2}\tr(j_{-}^{2})=-\kappa^{2}\,.
\label{Virasoro-SU(2)}
\end{equation}
By introducing the spectral parameter $x$\,, the Lax pairs are defined as
\begin{equation}
{\mathscr L}\eq \pa_{\sig}+\ko{\f{j_{+}}{1-x/g}-\f{j_{-}}{1+x/g}}\,,\qquad 
{\mathscr M}\eq \pa_{\tau}+\ko{\f{j_{+}}{1-x/g}+\f{j_{-}}{1+x/g}}\,,\qquad 
\label{L and M}
\end{equation}
which satisfy the flatness condition
\begin{equation}
[{\mathscr L}, {\mathscr M}]=0\,,
\label{flatness}
\end{equation}
as long as the string equations of motion are satisfied.
\begin{figure}[t]
\begin{center}
\vspace{0.5cm}
\includegraphics[scale=0.9]{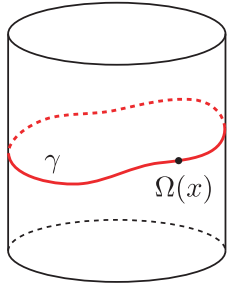}
\vspace{0.5cm}
\caption{\small Monodromy around a closed string.}
\label{fig:monodromy}
\end{center}
\end{figure}

We can now define the {\em monodromy matrix} of the Lax connection around the closed string.
It is independent of the path $\gamma(\sigma)$ around the string, but only depends on the starting (ending) point $\gamma(0)=\gamma(2\pi)$\,.
Therefore we can compute it as\footnote{The symbol $\overline{{\mathscr P}}$ denotes anti-path-ordered, {\em i.e.}, the values of $\sigma$ decrease from left to right.}
\begin{equation}
\Omega(x)=\overline{{\mathscr P}}\exp\int_{0}^{2\pi}d\sig\, \f{1}{2}\ko{\f{j_{+}}{1-x/g}-\f{j_{-}}{1+x/g}}
~ \in ~ SU(2)\,.
\label{Omega}
\end{equation}
It is a unimodular matrix, and thus can be diagonalised as $\Omega(x)\sim {\rm diag}(\lam_{+}(x), \lam_{-}(x))\eq{\rm diag}(e^{ip_{\rm s}(x)}, e^{-ip_{\rm s}(x)})$\,.
The physical quantity is the quasi-momentum $p_{\rm s}(x)$\,, which is defined through the trace of the monodromy matrix,
\begin{equation}
\tr\Omega(x)=2\cos p_{\rm s}(x)\,.
\label{def:Omega}
\end{equation}
Due to the flatness condition (\ref{flatness}), $\tr\Omega(x)$ does not depend on $\tau$\,.
Hence the quasi-momentum $p_{\rm s}(x)$ generates infinitely many local conserved charges by Taylor-expanding it around $x$\,.
The pole structure of the quasi-momentum can be read off from (\ref{Omega}) the Virasoro conditions (\ref{Virasoro-SU(2)}) as
\begin{equation}
p_{\rm s}(x)\mathrel{\mathop{\kern0pt\longrightarrow}\limits_{\hspace{3mm}x\, \to\, \pm 1\hspace{3mm}}}
-\frac{E/4g}{x/g\mp 1}+\sum_{n=2}^{\infty}\f{\mbox{($n$\,-th local charge)}}{\ko{x/g\mp 1}^{n}}\,.
\label{x -> pm 1}
\end{equation}
The asymptotic behaviors of the quasi-momentum in $x\to 0,\,\infty$ are determined as follows.
For $x\to \infty$\,, (\ref{def:Omega}) can be evaluated as
\begin{align}
2\cos p_{\rm s}(x)=\tr\Omega(x)\, &=
\tr\kko{{\bf 1}+\f{1}{x}\int_{0}^{2\pi}d\sig\, j_{0}(\sig)+\f{1}{x^{2}}\int_{0}^{2\pi}d\sig\int_{0}^{\sig}d\sig'\, j_{0}(\sig')j_{0}(\sig)+\dots}\no\\[2mm]
&\sim 2+\f{1}{2x^{2}}\tr\kko{\int_{0}^{2\pi}d\sig\, j_{0}(\sig)}^{2}
=2+\f{1}{x^{2}}\kko{Q_{\rm R}^{A}Q_{\rm R}^{B}\tr\sig^{A}\sig^{B}}\,.
\end{align}
Using $\tr(\sig^{A}\sig^{B})=2\delta^{AB}$ and assuming the classical string solution is the highest-weight state in $SU(2)_{\rm R}\times SU(2)_{\rm L}$\,, one can verify that $p_{\rm s}(x)$ behaves as
\begin{align}
p_{\rm s}(x)\mathrel{\mathop{\kern0pt\longrightarrow}\limits_{\hspace{3mm}x\, \to\, \infty\hspace{3mm}}}
-\f{Q_{\rm R}^{A=3}}{2x}+\ldots 
=-\f{J_{2}-J_{1}}{2x}+\ldots\,.
\label{x -> infty}
\end{align}
Similarly, for $x\to 0$\,, we have
\begin{align}
\tr\Omega(x)&=\tr\kko{{\mathcal G}(2\pi)^{-1}\overline{{\mathscr P}}\exp\ko{x\int_{0}^{2\pi}d\sig\, l_{0}+\dots}{\mathcal G}(0)}\no\\[2mm]
&\sim 2+\f{x^{2}}{2}\tr\kko{\int_{0}^{2\pi}d\sig\, l_{0}(\sig)}^{2}
=2+\f{x^{2}}{2}\kko{Q_{\rm L}^{A}Q_{\rm L}^{B}\tr\sig^{A}\sig^{B}}\,,\\[2mm]
\mbox{\em i.e.},\quad &p_{\rm s}(x)\mathrel{\mathop{\kern0pt\longrightarrow}\limits_{\hspace{3mm} x\, \to\, 0\hspace{3mm}}}
2\pi m+\f{Q_{\rm L}^{A=3}}{2}\,x+\ldots 
=2\pi m-\f{J_{1}+J_{2}}{2}\,x+\ldots\,.
\label{x -> 0}
\end{align}
Here the integer $m$ accounts for the fact $p(0)\in \mathbb Z$ that follows from the periodicity of the string $\Omega(0)={\mathcal G}^{-1}(2\pi){\mathcal G}(0)={\bf 1}$\,.

\subsubsection*{Classical string Bethe ansatz equations}

The quasi-momentum $p_{\rm s}(x)$ is defined on a complex plane with branch-cuts.
The two eigenvalues $\lam_{\pm}=e^{\pm i p_{\rm s}(x)}$ of the monodromy matrix, which are the two roots of the characteristic equation $\lam^{2}-\tr\Omega(x)\,\lam+1=0$\,, are interchanged on either side of the cut, up to a factor of $2\pi$\,.
It follows from the unimodularity of the monodromy matrix.
Denoting the branch-cuts as $\cC_{k}$ $(k=1,\dots, K)$\,, the condition can be expressed as
\begin{equation}
\ps{}_{\rm \hspace{-.1mm}s}(x)=2\pi n_{k}\,,\quad x\in\cC_{k}\,,\quad n_{k}\in\mathbb Z\,.
\label{SBA p}
\end{equation}
Due to the presence of the branch-cuts, the spectral parameter space becomes a two-sheeted Riemann-surface.
Both sheets exhibit the singular structure (\ref{x -> pm 1})\,.
By subtracting the poles from the quasi-momentum, we can define the resolvent $G_{\rm s}(x)$ which is regular on the whole physical sheet,
\begin{equation}
G_{\rm s}(x)=p_{\rm s}(x)+\f{E/4g}{x/g-1}+\f{E/4g}{x/g+1}\,.
\end{equation}
In addition to the branch-cuts $\cC_{k}$\,, there can also be so-called condensate cuts.
Let us denote them as $\cB_{j}$ and assume there are $K'$ such condensate cuts.
Then on the physical sheet, the resolvent can be expressed as 
\begin{equation}
G_{\rm s}(x)=\int_{\cC}dy\,\f{\sig_{\rm s}(y)}{x-y}+\int_{\cB}dy\,\f{\sig_{\rm s}(y)}{x-y}\,,
\end{equation}
where $\cC\eq \cC_{1}\cup\dots\cup\cC_{K}$ and $\cB\eq \cB_{1}\cup\dots\cup\cB_{K'}$\,, and $\sig_{\rm s}(x)$ plays the role of the density.
For a condensate cut $\cB_{k}$\,, the density is constant and given by $\sig_{\rm s}(x)=-in_{k}$\,.
The unimodularity condition (\ref{SBA p}) can be translated into
\begin{equation}
\Gs{}_{\rm \hspace{-.1mm}s}(x)=2\pint_{\cC\cup\cB} dy\,\f{\sig_{\rm s}(y)}{x-y}
=\f{xE}{x^{2}-g^{2}}+2\pi n_{k}\,,\quad x\in\cC_{k}\,.
\label{SBA G}
\end{equation}
Using the resolvent, or the density, the normalisation condition and the asymptotic behaviors (\ref{x -> infty}\,-\,\ref{x -> 0}) are cast into, respectively,\footnote{{\em C.f.}, the one-loop gauge theory analysis (\ref{1-loop BAE : x}\,-\,\ref{1-loop DR : x}) or (\ref{1-loop BAE : x 2}\,-\,\ref{1-loop DR : x 2}).}
\begin{alignat}{3}
-\oint \f{dx}{2\pi i}\, G_{\rm s}(x)&=\int dx\, \sig_{\rm s}(x) &&= J_{2}+\f{E-J}{2}\,,
\label{s-1}\\[2mm]
-\oint \f{dx}{2\pi i}\, \f{G_{\rm s}(x)}{x}&=\int dx\, \f{\sig_{\rm s}(x)}{x} &&= 2\pi m\,,
\label{s-2}\\[2mm]
-\oint \f{dx}{2\pi i}\, \f{G_{\rm s}(x)}{x^{2}}&=\int dx\, \f{\sig_{\rm s}(x)}{x^{2}} &&= \f{E-J}{2g^{2}}\,,
\label{s-3}
\end{alignat}
where $J\eq J_{1}+J_{2}$\,.
By substituting (\ref{s-3}), we can rewrite the integral equation (\ref{SBA G}) as,
\begin{equation}
2\pint dy\, \f{\sig_{\rm s}(y)}{x-y}=\f{xJ}{x^{2}-g^{2}}+2g^{2}x\int dy\, \f{\sig_{\rm s}(y)}{y^{2}(x^{2}-g^{2})}+2\pi n_{k}\,,\quad x\in\cC_{k}\,.
\label{SBA rho}
\end{equation}
This is the integral equation reflecting the integrability of the string sigma model, which is often referred to as the {\em classical string Bethe equation}.

\paragraph{}
To compare the integrable structures of gauge and string theory at the level of classical integral equations, let us derive the corresponding integral Bethe ansatz equations for the gauge theory side, by taking the thermodynamic limit of the BDS model studied in Section \ref{sec:BDS}.
As we saw, the model is correct up to the three-loop level in perturbation theory by construction.
In the thermodynamic limit, the BDS rapidity variable (\ref{rap}) scales as $u\sim \cO(L)$ as before, so let us rescale it as ${\tilde u}=u/L$ and also define ${\tilde g}=g/L$\,.
Then the thermodynamic limit of the BDS Bethe ansatz equations (\ref{BDS bethe eq}) become
\begin{equation}
2\pint_{\cC}d{\tilde u}'\,\f{\rho_{\rm g}({\tilde u}')}{{\tilde u}-{\tilde u}'}=\f{1}{\sqrt{{\tilde u}^{2}-4\tilde g^{2}}}+2\pi n_{k}\,,\quad {\tilde u}\in \cC_{k}\,.
\label{scaling limit BDS 2}
\end{equation}
It can be rewritten as, in terms of the spectral parameter, 
\begin{equation}
2\pint dy\, \f{\sig_{\rm g}(y)}{x-y}=\f{xL}{x^{2}-g^{2}}+2g^{2}x\int dy\, \f{\sig_{\rm g}(y)}{yx(yx-g^{2})}+2\pi n_{k}\,,\quad x\in\cC_{k}\,.
\label{BA sig}
\end{equation}
where we defined $\sig_{\rm g}(x)\eq\rho_{\rm g}(\tilde u)$\,.
Comparing this expression with (\ref{SBA rho}), one notices the only difference between the perturbative gauge theory and semi-classical string theory is the second term in the RHSs.
Also, in this form, agreement of all charges up to two-loops is manifest, and we see the discrepancy starts at the three-loop level.
This feature applies to all classical solutions in the $SU(2)$ sector.

So far we have seen algebraic curve description of classical strings on $\mathbb R\times S^{3}$ \cite{Kazakov:2004qf}.
This line of approach was later extended to $\mathbb R\times S^{5}$ sector \cite{Beisert:2004ag}, to $AdS_{5}\times S^{1}$ sector \cite{SchaferNameki:2004ik}, and to the full $AdS_{5}\times S^{5}$ sector \cite{Beisert:2005bm,Alday:2005gi}.

\subsubsection*{Toward quantum string Bethe ansatz equations}

There exist strong indications that string theory on $AdS_{5}\times S^{5}$ is integrable not only at the classical level but also at the quantum level.
There is an expectation that, just as for the perturbative gauge theory side, the quantum integrability can be also captured by a set of Bethe ansatz equations in a discrete form, namely the {\em quantum string Bethe equation}.
In order to obtain it, one needs to ``undo the thermodynamic limit'' in (\ref{SBA rho}), {\em i.e.}, to turn the classical string Bethe ansatz equation (\ref{SBA rho}) into a discretised form.
Such a set of quantum Bethe ansatz equations for string theory, if it exists, should yield the exact spectrum of the string theory on $AdS_{5}\times S^{5}$\,, whose quantisation has not achieved ever.
It should also describe the non-perturbative regime of $\cN=4$ SYM in the large-$N$ limit.

As we already briefly mentioned in the end of Section \ref{sec:discrepancy}, the S-matrix for the quantum string Bethe ansatz equations may be formulated by multiplying an additional phase factor, called the {\em dressing factor} to the gauge theory BDS S-matrix.
The first quantum string Bethe equations for $\mathbb R\times S^{3}$ string were proposed by Arutyunov, Frolov and Staudacher in \cite{Arutyunov:2004vx}.
The construction was extended to other rank-one sectors by Staudacher in \cite{Staudacher:2004tk}.
In \cite{Beisert:2005fw}, Beisert and Staudacher generalised the quantum string Bethe equation conjecture to the full $AdS_{5}\times S^{5}$ sector.
Their proposals are summarised in Chapter \ref{chap:dressing}, where we also report the current status of the issue.

\subsubsection*{Examples\,: BMN strings}

As the simplest example, let us see how the BMN energy formula (\ref{BMN sqrt formula}) can be reproduced \`a la KMMZ.
In the BMN limit, the cuts shrink to almost point-like, tiny cuts.
For the simplest two-impurity case $\al_{n}^{\dagger}{}^{j}\al_{-n}^{\dagger}{}^{j}\kket{0,p^{+}}$\,, the density function for the shrunken cuts are approximately given by 
\begin{equation}
\sig_{\rm s}(x)\simeq \f{A}{2}\ko{\delta(x-x_{0})+\delta(x+x_{0})}\,,
\end{equation}
where $\cC_{1}=[x_{0}-\ep,x_{0}+\ep]$ and $\cC_{2}=[-x_{0}-\ep,-x_{0}+\ep]$ are the two point-like cuts ($0<\ep\ll 1$) with mode numbers $n_{1}=-n$ and $n_{2}=n$ respectively, and $A$ is a constant to be determined.
The points $\pm x_{0}$ can be determined by solving the string Bethe equation (\ref{SBA G}).
In this BMN case, $\Gs{}_{\rm \hspace{-.1mm}s}(x)=0$ almost everywhere, so that we have
\begin{equation}
x_{0}=-\f{E}{4\pi n}\ko{1+\sqrt{1+\f{(4\pi n g)^{2}}{E^{2}}}}\,,
\label{BMN x0}
\end{equation}
and the conditions (\ref{s-1}) and (\ref{s-3}) become
\begin{equation}
A=J_{2}+\f{E-J}{2}\,,\qquad \f{A}{x_{0}^{2}}=\f{E-J}{2g^{2}}\,.
\label{BMN s-1,3}
\end{equation}
Here actually $J_{2}=2$ in our two-impurity case.
Note that (\ref{s-2}) is automatically satisfied with $m=0$\,.
From (\ref{BMN x0}) and (\ref{BMN s-1,3}), it is easy to show that
\begin{equation}
E-J_{1}=J_{2}\sqrt{1+\f{(4\pi n g)^{2}}{E^{2}}}
=2\sqrt{1+\f{n^{2}\lam}{E^{2}}}\,.
\end{equation}
This agrees with the previous BMN result (\ref{BMN sqrt formula}) up to the difference between $E$ and $J$\,, which is negligible in the first order approximation.

\subsubsection*{Other classical strings}

In the KMMZ formalism, every string solution is characterised by a spectral curve endowed with an Abelian integral called quasimomentum.
Several more examples are\,: elliptic folded/circular strings \cite{Beisert:2004hm}, (dyonic) giant magnons on $\mathbb R\times S^{3}$ and $AdS_{3}\times S^{1}$ \cite{Minahan:2006bd}, pulsating strings rotating \cite{Minahan:2006bd}, helical strings \cite{Vicedo:2007rp} and oscillating helical strings \cite{Hayashi:2007bq}.
In \cite{Dorey:2006zj, Dorey:2006mx}, general finite-gap solutions to the equations of motions on $\mathbb R\times S^{3}$ were constructed.

For the finite-gap interpretations of helical strings on $\mathbb R\times S^{3}$\,, see Section \ref{sec:FG} (see also the end of Section \ref{sec:DGM} for the finite-gap interpretation of dyonic giant magnons).

\section[Appendix for Chapter \ref{chap:FT}]{Appendix for Chapter \ref{chap:FT}\,: Effective sigma-model approach to spinning-strings/spin-chains\label{sec:Kruczenski}}

In the main text we have discussed how one can compare the energies of particular string states with the conformal dimensions of particular SYM operators ({\em i.e.}, particular configurations of Bethe roots in the SYM spin-chain) in the far-from-BPS sector, and have seen the one-, two-loop agreements and the three-loop discrepancy.

Actually we can see the one-loop agreement of the spectra at the level of effective actions, as demonstrated by Kruczenski \cite{Kruczenski:2003gt}.
This provides a direct map between concrete solutions of both theories, bypassing the need to compute conformal dimensions for explicit solutions via the Bethe ansatz method.
In this appendix, we will discuss Kruczenski's original effective sigma-model approach briefly.
We will restrict our argument to the one-loop level for simplicity, but the matching in this formalism can be shown to be successful even at two loops \cite{Kruczenski:2004kw}, which is of course consistent with what we found in Chapter \ref{chap:FT}.

\paragraph{}
Let us start with the $SU(2)$ dilatation operator at the one-loop ({\em i.e.}, the Heisenberg XXX${}_{1/2}$ spin-chain Hamiltonian) (\ref{1-loop Ham}),
\begin{equation}
H=\sum_{\ell=1}^{L}\cH_{\ell,\ell+1}\,,\qquad 
\cH_{\ell,\ell+1}=\f{\lambda}{16\pi^{2}}\ko{I_{\ell}\cdot I_{\ell+1}-\vec \sigma_{\ell}\cdot \vec \sigma_{\ell+1}}\,.
\end{equation}
Following \cite{Kruczenski:2003gt, Fradkin:1991:FTC}, let us perform a so-called ``coherent state path integral'' to obtain an effective action for the spin-chain system. First we consider the path integral for one spin, that is, for one site in the chain.
A coherent state $\ket{n}$ is defined as 
\begin{equation}\label{coh-vec}
\ket{n}\eq \ket{n(\theta,\phi)}\eq \exp\kko{-i\theta\ko{\sin\phi\, \sigma^{x}-\cos\phi\, \sigma^{y}}/2}\ket{0}\,,
\end{equation}
where $\ket{0}$ is the highest weight state of the spin-$\hf{1}{2}$ representation.  
The coherent state is defined to have the following remarkable properties.
First, the expectation value of Pauli matrices in a coherent state (\ref{coh-vec}) gives an unit three-vector parametrized by $\theta$ and $\phi$\,, {\em i.e.},
\begin{equation}
\vec n_{\ell}\eq \bra{n}\vec\sigma_{\ell}\ket{n}
=\ko{\sin\theta_{\ell}\cos\phi_{\ell},\, \sin\theta_{\ell}\sin\phi_{\ell},\, \cos\theta_{\ell}}\,.
\end{equation}
The ``northern pole'' of the two-sphere would be represented as $\vec n_{0}\eq \ko{0,0,1}$\,.
Second, the inner product of two coherent states $\ket{n_{1}}$ and $\ket{n_{2}}$ is given by
\begin{equation}
\langle n_{1} | n_{2}\rangle
={\ko{\hf{1+\vec n_{1}\cdot \vec n_{2}}{2}}^{1/2}} e^{i{\cA}\ko{\vec n_{1},\vec n_{2},\vec n_{0}}}\,,
\end{equation}
where ${\cA}\ko{\vec n_{1},\vec n_{2},\vec n_{0}}$ denotes the oriented area of the spherical triangle with vertices at $\vec n_{1}$\,, $\vec n_{2}$ and $\vec n_{0}$\,.  When performing the coherent state path integral, the factor containing $\vec n_{1}\cdot \vec n_{2}$ does not contribute to the final expression, and the rest $e^{i{\cA}\ko{\vec n_{1},\vec n_{2},\vec n_{0}}}$ produces the following so-called ``Wess-Zumino'' term,
\begin{equation}
\cS_{\rm WZ}[\vec n_{\ell}]=\int {\cal A}_{\ell}=\int_{0}^{1}d\rho\int dt\, \vec{n}_{\ell}(t,\rho)\cdot \kko{\pa_{t}\vec{n}_{\ell}(t,\rho)\times \pa_{\rho}\vec{n}_{\ell}(t,\rho)}\,.
\end{equation}
Here $\vec{n}(t,\rho)$ with $\rho \in[0,1]$ is an extension of $\vec{n}(t)$ defined such that $\vec{n}(t,0)\eq \vec{n}(t)$
and $\vec{n}(t,1)\eq \vec{n}_{0}$\,.  
As we are considering a classical solution, the Wess-Zumino term can be partially integrated to give a $\rho$\,-independent expression like $\pa_{t}\phi \cos\theta$\,.

The total action is the sum of the Wess-Zumino term and the expectation value of the Hamiltonian in the coherent states,
\begin{equation}\label{coh-action-def}
\cS [\vec n_{\ell}] = \f{1}{2}\,\cS_{\rm WZ}[\vec n_{\ell}]+\int dt \bra{n_{\ell}} \cH \ket{n_{\ell}}\,.
\end{equation}
In the scaling limit where $\lambda/L^{2}$ is fixed finite but $L$ is large,
the expectation value of the Hamiltonian in (\ref{coh-action-def}) can be evaluated as
\begin{align}
\bra{n} \ko{I_{\ell}\cdot I_{\ell+1}-\vec\sigma_{l}\cdot \vec\sigma_{\ell+1} } \ket{n}
=\hf{1}{2}\ko{\vec n_{\ell+1}-\vec n_{\ell}}^{2}
~\sim~{\hf{1}{2}}\ko{\theta'{}^{2}+\phi'{}^{2}\sin^{2}\theta} \cdot{\ko{\hf{2\pi}{L}}}^{2}\,,
\end{align}
where we defined $\vec n(\hs)=\ko{\sin\theta (\hs)\cos\phi (\hs),\sin\theta (\hs)\sin\phi (\hs),\cos\theta (\hs)}$ with the identification $\vec n(\komoji{\f{2\pi \ell}{L}})\eq \vec n_{\ell}$\,, and the prime $(\, {}'\, )$ denotes a derivative with respect to such defined $\hs$\,.
Plugging these into (\ref{coh-action-def}), we arrive at the following
effective action of the spin-chain,
\begin{align}\label{coh-action-eff}
S_{\rm eff} &=\sum_{\ell=1}^{L}\cS [\vec n_{\ell}]
= \f{L}{2\pi}\int_{0}^{2\pi} d\hs\, \kko{\f{1}{2}\,\cS_{\rm WZ}[\vec n(\hs)]+\int dt \bra{n(\hs)} \cH \ket{n(\hs)}}\no\\[2mm]
&= \f{L}{4\pi}\int_{0}^{2\pi} d \hs \int dt\, \pa_{t}\phi \cos\theta
-\f{\lambda}{16\pi L}\int_{0}^{2\pi} d \hs \int dt \kko{\theta'{}^{2}+\phi'{}^{2}\sin^{2}\theta}\,,
\end{align}
which turns out to be the so-called Landau-Lifshitz action.
The equations of motion that follow from the action (\ref{coh-action-eff}) becomes 
\begin{equation}
\dot n_{i}=\f{\lam}{2L^{2}}\,\ep^{ijk}n_{j}n_{k}''\,,
\label{LL eq}
\end{equation}
which is known as the Landau-Lifshitz equation.
It describes the time-evolution of magnetization vector of a classical ferromagnet.

\paragraph{}
We can derive the same expression (\ref{coh-action-eff}) on the string sigma model side.
Let us start with the $\mathbb R\times S^{3}$ string action (\ref{RtS3_action}),\footnote{Here we use $t$ instead of $\eta_{0}$ to denote the AdS-time to avoid the confusion with another variable $\eta$\,.} and define new angle variables by 
\begin{equation}
\zeta\eq\f{\vp_{1}+\vp_{2}}{2}\,,\qquad 
\eta\eq\f{\vp_{1}-\vp_{2}}{2}\,.
\end{equation}
In terms of these angles, the two-sphere can be parametrized by $U_{j}e^{i\zeta}$ $(j=1,2)$\,, where $U_{1}=\cos\psi\,e^{i\eta}$ and $U_{2}=\sin\psi\,e^{-i\eta}$ are ${\mathbb{CP}}^{1}$ coordinates.  
Note that $t$ and $\zeta$ are ``fast'' variables that have no
counterparts in gauge theory side, they should therefore be gauged away
through appropriate constraints 
so that the sigma model action reduces to the one written in terms of only the ``slow'' variables $\psi$ and $\eta$\,.
The Lagrangian then takes the form
\begin{align}
\cL_{\mathbb R\times {S}^{3}}&=-\f{\sqrt{\lambda}}{2}\kkko{
\ga^{\al\be}\kko{-\pa_{\al}t\pa_{\be}t+\pa_{\al}\psi\pa_{\be}\psi+\ko{\pa_{\al}\zeta\pa_{\be}\zeta+\pa_{\al}\eta\pa_{\be}\eta}+2\cos\ko{2\psi}\pa_{\al}\zeta\pa_{\be}\eta}}
\end{align}
As usual, we gauge-fix the AdS-time as $t=\kappa\tau$\,, which solves the equation of motion $\pa^{2} t=0$\,.  
We make one more change of variables as $u\eq \zeta-t$ so that $\dot
u$ behaves as $\kappa^{-1}+\cO(\kappa^{-3})$\,.  
Then the Virasoro constraints are written as
\begin{align}
0&= \kappa^{2}+\dot\psi^{2}+\psi'{}^{2}+\kko{-\kappa^{2}+\dot u^{2}+ u '{}^{2}\dot\eta^{2}+\eta'{}^{2}+2\kappa\dot u+2\cos\ko{2\psi}\ko{\kappa\dot\eta+\dot u\dot\eta+u'\eta'}}\,,\\
0&= \dot\psi \psi'+\kko{\dot u u'+\dot\eta \eta'+2\kappa u'+2\cos\ko{2\psi}\ko{\kappa \eta'+\dot u \eta'+\dot\eta u'}}\,.
\end{align}
Here, as usual, the dots $(\, \dot{} \,)$ and the primes $(\, {}' \,)$ denote the derivatives with respect to the worldsheet time- $(\tau)$ and the space- $(\sigma)$ coordinate.

To obtain the string solutions whose energy behaves as $E \sim
J+\dots$ in the large-spin limit $J\to \infty$\,, 
we need to rescale the worldsheet time variable and to take a
special limit.
Following the original paper \cite{Kruczenski:2003gt}, we adopt the following limit\,:
\begin{equation}\label{Kruczenski_limit}
\kappa\to \infty\,,\quad 
\dot X\to 0\,,\quad 
\kappa \dot X~:~\mbox{fixed}\,,\quad 
X'~:~\mbox{fixed}\qquad 
\mbox{for}\quad 
X=\psi,~u,~\eta\,.
\end{equation}
To match the string action with the effective action of gauge theory side, it is needed to use reduced Virasoro constraints in the limit (\ref{Kruczenski_limit}) and also remove the total derivative term.
Further we should change variables such that $-2\eta\mapsto \phi$ $(\in [0,2\pi))$ 
and $2\psi\mapsto \theta$ $(\in [0,\pi])$\,, and rescale the
worldsheet variables as $\ttau=\tau/\kappa$ 
and $\tsig=\sqrt{\lambda}\,\kappa\sigma/J$\,.
Then the action finally takes the form,
\begin{align}\label{Lagrangian}
S_{\mathbb R\times {S}^{3}}
=\f{J}{4\pi}\int d\ttau d\tsig\, \dot\phi\cos\theta-\f{\lambda}{16\pi
  J}
\int d\ttau d\tsig\kko{\theta'{}^{2}+\phi'{}^{2}\sin^{2}\theta}\,.
\end{align}
Here we have redefined the notations of dots and primes so that $\, \dot{}\, =\pa_{\ttau}$ and ${}'\, =\pa_{\tsig}$\,.
This is the same Landau-Lifshitz effective action as we saw in the
gauge theory side, (\ref{coh-action-eff}), 
under the identifications $J\eq L$\,, $\ttau\eq t$ and $\tsig\eq \hs$\,.\footnote{We should note that the above procedure based on the conformal gauge choice is applicable only for the one-loop analysis.
Beyond the one-loop order, we should take the two-dimensional (2D) ``T-dual'' \cite{Kruczenski:2004kw} along $\zeta (\sigma)$ and
introduce the T-dualised field $\widetilde \zeta(\sigma)$\,, 
then gauge-fix as $\widetilde \zeta(\sigma)=J\sigma/\sqrt{\lambda}$\,.
For more details, see \cite{Kruczenski:2004kw,Kruczenski:2004cn,Stefanski:2004cw} and Tseytlin's review \cite{Tseytlin:2004xa}.}

\begin{figure}[t]
\begin{center}
\vspace{.3cm}
\includegraphics[scale=0.85]{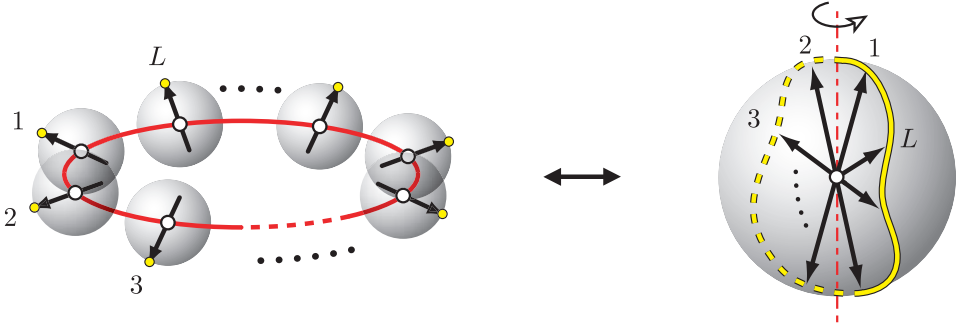}
\vspace{.3cm}
\caption{\small Coherent spin-chain states can be directly identified with string states.}
\label{fig:coherent}
\end{center}
\end{figure}

\paragraph{}
In terms of $\theta$ and $\phi$\,, the Landau-Lifshitz equation (\ref{LL eq}) can be rephrased as
\begin{alignat}{3}
\mbox{for}~~\theta\,,\qquad 
0&=\dot\phi\sin\theta+\f{\lambda}{2J^{2}}\kko{\theta''-\sin\theta\cos\theta\phi'{}^{2}}\,,
\label{EOM-theta}\\[2mm]
\mbox{for}~~\phi\,,\qquad 
0&=\dot\theta\sin\theta+\f{\lambda}{2J^{2}}\kko{\phi'\sin^{2}\theta}'\,.
\label{EOM-eta}
\end{alignat}
The conserved sigma model charges such as the $z$-component of the spin $S_{z}=(J_{1}-J_{2})/2$ (angular momentum along $\phi$\,-direction), the total spin $J=J_{1}+J_{2}$ and the energy $E$ are calculated as
\begin{align}
S_{z}&=\f{J}{4\pi}\int_{0}^{2\pi} d\hs\int_{0}^{1}d\rho\, \pa_{\rho}\theta\sin\theta
=\f{J}{4\pi}\int^{2\pi}_{0} d\hs\, \cos\theta\,,\\[2mm]
J&=\f{J}{2\pi}\int_{0}^{2\pi} d\hs\,,\\[2mm]
E&=\f{\lambda}{16\pi J}\int_{0}^{2\pi} d\hs\ko{\theta'{}^{2}+\phi'{}^{2}\sin^{2}\theta}\,,
\end{align}
respectively.  
The filling-fraction $\ag$\,, or the spin-fraction $\as$\,, which we denote $\al$ here, is related to the sum of the $z$-component of the spin as $S_{z}/J=\hf{1}{2}-\al$\,.

Particular solutions are obtained by making suitable ansatz on the sigma model.
For example, the ansatz $\phi'=0$\,, which implies $\dot\theta=0$ and $\ddot \phi=0$ in view of the equations of motion (\ref{EOM-theta}, \ref{EOM-eta}), leads to a single sine-Gordon equation for $\theta=\theta(\hs)$\,.
Following the same argument as we made in Section \ref{sec:elliptic}, we again reach the ``folded'' (``double-contour'') and the ``circular'' (``imaginary-root'') spectra, reproducing the one-loop results for both types of solutions.
As another example, the ansatz of the form $\theta={\rm const.}$ and $\phi=m\hs$ yields a rational circular solution with winding number $m$\,.

A string solution called dyonic giant magnon, which will be investigated in detail later in Section \ref{chap:DGM}, also corresponds to a special solution of the Landau-Lifshitz equation.
There is a localised soliton solution known in the Landau-Lifshitz model on an infinite line, whose dispersion relation agrees with that of a dyonic giant magnon (\ref{res}) to the leading order of the expansion in powers of $\lam/J_{2}^{2}$\,.

\chapter[Helical Strings]
	{Helical Strings\label{chap:OS}}

\section{Overview}

In the previous chapters we saw two concrete examples of the spinning-string/spin-chain duality in the $SU(2)$ subsector, namely the ``folded-string/double-contour'' and ``circular-string/imaginary-root'' correspondence.
Both of them are related to particular periodic solutions of the integrable Neumann model, or more precisely, periodic soliton solutions of the sine-Gordon equation.
It would then be natural to seek for more generic two-spin string solutions that are related with more general (periodic) solutions of the integrable model.
They would shed more light on the integrable structures of string theory, or in light of the AdS/CFT, on the gauge theory as well.

\paragraph{}
In \cite{Okamura:2006zv}, we explicitly constructed a family of classical string solutions with large spins on $\mathbb R\times {S}^{3}\subset {AdS}_{5}\times {S}^{5}$\,, which are related to periodic soliton solutions of the Complex sine-Gordon equation via the so-called Pohlmeyer reduction procedure.
It was shown that they interpolate between the spinning/rotating strings of Frolov and Tseytlin \cite{Frolov:2003xy} and the so-called dyonic giant magnons \cite{Chen:2006gq}.\footnote{In this chapter we use terminology such as (dyonic) giant magnons.
Definitions as well as a recipe for their construction (the Pohlmeyer reduction procedure) will be introduced in great detail later in Chapter \ref{chap:DGM}.
We decided to do so for the coherency of the thesis.}
In the succeeding work \cite{Hayashi:2007bq}, another family of classical string solutions with large windings on $\mathbb R\times {S}^{3}$ are constructed, which are related to the ones with large spins via the worldsheet $\tau\leftrightarrow\sig$ flip on the sphere side (while keeping the gauge $t=\kappa\tau$).
They interpolate the pulsating strings and the so-called ``single-spike'' solutions of \cite{Ishizeki:2007we}.
Those solutions presented in \cite{Okamura:2006zv,Hayashi:2007bq} turn out to fill the most general elliptic classical strings on $\mathbb R\times {S}^{3}$\,, and we named them ``helical strings'', after their appearance in spacetime.
Their profiles can be written in terms of combination of elliptic theta functions, which favours a finite-gap interpretation, and we will also explain it in Section \ref{sec:FG}.

\paragraph{}
Let us take a close look at them.
There are four types of helical strings in the $SU(2)$ sector\,; two of which (the types $(i)$ and $(ii)$) in the large spin sector and the rest two (the types $(i)'$ and $(ii)'$) in the large winding sector.

The type $(i)$ helical string includes a (two-spin) folded string and a so-called giant magnon solution as its particular limits, while the type $(ii)$ helical string interpolates between a (two-spin) circular string and a giant magnon in a similar manner.
They have different topology\,; the type $(i)$ stays on only one of the hemispheres about the equator (see Figure \ref{fig:type-I}), while type $(ii)$  sweeps both hemispheres, crossing the equator twice in a period (Figure \ref{fig:type-II}).
They are elliptic solutions and are related by an analytic continuation of the moduli parameter, just as was the case with the GKP \cite{Gubser:2002tv} (single-spin) or the Frolov-Tseytlin \cite{Frolov:2003xy} (two-spin) folded/circular strings.\footnote{Recall that the elliptic moduli parameter $C$ defined in (\ref{def:C}) controlled the topology of the string, classifying the solutions into to of folded type ($C<1$) and of circular type ($C>1$).}

The other two helicals are called type $(i)'$ and $(ii)'$\,.
They are of oscillatory nature rather than rotating (although they have finite spins in general).
They are related to the type $(i)$ and $(ii)$ solutions, respectively, by an interchange of worldsheet variables $\tau\leftrightarrow\sig$ on the sphere side (while leaving the AdS side intact).
Throughout this thesis, we will refer to this transformation as the ``$\ts$ transformation'', or just ``2D transformation''.
The difference between the type $(i)'$ and $(ii)'$ lies in the region they oscillate, see Figures \ref{fig:T-helical} and \ref{fig:pulsating}.

It will be also shown that, in the finite-gap language, the two classes of string solutions \---- rotating/spinning with large-spins (types $(i)$ and $(ii)$) on one hand, and oscillating strings with large windings (types $(i)'$ and $(ii)'$) on the other \---- correspond to two equivalence classes of representations of a generic algebraic curve with two cuts.
The $\ts$ operation turns out to correspond to rearranging the configuration of cuts with respect to two singular points on the real axis of the spectral parameter plane, see Figure \ref{fig:2-cuts}.\footnote{An alternative description of $\ts$ operation is to swap the definition of quasi-momentum and quasi-energy.
We will clarify this point later.}

\paragraph{}
We will revisit the helical strings in Section \ref{sec:helical2}, where we explain how to construct them.
In this chapter, we will only display the results and see the properties.
We employ the so-called Pohlmeyer reduction procedure to construct the helical strings, and the method will be first discussed for the construction of (dyonic) giant magnons.

\section{Large spin sector\label{sec:helical large-spin}}

As advertised, there are two helical strings in the large-spin sector; the type $(i)$ and type $(ii)$ strings.

\subsection{Type $\bmt{(i)}$ helical string}

\begin{figure}[p]
\begin{center}
\vspace{1.5cm}
\includegraphics[scale=0.9]{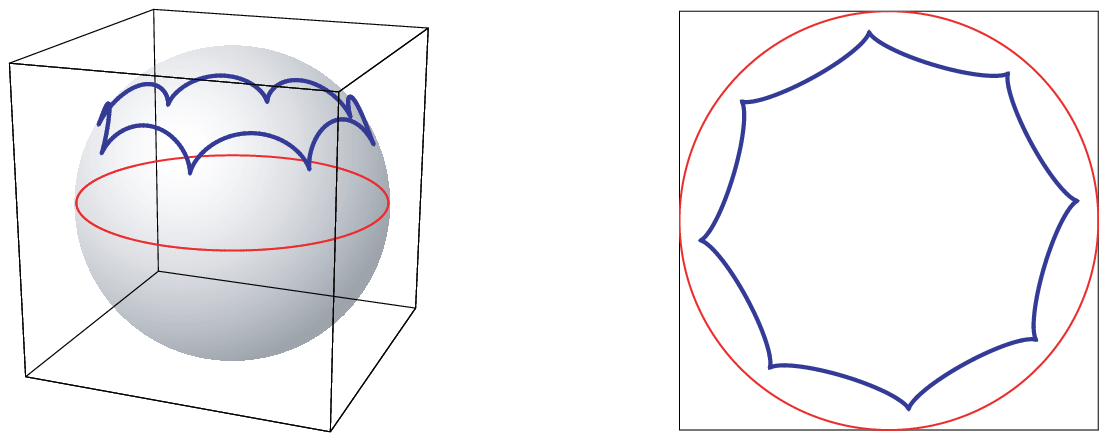}

\vspace{1.5cm}
\includegraphics[scale=0.9]{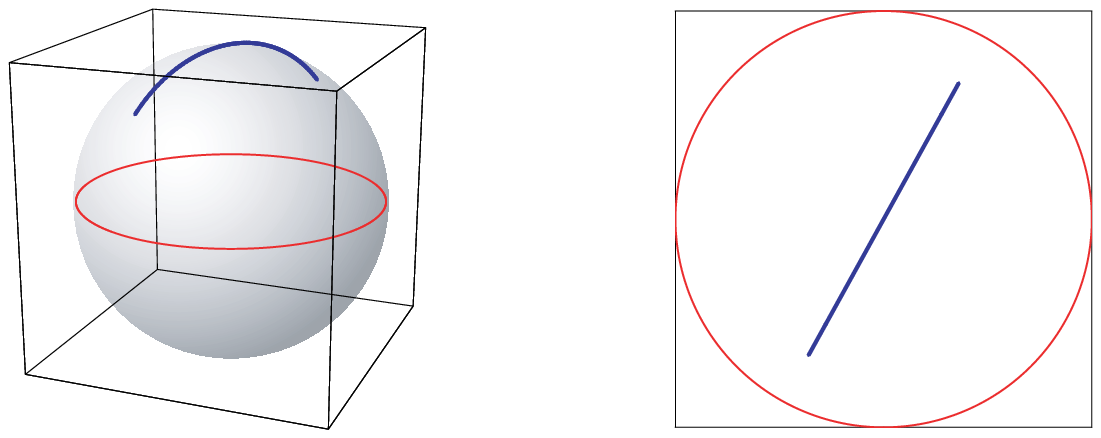}

\vspace{1.5cm}
\hspace{1.8mm}\includegraphics[scale=0.9]{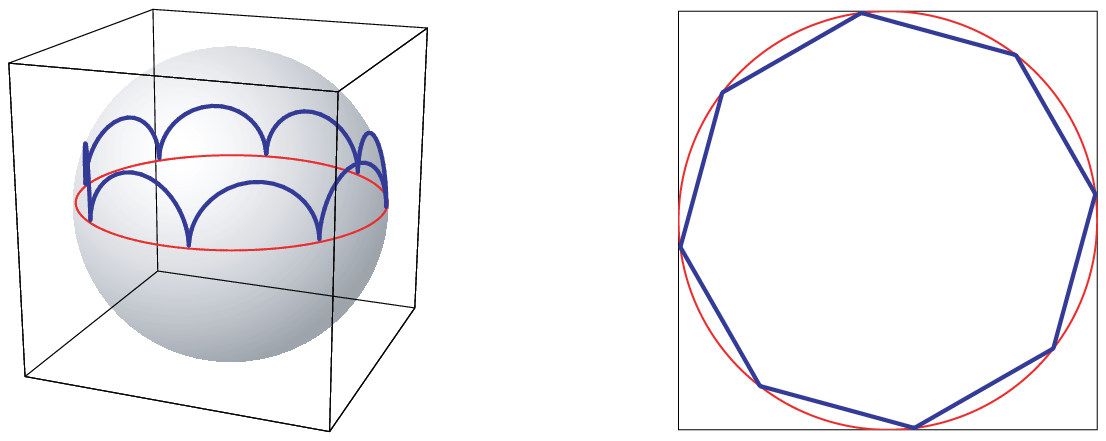}
\vspace{0.5cm}
\caption{\small {\sc Top}\,: Type $(i)$ helical string with a single spin ($k=0.68$\,, $n=8$), projected onto $S^{2}$\,.
Each turning points are located away from the equator, and each segment curves {\it inwards}.
{\sc Middle}\,: The $\om=0$ case ($k=0.75$).
{\sc Bottom}\,: The $k\to 1$ limit.
The diagram shows $n=8$ case, which can be viewed as an array of $n=8$ giant magnons.
All three diagrams show the single-spin cases $(u_{2}=\om_{2}=0)$\,.}
\label{fig:type-I}
\end{center}
\end{figure}

The type $(i)$ string is the branch which includes the folded string as a special case.
See Figure \ref{fig:type-I} for the diagram.
The generic profile is given by \cite{Okamura:2006zv}\footnote{Throughout this chapter, we often
omit the elliptic moduli $k$ from expressions of elliptic
functions. For example, we will often write $\bTh_{\nu}(z)$ or
$\eK$ instead of $\bTh_{\nu}(z|k)$ or $\eK(k)$\,. }
\begin{align}
\eta_{0}(\tau, \sig)&=aT+bX\,,\label{zf0}\\
\xi_{1}(\tau, \sig)&=C\f{\bTh_{0}(0|k)}{\sqrt{k}\, \bTh_{0}(i\om_{1}|k)}\f{\bTh_{1}(X-i\om_{1}|k)}{\bTh_{0}(X|k)}\,\exp\kko{\eZ_{0}(i\om_{1}|k)X+i u_{1}T}\,,\label{zf1}\\
\xi_{2}(\tau, \sig)&=C\f{\bTh_{0}(0|k)}{\sqrt{k}\, \bTh_{2}(i\om_{2}|k)}\f{\bTh_{3}(X-i\om_{2}|k)}{\bTh_{0}(X|k)}\,\exp\kko{\eZ_{2}(i\om_{2}|k)X+i u_{2}T}\,,\label{zf2}
\end{align}
Here $\omega_1$ and $\omega_2$ are real parameters, and $(T,X)$ are boosted worldsheet variables with boost parameter $v$\,,
\begin{equation}
T(\tau,\sig) \equiv \frac{\tilde \tau - v \tilde \sigma}{\sqrt{1 - v^2}} \,,\qquad
X(\tau,\sig) \equiv \frac{\tilde \sigma - v \tilde \tau}{\sqrt{1 - v^2}} \,,\qquad 
(\tilde \tau, \tilde \sigma)\equiv (\mu\tau,\mu\sigma)\,,\quad \mu\in\mathbb R\,.
\label{def:X,T}
\end{equation}
When we clarify the map between string solutions and Complex sine-Gordon solitons later, the boost parameter $v$ turns out to be the very velocity of the solitons.
The normalisation factor $C$ is determined by the sigma model condition $\left| \xi_{1} \right|^2 + \left| \xi_{2} \right|^2 = 1$ as
\begin{equation}
C=\kko{\f{\dn^{2}(i\om_{2})}{k^{2}\cn^{2}(i\om_{2})}-\sn^{2}(i\om_{1})}^{-1/2}\,.
\label{normalisation i}
\end{equation}
The parameters $a$ and $b$ in (\ref{zf0}) are fixed by the Virasoro conditions, which imply
\begin{alignat}{2}
&a^2 + b^2 &\ &= k^2 - 2 k^2 \sn^2 (\iomm{1}) - U + 2 \ssp u_2^2 \,,
\label{a,b-f1}  \\
&\quad a \ssp b &\ &= - i \, C ^2
\pare{u_1 \sn (\iomm{1}) \cn (\iomm{1}) \dn (\iomm{1}) - u_2 \, \frac{1-k^2}{k^2} \, \frac{\sn (\iomm{2}) \dn (\iomm{2})}{\cn^3 (\iomm{2})} }\label{a,b-f2}\,.
\end{alignat}
We can adjust the soliton velocity $v$ so that the AdS-time is proportional to the worldsheet time variable. 
It then follows that $v \equiv b/a \le 1$ and $\eta_{0} = \sqrt{a^2 - b^2} \, \tilde\tau $\,.
The two ``angular velocities'' $u_{1}$ and $u_{2}$ are constrained as 
\begin{equation}
u_1^2 = U + \dn^2 (\iomm{1}) \,,\qquad
u_2^2 = U - \frac{(1 - k^2) \sn^2 (\iomm{2})}{\cn^2 (\iomm{2})} \,,
\label{u_1 and u_2}
\end{equation}
where $U$ is a parameter that controls the type of solution.
This parameter actually corresponds to the eigenvalue of the so-caled Lam\'{e} equation, to which the string equation of motion can be translated (see (\ref{reduced_eom})).
From \eqref{u_1 and u_2} we find the two angular velocities satisfy
\begin{equation}
u_1^2 - u_2^2 = \dn^2 (\iomm{1}) + \frac{(1 - k^2) \sn^2 (\iomm{2})}{\cn^2 (\iomm{2})} \,.
\label{u1-u2:f}
\end{equation}

We are interested in closed string solutions, which means we need to consider periodicity conditions.
The period in $\sigma$\,-direction is defined such that it leaves the theta functions in \eqref{zf1} and \eqref{zf2} invariant, namely it is given by
\begin{equation}
-\ell \le \sigma \le \ell,\qquad \ell = \frac{\eK \sqrt{1-v^2}}{\ssp \mu} \,.
\label{period I}
\end{equation}
Then the periodicity of the string requires
\begin{alignat}{2}
\Delta \sigma \Big|_{\mbox{\scriptsize one-hop}} &\equiv \frac{2 \pi}{n} = \frac{2 \eK \sqrt{1 - v^2}}{\mu} \,,  \label{Dsigma_cl_dy} \\[2mm]
\Delta \varphi_1 \Big|_{\mbox{\scriptsize one-hop}} &\equiv \frac{2 \pi N_1}{n} = 2 \eK \pare{ -i \eZ_0 (\iomm{1}) - v \ssp u_1 } + (2 \ssp n'_1 + 1) \pi \,, \label{Dphi1_cl_dy} \\[2mm]
\Delta \varphi_2 \Big|_{\mbox{\scriptsize one-hop}} &\equiv \frac{2 \pi N_2}{n} = 2 \eK \pare{ -i \eZ_2 (\iomm{2}) - v \ssp u_2 } + 2 \ssp n'_2 \ssp \pi \,.  \label{Dphi2_cl_dy}
\end{alignat}
As $\sigma$ runs from $0$ to $2\pi$ in the worldsheet, the string hops $n$ times in the target space, winding $N_{1}$ and $N_{2}$ times in $\varphi_1$- and $\varphi_2$-direction, respectively.

The global conserved charges can be computed as usual.
The rescaled energy $\cE=(\pi/\sqrt{\lam})E$ and the spins $\cJ_{j}=(\pi/\sqrt{\lam})J_{j}$ $(j=1,2)$ are evaluated after a little algebra to give 
\begin{align}
\cE &= n \ssp a \pare{1 - v^2 }  \eK \,,\label{f E} \\[2mm]
\cJ_1  &= \frac{n \ssp C^2 \, u_1}{{k^2 }}\kko{ { - \eE + \left( {\dn^2 (\iomm{1}) + \frac{v \ssp k^2}{{u_1 }} \ssp i \sn(\iomm{1}) \cn(\iomm{1}) \dn (\iomm{1}) } \right)\eK} } \,, \label{f J1}\\[2mm]
{\cal J}_2  &= \frac{n \ssp C^2 \, u_2}{{k^2 }}\cpare{ \eE + (1-k^2) \left( 
\frac{\sn^2 (\iomm{2})}{\cn^2 (\iomm{2})} - \frac{v}{u_2} \frac{i \sn(\iomm{2})\dn(\iomm{2})}{\cn^3(\iomm{2})} \right)\eK}\,.\label{f J2}
\end{align}

\subsubsection*{$\bullet$ \bmt{\om_{1,2}\to 0} limit\,:~ Elliptic folded strings}

When both $\om_{1,2}$ vanish and the soliton velocity $v$ goes to zero, the boosted coordinates \eqref{def:X,T} become $(T, X) \to (\tilde \tau, \tilde \sigma)$\,.
In this case, the profile (\ref{zf0}\,-\,\ref{zf2}) reduces to
\begin{equation}
\eta_{0} = \sqrt{k^2 + u_2^2} \ \tilde \tau \,, \qquad 
\xi_{1} = k \sn(\tilde \sigma | k) \, e^{i u_1 \tilde \tau} \,, \qquad 
\xi_{2} = \dn(\tilde \sigma | k) \, e^{i u_2 \tilde \tau} \,,
\label{helical->fold}
\end{equation}
with the constraint $u_{1}^{2}-u_{2}^{2}=1$\,.
This is the folded spinning/rotating string of \cite{Frolov:2003xy} studied in Section \ref{sec:elliptic}.
To compare (\ref{helical->fold}) with the one presented in \cite{Frolov:2003xy}, we should relate the parametrisation as
\begin{equation}
\tilde \tau = \mu_{0} \ssp \tau_{\ssp \rm FT} \,,\quad 
\tilde \sig = \mu_{0} \ssp \sigma_{\ssp \rm FT} \,,\quad 
\kappa_{\ssp \rm FT} = \mu_{0} \sqrt{k^2 + u_2^2}\,,\quad
w_{i} = \mu_{0} \ssp u_{i}\qquad 
\mbox{with}\quad 
\mu_{0} = \sqrt{w_{1}^{2}-w_{2}^{2}}\,.\no
\end{equation}
The conserved charges take the following simple form,
\begin{align}
\cE = n\sqrt{k^2 + u_2^2} \; \eK \,, \quad 
\cJ_1 =n u_1 \pare{\eK - \eE} \,,  \quad 
\cJ_2 =n u_2 \, \eE \,,
\label{charges-fold}
\end{align}
with the hopping number $n$ now represents the folding number.
{\em C.f.}, (\ref{spins folded}, \ref{energy folded}).

\subsubsection*{$\bullet$ \bmt{k\to 1} limit\,:~ Dyonic giant magnons}

When the moduli parameter $k$ goes to unity, the string becomes an array of so-called dyonic giant magnons.\footnote{We will investigate the giant magnon and its two-spin extension, the dyonic giant magnon, in great detail later in Section \ref{sec:DGM}.}
The relation (\ref{u1-u2:f}) (or (\ref{u1-u2:c})) implies that the $\omega_2$-dependence of the solutions disappears in this limit, so here we write $\omega$ instead of $\omega_1$\,.
The relation $u_{1}^{2}-u_{2}^{2}=1+\tan^{2}\om$ implies $a=u_{1}$ and $b=\tan\om$ in view of (\ref{a,b-f1}, \ref{a,b-f2}) (or (\ref{a,b-c1}, \ref{a,b-c2})), and the profile becomes
\begin{align}
\eta_{0} = \sqrt{1 + u_2^2}\, \tilde \tau  \,,\qquad 
\xi_{1} = \frac{\sinh(X  - \iom)}{\cosh(X )}
\, e^{i \tan(\omega)X + i u_1 T}\,,\qquad 
\xi_{2} = \frac{\cos (\omega)}{\cosh(X )} \;
\, e^{i u_2 T} \,.
\end{align}
The boundary conditions are imposed as 
\begin{equation}
\xi_{1} \to \exp\pare{\pm i \Delta \varphi_1/2 + i \tilde \tau},\quad 
\xi_{2} \to 0 \qquad {\rm as} \quad \tilde \sigma \to \pm \infty \,,
\label{bc}
\end{equation}
which requires $\mu\to \infty$ as well as the relation $\Delta \varphi_1 = \pi - 2 \omega$\,.
The conserved charges for one-hop ({\it i.e.}, single giant magnon) are given by
\begin{eqnarray}
&\ds \cE ={u_1 \pare{1 - \frac{\tan^2 \omega}{u_1^2}} \eK (1)} \,,& \\[2mm]
&\ds \cJ_1 ={u_1 \kko{ \pare{1 - \frac{\tan^2 \omega}{u_1^2}} \eK (1) - \cos^2 \omega }}  \,, \qquad
\cJ_2 = u_2 \cos^2 \omega \,,&
\label{charges DGM}
\end{eqnarray}
where $\eK (1)$ is a divergent constant (hence $\cE$\,, $\cJ_1\to \infty$).
The energy-spin relation then becomes
\begin{equation}
\cE - \cJ_1 = \sqrt{ \cJ_2^2 + \cos^{2}\omega }\,,\qquad 
\mbox{\em i.e.}, \qquad 
E-J_1 = \sqrt{J_2^2 + \frac{\lambda}{\pi^2} \sin^2\ko{\f{\Delta\varphi_{1}}{2}}}\,.
\label{energy_spin_CDO}
\end{equation}
It would be useful to note that one can match the expressions above with the ones used in \cite{Chen:2006gq} by redefining the parameters as
\begin{equation}
T = (\cos \alpha) \, \widetilde T\,, \quad
X = (\cos \alpha) \, \widetilde X 
\qquad {\rm and} \qquad
u_2 \equiv \tan \alpha \,,
\end{equation}
where $\widetilde T$ and $\widetilde X$ are the boosted worldsheet variables used in \cite{Chen:2006gq}, and $\al$ is the ``rotational parameter'' for the corresponding CsG kink soliton (see Chapter \ref{sec:DGM}).

\subsubsection*{$\bullet$ \bmt{k\to 0} limit\,:~ point-like (BPS) strings}

Another interesting limit is to send $k$ to zero, where the elliptic functions reduce to rational functions.
The Virasoro conditions become
\begin{equation}
a^2 + b^2 = u_2^2 + \tanh^2 \omega_{2} \qquad
\mbox{and}\qquad
ab = \pm u_2 \tanh \omega_{2}\,,
\label{uniform Vir}
\end{equation}
where $u_2 = \sqrt{U + \tanh^2 \omega_{2}}$\,.
They can be solved by $(a,b) = (u_{2},\tanh \omega_{2})$ (assuming $U > 0$).
Then the string profile becomes
\begin{equation}
\eta_{0} = \sqrt{U}\,\tilde\tau\,,\qquad 
\xi_{1} =0\,,\qquad 
\xi_{2} =e^{i \sqrt{U}\,\tilde\tau}\,.
\end{equation}
The conserved charges for one-hop are $E=\sqrt{\lam}/2$\,, $J_{1}=0$ and $J_{2}=\sqrt{\lam}/2$\,.
This is a point-like, BPS ({\em i.e.}, $E-J_{2}=0$) string studied in Section \ref{sec:rot-string-ansatz2}, this time rotating along the great circle in the $X^{3}$\,-\,$X^{4}$ plane.

\subsection{Type $\bmt{(ii)}$ helical string}

\begin{figure}[p]
\begin{center}
\vspace{1.5cm}
\includegraphics[scale=0.9]{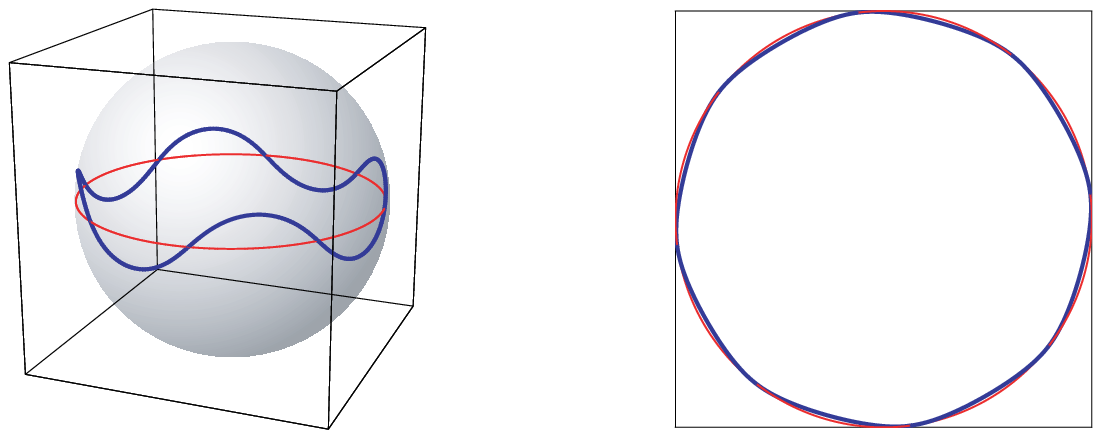}

\vspace{1.5cm}
\includegraphics[scale=0.9]{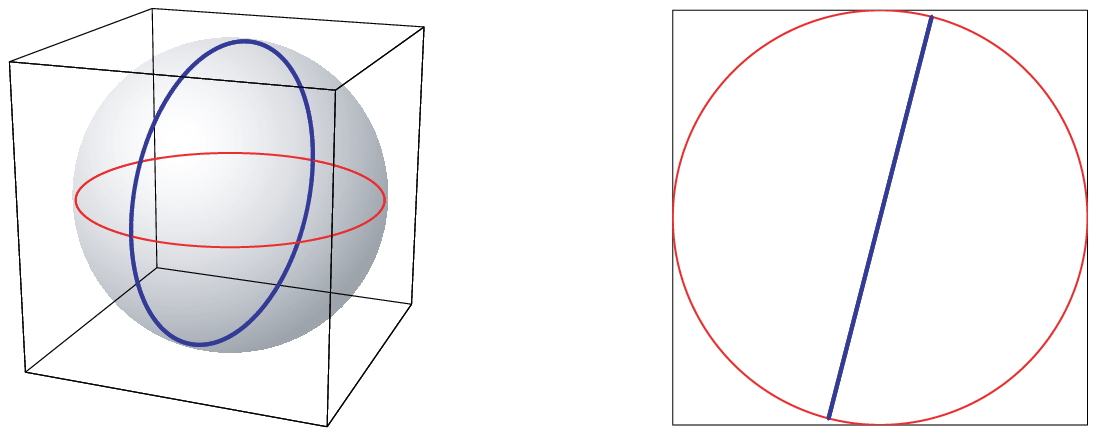}

\vspace{1.5cm}
\hspace{1.8mm}\includegraphics[scale=0.9]{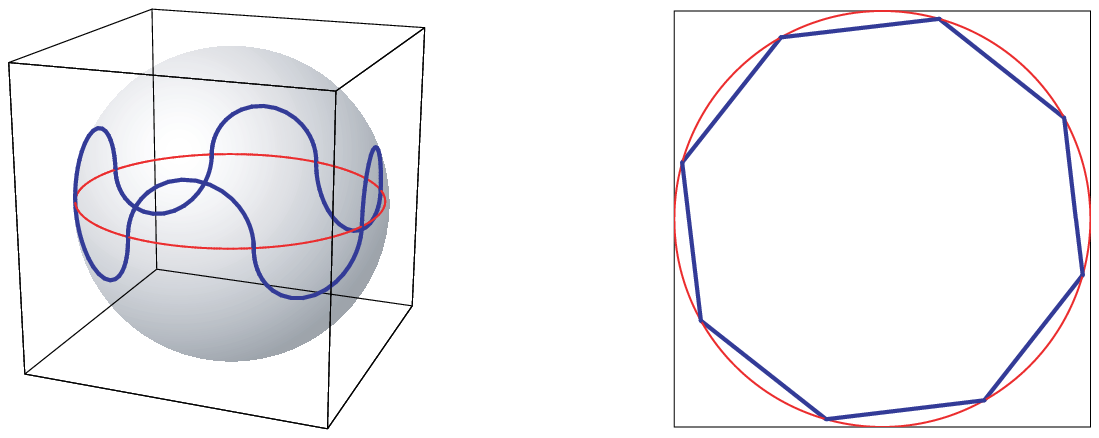}
\vspace{0.5cm}
\caption{\small {\sc Top}\,: Type $(ii)$ helical string with a single spin, projected onto $S^{2}$\,.
The diagram shows $k=0.68$ and $m=8$ case.
As compared to the type I case, each segment curves {\it outwards} about the ``northern pole''.
{\sc Middle}\,: The $\om=0$ case.
This can be regarded as a circular string of \cite{Frolov:2003xy}.
{\sc Bottom}\,: The $k\to 1$ limit.
The diagram shows $m=8$ case, and it can be realised as an array of four giant magnons and four flipped giant magnons by turns.
All three diagrams show the single-spin cases $(u_{2}=\om_{2}=0)$\,.}
\label{fig:type-II}
\end{center}
\end{figure}

In contrast to the type $(i)$ case, the type $(ii)$ string winds around the equator of ${S}^{2}$\,, waving up and down\,; see Figure \ref{fig:type-II}.
The profile is given by\footnote{We use a hat to indicate type $(ii)$ quantities.} \cite{Okamura:2006zv}
\begin{align}
\hat\eta_{0}(\tau, \sig)&=\hat aT+\hat bX\,,\label{zc0}\\
\hat\xi_{1}(\tau, \sig)&=\hat C\f{\bTh_{0}(0|k)}{\sqrt{k}\, \bTh_{0}(i\om_{1}|k)}\f{\bTh_{1}(X-i\om_{1}|k)}{\bTh_{0}(X|k)}\,\exp\kko{\eZ_{0}(i\om_{1}|k)X+i u_{1}T}\,,\label{zc1}\\
\hat\xi_{2}(\tau, \sig)&=\hat C\f{\bTh_{0}(0|k)}{\sqrt{k}\, \bTh_{3}(i\om_{2}|k)}\f{\bTh_{2}(X-i\om_{2}|k)}{\bTh_{0}(X|k)}\,\exp\kko{\eZ_{3}(i\om_{2}|k)X+i u_{2}T}\,,\label{zc2}
\end{align}
where $\omega$ is again a real parameter. 
The AdS-time can be written as $\hat \eta_{0} = \tilde \tau$\,, and the normalisation factor $\hat C$ is given by
\begin{equation}
\hat C = \pare{ \frac{\cn^2 (\iomm{2})}{\dn^2 (\iomm{2})} - \sn^2 (\iomm{1}) }^{-1/2}\,.
\label{normalisation ii}
\end{equation}
The Virasoro conditions constrain the coefficients $\hat a$ and $\hat b$ as
\begin{alignat}{2}
&\hat a^2 + \hat b^2 &\ &= \quad k^2 - 2 k^2 \sn^2 (\iomm{1}) - U + 2 \ssp u_2^2 \,, \label{a,b-c1}\\
&\quad \hat a \, \hat b &\ &= - i \, \hat C^2
\pare{u_1 \sn (\iomm{1})\cn(\iomm{1}) \dn(\iomm{1}) + u_2 \pare{1-k^2} \frac{\sn (\iomm{2})\cn(\iomm{2})}{\dn^3(\iomm{2})}} \,.
\label{a,b-c2}
\end{alignat}
The soliton velocity is given by $\hat v \equiv \hat b/\ssp \hat a \le 1$ so that we have $\hat \eta_{0} = \sqrt{\hat a^2 - \hat b^2}\,\tilde \tau$\,.
The angular velocities $u_{1}$ and $u_{2}$ satisfy
\begin{equation}
u_1^2 = U + \dn^2 (\iomm{1}) \,,\qquad
u_2^2 = U + \frac{1 - k^2}{\dn^2 (\iomm{2})} \,,
\end{equation}
and are constrained as
\begin{equation}
u_1^2 - u_2^2 = \dn^2 (\iomm{1}) - \frac{1 - k^2}{\dn^2 (\iomm{2})} \,.
\label{u1-u2:c}
\end{equation}
The periodicity conditions for the type $(ii)$ solution are given by
\begin{alignat}{2}
\Delta \sigma \Big|_{\rm one\mbox{\tiny\,-\,}hop} &\equiv \frac{2 \pi}{m} = \frac{2 \eK \sqrt{1 - v^2}}{\mu} \,,  \label{Dsigma2_cl_dy} \\[2mm]
\Delta \varphi_1 \Big|_{\rm one\mbox{\tiny\,-\,}hop} &\equiv \frac{2 \pi M_1}{m} = 2 \eK \pare{ \eZ_0 (\iomm{1}) - \hat v \ssp u_1 } + (2 \ssp m'_1 + 1) \pi  \,,  \label{Dphi3_cl_dy} \\[2mm]
\Delta \varphi_2 \Big|_{\rm one\mbox{\tiny\,-\,}hop} &\equiv \frac{2\pi M_2}{m} = 2 \eK \pare{ \eZ_3 (\iomm{2}) - \hat v \ssp u_2 } + (2 \ssp m'_2 + 1) \pi  \,, \label{Dphi4_cl_dy}
\end{alignat}
where $m=1,2,\dots$ is again the number of hops for $0\leq \sigma \leq 2\pi$\,, and $M_{1}$ and $M_{2}$ are winding numbers for $\varphi_1$- and $\varphi_2$-direction, respectively.

The conserved charges of the type $(ii)$ string with $m$ hops can be evaluated as
\begin{align}
\hat \cE &= m \ssp \hat a \pare{1 - \hat v^2 } \; \eK \,, \\[2mm]
\hat \cJ_1  &= \frac{m \ssp \hat C^2 \, u_1 }{{k^2 }}\kko{ { - \eE + \left( {\dn^2 (\iomm{1}) + \frac{\hat v \ssp k^2}{{u_1 }} \ssp i \sn (\iomm{1}) \cn (\iomm{1}) \dn (\iomm{1}) } \right)\eK} } \,, \\[2mm]
\hat \cJ_2  &= \frac{m \ssp \hat C^2 \, u_2 }{{k^2 }}\kko{ \eE - (1-k^2) \pare{ \frac{1}{\dn^2 (\iomm{2})} - \frac{\hat v \ssp k^2}{u_2} \, \frac{i \sn (\iomm{2})\cn(\iomm{2})}{\dn^3(\iomm{2})}} \eK }\,.
\end{align}

\subsubsection*{$\bullet$ \bmt{\om_{1,2}\to 0} limit\,:~ Elliptic circular strings}

Elliptic circular strings of Frolov-Tseytlin \cite{Frolov:2003xy} are reproduced by taking the stationary limit $v\to 0$ for the type $(ii)$ solutions.
In this case (\ref{zc0}\,-\,\ref{zc2}) reduce to
\begin{equation}
\hat \eta_{0} = \sqrt{1 + u_2^2} \ \tilde \tau \,, \qquad
\hat \xi_{1} = \sn(\tilde \sigma | k) \, e^{i u_1 \tilde \tau} \,, \qquad
\hat \xi_{2} = \cn(\tilde \sigma | k) \, e^{i u_2 \tilde \tau} \,,
\end{equation}
with a constraint $u_{1}^{2}-u_{2}^{2}=k^{2}$\,.
The conserved charges are given by
\begin{equation}
\cE = m\sqrt{1 + u_2^2} \; \eK \,, \quad 
\cJ_1 = \frac{m u_1}{k^2} \pare{\eK - \eE} \,, \quad 
\cJ_2 = \frac{m u_2}{k^2} \pare{\eE - (1-k^2) \eK} \,,
\end{equation}
with $m$ now representing the winding number in $\theta$-direction.
{\em C.f.}, (\ref{spins circular}, \ref{energy circular}).

\subsubsection*{$\bullet$ \bmt{k\to 1} limit\,:~ Dyonic giant magnons}

The $k\to 1$ limit takes the type $(ii)$ solution to an array of dyonic giant magnons as in the type $(i)$ case.
The only difference to the type $(i)$ case is that while all the ``hop''s are on one side of the hemisphere in the type $(i)$ case, in this type $(ii)$ case they appear alternatively on both sides of the equator, see the third diagram in Figure \ref{fig:type-II}.
However, this is not an essential difference, since the two configurations can switch to each other without energy cost.
In this sense they are essentially the same configuration.\footnote{This can be also understood from the viewpoint of finite-gap solutions.
Both solutions can be represented by the same condensate cut (when the associated periods are properly normalised).}

\subsubsection*{$\bullet$ \bmt{k\to 0} limit\,:~ Rational circular strings}

In the $k\to 0$ limit, the profile becomes
\begin{align}
\hat \eta_{0} = \sqrt{\hat a^2 - \hat b^2} \; \tilde \tau\,,\quad
\hat \xi_{1} = \hat C \; \sin (X - \iomm{1}) \, e^{i u_1 T}\,,\quad
\hat \xi_{2} = \hat C \; \cos (X - \iomm{2}) \, e^{i u_2 T}\,,
\end{align}
where $\hat C = \pare{ \cosh^2 \omega_2 + \sinh^2 \omega_1 }^{-1/2}$\,.
The angular velocities satisfy $u_1^2 = u_2^2 = U + 1$\,.
The parameters $\hat a$ and $\hat b$ (with $\hat a \ge \hat b$) are determined by
\begin{equation}
\hat a^2 + \hat b^2 = U+2  \qquad
\mbox{and}\qquad
\hat a \, \hat b = \hat C^2 \sqrt{U + 1}
\pare{\sinh\omega_1 \cosh \omega_1 \mp \sinh\omega_2 \cosh \omega_2 }
\label{pm}\,,
\end{equation}
where $\mp$ reflects the sign ambiguity of the angular momenta.
The conserved charges for 
one-hop are evaluated as
\begin{eqnarray}
\hat \cE = \frac{\pi \hat a \pare{1 - \hat  v^2}}{2}  \,,\quad 
\hat \cJ_1  = - \frac{\pi \ssp \hat C^2 \hat v}{2} \, \sinh \omega_1 \cosh \omega_1 \,, \quad 
\hat \cJ_2  = \frac{\pi \ssp \hat C^2 \hat v}{2} \, \sinh \omega_2 \cosh \omega_2 \,.
\label{charges unif IIb}
\end{eqnarray}
As we are assuming $\hat a \ge \hat b \ge 0$\,, the situation $\hat b =0$ can be realised only when $\om_{1}=\om_{2}$ with ``$-$'' sign in (\ref{pm}), or when $\om_{1} = - \om_{2}$ with ``$+$'' sign.
In both cases, the soliton velocity $\hat v = \hat b/ \ssp \hat a$ vanishes, which then implies the equal spin relation $J_{1}=J_{2}$ in view of \eqref{charges unif IIb}.
This equal two-spin (rational) solution can also be realised as the $J_{1}=J_{2}$ case of the rational (``constant-radii'') solution we studied in Section \ref{sec:Rational circular}.
From the viewpoint of a SYM finite-gap problem, the equal two-spin case corresponds to the single-cut limit of the symmetric two-cut imaginary root solution (\ref{IR solution}), that is, the limit when the outer two branch points of the cuts go to infinity, thus reducing it a single-cut.\footnote{To be more precise, the outer branch points $\pm i t$ go to $\pm i\infty$ while the inner ones $\pm i s$ go to $\pm i/2\pi m$\,.}
This situation can also be realised as the $\ag\to \hf{1}{2}$ limit of the single cut distribution of Bethe roots studied in Section \ref{sec:rational 1-loop}.

\section{Large winding sector\label{sec:helical large-winding}}

In the previous section we investigated helical strings with large spins.
In this section, we explore another branch of the helical string, which are obtained by performing the worldsheet $\ts$ transformation (``2D transformation'') to the sphere side.
The resulting solutions again exhibit profiles in terms of elliptic theta functions, but this time they show oscillating behavior rather than fast-rotation.

Let us look at the string equations of motion \eqref{string_eom} and the Virasoro constraints \eqref{string_Virasoro}. 
Since they are invariant under the $\tau \leftrightarrow \sigma$ flip, any string solution is mapped to another solution under this map.
Upon closer inspection of the Virasoro constraints \eqref{string_Virasoro}, one finds that the $\ts$ operation can be applied independently to the $\bb{R} \subset AdS_5$ and $S^3 \subset S^5$ part.
We will use this observation to generate new string solutions from known solutions on $\mathbb R\times S^{3}$\,, by transforming only the $S^{3}$ part while retaining the gauge $t=\kappa \tau$\,.
In order to satisfy other consistency conditions such as closedness of the string, one needs to reconsider the periodicity in the new $\sigma$ direction (that used to be the $\tau$ direction before the flip).
The 2D transformed versions of type $(i)$ and $(ii)$ strings are called type $(i)'$ and $(ii)'$ strings, respectively.

\paragraph{}
This kind of ``2D duality'' is actually well-known in the context of rotating strings and pulsating string solutions, both of which are characterised by the same special Neumann-Rosochatius integrable system \cite{Arutyunov:2003uj,Arutyunov:2003za}.
In contrast to the ansatz for rotating strings (\ref{rotating_string_ansatz}), the one for pulsating strings reads
\begin{equation}
\xi _i\ko{\tau,\sigma} = r_i\ko{\tau}\,\exp\kkko{i \varphi_{i}(\tau,\sig)}=r_i\ko{\tau}\,\exp\kkko{i \kko{ m_i\tau+\alpha_i\ko{\tau} }}\,.\label{pulsating_string_ansatz}
\end{equation}
Here $m_{j}$ play the role of the integer winding numbers ({\em c.f.}, the counterparts $w_{j}$ in (\ref{rotating_string_ansatz}) as angular velocities, which are continuous parameters in classical string theory but should take discrete values in quantum theory).
It is reminiscent of the ordinary T-duality that the angular momenta (spins) and winding numbers are interchanged, however, one should
also take notice that in our 2D transformation case not only the angular part $\varphi_{j}$ but also the radial part $r_{j}$ are transformed (as in (\ref{rotating_string_ansatz}) $\leftrightarrow$ (\ref{pulsating_string_ansatz})).
Consequently, there are two effects of this $\ts$ map\,:
\begin{itemize}
\item Large spin states become large winding states.
\item Rotating/spinning states become oscillating states.
\end{itemize}
We will see these features for the case of 2D transformed helical strings, and see how they interpolate between particular pulsating strings ($\ts$ transformed folded/circular strings \cite{Frolov:2003xy}) and the ``single-spike'' strings \cite{Ishizeki:2007we} ($\ts$ transformed dyonic giant magnons).

\paragraph{}
We first study the type $(i)'$ case in the following Section \ref{sec:type (i)'}.
The results on the type $(ii)'$ solutions will be collected in Section \ref{sec:type (ii)'}.

\subsection{Type $\bmt{(i)'}$ Helical Strings\label{sec:type (i)'}}

Starting from (\ref{zf0}\,-\,\ref{zf2}), by
swapping $\tau$ and $\sigma$ in the sphere coordinates $\xi_{1,2}(\tau, \sigma)$ while keeping the relation $\eta_{0}(\tau, \sigma)=aT+bX$ for the AdS side as it
is, we obtain the 2D transformed version of the type $(i)$
two-spin helical strings, which we call type $(i)'$ helical
strings \cite{Hayashi:2007bq}
\begin{align}
\xi_{1} &= C \frac{\bTh _0 (0|k)}{\sqrt{k} \, \bTh _0 (\iomm{1}|k)} \frac{\bTh _1 (T   - \iomm{1}|k)}{\bTh _0 (T|k)} \,
\exp \Big(  \eZ_0 (\iomm{1}|k)  T+ i  u_1 X\Big)\,,
\label{zf1-T} \\[2mm]
\xi_{2} &= C \frac{\bTh _0 (0|k)}{\sqrt{k} \,\bTh _2 (\iomm{2}|k)} \frac{\bTh _3 (T   - \iomm{2}|k)}{\bTh _0 (T|k)} \, \exp \Big( \eZ_2 (\iomm{2}|k)T + i   u_2 X\Big)\,,
\label{zf2-T}
\end{align}
with the same normalisation constant as (\ref{normalisation i}).
The diagrams are shown in Figures \ref{fig:T-helical} and \ref{fig:pulsating}.
The Virasoro constraints (\ref{string_Virasoro}) force the parameters $a$ and $b$ to satisfy the same relations as the ones for the type $(i)$ case, (\ref{a,b-f1}, \ref{a,b-f2}).
The same relations as (\ref{u_1 and u_2}) are also satisfied.
We can adjust the parameter $v$ such that the AdS-time is proportional to the worldsheet time variable, namely $\eta_{0} = \sqrt{a^2 - b^2} \, \tilde\tau $ with $v \equiv b/a \le 1$\,.

\paragraph{}
We are interested in closed string solutions, so we need to reconsider the periodicity conditions.
The period in the $\sigma$\,-direction is again defined such that it leaves the theta functions in (\ref{zf1-T}, \ref{zf2-T}) invariant.
Reflecting the $v\leftrightarrow 1/v$ feature between before and after the $\ts$ map, in contrast to (\ref{period I}), the period now becomes
\begin{equation}
-\ell \le \sigma \le \ell,\qquad \ell = \frac{\eK \sqrt{1-v^2}}{v \ssp \mu}~~(>0)\,,
\end{equation}
hence the periodicity of the string requires
\begin{alignat}{2}
\Delta \sigma &\equiv \frac{2 \pi}{n} & &= \frac{2 \eK \sqrt{1 - v^2}}{v \ssp \mu} \,,
\label{Dsigma_cl f} \\[2mm]
\Delta \varphi_1 &\equiv \frac{2 \pi N_1}{n} & &= 2 \eK \pare{ \frac{u_1}{v} + i \eZ_0 (\iomm{1}) } + (2 \ssp n'_1 + 1) \pi  \,,
\label{Dphi1_cl f} \\[2mm]
\Delta \varphi_2  &\equiv \frac{2 \pi N_2}{n} & &= 2 \eK \pare{ \frac{u_2}{v} + i \eZ_2 (\iomm{2}) } + 2 \ssp n'_2 \ssp \pi \,.
\label{Dphi2_cl f}
\end{alignat}
The integers $n=1,2,\dots$ counts the number of periods in $0\leq \sigma \leq 2\pi$\,, and $N_{1, 2}$ are the winding numbers in $\varphi_{1,2}$\,-directions respectively.
The integers $n'_{1,2}$ specify the ranges of $\om_{1,2}$ respectively.\footnote{\,
When $\om_{i}$ are shifted by $2\eK'$\,, the integers $n'_{i}$ change by one while $\xi_{i}$ and $\Delta \varphi_{i}$ are unchanged.}

The energy ${\cal E}$ and spins ${\cal J}_{i}$ of the type $(i)'$ string
with $n$ periods are evaluated as
\begin{align}
{\cal E} &= \frac{n a (1 - v^2)}{v} \, \eK = \frac{n (a^2 - b^2)}{b} \, \eK \,,
\label{t-f E} \\[2mm]
{\cal J}_1  &= \frac{n \ssp C^2 \, u_1}{k^2 }\cpare{ \eE - \left( {\dn^2 (\iomm{1}) + \frac{i \ssp k^2}{v \ssp u_1} \sn(\iomm{1}) \cn(\iomm{1}) \dn (\iomm{1}) } \right) \eK } \,,
\label{t-f J1} \\[2mm]
{\cal J}_2  &= \frac{n \ssp C^2 \, u_2}{k^2 } \cpare{ - \eE - (1-k^2) \left(
\frac{\sn^2 (\iomm{2})}{\cn^2 (\iomm{2})} - \frac{i}{v \ssp u_2} \frac{\sn(\iomm{2})\dn(\iomm{2})}{\cn^3(\iomm{2})} \right)\eK}\,.
\label{t-f J2}
\end{align}
Notice that in the limit $v \to 0$ (or $\om_{1,2}\to 0$)\,, all the
winding numbers in (\ref{Dsigma_cl f}\,-\,\ref{Dphi2_cl f}) become
divergent (and so ill-defined), due to the fact that the $\theta$\,-angle defined in (\ref{S-coord}) becomes independent of $\sigma$\,.
Therefore, in this limiting case, we may choose $\mu$ arbitrarily
without the need of solving \eqref{Dsigma_cl f}, provided that
$N_1$ and $N_2$ are both integers.

It is meaningful to compare the above expressions with (\ref{f E}\,-\,\ref{f J2}) for the original type $(i)$ helical strings.
The type $(i)'$ charges ${\cal E}^{(i)'}$ and ${\cal J}_{1,2}^{(i)'}$ as functions of $v =
b/a$ are related
to the type $(i)$ charges by ${\cal E}^{(i)'} (a,b) = - {\cal E}^{(i)}
(b,a)$ and ${\cal J}_{1,2} ^{(i)'}(v) = - {\cal J}_{1,2}^{(i)} (-1/v)$\,.
Similar relations also hold for the winding numbers given in
\eqref{Dphi1_cl f} and \eqref{Dphi2_cl f} as $N_{1,2} ^{(i)'}(v) = - N_{1,2}^{(i)}(-1/v)$\,. They are simply the consequence of the symmetry $a\leftrightarrow b$ of the Virasoro constraints.
For example, if $(a, b) = (a_0, b_0)$ solves (\ref{a,b-f1}, \ref{a,b-f2}), then $(a, b) = (b_0, a_0)$ gives another solution.

\paragraph{}
The type $(i)'$ helical strings contain pulsating strings and single-spike strings in particular limits.
Below we will consider various limits including them.

\subsubsection*{$\bullet$ \bmt{\om_{1,2}\to 0} limit\,:~ Pulsating strings}

Let us first consider the $\om_{1,2}\to 0$ limit. In this limit,
the boosted coordinates (\ref{def:X,T}) reduce to $(T,
X) \to (\tilde \tau, \tilde \sigma)$\,, and (\ref{zf0}, \ref{zf1-T}, \ref{zf2-T}) become
\begin{equation}
\eta_{0} = \sqrt{k^2 + u_2^2} \ \tilde \tau \,, \qquad
\xi_{1} = k \sn(\tilde \tau | k) \, e^{i u_1 \tilde \sigma} \,, \qquad
\xi_{2} = \dn(\tilde \tau | k) \, e^{i u_2 \tilde \sigma} \,,
\label{profiles stat-f}
\end{equation}
with the constraint $u_{1}^{2}-u_{2}^{2}=1$\,. Since the radial
direction is independent of $\sigma$\,, we may regard $\mu$ as a
free parameter satisfying $N_1 = \mu u_1$ and $N_2 = \mu u_2$\,.
Then the conserved charges for a single period become
\begin{align}
{\cal E} = \pi \sqrt{k^{2}N_1^2 + \ko{1-k^{2}} N_2^2} \,, \qquad
{\cal J}_1 = {\cal J}_2 = 0\,.
\label{charges stat-f}
\end{align}
Left of Figure \ref{fig:pulsating} shows the time evolution of the
type $(i)'$ pulsating string. It stays in the northern hemisphere, and
sweeps back and forth between the north pole $(\theta=\pi/2)$
and the turning latitude determined by $k$\,.

When set $u_{2}=0$\,, this string becomes identical to the
simplest pulsating solution studied in \cite{Minahan:2002rc} (the
zero-rotation limit of ``rotating and pulsating'' strings studied in
\cite{Engquist:2003rn, Kruczenski:2004cn}).\footnote{\, The type
$(i)'$ pulsating solution studied here and also the type $(ii)'$
pulsating string discussed later are qualitatively different
solutions from the ``rotating and pulsating'' string of
\cite{Engquist:2003rn}, so that the finite-gap interpretation and
the gauge theory interpretation of type $(i)'$ and $(ii)'$ are
also different from those of \cite{Engquist:2003rn}. }

\subsubsection*{$\bullet$ \bmt{k\to 1} limit\,:~ Single-spike strings}

When the moduli parameter $k$ goes to unity, type $(i)'$ helical
string becomes an array of single-spike strings studied in
\cite{Ishizeki:2007we, Mosaffa:2007ty}. The dependence on $\omega_2$
drops out in this limit, so we write $\omega$ instead
$\omega_1$\,. The Virasoro constraints can be explicitly solved by
setting $a=u_{1}$ and $b=\tan\omega$\,. The profile of the string
then becomes
\begin{align}
\eta_{0} = \sqrt{1 + u_2^2}\, \tilde \tau  \,,\qquad
\xi_{1} = \frac{\sinh(T  - \iom)}{\cosh(T)} \, e^{i \tan(\omega)T + i u_1 X}\,,\qquad
\xi_{2} = \frac{\cos (\omega)}{\cosh(T)} \; \, e^{i u_2 X} 
\label{profiles spk}
\end{align}
with the constraint
$u_{1}^{2}-u_{2}^{2}=1+\tan^{2}\omega$\,.\footnote{\, Here
$u_{1,2}$ and $\om$ are related to $\gamma$ used in
\cite{Ishizeki:2007we} (see their equation (6.23)) by
$u_{1}=1/\cos\gamma\cos\om$ and
$u_{2}=\tan\gamma/\cos\om$\,. } The conserved charges are
computed as
\begin{equation}
{\cal E} = \pare{\frac{u_1^2 - \tan^2 \omega}{\tan \omega}} \eK (1) \,, \qquad
{\cal J}_1 = u_1 \cos^2 \omega \,, \qquad
{\cal J}_2 = u_2 \cos^2 \omega \,,
\label{charges spk}
\end{equation}
where $\eK (1)$ is a divergent constant.
For $n=1$ case ({\em i.e.}, single spike), the expressions \eqref{charges spk} result in
\begin{equation}
{\cal J}_1 = \sqrt{ {\cal J}_2^2 + \cos^{2}\omega }\,,\qquad
{\it i.e.},\quad
J_{1}=\sqrt{J_{2}^{2}+\f{\lam}{\pi^{2}}\cos^{2}\omega}\,.
\label{spin spin}
\end{equation}

Since the winding number $\Delta \varphi_{1}$ also diverges as
$k\to 1$\,, this limit can be referred to as the ``infinite
winding'' limit,\footnote{\, Notice, however, that the string
wraps very close to the equator but touches it only once every
period (every ``cusp''). } which can be viewed as the
2D transformed version of the ``infinite spin'' limit of
\cite{Hofman:2006xt}. By examining the periodicity condition
carefully, one finds that both of the divergences come from the
same factor $\eK(k)\big|_{k\to 1}$\,. Using the formula
(\ref{leading Zeta}), one can deduce that
\begin{equation}
\left. {\mathcal E}-\f{\Delta\varphi_{1}}{2}\right|_{k\to 1}
=-\ko{\omega - \frac{\pare{2 \ssp n'_1 + 1}\pi}{2} } \equiv \bar \theta \,.
\label{ene-wind}
\end{equation}
Using the $\bar \theta$ variable introduced above, which is the same definition as used in \cite{Ishizeki:2007we}, one can see \eqref{spin spin} precisely reproduces the relation between spins obtained in \cite{Ishizeki:2007we}.\footnote{Let us comment on a subtlety about $v \to 0$ (or equivalently $\om\to 0$) limit of a single spike string. It is easy to see that the profile of single-spike solution \eqref{profiles spk} with $\omega=0$ agrees with that of pulsating string solution \eqref{profiles stat-f} with $k=1$\,.
However, due to a singular nature of the $v \to 0$ limit, the angular momenta of both solutions \eqref{spin spin} and \eqref{charges stat-f} do not agree if we just naively take the limits on both sides.}

\subsubsection*{$\bullet$ \bmt{k\to 0} limit\,:~ Rational circular (static) strings}

The constraints among the parameters are the same as the $k\to 0$ limit of the type $(i)$ string, and the profile becomes
\begin{equation}
\eta_{0} = \sqrt U \tilde \tau\,,\qquad
\xi_{1} =0\,,\qquad
\xi_{2} =e^{i \sqrt U \tilde \sigma}\,.
\label{rational static prof}
\end{equation}
This is an unstable string that has no spins and just wraps around one of the great circles, and can be viewed as the $\ts$ transformed version of a point-like, BPS string with $E-(J_{1}+J_{2})=0$\,.
The conserved charges for a single period reduce to $\cE= \pi\mu \sqrt{U}$ and $\cJ_1=\cJ_2=0$\,.
The winding number for the $\varphi_{2}$\,-direction becomes
$N_{2}=\mu \sqrt{U}$\,, so the energy can also be written as 
\begin{equation}
E=N_{2}\sqrt{\lam}\,.
\label{winding energy}
\end{equation}
This result will be suggestive when we
discuss gauge theory in Appendix \ref{sec:gauge theory},
since it predicts that the conformal dimension of SYM dual
operator, which should be an $SO(6)$ singlet state, is also given
by ${\rm (integer)}\times \sqrt{\lam}$ in this limit.
Note also that in the limit $\mu \sqrt{U} \to \infty$\,, the profile \eqref{rational static prof} agrees with the $\omega = \pi/2$ case of the single-spike string after the interchange $\xi_1 \leftrightarrow \xi_2$\,.
We will also refer to this fact in the gauge theory discussion in Appendix \ref{sec:gauge theory}.

\subsection{Type \bmt{(ii)'} Helical Strings\label{sec:type (ii)'}}

\begin{figure}[p]
\begin{center}
\vspace{1.5cm}
\includegraphics[scale=0.9]{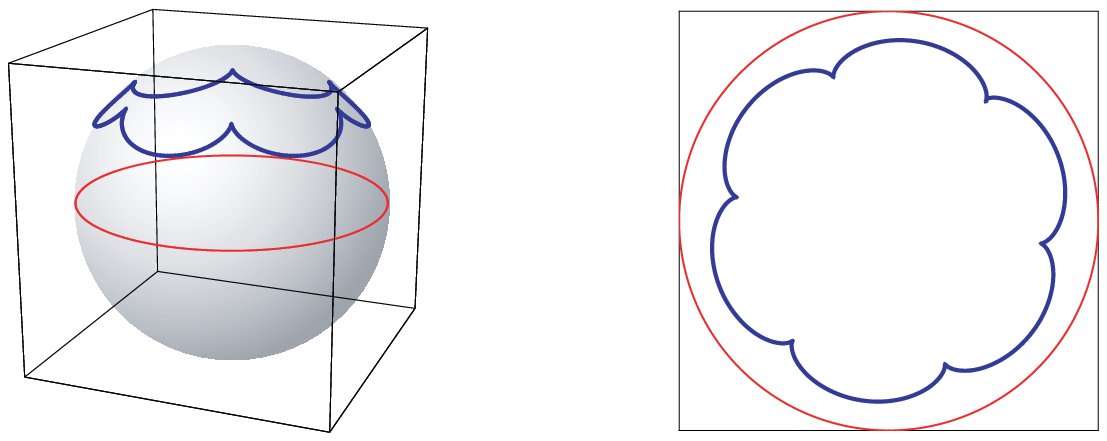}

\vspace{1.5cm}
\includegraphics[scale=0.9]{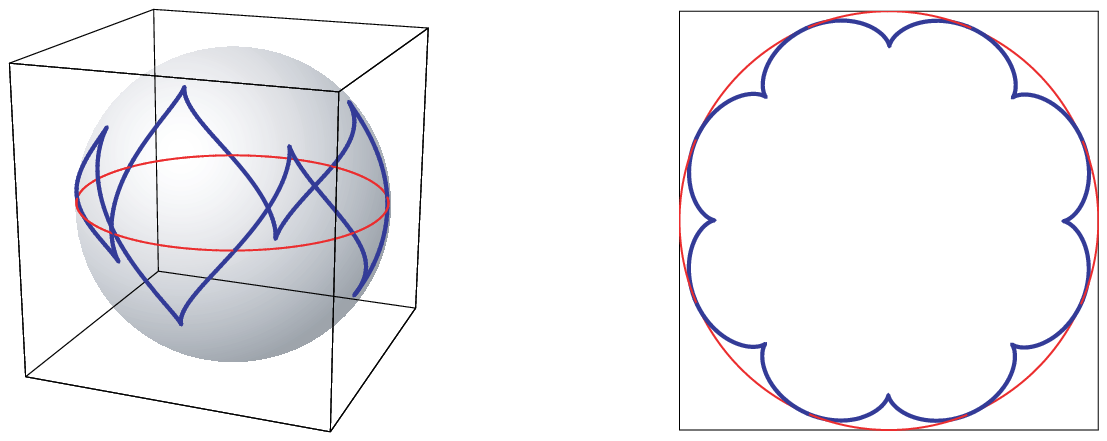}

\vspace{1.5cm}
\includegraphics[scale=0.9]{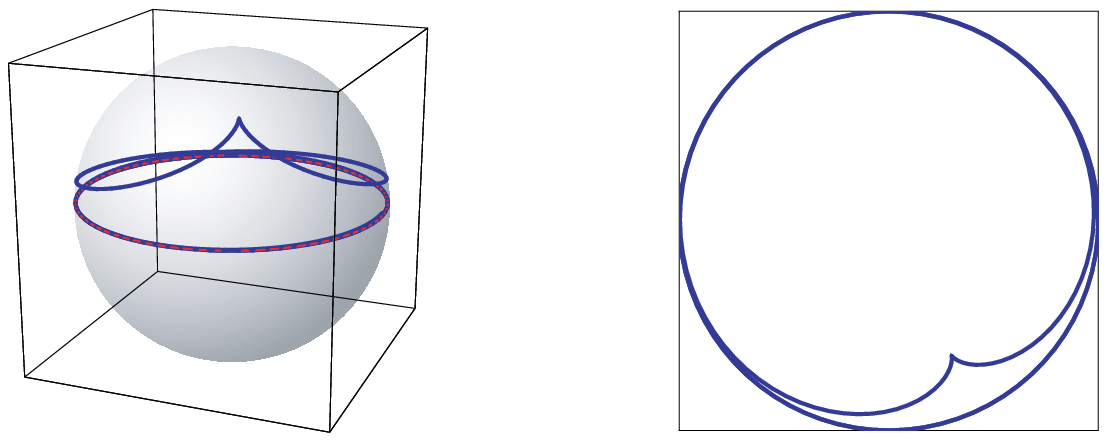}
\vspace{0.5cm}
\caption{\small {\sc Top}\,: Type $(i)'$ helical string
$(k=0.68\,, n=6)$\,, projected onto $S^{2}$\,. 
The (red) circle indicates
the $\theta=0$ line (referred to as the ``equator'' in the main
text).
{\sc Middle}\,: Type $(ii)'$ helical string $(k=0.40\,, m=8)$\,.
{\sc Bottom}\,: The $k\to 1$ limit of type $(i)'$ helical string\,: single-spike string $(\omega=0.78)$\,.
All three diagrams show the single-spin cases $(u_{2}=\om_{2}=0)$\,.}
\label{fig:T-helical}
\end{center}
\end{figure}

\begin{figure}[tp]
\begin{center}
\vspace{0.5cm}
\includegraphics{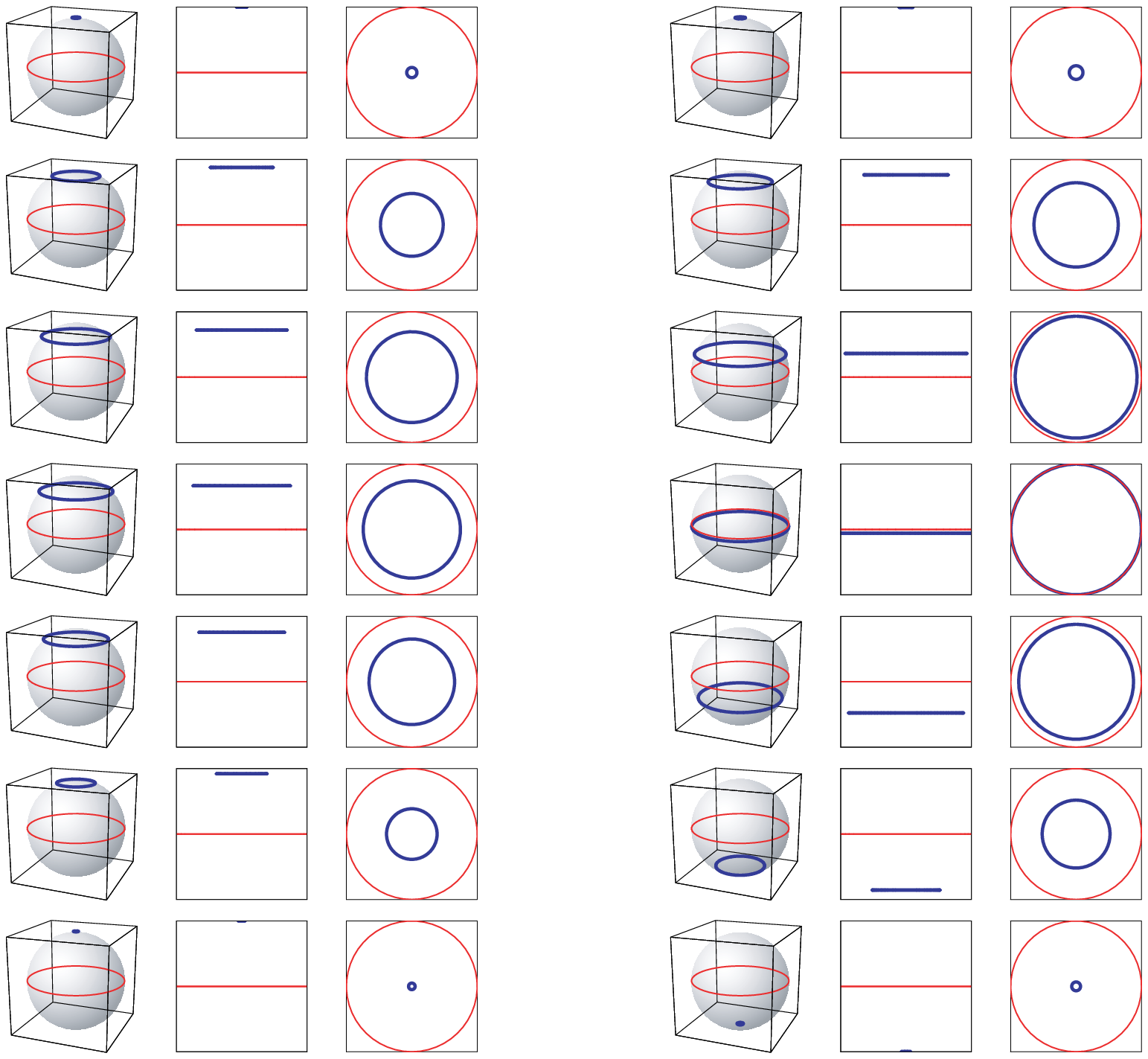}
\vspace{0.5cm}
\caption{\small In the $\om_{1,2}\to 0$ limit, the type
$(i)'$ (Left figures) and type $(ii)'$ (Right figures) helical
strings reduce to different types of pulsating strings. Their
behaviors are different in that the type $(i)'$ sweeps back and
forth only in the northern hemisphere with turning latitude controlled
by the elliptic modulus $k$\,, while the type $(ii)'$ pulsates on the
entire sphere, see Section \ref{sec:type (ii)'}. For the type
$(ii)'$ case, we only showed half of the oscillation period (for
the other half, it sweeps back from the south pole to the northern pole). }
\label{fig:pulsating}
\end{center}
\end{figure}

The type $(ii)'$ solution can be obtained from the type $(i)'$ solutions, either by shifting $\omega_2\mapsto \omega_2+\eK'$ or by transforming $k$ to $1/k$\,.
The profile is given by\footnote{\,
We use a hat to indicate type $(ii)'$ variables (not to be confused with the one for type $(ii)$ variables).}
\begin{align}
\hat \xi_{1} &= \hat C \frac{\bTh _0 (0|k)}{\sqrt{k} \, \bTh _0 (\iomm{1}|k)} \frac{\bTh _1 (T - \iomm{1}|k)}{\bTh _0 (T|k)} \, \exp \Big( \eZ_0 (\iomm{1}|k) T + i u_1 X\Big)\,,
\label{zc1-T} \\[2mm]
\hat \xi_{2} &= \hat C \frac{\bTh _0 (0|k)}{\sqrt{k} \,\bTh _3 (\iomm{2}|k)} \frac{\bTh _2 (T - \iomm{2}|k)}{\bTh _0 (T|k)} \, \exp \Big( \eZ_3 (\iomm{2} |k) T + i u_2 X \Big)\,,
\label{zc2-T}
\end{align}
where $\hat C$ is the same normalisation constant as (\ref{normalisation ii}).
The AdS coordinate is the same as the type $(ii)$ case, (\ref{zc0}).
See Figures \ref{fig:T-helical} and \ref{fig:pulsating} for the diagrams.

The equations of motion force $u_{1}$ and $u_{2}$ to satisfy the same relation as (\ref{u1-u2:c}),
and the Virasoro conditions impose the same constraints as (\ref{a,b-c1}, \ref{a,b-c2}).
As in the type $(i)'$ case, we can set the AdS-time to be $\hat \eta_{0} = \sqrt{\hat a^2 - \hat b^2}\,\tilde \tau$ with $\hat v \equiv \hat b/\ssp \hat a \le 1$\,.
The periodicity conditions for the type $(ii)'$ solutions become
\begin{alignat}{2}
\Delta \sigma &\equiv \frac{2 \pi}{m} = \frac{2 \eK \sqrt{1 - \hat v^2}}{\hat v \ssp \mu} \,,
 \\[2mm]
\Delta \varphi_1 &\equiv \frac{2 \pi M_1}{m} = 2 \eK \pare{ \frac{u_1}{\hat v} + i \eZ_0 (\iomm{1}) } + \pare{2 \ssp m'_1 + 1} \pi  \,,
 \\[2mm]
\Delta \varphi_2 &\equiv \frac{2\pi M_2}{m} = 2 \eK \pare{ \frac{u_2}{\hat v} + i \eZ_3 (\iomm{2}) } + \pare{2 \ssp m'_2 + 1} \pi  \,,
\end{alignat}
where $m=1,2,\dots$ counts the number of periods in $0\leq \sigma
\leq 2\pi$\,, and $M_{1, 2}$ are the winding numbers in the
$\varphi_{1,2}$-directions respectively, and $m'_{1,2}$ are
integers. The conserved charges are given by
\begin{align}
\hat {\cal E} &= \frac{m a (1 - v^2)}{v} \, \eK = \frac{n (a^2 - b^2)}{b} \, \eK \,,
\label{t-c E} \\[2mm]
\hat {\cal J}_1  &= \frac{m \ssp \hat C^2 \, u_1 }{k^2} \cpare{ \eE - \left( {\dn^2 (\iomm{1}) + \frac{i \ssp k^2}{\hat v \ssp u_1} \, \sn (\iomm{1}) \cn (\iomm{1}) \dn (\iomm{1}) } \right) \eK } \,,
\label{t-c J1} \\[2mm]
\hat {\cal J}_2  &= \frac{m \ssp \hat C^2 \, u_2 }{k^2} \cpare{ - \eE + (1-k^2) \pare{ \frac{1}{\dn^2 (\iomm{2})} - \frac{i \ssp k^2}{\hat v \ssp u_2} \, \frac{\sn (\iomm{2})\cn(\iomm{2})}{\dn^3(\iomm{2})}} \eK }\,.
\label{t-c J2}
\end{align}

Just as in the type $(i)\leftrightarrow (i)'$ case, the winding
numbers and the conserved charges of the type $(ii)$ and
$(ii)'$ strings are related by ${\cal \hat E}^{(ii)'} (\hat a, \hat b) = - {\cal \hat
E}^{(ii)} (\hat b, \hat a)$\,, ${\cal \hat J}_{1,2}^{(ii)'}(\hat v) = - {\cal
\hat J}_{1,2}^{(ii)} (-1/\hat v)$ and $M_{1,2}^{(ii)'}(\hat v) = - M_{1,2}^{(ii)}(-1/\hat v)$\,.

\paragraph{}
As in the type $(i)'$ case, we can take various limits.

\subsubsection*{$\bullet$ \bmt{\om_{1,2}\to 0} limit\,:~ Pulsating strings}

When both $\om_{1}$ and $\om_{2}$ go to zero, the type $(ii)'$ string becomes a pulsating string.
In contrast to the type $(i)'$ pulsating string, this type $(ii)'$ pulsating string pulsates on the entire sphere.
See the right diagrams of Figure \ref{fig:pulsating} for its time evolution. 

In this limit, the profile (\ref{zc0}, \ref{zc1-T}, \ref{zc2-T}) reduces to
\begin{equation}
\hat \eta_{0} = \sqrt{1 + u_2^2} \ \tilde \tau \,, \qquad
\hat \xi_{1} = \sn(\tilde \tau |k) \, e^{i u_1 \tilde \sigma} \,, \qquad
\hat \xi_{2} = \cn(\tilde \tau |k) \, e^{i u_2 \tilde \sigma}
\end{equation}
with the constraint $u_{1}^{2}-u_{2}^{2}=k^{2}$\,.
The conserved charges for a single period become
\begin{equation}
{\cal E} = \f{\pi}{k} \sqrt{M_1^2 + \pare{k^2 - 1} M_2^2} \,, \qquad
{\cal J}_1 = {\cal J}_2 = 0 \,.
\label{charges stat-c}
\end{equation}
Again, when set $u_{2}=0$\,,
this string reduces to the simplest pulsating solution studied in
\cite{Minahan:2002rc}.

\subsubsection*{$\bullet$ \bmt{k\to 1} limit\,:~ Single-spike strings}

The $k\to 1$ limit results in essentially the same solution as the type
$(i)'$ case, that is an array of single-spike strings. The only
difference is that while in the type $(i)'$ case every cusp
appears on the same side about the equator, say the northern
hemisphere, in the type $(ii)'$ case cusps appear in both the
northern and southern hemispheres in turn, each after an infinite
winding.

\subsubsection*{$\bullet$ \bmt{k\to 0} limit\,:~ Rational circular strings}

In the $k\to 0$ limit, the profile becomes
\begin{align}
\hat \eta_{0} = \sqrt{\hat a^2 - \hat b^2} \; \tilde \tau\,,\quad
\hat \xi_{1} = \hat C \; \sin (T - \iomm{1}) \, e^{i u_1 X}\,,\quad
\hat \xi_{2} = \hat C \; \cos (T - \iomm{2}) \, e^{i u_2 X}
\end{align}
with $\hat C = \pare{ \cosh^2 \omega_2 + \sinh^2 \omega_1
}^{-1/2}$ and $u_1^2 = u_2^2 = U + 1$\,. 
The Virasoro constraints imply the same set of relations as (\ref{pm}).
The
periodicity conditions become
\begin{alignat}{2}
\Delta \sigma &\equiv \frac{2 \pi}{m} = \frac{\pi \sqrt{1 - \hat v^2}}{\hat v \ssp \mu} \,, \\[2mm]
\Delta \varphi_1 &\equiv \frac{2 \pi M_1}{m} = \frac{\pi \ssp u_1}{\hat v} + \pare{2 \ssp m'_1 + 1} \pi \,, \\[2mm]
\Delta \varphi_2 &\equiv \frac{2\pi M_2}{m} = \frac{\pi \ssp u_2}{\hat v} + \pare{2 \ssp m'_2 + 1} \pi \,.
\end{alignat}
The conserved charges for a single period are evaluated as
\begin{equation}
\hat {\cal E} = \frac{\pi \hat a \pare{1 - \hat  v^2}}{2 \ssp \hat v}  \,, \qquad
\hat {\cal J}_1  = \frac{\pi \ssp \hat C^2}{2 \hat v} \, \sinh \omega_1 \cosh \omega_1 \,, \qquad
\hat {\cal J}_2  = - \frac{\pi \ssp \hat C^2}{2 \hat v} \, \sinh \omega_2 \cosh \omega_2 \,.
\end{equation}

\subsubsection*{Summary\, \----\, The complete catalogue of elliptic strings on $\bmt{\mathbb R\times S^{3}}$\, \----}

In \cite{Okamura:2006zv}, we constructed the most general elliptic (``two-cut'') classical string solutions on $\mathbb R\times S^{3}\subset AdS_{5}\times S^{5}$\,, called helical strings.
They were shown to include various strings studied in the {\em large-spin} sector. Schematically, the family tree reads
\begin{alignat}{5}
&\, {\rm I}&{}:\quad &{}
\begin{array}{l}
\mbox{Type $(i)$ helical string} \\
\mbox{with generic $k$ and $\om_{1,2}$}
\end{array}
~ &\longrightarrow~ &
\left\{
\begin{array}{ll}
\mbox{- Point-like (BPS), rotating string} & (k\to 0) \\
\mbox{- Array of dyonic giant magnons} & (k\to 1) \\
\mbox{- Elliptic, spinning folded string} & (\om_{1,2}\to 0)
\end{array}
\right.\,,\no\\[2mm]
&{\rm II}&{}:\quad &{}
\begin{array}{l}
\mbox{Type $(ii)$ helical string} \\
\mbox{with generic $k$ and $\om_{1,2}$}
\end{array}
~ &\longrightarrow~ &
\left\{
\begin{array}{ll}
\mbox{- Rational, spinning circular string} & (k\to 0) \\
\mbox{- Array of dyonic giant magnons} & (k\to 1) \\
\mbox{- Elliptic, spinning circular string} & (\om_{1,2}\to 0)
\end{array}
\right.\,.\no
\end{alignat}
Moreover, the single-spin limit of the type $(i)$ helical strings agrees with so-called ``spiky strings'' studied in \cite{Ryang:2005yd, Arutyunov:2006gs, Astolfi:2007uz}.\footnote{The two-spin helical strings are different from the so-caled ``spiky strings'' in that they have no singular points in spacetime. 
When embedded in $\mathbb R\times S^{3}$\,, the singular ``cusps'' of the spiky
strings that apparently existed on $\mathbb R\times S^{2}$ are all smoothed out to result in non-spiky profiles.}

\paragraph{}
In contrast, in \cite{Hayashi:2007bq}, we explored another branch of helical strings.
This includes a {\em large-winding} sector where $m\sqrt{\lam}$ becomes of the same order as the energy which diverges ($m$ being the winding number).
We saw that when classical strings on $\mathbb R \times S^3 \subset AdS_5\times S^5$ are considered in conformal gauge, an operation of interchanging $\tau$ and $\sigma$\,, as well as keeping temporal gauge $t \propto \tau$\,, maps the original helical strings to
another type of helical strings.
Roughly speaking, rotating/spinning solutions with large spins became oscillating
solution with large windings.
Again, schematically, we found\,:
\begin{alignat}{5}
&\, {\rm I}'&{}:\quad &{}
\begin{array}{l}
\mbox{Type $(i)'$ helical string} \\
\mbox{with generic $k$ and $\om_{1,2}$}
\end{array}
~ &\longrightarrow~ &
\left\{
\begin{array}{ll}
\mbox{- Rational, static circular string} & (k\to 0) \\
\mbox{- Array of single-spike strings} & (k\to 1) \\
\mbox{- Elliptic, type $(i)'$ pulsating string} & (\om_{1,2}\to 0)
\end{array}
\right.\,,\no\\[2mm]
&{\rm II}'&{}:\quad &{}
\begin{array}{l}
\mbox{Type $(ii)'$ helical string} \\
\mbox{with generic $k$ and $\om_{1,2}$}
\end{array}
~ &\longrightarrow~ &
\left\{
\begin{array}{ll}
\mbox{- Rational circular string} & (k\to 0) \\
\mbox{- Array of single-spike strings} & (k\to 1) \\
\mbox{- Elliptic, type $(ii)'$ pulsating string} & (\om_{1,2}\to 0)
\end{array}
\right.\,.\no
\end{alignat}

In Appendix \ref{app:AdS helicals}, similar helical solutions on $AdS_{3}\times S^{1}$ are discussed, which are also constructed in \cite{Hayashi:2007bq}.
They include the ``$SL(2)$ (dyonic) giant magnons'' and the two-spin folded strings of \cite{Frolov:2002av} as special limiting cases.

\section{Finite-gap Interpretations\label{sec:FG}}

\subsubsection*{Overview}

As is clear to those who are fluent in the finite-gap language, the profiles of type $(i)$\,, $(ii)$\,, $(i)'$ and $(ii)'$ strings, (\ref{zf0}\,-\,\ref{zf2}), (\ref{zc0}\,-\,\ref{zc2}), (\ref{zf0}, \ref{zf1-T}, \ref{zf2-T}) and (\ref{zc0}, \ref{zc1-T}, \ref{zc2-T}) respectively, are closely related to the so-called Baker-Akhiezer function\,; see \cite{Dorey:2006zj} and references therein.
In our case, the typical form of the profiles tells us that they are described by two cuts in the spectral parameter plane, just as are the well-known cases with folded and circular strings.
In order for the charges to be real, the four branch-points of the cuts must satisfy the so-called reality constraint, that is, they must locate symmetrically with respect to the real axis.
Then the most general ansatz for the location of the branch-points $\{ x_{k} \}_{k=1,\dots,4}\in {\mathbb C}$ for our helical solutions can be written as $x_{1}=i\rho\,e^{-i(\al+\delta)}$\,, $x_{2}=x_{1}^{*}$\,, $x_{3}=-x_{2}\,e^{-i\delta}/\rho$ and $x_{4}=x_{3}^{*}$\,, where $\rho$\,, $\alpha$ and $\delta$ are real parameters.
As compared to the folded or circular string cases, there are now extra degrees of freedom $\rho$ and $\al$\,.
Another remark is that, in the $k\to 1$ limit, the type $(i)$ solutions can be switched to the type $(ii)$ branch without energy costs, changing the number of ``spikes'' ($n$) into the number of crossing the equator ($m$).
This can be understood as changing of the periods for $\mathcal A$- and $\mathcal B$-cycles defined for the two cuts, according to particular modular transformations of the elliptic functions. 

Indeed in \cite{Vicedo:2007rp}, the type $(i)$ (and $(ii)$) helical strings were reconstructed as finite-gap solutions (see also \cite{Dorey:2006zj}).
The type $(i)$ and $(ii)$ helical strings (\ref{zf1}, \ref{zf2}) were shown to be equivalent to the most general elliptic (two-cut) finite-gap solution for $\mathbb R\times S^{3}\subset AdS_{5}\times S^{5}$ sigma model, with both cuts intersecting the real axis within the interval $(-1,1)$ (see Figure \ref{fig:2-cuts} (a)).\footnote{Our convention for the spectral parameter used in this section is explained in the next footnote.}
On the other hand, the $\ts$ transformed helical string (\ref{zf1-T}, \ref{zf2-T}) of type $(i)'$ $(ii)'$ strings are again described as the most general two-cut finite-gap solutions, the only difference from the type $(i)$ and $(ii)$ cases being the asymptotic behaviors of differentials at $x\to \pm 1$ (or equivalently, different
configurations of cuts with respect to interval $(-1, 1)$\,, see Figure \ref{fig:2-cuts} (b)). 
Below we will see those features in more detail.

\subsubsection*{Preliminaries}

Recall first from \cite{Vicedo:2007rp} that the $(\sigma,\tau)$-dependence of the general finite-gap solution enters solely through the differential form
\begin{equation} \label{dQ}
d \mathcal{Q}(\sigma, \tau) = \frac{1}{2 \pi} \left( \sigma dp +
\tau dq \right)\,,
\end{equation}
where $dp$ and $dq$ are the differentials of the quasi-momentum and quasi-energy defined below by their respective asymptotics near the points $x = \pm 1$\,.\footnote{The spectral parameter $x$ used in this section is the ``rescaled'' one, which is more convenient in studying the strong coupling region $g\gg 1$\,.
It is related to the spectral parameter $x_{\rm old}$ used in Section \ref{sec:SBAE} as $gx=x_{\rm old}$\,.
Therefore, the singularities at $x_{\rm old}=\pm g$ the string Bethe ansatz equations (\ref{SBA rho}) (originated from (\ref{L and M})) had are now translated to $x=\pm 1$\,.
Notice also that the $x^{\pm}$ spectral parameters introduced in (\ref{x(u)}) were also ``rescaled'' ones.}
The differential multiplying $\sigma$ in $d\mathcal{Q}(\sigma,\tau)$ (namely $dp$) is related to the eigenvalues of the monodromy matrix (\ref{Omega}), which by definition is the parallel transporter along a closed loop $\sigma \in [ 0, 2 \pi)$ on the worldsheet. This is because the Baker-Akhiezer vector $\bmt{\psi}(P,\sigma,\tau)$\,, whose $(\sigma,\tau)$-dependence also enters solely through the differential form $d\mathcal{Q}(\sigma, \tau)$ in \eqref{dQ}, satisfies \cite{Dorey:2006zj}
\begin{equation*}
\bmt{\psi}(P,\sigma + 2 \pi,\tau) = \exp \left\{ i \int_{\infty^+}^P dp
\right\} \bmt{\psi}(P,\sigma,\tau)\,.
\end{equation*}
Now it is clear from \eqref{dQ} that the $\sigma \leftrightarrow \tau$ operation can be realised on the general
finite-gap solution by simply interchanging the quasi-momentum with the so-called quasi-energy,\footnote{For the definitions of $dp$ and $dq$\,, see below.}
\begin{equation} \label{dp-dq}
dp ~\leftrightarrow~ dq\,.
\end{equation}
However, since we wish $dp$ to always denote the differential related to the eigenvalues of the monodromy matrix, by the above argument it must always appear as the coefficient of $\sigma$ in $d\mathcal{Q}(\sigma, \tau)$\,. Therefore equation \eqref{dp-dq} should be interpreted as saying that the respective definitions of the
differentials $dp$ and $dq$ are interchanged, but $d\mathcal{Q}(\sigma, \tau)$ always takes the same form as in
\eqref{dQ}.

\begin{figure}[htbp]
\begin{center}
\vspace{0.5cm}
\includegraphics{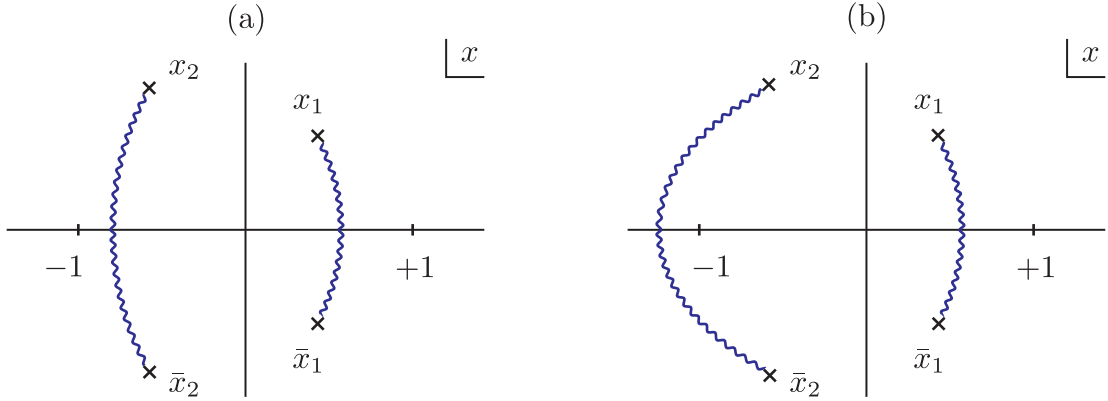}
\vspace{0.5cm}
\caption{\small Different possible arrangements of cuts relative to $x
= \pm 1$\,: (a) corresponds to the helical string, (b) corresponds to
the $\ts$ transformed helical string.}
\label{fig:2-cuts}
\end{center}
\end{figure}

Before proceeding let us note the precise definitions of these differentials $dp$ and $dq$\,. Consider an algebraic curve $\Sigma$\,, which admits a hyperelliptic representation with cuts.
For what follows it will be important to specify the position of the different cuts relative to the points $x = \pm 1$\,, {\em i.e.}, Figures \ref{fig:2-cuts} (a) and \ref{fig:2-cuts} (b) are to be distinguished for the purpose of defining $dp$ and $dq$\,.
We could make this distinction by specifying an equivalence relation on representations of $\Sigma$ in terms of cuts, where two representations are equivalent if the cuts of one can be deformed into the cuts of the other within $\mathbb{C}\!\setminus\!\{ \pm 1\}$\,. It is straightforward to see that there are only two such equivalence classes for a general algebraic curve $\Sigma$\,. 
For example, in the case of an elliptic curve $\Sigma$ the representatives of these two equivalence classes are given in Figures \ref{fig:2-cuts} (a) and \ref{fig:2-cuts} (b).
Now with respect to a given equivalence class of cuts, the differentials $dp$ and $dq$ can be uniquely defined on $\Sigma$ as in \cite{Dorey:2006zj} by the following conditions\,:
\begin{itemize}
\item[$(1)$] their $\mathcal A$-period vanishes.
\item[$(2)$] their respective poles at $x = \pm 1$ are of the following form, up to a trivial overall change of sign
(see \cite{Vicedo:2007rp}),\footnote{Recall the asymptotics (\ref{x -> pm 1}) and the relation $E=\sqrt{\lam}\,\kappa=4\pi g\kappa$\,.}

\begin{equation} \label{dp asymptotics at pm 1}
dp(x^{\pm}) \underset{x \rightarrow +1}\sim \mp \frac{\pi \kappa dx}{(x -
1)^2}\,, \qquad dp(x^{\pm}) \underset{x \rightarrow -1}\sim \mp
\frac{\pi \kappa dx}{(x + 1)^2}\,,
\end{equation}
\begin{equation} \label{dq asymptotics at pm 1}
dq(x^{\pm}) \underset{x \rightarrow +1}\sim \mp \frac{\pi \kappa dx}{(x -
1)^2}\,, \qquad dq(x^{\pm}) \underset{x \rightarrow -1}\sim \pm
\frac{\pi \kappa dx}{(x + 1)^2}\,,
\end{equation}
where $x^{\pm} \in \Sigma$ denotes the pair of points above $x$\,, with $x^{+}$ being on the physical sheet, and $x^{-}$ on the other sheet.\footnote{They should not be confused with AdS/CFT spectral parameters (\ref{x(u)}).}
\end{itemize}
Once the differentials $dp$ and $dq$ have been defined by (\ref{dp asymptotics at pm 1}, \ref{dq asymptotics at pm 1}) with respect to a given equivalence class of cuts, one can move the cuts around into the other equivalence class (by crossing say $x = - 1$ with a single cut) to obtain a representation of $dp$ and $dq$ with respect to the other equivalence class of cuts.
So for instance, if we define $dp$ and $dq$ by \eqref{dp asymptotics at pm 1} and \eqref{dq asymptotics at pm 1} with respect to the equivalence class of cuts in Figure \ref{fig:2-cuts} (a), then with respect to the equivalence
class of cuts in Figure \ref{fig:2-cuts} (b) the definition of $dp$ will now be \eqref{dq asymptotics at pm 1} and that of $dq$ will now be \eqref{dp asymptotics at pm 1}.

In summary, both equivalence classes of cuts represents the very same algebraic curve $\Sigma$\,, but each equivalence class gives rise to a different definition of $dp$ and $dq$\,.
So the two equivalence classes of cuts give rise to two separate finite-gap solutions but which can be related by a $\ts$ transformation \eqref{dp-dq}.
Indeed, if in the construction of \cite{Vicedo:2007rp}, the generic configuration of cuts is assumed to be the one given in Figure \ref{fig:2-cuts} (b), instead of Figure \ref{fig:2-cuts} (a) as was done in originally, then the resulting solution becomes the generic helical string but with
\begin{equation*}
X ~\leftrightarrow~ T\,,
\end{equation*}
namely the 2D transformed helical string (\ref{zf1-T}, \ref{zf2-T}). Therefore, with $dp$ and $dq$ defined as above by their respective asymptotics (\ref{dp asymptotics at pm 1}, \ref{dq asymptotics at pm 1}) at $x = \pm 1$\,, the type $(i)$ and $(ii)$ helical string of \cite{Okamura:2006zv,Vicedo:2007rp} is the general finite-gap solution corresponding to the class represented by Figure \ref{fig:2-cuts} (a), whereas the 2D transformed versions, type $(i)'$ and $(ii)'$ helical string corresponds to the most general elliptic finite-gap solution on $\mathbb R\times S^{3}$ with cuts in the other class represented in Figure \ref{fig:2-cuts} (b).

\subsubsection*{Type $\bmt{(i)}$ and $\bmt{(ii)}$ helical strings}

We can obtain expressions for the global charges $J_1=(J_{\rm L}+J_{\rm R})/2$\,, $J_2 = (J_{\rm L}-J_{\rm R})/2$ along the same lines as in \cite{Vicedo:2007rp}.
In terms of the differential form
\begin{equation}
\alpha \equiv \frac{\sqrt{\lambda}}{4 \pi} \left( x + \frac{1}{x}
\right) dp\,, \qquad \tilde{\alpha} \equiv \frac{\sqrt{\lambda}}{4
\pi}\left( x - \frac{1}{x} \right) dp\,,
\end{equation}
we can write
\begin{align}
\begin{split}
J_1 &= -\text{Res}_{0^+} \alpha + \text{Res}_{\infty^+} \alpha =
\text{Res}_{0^+} \tilde{\alpha} + \text{Res}_{\infty^+}
\tilde{\alpha}\,, \\
J_2 &= -\text{Res}_{0^+} \alpha - \text{Res}_{\infty^+} \alpha\,. 
\label{J1J2 res}
\end{split}
\end{align}
Note that $\alpha$ and $\tilde{\alpha}$ both have simple poles at $x = 0$\,, $\infty$\,, but $\tilde{\alpha}$ also has simple poles at $x = \pm 1$ coming from the double poles in $dp$ at $x = \pm 1$\,.
It then follows that we can rewrite \eqref{J1J2 res} as
\begin{align}
\begin{split}
J_1 &= - \sum_{I=1}^2 \frac{1}{2 \pi i} \int_{{\mathcal A}_I} \tilde{\alpha} -
\text{Res}_{(+1)^+} \tilde{\alpha} - \text{Res}_{(-1)^+} \tilde{\alpha}\,,\\
J_2 &= \sum_{I=1}^2 \frac{1}{2 \pi i} \int_{{\mathcal A}_I} \alpha\,,
\label{J1J2 fg}
\end{split}
\end{align}
where ${\mathcal A}_I$ is the $\mathcal A$-cycle around the $I$-th cut.
The residues of $\tilde{\alpha}$ at $x= \pm 1$ are of the same sign (as a consequence of $p(x)$ having equal residues at $x = \pm 1$) so that their sum give the energy $E$ of the string, hence we have
\begin{equation} \label{J_1,J_2 periods}
E- J_1 = \sum_{I=1}^2 \frac{1}{2 \pi i} \int_{{\mathcal A}_I} \tilde{\alpha}\,,
\qquad J_2 = \sum_{I=1}^2 \frac{1}{2 \pi i} \int_{{\mathcal A}_I} \alpha\,.
\end{equation}

\begin{figure}[htbp]
\begin{minipage}{0.5\hsize}
\begin{center}
\vspace{0.5cm}
\includegraphics{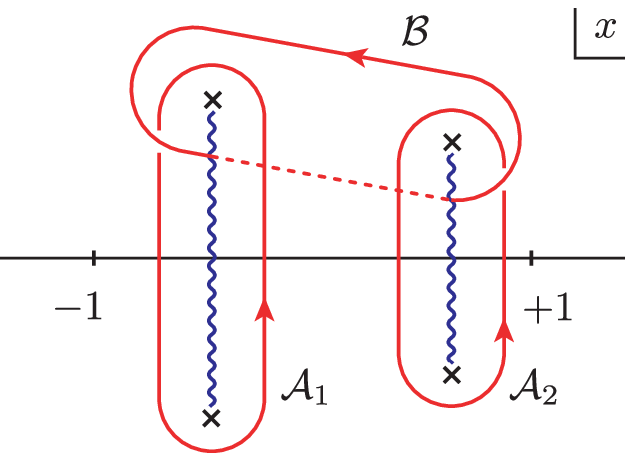}
\vspace{0.5cm}
\caption{\small Definitions of cycles.}
\label{fig:cycles2}
\end{center}
\end{minipage}
\begin{minipage}{0.5\hsize}
\begin{center}
\vspace{0.5cm}
\includegraphics{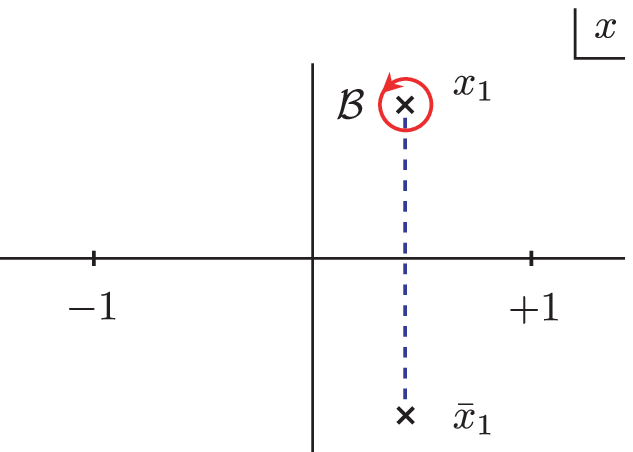}
\vspace{0.5cm}
\caption{\small $k\to 1$ limit of cuts.}
\label{fig:ss-FG}
\end{center}
\end{minipage}
\vspace{0.5cm}
\end{figure}

\paragraph{}
We would like to consider two types of limits here\,: (a) the symmetric cut limit (where the curve acquires the extra symmetry $x \leftrightarrow -x$) which corresponds to taking $\omega_{1,2} \rightarrow 0$ in the finite-gap solution, and (b) the singular curve limit which corresponds to taking the moduli of the curve to one, $k \rightarrow 1$\,.
In the symmetric cut limit, the discussion is identical to that in \cite{Vicedo:2007rp} (when working with the configuration of cuts in Figure \ref{fig:2-cuts} (a)), in particular there are two possibilities corresponding to the type $(i)$ and type $(ii)$ cases, for which the cuts are symmetric with $x_1 = -\bar{x}_2$ and imaginary with $x_1 = -\bar{x}_1$\,, $x_2 = -\bar{x}_2$ respectively (see Figure 2 of \cite{Vicedo:2007rp}).
Here we parametrised the elliptic curve $\Sigma$ as 
\begin{equation}
\Sigma\,:~y(x)=(x-x_{1})(x-\bar{x}_{1})(x-x_{2})(x-\bar{x}_{2})\,.
\end{equation}

In the singular limit $k \to 1$ where both cuts merge into a pair of singular points at $x = x_1$\,, $\bar{x}_1$ \cite{Vicedo:2007rp}, the
sum of $\mathcal A$-cycles turns into a sum of cycles around the points $x_1$\,, $\bar{x}_1$\,, so that \eqref{J_1,J_2 periods} yields in this limit
\begin{equation} \label{J_1,J_2 res}
E- J_1 = \text{Res}_{x_1} \tilde{\alpha} + \overline{\text{Res}_{x_1}
\tilde{\alpha}}\,, \qquad
J_2 = \text{Res}_{x_1} \alpha + \overline{\text{Res}_{x_1} \alpha}\,.
\end{equation}
Moreover, in the singular limit $dp$ acquires simple poles at $x =
x_1$\,, $\bar{x}_1$ so that the periodicity condition about the $\mathcal B$-cycle, $\int_{\mathcal B} dp = 2 \pi n$\,, implies
\begin{equation}
\text{Res}_{x_1} dp = \frac{n}{i}\,.
\label{B fg}
\end{equation}
Let us set $n=1$ ($n$ can be easily recovered at any moment).
Then \eqref{J_1,J_2 res} simplifies to
\begin{align}
E- J_1 &= \frac{\sqrt{\lambda}}{4 \pi} \left| \left( x_1 -
\frac{1}{x_1} \right) -\left( \bar{x}_1 - \frac{1}{\bar{x}_1}
\right) \right|\,, \label{J_1 res-2}\\ J_2 &= \frac{\sqrt{\lambda}}{4 \pi}
\left| \left( x_1 + \frac{1}{x_1} \right) - \left( \bar{x}_1 +
\frac{1}{\bar{x}_1} \right) \right|\,.\label{J_2 res-2}
\end{align}
Then if we relate $\Delta\varphi_{1}$ for a dyonic giant magnon (see (\ref{bc}) for the definition) to the spectral data $x_1$ of the singular curve by the identification
\begin{equation}
\Delta\varphi_{1} = \f{1}{i} \ln
\left(\frac{x_1}{\bar{x}_1}\right), \label{p fg}
\end{equation}
the expressions (\ref{J_1 res-2}\,-\,\ref{p fg}) together
imply the relation
\begin{equation}
E-J_1 = \sqrt{J_2^2 + \frac{\lambda}{\pi^2} \sin^2
 \ko{\f{\Delta\varphi_{1}}{2}}}\,.
\end{equation}
This is the energy-spin relation for the dyonic giant magnons (\ref{energy_spin_CDO}).

\subsubsection*{Type $\bmt{(i)'}$ and $\bmt{(ii)'}$ helical strings}

As discussed before, a given finite-gap solution is not associated with a particular equivalence class of cuts; since $dp$ and $dq$ are defined relative to an equivalence class of cuts, one can freely change equivalence class provided one also changes the definitions of $dp$ and $dq$ with respect to this new equivalence class according to \eqref{dp-dq}, so that in the end $dp$ and $dq$ define the same differentials on $\Sigma$ in either representation.
For example, we can describe the 2D transformed helical string in two different ways: either we take the  configuration of cuts in Figure \ref{fig:2-cuts} (b) with $dp$ and $dq$ defined as usual by their asymptotics (\ref{dp asymptotics at pm 1}, \ref{dq asymptotics at pm 1}) at $x = \pm 1$\,, or we take the configuration of cuts in Figure \ref{fig:2-cuts} (a) but need to swap the definitions of $dp$ and $dq$ in (\ref{dp
asymptotics at pm 1}, \ref{dq asymptotics at pm 1}).
In the following we will use the latter description of Figure \ref{fig:2-cuts} (a) in order to take the singular limit $k \to 1$ where the cuts merge into a pair of singular points.

\paragraph{}
The discussion up to (\ref{J1J2 fg}) is the same the type $(i)$ and $(ii)$ cases.
The difference come about when we consider the residues of $\tilde{\alpha}$ at $x= \pm 1$\,.
They were of the same sign in the type $(i)$ and $(ii)$ cases, but after the $\ts$ transformation, the residues of $\tilde{\alpha}$ at $x = \pm 1$ become opposite (since $p(x)$ now has opposite residues at $x = \pm 1$) and therefore cancel in the expression for $J_1$ in (\ref{J1J2 fg}), resulting in the following expressions
\begin{equation} \label{J_1,J_2 periods-T}
- J_1 = \sum_{I=1}^2 \frac{1}{2 \pi i} \int_{{\mathcal A}_I} \tilde{\alpha}\,,
\qquad J_2 = \sum_{I=1}^2 \frac{1}{2 \pi i} \int_{{\mathcal A}_I} \alpha\,.
\end{equation}
In parallel to the discussion of the type $(i)$ and $(ii)$ helical string cases, there are two types of limits one can consider: (a) the symmetric cut limit $\om_{1,2}\to 0$\,, and (b) the singular curve limit $k \rightarrow 1$\,.
The symmetric cut limit $\om_{1,2}\to 0$ leads to the same situation as the type $(i)$ and $(ii)$ cases except the equivalence class.
In the $k\to 1$ limit, both cuts merge into a pair of singular points at $x = x_1$\,, $\bar{x}_1$ \cite{Vicedo:2007rp}, the sum of $\mathcal A$-cycles turns into a sum of cycles around the points $x_1$\,, $\bar{x}_1$\,, so that \eqref{J_1,J_2 periods} yields in this limit, setting $n=1$ again, 
\begin{align}
- J_1 &= \frac{\sqrt{\lambda}}{4 \pi} \left| \left( x_1 -
\frac{1}{x_1} \right) -\left( \bar{x}_1 - \frac{1}{\bar{x}_1}
\right) \right|\,, \label{J_1 res-2-T}\\[2mm]
J_2 &= \frac{\sqrt{\lambda}}{4 \pi}
\left| \left( x_1 + \frac{1}{x_1} \right) - \left( \bar{x}_1 +
\frac{1}{\bar{x}_1} \right) \right|\,.\label{J_2 res-2-T}
\end{align}
The periodicity condition in the singular limit is the same as (\ref{B fg}).
The energy $E = \sqrt{\lambda}\,\kappa = (n \sqrt{\lambda}/\pi)
\,\mathcal{E}$ diverges in the singular limit $k \to 1$\,, but
this divergence can be related to the one in $\Delta \varphi_1$ for the single-spike string (see \eqref{Dphi1_cl f}).
In the present case the $\sigma$-periodicity condition
$\int_{\mathcal B} dp \in 2 \pi \mathbb{Z}$ can be written as
({\em c.f.}, equation $(2.23)$ in \cite{Vicedo:2007rp})
\begin{equation*}
- \frac{2 \eK(k) \sqrt{1 - v^2}}{v} = 
\frac{2 \pi\kappa |x_1 - \bar{x}_2|}{n \sqrt{y_+ y_-}}\,,
\end{equation*}
where $y_{\pm} = y(x)\big|_{x = \pm 1} > 0$ and $v$ can be
expressed in the present setup as $v = \mbox{\large $\frac{y_+ - y_-}{y_+ + y_-}$}$
(see \cite{Vicedo:2007rp}). Using this $\sigma$-periodicity condition,
the energy can be expressed in the $k \to 1$ limit as
\begin{equation*}
\mathcal{E} = \frac{u_1}{v} (1 - v^2) \,\eK(1)\,.
\end{equation*}
We can relate this divergent expression with $\Delta \varphi_1$ in \eqref{Dphi1_cl f} which also diverge in the
limit $k \to 1$\,, by making use of the relation $u_1 v = \tan \omega_1$ (see
\cite{Vicedo:2007rp} where the notation is $u_1 = v_-$ and $\omega_1 =
\tilde{\rho}_-$), and find
\begin{equation}
\mathcal{E} - \frac{\Delta \varphi_1}{2} = - \left( \omega_1 -
\frac{(2 n'_1 + 1) \pi}{2} \right) \equiv \bar{\theta}\,. \label{E fg}
\end{equation}
Comparing this scenario with the one for the original helical strings in
\cite{Vicedo:2007rp}, we can write an expression for $\bar{\theta}$ in
terms of the spectral data $x_1$ of the singular curve. Identifying
\begin{equation}
\bar{\theta} = -\frac{i}{2} \ln
\left(\frac{x_1}{\bar{x}_1}\right), \label{theta fg}
\end{equation}
the expressions (\ref{J_1 res-2-T}, \ref{J_2 res-2-T}) and \eqref{theta fg} together
imply the relation
\begin{equation}
-J_1 = \sqrt{J_2^2 + \frac{\lambda}{\pi^2} \sin^2
 \bar{\theta}}\,.
\end{equation}
This is the same relation as (\ref{spin spin}).\footnote{The sign difference between (\ref{spin spin}) and here is not essential.}

\paragraph{}
Let us summarise.
In this section, we investigated both types (large-spin and large-winding) of helical
strings from the finite-gap perspective. We were able to
understand the effect of the $\tau \leftrightarrow \sigma$
operation as an interchange of quasi-momentum and quasi-energy.
The transformed helical strings were described as general two-cut
finite-gap solutions as in the original case \cite{Vicedo:2007rp},
the only difference being the asymptotic behaviors of
differentials at $x\to \pm 1$ (or equivalently, different
configurations of cuts with respect to interval $(-1, 1)$). By
expressing the charges in terms of spectral parameters
(branch-points of the cuts), the charge relations for single
spikes were also reproduced.

\section[Appendix for Chapter \ref{chap:OS}]{Appendix for Chapter \ref{chap:OS}\,: Single-spin limit}

In this appendix we collect some results of single-spin limits of various helical strings for readers' convenience.

\subsubsection*{Type $\bmt{(i)}$ single-spin helical string}

The single spin-limit corresponds to $u_{2} = \om_2 = 0$\,.
For the type $(i)$ string, the profile becomes
\begin{align}
\eta_{0} (T,X)&=a T+b X\qquad 
\mbox{with}\quad a =k\cn(\iom | k)\,,\quad b =-ik\sn(\iom | k)\,,
\label{eta0-f2}\\
\xi_{1} (T,X)&=\f{\sqrt{k}}{\dn(\iom | k)}\f{\bTh_{0}(0 | k)}{\bTh_{0}(\iom | k)}\f{\bTh_{1}(X-\iom | k)}{\bTh_{0}(X | k)}\, 
\exp\kko{\eZ_{0}(\iom | k)X+i\dn(\iom | k)T}\,,
\label{xi1-f2}\\
\xi_{2} (T,X)&=\f{\dn (X | k)}{\dn(\iom | k)}\,,
\label{xi2-f2}
\end{align}
with the periodicity conditions
\begin{alignat}{2}
\Delta \sigma \Big|_{\rm one\mbox{\tiny\,-\,}hop} &\equiv \frac{2 \pi}{n} = \frac{2 \eK \sqrt{1 - v^2}}{\mu} \,,
\label{Dsigma_cl} \\[2mm]
\Delta \varphi_1 \Big|_{\rm one\mbox{\tiny\,-\,}hop} &\equiv \frac{2 \pi N_{1}}{n} = 2 \eK \pare{ - i \eZ_0 (\iom) + \frac{i \sn (\iom) \dn (\iom)}{\cn (\iom)} } + \pare{2 \ssp n'_1 + 1}\pi \,,
\label{Dphi1_cl}
\end{alignat}
where $n = 1, 2, \ldots$\,, and $N_1\,, n'_1$ are integers.
Here $\varphi_{1}$ is the azimuthal angle defined in (\ref{S-coord}).
When $\sigma$ runs from $0$ to $2\pi$\,, an array of $n$ hops winds $N_{1}$ times in $\varphi_{1}$-direction in the target space, thus making the string closed.
The energy $E$ and the spin $J_{1}$ are computed as
\begin{equation}
{\cal E} = \frac{n k \, \eK}{\cn (\iom)} \,, \qquad
{\cal J}_{1} = \frac{n(\eK - \eE)}{\dn (\iom)} \,.
\label{1ch-f}
\end{equation}

\paragraph{}
In the $\om\to 0$ limit, the type $(i)$ single-spin solution reduces to a folded string solution studied in \cite{Gubser:2002tv}.
In this limit, boosted worldsheet coordinates become 
$(T, X) \to (\tilde \tau, \tilde \sigma)$ defined in \eqref{def:X,T},
and the fields (\ref{eta0-f2}\,-\,\ref{xi2-f2}) reduce to, respectively,
\begin{align}
\eta_{0}\to k \tilde \tau\,,\qquad 
\xi_{1}\to k \sn(\tilde \sig | k)\, e^{i \tilde\tau}\,,\qquad 
\xi_{2}\to \dn(\tilde \sig | k)\,.
\end{align}
This solution corresponds to a kink-array of sG equation {at rest} ($v=0$), and it spins around the northern pole of an ${S}^2$ with its centre of mass fixed at the pole.
The integer $n$ counts the number of folding, which is related to $\mu$ via the boundary condition (\ref{Dsigma_cl}).

\paragraph{}
The limit $k\to 1,\ \mu \to \infty$ takes the type $(i)$ single-spin solution to an array of giant magnons, each of which having the same soliton velocity of the sG system \cite{Hofman:2006xt}.
The endpoints of the string move on the equator $\theta=\pi/2$ at the speed of light.
In this limit, boosted worldsheet coordinates become $T\to \tilde \tau/\cos\om-\ko{\tan\om}\tilde \sig$ and $X\to \tilde \sig/\cos\om-\ko{\tan\om} \tilde \tau$\,, 
and the fields (\ref{eta0-f2}\,-\,\ref{xi2-f2}) reduce to
\begin{align}
\eta_{0}\to \tilde \tau\,,\quad 
\xi_{1}\to{\kko{\tanh\ko{\mbox{\large $\f{\tilde\sig-\ko{\sin\om}\tilde \tau}{\cos\om}$}}\cos\om-i\sin\om}}\, e^{i \tilde \tau}\,,\quad 
\xi_{2}\to \f{\cos\om}{\cosh\ko{\mbox{\large $\f{\tilde \sig-\ko{\sin\om}\tilde \tau}{\cos\om}$}}}\,.
\label{one_giant_magnon}
\end{align}
The following boundary conditions are imposed at each end of hops\,:
\begin{equation}
\xi_{1} \to \exp\pare{\pm i \Delta \varphi_1/2 + i \tilde \tau},\quad 
\xi_{2} \to 0 \qquad \mbox{as} \quad \tilde \sigma \to \pm \infty \,,
\end{equation}
in place of \eqref{Dsigma_cl} and \eqref{Dphi1_cl}. 
One can see $\Delta \varphi_1$ is determined only by $\omega$\,, which is further related to the magnon momentum $p$ of the gauge theory as $\Delta \varphi_1 = p = \pi - 2 \omega$ in view of the AdS/CFT \cite{Hofman:2006xt}.

\subsubsection*{Type $\bmt{(ii)}$ single-spin helical string}

The profile is given by
\begin{align}
\hat \eta_{0} (T,X)&= \hat a \, T + \hat b X \,,\qquad 
\mbox{with}\quad \hat a =\dn(\iom | k)\,,\quad \hat b =-ik\sn(\iom | k) \,,
\label{eta0-c2}\\
\hat \xi_{1} (T,X)&=\f{1}{\sqrt{k}\,\cn(\iom | k)}\f{\bTh_{0}(0 | k)}{\bTh_{0}(\iom | k)}\f{\bTh_{1}(X-\iom | k)}{\bTh_{0}(X | k)}\, 
\exp\kko{\eZ_{0}(\iom | k)X+ik\cn(\iom | k)T}\,,
\label{xi1-c2}\\
\hat \xi_{2} (T,X)&=\f{\cn (X | k)}{\cn(\iom | k)}\,,
\label{xi2-c2}
\end{align}
where $\omega$ is again a real parameter, and the soliton velocity is given by $\hat v \equiv \hat b/\ssp \hat a$\,. 
In this type $(ii)$ case, the AdS-time can be written as $\hat \eta_{0} = \tilde \tau$\,.
The following periodic boundary conditions are imposed\,:
\begin{alignat}{2}
\Delta \sigma \Big|_{\rm one\mbox{\tiny\,-\,}hop} &\equiv \frac{2 \pi}{m} = \frac{2 \eK \sqrt{1 - v^2}}{\mu} \,,  \label{Dsigma2_cl} \\[2mm]
\Delta \varphi_1 \Big|_{\rm one\mbox{\tiny\,-\,}hop} &\equiv \frac{2 \pi M_1}{m} = 2 \eK \pare{ - i \eZ_0 (\iom) + \frac{i k^2 \sn (\iom) \cn (\iom)}{\dn (\iom)} } + (2 \ssp m'_1 + 1) \pi  \,,  \label{Dphi2_cl}
\end{alignat}
where $m=1,2,\dots$ is the number of hops, $M_1$ is the winding number in $\varphi_{1}$-direction, and $m'_1$ is an integer.
The conserved charges are given by
\begin{equation}
\hat {\cal E} = \frac{m \eK}{\dn (\iom)} \,, \qquad
\hat {\cal J} = \frac{m(\eK - \eE)}{k \cn (\iom)} \,. \label{1ch-c} 
\end{equation}

\paragraph{}
In the $\omega \to 0$ limit, the type $(ii)$ solution reduces to a circular string studied in \cite{Gubser:2002tv}. 
Again, the boosted coordinates \eqref{def:X,T} become $(T, X) \to (\tilde \tau, \tilde \sigma)$\,, and the profile reduces to
\begin{equation}
\hat \eta_{0} \to \tilde \tau\,,\qquad 
\hat \xi_{1} \to \sn(\tilde \sig | k)\, e^{i \tilde \tau}\,,\qquad 
\hat \xi_{2} \to \cn(\tilde \sig | k)\,.
\end{equation}
The integer $m$ counts the number of winding, which is related to $\mu$ via the boundary condition (\ref{Dsigma2_cl}).

\paragraph{}
The limit $k\to 1$ with $\mu \to \infty$ reduce the type $(ii)$ solution to an array of giant magnons and flipped giant magnons, one after the other.

\subsubsection*{Type $\bmt{(i)'}$ single-spin helical string}

A single-spin type $(i)'$ helical string is obtained by setting $u_{2}=\om_{2}=0$\,, which results in $J_2 = N_2 = 0$\,.\footnote{\,
It turns out the other single-spin limit $u_1\,,\omega_1 \to 0$\,, which gives $J_1 = 0$\,,  does not result in a real solution for the type $(i)'$ solution.}
In view of \eqref{u_1 and u_2}, the condition $u_2 = \omega_2 = 0$ requires $U = 0$\,, $u_1 = \dn (\iom)$ and $C = \sqrt{k}/\dn (\iom)$\,, and the Virasoro constraints (the same as (\ref{a,b-f1}, \ref{a,b-f2})) are solved by setting $a = k \cn (\iom)$\,, $b = -ik \sn (\iom)$ and $v = - i \,\sn (\iom)/\cn (\iom)$\,.
The profile is given by just swapping $T$ and $X$ in (\ref{eta0-f2}\,-\,\ref{xi2-f2}), which is the same as the single-spin single-spike solution first obtained in \cite{Ishizeki:2007we}.
The periodicity conditions become
\begin{alignat}{3}
\Delta \sigma &= \ \frac{2 \pi}{n} & &= \frac{2 i \eK}{\mu \sn(\iom)} \,, \qquad
\frac{2 \pi N_2}{n} = 0\,, \\[2mm]
\Delta \varphi_1 &= \frac{2 \pi N_1}{n} & &= 2 i \eK \pare{ \frac{\cn(\iom) \dn (\iom)}{\sn (\iom)} + \eZ_0 (\iom) } + \pare{2 \ssp n'_1 + 1} \pi \,,
\label{closed single phi-i}
\end{alignat}
and the conserved charges for a single period are
\begin{equation}
{\cal E} = \frac{i k}{\sn(\iom)} \, \eK \,,\qquad
{\cal J}_1 = \frac{1}{k \dn (\iom)} \Big[ \eE - \pare{1 - k^2} \eK \Big] \,,\qquad
{\cal J}_2 = 0\,.
\label{single ch-i}
\end{equation}

\subsubsection*{Type $\bmt{(ii)'}$ single-spin helical string}

As in the type $(i)'$ case, we obtain the type $(ii)'$ helical
strings with $J_2 = M_2 = 0$ by setting $u_2 = \omega_2 =
0$\,.\footnote{\,For the type $(ii)'$ case, the other single-spin
limit $u_1 = \omega_1 = 0$ results in $U = - 1$\,, $u_2^2 = - 1 +
(1-k^2)/\dn^2 (\iomm{2})$ and $\hat C = \dn (\iomm{2})/\cn
(\iomm{2})$\,. It turns out equivalent to the $\om_{1,2}\to 0$
limit, because $u_2$ must be real, and thus
the second condition implies $\omega_2 = 0$\,.} Then we find $U =
- 1 + k^2$\,, $u_1 = k \cn (\iom)$ and $\hat C = 1/\cn (\iom)$\,.
The Virasoro conditions require $\hat a = \dn (\iom)$\,, $\hat b =
- i k \sn (\iom)$ and $\hat v = - i k \sn (\iom)/\dn (\iom)$\,.
The profile is given by just swapping $T$ and $X$ in (\ref{eta0-c2}\,-\,\ref{xi2-c2}), and the periodicity conditions become
\begin{alignat}{3}
\Delta \sigma &= \ \frac{2 \pi}{m} & &= \frac{2 i \eK}{\mu k \sn(\iom)} \,,\qquad  \frac{2\pi M_2}{m} = 0\,,  \\[2mm]
\Delta \varphi_1 &= \frac{2 \pi M_1}{m} & &= 2 i \eK \pare{ \frac{\cn (\iom) \dn (\iom)}{\sn (\iom)} + \eZ_0 (\iom) } + \pare{2 \ssp m'_1 + 1} \pi  \,,
\label{closed single phi-ii}
\end{alignat}
and the conserved charges for a single period are given by
\begin{equation}
\hat {\cal E} = \frac{i}{k \sn(\iom)} \, \eK \,,\qquad
\hat {\cal J}_1 = \frac{1}{k \cn (\iom)} \, \eE \,,\qquad
\hat {\cal J}_2 = 0\,.
\end{equation}

\part[The AdS/CFT S-matrix]
	{The AdS/CFT S-matrix\label{part:5}}

\chapter[The Asymptotic Spectrum of $\cN=4$ SYM Spin-Chain]
	{The Asymptotic Spectrum of $\boldsymbol{\cN=4}$ SYM Spin-Chain\label{chap:Asymptotic}}

In this Part \ref{part:5}, we will mostly investigate ``asymptotic'' states in $AdS_{5}\times S^{5}$ string/$\cN=4$ SYM.
As we already emphasised, since the work of \cite{Staudacher:2004tk}, it has been noticed that the S-matrix\footnote{The ``internal'' S-matrix we are discussing here (and throughout this thesis) should not be confused with the ``external'' S-matrix of $\cN=4$ SYM, {\em i.e.}, the multi-gluon scattering in four-dimensional spacetime.
It is interesting to note that there has been a remarkable progress in the study of the multi-gluon scattering amplitude recently, where the ``internal'' (worldsheet) and ``external'' (spacetime) S-matrices dramatically met; 
the ``cusp'' anomalous dimension computed in \cite{Bern:2006ew,Cachazo:2006az} agreed (numerically) with the prediction from the conjectured Bethe ansatz equations \cite{Eden:2006rx,Beisert:2006ez}.
} defined for both gauge and string theory should be the clue to unify the integrability which both theories are believed to possess.
Our main interest will be then how we can define and construct the (worldsheet) S-matrix of AdS/CFT.
Remarkably, the matrix structure of the asymptotic S-matrix can be completely determined by symmetry argument only, and the only remaining problem is to fix the overall scalar phase factor.
Before seeing this in Chapter \ref{chap:dressing}, in this chapter, we explain what is the asymptotic states, which representation they belong to, and how the asymptotic spectrum can be determined.

\section{The asymptotic states}

The asymptotic state corresponds to local excitations above the ferromagnetic groundstate of the spin-chain.
The latter state corresponds to the gauge theory operator $\tr(\cZ^{J_{1}})$ where $\cZ$ is a complex adjoint scalar field with R-charge $J_{1}=1$\,.
The ferromagnetic groundstate is not invariant under the full superconformal algebra $PSU(2,2|4)$\,, but instead is only preserved by the subalgebra $\left(PSU(2|2)\times PSU(2|2)\right)\ltimes {\mathbb{R}}$\,.
The residual symmetry algebra can also be understood as two copies of $SU(2|2)$ with their central charges identified.
This common central charge will play the role of Hamiltonian for the associated spin-chain whose eigenvalue is identified with the combination $\Delta-J_{1}$\,.

Moreover as noted in \cite{Beisert:2005tm}, an important subtlety arising is that this symmetry algebra needs to be further extended by two additional central charges in order to describe excitations of non-zero momenta.
The central extension is necessary also because otherwise anomalous dimensions cannot vary continously with the coupling $g$ (note that the only possible fundamental $({\bf 2}|{\bf 2})$ representations of $SU(2|2)$ have central charges $\pm \hf{1}{2}$).

This extended unbroken symmetry is linearly realised on excitations above the groundstate which consequently form representations of the corresponding non-abelian symmetry group $(PSU(2|2)\times PSU(2|2))\ltimes {\mathbb{R}}^{3}$\,.
In this chapter we will determine which representations appear in the spectrum of asymptotic states, and describe the minimal possibility for the complete spectrum of asymptotic states of the spin-chain.

\paragraph{}
As we explained in Part \ref{part:1}, a magnon is the fundamental excitation of the spin-chain, which corresponds to an insertion of a single impurity, with definite momentum $p$\,, into the groundstate operator $\tr(\cZ^{J_{1}})$\,.
There are a total of sixteen possible choices for the impurity corresponding to the various scalars and spinor fields and covariant derivatives of the ${\cal N}=4$ theory \cite{Berenstein:2002jq}.
As we review below, these excitations fill out a multiplet in the bifundamental representation of $(PSU(2|2)\times PSU(2|2))\ltimes {\mathbb{R}}^{3}$\,. In terms of the centrally-extended algebra described above, these are short representations with an exact BPS dispersion relation which is uniquely given by the closure of the algebra to be \cite{Beisert:2005tm,Beisert:2004hm,Rej:2005qt},   
\begin{equation}
\Delta-J_{1}=\sqrt{1+f(\lam)\sin^{2}\left(\frac{p}{2}\right)}\,.
\label{Magnondispersion}
\end{equation}
Here the coupling dependent function $f(\lam)$ is determined to be $f(\lam)=\lam/\pi^{2}=16g^{2}$ by considering the BMN limit \cite{Berenstein:2002jq}.
As the residual symmetry generators commute with the Hamiltonian of the spin-chain, each state in the multiplet has the same dispersion relation 
(\ref{Magnondispersion}).
With this in mind, we can think of the sixteen states in the bifundamental multiplet as distinct ``polarisations'' of a single excitation.

We restrict our attention to magnons of a single polarisation, namely, we consider an $SU(2)$ sector.
Within this subsector, it is known that the asymptotic spectrum also includes an infinite tower of magnon boundstates \cite{Dorey:2006dq}.
These excitations are labelled by a positive integer $Q$\,, which corresponds to the number of constituent magnons of different flavours, as well as their conserved momenta $P$\,.
The location of the corresponding poles in the exact magnon S-matrix indicates that these states have an exact dispersion relation of the form,  
\begin{equation}
\Delta-J_{1}=\sqrt{Q^{2}+16g^{2}\sin^{2}\left(\frac{P}{2}\right)}
\label{Magnondispersion2}
\end{equation}
which generalises (\ref{Magnondispersion}).
For $Q>1$\,, this formula is a generalisation of the exact \cite{Beisert:2005tm} magnon dispersion relation obtained in \cite{Beisert:2004hm,Staudacher:2004tk,Beisert:2005fw} (see also \cite{Santambrogio:2002sb, Berenstein:2005jq}). 
The corresponding classical string solution which precisely reproduces (\ref{Magnondispersion2}) will be discussed in Chapter \ref{chap:DGM}.
Scattering matrices for these states will be discussed in Chapter \ref{chap:S-matrices in HM}.

In the context of the full model, these asymptotic states in the $SU(2)$ sector should be particular representatives from complete representations of the symmetry group $(PSU(2|2)\times PSU(2|2))\ltimes {\mathbb{R}}^{3}$\,.
In fact, we will see below that the $Q$-magnon boundstate lies in a short irreducible representation of dimension $16Q^{2}$ \cite{Beisert:2006qh}.
The representation in question can be thought of as a supersymmetric extension of the rank-$Q$ traceless symmetric tensor representation of the unbroken $SO(4)\simeq SU(2)\times SU(2)$ R-symmetry which is a subgroup of $(PSU(2|2)\times PSU(2|2))\ltimes{\mathbb{R}}^{3}$\,.
This particular representation includes the known BPS boundstates of magnons in the $SU(2)$ sector. 
An important consistency check is that the representation does not lead to boundstates in any of the other rank one subsectors which are known to be absent \cite{Hofman:2006xt}. 
In \cite{Beisert:2005tm}, the dispersion relation (\ref{Magnondispersion}) for excitations transforming in the bifundamental representation of $(PSU(2|2)\times PSU(2|2))\ltimes {\mathbb{R}}^{3}$ was derived by purely group theoretical means. 
As an additional test of our results, we will extend the analysis to the symmetric tensor representations relevant for the boundstates described above 
to provide a parallel group theoretic derivation of the dispersion relation (\ref{Magnondispersion2}).

\section{The representation theory}

\subsection{The algebra}

To begin, let us first focus on a single copy of $SU(2|2)\subset\left(PSU(2|2)\times PSU(2|2)\right)\ltimes {\mathbb{R}}$ and review some associated basic facts following \cite{Beisert:2005tm}.
The algebra consists of two three-component bosonic generators $\fL^{\al}{}_{\be}$ and $\fR^{a}{}_{b}$ which generate $SU(2)_{\rm L}\times SU(2)_{\rm R}$ rotations;
two four-component fermionic supersymmetry generators $\fQ^{\al}{}_{b}$ and $\fS^{a}{}_{\be}$\,, and finally the algebra also contains a central charge $\fC$ which is shared with the other $SU(2|2)$\,.
These generators obey the following (anti-)commutation relations:
\begin{align}
[\fR^{a}{}_{b},
  \fJ^{c}]&=\delta^{c}_{b}\fJ^{a}-\half\delta^{a}_{b}\fJ^{c}\,,
\label{comm1}\\
[\fL^{\al}{}_{\be},
  \fJ^{\ga}]&=\delta^{\ga}_{\be}\fJ^{\al}-\half\delta^{\al}_{\be}\fJ^{\ga}\,,
\label{comm2}\\
\{ \fQ^{\al}{}_{a}, \fS^{b}{}_{\be}
\}&=\delta^{b}_{a}\fL^{\al}{}_{\be}+\delta^{\al}_{\be}\fR^{b}{}_{a}+
\delta^{b}_{a}
\delta^{\al}_{\be}\fC\,,
\label{comm3}
\end{align}
where $\fJ$ stands for any generator with appropriate indices.

In addition, as discussed in \cite{Beisert:2005tm}, the $SU(2|2)$ algebra is too restrictive for the discussion of excitations with non-zero momentum and it is necessary to enlarge it to $SU(2|2)\ltimes\mathbb R^{2}\cong PSU(2|2)\ltimes {\mathbb{R}}^{3}$\,, with two extra central charges $\fP$ and $\fK$ satisfying the anti-commutation relations, 
\begin{alignat}{3}
\{ \fQ^{\al}{}_{a}, \fQ^{\be}{}_{b} \}&=\ep^{\al\be}\ep_{ab}\fP\,,\qquad 
&\{ \dot\fQ^{\dot\al}{}_{\dot a}, \dot\fQ^{\dot\be}{}_{\dot b}
\}&=\ep^{\dot \al\dot \be}\ep_{\dot a\dot b}\fP\,,
\label{comm4}\\
\{ \fS^{a}{}_{\al}, \fS^{b}{}_{\be} \}&=\ep^{ab}\ep_{\al\be}\fK\,,\qquad 
&\{ \dot\fS^{\dot a}{}_{\dot \al}, \dot\fS^{\dot b}{}_{\dot \be} \}&=
\ep^{\dot a\dot b}\ep_{\dot \al\dot \be}\fK\label{comm5}\,.
\end{alignat}
The two extra central charges $\fP$ and $\fK$ are unphysical in the sense that they vanish when the constraint of vanishing total momentum is imposed.
These two extra central charges $\fP$ and $\fK$ in fact combine with $\fC$ to give a vector under group $SO(1,2)$ \cite{Beisert:2005tm,Hofman:2006xt}.
The full extended subalgebra is then obtained by taking direct product between two copies of $PSU(2|2)\ltimes{\mathbb{R}}^{3}$ and identifying their central charges (both physical and unphysical ones), which extend the residual symmetry algebra from $\left(PSU(2|2)\times PSU(2|2)\right)\ltimes{\mathbb{R}}$ to $\left(PSU(2|2)\times PSU(2|2)\right)\ltimes{\mathbb{R}}^{3}$\,.
Under the extended residual symmetry algebra, the central charge $\fC$ can be identified with the Hamiltonian for the spin-chain, whereas the two extra central charges play the role of gauge transformation generators which insert or remove a background chiral field $\cZ$ \cite{Beisert:2005tm}.

\subsection{Elementary magnon case}

The fundamental representation of $PSU(2|2)\ltimes {\mathbb{R}}^{3}\simeq SU(2|2)\ltimes{\mathbb{R}}^{2}$ corresponds to a ${\bf 2}|{\bf 2}$ dimensional superspace given by the basis
\begin{equation}
\sbx \equiv\left(\begin{array}{c}\phi^{a}\\
\hline
  \psi^{\alpha}\end{array}\right)
\,,\qquad a=1\,,2\,,\quad \alpha=1\,,2\,.
\end{equation}
Here we have adopted the notation for super Young diagrams introduced in \cite{BahaBalantekin:1980pp,Bars:1982ps,Bars:1984rb}.
The fields $\phi^{a}$ and $\psi^{\alpha}$ are bosonic and fermionic, respectively.
The group generators acting on this space can be written in the following $4\times 4$ supermatrix form:
\begin{equation} 
\left(
\begin{array}{c|c}
\fR^{a}{}_{b} & \fQ^{\al}{}_{b}\\
\hline
\fS^{a}{}_{\be} & \fL^{\alpha}{}_{\beta}\\
\end{array}
 \right)
\end{equation}
where $\fR$ and $\fL$ are the internal and spacetime $SU(2)$ rotation generators, respectively, while $\fQ$ and $\fS$ are the supersymmetry and superboost generators, respectively.
We can decompose the fundamental representation $\sbx$ under the maximal bosonic subgroup $SU(2)\times SU(2)$ as, 
\begin{equation}
\sbx =\overbrace{(\bx, {\bf{1}})}^{\mbox{\small $\vphantom{\f{\f{}{}}{\f{}{}}}\,\, \phi^{a}$}}\, \oplus\, 
\overbrace{({\bf{1}}, \bx)}^{\mbox{\small $\vphantom{\f{\f{}{}}{\f{}{}}}\,\, \psi^{\alpha}$}}\,.
\label{elemag1}
\end{equation}
The canonical action of the $PSU(2|2)\ltimes {\mathbb{R}}^{3}$ generators on the components $\phi^{a}$ and $\psi^{\alpha}$ is then given by \cite{Beisert:2005tm}
\begin{alignat}{3}
&\fQ^{\al}{}_{a}\kket{\phi^{b}}&&=\a\,\delta^{b}_{a}\kket{\psi^{\al}}\,,
\label{Q-phi}\\
&\fQ^{\al}{}_{a}\kket{\psi^{\be}}&&=\b\,\ep^{\al\be}\ep_{ab}\kket{\phi^{b}\cZ^{+}}\,,
\label{Q-psi}\\
&\fS^{a}{}_{\al}\kket{\phi^{b}}&&=\c\,\ep^{ab}\ep_{\al\be}\kket{\psi^{\be}\cZ^{-}}\,,
\label{S-phi}\\
&\fS^{a}{}_{\al}\kket{\psi^{\be}}&&=\d\,\delta^{\be}_{\al}\kket{\phi^{a}}
\label{S-psi}\,,
\end{alignat}
whereas the $SU(2)$ generators $\fR$ and $\fL$ act on bosonic and fermionic components as
\begin{alignat}{5}
&\fR^{a}{}_{b}\kket{\phi^{c}}&&=\delta^{c}_{b}\kket{\phi^{a}}-\half \delta^{a}_{b}\kket{\phi^{c}}\,,&\qquad
&\fR^{a}{}_{b}\kket{\psi^{\ga}}=0\,,\label{R-phi}\\
&\fL^{\al}{}_{\be}\kket{\psi^{\ga}}&&=\delta^{\ga}_{\be}\kket{\psi^{\al}}-\half \delta^{\al}_{\be}\kket{\psi^{\ga}}\,,&\qquad 
&\fL^{\al}{}_{\be}\kket{\phi^{c}}=0\,.\label{L-phi}
\end{alignat}
Finally the central charges $\fC$\,, $\fP$ and $\fK$ act as
\begin{alignat}{5}
\fC\kket{\xi^{A}}=\cC\kket{\xi^{A}}\,, \qquad \fP\kket{\xi^{A}}=\cP\kket{\xi^{A}}\,, \qquad \fK\kket{\xi^{A}}=\cK\kket{\xi^{A}}\,,
\end{alignat}
where we have also introduced $\xi^{A_{i}}=\{\phi^{a_{i}};\psi^{\al_{i}}\}$ a generalised vector and $A_{i}=\{a_{i},\alpha_{i}\}$ a generalised index for notational conveniences. 
The coefficients $\a$\,, $\b$\,, $\c$ and $\d$ in (\ref{Q-phi}\,-\,\ref{S-psi}) can be expressed as functions of the magnon spectral parameters $x^{+}$ and $x^{-}$\,, which in turns are related to individual magnon momentum $p$ by (see (\ref{p}))
\begin{equation}
\exp(ip)=\frac{x^{+}}{x^{-}}\,.\label{momentum}
\end{equation}
The symbols $\cZ^{\pm}$ in (\ref{Q-psi}) and (\ref{S-phi}) denote an inserting $(+)$ or a removing $(-)$ of a background $\cZ$ field on the right of the excitation $\phi^{a}$ or $\psi^{\alpha}$\,, respectively.
In other words, they are length-changing operators which make the spin-chain dynamic.
It is important to note that the fundamental representation $\sbx$ is in fact a short (or atypical) representation of $PSU(2|2)\ltimes {\mathbb{R}}^{3}$\,, and it satisfies the shortening condition which for this case is given in terms of the three central charges $\cC(x^{\pm})$\,, $\cP(x^{\pm})$ and $\cK(x^{\pm})$ as \cite{Beisert:2006qh}
\begin{equation}
\cC^{2}-\cP\cK=\frac{1}{4}\,.
\label{shortening1}
\end{equation}
Using the explicit parameterisations for the central charges in terms of spectral parameters given in \cite{Beisert:2005tm}, the shortening condition is equivalent to the constraint on the magnon spectral parameters ({\em c.f.}, (\ref{constraint xpm}))\,:
\begin{equation}
x^{+}+\frac{1}{x^{+}}-x^{-}-\frac{1}{x^{-}}=\f{i}{g}\,.
\label{constraint1}
\end{equation}
The exact magnon dispersion relation (\ref{Magnondispersion}) then arises from the protected central charge $\cC$ carried by the fundamental representation $\sbx$\,.

Let us recall here that, in terms of ${\cal N}=4$ SYM, the elementary excitation of the spin-chain corresponds to the insertion of an impurity field with\footnote{Here $\Delta_{0}$ denotes the bare dimension of the inserted field.} $\Delta_{0}-J_{1}=1$ into $\mathrm{Tr}\left(\cZ^{J_{1}}\right)$\,.
In the limit $J_{1}\to \infty$\,, this corresponds to a single magnon propagating over the ferromagnetic groundstate of the infinite chain.
There are eight bosonic and eight fermionic impurities which correspond to sixteen different possible polarisations of the magnon. 
Explicitly, they correspond to different elements of the set $\{ \Phi_{i}, D_{\mu}, \Psi_{\al \be}, \Psi_{\dot\al\dot \be} \}$\,.
Here $i$\,, $\mu=1,\dots,4$ are indices in the vector representation of the two $SO(4)$ factors left unbroken by the ferromagnetic groundstate. 
The former is the unbroken R-symmetry of the ${\cal N}=4$ theory while the latter corresponds to conformal spin. In view of their interpretation as rotations in the dual string geometry, we denote these $SO(4)_{S^{5}}$ and $SO(4)_{AdS_{5}}$\,, respectively.
The scalars $\Phi_{i}$ and covariant derivatives $D_{\mu}$ form a vector representation of each group.
We also use the standard isomorphism $SO(4)\simeq SU(2)_{\rm L}\times SU(2)_{\rm R}$ to introduce dotted and undotted spinor indices for each factor.
The fermionic fields of the ${\cal N}=4$ theory, denoted $\Psi_{\al \be}$\,, $\Psi_{\dot\al\dot\be}$ $(\al, \dot\al=1,2)$ transform in the appropriate bispinor representations. 
The quantum numbers of the ${\cal N}=4$ fields under the bosonic symmetries are summarised in Table \ref{tab1} (for more details, see for example \cite{Sadri:2003pr}).

\begin{table}[tbph]
\caption{\small $SU(2)^{4}$ representations of $\cN=4$ fields.}
\begin{center}
\begin{tabular}{|l|lcccccccr|c|c|}
\hline
Fields	&	& $\!\!\!\!SU(2)_{{S}^{5},{\rm R}}\!\!\!\!$&$\!\!\!\!
\times\!\!\!\!$&$\!\!\!\!SU(2)_{{AdS}_{5},{\rm R}}
\!\!\!\!$&$\!\!\!\!\times\!\!\!\!$&$\!\!\!\!SU(2)_{{S}^{5},{\rm L}}
\!\!\!\!$&$\!\!\!\!\times\!\!\!\!$&$\!\!\!\!
SU(2)_{{AdS}_{5},{\rm L}}\!\!\!\!$
& &	$\Delta_{0}-J_{1}$	&	$\Delta_{0}+J_{1}$\\
\hline\hline
\, $\cZ$	&	$($&$ {\bf 1}$&$,$&${\bf 1}$&$;$&${\bf 1}$&$,$&${\bf 1}$&$)$	&	$0$	&$2$\\
\, $\bar \cZ$	&	$($&$ {\bf 1}$&$,$&${\bf 1}$&$;$&${\bf 1}$&$,$&${\bf 1}$&$)$	&	$2$	&$0$\\
\hline
\, $\Phi_{i}$	&	$($&$ \bx$&$,$&$\ichi$&$;$&$\bx$&$,$&$\ichi$&$)$	&	$1$	&$1$\\
\, $D_{\mu}$	&	$($&$ {\bf 1}$&$,$&$\bx$&$;$&$\ichi$&$,$&$\bx$&$)$	&	$1$	&$1$\\
\, $\Psi_{\al{\be}}$	&	$($&$ \bx$&$,$&$\ichi$&$;$&$\ichi$&$,$&$\bx$&$)$	&	$1$	&$2$\\
\, $\Psi_{{\dot\al}{\dot\be}}$	&	$($&$ {\bf 1}$&$,$&$\bx$&$;$&$\bx$&$,$&$\ichi$&$)$	&	$1$	&$2$\\
\hline
\, $\Psi_{\al{\dot\be}}$	&	$($&$ {\bf 1}$&$,$&${\bf 1}$&$;$&$\bx$&$,$&$\bx$&$)$	&	$2$	&$1$\\
\, $\Psi_{{\dot\al}\be}$	&	$($&$ \bx$&$,$&$\bx$&$;$&${\bf 1}$&$,$&${\bf 1}$&$)$	&	$2$	&$1$\\
\hline
\end{tabular}
\end{center}
\label{tab1}
\end{table}

\paragraph{}
In order to interpret the impurities described above in terms of the supergroup $(PSU(2|2)\times PSU(2|2))\ltimes{\mathbb{R}}^{3}$\,, we note that the bifundamental representation is given by the direct product between two copies of fundamental $\sbx$ described above,   
\begin{equation}
\left(\sbx;\sbx\right)=\left(\bx,\ichi;\bx,\ichi\right)\oplus
\left(\bx,\ichi;\ichi,\bx\right)\oplus
\left(\ichi,\bx;\bx,\ichi\right)
\oplus\left(\ichi,\bx;\ichi,\bx\right)\,.\label{elemag2}
\end{equation}  
Here we have also decomposed $\left(\sbx;\sbx\right)$ in terms of representations of the $SU(2)^{4}$ bosonic subgroup of $(PSU(2|2)\times PSU(2|2))\ltimes {\mathbb{R}}^{3}$\,.
There are again sixteen components within this decomposition, precisely what one needs to incorporate the elementary excitations listed in Table \ref{tab1}.
By identifying the four $SU(2)$ factors in (\ref{elemag2}), column by column, with the other four in Table \ref{tab1}, we can identify each term in (\ref{elemag2}) with an impurity $\cN=4$ theory according to,  
\begin{equation}
\begin{CD}
\Phi_{i}\equiv \left(\bx,\ichi;\bx,\ichi\right) 
	@>{\mbox{$\vphantom{\f{}{}}$ \footnotesize $PSU(2|2)_{\rm L}$}}>> 
\Psi_{\al\be}\equiv\left(\bx,\ichi;\ichi,\bx\right)\\
	@V{\mbox{$\vphantom{\f{}{}}$ \footnotesize $PSU(2|2)_{\rm R}$}}VV 
	@VV{\mbox{$\vphantom{\f{}{}}$ \footnotesize $PSU(2|2)_{\rm R}$}}V \\
\Psi_{\dot\al \dot\be}\equiv\left(\ichi,\bx;\bx,\ichi\right) 
	@>{\mbox{$\vphantom{\f{}{}}$ \footnotesize $PSU(2|2)_{\rm L}$}}>> 
D_{\mu}\equiv \left(\ichi,\bx;\ichi,\bx\right)
\end{CD}
\end{equation}
Therefore the sixteen elementary excitations completely fill up the bifundamental representation of $SU(2|2)\times SU(2|2)$\,.

\subsection{Magnon boundstate case}

Having treated the case of the elementary magnon, we now proceed to determine the corresponding representations of $(PSU(2|2)\times PSU(2|2))\ltimes {\mathbb{R}}^{3}$ relevant for the magnon boundstates discovered in \cite{Dorey:2006dq}.
The natural starting point for the $Q$-magnon boundstate is to consider the tensor product between $Q$ copies of the elementary magnon representation $(\sbx;\sbx)$ as given in (\ref{elemag2}).
In particular the magnon boundstates should transform in the short irreducible representations under the residual symmetry algebra $(PSU(2|2)\times PSU(2|2))\ltimes {\mathbb{R}}^{3}$\,.

As above we will begin by considering a single copy of $PSU(2|2)\ltimes{\mathbb{R}}^{3}$ and will start with the simplest case taking the tensor product between two fundamentals $\sbx$ as described in (\ref{elemag1}).
In the usual experience of dealing with Lie algebra, one expects that tensoring two or more irreducible representations ({\em e.g.}, the fundamental representation) would yield direct sum of irreducible representations (including both long and short).
However, as pointed out in \cite{Beisert:2006qh}, such multiplet splitting does not happen generally for $PSU(2|2)\ltimes{\mathbb{R}}^{3}$\,.
In particular, for the tensor product of two fundamental representations, the splitting into irreducible representations of lower dimensions can only happen if the central charges $\cC_{i}$\,, $\cP_{i}$ and $\cK_{i}$ carried by the two constituent magnons ($i=1,2$) satisfy the ``splitting condition''
\begin{equation}
(\cC_{1}+\cC_{2})^{2}-(\cP_{1}+\cP_{2})(\cK_{1}+\cK_{2})=1\quad 
\Rightarrow \quad 
2\cC_{1}\cC_{2}-\cP_{1}\cK_{2}-\cK_{1}\cP_{2}=\frac{1}{2}\,.
\label{splitting1}
\end{equation}
Clearly for arbitrary combinations of the central charges, (\ref{splitting1}) would not be satisfied, hence tensoring two fundamental representations generically gives us a long irreducible representation of sixteen dimensions. 
Interestingly, the splitting condition (\ref{splitting1}) can be satisfied when the spectral parameters obey the boundstate pole condition established in \cite{Dorey:2006dq,Chen:2006gq}, that is 
\begin{equation}
x^{-}_{1}=x^{+}_{2}\,.
\label{polecondition}
\end{equation}
This can be shown by explicitly calculating the expression in (\ref{splitting1}) using the spectral parameters.

In this special case, the long multiplet of sixteen dimensions splits into direct sum of two short representations of eight dimensions, and we can label them using the branching rules for super Young diagrams worked out in \cite{BahaBalantekin:1980pp}, 
\begin{equation}
\sbx\otimes\sbx = \twosbx \oplus\vtwosbx\,.\label{ten2sbx}
\end{equation}
The two terms on the RHS 
represent distinct irreducible representations of $PSU(2|2)\ltimes{\mathbb{R}}^{3}$\,.
The first irreducible representation, denoted $\twosbx$\,, corresponds to a symmetrisation of indices for the bosonic components $\phi^{a}$s of each fundamental representation and anti-symmetrisation of indices for the corresponding Grassmann components $\psi^{\al}$s.
We will call this the ``super-symmetric'' representation.
In contrast, the second term $\vtwosbx$ corresponds to a ``super-antisymmetric'' representation where the bosonic/fermionic indices are antisymmetrised/symmetrised, respectively.
Both of them are in fact short irreducible representations of $PSU(2|2)\ltimes {\mathbb{R}}^{3}$\,, satisfying the shortening condition (\ref{splitting1}) and carrying the protected central charges.\footnote{The splitting condition (\ref{splitting1}) is satisfied either by $x^{-}_{1}=x^{+}_{2}$ or $x^{+}_{1}=x^{-}_{2}$\,, in which the S-matrix becomes a projector onto the super-antisymmetric and super-symmetric representation, respectively \cite{Beisert:2006qh}.}

We can further decompose these short representations into representations under its $SU(2)\times SU(2)$ bosonic subgroup\footnote{In fact, our situation is further simplified as the subgroups only involve $SU(2)$s, whose Young diagrams only contain single rows.}.
In terms of standard $SU(2)$ Young diagrams the decompositions are  
\begin{align}
\twosbx & =(\twobx, {\bf 1})\oplus(\bx, \bx)\oplus({\bf 1}, {\bf 1})\,,
\label{twosbx}\\
\vtwosbx & =({\bf 1}, {\bf 1})\oplus(\bx, \bx)\oplus({\bf 1}, \twobx)\,.
\label{vtwosbx}
\end{align}

The generalisation to the physical case with two factors of $PSU(2|2)\ltimes {\mathbb{R}}^{3}$ with their central charges identified is straightforward. Combining (\ref{elemag2}) and (\ref{ten2sbx}), the tensor product of two bifundamental representations can be decomposed
as
\begin{equation}
(\sbx;\sbx) \otimes (\sbx;\sbx) = (\twosbx;\twosbx) \oplus
  (\twosbx;\vtwosbx) \oplus (\vtwosbx;\twosbx) 
\oplus (\vtwosbx;\vtwosbx)\,.
\label{sbx2-x-sbx2}
\end{equation}
Each irreducible representation in the decomposition in (\ref{sbx2-x-sbx2}) is manifestly supersymmetric, containing equal number of bosonic and 
fermionic components.
To identify the nature of the corresponding states, it is convenient to further decompose each term in the decomposition (\ref{sbx2-x-sbx2}) into the irreducible representations of the four $SU(2)$ subgroups. For example, the first term yields, 
\begin{align}
(\twosbx;\twosbx) 
&= (\twobx,\ichi;\twobx,\ichi)\oplus (\twobx,\ichi;\ichi,\ichi)\oplus(\ichi,\ichi;\twobx,\ichi)\oplus(\ichi,\ichi;\ichi,\ichi)\cr
&\qquad {}\oplus(\bx,\bx;\ichi,\ichi)\oplus (\bx,\bx;\twobx,\ichi)\cr
&\qquad {}\oplus(\ichi,\ichi;\bx,\bx)\oplus (\twobx,\ichi;\bx,\bx)\cr
&\qquad {}\oplus(\bx,\bx;\bx,\bx)\,.
\label{decten2bx}
\end{align}
As each state in the constituent bifundamental multiplet corresponds to an insertion of a particular impurity in the ${\cal N}=4$ SYM theory, we can identify the terms on the RHS of (\ref{decten2bx}) with appropriate bilinears in the ${\cal N}=4$ fields.
In Appendix \ref{app:rep}, the $SU(2)^{4}$ quantum numbers of arising from each product of two ${\cal N}=4$ impurities are listed.
Comparing (\ref{decten2bx}) with the results in the appendix, we identify the relevant bilinears as, 
\begin{equation}
(\twosbx;\twosbx)\equiv ~(\Phi_{i}\otimes
  \Phi_{j})~\oplus~(\Phi_{i}\otimes \Psi_{\al\be})~ \oplus
  ~(\Phi_{i}\otimes \Psi_{\dot\al\dot\be})~ 
\oplus~(D_{\mu}\otimes\Phi_{i})~\,.
\label{2N4fields}
\end{equation}
where appropriate (anti-)symmetrisations over indices is understood. 

As explained above, the two magnon boundstates in the $SU(2)$ sector must correspond to (at least) one of the short representations of $(PSU(2|2)\times PSU(2|2))\ltimes{\mathbb{R}}^{3}$ appearing in the decomposition (\ref{decten2bx}).
To identify the relevant representation we note that each magnon of the $SU(2)$ sector carries one unit of a second $U(1)$ R-charge denoted $J_{2}$ in \cite{Dorey:2006dq}.
The charge $J_{2}$ corresponds to one Cartan generator of the unbroken R-symmetry group $SO(4)\simeq SU(2)\times SU(2)\subset (PSU(2|2)\times PSU(2|2))\ltimes {\mathbb{R}}^{3}$ normalised to that states in the bifundamental representation of $SU(2)\times SU(2)$ have charges $-1\leq J_{2}\leq +1$\,.
The two-magnon boundstate has charge $J_{2}=2$\,.
It is straightforward to check that this value is realised in the term $(\twobx,1;\twobx,1)$ appearing in the decomposition (\ref{decten2bx}) of the ``bi-super-symmetrised'' representation $(\twosbx;\twosbx)$ of $(PSU(2|2)\times PSU(2|2))\ltimes {\mathbb{R}}^{3}$\,.
One may also check that the remaining irreducible representations in the decomposition (\ref{sbx2-x-sbx2}) of the tensor product do not contain states with $J_{2}=2$\,.

Summarising the above discussion we deduce that the two magnon boundstate discovered in \cite{Dorey:2006dq} is one component of a multiplet of states in the $(\twosbx;\twosbx)$ of $(PSU(2|2)\times PSU(2|2))\ltimes {\mathbb{R}}^{3}$\,.
The dimension of this representation is sixty-four, which corresponds to the number of independent polarisations of the two magnon boundstate.
The various bilinear impurities corresponding to these polarisations appear in (\ref{2N4fields}).   
A check on the identification described above is that there are no bilinears involving only either two fermions or two derivatives.
This agrees with the known absence of two magnon boundstates in the $SU(1|1)$ and $SL(2,{\mathbb{R}})$ sectors, respectively \cite{Hofman:2006xt,Minahan:2006bd,Roiban:2006gs}.

\paragraph{}
It is straightforward to extend the discussion to the case of general $Q$-magnon scattering, now the multiplet splitting condition can be given by
\begin{equation}
\bcC_{Q}^{2}-\bcP_{Q}\bcK_{Q}=\frac{Q^{2}}{4}\,,
\label{splitting2}
\end{equation}
where $\bcC_{Q}$\,, $\bcP_{Q}$ and $\bcK_{Q}$ are the central charges carried by the generic long irreducible representation formed by tensor product between $Q$ fundamentals. 
This can be satisfied when we impose the boundstate condition
\begin{equation}
x_{i}^{-}=x_{i+1}^{+}\,,~~~i=1\,,2\,,\dots\,,Q-1\,.\label{polecond2}
\end{equation}
The tensor product between $Q$ fundamental representations generally consists of direct sum of long representations \cite{Beisert:2006qh}.
In this special limit (\ref{polecond2}), it can be further decomposed into direct sum of short representations and labelled by the branching rules as
\begin{equation}
\underbrace{(\sbx;\sbx) \otimes\dots \otimes (\sbx;\sbx)}_{Q}
=(\, \underbrace{\vphantom{()}\Qsbx}_{Q}\,
;\,\underbrace{\vphantom{()}
\Qsbx}_{Q}\, )\oplus\cdots\,,
\label{brule}
\end{equation}
where the dots represents the direct sum of other irreducible representations.
In particular, the representation $\Qsbx$ being again a short representation under $PSU(2|2)\ltimes {\mathbb{R}}^{3}$ satisfies the shortening condition in \cite{Beisert:2006qh} and carries protected central charges.
Furthermore, by considering the multi-magnon boundstates in the $SU(2)$ spin-chain, we can conclude that the most general $Q$-magnon boundstate should be contained in the first term of the decomposition (\ref{brule}), as such term contains a state of highest weight $Q$\,.
It should be a straightforward but tedious excercise to decompose $(\Qsbx ;\Qsbx)$ into the irreducible representations of $SU(2)^{4}$\,, and rewrite the various terms in the decomposition in terms of the $\cN=4$ SYM fields as we did for the case of $Q=2$\,.
It would also be interesting to identify these different species of boundstates from the poles in their associated scattering matrices \cite{Beisert:2005tm,Beisert:2005fw}.
Even though the classification here does not completely rule out the possibility of having boundstates in other irreducible representation at larger $Q$\,, the states in $(\Qsbx\,;\Qsbx)$ should be regarded as the minimal set of boundstates in the asymptotic spectrum.

\paragraph{}
Here we would like to discuss the number of the possible polarisations for a $Q$-magnon boundstate. 
In decomposing the irreducible representations of $SU(2|2)$ into those of the $SU(2)\times SU(2)$ subgroup, the valid Young diagrams involved should only contain single rows to comply with the usual rules.
As the result the decomposition for irreducible representation of our interests terminates after three terms: 
\begin{equation}
\underbrace{\vphantom{()}\Qsbx}_{Q}=(\underbrace{\vphantom{()}\Qbx}_{Q}\, ,{\bf
  1})+(\underbrace{\vphantom{()}\Qbx}_{Q-1}\, ,\bx)+
(\underbrace{\vphantom{()}\Qbx}
_{Q-2}\, ,{\bf1})\,.
\label{Qsbx}
\end{equation}
Simple counting shows that there are $4Q$ states in this decomposition, therefore, there are $(4Q)^{2}=16Q^{2}$ states for $(\Qsbx\,;\Qsbx)$ which contains all possible polarisations for $Q$-magnon boundstates.
This is the degeneracies for a given boundstate charge $Q$ and it is drastically different from the number of possible out-going states for $Q$-magnon scatterings, which goes exponentially with $Q$\,.
This concludes our discussion on the representation of the magnon boundstates.

\section{The asymptotic spectrum\label{sec:asymptotic spectrum}}

Having worked out the representation, it is rather straightforward to obtain an exact dispersion relation for the general $Q$-magnon boundstates by 
extending the arguments in \cite{Beisert:2005tm}.
The idea is that, as we discussed earlier, the energy $\Delta-J_{1}$ of the magnon boundstate should again be the physical central charge $\bcC_{Q}$ carried by the associated irreducible representation $(\Qsbx\,;\Qsbx)$ under the extended residual symmetry algebra. 
Recall that this central charge (along with the two extra ones $\bcP_{Q}$ and $\bcK_{Q}$) is shared between the two $SU(2|2)$s in the extended algebra, in addition, the magnon boundstate transforms under identical short irreducible representation with respect to each $SU(2|2)$\,.
We conclude that it is sufficient to consider the action of only a single $SU(2|2)$ (with two extra central charges) on the boundstate, and treat the components transforming under the other $SU(2|2)$ as spectators, just like the infinite number of background $\cZ$ fields. 
Moreover, as $\fC$ should commute with other group generators which relate all $16Q^{2}$ different polarisations for magnon boundstate of charge $Q$\,, the dispersion relation deduced here should be identical for all of them.

Moving on to a $Q$-magnon boundstate which transforms as $\Qsbx$ under $PSU(2|2)\ltimes {\mathbb{R}}^{3}$\,, we have
\begin{equation}
\underbrace{\vphantom{()}\Qsbx}_{Q}\, ~:\quad 
\kket{\Xi_{Q}}\eq\kket{\xi^{(A_{1}}\xi^{A_{2}}\dots\xi^{A_{Q-1}}\xi^{A_{Q})}}\,,
\label{Qbound}
\end{equation}
where we have omitted the infinite number of background $\cZ$ fields as before.
We are interested in the central charge $\bcC_{Q}$ carried by such a state, which would in turn give us the required dispersion relation. 
This can be obtained by considering the actions from both sides of the commutator (\ref{comm3}) on the higher tensor representations, and combining with the algebraic relations (\ref{Q-phi}\,-\,\ref{L-phi}).

We shall give our calculational details in the generalised indices $A_{i}$ and only focus on the algebraic structures, the explicit conversion into bosonic and fermionic indices, $a_{i}$ and $\alpha_{i}$ respectively, should be obvious. 
First let us act the LHS of (\ref{comm3}) on $\kket{\Xi_{Q}}$ using (\ref{Q-phi}\,-\,\ref{S-psi}) to obtain
\begin{align}
\sum_{i=1}^{Q}\{ \fQ, \fS \}_{A_{i}}^{B_{i}}\kket{\Xi_{Q}^{C_{i}}}
&=\sum_{i=1}^{Q}\ko{\a_{i}\d_{i}-\b_{i}\c_{i}}\delta^{C_{i}}_{A_{i}}\kket{\Xi_{Q}^{B_{i}}}
+\sum_{i=1}^{Q}\b_{i}\c_{i}\delta^{B_{i}}_{A_{i}}\kket{\Xi_{Q}^{C_{i}}}\,.
\label{LHS}
\end{align}
The notation here means that $\{\fQ,\fS\}^{B_{i}}_{A_{i}}$ only acts on the $i$\,-th fundamental representation in the tensor representation and the superscript $C_{i}$ in $\kket{\Xi_{Q}^{C_{i}}}$ is also for highlighting such fact.

On the other hand, the action of the RHS of (\ref{comm3}) on $\kket{\Xi_{Q}}$ gives.
\begin{equation}
\sum_{i=1}^{Q}\left(\fL+\fR+\fC\right)^{B_{i}}_{A_{i}}\kket{\Xi_{Q}^{C_{i}}}
=\sum_{i=1}^{Q}\left\{\delta^{C_{i}}_{A_{i}}\kket{\Xi_{Q}^{B_{i}}}
+\left(\cC_{i}-\hf{1}{2}\right)\delta^{B_{i}}_{A_{i}}\kket{\Xi_{Q}^{C_{i}}}\right\}\,,
\label{RHS}
\end{equation}
where we have used $\cC_{i}$ to denote the central charge carried by the $\xi^{A_{i}}$\,, that is $\fC\kket{\xi^{A_{i}}}=\cC_{i}\kket{\xi^{A_{i}}}$\,.
From (\ref{LHS}, \ref{RHS}), we can deduce the closure of the symmetry algebra requires 
\begin{equation}
\a_{i}\d_{i}-\b_{i}\c_{i}=1\quad \mbox{and}\quad {\cC}_{i}=\ko{\hf{1}{2}+\b_{i}\c_{i}}\,,\qquad i=1\,,\dots\,,Q\,,
\end{equation}
which then implies that $\cC_{i}=\hf{1}{2}(\a_{i}\d_{i}+\b_{i}\c_{i})$\,.
The central charge $\bcC_{Q}$ of $\kket{\Xi_{Q}}$ is given by sum of the individual central charges, hence we have
\begin{equation}
\bcC_{Q}=\sum_{i=1}^{Q}\cC_{i}=\hf{1}{2}\sum_{i=1}^{Q}\ko{\a_{i}\d_{i}+\b_{i}\c_{i}}=\hf{1}{2}\ko{\A\D+\B\C}\,.
\label{cC}
\end{equation}
This is the central charge of the $Q$-magnon boundstate in terms of $\a_{i}$\,, $\b_{i}$\,, $\c_{i}$ and $\d_{i}$\,, 
and here we have also introduced $\A$\,, $\B$\,, $\C$ and $\D$ which should be functions of the spectral parameters for the boundtstates $X^{\pm}$\,.
To proceed obtaining the explicit expression for $\bcC_{Q}$\,, we need to work out $\A$\,, $\B$\,, $\C$ and $\D$ or at least some combinations of them in terms of the magnon boundstate spectral parameters, this is where the two extra central charges $\fP$ and $\fR$ in (\ref{comm4}, \ref{comm5}) come in. 
First consider the actions of (\ref{comm4}, \ref{comm5}) on $\kket{\Xi_{Q}}$\,, one can deduce that 
\begin{equation}
\fP\kket{\Xi_{Q}}
=\sum_{i=1}^{Q}\a_{i}\b_{i}\prod^{Q}_{j=i+1}e^{-ip_{j}}\kket{\Xi_{Q}^{C_{i}}\cZ^{+}}\,,\qquad
\fK\kket{\Xi_{Q}}
=\sum_{i=1}^{Q}\c_{i}\d_{i}\prod^{Q}_{j=i+1}e^{ip_{j}}\kket{\Xi_{Q}^{C_{i}}\cZ^{-}}\,.\label{PKQbound}
\end{equation}
In deducing (\ref{PKQbound}), we have also used the consistency relation 
$\kket{\cZ^{\pm}\xi^{A_{i}}}=e^{\mp i p_{i}}\kket{\xi^{A_{i}}\cZ^{\pm}}$ to shift the insertion/removal of $\cZ$ field to the far right.
Actually the additional central charges $\fP$ and $\fK$ generate gauge transformations \cite{Beisert:2005tm}.
Writing the actions on arbitrary fundamental excitation $\xi$ as $\fP\kket{\xi}=\al\kket{[\cZ^{+},\xi]}$ and $\fK\kket{\xi}=\be\kket{[\cZ^{-},\xi]}$\,, where $\al$ and $\be$ are some constants which are common to all excitations, we have $\a_{i}\b_{i}=\alpha(e^{-ip_{i}}-1)$ and $\c_{i}\d_{i}=\beta(e^{ip_{i}}-1)$\,.
The two additional central charges carried by the magnon boundstate are then given by
\begin{alignat}{3}
\A\B&={\bcP}_{Q}&&=\alpha(e^{-iP}-1)&&=\alpha\left(\frac{X^{-}}{X^{+}}-1\right)\,,\label{ABCD1}\\[2mm]
\C\D&={\bcK}_{Q}&&=\beta(e^{iP}-1)&&=\beta\left(\frac{X^{+}}{X^{-}}-1\right)\,,\label{ABCD2}
\end{alignat}
where $P=\sum_{i=1}^{Q}p_{i}$ is the momentum carried by the $Q$-magnon boundstate that is given by the sum of constituent momenta.
When we restrict to the physical states living in $PSU(2|2)\ltimes {\mathbb{R}}^{3}$\,, both extra central charges should vanish.

Moreover, as the $Q$-magnon boundstates transform in the short representation $\Qsbx$ of $PSU(2|2)\ltimes{\mathbb{R}}^{3}$\,, in terms of their central charges $\bcC_{Q},\bcP_{Q}$ and $\bcK_{Q}$\,, the shortening condition reads
\begin{equation}
\bcC_{Q}^{2}-\bcP_{Q}\bcK_{Q}=\frac{Q^{2}}{4}\,.\label{shortening2}
\end{equation}
In the light of (\ref{shortening1}) and (\ref{constraint1}),
this should in turn provide a constraint on the boundstate spectral parameters $X^{\pm}$ as
\begin{equation}
X^{+}+\frac{1}{X^{+}}-X^{-}-\frac{1}{X^{-}}=\f{iQ}{g}\,.
\label{constraint2}
\end{equation}
This can be guaranteed and reduced correctly to trivial $Q=1$ case if we set 
\begin{equation}
\A\D=\sum^{Q}_{i=1}\a_{i}\d_{i}=-i(X^{+}-X^{-})\,,
\qquad 
\B\C=\sum^{Q}_{i=1}\b_{i}\c_{i}=i\left(\frac{1}{X^{+}}-\frac{1}{X^{-}}\right)\,.\label{XpXm}
\end{equation}
Using the explicit expressions for $\a_{i},\b_{i},\c_{i}$ and $\d_{i}$ in terms of the magnon spectral parameters given in \cite{Beisert:2005tm}, we deduce that
\begin{equation}
X^{+}-X^{-}=\sum^{Q}_{i=1}(x_{i}^{+}-x_{i}^{-})\,,
\qquad 
\frac{1}{X^{+}}-\frac{1}{X^{-}}=\sum^{Q}_{i=1}\left(\frac{1}{x_{i}^{+}}-\frac{1}{x_{i}^{-}}\right)\,.\label{XpXm2}
\end{equation}
Combining (\ref{XpXm2}) with (\ref{ABCD1}, \ref{ABCD2}), they give three constraints on $\{x^{\pm}_{1}\,,\dots\,,x^{\pm}_{Q}\}$ in terms of $X^{\pm}$ which can be satisfied by the combination
\begin{equation}
X^{+}=x^{+}_{1}\,,\qquad X^{-}=x^{-}_{Q}\,,\qquad
x_{i}^{-}=x_{i+1}^{+}\,,\quad (i=1\,,\dots\,, Q-1)\label{constraint4}\,.
\end{equation} 
The last set of equations in (\ref{constraint4}) is identical to the multiplet splitting condition given earlier (\ref{splitting2}), as the ``super-symmetric'' representation $\Qsbx$ can only arise from the decomposition of general $Q$-magnon tensor product after (\ref{splitting2}) is imposed.

From (\ref{ABCD1}, \ref{ABCD2}) and (\ref{shortening2}) (or (\ref{constraint2})), we can also deduce $\bcC_{Q}$ for the magnon boundstate,\footnote{There are two choices for the sign in front of the square root in (\ref{bcC}), and here we took the plus sign.
These two possibilities correspond to the spectra of a ``particle'' and an ``antiparticle''.}
\begin{equation}
\bcC_{Q}=\f{1}{2}\sqrt{\ko{\A\D-\B\C}^{2}+4\A\B\C\D}
=\f{1}{2}\sqrt{Q^{2}+16\al \be \sin^{2}\ko{\f{P}{2}}}\,.
\label{bcC}
\end{equation}
The product $\al\be$ is in general a function of the 't Hooft coupling $\lambda$\,. 
As we already mentioned, for the case of single magnon, it was set to $\al\be=g^{2}=\lambda/16\pi^{2}$ by considering the BMN limit \cite{Berenstein:2002jq}.
This dependence should interpolate to case of $Q> 1$\,, and indeed one can confirm that for example by considering the Frolov-Tseytlin limit (\ref{thermodynamic}) (or (\ref{Frolov-Tseytlin limit})) \cite{Frolov:2003xy} as in \cite{Dorey:2006dq}. In any case, we deduce that the dispersion relation for the magnon boundstate from the group theoretical means is 
\begin{equation}
\Delta-J_{1}\eq 2\bcC_{Q}=\sqrt{Q^{2}+\f{\lambda}{\pi^{2}}\sin^{2}\ko{\f{P}{2}}}\,.
\label{kdisprel}
\end{equation}
This formula reduces to the one proposed in \cite{Dorey:2006dq} for single magnon boundstate of charge $Q=1$\,, with $\Delta-J_{1}$ coincides with (\ref{Magnondispersion}). It is also important to note that, as discussed earlier, there will be $16Q^{2}$\,-fold degeneracies which correspond to the all possible polarisations of a $Q$-magnon boundstate, all share the same dispersion relation (\ref{Magnondispersion}).

Let us make a comment on the situation where there are more than one boundstate in the asymptotic spin-chain, namely a state of the form $\kket{\Xi_{Q_{1}}\dots\Xi_{Q_{M}}}$\,.
Here $M$ is the number of the boundstates each of which are well-separated, and $Q_{k}$ is the number of constituent magnons in the $k$\,-th boundstate.
In this case the dispersion relation (\ref{kdisprel}) is simply generalised to give
\begin{equation}
\Delta-J_{1}\eq \sum_{k=1}^{M}2\bcC_{Q_{k}}\quad \mbox{with}\quad  
\bcC_{Q_{k}}\eq \f{1}{2} \sqrt{Q_{k}^{2}+\f{\lambda}{\pi^{2}}\sin^{2}\ko{\f{P_{k}}{2}}}\,,
\end{equation}
where $P_{k}$ is the total momentum of the $k$-th boundstate in the
asymptotic spin-chain.

\paragraph{}
We have so far identified the representations under which the infinite tower of BPS boundstates in the asymptotic spectrum of $\cN=4$ SYM transforms.
As these are short representations we expect that these states are present for all values of the 't Hooft coupling, $\lambda$\,.
Indeed, as discussed in \cite{Dorey:2006dq,Chen:2006ge}, the representatives of the boundstate multiplets lying in a given $SU(2)$ sector are directly visible both in one-loop gauge theory and in semiclassical string theory which correspond to small and large $\lambda$ respectively.
An obvious question is whether additional asymptotic states are also present.
At this point we cannot rule out the possibility that some of the additional short representations, which appear in the tensor product of bi-fundamentals when the shortening condition is obeyed, also correspond to BPS boundstates in the spectrum.
However, the representations which can occur are certainly constrained by the known absence of boundstates in the remaining rank one sectors.
In particular this rules out additional boundstates with $Q=2$\,.

We should also note that there are two classes of states which we have not included in our discussion.
First, the semiclassical string theory analysis of \cite{Hofman:2006xt} suggests the presence of an infinite tower of neutral boundstates appearing as poles in the two-magnon S-matrix. 
These poles should appear at values of the kinematic variables which do not satisfy the shortening condition.
In fact, for such generic values of the momenta the tensor product of two bi-fundamentals actually consists of a single irreducible long multiplet \cite{Beisert:2006qh}. 
Each of the neutral boundstates of \cite{Hofman:2006xt} must therefore fill out such a multiplet.
As the energies of these states are not protected, their behaviour away from the region of large $\lam$ is still unclear.
Finally we recall that the ${\cal N}=4$ SYM spin-chain also contains a singlet state of zero energy \cite{Beisert:2005tm}.
In a crossing invariant theory, however, this state is indistinguishable from the vacuum.

\newpage
\section[Appendix for Chapter \ref{chap:Asymptotic}]{Appendix for Chapter \ref{chap:Asymptotic}\,: Decomposition of \bmt{\cN=4} fields into \bmt{SU(2)^{4}} Representations\label{app:rep}}

Here we list all possible tensor decompositions between two $\cN=4$ SYM excitations.
They are useful in interpreting the terms on the RHS of (\ref{decten2bx}) in terms of appropriate bilinears in the ${\cal N}=4$ fields.
\begin{alignat}{3}
\Phi_{i}&\otimes \Phi_{j}\quad &:\quad &(\bx,{\bf 1};\bx,{\bf 1})\otimes (\bx,{\bf 1};\bx,{\bf 1})\cr
&&&\quad =(\twobx,{\bf 1};\twobx,{\bf 1})
\oplus(\twobx,{\bf 1};{\bf 1},{\bf 1})
\oplus({\bf 1},{\bf 1};\twobx,{\bf 1})
\oplus({\bf 1},{\bf 1};{\bf 1},{\bf 1})\,,\\[2mm]
D_{\mu}&\otimes D_{\nu}\quad &:\quad &({\bf 1},\bx;{\bf 1},\bx)\otimes ({\bf 1},\bx;{\bf 1},\bx)\cr
&&&\quad =({\bf 1},\twobx;{\bf 1},\twobx)
\oplus({\bf 1},\twobx;{\bf 1},{\bf 1})
\oplus({\bf 1},{\bf 1};{\bf 1},\twobx)
\oplus({\bf 1},{\bf 1};{\bf 1},{\bf 1})\,,\\[2mm]
D_{\mu}&\otimes \Phi_{i}\quad &:\quad &({\bf 1},\bx;{\bf 1},\bx)\otimes (\bx,{\bf 1};\bx,{\bf 1})
=(\bx,\bx;\bx,\bx)\,,\\[2mm]
\Psi_{{\dot\al}\dot\be}&\otimes \Psi_{\ga{\delta}}\quad &:\quad &({\bf 1},\bx;\bx,{\bf 1})\otimes (\bx,{\bf 1};{\bf 1},\bx)=(\bx,\bx;\bx,\bx)\,,\\[2mm]
\Psi_{\al{\be}}&\otimes \Psi_{\ga{\delta}}\quad &:\quad &(\bx,{\bf 1};{\bf 1},\bx)\otimes (\bx,{\bf 1};{\bf 1},\bx)\cr
&&&\quad =(\twobx,{\bf 1};{\bf 1},\twobx)
\oplus(\twobx,{\bf 1};{\bf 1},{\bf 1})
\oplus({\bf 1},{\bf 1};{\bf 1},\twobx)
\oplus({\bf 1},{\bf 1};{\bf 1},{\bf 1})\,,\\[2mm]
\Psi_{{\dot\al}\dot\be}&\otimes \Psi_{{\dot\ga}\dot\delta}\quad &:\quad &({\bf 1},\bx;\bx,{\bf 1})\otimes ({\bf 1},\bx;\bx,{\bf 1})\cr
&&&\quad =({\bf 1},\twobx;\twobx,{\bf 1})
\oplus({\bf 1},\twobx;{\bf 1},{\bf 1})
\oplus({\bf 1},{\bf 1};\twobx,{\bf 1})
\oplus({\bf 1},{\bf 1};{\bf 1},{\bf 1})\,,\\[2mm]
\Phi_{i}&\otimes \Psi_{\al{\be}}\quad &:\quad &(\bx,{\bf 1};\bx,{\bf 1})\otimes (\bx,{\bf 1};{\bf 1},\bx)
=({\bf 1},{\bf 1};\bx,\bx)
\oplus(\twobx,{\bf 1};\bx,\bx)\,,\\[2mm]
\Phi_{i}&\otimes \Psi_{{\dot\al}\dot\be}\quad &:\quad &(\bx,{\bf 1};\bx,{\bf 1})\otimes ({\bf 1},\bx;\bx,{\bf 1})
=(\bx,\bx;{\bf 1},{\bf 1})
\oplus(\bx,\bx;\twobx,{\bf 1})\,,\\[2mm]
D_{\mu}&\otimes \Psi_{\al{\be}}\quad &:\quad &({\bf 1},\bx;{\bf 1},\bx)\otimes (\bx,{\bf 1};{\bf 1},\bx)
=(\bx,\bx;{\bf 1},{\bf 1})
\oplus(\bx,\bx;{\bf 1},\twobx)\,,\\[2mm]
D_{\mu}&\otimes \Psi_{{\dot\al}\dot\be}\quad &:\quad &({\bf 1},\bx;{\bf 1},\bx)\otimes ({\bf 1},\bx;\bx,{\bf 1})
=({\bf 1},{\bf 1};\bx,\bx)
\oplus({\bf 1},\twobx;\bx,\bx)\,.
\end{alignat}

\chapter[The Conjectured AdS/CFT S-matrix]
	{The Conjectured AdS/CFT S-matrix\label{chap:dressing}}

\section[The $SU(2|2)$ dynamic S-matrix and the dressing phase]
	{The \bmt{SU(2|2)} dynamic S-matrix and the dressing phase}

We have so far discussed the dispersion relation for the asymptotic SYM spin-chain.
We are now going to investigate another physical object of interest, that is the S-matrix.
As emphasised in \cite{Staudacher:2004tk}, the S-matrix is a powerful tool to indirectly compare the respective spectra of gauge theory and string theory via the AdS/CFT correspondence.
Assuming integrability, it allows us to obtain spectral information despite the absence of detailed knowledge of the underlying exact dilatation operator.
Actually the S-matrix appears to be much simpler than the dilatation operator of gauge theory or the (quantum) Hamiltonian of string theory.
Our aim here is to construct an S-matrix which describes either side of the (free-)AdS/(planar-)CFT correspondence, which we call {\em the AdS/CFT S-matrix}.

\begin{figure}[t]
\begin{center}
\vspace{.3cm}
\includegraphics[scale=0.7]{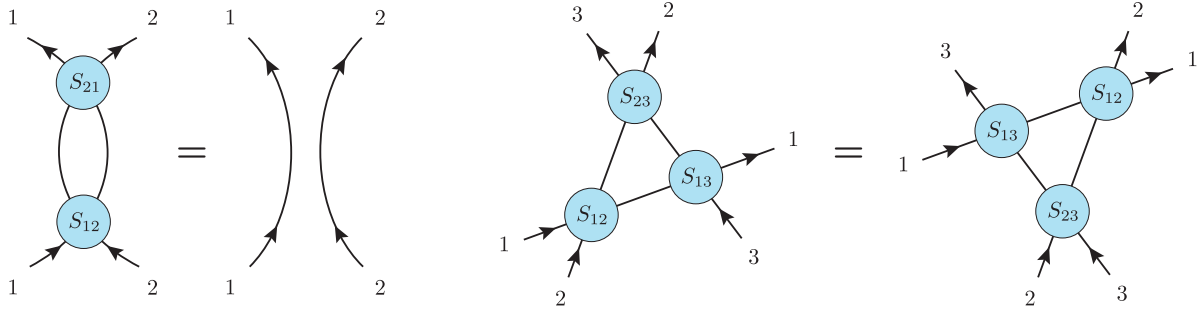}
\vspace{.3cm}
\caption{\small Unitarity condition (\ref{unitarity}) and Yang-Baxter relation (\ref{YBE}).}
\label{fig:unitarity-YB}
\end{center}
\end{figure}

\paragraph{}
The S-matrix of an integrable system should satisfy fundamental symmetry requirements such as {\em unitarity} and the {\em Yang-Baxter relation},
\begin{align}
&S_{12}S_{21}=1\,,\label{unitarity}\\
&S_{12}S_{13}S_{23}=S_{23}S_{13}S_{12}\,,\label{YBE}
\end{align}
respectively, see Figure \ref{fig:unitarity-YB} for corresponding diagrams.
Here we used shorthand notations $S_{ij}=S(p_{i},p_{j})$\,.
We saw in Chapter \ref{chap:Asymptotic} that for the asymptotic $\cN=4$ SYM spin-chain, the symmetry algebra should be enlarged to $(PSU(2|2)\times PSU(2|2))\ltimes {\mathbb{R}}^{3}$\,.
Remarkably, as shown by Beisert \cite{Beisert:2005tm}, the matrix structure of the S-matrix with $(PSU(2|2)\times PSU(2|2))\ltimes {\mathbb{R}}^{3}$ symmetry satisfying the conditions (\ref{unitarity}) and (\ref{YBE}) is completely obtained by symmetry considerations alone.
The most general ansatz for the action of the S-matrix on two-particle states $\kket{\cX_{1}\cX_{2}}$\,, where each $\cX$ is a fundamental excitation, $\cX\in \{ \phi^{1},\phi^{2}|\psi^{1},\psi^{2} \}$\,, is given by
\begin{align}
&S_{12}\kket{\phi_{1}^{a}\phi_{2}^{b}}
=A_{12}\kket{\phi_{2}^{\{a}\phi_{1}^{b\}}}
+B_{12}\kket{\phi_{2}^{[a}\phi_{1}^{b]}}
+\half C_{12}\ep^{ab}\ep_{\al\be}\kket{\psi_{2}^{\al}\psi_{1}^{\be}\cZ^{-}}\,,\label{S12-1}\\
&S_{12}\kket{\psi_{1}^{\al}\psi_{2}^{\be}}
=D_{12}\kket{\psi_{2}^{\{\al}\psi_{1}^{\be\}}}
+E_{12}\kket{\psi_{2}^{[\al}\psi_{1}^{\be]}}
+\half F_{12}\ep^{\al\be}\ep_{ab}\kket{\phi_{2}^{a}\phi_{1}^{b}\cZ^{+}}\,,\\
&S_{12}\kket{\phi_{1}^{a}\psi_{2}^{\be}}
=G_{12}\kket{\psi_{2}^{\be}\phi_{1}^{a}}
+H_{12}\kket{\phi_{2}^{a}\psi_{1}^{\be}}\,,\\
&S_{12}\kket{\psi_{1}^{\al}\phi_{2}^{b}}
=K_{12}\kket{\psi_{2}^{\al}\phi_{1}^{b}}
+L_{12}\kket{\phi_{2}^{b}\psi_{1}^{\al}}\,.\label{S12-2}
\end{align}
The ten coefficients $A_{12}\,,\dots, L_{12}$ are functions of $x_{1}^{\pm}$ and $x_{2}^{\pm}$\,.
Requiring the invariance of the S-matrix under the symmetry algebra $PSU(2|2)\ltimes \mathbb R^{3}$\,, {\em i.e.}, $[\fJ_{1}+\fJ_{2},S_{12}]=0$ with any generators $\fJ_{1,2}$\,, gives constraints among the ten coefficients.
Actually, for bosonic generators $\fR$\,, $\fL$ and for central charges $\fC$\,, $\fP$ and $\fK$\,, the commutation relation is trivially satisfied, and so we only need to care about the fermionic generators.
For example, when both $\fJ_{j}$ are $\fQ_{j}$\,, acting both sides of the commutation relation $[(\fQ_{1}+\fQ_{2})^{\ga}_{c},S_{12}]=0$ on the state $\kket{\phi_{1}^{a}\phi_{2}^{b}}$ yields 
\begin{align}
0= ~ &\kko{\a_{1}K_{12}-\f{1}{2}\ko{A_{12}+B_{12}}\a_{2}-\f{1}{2} C_{12}\b_{1}}\kket{\psi_{2}^{\ga}\phi_{1}^{b}}+{}\no\\[2mm]
+{}&\kko{\a_{1}L_{12}-\f{1}{2}\ko{A_{12}-B_{12}}\a_{1}+\f{1}{2} C_{12}\b_{2}\ko{\f{x_{1}^{-}}{x_{1}^{+}}}}\kket{\phi_{2}^{b}\psi_{1}^{\ga}}\no\\[2mm]
+{}&\kko{\a_{2}G_{12}-\f{1}{2}\ko{A_{12}-B_{12}}\a_{2}+\f{1}{2} C_{12}\b_{1}}\kket{\psi_{2}^{\ga}\phi_{1}^{a}}\no\\[2mm]
+{}&\kko{\a_{2}H_{12}-\f{1}{2}\ko{A_{12}+B_{12}}\a_{1}-\f{1}{2} C_{12}\b_{2}\ko{\f{x_{1}^{-}}{x_{1}^{+}}}}\kket{\phi_{2}^{b}\psi_{1}^{\ga}}\no\,,
\end{align}
which implies
\begin{align}
\begin{split}
&\a_{1}K_{12}+\a_{2}G_{12}=\a_{2}A_{12}\,,\qquad 
\a_{1}K_{12}-\a_{2}G_{12}=\a_{2}B_{12}+\b_{1}C_{12}\,,\\
&\a_{1}L_{12}+\a_{2}H_{12}=\a_{1}A_{12}\,,\qquad 
\a_{1}L_{12}-\a_{2}H_{12}=-\a_{1}B_{12}-\b_{2}\ko{\f{x_{1}^{-}}{x_{1}^{+}}}C_{12}\,.
\end{split}
\end{align}
In the intermediate calculation, relations such as $\kket{\phi_{2}\cZ^{+}\psi_{1}\cZ^{-}}=\mbox{\large $\f{x_{1}^{-}}{x_{1}^{+}}$}\kket{\phi_{2}\psi_{1}\cZ^{+}\cZ^{-}}=\mbox{\large $\f{x_{1}^{-}}{x_{1}^{+}}$}\kket{\phi_{2}\psi_{1}}$ and an identity of the form $\ep^{ij}\ep_{kl}=\delta^{i}_{l}\delta^{j}_{k}-\delta^{i}_{k}\delta^{j}_{l}$ are employed.
In this way, we can obtain constraints among the ten coefficients.
By solving the resulting set of equations, they are uniquely determined {up to an unknown overall scalar phase $S^{0}_{12}$} as \cite{Beisert:2005tm}\footnote{There is an alternative systematic derivation of the $SU(2|2)$ S-matrix based on the $SU(1|2)$ symmetric formulation \cite{Beisert:2005tm,Janik:2006dc}, in which the length of the spin-chain does not fluctuate.}
\begin{align}
&A_{12}=S^{0}_{12}\,\f{x_{2}^{+}-x_{1}^{-}}{x_{2}^{-}-x_{1}^{+}}\,,\label{A12}\\[2mm]
&B_{12}=S^{0}_{12}\,\f{x_{2}^{+}-x_{1}^{-}}{x_{2}^{-}-x_{1}^{+}}\ko{1-2\,\f{1-1/x_{2}^{-}x_{1}^{+}}{1-1/x_{2}^{-}x_{1}^{-}}\f{x_{2}^{+}-x_{1}^{+}}{x_{2}^{+}-x_{1}^{-}}}\,,\\[2mm]
&C_{12}=S^{0}_{12}\,\f{2\sqrt{i(x_{2}^{-}-x_{2}^{+})}\sqrt{i(x_{1}^{-}-x_{1}^{+})}}{x_{2}^{-}x_{1}^{-}-1}\f{x_{2}^{+}-x_{1}^{+}}{x_{2}^{-}-x_{1}^{+}}\,,\\[2mm]
&D_{12}=-S^{0}_{12}\,,\\[2mm]
&E_{12}=-S^{0}_{12}\ko{1-2\,\f{1-1/x_{2}^{+}x_{1}^{-}}{1-1/x_{2}^{+}x_{1}^{+}}\f{x_{2}^{-}-x_{1}^{-}}{x_{2}^{-}-x_{1}^{+}}}\,,\\[2mm]
&F_{12}=S^{0}_{12}\,\f{2\sqrt{i(x_{2}^{-}-x_{2}^{+})}\sqrt{i(x_{1}^{-}-x_{1}^{+})}}{x_{2}^{+}x_{1}^{+}-1}\f{x_{2}^{-}-x_{1}^{-}}{x_{2}^{-}-x_{1}^{+}}\,,\\[2mm]
&G_{12}=S^{0}_{12}\,\f{x_{2}^{+}-x_{1}^{+}}{x_{2}^{-}-x_{1}^{+}}\,,\\[2mm]
&H_{12}=S^{0}_{12}\,\sqrt{\f{x_{1}^{-}-x_{1}^{+}}{x_{2}^{-}-x_{2}^{+}}}\f{x_{2}^{+}-x_{2}^{-}}{x_{2}^{-}-x_{1}^{+}}\,,\\[2mm]
&K_{12}=S^{0}_{12}\,\sqrt{\f{x_{2}^{-}-x_{2}^{+}}{x_{1}^{-}-x_{1}^{+}}}\f{x_{1}^{+}-x_{1}^{-}}{x_{2}^{-}-x_{1}^{+}}\,,\\[2mm]
&L_{12}=S^{0}_{12}\,\f{x_{2}^{-}-x_{1}^{-}}{x_{2}^{-}-x_{1}^{+}}\,.\label{L12}
\end{align}
In writing down the coefficients, we took a Hermitian representation, where $x^{+}$ and $x^{-}$ are complex conjugates, and the four parameters $\a$\,, $\b$\,, $\c$ and $\d$ employed in Section \ref{sec:asymptotic spectrum} are explicitly given by
\begin{eqnarray}
&\ds \a=\sqrt{g}\,\ga\,,\quad
\b=-\f{\sqrt{g}}{\ga}\ko{1-\f{x^{-}}{x^{+}}}\,,\quad 
\c=i\sqrt{g}\,\ga\,\f{1}{x^{-}}\,,\quad
\d=-\f{i\sqrt{g}}{\ga}(x^{+}-x^{-})&\no\\[2mm]
&\ds \mbox{with}\quad \ga=\sqrt{i(x^{-}-x^{+})}\,, \quad \al=\be=g\,.&\no
\end{eqnarray}
Notice that the S-matrix is not of a familiar difference form, $S(u_{1},u_{2})\neq S(f(u_{1})-f(u_{2}))$\,.

Recall that we needed the so-called dressing phase factor \cite{Arutyunov:2004vx} to match the gauge theory Bethe ansatz in the thermodynamic limit (\ref{BA sig}) to the string theory integral equation (\ref{SBA rho}).
The unfixed degree of freedom, that is the overall scalar phase factor $S^{0}_{12}$ in the S-matrix (\ref{A12}\,-\,\ref{L12}), just corresponds to the dressing factor.
The relation between the scalar phase $S^{0}_{jk}$ and the dressing phase $\sig_{jk}$ is given by
\begin{equation}
(S^{0}(x_{j},x_{k}))^{2}=\f{x_{k}^{+}-x_{j}^{-}}{x_{k}^{-}-x_{j}^{+}}\f{1-1/x_{k}^{-}x_{j}^{+}}{1-1/x_{k}^{+}x_{j}^{-}}\,\sig^{2}(x_{j},x_{k})\,,\qquad 
\sig(x_{j},x_{k})=\exp(i\theta(x_{j},x_{k}))\,.
\label{S0}
\end{equation}
Indeed symmetry arguments allow us to include such a degree of freedom.

\begin{figure}[t]
\begin{center}
\vspace{.3cm}
\includegraphics[scale=0.75]{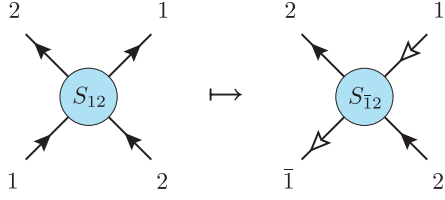}
\vspace{.3cm}
\caption{\small Crossing transformation.}
\label{fig:crossing}
\end{center}
\end{figure}

Moreover, there are theoretical arguments which favour the existence of the dressing factor\,; a non-trivial dressing factor is needed for the S-matrix to possess {\em crossing symmetry}, which S-matrices of integrable models are commonly expected to obey.
Via the crossing transformation, one of the scattering particles, say particle $j$ with $x_{j}^{\pm}$\,, is replaced with its conjugate particle $\bar j$ (therefore it is an antiparticle propagating backwards in space and time).
This corresponds to replacing $x_{j}^{\pm}$ with $x_{\bar j}^{\pm}=1/x_{j}^{\pm}$\,, see Figure \ref{fig:crossing}.
Janik derived a crossing equation \cite{Janik:2006dc}
\begin{equation}
S^{0}(x_{j},x_{k})S^{0}(1/x_{j},x_{k})=f(x_{j},x_{k})^{-1}
\label{crossing S0}
\end{equation}
where the function $f(x_{j},x_{k})$ is given by
\begin{equation}
f(x_{j},x_{k})=\f{x_{k}^{+}-x_{j}^{+}}{x_{k}^{+}-x_{j}^{-}}\f{1-1/x_{k}^{-}x_{j}^{+}}{1-1/x_{k}^{-}x_{j}^{-}}\,.
\end{equation}
The equation (\ref{crossing S0}) leads to a constraint 
\begin{align}
&\bth(x_{j}^{\pm},x_{k}^{\pm})+\bth(1/x_{j}^{\pm},x_{k}^{\pm})
=-i\ln h(x_{j}^{\pm},x_{k}^{\pm})\no\\[2mm]
&\qquad \mbox{with}\quad 
h(x_{j}^{\pm},x_{k}^{\pm})=\f{x_{k}^{-}}{x_{k}^{+}}\,\f{x_{j}^{-}-x_{k}^{+}}{x_{j}^{+}-x_{k}^{+}}\,\f{1-1/x_{j}^{-}x_{k}^{-}}{1-1/x_{j}^{+}x_{k}^{-}}\,.
\end{align}
Obviously, the trivial phase $\sig_{jk}=1$ is not consistent with the crossing symmetry.
On the other hand, we know from the gauge theory side, that $\bsig_{jk}$ must be one at the first three orders in $\lam$ \cite{Serban:2004jf}\,: $\bsig_{jk}=1+\cO(g^{6})$\,, and from the string theory side, that $\bsig_{jk}$ must be non-trivial around $\lam\sim \infty$ \cite{Arutyunov:2004vx}.
It is also known that there is an essential singularity at $\lam=\infty$\,.
A recent study on the gauge theory side \cite{Bern:2006ew} has provided important evidence for the existence of the dressing phase, where a first non-vanishing term in the dressing phase was indeed indirectly observed by a direct perturbative computation.
We will investigate the structure of the dressing phase in more detail in Section \ref{sec:dressing phase}.

\section[The Conjectured S-matrix of AdS/CFT]
	{The Conjectured S-matrix of AdS/CFT\label{sec:full S-matrix}}

The full S-matrix for all sixteen magnons can be obtained by straightforwardly generalising the symmetry algebra to $(PSU(2|2)\times PSU(2|2))\ltimes \mathbb R^{3}$\,.
It is given by
\begin{equation}
S^{\rm full}_{12}=(S^{0}_{12})^{2}\kko{S_{12}^{SU(2|2)_{\rm L}}\otimes S_{12}^{SU(2|2)_{\rm R}}}\,,
\end{equation}
where each $S_{12}^{SU(2|2)_{\rm L}}$ and $S_{12}^{SU(2|2)_{\rm R}}$ is given by the $SU(2|2)$ S-matrix appearing in (\ref{S12-1}\,-\,\ref{S12-2}) with coefficients (\ref{A12}\,-\,\ref{L12}).
The terms in the bracket are determined by the symmetries (\ref{unitarity}) and (\ref{YBE}), while the scalar phase cannot be determined by such symmetry requirements.

We discussed the Beisert-Staudacher complete asymptotic Bethe ansatz for the full $PSU(2,2|4)$ super spin-chain at the one-loop in Section \ref{sec:full one-loop}.
Generalising the results, a conjecture for the full model at all-loop was made in \cite{Beisert:2005fw} by the same authors.
A convenient choice of the super Dynkin diagram for the representation is the following one\,:
\newsavebox{\boxSdynkinn}
\sbox{\boxSdynkinn}{\includegraphics{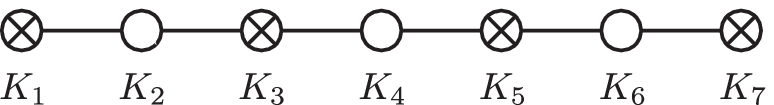}}
\newlength{\bwSdynkinn}
\settowidth{\bwSdynkinn}{\usebox{\boxSdynkinn}} 
\begin{center}
\vspace{.5cm}
\parbox{\bwSdynkinn}{\usebox{\boxSdynkinn}}\,,
\vspace{.5cm}
\end{center}
where $0\leq K_{2}\leq K_{1}+K_{3}\leq K_{4}\geq K_{5}+K_{7}\geq K_{6}\geq 0$\,.\footnote{Note that the condition for the filling numbers $K_{i}$ is different from the one for the one-loop case (which is discussed in page \pageref{1-loop K}).}
In this assignment, the Bethe ansatz equations for the full $\cN=4$ model take the form
\begin{align}
1&=\prod_{j=1}^{K_{4}}\f{x_{4,j}^{+}}{x_{4,j}^{-}}\,,\\[2mm]
1&=\prod_{j=1}^{K_{2}}\f{u_{1,k}-u_{2,j}+i/2}{u_{1,k}-u_{2,j}-i/2}
	\prod_{j=1}^{K_{4}}\f{1-g^{2}/x_{1,k}x_{4,j}^{+}}{1-g^{2}/x_{1,k}x_{4,j}^{-}}\,,\\[2mm]
1&=\prod_{\mbox{$j=1\atop j\neq k$}}^{K_{2}}\f{u_{2,k}-u_{2,j}-i}{u_{2,k}-u_{2,j}+i}
	\prod_{j=1}^{K_{3}}\f{u_{2,k}-u_{3,j}+i/2}{u_{2,k}-u_{3,j}-i/2}
	\prod_{j=1}^{K_{1}}\f{u_{2,k}-u_{1,j}+i/2}{u_{2,k}-u_{1,j}-i/2}\,,\\[2mm]
1&=\prod_{j=1}^{K_{2}}\f{u_{3,k}-u_{2,j}+i/2}{u_{3,k}-u_{2,j}-i/2}
	\prod_{j=1}^{K_{4}}\f{x_{3,k}-x_{4,j}^{+}}{x_{3,k}-x_{4,j}^{-}}\,,\\[2mm]
\ko{\f{x_{4,k}^{+}}{x_{4,k}^{-}}}^{J}&=\prod_{\mbox{$j=1\atop j\neq k$}}^{K_{4}}\ko{\f{u_{4,k}-u_{4,j}+i}{u_{4,k}-u_{4,j}-i}\,\bsig^{2}(u_{4,k},u_{4,j})}\times {}\no\\
	&\quad {}\times \prod_{j=1}^{K_{1}}\f{1-g^{2}/x_{4,k}^{-}x_{1,j}}{1-g^{2}/x_{4,k}^{+}x_{1,j}}
	\prod_{j=1}^{K_{3}}\f{x_{4,k}^{-}-x_{3,j}}{x_{4,k}^{+}-x_{3,j}}
	\prod_{j=1}^{K_{7}}\f{1-g^{2}/x_{4,k}^{-}x_{7,j}}{1-g^{2}/x_{4,k}^{+}x_{7,j}}
	\prod_{j=1}^{K_{5}}\f{x_{4,k}^{-}-x_{5,j}}{x_{4,k}^{+}-x_{5,j}}\,,\label{u4}\\[2mm]
1&=\prod_{j=1}^{K_{6}}\f{u_{5,k}-u_{6,j}+i/2}{u_{5,k}-u_{6,j}-i/2}
	\prod_{j=1}^{K_{4}}\f{x_{5,k}-x_{4,j}^{+}}{x_{5,k}-x_{4,j}^{-}}\,,\\[2mm]
1&=\prod_{\mbox{$j=1\atop j\neq k$}}^{K_{6}}\f{u_{6,k}-u_{6,j}-i}{u_{6,k}-u_{6,j}+i}
	\prod_{j=1}^{K_{5}}\f{u_{6,k}-u_{5,j}+i/2}{u_{6,k}-u_{5,j}-i/2}
	\prod_{j=1}^{K_{1}}\f{u_{6,k}-u_{7,j}+i/2}{u_{6,k}-u_{7,j}-i/2}\,,\\[2mm]
1&=\prod_{j=1}^{K_{6}}\f{u_{7,k}-u_{6,j}+i/2}{u_{7,k}-u_{6,j}-i/2}
	\prod_{j=1}^{K_{4}}\f{1-g^{2}/x_{7,k}x_{4,j}^{+}}{1-g^{2}/x_{7,k}x_{4,j}^{-}}\,,
\end{align}
where the length $L$ of the super spin-chain and the quantum number $J$ in (\ref{u4}) are related as
\begin{equation}
J=L+K_{4}+\mbox{\large $\f{1}{2}$}\ko{K_{1}-K_{3}-K_{5}+K_{7}}\,.
\end{equation}
The set of Bethe ansatz equations describes the scattering of string states in the usual temporal gauge $p_{\varphi}=J$ in the limit $J\to \infty$ while $\lam$ fixed (the ``decompactifying limit''), so that the gauge-fixed string sigma model becomes a two-dimensional field theory defined on a plane (rather than on a cylinder).

In particular, for the three simplest closed rank-one sectors, the $SU(2)$\,, $SU(1|1)$ and $SL(2)$ sectors, the Bethe ansatz equations reduce to the following forms,
\begin{equation}
\ko{\f{x_{k}^{+}}{x_{k}^{-}}}^{L}=\prod_{\mbox{$j=1\atop j\neq k$}}^{M}\ko{\f{x_{j}^{+}-x_{k}^{-}}{x_{j}^{-}-x_{k}^{+}}}^{\eta}
\f{1-1/(x_{j}^{+}x_{k}^{-})}{1-1/(x_{j}^{-}x_{k}^{+})}\,\bsig^{2}(x_{k}^{\pm},x_{j}^{\pm})\,,\qquad 
L=J+\f{\eta+1}{2}M\,,
\label{all-loop rank-1}
\end{equation}
where $\eta$ specifies the sectors as\footnote{It is interesting to note that the all-loop $SL(2)$ S-matrix is identical to the scalar phase factor (\ref{S0}).}
\begin{equation}
\eta=\left\{
\begin{array}{cc}
+1 \quad &\mbox{$SU(2)$}\,,\\
0 \quad &\mbox{$SU(1|1)$}\,,\\
-1 \quad &\mbox{$SL(2)$}\,.\end{array}\right.
\end{equation}
The all-loop Bethe ansatz equations (\ref{all-loop rank-1}) generalise the one-loop results (\ref{1-loop BAE : u}), (\ref{1-loop BAE : u SU(1|1)}) and (\ref{1-loop BAE : u SL(2)}) for the $SU(2)$\,, $SU(1|1)$ and $SL(2)$ sectors, respectively.
All the three sectors have the dispersion relation $\ga=ig\sum_{k=1}^{M}\ko{1/x_{k}^{+}-1/x_{k}^{-}}$ and the momentum condition $\prod_{k=1}^{M}x_{k}^{+}/x_{k}^{-}=1$ in common.
The higher charges are given by $\cQ_{r}=\sum_{k=1}^{K}q_{r}(x_{k}^{\pm})$ with $q_{r+1}(x^{\pm})=(i/r)\kko{1/(x_{k}^{+})^{r}-1/(x_{k}^{-})^{r}}$ $(r=1,2,\dots)$\,.

\section{Structure of the dressing phase\label{sec:dressing phase}}

We are now going to refine the structure of the dressing phase $\bsig^{2}(x_{k}^{\pm},x_{j}^{\pm})$\,, which is the last piece to be determined.
Let us recall our notation
\begin{equation}
\bsig(x_{k}^{\pm},x_{j}^{\pm};g)=\exp\kko{i\bth(x_{k}^{\pm},x_{j}^{\pm};g)}\,,
\end{equation}
and study the phase function $\theta_{kj}(g)$ in both the strong and the weak coupling limits.

\subsection{The strong coupling expansion}

\paragraph{Classical string level.}

In \cite{Arutyunov:2004vx}, the leading order structure of the phase was determined by Arutyunov, Frolov and Staudacher, originally as a way to explain the three-loop discrepancy.
This result is known as the {\em AFS phase} \cite{Arutyunov:2004vx}, which is given by
\begin{equation}
\bth(x_{k}^{\pm},x_{j}^{\pm};g)=g\sum_{r=2}^{\infty}\kko{q_{r}(x_{k}^{\pm})q_{r+1}(x_{j}^{\pm})-q_{r}(x_{j}^{\pm})q_{r+1}(x_{k}^{\pm})}\,.
\label{AFS phase}
\end{equation}
The charges $q_{r}(x^{\pm})$ are defined as in (\ref{higher charges}).
The phase is obtained by discretising the integral string Bethe ansatz equations (\ref{SBA rho}), that is to ``undo the thermodynamic limit'' to ``come back'' to its quantum form.
By construction, the AFS phase has a well-defined classical limit, and also yields the correct energies of massive states in the strict strong-coupling limit $g\to \infty$ \cite{Arutyunov:2004vx}.

However, (\ref{AFS phase}) captures only the leading order, and there are additional (infinitely many) quantum corrections.
At the quantum level, the phase can be expanded as
\begin{equation}
\bth(x_{k}^{\pm},x_{j}^{\pm};g)=\sum_{r=2}^{\infty}\sum_{s=r+1}^{\infty}c_{r,s}(g)\kko{q_{r}(x_{k}^{\pm})q_{s}(x_{j}^{\pm})-q_{r}(x_{j}^{\pm})q_{s}(x_{k}^{\pm})}\,.
\label{theta kj}
\end{equation}
It is a function of the coupling constant $g$ and the infinite tower of conserved higher charges $q_{r}$\,.
Actually this is the most general long-range integrable deformation of the Heisenberg spin-chain \cite{Beisert:2005wv}.
An infinite tower of unknown coefficients $c_{r,s}$ depending on the coupling constant $g$ remains to be determined.

\paragraph{First quantum corrections.}

Let us further expand the coefficients into ``modes'' at strong coupling,
\begin{align}
c_{r,s}(g)=\sum_{n=0}^{\infty}c_{r,s}^{(n)}g^{1-n}\,.
\label{c_{r,s}}
\end{align}
As we have just seen in (\ref{AFS phase}), agreement with classical string theory uniquely determines the first term, namely the AFS phase \cite{Arutyunov:2004vx} as $c_{r,s}^{(0)}=\delta_{r+1,s}$\,.
The leading corrections in $1/g$ ({\em i.e.}, the second modes $n=1$) were determined by Hernandez and Lopez \cite{Hernandez:2006tk} (see also \cite{Beisert:2005cw,SchaferNameki:2005is,Arutyunov:2006iu,Freyhult:2006vr}) by comparison with the one-loop corrections in the string sigma model, which are obtained from the spectrum of quadratic fluctuations around a classical solution \cite{Frolov:2003tu,Frolov:2004bh,Park:2005ji,Beisert:2005mq,Fuji:2005ry}.
This leads to
\begin{align}
c_{r,s}^{(1)}=\f{(-1)^{r+s}-1}{\pi}\f{(r-1)(s-1)}{(s+r-2)(s-r)}\,.
\label{HL 2}
\end{align}
A careful analysis of the one-loop sums over bosonic and fermionic string frequencies was performed in \cite{SchaferNameki:2005tn,Beisert:2005cw,SchaferNameki:2006gk,Hernandez:2006tk,Freyhult:2006vr}, which reproduced the Hernandez-Lopez phase (\ref{HL 2}).
The phase can be also derived by algebraic means \cite{Gromov:2007cd}.
Remarkably, it was shown that the AFS plus Hernandez-Lopez phase obey the crossing relation \cite{Arutyunov:2006iu} to one-loop order.

\paragraph{Higher corrections.}

An important feature of the general form of the dressing phase is that it is bilinear in the conserved charges $q_{r}$\,.
As a consequence, the scattering phase $\theta(x_{j}^{\pm},x_{k}^{\pm})$ is separately odd under the interchange of $x_{j}^{+}$ and $x_{j}^{-}$ and under the interchange of $x_{k}^{+}$ and $x_{k}^{-}$\,.
Because of unitarity, $\theta_{jk}$ is also odd under the interchange $x_{j}^{\pm}\leftrightarrow x_{k}^{\pm}$\,.
In fact, the general form (\ref{theta kj}), together with (\ref{higher charges}), implies that we can cast the dressing factor into the following form \cite{Arutyunov:2006iu},
\begin{equation}
\sigma(x_{j}^{\pm}, x_{k}^{\pm})^{2}=\ko{\f{R(x_{j}^{+},
    x_{k}^{+})R(x_{j}^{-}, x_{k}^{-})}
{R(x_{j}^{+}, x_{k}^{-})R(x_{j}^{-}, x_{k}^{+})}}^{2}\,,\qquad 
R(x_{j}, x_{k})=\exp\kko{i\chi(x_{[j}, x_{k]})}
\label{BEHLS}
\end{equation}
with $\chi(x_{[j}, x_{k]})\eq \chi(x_{j}, x_{k})-\chi(x_{k}, x_{j})$ an antisymmetric combination.
Defining $gk(x_{1},x_{2})\eq \chi(x_{[1},x_{2]})$\,, we can write it as
\begin{equation}
\theta(x_{1}^{\pm},x_{2}^{\pm})=2g\left[k(x_{1}^{+},x_{2}^{+})+k(x_{1}^{-},x_{2}^{-}) -k(x_{1}^{+},x_{2}^{-})-k(x_{1}^{-},x_{2}^{+})\right]\,.
\label{factorizedAFS}
\end{equation}
The function $k(x,y)$ admits an expansion of the form,
\begin{align}
k(x,y)&=-\sum_{r=2}^{\infty}\sum_{s=r+1}^{\infty}\f{c_{r,s}(g)}{(r-1)(s-1)}\f{1}{x^{r-1}y^{s-1}}\\
&=k_{0}(x,y)+ g^{-1}\,k_{1}(x,y)+{\mathcal{O}}(g^{-2}) 
\end{align}
The leading term can be deduced from the AFS phase $c_{r,s}^{(0)}=\delta_{r+1,s}$ to be 
\begin{equation}
k_{0}(x,y)=-\left[\left(x+\frac{1}{x}\right)-\left(y+\frac{1}{y}\right)\right]\ln\left(1-\frac{1}{xy}\right)
\label{elemetarychi0}\,.
\end{equation} 
The Hernandez-Lopez phase can be written in terms of the dilogarithm function $\mathrm{Li}_{2}(z)=\sum_{k=1}^{\infty}z^{k}/k^{2}$ as \cite{Arutyunov:2006iu,Beisert:2006ib}
\begin{align}
k_{1}(x,y)&=\kappa_{1}(x,y)-\kappa_{1}(y,x)\,,\label{HL-strong}\\[2mm]
\kappa_{1}(x,y)&=\f{1}{\pi}\ln\ko{\f{y-1}{y+1}}\ln\ko{\f{x-1/y}{x-y}}+{}\no\\[2mm]
&\quad {}+\f{1}{\pi}\komoji{\kko{\mathrm{Li}_{2}\ko{\f{\sqrt{y}-1/\sqrt{y}}{\sqrt{y}-\sqrt{x}}}-\mathrm{Li}_{2}\ko{\f{\sqrt{y}-1/\sqrt{y}}{\sqrt{y}-\sqrt{x}}}+\mathrm{Li}_{2}\ko{\f{\sqrt{y}-1/\sqrt{y}}{\sqrt{y}-\sqrt{x}}}-\mathrm{Li}_{2}\ko{\f{\sqrt{y}-1/\sqrt{y}}{\sqrt{y}-\sqrt{x}}}}}\,.
\end{align}
\paragraph{}
Finally, a crossing symmetric, all-order conjecture was made by Beisert, Hernandez and Lopez \cite{Beisert:2006ib}. 
The modes in (\ref{c_{r,s}}) are proposed as
\begin{align}
c_{r,s}^{(n)}=\f{\ko{1-(-1)^{r+s}}\zeta(n)}{2(-2\pi)^{n}\Gamma(n-1)}\,(r-1)(s-1)\,\f{\Gamma[(s+r+n-3)/2]\Gamma[(s-r+n-1)/2]}{\Gamma[(s+r-n+1)/2]\Gamma[(s-r-n+3)/2]}\,.
\label{c_{r,s}^{n}}
\end{align}
This conjecture is based on the following inputs.
The even $n$ parts are determined such that they satisfy the crossing relation.
The odd $n$ parts are chosen ``naturally'' (in the sense of ``simplest guess''), see \cite{Beisert:2006ib} for the details.\footnote{It solves the homogeneous crossing relation, but the choice is not unique and there is a room to include additional homogeneous solutions.
Some additional conditions (such as the transcendantality principle we are discussing in the next subsection) are needed to constrain the structures of the dressing phase further.}

\subsection{The weak coupling expansion\label{sec:weak-expansion}}

The general weak-coupling expansion of the dressing phase is given by \cite{Arutyunov:2004vx, Beisert:2005wv}
\begin{equation}
\bth(x_{k}^{\pm},x_{j}^{\pm};g)=\sum_{r=2}^{\infty}\sum_{s=r+1}^{\infty}\beta_{r,s}(g)g^{2-r-s}\kko{q_{r}(x_{k}^{\pm})q_{s}(x_{j}^{\pm})-q_{r}(x_{j}^{\pm})q_{s}(x_{k}^{\pm})}\,.
\label{theta kj weak}
\end{equation}
The coefficients for the all-order expansion, 
\begin{equation}
\beta_{r,s}(g)=\sum_{\ell=s-1}^{\infty}\beta_{r,s}^{(\ell)}g^{2\ell}\,,
\label{beta_{r,s}}
\end{equation}
where $\ell$ counts the number of loops in gauge theory,\footnote{Note that, the other coupling-dependent factor $g^{2-r-s}$ in (\ref{theta kj weak}) comes from the rescaling of the higher-order charges $q_{r}(x^{\pm})$ according to the rescaling ${\rm x}^{\pm}/g= x^{\pm}$\,.
See the footnote \ref{foot:x} in page \pageref{foot:x}.} are proposed as \cite{Beisert:2006ez}
\begin{align}
\beta_{r,s}^{(\ell)}&=2(-1)^{(2\ell+r-s-1)/2}\f{(r-1)(s-1)}{2\ell-r-s+2}\times{}\no\\
&\qquad {}\times \bigg(\begin{array}{c}2\ell-r-s+2 \\ \ell-r-s+2 \end{array}\bigg)
\bigg(\begin{array}{c}2\ell-r-s+2 \\ \ell-s+1 \end{array}\bigg)
\zeta(2\ell-r-s+2)\label{beta_{r,s}l-1}\\[2mm]
&=2(-1)^{(m+r-1)/2}\f{(r-1)(s-1)}{2m+1}\times{}\no\\
&\qquad {}\times \bigg(\begin{array}{c}2m+1 \\ m-(r+s-3)/2 \end{array}\bigg)
\bigg(\begin{array}{c}2m+1 \\ m-(s-r-1)/2 \end{array}\bigg)
\zeta(2m+1)
\label{beta_{r,s}l-2}
\end{align}
with $\beta_{r,s}^{(\ell)}=0$ for $\ell<r+s-2$\,.
In the second expression, the integer $m$ is defined through the relation $2\ell=2m+r+s-1$\,.
The coefficients $\beta_{r,s}^{(\ell)}$ have degree of transcendantality\footnote{The degree $k$ of transcendantality is attributed to each constant $\pi^{k}$ or $\zeta(k)$\,.} $2\ell-r-s+2$\,.
This turns out to ensure the so-called {\em ``transcendantality principle''} \cite{Kotikov:2002ab}, which states the transcendantality should be preserved to arbitrary orders in the weak coupling expansion \cite{Beisert:2006ez}.
Notice also that $r\pm s$ must be an odd integer for the absence of contributions from fractional loop orders in gauge theory.

Dramatically, the expressions in the strong and the weak coupling region \---- (\ref{theta kj}, \ref{c_{r,s}}) with (\ref{c_{r,s}^{n}}) for the former, and (\ref{theta kj weak}, \ref{beta_{r,s}}) with (\ref{beta_{r,s}l-1}) (or (\ref{beta_{r,s}l-2})) for the latter \---- can be shown to match by using a trick developed in \cite{Beisert:2006ez}.
It is a kind of analytic continuation of the index parameters, $c_{r,s}(g)=\sum_{n=0}^{\infty}c_{r,s}^{(n)}g^{1-n}=-\sum_{n=0}^{\infty}\tilde c_{r,s}^{(-n)}g^{1+n}$ with $c_{r,s}^{(n)}=\tilde c_{r,s}^{(-n)}$\,, which implies $\beta_{r,s}^{(\ell)}=-c_{r,s}^{(-(2\ell-r-s+1))}$\,.

In the weak coupling regime, this conjecture was shown to be consistent with the study of general four- and five-loop long-range spin-chains, where it was found that the above structure of $\bsig(g)$ (with analytic $c_{r,s}(g)$) indeed appears \cite{Beisert:2005wv}.
Significantly, as we briefly mentioned before, it was observed \cite{Bern:2006ew}, following the strategy proposed in \cite{Eden:2006rx}, that the gauge theory four-loop computation requires a first non-vanishing piece in the dressing phase, and it matches with the proposal above.

\paragraph{}
The proposed dressing phase breaks BMN scaling explicitly.
This can be seen easily by first noticing that the phase is dominated by the first $(r,s)=(2,3)$ piece in the weak coupling limit, then evaluating it as
\begin{align}
\theta(u,u';g)&= \beta_{2,3}^{(3)}g^{6}g^{2-2-3}\ko{q_{2}q_{3}'-q_{2}'q_{3}}+\dots\no\\
&=g^{6}J^{-4}\times {\rm finite}+\dots =\tlambda^{3}J^{2}\times {\rm finite}+\dots \,,
\label{break BMN}
\end{align}
where $\tlambda= \lam/J^{2}$ is the BMN coupling.
Here we used the fact that each $q_{r}$ scales as $(g/J)^{r-1}$ in this limit, and the difference term $(q_{r}q_{s}'-q_{r}'q_{s})$ yields an extra factor of $1/J$\,.
The $\tlambda^{3}J^{2}$ term in (\ref{break BMN}) diverges in the BMN limit (\ref{BMN limit}), thus explicitly breaking the BMN scaling.\footnote{The breakdown of the BMN scaling was already observed earlier in the plane-wave matrix model \cite{Fischbacher:2004iu} as a simple toy model for $\cN=4$ SYM.}
In light of this feature, the ``three-loop discrepancy'' observed in the near-BMN and the Frolov-Tseytlin limits discussed in Section \ref{sec:discrepancy} turned out to have resulted from such ill-defined limits.

\subsubsection*{Closed integral form of the dressing factor}

Using the proposed all-order coefficients (\ref{c_{r,s}^{n}}), the function $\chi(x_{j}, x_{k})$ is obtained as \cite{Beisert:2006ez}
\begin{align}
\chi(x_{j}, x_{k})=-\sum_{r=2}^{\infty}\sum_{s=r+1}^{\infty}\f{2\cos\kko{(s-r-1)\pi/2}}{x_{j}^{r-1}x_{k}^{s-1}}\int_{0}^{\infty}\f{dt}{t}\f{{J}_{r-1}(2gt){J}_{s-1}(2gt)}{e^{t}-1}\,,
\label{chi 0}
\end{align}
where ${J}_{r}(z)=\sum_{m=0}^{\infty}\mbox{\large $\f{(-1)^{m}}{m!\Gamma(m+r+1)}$}\ko{\mbox{\large $\f{z}{2}$}}^{2m+r}$ are standard Bessel functions.
Starting with the expression (\ref{chi 0}), one can derive the following closed integral form \cite{Dorey:2007xn}\,:
\begin{equation}
\chi(x_{j},x_{k})=-i\oint_{\mathcal C}\f{dz_{1}}{2\pi}\oint_{\mathcal
  C}\f{dz_{2}}{2\pi}
\f{\ln\Gamma\ko{1+ig\ko{z_{1}+\hf{1}{z_{1}}-z_{2}-\hf{1}{z_{2}}}}}{(z_{1}-x_{j})(z_{2}-x_{k})}\,.
\label{chi}
\end{equation}
The contours in (\ref{chi}) are unit circles $|z_{1}|=|z_{2}|=1$\,.
In fact, (\ref{chi}) differs from (\ref{chi 0}) by terms symmetric under the interchange $x_{j}\leftrightarrow x_{k}$\,, but both (\ref{chi}) and (\ref{chi 0}) yield the same dressing phase since $\chi(x_{j},x_{k})$ enters in the antisymmetric form $\chi(x_{[j},x_{k]})=\chi(x_{j},x_{k})-\chi(x_{k},x_{j})$\,.
We will make use of the expression (\ref{chi}) in studying the singular structures of the AdS/CFT S-matrix later in the final chapter \ref{chap:Singularities}.

\chapter[Dyonic Giant Magnons]
	{Dyonic Giant Magnons\label{chap:DGM}}

\section{Solitons in AdS/CFT}

In \cite{Hofman:2006xt}, Hofman and Maldacena identified a particular limit where the problem of determining the spectrum simplifies considerably.
The limit is defined as
\begin{equation}
\Delta\,,~J_{1}\to \infty\,,\qquad 
\Delta-J_{1}\, :~\mbox{fixed}\,,\qquad 
\lam\, :~\mbox{fixed}\,.
\label{HM limit}
\end{equation}
In this Hofman-Maldacena limit, both the gauge theory spin-chain and the dual string effectively become infinitely long.
The spectrum can then be analysed in terms of asymptotic states and their scattering.
On both sides of the correspondence the limiting theory is characterised by a centrally-extended $SU(2|2)\times SU(2|2)$ supergroup we discussed in Chapter \ref{chap:Asymptotic}, which strongly constrains the spectrum and S-matrix \cite{Beisert:2005tm}.

\begin{figure}[thb]
\begin{center}
\vspace{.3cm}
\includegraphics[scale=0.9]{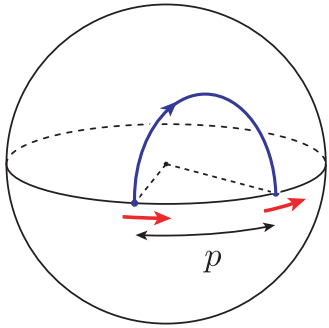}
\vspace{.3cm}
\caption{\small A giant magnon solution.
The endpoints of the string move on the equator 
$\theta=\pi/2$ at the speed of light.
The magnon momentum is given by $p=\Delta\varphi_{1}$\,, where
$\Delta\varphi_{1}$ is the angular distance between 
two endpoints of the string.}
\label{fig:GM}
\end{center}
\end{figure}

The basic asymptotic state carries a conserved momentum $p$\,, and lies in a short multiplet of supersymmetry.
States in this multiplet have different polarisations corresponding to transverse fluctuations of the dual string in different directions in $AdS_{5}\times S^{5}$\,. 
The BPS condition essentially determines the dispersion relation for all these states to be \cite{Beisert:2004hm, Staudacher:2004tk, Beisert:2005fw, Beisert:2005tm} (see also \cite{Santambrogio:2002sb, Berenstein:2005jq}), 
\begin{eqnarray}
\Delta-J_{1} & = &
 \sqrt{1+\frac{\lambda}{\pi^{2}}\sin^{2}\left(\frac{p}{2}\right)}\,,
\label{disp1}
\end{eqnarray}
as we showed in more general form in Chapter \ref{chap:Asymptotic}, see (\ref{kdisprel}).
In the spin-chain description, this multiplet corresponds to an elementary excitation of the ferromagnetic vacuum, namely a magnon.
The dual state in semiclassical string theory was identified in \cite{Hofman:2006xt}.
It corresponds to a localised classical soliton which propagates on an infinite string moving on an ${\mathbb{R}}\times 
S^2$ subspace of $AdS_{5}\times S^{5}$\,.
The conserved magnon momentum $p$ corresponds to a certain geometrical angle in the target space explaining the periodic momentum dependence appearing in (\ref{disp1}).
Following \cite{Hofman:2006xt}, this classical string configuration is referred to as a {\em giant magnon}.
For a diagram, see Figure \ref{fig:GM}.
The single-spin giant magnon was generalised to the two-spin giant magnon in \cite{Chen:2006ge}, which is called the {\em dyonic giant magnon}.
Below we are going to review those soliton solutions of string theory as well as their gauge theory duals, and explain the important roles they played in testing the conjectured AdS/CFT S-matrix.

\section{Giant magnons}

In static gauge, the string equations of motion are essentially those of a bosonic $O(3)$ sigma model supplemented by the Virasoro constraints. 
An efficient way to find the relevant classical solutions exploits the equivalence of this system to the sine-Gordon (sG) equation,
\begin{equation}
\partial _a \partial ^{\ssp a} \phi  - \sin \phi =0\,.
\label{sG eq}
\end{equation}
discovered many years ago by Pohlmeyer \cite{Pohlmeyer:1975nb} (see also \cite{Mikhailov:2005qv,Mikhailov:2005sy}).

\paragraph{}
We begin by discussing strings on ${\mathbb{R}}\times S^{2}$ in the Hofman-Maldacena limit, then explain the classical equivalence of the string equations to the sine-Gordon equation.
In the Hofman-Maldacena limit (\ref{HM limit}), $\lam$ is held fixed.
This fact allows us to interpolate between the regimes of small and large $\lambda$\,, where perturbative gauge theory and semiclassical string theory respectively are valid. 
It is convenient to implement the Hofman-Maldacena limit for the string by defining the following rescaled worldsheet coordinates, $(x,t)\equiv(\kappa\sigma,\kappa\tau)$\,, which are held fixed as $\kappa=\Delta/\sqrt{\lambda}\to\infty$\,.
Under this rescaling, the interval $-\pi \le \sigma \le \pi$ corresponding to the closed string is mapped to the real line $-\infty\le x \le \infty$ with the point $\sigma=\pm\pi$ mapped to $x=\pm\infty$\,.
As always, a consistent closed string configuration always involves at least two magnons with zero total momentum.
The condition that the total momentum vanishes (modulo $2\pi$) is then enforced by the closed string boundary condition.
However, after the above rescaling, the closed string boundary condition and thus the vanishing of the total momentum can actually be relaxed.
This allows us to focus on a single worldsheet excitation or magnon carrying non-zero momentum $p$\,.

The conserved charges of the system which remain finite in the Hofman-Maldacena limit
are given as, 
\begin{align}
\Delta-J_{1}&=\frac{\sqrt{\lambda}}{2\pi}\int^{+\infty}_{-\infty}dx\,
(1-{\mathrm{Im}}
[\xi_{1}^{*}\partial_{t}\xi_{1}])
\label{E-J1}\,,\\[2mm]
J_{2}&=\frac{\sqrt{\lambda}}{2\pi}\int^{\infty}_{-\infty}dx\,{\mathrm{Im}}
[\xi_{2}^{*}\partial_{t}\xi_{2}]
\label{J2*}\,.
\end{align}
It is important to note that these quantities will not necessarily be equal to their counter-parts (\ref{S-charges}) computed in the original worldsheet coordinates.
In general, the latter may include an additional contribution coming from a neighbourhood of the point $\sigma=\pm\pi$ which is mapped to $x=\infty$\,. 
As in the above discussion of worldsheet momentum, the extra contribution reflects the presence of additional magnons at infinity.

\paragraph{}
There is a (classical) equivalence between the $O(3)$ sigma model and the sG theory, which will be explained in the next section in detail, for the case of a more general equivalence between the $O(4)$ sigma model and the Complex sine-Gordon theory.
For the moment, let us just make use of the consequences of the equivalence.
Then one can construct classical string solutions on $\mathbb R\times S^{2}$ by the following recipe\,:\label{Phlmeyer's recipe}
Firstly, find a solution $\phi$ of sG equation (\ref{sG eq}).
Secondly, relate the sG solution $\phi$ with the sigma model fields as 
\begin{equation}
\pa_{a}\vec\xi\cdot\pa^{\ssp a}\vec\xi^{*}\equiv \cos \phi \,.
\label{def_cos_phi}
\end{equation}
Thirdly, plug (\ref{def_cos_phi}) into the string equation of motion (\ref{string_eom})\,, which leads to a Schr\"{o}dinger type differential equation
\begin{equation}
\pa_{a} \pa^{\ssp a} \vec \xi + (\cos \phi) \vec \xi = \vec 0 \,.
\label{phi<->del X del X}
\end{equation}
Fourthly, solve the differential equation (\ref{phi<->del X del X}) under appropriate boundary conditions.
Fifthly and finally, the resulting set of $\eta_{0}\,(=\kappa \tau)$ and $\vec\xi$ gives the corresponding string profile in $\mathbb R\times S^{2}$\,.

For our purpose of obtaining a giant magnon solution, we choose a sG kink soliton solution
\begin{equation}
\phi(x,t)=2\arcsin\kko{{1 \mathord{\left/ {\vphantom {\cosh\ko{\f{x - vt}{\sqrt{1-v^{2}}}}}} \right. \kern-\nulldelimiterspace} \cosh\ko{\f{x - vt}{\sqrt{1-v^{2}}}}}}
\label{sG kink}
\end{equation}
as the solution $\phi$ in the first step, and impose the boundary conditions
\begin{equation}
\xi_{1}\to \exp\left(it\pm i\Delta\varphi_{1}/2\right)\,,\qquad 
\xi_{2}\to 0,
\quad {\mathrm{as}}\quad x\to\pm\infty
\label{b.c.}
\end{equation}
in the fourth step.
The boundary condition can be fulfilled if and only if $v=\cos(\Delta\varphi_{1}/2)$\,, when the differential equation is solved to give
\begin{align}
\xi_{1}&={\kko{\sin\ko{\mbox{\large $\f{\Delta\varphi_{1}}{2}$}}\tanh\ko{\mbox{\large $\f{x-\cos(\Delta\varphi_{1}/2)\, t}{\sin(\Delta\varphi_{1}/2)}$}}-i\cos\ko{\mbox{\large $\f{\Delta\varphi_{1}}{2}$}}}}\exp\ko{i t}\,,\label{GM1}\\[2mm]
\xi_{2}&= \left.\sin\ko{\mbox{\large $\f{\Delta\varphi_{1}}{2}$}}\right/\cosh\ko{\mbox{\large $\f{x-\cos(\Delta\varphi_{1}/2)\, t}{\sin(\Delta\varphi_{1}/2)}$}}
\,.\label{GM2}
\end{align}
This is the profile of the giant magnon, and is equivalent to the one given as (2.16) in the original paper \cite{Hofman:2006xt}.
We will rederive it as a special case of the more general solution presented below.
One may check that while the energy $\Delta$ and the angular momentum $J_{1}$ of the solution diverge, the combination $\Delta-J_{1}$ remains finite and is given by
\begin{equation}
\Delta-J_{1}=\frac{\sqrt{\lambda}}{\pi}\left|\sin\left(\frac{p}{2}\right)
\right|\label{hmE-J1}
\end{equation}
in agreement with the large-$\lambda$ limit of (\ref{disp1}).
Moreover the solution (\ref{GM1}), (\ref{GM2}) carries only one non-vanishing angular momentum, having $J_{2}=0$\,.

\section{Dyonic giant magnons\label{sec:DGM}}

We saw in Chapter \ref{chap:Asymptotic} that in addition to the elementary magnon, the asymptotic spectrum of the $\cN=4$ SYM spin-chain also contains an infinite tower of boundstates
\cite{Dorey:2006dq}.
Magnons with polarisations in an $SU(2)$ subsector carry a second conserved $U(1)$ R-charge, denoted $J_{2}$\,, and form 
boundstates with the exact dispersion relation (see (\ref{kdisprel})),
\begin{equation}
\Delta-J_{1} = 
 \sqrt{J_{2}^{2}+\frac{\lambda}{\pi^{2}}\sin^{2}\left(\frac{p}{2}\right)}\,.
\label{disp3}
\end{equation}
The elementary magnon in this subsector has charge $J_{2}=1$ and states with $J_{2}=Q$ correspond to $Q$-magnon boundstates.
These states should exist for all integer values of $J_{2}$ and for all values of the 't Hooft coupling \cite{Dorey:2006dq}.
In particular we are free to consider states where $J_{2}\sim \sqrt{\lambda}$\,.
For such states the dispersion relation (\ref{disp3}) has the appropriate scaling for a classical string carrying a second large classical angular momentum $J_{2}$\,.

In this section we will identify the corresponding classical solutions of the worldsheet theory and determine some of their properties.
In particular, we will reproduce the exact BPS dispersion relation (\ref{disp3}) from a purely classical calculation in string theory.
We also briefly discuss semiclassical quantisation of these objects which simply has the effect of restricting the R-charge $J_{2}$ to integer values.

\subsubsection*{Pohlmeyer reduction procedure for the $\bmt{O(4)}$ string sigma model}

The minimal string solutions carrying two independent angular momenta, $J_{1}$ and $J_{2}$ correspond to strings moving on an 
${\mathbb{R}}\times S^{3}$ subspace of $AdS_{5}\times S^{5}$\,.
Again we can make use of the classical equivalence between the $O(4)$ string sigma model in conformal gauge and the Complex sine-Gordon (CsG) model in order to construct the string solutions.
The CsG equation is completely integrable and has a family of soliton solutions \cite{Lund:1976ze, Lund:1977dt, Getmanov:1977nf, Getmanov:1980cq, de Vega:1981ka, de Vega:1982sh, Dorey:1994mg, Bowcock:2002vz}.
In addition to a conserved momentum the soliton also carries an additional conserved charge associated with rotations in an internal space.
The problem of reconstructing the corresponding string motion, while still non-trivial, involves solving linear differential equations only (as we already demonstrated for the case of an elementary giant magnon).
We construct the two-parameter family of string solutions corresponding to a single CsG soliton and show that they have all the expected properties of giant magnons.
In particular they carry non-zero $J_{2}$ and obey the BPS dispersion relation (\ref{disp3}).
It is quite striking that we obtain the {\em exact} BPS formula, for all values of $J_{2}$\,, from a classical calculation. 
This situation seems to be very analogous to that of BPS-saturated Julia-Zee dyons in ${\cal N}=4$ SYM \cite{Julia:1975ff, TW:1976:SQG}.
These objects also have a classical BPS mass formula which turns out to be exact.
It seems appropriate to call our new two-charge configurations {\em dyonic giant magnons}.

\paragraph{}
We begin by discussing strings on ${\mathbb{R}}\times S^{3}$ in the Hofman-Maldacena limit, and explain the classical equivalence of the string equations to the CsG equation.
Then we construct the required string solutions from the CsG solitons and determine their properties. 

\begin{figure}[htb]
\begin{center}
\vspace{.3cm}
\hspace{-.0cm}\includegraphics[scale=0.9]{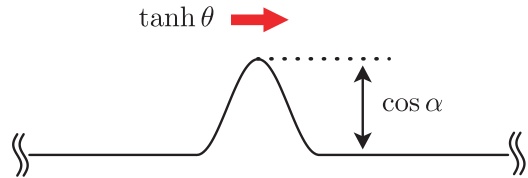}
\vspace{.3cm}
\caption{\small A kink soliton solution of CsG equation.}
\label{fig:kink CsG}
\end{center}
\end{figure}

The equation of the motion for the target space coordinate $\vec{X}(x,t)$
can be written in terms of light-cone coordinates $x_{\pm}=(t\pm x)/2$
as
\begin{equation}
\partial_{+}\partial_{-}\vec{X}+(\partial_{+}\vec{X}\cdot\partial_{-}\vec{X})
\vec{X}=0
\label{eomX}\,.
\end{equation}
A physical string solution must also satisfy the
Virasoro constraints. In terms of the rescaled coordinates these
become,  
\begin{equation}
\partial_{+}\vec{X}\cdot\partial_{+}\vec{X}=
\partial_{-}\vec{X}\cdot\partial_{-}\vec{X}=1
\label{virasoro}\,.
\end{equation}
To solve the string equations of motion (\ref{eomX}) in the general
case, together with the
Virasoro conditions (\ref{virasoro}), we will exploit the 
equivalence of this system with the CsG
equation. Following \cite{Pohlmeyer:1975nb}, we will begin by
identifying the $SO(4)$ 
invariant combinations of the worldsheet fields $\vec{X}$ and their
derivatives. As the first derivatives $\partial_{\pm}\vec{X}$ are unit 
vectors, we can define a real scalar field $\phi(x,t)$ via the 
relation,    
\begin{equation}
\cos\phi=\partial_{+}\vec{X}\cdot\partial_{-}\vec{X}\label{cosphi}\,.
\end{equation}
Taking into account the constraint $|\vec{X}|^{2}=1$\,, we see that 
there are no other independent $SO(4)$ invariant quantities that can be
constructed out of the fields and their first derivatives. At the
level of second derivatives we can construct two additional
invariants;     
\begin{equation}
u\sin\phi =\partial_{+}^{2}\vec{X}\cdot \vec{K}\,,\;\;\; 
v\sin\phi =\partial_{-}^{2}\vec{X}\cdot\vec{K}\label{uv}\,,
\end{equation}
where the components of vector $\vec{K}$ are given by 
$K_{i}=\epsilon_{ijkl}X_{j}\partial_{+}X_{k}\partial_{-}X_{l}$\,.
The connection with the CsG model arises from the equations of motion for
$u$\,, $v$ and $\phi$ derived in \cite{Pohlmeyer:1975nb}. 
In fact the resulting equations imply 
that $u$ and $v$ are not independent and can be
eliminated in favour of a new field $\chi(x,t)$ as, 
\begin{equation}
u=\partial_{+}\chi\tan\left(\frac{\phi}{2}\right)\,,\;\;\;\;\;
v=-\partial_{-}\chi\tan\left(\frac{\phi}{2}\right)\label{newuv}\,.
\end{equation}
The equations of motion for $\chi$ and $\phi$ can then be written as
\begin{align}
0&=\partial_{+}\partial_{-}\phi+\sin\phi-\frac{\tan^{2}(\phi/2)}{\sin\phi}\,
\partial_{+}\chi\partial_{-}\chi
\label{eomphi}\,,\\
0&=\partial_{+}\partial_{-}\chi+\frac{1}{\sin\phi} 
(\partial_{+}\phi\,\partial_{-}\chi+
\partial_{-}\phi\,\partial_{+}\chi)
\label{eomchi}\,.
\end{align}
In the special case of constant $\chi$ they reduce to the usual
sine-Gordon equation for $\phi(x,t)$\,. 
Finally we can combine the real fields $\phi$ and $\chi$ to form a  
complex field
$\psi=\sin\left(\phi/2\right)\exp(i\chi/2)$\,, which
obeys the equation,
\begin{equation} 
\partial_{+}\partial_{-}\psi+
\psi^{*}
\,\frac{\partial_{+}\psi\,\partial_{-}\psi}{1-|\psi|^{2}}+\psi(1-|\psi|^{2})=0\,.
\label{csg}
\end{equation}
Equation (\ref{csg}) is known as the {\em Complex sine-Gordon equation}.
Like the
ordinary sG equation, it is completely integrable and has 
localised soliton solutions which undergo factorised scattering. 
The CsG equation is invariant under a global rotation of the phase of
the complex field\,: $\psi\rightarrow \exp(i\nu)\psi$ and
$\psi^{*}\rightarrow \exp(-i\nu)\psi^{*}$\,. In addition to momentum and
energy, CsG solitons carry the corresponding conserved $U(1)$ 
Noether charge\footnote{Note that there is no simple relation 
between the $U(1)$ 
charge of the CsG soliton and the string angular momentum $J_{2}$\,.}.  
The most general one-soliton solution to (\ref{csg}) is given by (see
eg \cite{Dorey:1994mg}), 
\begin{equation}
\psi_{\rm 1\mbox{\scriptsize -}soliton} =  
\frac{(\cos\alpha)\,e^{i\mu}}{\cosh[(\cos\alpha)(X-X_{0})]}\exp[i(\sin\alpha)
  T]
\label{onesol}
\end{equation}
with
\begin{equation}
\left[\begin{array}{c}
X \\
T\end{array}\right]
=
\left[\begin{array}{cc}
\cosh\th & -\sinh\th \\
-\sinh\th & \cosh\th \end{array}\right]
\left[\begin{array}{c}
x \\
t\end{array}\right]\,.
\label{XandT}
\end{equation}
The constant phase $\mu$ is irrelevant for our purposes as only the
derivatives of the field $\chi$ affect the corresponding string
solution. The parameter $X_{0}$ can be absorbed by a constant 
translation of the world-sheet coordinate $x$ and we will set it to
zero. The two remaining parameters of the solution 
are the rapidity $\theta$ of the soliton and an additional  
real number $\alpha$ which determines the $U(1)$ charge carried by the
soliton.
See Figure \ref{fig:kink CsG} for the diagram.

\subsubsection*{Dyonic giant magnons from CsG kinks}
In order to obtain a dyonic giant magnon via Pohmeyer's reduction procedure, we have only to modify the steps 1 and 2 of the recipe for an elementary giant magnon case (see page \pageref{Phlmeyer's recipe}) such that $\cos\phi\eq \partial_{a} \vec\xi \, {}^* \cdot \partial^{\ssp a} \vec \xi$\,, where $\phi$ appears in the real part of a solution $\psi$ of CsG equation (\ref{csg}).
Taking the limit $\alpha\rightarrow 0$\,, the field $\psi$ corresponding to the one-soliton solution (\ref{onesol}) reduces to the kink
solution of the ordinary sG equation.
As it is the only known solution of the CsG equation with this property, it is the unique candidate for the dyonic giant magnon solution we seek.
It remains to reconstruct the corresponding configuration of the string worldsheet fields $\vec{X}$ (or equivalently $\xi_{1}$ and $\xi_{2}$) corresponding to (\ref{onesol}) for general values of the rapidity $\theta$ and rotation parameter $\alpha$\,.
In this case we have, 
\begin{equation}
\partial_{+}\vec{\xi}\cdot\partial_{-}\vec{\xi}^{*}=\cos\phi=1-
\frac{2\cos^{2}\alpha}{\cosh^{2}\left[(\cos\alpha)X\right]}\,.
\label{sc1}
\end{equation}
Hence the complex coordinates $\xi_{1}$ and $\xi_{2}$ must both solve 
the linear equation, 
\begin{eqnarray}
\frac{\partial^{2}\vec \xi}{\partial t^{2}}-
\frac{\partial^{2}\vec \xi}{\partial x^{2}}+\left(1-
\frac{2\cos^{2}\alpha}{\cosh^{2}\left[(\cos\alpha) X\right]}\right)\vec \xi =\vec 0
\label{xt1}
\end{eqnarray}
where, as above $X=(\cosh\theta)x-(\sinh\theta)t$ and we impose the boundary conditions appropriate for a giant magnon with momentum $P$\,,
\begin{equation}
\xi_{1}\to \exp\left(it\pm iP/2\right)\,,\quad \xi_{2}\to 0\,,
\quad 
{\mathrm{as}}\quad 
x\to\pm\infty\label{b.c.2}\,.
\end{equation} 
As always the two complex fields obey the 
constraint $|\,\vec \xi\,|^{2}=|\xi_{1}|^{2}+|\xi_{2}|^{2}=1$\,.
We will find unique solutions of the linear equation (\ref{xt1})
obeying these conditions and then, for self-consistency, check that
they correctly reproduce (\ref{sc1}).   
\paragraph{}
It is convenient to express the solution of (\ref{xt1}) 
in terms of the boosted coordinates $X$ and $T$\,. In terms of these
variables $\vec \xi =\vec \xi (X,T)$ obeys, 
\begin{equation}
\frac{\partial^{2}\vec \xi}{\partial T^{2}}-
\frac{\partial^{2}\vec \xi}{\partial X^{2}}+\ko{1-
\frac{2\cos^{2}\alpha}{\cosh^{2}\kko{(\cos\alpha)X}}}\vec \xi = \vec 0\,. 
\label{xt2}
\end{equation}
The problem now has the form of an ordinary Klein-Gordon equation describing the scattering of a relativistic particle in one spatial dimension incident on a static potential well.
We give the detailed derivation of the solution in Appendix \ref{app:derivation of DGM}, and here we just display the final result as
\begin{align}
\begin{split}
\xi_{1} & = 
\kko{\sin\ko{\hf{P}{2}}\tanh\left[(\cos\alpha)X\right]-i\cos\ko{\hf{P}{2}}}\exp(it)\,, \\[2mm]
\xi_{2} & = 
\kko{\left.\sin\ko{\hf{P}{2}}\right/\cosh\left[(\cos\alpha)X\right]}
\exp\left(i(\sin\alpha)T\right)\,,
\label{soln}
\end{split}
\end{align}
where $X$\,, $T$ are given by
\begin{align}
\left[\begin{array}{c}
X \\
T\end{array}\right]
=
\f{1}{\sin(P/2)}
\left[\begin{array}{cc}
\sqrt{1-\cos^{2}(P/2)\sin^{2}\al} & -\cos(P/2)\cos\al \\
-\cos(P/2)\cos\al & \sqrt{1-\cos^{2}(P/2)\sin^{2}\al}\end{array}\right]
\left[\begin{array}{c}
x \\
t\end{array}\right]\,.
\end{align}
One may easily check that this solution, in addition to obeying the string equation of motion (\ref{xt1}) and boundary conditions (\ref{b.c.2}), obeys the Virasoro constraints and satisfies the self-consistency condition (\ref{sc1}). 
It also reduces to the Hofman-Maldacena solution (\ref{GM1}), (\ref{GM2}) in the non-rotating case $\alpha=0$\,.
Setting $P=\pi$\,, we obtain one-half of the folded string configuration discussed in \cite{Dorey:2006dq}.\footnote{Recall that in the Hofman-Maldacena limit we have relaxed the closed string boundary condition.
To obtain a consistent closed string configuration we should add a second magnon at infinity in the coordinate $x$\,.
In spacetime this corresponds to adding a second open string to form a folded closed string.}

\paragraph{}
The solution (\ref{soln}) depends on two parameters: $P$ and $\alpha$\,.
We can now evaluate the conserved charges (\ref{E-J1}) and (\ref{J2*}) as a function of these parameters.
\begin{align}
E-J_{1} & = \frac{\sqrt{\lambda}}{\pi}\, \sin\ko{\f{P}{2}}\, 
\frac{\sqrt{1-\cos^{2}(P/2)\sin^{2}\al}}{\cos\alpha}\,, \nonumber \\[2mm]
J_{2} & = \frac{\sqrt{\lambda}}{\pi}\, \sin\ko{\f{P}{2}}\tan\alpha\,. 
\label{param}
\end{align}
Eliminating $\alpha$ we obtain the dispersion relation, 
\begin{equation}
E-J_{1} = \sqrt{J_{2}^{2}+\frac{\lambda}{\pi^{2}}\sin^{2}
\left(\frac{P}{2}\right)}\,,
\label{res}
\end{equation}
which agrees precisely with the BPS dispersion relation (\ref{disp3}) for
the magnon boundstates obtained in \cite{Dorey:2006dq}.  

\subsubsection*{The action variable}

The time dependence of the solution (\ref{soln}) is also of interest. 
As in the orginal Hofman-Maldacena solution the 
constant phase rotation of $\xi_{1}$ with exponent $it$
ensures that the endpoints of the string move on an equator of the
three-sphere at the speed of light. We can remove this dependence by
changing coordinates from $\xi_{1}$ to $\tilde{\xi}_{1}=\exp(-it)\xi_{1}$\,. 
In the new frame, the string configuration 
depends periodically on time through 
the $t$\,-dependence of $\xi_{2}$\,. The period, $T_{0}$\,, 
for this motion is the time 
for the solution to come back to itself up to a translation of
the worldsheet coordinate $x$\,. From (\ref{soln}) we find
\begin{equation}
T_{0}=2\pi\,\frac{\cosh\theta}{\sin\alpha}\,.
\label{t}
\end{equation}

As we have a periodic classical solution it is natural to define a corresponding action variable.
A leading-order semiclassical quantization can then be performed by restricting the action variable to integral values according to the Bohr-Sommerfeld condition.
Following \cite{Hofman:2006xt}, the action variable ${\mathcal I}$ is defined by the equation, 
\begin{equation} 
d{\mathcal I}=\frac{T_{0}}{2\pi}\, d(E-J_{1})\Big|_{P} \,,
\end{equation} 
where the subscript $P$ indicates that the differential is taken with
fixed $P$\,.
Using (\ref{param}), (\ref{res}) and (\ref{t}) we obtain simply $d{\mathcal I}=dJ_{2}$ which is consistent with the identification ${\mathcal I}=J_{2}$\,.
This is very natural as we expect the angular momentum $J_{2}$ to be integer valued in the quantum theory.
It is also consistent with the semiclassical quantization of finite-gap solutions discussed in \cite{Dorey:2006zj} where the action variables correspond to the filling fractions.
These quantities are simply the number of units of $J_{2}$ carried by each worldsheet excitation.

\subsubsection*{Finite-gap interpretation}

Let us see how the dyonic giant magnon can be described as a finite-gap solution (see Section \ref{sec:SBAE} for the general setup).
Such a classical string solution living in the Hofman-Maldacena sector turns out to be described simply by a condensate cut with constant density $\rho(x)=-i$ with endpoints $x=X^{\pm}$ (the spectral parameter $x$ used here is related to the one $x_{\rm old}$ used in Section \ref{sec:SBAE} as $x=x_{\rm old}/g$).
See Figure \ref{fig:ss-FG} for the diagram.
In this case, the equations (\ref{s-1}), (\ref{s-2}) and (\ref{s-3}) reduce to, respectively,
\begin{alignat}{3}
&-i\int_{X^{-}}^{X^{+}}dx&{}&=-i\ko{X^{+}-X^{-}}&{}&=\f{1}{g}\kko{J_{2}+\f{E-J}{2}}\,,\label{FG-DGM-1}\\[2mm]
&-i\int_{X^{-}}^{X^{+}}\f{dx}{x}&{}&=-i \ln\ko{\f{X^{+}}{X^{-}}}&{}&=P\,,\label{FG-DGM-2}\\[2mm]
&-i\int_{X^{-}}^{X^{+}}\f{dx}{x^{2}}&{}&=+i\ko{\f{1}{X^{+}}-\f{1}{X^{-}}}&{}&=\f{E-J}{2g}\,.\label{FG-DGM-3}
\end{alignat}
Note that $J=J_{1}+J_{2}$\,, and also that we have replaced $2\pi m$ in (\ref{s-2}) with generic $P$ since the closedness condition can be relaxed in the Hofman-Maldacena limit.
The classical string Bethe ansatz equation (\ref{SBA rho}) together with the relations (\ref{FG-DGM-1})\,-\,(\ref{FG-DGM-3}) implies the mode numbers $n_{k}$ in (\ref{SBA rho}) must be infinite.
From (\ref{FG-DGM-1})\,-\,(\ref{FG-DGM-3}), we can immediately obtain
\begin{align}
E-J_{1}&=\f{g}{i}\kko{\ko{X^{+}-\f{1}{X^{+}}}-\ko{X^{-}-\f{1}{X^{-}}}}\,,\label{E DGM}\\
J_{2}&=\f{g}{i}\kko{\ko{X^{+}+\f{1}{X^{+}}}-\ko{X^{-}+\f{1}{X^{-}}}}\label{Q(X) DGM}\,,\\
P&=\f{1}{i}\ln\ko{\f{X^{+}}{X^{-}}}\,.\label{P(X) DGM}
\end{align}
By eliminating the dependence on $X^{\pm}$ in (\ref{FG-DGM-1})\,-\,(\ref{FG-DGM-3}), we can reproduce the energy-spin relation (\ref{res}).

\section{Helical strings revisited\label{sec:helical2}}

\begin{figure}[thb]
\begin{center}
\vspace{.3cm}
\includegraphics[scale=0.9]{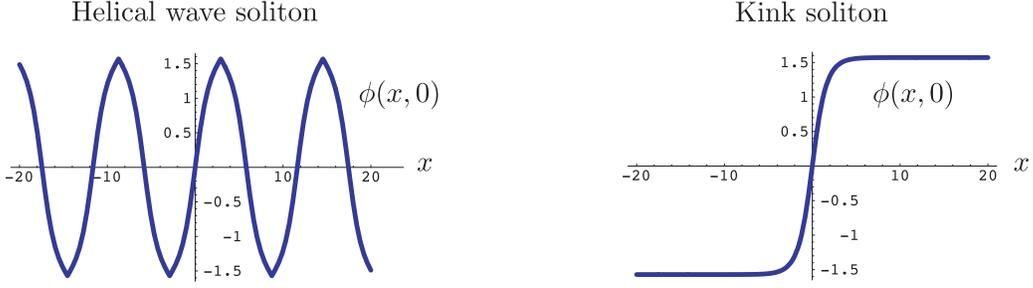}
\vspace{.3cm}
\caption{\small The helical wave soliton (\ref{helical wave : 1-spin}) (Left) and the kink soliton (\ref{sG kink}) (Right).
The latter is obtained from the former by taking the limit $k\to 1$\,.}
\label{fig:helical-trans}
\end{center}
\end{figure}

We saw in this chapter that the dyonic giant magnons can be mapped from kink soliton solutions of the CsG equation.
In fact, there is a more general, periodic soliton solution in CsG theory, which is known as a ``helical wave''.
It is a rigid array of kinks, see the left diagram of Figure \ref{fig:helical-trans}.
Starting with this soliton and re-exploiting the Pohlmeyer reduction procedure that relate the classical CsG system and the $O(4)$ string sigma model, we can obtain more general classical string solution.
This was achieved in \cite{Okamura:2006zv}, where a family of classical string solutions that interpolate dyonic giant magnons and Frolov-Tseytlin folded/circular strings \cite{Frolov:2003xy} was constructed.
They are called {\em helical strings}, and they turn out to be the most general elliptic classical string solutions on $\mathbb R\times S^{3}$\,.
For more details, see Chapter \ref{chap:OS}, where another family of helical strings in the large-winding sector is also investigated.

In this section we will explain how to obtain those helical strings by using simple single-spin cases as an example.
For more details, see \cite{Okamura:2006zv}.
An example of helical wave solution of sG equation is given by\footnote{The initial values for $(t,x)$ are set to be zero.}
\begin{equation}
\phi(t,x)=2\arcsin\kko{\cn\ko{\left. \f{x-vt}{k\sqrt{1-v^{2}}}\right| k}}
\label{helical wave : 1-spin}
\end{equation}
where $v$ is the soliton velocity, and cn is the Jacobian cn function.
The parameter $k$ determines the spatial period (or gwavelengthh) of $\phi$ field with respect to $x-vt$ as $4k\eK(k)\sqrt{1-v^{2}}$\,.
In the limit $k\to 1$\,, (\ref{helical wave : 1-spin}) reduces to an ordinary single-kink soliton with velocity $v$\,, (\ref{sG kink}).

Apart from (\ref{helical wave : 1-spin}), there are other types of helical waves exist in sG theory.
Substituting those helical waves into the string equation of motion (\ref{phi<->del X del X}) yields in general the following Schr\"{o}dinger equation, now with a potential in terms of the elliptic function,
\begin{equation}
\kko{- \partial_\tau^2 + \pa_{\sigma}^{2} - \mu^2 k^{2}\ko{2\sn^{2} \pare{\left.\frac{\mu (\sigma - v \tau)}{\sqrt{1 - v^2}}\right|k} - 1}} \vec \xi = \mu^2 U \, \vec \xi \,,  \label{lame_0}
\end{equation}
where $(k \mu \ssp \tau, k \mu \ssp \sigma) \equiv (t,x)$\,.
In particular, $U=0$ corresponds to the cn-type helical soliton (\ref{helical wave : 1-spin}).
We will keep $U$ so that we can keep various possibilities.
Introducing the boosted worldsheet coordinates (\ref{def:X,T}), we can rewrite the string equation of motion \eqref{lame_0} as 
\begin{equation}
\kko{-\pa_{T}^{2}+\pa_{X}^{2} - k^{2}\ko{2\sn^{2}(X)-1}} \vec\xi = \ssp U \,\vec\xi\,.
\label{reduced_eom}
\end{equation}
The consistency condition \eqref{def_cos_phi} is indeed satisfied as
\begin{equation}
\frac{1}{\mu^2} \kko{ \left| \partial_\sigma \vec\xi \,\right|^2
- \left| \partial_\tau \vec\xi \,\right|^2 } = k^2 
- 2 k^2 \sn^2 (X ) - U\,,
\end{equation}
from which we can deduce the equation of motion \eqref{reduced_eom}.

We can solve this equation under an ansatz
\begin{equation}
\xi_{j} (T,X;w_{j})=  {\mathcal Y}_{j}(X;w_{j}) \, e^{i u_{j}(w_{j}) T}\qquad (j=1,2)\,.
\label{ansatz xi}
\end{equation}
Here $w_{j}$ are complex parameters and ${\mathcal Y}_{j}$ are independent of $T$\,. 
As for constraints on $w$\,, see appendix \ref{app:Details of Calculations}.
The differential equation satisfied by ${\mathcal Y}_{j}$ then takes the form
\begin{equation}
\cpare{ \,\frac{d^2}{d X^2} - k^{2}\Big( 2 \sn^{2} \pare{X|k} - 1 \Big) +u_{j}^{2}}{\mathcal Y}_{j} = U \, {\mathcal Y}_{j} \,,  \label{Lame}
\end{equation}
which is known as the {Lam\'{e} equation}.
For general eigenfunctions of Lam\'{e} equations, see Chapter 23.7 of \cite{WhWa27} for details.
They are given by
\begin{equation}
{\mathcal Y}(X; w)\,  \propto\,\f{\bTh_1 (X-w|k)}{\bTh_{0}(X|k)}\,\exp\ko{\eZ_{0}(w|k) X}
\quad 
\mbox{with}\quad 
u^2 = \dn^2 (w|k) + U \,,
\label{Lame_eig_fn}
\end{equation}
where $\Theta_\nu$\,, $Z_\nu$ are the Jacobian theta and zeta functions defined in Appendix \ref{app:Elliptic Functions}, respectively.

We would like to find solutions that satisfy the string equation of motion (\ref{lame_0}), the consistency condition for Pohlmeyer reduction (\ref{def_cos_phi}) and the Virasoro conditions \eqref{string_Virasoro}.
Actually it turns out that, corresponding to several possibilities of choosing a helical soliton solution of (C)sG equation, there can be as many consistent string solutions.

\paragraph{}
The periodic sG soliton (\ref{helical wave : 1-spin}) is mapped to a simple single-spin helical string that interpolate GKP folded/circular strings and a Hofmann-Maldacena giant magnon.
For example, the profile of the single-spin, type $(i)$ helical string is given by
\begin{align}
\eta_{0} (T,X)&=a T+b X\qquad 
\mbox{with}\quad a =k\cn(\iom | k)\,,\quad b =-ik\sn(\iom | k)\,,
\label{eta0-f}\\[2mm]
\xi_{1} (T,X)&=\f{\sqrt{k}}{\dn(\iom | k)}\f{\bTh_{0}(0 | k)}{\bTh_{0}(\iom | k)}\f{\bTh_{1}(X-\iom | k)}{\bTh_{0}(X | k)}\, 
\exp\kko{\eZ_{0}(\iom | k)X+i\dn(\iom | k)T}\,,
\label{xi1-f}\\[2mm]
\xi_{2} (T,X)&=\f{\dn (X | k)}{\dn(\iom | k)}\,,
\label{xi2-f}
\end{align}
with $\omega$ a real parameter.
The soliton velocity $v$\,, which appeared in the definitions of $T$ and $X$ \eqref{def:X,T}, is related to the parameters $a$ and $b$ in \eqref{eta0-f} as $v\eq b/a$\,.
Using various properties and identities listed in Appendices \ref{app:Elliptic Functions} and \ref{app:Details of Calculations}, one can check the proposed set of solutions (\ref{eta0-f})\,-\,(\ref{xi2-f}) indeed satisfy the required physical constraints.
Note that the AdS-time variable $\eta_{0}$ can be rewritten as $\eta_{0} = k \tilde \tau$\,.
This solution must satisfy the periodic boundary conditions for the spacetime coordinate for them to represent a closed string, which constrain the parameters $v$\,, $k$ and $\om$ appearing in the profile.
For more details, see Chapter \ref{chap:OS} and the original paper \cite{Okamura:2006zv}, where more general two-spin cases are discussed.

\subsubsection*{Boundstates of (dyonic) giant magnons?}

In \cite{Hofman:2006xt}, a sort of non-rigid string with time-dependent profile was constructed.
It was obtained as a ``boundstate'' of two giant magnons with complex conjugate momenta.
In \cite{Spradlin:2006wk}, the same solution was re-examined and also generalised to two-spin cases.
It was realised in \cite{Dorey:2007xn} that the ``boundstate'' of two (dyonic) giant magnons found in \cite{Hofman:2006xt, Spradlin:2006wk} was not really a boundstate literally, but rather turned out to be a superposition (scattering state) of two BPS boundstates.
We will discuss it in Chapter \ref{chap:Singularities}.\\

Finally, for further literature on giant magnons and the physics in the Hofman-Maldacena sector, see \cite{Spradlin:2006wk,Bobev:2006fg,Kruczenski:2006pk,Kalousios:2006xy,Hirano:2006ti,Roiban:2006gs,Chu:2006ae,Okamura:2006zv,Ryang:2006yq} (See also \cite{Ryang:2005yd,Berenstein:2005jq,Vazquez:2006hd,Hatsuda:2006ty,Berenstein:2007zf}).

\section[Appendix for Chapter \ref{chap:DGM}]{Appendix\,: Solving the differential equation (\ref{xt2})\label{app:derivation of DGM}}

As usual the general solution of this equation can be written as a linear
combination of ``stationary states'' of the form, 
\begin{equation} 
\xi (X,T)=F(X_{\al})\exp(i\om T)\,,\qquad X_{\al}\eq (\cos\alpha)X\,.
\end{equation} 
We find that the function $F(X_{\al})$ obeys the equation, 
\begin{equation}
-\frac{d^{2}F}{dX_{\al}^{2}}-\frac{2}{\cosh^{2}X_{\al}}\, F=\varepsilon^{2}F\,.
\label{RM}
\end{equation}
\paragraph{}
Equation (\ref{RM}) coincides with the time-independent
Schr\"{o}dinger equation for a particle in (a special case of) 
the Rosen-Morse potential 
\cite{PhysRev.42.210}, 
\begin{equation}
V(x)= \frac{-2}{\cosh^{2}x}\,.
\label{RM2}
\end{equation}
The exact spectrum of this problem is known (see {\em e.g.}, \cite{LL:1965:QM}). There is a single normalisable boundstate with 
energy $\varepsilon^{2}=-1$ and wavefunction, 
\begin{equation}
F_{-1}(X_{\al})=\frac{1}{\cosh X_{\al}}
\label{bs}  
\end{equation}
and a continuum of scattering states with $\varepsilon^{2}=k^{2}$ for 
$k>0$ and wavefunctions, 
\begin{equation}
F_{k^{2}}(X_{\al})=\left(\tanh X_{\al}-ik\right)\exp(ik X_{\al})
\label{scat}
\end{equation}
with asymptotics $F_{k^{2}}(X_{\al})\sim \exp\left(ik X_{\al}\pm i\delta/2\right)$\,, where the scattering phase-shift is given as
$\delta=2\tan^{-1}(1/k)$\,.
The general solution to the original linear equation (\ref{xt1}) can be constructed as a linear combination of these boundstate and scattering wavefunctions.
The particular solutions corresponding to the worldsheet fields $\xi_{1}$ and $\xi_{2}$ are singled out by the boundary conditions (\ref{b.c.2}).
In particular, the boundary condition (\ref{b.c.2}) can only be matched by a solution corresponding to a single scattering mode $F_{k^{2}}(X_{\al})$\,;
\begin{equation} 
\xi_{1}=c_{1}F_{k^{2}}(X_{\al})\,\exp(i \omega_{k^{2}} T)\,,\quad 
\mbox{where}\quad \omega_{k^{2}}=\sqrt{k^{2}\cos^{2}\alpha+1}\,.
\end{equation}
We find that (\ref{b.c.2}) is obeyed provided we set, 
\begin{equation}
k=\frac{\sinh\theta}{\cos\alpha} 
\label{k}
\end{equation}
which yields the magnon momentum $P=\delta=2\tan^{-1}(1/k)$\,.
The boundary condition (\ref{b.c.2}) dictates that $\xi_{2}$ decays at left and right infinity.
This is only possible if we identify it with the solution corresponding to the unique normalisable boundstate of the potential (\ref{RM}), 
\begin{equation}  
\xi_{2}=c_{2}F_{-1}(X_{\al})\exp(i\omega_{-1}T)\quad 
\mbox{with}\quad \omega_{-1}=\sin\alpha\,.
\end{equation} 
Without loss of generality we can choose the
constants $c_{1}$ and $c_{2}$ to be real. The 
condition $|\xi_{1}|^{2}+|\xi_{2}|^{2}=1$ then yields, 
\begin{equation}  
c_{1}=c_{2}=\frac{1}{\sqrt{1+k^{2}}}=\sin\ko{\f{P}{2}}\,,
\end{equation}
which reproduces the profile (\ref{soln}) displayed in the main text.

\chapter[ Scattering of AdS/CFT Solitons]
	{Scattering of AdS/CFT Solitons\label{chap:S-matrices in HM}}

\section{Infinite Spin/R-charge Limit of the AdS/CFT S-matrices}

In the Hofman-Maldacena limit, the spin-chain/string becomes infinitely long and the spectrum consists of local excitations which propagate freely apart from pairwise scattering.
The physical content of the limiting theory is the spectrum of asymptotic states and their S-matrix and the main problem is to compare the spectrum and S-matrix which appear on both sides of the correspondence.
We compared the spectrum in the previous chapters, where we found exact matching between both sides, (\ref{disp3}) and (\ref{res}).
In this chapter, we will study the scattering of the BPS states (magnon boundstates or dyonic giant magnons).

\paragraph{}
As we saw in Chapter \ref{chap:dressing}, the exact S-matrix for the magnons themselves is known up to a single overall phase.
In the $SU(2)$ sector, the remaining ambiguity corresponds to the dressing factor first introduced in \cite{Arutyunov:2004vx, Beisert:2005wv}. 
As we already saw in (\ref{theta kj}), the dressing factor takes a very specific form as a function of the conserved charges of the theory but still involves an infinite number of undetermined coefficients.
In an integrable theory, the scattering of boundstates is uniquely determined by the scattering of their constituents \cite{Zamolodchikov:1978xm,Karowski:1978ps,Faddeev:1996iy}. 
In this section, we will take the exact magnon S-matrix, including the dressing factor, as a starting point and derive the corresponding S-matrix for the scattering of magnon boundstates in the $SU(2)$ sector.
The resulting S-matrix has an interesting analytic structure with simple poles corresponding to boundstate contributions in the s- and t- channels as well as double-poles corresponding to anomalous thresholds.
The boundstate S-matrix also includes a dressing factor which is functionally identical to the one appearing in the fundamental magnon S-matrix.
We note that this universality of the dressing factor is essentially equivalent to its conjectured form as a function of the conserved charges mentioned above.

\begin{figure}[t]
\begin{center}
\vspace{.3cm}
\includegraphics[scale=0.9]{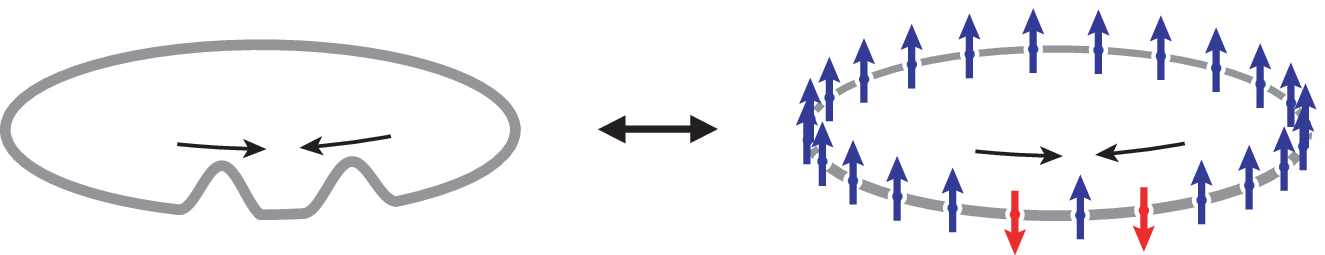}
\vspace{.3cm}
\caption{\small The AdS/CFT S-matrix\,: scattering in string theory (left) and in gauge theory (right).}
\label{fig:spin-chain}
\end{center}
\end{figure}

\paragraph{} 
On the string theory side, the fundamental magnons and their boundstates correspond to solitons of the worldsheet theory which can be studied using semiclassical methods for $g\gg 1$ \cite{Hofman:2006xt}.
In particular, $SU(2)$ sector boundstates with values of $Q$ which scale linearly with $g$\,, are identified with dyonic giant magnons studied in the previous chapter.
We found that in string theory, $Q$ corresponds to the second conserved angular momentum $J_{2}$ on $S^{5}$ and the exact dispersion relation (\ref{kdisprel}) is already obeyed at the classical level, see (\ref{res}).

In the following we will again utilise the CsG description to obtain a semiclassical approximation to the S-matrix for the dyonic giant magnons.
Multi-soliton solutions of the CsG equation are also available in the literature \cite{Lund:1976ze, Lund:1977dt, Getmanov:1977nf, Getmanov:1980cq, deVega:1981ka, deVega:1982sh, Dorey:1994mg, Bowcock:2002vz}.
In the classical theory these objects undergo factorised scattering with a known {time-delay}.
This is precisely the information required to calculate the semiclassical approximation to the S-matrix.

Our main result is that the so-obtained string theory S-matrix precisely matches the large-$g$ limit of the magnon boundstate (gauge theory) S-matrix described above.
A similar comparison in the $Q=1$ case of fundamental magnons was performed in \cite{Hofman:2006xt}.
A new feature of the present case is that both the dressing factor and the remaining factor which originates from the all-loop gauge theory Bethe ansatz of \cite{Beisert:2004hm} contribute at leading order in the $g\to \infty$ limit and therefore both parts are tested by the comparison.

\section{Scattering of magnon boundstates}

A generic asymptotic state in the $SU(2)$ sector has two independent quantum numbers $P$ and $Q$\,.
It will be convenient to use an alternative parametrisation in terms of two complex variables $X^{\pm}$ introduced in Section \ref{sec:asymptotic spectrum}.
For convenience let us here brush up our notations about the magnon boundstate parameters.

In term of the spectral parameters $x^{\pm}$\,, the BDS piece of the S-matrix takes the form, (re-displaying (\ref{S in x}) here for convenience,)
\begin{equation}
S_{\rm BDS}(x_{j}^{\pm}, x_{k}^{\pm})=\f{x_{j}^{+}-x_{k}^{-}}{x_{j}^{-}-x_{k}^{+}}\cdot
\f{1-1/(x_{j}^{+}x_{k}^{-})}{1-1/(x_{j}^{-}x_{k}^{+})}\,.\no
\end{equation}
We note the presence of a simple pole at $x_{j}^{-}=x_{k}^{+}$\,. 
As explained in \cite{Dorey:2006dq}, this pole indicates the formation of a normalisable BPS boundstate of two magnons. In fact the theory also contains a $Q$-magnon boundstate for each value of $Q>1$\,, related to a corresponding pole of the multi-particle S-matrix which can be expressed as the product of two body factors by virtue of integrability.
The spectral parameters of the constituent magnons in a $Q$-magnon boundstate are,    
\begin{equation}
x_{j}^{-} = x_{j+1}^{+}
\qquad \mbox{for}\quad 
j=\komoji{1,\dots, Q-1}\,.
\end{equation}
The resulting boundstate of rapidity $U$ is described by introducing spectral parameters 
\begin{equation}
X^{\pm}(U;Q)=x\ko{U\pm\mbox{\large $\f{iQ}{2g}$}}\,,\qquad 
\mbox{\em i.e.},\quad 
X^{+}\equiv x_{1}^{+}\,,\quad 
X^{-}\equiv x_{Q}^{-}\,,
\end{equation}
The total momentum $P$ and $U(1)$ charge $Q$ of the boundstate are then expressed as
\begin{align}
P(X^{\pm})&=\f{1}{i}\ln\ko{\f{X^{+}}{X^{-}}}\,,\label{P(X)}\\
Q(X^{\pm})&=\f{g}{i}\kko{\ko{X^{+}+\f{1}{X^{+}}}-\ko{X^{-}+\f{1}{X^{-}}}}\label{Q(X)}\,.
\end{align}
One can also show the rapidity $U$ and energy $E=\sum_{k=1}^{Q}\ep_{k}$ for the boundstate are related to the spectral parameters $X^{\pm}$ through the expressions
\begin{alignat}{3}
\hspace{-0.2cm}U(X^{\pm})	&=\f{1}{2}\kko{\ko{X^{+}+\f{1}{X^{+}}}+\ko{X^{-}+\f{1}{X^{-}}}}
=\f{1}{2g}\cot\ko{\f{P}{2}}\sqrt{Q^{2}+16g^{2}\sin^{2}\ko{\f{P}{2}}}\,,\label{U(X)}\\[2mm]
\hspace{-0.2cm}E(X^{\pm})	&=\f{g}{i}\kko{\ko{X^{+}-\f{1}{X^{+}}}-\ko{X^{-}-\f{1}{X^{-}}}}
=\sqrt{Q^{2}+16g^{2}\sin^{2}\ko{\f{P}{2}}}\,.
	\label{E}
\end{alignat}
If $Q$ and $E$ are regarded as free complex parameters then $X^{+}$ and $X^{-}$ are unconstrained complex variables. 
The condition of fixed integer charge $Q$ provides a cubic constraint on $X^{\pm}$ which defines a complex torus \cite{Janik:2006dc}.
The case $Q=1$ corresponds to the fundamental magnon and the variables $X^{\pm}$ coincide with the usual spectral parameters $x^{\pm}$\,.
As before, we will reserve the use of lower-case variables $x^{\pm}$ for this special case.
For any positive integer $Q$\,, physical states with real momentum and positive energy are obtained by imposing the conditions $X^{-}=(X^{+})^{*}$ and $|X^{\pm}|>1$\,.

Notice that the expressions (\ref{P(X)}), (\ref{Q(X)}) and (\ref{E}) precisely agree with (\ref{P(X) DGM}), (\ref{Q(X) DGM}) and (\ref{E DGM}), respectively, when we naturally identify $(P,Q,E)$ in the former expressions with $(P,J_{2},E-J_{1})$ appearing in the latter.\footnote{The symbol $E$ used in this section to represent the spin-chain excitation energy (=$\Delta-J_{1}$) should not be confused with $E$ used in Section \ref{sec:DGM} to represent the string energy.}

\subsection{The boundstate S-matrix}

As a consequence of integrability, the asymptotic states described above undergo factorised scattering. In other words the S-matrix for the scattering of an arbitrary number of excitations can be written consistently as a product of two-body factors.
In the case of fundamental magnons in the $SU(2)$ sector, with spectral parameters $x_{1}^{\pm}$ and $x_{2}^{\pm}$ respectively, the exact two body S-matrix can be written as,   
\begin{align}
s(x_{1}^{\pm},x_{2}^{\pm})&\equiv \hat{s}(x_{1}^{\pm},x_{2}^{\pm})\times\sigma^{2}(x_{1}^{\pm},x_{2}^{\pm})\no\\
&\equiv \hat{s}(x_{1}^{\pm},x_{2}^{\pm})\times\exp\kko{2\th(x_{1}^{\pm},x_{2}^{\pm})}\,.
\label{elementarySmatrix}
\end{align}
The first factor,  
\begin{eqnarray}
\hat{s}(x_{1}^{\pm},x_{2}^{\pm})&=&\frac{x_{1}^{+}-x_{2}^{-}}{x_{1}^{-}-x_{2}^{+}}\frac{1-1/x_{1}^{+}x_{2}^{-}}{1-1/x_{1}^{-}x_{2}^{+}}=
\frac{u(x_{1}^{\pm})-u(x_{2}^{\pm})+i/g}{u(x_{1}^{\pm})-u(x_{2}^{\pm})-i/g}
\label{hatsBDS}
\end{eqnarray}
originates in the all-loop asymptotic Bethe ansatz of BDS \cite{Beisert:2004hm}.
This factor has a pole in the physical region of the spectral plane at the point $x_{1}^{-}=x_{2}^{+}$ which corresponds to the formation of the $Q=2$ BPS boundstate in the s-channel \cite{Dorey:2006dq}.
The second term $\exp\kko{2\th(x_{1}^{\pm},x_{2}^{\pm})}$ is the dressing phase factor we discussed in Section \ref{sec:dressing phase}.

\begin{figure}[t]
\begin{center}
\vspace{.5cm}
\hspace{-.0cm}\includegraphics[scale=1.0]{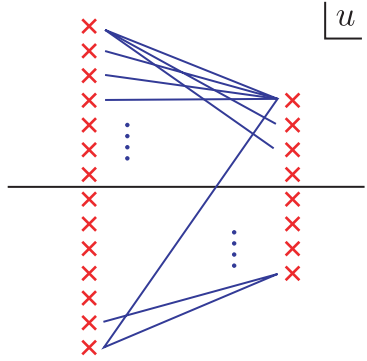}
\vspace{.0cm}
\caption{\small Constructing boundstate S-matrix by fusion.
Each boundstate is represented by an equally spaced sequence of Bethe roots (Bethe string).}
\label{fig:fusion}
\end{center}
\end{figure}

\paragraph{}
Let us now consider two magnon boundstates with charges $Q_{1}$ and $Q_{2}$ and momenta $P_{1}$ and $P_{2}$ respectively.
We assume $Q_{1}\geq Q_{2}$\,.
Equivalently we can describe these states with spectral parameters $X_{1}^{\pm}$ and $X_{2}^{\pm}$ with, 
\begin{equation}
\exp(iP_{1})=\frac{X_{1}^{+}}{X_{1}^{-}}\,, \qquad \exp(iP_{2})=\frac{X_{2}^{+}}{X_{2}^{-}}\,,
\end{equation}
where $X_{1}^{\pm}$ satisfies (\ref{Q(X)}) with $Q=Q_{1}$ and a similar equation holds for $X_{2}^{\pm}$ with $Q=Q_{2}$\,. 
Our goal is to find the S-matrix $S(X_{1}^{\pm},X_{2}^{\pm})$ describing the scattering of these two boundstates states. In an integrable quantum theory, the S-matrix for the scattering of boundstates is uniquely determined by the S-matrix of their constituents.
Thus, in the present case, we begin by considering the scattering of $Q_{1}+Q_{2}$ fundamental magnons with individual spectral parameters,   
\begin{equation}
\{x_{j_{1}}^{\pm}\}\,,\quad \{y_{j_{2}}^{\pm}\}\quad \mbox{with}\quad j_{1}=1,\dots,Q_{1}\,,\quad j_{2}=1,\dots,Q_{2}\,.
\label{xj1xj2}
\end{equation}
As above the spectral parameters for fundamental magnons satisfy the constraints, 
\begin{alignat}{3}
\frac{i}{g}&=\bigg(x_{j_{1}}^{+}+\frac{1}{x_{j_{1}}^{+}}\bigg)&-&\bigg(x_{j_{1}}^{-}+\frac{1}{x_{j_{1}}^{-}}\bigg)
\,,&\qquad &j_{1}=1,\dots,Q_{1}\label{xj1id}\,;\\[2mm]
\frac{i}{g}&=\bigg(y_{j_{2}}^{+}+\frac{1}{y_{j_{2}}^{+}}\bigg)&-&\bigg(y_{j_{2}}^{-}+\frac{1}{y_{j_{2}}^{-}}\bigg)
\,,&\qquad &j_{2}=1,\dots,Q_{2}\label{yj2id}\,.
\end{alignat}
By factorisability, the S-matrix for the scattering of the constituent magnons is simply a product of two-body factors.
The formation of two boundstates of charges $Q_{1}$ and $Q_{2}$ corresponds to the pole in this multi-particle S-matrix appearing at,  
\begin{alignat}{3}
\xm_{j_{1}}&=\xp_{j_{1}+1}\,,&\qquad &j_{1}=1,\dots\, , Q_{1}-1\,;\label{polexj1}\\
\ym_{j_{2}}&=\yp_{j_{2}+1}\,,&\qquad &j_{2}=1,\dots\, , Q_{2}-1\,.\label{poleyj2}
\end{alignat}
The resulting boundstate spectral parameters $X_{1}^{\pm}$ and $X_{2}^{\pm}$ can then be
identified as:
\begin{equation}
X_{1}^{+}=\xp_{1}\,,\qquad 
X_{1}^{-}=\xm_{Q_{1}}\,;\qquad 
X_{2}^{+}=\yp_{1}\,,\qquad 
X_{2}^{-}=\ym_{Q_{2}}\,.
\end{equation}
where it is easy to check that the appropriate constraint equation for $X_{1}^{\pm}$ ({\em i.e.}, Eqn (\ref{Q(X)}) with $Q=Q_{1}$) is obeyed by virtue of (\ref{xj1id}) and (\ref{polexj1}) and similarly for $X_{2}^{\pm}$\,.
Consistency of scattering in such that an integrable theory provides a simple recipe for extracting the boundstate S-matrix: it is simply the
residue of the multi-particle scattering matrix of the constituent magnons at the pole specified above.
This prescription is most familar in the context of relativistic field theories in $(1+1)$-dimensions \cite{Zamolodchikov:1978xm,Karowski:1978ps}, but has also been applied successfully in the context of integrable spin-chains \cite{Faddeev:1996iy}.
Starting with the S-matrix (\ref{elementarySmatrix}) for elementary magnons, it is straightforward to obtain the corresponding S-matrix for magnon boundstates of arbitrary charges by fusion.
Because of factorization the multi-particle S-matrix, the boundstate S-matrix is nothing other than the product of two-body S-matrices describing all possible pair-wise scatterings between the consitituent magnons.  
The proceedure is illustrated schematically in Figure \ref{fig:fusion}.
Thus in terms of the single magnon S-matrix $s(x_{1}^{\pm},x_{2}^{\pm})$ given in (\ref{elementarySmatrix}), the boundstate S-matrix is,    
\begin{equation}
S\left(Q_{1},Q_{2};P_{1},P_{2}\right)
=S(X_{1}^{\pm},X_{2}^{\pm})
\eq \prod_{j_{1}=1}^{Q_{1}}\prod_{j_{2}=1}^{Q_{2}}s(x^{\pm}_{j_{1}},y^{\pm}_{j_{2}})
\label{BSSmatrix}\,.
\end{equation}
It will be convenient to write $S$ as the product of two factors,  
\begin{align}
S(X_{1}^{\pm},X_{2}^{\pm})&\eq \hat{S}(X_{1}^{\pm},X_{2}^{\pm})\times \Sigma(X_{1}^{\pm},X_{2}^{\pm})\no\\
&\equiv \hat{S}(X_{1}^{\pm},X_{2}^{\pm})\times\exp\kko{2\Th(X_{1}^{\pm},X_{2}^{\pm})}
\label{BSSmatrix2parts}\,.
\end{align}
Here $\hat{S}$ is the contribution coming from the BDS factor $\hat{s}$ in the single magnon S-matrix, which is defined in (\ref{hatsBDS}).
The remaining piece $\Sigma$ originates from the dressing factor $\sigma$ in the single magnon  S-matrix, as defined in (\ref{theta kj}).
We will consider these two factors in turn. 

\paragraph{The BDS part.}

The BDS piece of the boundstate S-matrix is straightforwardly obtained by direct evaluation of the product (\ref{BSSmatrix}).
The corresponding calculation for the XXX${}_{1/2}$ Heisenberg spin-chain is reviewed in \cite{Faddeev:1996iy}.
The pole conditions (\ref{polexj1}), (\ref{poleyj2}) lead to numerous cancellations between the $Q_{1}Q_{2}$ factors in the product.
The remaining factors can be conveniently presented as,  
\begin{equation}
 \hat{S}\left(Q_{1},Q_{2},p_{1},p_{2}\right)
=G\left(Q_{1}-Q_{2}\right)\left[\prod_{l=1}^{Q_{2}-1}
G\left(Q_{1}-Q_{2}+2l\right)\right]^{2}G\left(Q_{1}+Q_{2}\right)
\label{BDSBSSmatrix}\,,
\end{equation}
where 
\begin{equation}
G(q)=\frac{U(P_{1},Q_{1})-U(P_{2},Q_{2})+i q/2g}{U(P_{1},Q_{1})-U(P_{2},Q_{2})-i q/2g}\,.
\end{equation}
The singularities of the final answer (\ref{BDSBSSmatrix}) have a natural interpretation in terms of on-shell intermediate states.
First, the simple pole of the factor $G(Q_{1}+Q_{2})$ corresponds to the formation of a boundstate with $Q=Q_{1}+Q_{2}$ in the s-channel.
This is a direct generalisation of the $Q=2$ pole in the S-matrix of two elementary magnons mentioned above.
Similarly, the other simple pole in $\hat{S}$\,, which comes from the factor $G(Q_{1}-Q_{2})$\,, precisely corresponds to the exchange of a boundstate with $Q=Q_{1}-Q_{2}>0$ in the t-channel.
The set of $Q_{2}-1$ double poles in the boundstate S-matrix also have a standard explanation in $(1+1)$-dimensional scattering theory
\cite{Coleman:1978kk}: they correspond to anomalous thresholds.
Specifically, the positions of the double poles are consistent with the kinematics of an intermediate state consisting of two on-shell boundstates with 
$Q=Q_{1}+l$ and $Q=Q_{2}-l$ respectively for $l=1,2,\ldots ,Q_{2}-1$\,.
We will investigate those simple and double poles (and also the poles in the dressing part as well) in greater detail later in Chapter \ref{chap:Singularities}, where we account for them by physical processes involving on-shell intermediate particles belonging to the BPS spectrum (\ref{E}).

\paragraph{The dressing part.}

The second contribution to the boundstate scattering matrix, denoted $\Sigma$ in (\ref{BSSmatrix2parts}) comes from the dressing factors of the elementary magnon S-matrices appearing in the product (\ref{BSSmatrix}).
Here we find an even more complete cancellation of factors appearing in the product.
In fact the final answer is simply that $\Sigma$ is identical as function of the higher conserved charges to the fundamental magnon dressing
factor $\sigma$\,.
Thus we have,
\begin{equation}
\Th(X_{1}^{\pm},X_{2}^{\pm})=\th(X_{1}^{\pm},X_{2}^{\pm})\,,
\label{exactThetaAFS}
\end{equation}
where $\th$ is the same function as defined in (\ref{theta kj}).
Thus the factor $\Sigma$ appearing in the boundstate S-matrix is equal to a universal function of the higher conserved charges.
Equivalently we have 
\begin{equation}
\Theta(X_{1}^{\pm},X_{2}^{\pm})=g\left[k(X_{1}^{+},X_{2}^{+})+k(X_{1}^{-},X_{2}^{-})
-k(X_{1}^{+},X_{2}^{-})-k(X_{1}^{-},X_{2}^{+})\right]
\label{factorizedAFS2}\,,
\end{equation} 
where $k(X_{1}^{\pm},X_{2}^{\pm})$ is the same function appearing in (\ref{factorizedAFS}).
As in the case of the single magnon S-matrix our knowledge of this function (or, equivalently of the coefficients $c_{r,s}(g)$) is limited to the first two orders in the strong coupling expansion.
As mentioned before, the general form (\ref{theta kj}) for the dressing factor originally arose as the most general integrable long-range deformation of the Heisenberg spin-chain \cite{Beisert:2005wv}.
In the present context it is interesting to note that it is essentially equivalent to the condition that the dressing factor should be the same universal function of the conserved charges for all BPS states in the theory.
Indeed one could start by imposing this universality as a requirement and, after also taking account of unitarity and parity invariance, one would immediately be lead to the general form (\ref{theta kj}).

\subsection{Strong coupling limit}

So far we have been considering the exact analytic expressions for the boundstate S-matrix.
To compare our results with those of semiclassical string theory we need to take the strong coupling limit $g\to \infty$\,.
As we discussed in \cite{Chen:2006ge}, the natural limit to take is one where the charges $Q_{1}$ and $Q_{2}$ also scale linearly with $g$\,.
As a consequence, both terms under the square root in the dispersion relation (\ref{E}) scale like $g^{2}$ and thus the energy $E$
has the appropriate coupling dependence for a semiclassical string state.
Conveniently, the spectral parameters $X_{1}^{\pm}$ and $X_{2}^{\pm}$ for boundstates with $Q=Q_{1}$ and $Q=Q_{2}$ respectively remain fixed in this limit.
Our next goal is to calculate the leading asymptotics of the boundstate S-matrix as a function of the spectral parameters.
As above we consider the two factors $\hat{S}$ and $\Sigma$ appearing in (\ref{BSSmatrix2parts}) in turn.

\paragraph{The BDS part in \bmt{g\to \infty}.} 
To take the strong coupling limit of $\hat{S}$\,, we begin by
exponentiating the product appearing in (\ref{BDSBSSmatrix}) to obtain
a sum in the exponent. As $g\rightarrow \infty$ this sum goes over to
an integral, with the integration limits depending only on the sum and
the difference between the charges $Q_{1}$ and $Q_{2}$\,. 
Interestingly, the leading contribution to $\hat{S}$ has the same
general form (\ref{factorizedAFS2}) as that of the dressing factor. 
In particular, the final result can be given as,  
\begin{equation}
\hat{S}(X_{1}^{\pm}, X_{2}^{\pm}) \simeq \exp[i\hat{\Theta}(X_{1}^{\pm}, X_{2}^{\pm})]
\label{SmatrixBSb}\,,
\end{equation}
where 
\begin{equation}
\hat{\Theta}(X_{1}^{\pm},X_{2}^{\pm})
=2g\left[\hat{k}(X_{1}^{+},X_{2}^{+})+\hat{k}(X_{1}^{-},X_{2}^{-})
-\hat{k}(X_{1}^{+},X_{2}^{-})-\hat{k}(X_{1}^{-},X_{2}^{+})\right]\,.
\label{Deltatotalb}
\end{equation}
Here the function $\hat{k}$ is given by, 
\begin{equation}
\hat{k}(X,Y)=
\left[\left(X+\frac{1}{X}\right)-\left(Y+\frac{1}{Y}\right)\right]\ln\left[(X-Y)\left(1-\frac{1}{XY}\right)\right]
\label{BDS2}\,.
\end{equation}

\paragraph{The dressing part in \bmt{g\to \infty}.} 
The strong-coupling limit of the dressing factor $\Sigma$ is simply given by replacing the function $k(X,Y)$ appearing in (\ref{factorizedAFS2}) by the function $k_{0}(X,Y)$ given in (\ref{elemetarychi0}):
\begin{align}
\Theta(X_{1}^{\pm},X_{2}^{\pm})\simeq 2g\left[k_{0}(X_{1}^{+},X_{2}^{+})+k_{0}(X_{1}^{-},X_{2}^{-})
-k_{0}(X_{1}^{+},X_{2}^{-})-k_{0}(X_{1}^{-},X_{2}^{+})\right]\,.
\end{align}

\paragraph{The total phase-shift.}
Collecting the results for the two factors we
find the final result for the strong coupling limit of the boundstate
S-matrix can be given as, 
\begin{equation}
S(X_{1}^{\pm}, X_{2}^{\pm}) \simeq \exp[i\Theta_{\rm gauge}(X_{1}^{\pm}, X_{2}^{\pm})]
\label{SmatrixBS}\,,
\end{equation}
where 
\begin{equation}
\Theta_{\rm gauge}(X_{1}^{\pm},X_{2}^{\pm})
=2g\left[K_{0}(X_{1}^{+},X_{2}^{+})+K_{0}(X_{1}^{-},X_{2}^{-})
-K_{0}(X_{1}^{+},X_{2}^{-})-K_{0}(X_{1}^{-},X_{2}^{+})\right]\,.
\label{Deltatotal}
\end{equation}
Here the function $K_{0}(X,Y)$ is given by, 
\begin{equation}
K_{0}\left(X,Y\right)=\hat{k}\left(X,Y\right)+k_{0}\left(X,Y\right)
=\left[\left(X+\frac{1}{X}\right)-\left(Y+\frac{1}{Y}\right)\right]\ln\left(X-Y\right)\,.\label{Kfunction}
\end{equation}
Note that $K\left(X,Y\right)$ is functionally different from $k_{0}(x,y)$ in (\ref{elemetarychi0}).\footnote{Note also, at the next order in the large-$g$ expansion of the phase, we have $K_{1}(X,Y)=k_{1}(X,Y)$ where $\Theta_{\rm gauge}(X,Y)=gK_{0}\left(X,Y\right)+K_{1}\left(X,Y\right)+\cO(g^{-1})$ and $k_{1}(x,y)$ is as given in (\ref{HL-strong}).}

\section{Scattering of dyonic giant magnons}

\begin{figure}[htb]
\begin{center}
\vspace{.3cm}
\hspace{-.0cm}\includegraphics[scale=0.9]{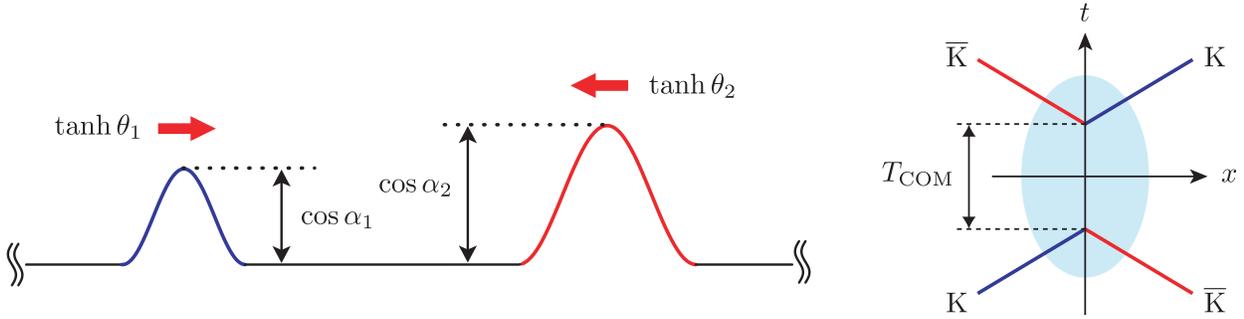}
\vspace{.3cm}
\caption{\small Scattering of CsG kink ($\rm K$) and antikink ($\rm \overline{K}$) solitons.
The right diagram shows the $(1+1)$-dimensional spacetime picture of the scattering (with interaction region shaded).}
\label{fig:2kinks CsG}
\end{center}
\end{figure}

In \cite{Chen:2006ge} we showed that the magnon boundstates in the $SU(2)$ sector described above appear in string theory on $AdS_{5}\times S^{5}$ as classical solitons of the worldsheet action, namely dyonic giant magnons.
The corresponding equations of motion together with the Virasoro constraint could be mapped onto the CsG equation. 
Under this equivalence, the classical string solution corresponding to a magnon boundstate of charge $Q$ and momentum $P$ is mapped to a certain one-soliton solution of the CsG equation.
The soliton in question has two parameters: a rapidity\footnote{Not to be confused with the magnon rapidity $U(X_{1}^{\pm})$ introduced above.} $\theta$ and an additional rotation parameter $\alpha$\,.
The dictionary between these parameters and the conserved quantities $E\, (=\Delta-J_{1})$\,, $Q$ and $P$ is,
\begin{align}
E&=\frac{4g\cos\al\cosh\theta}{\cos^{2}\al+\sinh^{2}\theta}
\label{Eithetaalpha}\,,\\[2mm]
Q&=\frac{4g\cos\al\sin\al}{\cos^{2}\al+\sinh^{2}\theta}
\label{Qithetaalpha}\,,
\end{align}
and  
\begin{equation}
\cot\left(\frac{P}{2}\right)=\frac{\sinh\theta}{\cos\al}\,.
\end{equation}
The CsG equation is completely integrable and has multi-soliton scattering solution which can be constructed explicitly via inverse scattering \cite{deVega:1982sh} or by the Hirota method \cite{Getmanov:1977nf, Getmanov:1980cq}.
The only effect of scattering is to induce a time delay for each soliton relative to free motion.
For two solitons with rapidities $\theta_{1}$ and $\theta_{2}$ and rotation parameters $\alpha_{1}$ and $\alpha_{2}$ the centre-of-mass (COM) frame is defined by the condition, $\cos\alpha_{1}\sinh\theta_{1}=-\cos\alpha_{2}\sinh\theta_{2}$ (see Figure \ref{fig:2kinks CsG}).

In this Lorentz frame the two solitons experience an equal time delay $\Delta T_{1}=\Delta T_{2}=\Delta T_{{\mathrm{COM}}}$ with \cite{Dorey:1994mg}, 
\begin{equation}
\Delta T_{{\mathrm{COM}}}=\frac{1}{\cos\al_{1}
\sinh\theta_{1}}\ln F\left(\Delta\theta,\Delta\alpha,\bar{\alpha}\right)\,,
\label{CsGtimedelay}
\end{equation}  
where we define $\Delta\theta=(\theta_{1}-\theta_{2})/2$\,, $\Delta\alpha=(\al_{1}-\al_{2})/2$ and $\bar{\alpha}=(\al_{1}+\al_{2})/2$ and the function $F$ is given by, 
\begin{equation}
F\left(\Delta\theta,\Delta\alpha,\bar{\alpha}\right)
=\frac{\sinh\left(\Delta\theta+i\Delta\al\right)\sinh\left(\Delta\theta-i\Delta\al\right)}
{\cosh\left(\Delta\theta+i\bar{\al}\right)\cosh\left(\Delta\theta-i\bar{\al}\right)}\label{Ffunction}\,.
\end{equation}
Time delays due to multiple soliton scattering are simply given by the sum of the delays experienced in each two-body collision.
This is a consequence of integrability, and is a classical analog of the factorisability of the S-matrix.
Indeed, the time delays determine the semiclassical approximation to the worldsheet S-matrix $S_{\rm string}=\exp(i\Theta_{\rm string})$\,.
In particular, if we express the S-matrix as a function of the energies $E_{1}$ and $E_{2}$ of the two excitations and their charges $Q_{1}$ and $Q_{2}$ we have
\cite{Jackiw:1975im},
\begin{equation}
\Delta T_{j}=\frac{\partial \Theta_{\rm string}}{\partial E_{j}}\,,\qquad 
j=1,\,2\,.
\label{DeltaT1T2b}
\end{equation}

\section{Comparison of the S-matrices}

Our aim here is to compare $S_{\rm string}$ with the semiclassical limit of the magnon boundstate S-matrix computed above.
Equivalently we can use the boundstate S-matrix to compute the time delay in boundstate scattering directly and compare with the COM frame
expression for $\Delta T_{1}$ and $\Delta T_{2}$ presented in (\ref{CsGtimedelay}) above.
To do so, one has to first express $\Theta(X_{1}^{\pm},X_{2}^{\pm})$ in terms of the charges $Q_{1}$\,, $Q_{2}$ and the energies $\ga_{1}$\,, $\ga_{2}$ using the relations:
\begin{align}
E_{j}&=\frac{g}{i}\left[\left(X_{j}^{+}-\frac{1}{X_{j}^{+}}\right)-\left(X_{j}^{-}-\frac{1}{X_{j}^{-}}\right)\right]\,,\label{E1}\\[2mm]
Q_{j}&=\frac{g}{i}\left[\left(X_{j}^{+}+\frac{1}{X_{j}^{+}}\right)-\left(X_{j}^{-}+\frac{1}{X_{j}^{-}}\right)\right]\,.\label{Q2}
\end{align}
We then define, 
\begin{equation}
\Delta {\tau}_{j}=\frac{\partial \Theta_{\rm gauge}}{\partial E_{j}}\,,
\label{DeltaT1T2}
\end{equation} 
while keeping the charges $Q_{1}$ and $Q_{2}$ fixed. 
Here we present the results of explicit differentiations exclusively in terms of spectral parameters:
\begin{align}
\Delta {\tau}_{1}&=ih_{1}(X_{1}^{\pm})\ln H(X_{1}^{\pm},X_{2}^{\pm})
-h_{2}(X_{1}^{\pm})\,,\label{DeltaT1exp}\\
\Delta {\tau}_{2}&=-ih_{1}(X_{2}^{\pm})\ln H(X_{1}^{\pm},X_{2}^{\pm})
+h_{2}(X_{2}^{\pm})
\,.\label{DeltaT2exp}
\end{align}
where the functions $H$ and $h_{1,2}$ are defined as
\begin{eqnarray}
&\ds H(x^{\pm},y^{\pm})=\frac{\xp-\yp}{\xp-\ym}\frac{\xm-\ym}{\xm-\yp}\,,&\\[2mm]
&\ds h_{1}(x^{\pm})=\f{((x^{+})^{2}-1)((x^{-})^{2}-1)}{(x^{+})^{2}-(x^{-})^{2}}\,,\qquad 
h_{2}(x^{\pm})=\ko{\f{1}{x^{+}}-\f{1}{x^{-}}}\f{x^{+}x^{-}+1}{x^{+}+x^{-}}\,.&
\end{eqnarray}

\paragraph{}
All that remains is to compare with the CsG time delays.
Combining the identities (\ref{E1})-(\ref{Q2}) and the relations (\ref{Eithetaalpha}) and (\ref{Qithetaalpha}), one can express the spectral parameters of the magnon boundstates in terms of\footnote{
In obtaining these expressions one needs to solve quadratic equations.
The appropriate root of the quadratic is selected by demanding that the corresponding state has positive energy.}
$\theta_{j}$ and $\alpha_{j}$\,,
\begin{equation}
X_{j}^{\pm}=\coth\kko{\frac{\theta_{j}}{2}\pm i\left(\frac{\al_{j}}{2}-\frac{\pi}{4}\right)}\,,
\label{X<->CsG}
\end{equation}
These expressions in turn yield
\begin{equation}
F(\Delta\theta,\Delta\alpha,\bar{\alpha})\equiv
F(X_{1}^{\pm},X_{2}^{\pm})=
H(X_{1}^{\pm},X_{2}^{\pm})\,.
\label{Fexplicit}
\end{equation}
Now comparing (\ref{CsGtimedelay}) with (\ref{DeltaT1exp}) and (\ref{DeltaT2exp}), taking into account the COM frame condition, we can see that the time-delays for boundstate scattering agree with those of CsG solitons up to a specific non-logarithmic term, 
\begin{equation}
\Delta \tau_{1}=
\Delta T_{1}+\left(\frac{1}{X_{2}^{+}}-\frac{1}{X_{2}^{-}}\right)\frac{X_{1}^{+}X_{1}^{-}+1}{X_{1}^{+}+X_{1}^{-}}\,.
\end{equation}
Upon integration with respect to $\ep_{1}$\,, we can obtain the relation between the scattering phases:
\begin{equation}
\Theta_{\rm string}=\Theta_{\rm gauge }+\left(\ep_{2}-Q_{2}\right)P_{1}\,.
\label{phaserelation}
\end{equation}
The non-logarithmic term in (\ref{DeltaT1exp}) integrates up to give the difference term above that is a direct generalisation of the one in eqn. (3.33)
of \cite{Hofman:2006xt}.
As in that case, the difference can be accounted for by taking into account the different effective length of the excitation on the both sides of the correspondence.
On the string theory side, we set $(t,x)=\kappa(\tau,\sig)$ in conformal gauge, and so the density of $E$ is constant, while on the gauge theory side a unit length is assigned to each site $\cZ$ or $\cW$\,.
Hence when there $J_{1}$ $\cZ$s and $Q$ $\cW$s in the spin-chain, we have $\Delta \ell_{\rm gauge}=\int d(J_{1}+Q)=\int dE-\int d\kko{(E-J_{1})-Q}=\Delta x_{\rm string}-(\ep-Q)$\,.
By exponentiating it, we see $S_{\rm string}=S_{\rm gauge}\,e^{i(\Delta x_{\rm string}-\Delta \ell_{\rm gauge})_{2}P_{1}}=S_{\rm gauge}\,e^{i(\ep_{2}-Q_{2})P_{1}}$ as expected from (\ref{phaserelation}).
One can also check that the expressions in (\ref{DeltaT1exp}) and (\ref{DeltaT2exp}) correctly satisfy $\Delta \tau_{1}=\Delta \tau_{2}$
in the COM frame.

\chapter[ Singularities of the AdS/CFT S-matrix]
	{Singularities of the AdS/CFT S-matrix\label{chap:Singularities}}

The correspondence between singularities of the S-matrix and on-shell
intermediate states is a standard feature of quantum field
theory. It can be understood as a consequence of the 
analyticity and unitarity of the S-matrix\footnote{More precisely, only those
singularities in a suitably defined ``physical region'' need to have an
explanation in terms of on-shell states.}. In \cite{Dorey:2007xn}, this
correspondence was investigated in the context of the spin-chain
description of planar ${\cal N}=4$ SYM.
In particular, the 
poles of the conjectured exact S-matrix for magnon scattering 
were precisely accounted for by considering processes involving the
exchange of one or more BPS magnon boundstates. The goal of the
present chapter is to extend this investigation to the corresponding
S-matrix for the scattering of the boundstates themselves.

\begin{figure}[htb]
\begin{center}
\vspace{.3cm}
\includegraphics[scale=0.8]{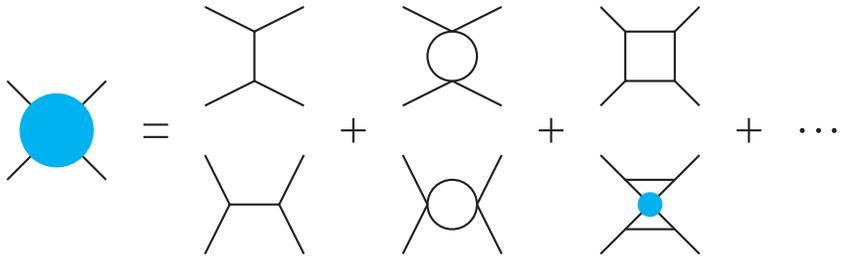}
\vspace{.3cm}
\caption{\small Landau diagrams associated with singularities of the two-body S-matrix.}
\label{fig:singularities}
\end{center}
\end{figure}

\paragraph{}
Initially we will focus on the $SU(2)$ sector of the ${\cal N}=4$
theory. 
The asymptotic Bethe ansatz equation \cite{Beisert:2005fw} for the $SU(2)$
sector of the theory reads, 
\begin{equation}
e^{i p_{j}}=\prod_{\mbox{$k=1\atop k\neq j$}}^{M} s(p_{j},p_{k})
\qquad \mbox{for}\quad 
j=1,\dots, M\,,
\end{equation}
where $M$ is the number of magnons, and the S-matrix 
is given by
\begin{equation}
s(p_{j},p_{k})
=S_{\rm BDS}^{-1}(p_{j},p_{k})\cdot
\sigma(p_{j}, p_{k})^{-2}\,,\qquad 
S_{\rm BDS}^{-1}(p_{j},p_{k})=\f{u(p_{j})-u(p_{k})-(i/g)}{u(p_{j})-u(p_{k})+(i/g)}\,,
\label{S-matrix 2}
\end{equation}
where the rapidity function $u$ is given in 
terms of the momentum by (\ref{rap}).
It is also convenient to introduce complex spectral parameters $x^{\pm}$\,, related to the rapidity (\ref{rap}) via (\ref{x(u)}).
The factor $\sigma^{-2}$ appearing in (\ref{S-matrix 2}) is the dressing factor,
and the cojectured exact expression for this function \cite{Beisert:2006ez} is
conveniently given as \cite{Dorey:2007xn} (\ref{BEHLS}).

For each solution of the Bethe
ansatz equations, the energy of the corresponding state is simply the
sum of the energies of the individual magnons. The energy of each
magnon is determined by the BPS dispersion relation (\ref{disp}).
In terms of the spectral parameters, the magnon momenta and energies
are expressed as (\ref{Delta}).

\section{Boundstates and their S-matrix\label{sec:S-matrix}}

In term of the spectral parameters BDS piece of the S-matrix takes the
form,
\begin{equation}
S_{\rm BDS}(x_{j}^{\pm}, x_{k}^{\pm})=\f{x_{j}^{+}-x_{k}^{-}}{x_{j}^{-}-x_{k}^{+}}\cdot
\f{1-1/(x_{j}^{+}x_{k}^{-})}{1-1/(x_{j}^{-}x_{k}^{+})}\,.
\label{S in x 2}
\end{equation}
We note the presence of a simple pole at $x_{j}^{-}=x_{k}^{+}$\,. 
As explained in \cite{Dorey:2006dq}, this pole indicates the formation
of a normalisable BPS boundstate of two magnons. In fact the theory
also contains a $Q$-magnon boundstate for each value of $Q>1$\,, 
related to a corresponding pole of the multi-particle S-matrix which can
be expressed as the product of two body factors by virtue of
integrability. These states were studied in detail in 
\cite{Dorey:2006dq, Chen:2006gq, Chen:2006gp}.
The spectral parameters of the constituent magnons in a $Q$-magnon
boundstate are,    
\begin{equation}
x_{j}^{-} = x_{j+1}^{+}
\qquad \mbox{for}\quad 
j=\komoji{1,\dots, Q-1}\,.
\end{equation}
The resulting boundstate of rapidity $U$ is described by introducing spectral
parameters 
\begin{equation}
X^{\pm}(U;Q)=x\ko{U\pm\mbox{\large $\f{iQ}{2g}$}}\,,\qquad 
\mbox{\em i.e.},\quad 
X^{+}\equiv x_{1}^{+}\,,\quad 
X^{-}\equiv x_{Q}^{-}\,,
\end{equation}
The total momentum $P$ and $U(1)$ charge $Q$ of the state are
then expressed as (\ref{P(X)}) and (\ref{Q(X)}).
The rapidity $U$ and energy $E=\sum_{k=1}^{Q}\ep_{k}$ for the boundstate are related to the spectral parameters $X^{\pm}$ as (\ref{U(X)}) and (\ref{E}).
It is useful to note the following properties of those functions of $X^{\pm}$\,.
\begin{enumerate}
\item By an interchange $X^{+}\leftrightarrow X^{-}$\,, $P$\,, $Q$\,, $E$ change signs and only $U$ remains the same:
\begin{equation}
    \begin{aligned}
	U(X^{\pm})&=U(X^{\mp})\,,\quad &
	P(X^{\pm})&=-P(X^{\mp})\,,\cr
	Q(X^{\pm})&=-Q(X^{\mp})\,,\quad &
	E(X^{\pm})&=-E(X^{\mp})\,.
    \end{aligned}
    \label{interchange}
\end{equation}
\item By an inversion $X^{\pm}\leftrightarrow 1/X^{\pm}$ known as
  crossing transformation \cite{Janik:2006dc,Arutyunov:2006iu}, 
$P$ and $E$ change signs, while $U$ and $Q$ remain the same:
\begin{equation}
    \begin{aligned}
	U(X^{\pm})&=U(1/X^{\pm})\,,\quad &
	P(X^{\pm})&=-P(1/X^{\pm})\,,\cr
	Q(X^{\pm})&=Q(1/X^{\pm})\,,\quad &
	E(X^{\pm})&=-E(1/X^{\pm})\,.
    \end{aligned}
    \label{crossing}
\end{equation}
\end{enumerate}

Note also the spectral parameters for the boundstates can be written as
\begin{equation}
X^{\pm}(P;Q)=R(P;Q)\,e^{\pm iP/2}
\quad \mbox{with}\quad R(P;Q)=\f{Q+\sqrt{Q^{2}+16g^{2}\sin^{2}\ko{P/2}}}{4g\sin\ko{P/2}}\,.
\label{radius}
\end{equation}

\paragraph{}
Starting with the S-matrix (\ref{S-matrix}) for elementary magnons, it is
straightforward to obtain the corresponding S-matrix for 
magnon boundstates of arbitrary charges 
by fusion, as worked out in \cite{Chen:2006gq,Roiban:2006gs}. Because of factorization of the multi-particle
S-matrix, the boundstate S-matrix is nothing other than the product of
two-body S-matrices describing all possible pair-wise scatterings
between the consitituent magnons.  
The proceedure is illustrated schematically
in Figure \ref{fig:fusion}.  
For two scattering boundstates (both in the same $SU(2)$ sector) with spectral parameters $Y_{1}^{\pm}$
and $Y_{2}^{\pm}$ and 
positive charges $Q_{1}\geq Q_{2}$ we find,  
\begin{align}
S(Y_{1}^{\pm},Y_{2}^{\pm})&=\prod_{j_{1}=1}^{Q_{1}}\prod_{j_{2}=1}^{Q_{2}}s(y_{j_{1}}^{\pm},y_{j_{2}}^{\pm})\no\\
&=G(Q_{1}-Q_{2})\kko{\prod_{n=1}^{Q_{2}-1}G(Q_{1}-Q_{2}+2n)^{2}}G(Q_{1}+Q_{2})\times {}\no\\[2mm]
& {}\qquad \times
\Sigma(Y_{1}^{\pm},Y_{2}^{\pm})^{-2}\,,\quad 
\mbox{where}\quad G(q)=\f{U_{1}-U_{2}-iq/2g}{U_{1}-U_{2}+iq/2g}\,.
\label{S-bounstates}
\end{align}
We note that $G(q)$ can be rewritten in terms of the spectral parameters as, 
\begin{equation}
G(q)
=\f{\ko{Y_{1}^{-}-Y_{2}^{+}}\ko{1-1/Y_{1}^{-}Y_{2}^{+}}+i(Q_{1}+Q_{2}-q)/2g}
	{\ko{Y_{1}^{+}-Y_{2}^{-}}\ko{1-1/Y_{1}^{+}Y_{2}^{-}}-i(Q_{1}+Q_{2}-q)/2g}
\label{G(q)}
\end{equation}
Here $\Sigma^{-2}$ stands for the appropriate dressing factor for
boundstates. As explained in \cite{Chen:2006gq}, its form as a 
function of the spectral
parameters is identical to that of the conjectured BES dressing factor \cite{Beisert:2006ez}
for elementary magnons. Explicitly we have, 
\begin{align}
\Sigma(Y_{1}^{\pm},
Y_{2}^{\pm})^{-2}&=\prod_{j_{1}=1}^{Q_{1}}\prod_{j_{2}=1}^{Q_{2}}\sigma(y_{j_{1}}^{\pm},
y_{j_{2}}^{\pm})^{-2}=\ko{\f{R(Y_{1}^{+}, Y_{2}^{+})R(Y_{1}^{-},
    Y_{2}^{-})}
{R(Y_{1}^{+}, Y_{2}^{-})R(Y_{1}^{-}, Y_{2}^{+})}}^{-2}\,.
\label{dressbs}
\end{align}
where the final equality arises after numerous cancellations are taken
into account.

\paragraph{}
From Eqn (\ref{S-bounstates}) and (\ref{G(q)}), a finite set of simple
and double poles of the boundstate S-matrix is apparent. In addition,
as we review below, the dressing factor (\ref{dressbs}) provides an
infinite sequence of additional double poles. 
Below we will investigate both simple and double poles of the
boundstate S-matrix (\ref{S-bounstates}), 
and discuss which of them are the physical singularities.
Then in Section \ref{sec:decode} we will interpret those
singularities as physical processes in terms 
of Landau diagrams.
We will also see their interpretation in terms Bethe root configurations in Section \ref{sec:Bethe String}.

\subsection{Simple poles}

Simple poles are found in $G(Q_{1}+Q_{2})$ and $G(Q_{1}-Q_{2})$ at
\begin{equation}
U_{1}-U_{2}=-\f{i}{2g}(Q_{1}+Q_{2})
\quad \mbox{and}\quad 
U_{1}-U_{2}=-\f{i}{2g}(Q_{1}-Q_{2})\,,
\end{equation}
As we discuss below it is natural to interpret these poles as due to 
exchange of BPS boundstates of charge $Q_{1}\pm
Q_{2}$ in 
s- and t-channel
processes respectively\footnote{In the special case $Q_{1}=Q_{2}$\, 
the t-channel process is absent.}. 
Note however that, in terms of the spectral parameters $Y_{1}^{\pm}$
and $Y_{2}^{\pm}$ of the two incoming particles, 
$G(Q_{1}\pm Q_{2})$ are written as,
\begin{align}
G(Q_{1}+Q_{2})&=\f{Y_{1}^{-}-Y_{2}^{+}}{Y_{1}^{+}-Y_{2}^{-}}\cdot\f{1-1/Y_{1}^{-}Y_{2}^{+}}{1-1/Y_{1}^{+}Y_{2}^{-}}\,,\label{sp1}\\[2mm]
G(Q_{1}-Q_{2})&=\f{Y_{1}^{-}-Y_{2}^{-}}{Y_{1}^{+}-Y_{2}^{+}}\cdot\f{1-1/Y_{1}^{-}Y_{2}^{-}}{1-1/Y_{1}^{+}Y_{2}^{+}}\,,\label{sp2}
\end{align}
and, in these variables, there are two simple poles originating in
{\em each} of $G(Q_{1}+Q_{2})$ and
$G(Q_{1}-Q_{2})$\,.  
For example, $G(Q_{1}+Q_{2})$ has simple poles at
$Y_{1}^{+}=Y_{2}^{-}$ and 
$Y_{1}^{+}=1/Y_{2}^{-}$\,.
The question of which, if any, of these poles correspond to a physical
processes will be investigated in Section \ref{sec:physical poles}.

\subsection{Double poles}
The singular structure of the dressing part is highly non-trivial, and
for scattering of elementary magnons, 
it was worked out in \cite{Dorey:2007xn}. This leads to an infinite
series of double poles in the magnon S-matrix. 
In the present
case of the 
boundstate S-matrix, there are two distinct sources of double poles 
which we will discuss in turn.

\subsubsection*{$\bullet$ BDS part}

The double poles of the BDS part of boundstate S-matrix
(\ref{S-bounstates}) 
locate at 
\begin{equation}
U_{1}-U_{2}=-\f{i}{2g}(Q_{1}+Q_{2}-2n)\,,\quad n=1,2,\dots, Q_{2}-1\,.
\label{u-u:3}
\end{equation}
As above, each of these, gives rise to a pair of double poles when
expressed in terms of the spectral parameters $Y_{1}^{\pm}$
and $Y_{2}^{\pm}$ at the two distinct roots of the equation\footnote{There are four ways to write down the condition (\ref{u-u:3}) in terms of the spectral parameters.
They all take the form $(Y_{1}^{\al}-Y_{2}^{\be})(1-1/Y_{1}^{\al}Y_{2}^{\be})+(i/g)\,n_{\al\be}=0$\,, where $(\al,\be)=(+,+), (+,-), (-,+), (-,-)$\,.
The integer $n_{\al\be}$ covers different region for these four choices, but the number of integers are the same, and is given by $\min\{Q_{1},Q_{2}\}-1$\,, namely $Q_{2}-1$ in our case.
The expression (\ref{u-u:3 x}) corresponds to $(\al,\be)=(+,-)$ case.}
\begin{equation}
Y_{1}^{+}+\f{1}{Y_{1}^{+}}-Y_{2}^{-}-\f{1}{Y_{2}^{-}}-\f{in}{g}=0\,,\quad n=1,2,\dots, Q_{2}-1\,.
\label{u-u:3 x}
\end{equation}

\subsubsection*{$\bullet$ Dressing part}

As the functional form of the dressing factor is essentially the same as the
elementary magnon scattering case we follow the analysis of 
\cite{Dorey:2007xn} .
In particular, we consider the derivative with respect to the coupling
$g$ of the function $\chi$ appearing in (\ref{dressbs}),  
\begin{align}
\pa_{g}\chi(Y_{1},Y_{2})&=-\oint_{\mathcal C}\f{dz_{1}}{2\pi}\oint_{\mathcal C}\f{dz_{2}}{2\pi}
\f{z_{1}+\hf{1}{z_{1}}-z_{2}-\hf{1}{z_{2}}}{(z_{1}-Y_{1})(z_{2}-Y_{2})}\,\komoji{\Psi\ko{1+ig\ko{z_{1}+\hf{1}{z_{1}}-z_{2}-\hf{1}{z_{2}}}}}\no\\[2mm]
&=-\oint_{\mathcal C}\f{dz_{1}}{2\pi}\oint_{\mathcal C}\f{dz_{2}}{2\pi}
\sum_{n=1}^{\infty}
\f{n/g^{2}}{(z_{1}-Y_{1})(z_{2}-Y_{2})}
\f{1}{z_{1}+\hf{1}{z_{1}}-z_{2}-\hf{1}{z_{2}}-\hf{in}{g}}\,,
\label{del-g-chi}
\end{align}
where we used the definition of digamma function $\Psi(x)$ and its asymptotic expansion,
\begin{equation}
\f{d}{dx}\ln \Gamma(x)=\Psi(x)
=-\gamma_{\rm E}-\sum_{n=1}^{\infty}\kko{\f{1}{x+n-1}-\f{1}{n}}\,,
\end{equation}
($\gamma_{\rm E}$ is Euler's constant).
Let $z_{2}=F_{n}(z_{1})$ be the root of the quadratic equation
\begin{equation}
z_{1}+\f{1}{z_{1}}-z_{2}-\f{1}{z_{2}}-\f{in}{g}=0\,,
\end{equation}
which satisfies $|F_{n}(z_{1})|<1$\,. 
By the same argument as in \cite{Dorey:2007xn}, singularities arise
when poles of the integrand pinch the integration contour. 
As explained in Section 5 of\cite{Dorey:2007xn} the only case where we
pinch the contour is when 
$Y_{1}=Y_{1}^{-}$ and $Y_{2}=Y_{2}^{+}$\,.
Plugging this into (\ref{del-g-chi}) and performing the double contour integrals, we reach the expression
\begin{align}
\pa_{g}\chi(Y_{1}^{-},Y_{2}^{+})=
-\f{n}{g^{2}}\sum_{n=1}^{\infty}\f{F_{n}(Y_{1}^{-})}{\kko{Y_{2}^{+}F_{n}(Y_{1}^{-})-1}\kko{1-F_{n}(Y_{1}^{-})^{2}}}\,.
\end{align}
This can be easily integrated, giving
\begin{equation}
\chi(Y_{1}^{-},Y_{2}^{+})=-i\sum_{n=1}^{\infty}\ln\ko{Y_{2}^{+}-F_{n}(Y_{1}^{-})^{-1}}\,.
\end{equation}
Then we see the relevant parts of our poles/zeros analysis become
\begin{equation}
\Sigma(Y_{1}^{\pm}, Y_{2}^{\pm})^{-2}
\sim e^{2i\kko{\chi(Y_{1}^{-}, Y_{2}^{+})-\chi(Y_{2}^{-}, Y_{1}^{+})}}
=\prod_{n=1}^{\infty}\kko{\f{Y_{2}^{+}-F_{n}(Y_{1}^{-})^{-1}}{Y_{1}^{+}-F_{n}(Y_{2}^{-})^{-1}}}^{2}\,.
\label{zero,pole}
\end{equation}
From (\ref{zero,pole}), we see the double poles lie at $Y_{1}^{+}=F_{n}(Y_{2}^{-})^{-1}$\,.
In view of $F_{n}(x)+F_{n}(x)^{-1}=x+x^{-1}-(in/g)$\,, this condition turns out 
\begin{equation}
Y_{1}^{+}+\f{1}{Y_{1}^{+}}-Y_{2}^{-}-\f{1}{Y_{2}^{-}}=-\f{in}{g}\,,
\label{u-u:2 x}
\end{equation}
which is one of the roots of the equation,  
\begin{equation}
U_{1}-U_{2}=-\f{i}{2g}(Q_{1}+Q_{2}+2n)\,,\quad n=1,2,\dots,\qquad U_{j}\eq u(Y_{j}^{\pm})\,.
\label{u-u:2}
\end{equation}
In the special case $Q_{1}=Q_{2}=1$\,, we reproduce the results of 
\cite{Dorey:2007xn}.

\section{Physicality conditions\label{sec:physical poles}}

In general S-matrix singularities occur at complex values of the external momenta and energies.
Only those singularities suitably close to the real axis (with positive energy) require a physical explanation.
In relativistic scattering there is a well-established notion of a ``physical sheet''.
In the present case, where the dynamics is non-relativistic, the
extent of the physical region is unclear. However, for each scattered
particle there are three distinct limits in which it is possible to
analyse the situation precisely.\footnote{There is yet another strong coupling limit of much interest, which is known as the near-flat-space limit \cite{Maldacena:2006rv}.
The momentum scales as $P\sim 1/\sqrt{g}$ in this limit, and it interpolates between the giant magnon limit and the plane-wave limit smoothly.
We will not dwell on this limit in this thesis, but it would be interesting to consider it in addition to those three limits described below, since in the near-flat-space region the physical poles might be identifiable.
We thank the referee of the paper \cite{Dorey:2007an} for pointing this out.}
These are, 
\begin{description}
\item[\bmt{(i)}] {\bf The Giant Magnon limit}\,:~ $g\to \infty$ while $P$ kept fixed, where 
\begin{equation}
Y^{+}\simeq 1/Y^{-}\simeq e^{iP/2}\,,\quad 
U\simeq 2\cos\ko{\f{P}{2}}\,,\quad 
E\simeq 4g\sin\ko{\f{P}{2}}\,.
\end{equation}
In this limit the particles with arbibtrary charge $Q$ become heavy solitonic states of the
string worldsheet theory. 

\item[\bmt{(ii)}] {\bf Plane-Wave limit}\,:~ $g\to \infty$ with $k\eq 2gP$ kept fixed, where 
\begin{equation}
Y^{+}\simeq Y^{-}\simeq \f{Q+\sqrt{Q^{2}+k^{2}}}{k}\in {\mathbb R}\,,\quad 
U\simeq \f{2}{k}\sqrt{Q^{2}+k^{2}}\,,\quad 
E\simeq \sqrt{Q^{2}+k^{2}}\,.
\label{pp-wave region}
\end{equation}
In this limit the magnon reduces to an elementary excitation of the
worldsheet theory. As before states with $Q>1$ are interpreted as
boundstates of the elementary $Q=1$ excitation. 
Notice one can also express 
\begin{equation}
E=\f{\xi^{2}+1}{\xi^{2}-1}\,Q\,,\quad 
k=\f{2\xi}{\xi^{2}-1}\,Q\,,\quad 
\xi\,e^{\pm i\delta/2} \eq Y^{\pm}\quad (\xi\in\mathbb R\,,~ 0<\delta\ll 1)\,.
\label{pp-wave region 2}
\end{equation}

\item[\bmt{(iii)}] {\bf Heisenberg spin-chain limit}\,:~ $g\ll 1$ limit, where 
\begin{equation}
Y^{\pm}\mp \f{iQ}{2g}\simeq U\simeq \f{1}{2g}\cot\ko{\f{P}{2}}\,,\quad 
E\simeq Q+\f{8g^{2}}{Q}\sin^{2}\ko{\f{P}{2}}\,.
\end{equation}
In this limit, the gauge theory can be studied in the one-loop
approximation where the dilatation operator in the $SU(2)$ sector is
precisely the Heisenberg Hamiltonian. 

\end{description}

In the following, as in \cite{Dorey:2007xn}, we will focus on
singularities which lie 
parametrically close to the positive real axis 
for both external energies.
In particular, this includes those singularities which come close to
the positive real axis in any of the three limits described above. 
Below we will identify which poles of the boundstate S-matrix fall
into this category. We will refer to them as physical poles. 

\subsection{Physical simple poles}

In (\ref{sp1}), (\ref{sp2}), we saw there are two simple poles for each
of $G(Q_{1}+Q_{2})$ and $G(Q_{1}-Q_{2})$ 
when written in terms of the spectral parameters. We will study the
behaviour of these poles in the limits described above. The key
question is whether, in any of these limits, the pole approaches 
a point where the energies of both particles scattering are real and
positive. If this is the case in at least one of the limits considered
then we will accept the pole as physical.

\paragraph{$\bullet$ Simple poles in \bmt{G(Q_{1}+Q_{2})}.}

First consider the case where the momenta of the two external particles are in the plane-wave region.
This means we have $Y_{i}^{+}\simeq Y_{i}^{-}$ $(i=1,2)$\,.
Let us suppose the first particle ($i=1$) is in the physical region
(so its energy $E_{1}$ is positive), 
and see whether one of the pole conditions $Y_{1}^{+}=Y_{2}^{-}$ $(\eq e^{i\delta /2})$ implies the second particle ($i=2$) is also physical.
The answer can be found by looking at the relative sign of energies 
between the two particles.
Since $e^{-i\delta /2}\simeq Y_{1}^{-}\simeq Y_{1}^{+}=e^{i\delta
  /2}=Y_{2}^{-}\simeq Y_{2}^{+}\simeq e^{-i\delta /2}$\,, 
the energy of the second particle is evaluated as
\begin{align}
E_{2}&=\f{g}{i}\kko{\ko{Y_{2}^{+}-\f{1}{Y_{2}^{+}}}-\ko{Y_{2}^{-}-\f{1}{Y_{2}^{-}}}}\no\\[2mm]
&\simeq\f{g}{i}\kko{\ko{Y_{1}^{-}-\f{1}{Y_{1}^{-}}}-\ko{Y_{1}^{+}-\f{1}{Y_{1}^+}}}
= -E_{1} ~ <0\,,
\end{align}
thus the second particle does not live near the physical region in
this limit.
On the other hand, for the other pole at 
$Y_{1}^{+}=1/Y_{2}^{-}$\, we have $Y_{1}^{-}\simeq 1/Y_{2}^{+}$
in the plane-wave region which leads to $E_{2}>0$\, 
thus corresponding to physical pole.

Next let us consider the case of the scattering of two dyonic giant magnons.
In this case, the spectral parameters are related as 
$Y_{i}^{+}\simeq 1/Y_{i}^{-}$ $(i=1,2)$\,.
By insisting that the energy of both particles is positive in this
limit, we again select the pole at $Y_{1}^{+}=1/Y_{2}^{-}$\,. The other
pole at $Y_{1}^{+}=Y_{2}^{-}$ again violates this criterion.

Finally let us consider the Heisenberg spin-chain limit.
By noticing the $g$\,-dependence of the spectral parameters, RHS of (\ref{sp1}) becomes in this limit,
\begin{equation}
G(Q_{1}+Q_{2})=\f{Y_{1}^{-}-Y_{2}^{+}}{Y_{1}^{+}-Y_{2}^{-}}\cdot\ko{1+\ord{g^{2}}}\,,
\label{Heisenberg limit}
\end{equation}
so one finds, contrast to the strong coupling results, it is the pole
$Y_{1}^{+}=Y_{2}^{-}$ that 
should be regarded as physical pole in the weak coupling region. 
In fact we can write down the wavefunction of the corresponding
boundstate explicitly in this limit. 

In conclusion we have found at least one limit in which each of the
two simple poles (at $Y_{1}^{+}=Y_{2}^{-}$ and at
$Y_{1}^{+}=1/Y_{2}^{-}$) occurs near the region of positive real
energies. Thus we will accept both poles as physical and seek an
expanation in terms of on-shell intermediate states. 

\paragraph{$\bullet$ Simple poles in \bmt{G(Q_{1}-Q_{2})}.}

We can apply the same line of reasoning to this case.
In particular can show that the pole at $Y_{1}^{+}=Y_{2}^{+}$\,, 
occurs near the region where both external particles have real
positive energies, in all the three of the limits discussed above 
(giant magnon, plane wave and Heisenberg spin-chain). In contrast one may
check that the remaining pole at $Y_{1}^{+}=1/Y_{2}^{+}$ stays away
from the physical region in each of the limits considered. For this
reason we will accept the first pole as physical but not the second. 

\paragraph{}
In summary, three of the four poles, $Y_{1}^{+}=Y_{2}^{-}$\,, $Y_{1}^{+}=1/Y_{2}^{-}$ and $Y_{1}^{+}=Y_{2}^{+}$ are identified with physical poles, giving $E_{i}>0$ and $Q_{i}>0$ for both external particles $Y_{i}$ ($i=1,2$) at least one of the three $(i)$-$(iii)$ regions.
They are summarised in Table \ref{tab:physical simple poles}.
Entries with checks ``\check\,'' indicate the pole result in $E_{i}>0$ and $Q_{i}>0$ in the region for both $i=1,2$\,.

\begin{table}[htbp]
\caption{\small The first three poles ($Y_{1}^{+}=Y_{2}^{-}$\,, $Y_{1}^{+}=1/Y_{2}^{-}$\,, $Y_{1}^{+}=Y_{2}^{+}$) are physical while the other one ($Y_{1}^{+}=1/Y_{2}^{+}$) is unphysical.}
\begin{center}
\begin{tabular}{|cl||c|c||c|c|}\hline
{} &  & \multicolumn{2}{c||}{$G(Q_{1}+Q_{2})$} & \multicolumn{2}{c|}{$G(Q_{1}-Q_{2})$}\\ \cline{3-6}
{} & {} & $Y_{1}^{+}=Y_{2}^{-}$ & $Y_{1}^{+}=1/Y_{2}^{-}$ & $Y_{1}^{+}=Y_{2}^{+}$ & $Y_{1}^{+}=1/Y_{2}^{+}$ \\ \hline
$(i)$ & Giant Magnon limit & \batsu & \check & \check & \batsu \\
$(ii)$ & Plane-Wave limit & \batsu & \check & \check & \batsu \\
$(iii)$ & Heisenberg spin-chain limit & \check & \batsu & \check & \batsu \\ \hline
\end{tabular}
\end{center}
\label{tab:physical simple poles}
\end{table}

\subsection{Physical double poles}

As we saw in the previous sections, double poles exist in two regions; one is in a finite interval (\ref{u-u:3 x}) that comes from the BDS part, and the other is an infinite interval (\ref{u-u:2 x}) from the dressing part.

\paragraph{$\bullet$ The BDS part.}
We start with investigating the first region originated from the BDS part.
Written in terms of the spectral parameters, there are two double poles in the BDS part of boundstate S-matrix (\ref{S-bounstates}), which are the two roots of the equation (\ref{u-u:3 x}).
For each $n$\,, one can solve the constraint for $Y_{1}^{+}$ to find the two roots
\begin{equation}
Y_{1}^{+}=y_{\pm}^{(n)}\eq \f{g(Y_{2}^{-})^{2}+i n Y_{2}^{-}+g\pm\sqrt{\ko{g(Y_{2}^{-})^{2}+i n Y_{2}^{-}+g}^{2}-4g^{2}(Y_{2}^{-})^{2}}}{2gY_{2}^{-}}\,.
\label{y pm}
\end{equation}
Here the subscripts in $y_{\pm}^{(n)}$ refers to the signs in front of the square root in (\ref{y pm}), and the integer $n$ runs $n=1,\dots,Q_{2}-1$\,.
We would like to find out which of the two roots corresponds to a physical double pole that satisfy the criteria established in the previous section.
We will do so for $n\sim 1$ and $n\sim Q_{2}$ regions, separately, when the identification becomes transparent due to that we already know which are the closest physical simple poles to each regions.
For the latter case with $n\sim Q_{2}$\,, we will further divide the case into two according to whether $Q_{2}\ll g$ or $Q_{2}\gg g$\,.

\paragraph{}
When $n$ is much smaller than $g$\,, the two roots in (\ref{y pm}) approach $y_{+}\to Y_{2}^{-}$ and $y_{-}\to 1/Y_{2}^{-}$\,.
In this region, since $n$ is close to zero, we can expect the physical double poles exist near the physical simple pole in $G(Q_{1}+Q_{2})$\,, which is $Y_{1}^{+}=1/Y_{2}^{-}$ as we identified in the previous section (see Table \ref{tab:physical simple poles}).
Hence we conclude that in both the plane wave and the giant magnon regions, it is the root $y_{-}$ that corresponds to a physical double pole, while in the Heisenberg spin-chain limit $g\ll 1$\,, the other root $y_{+}$ is physical since $y_{-}$ disappears in this limit, just as we saw in (\ref{Heisenberg limit}).

Next let us turn to the other side of the BDS double pole spectrum, $n=Q_{2}-1,\,Q_{2}-2,\dots$\,, which is near the physical simple pole in $G(Q_{1}-Q_{2})$\,.
First notice that when $n$ is close to $Q_{2}$\,, the two roots in (\ref{y pm}) approach to either $Y_{2}^{+}$ or $1/Y_{2}^{+}$\,.
To see this, let us solve the constraint $Y_{2}^{+}+1/Y_{2}^{+}-iQ_{2}/2g=Y_{2}^{-}+1/Y_{2}^{-}+iQ_{2}/2g$ for $Y_{2}^{+}$ \,, which leads to
\begin{equation}
Y_{2}^{+}=y'{}_{\pm}\eq \f{g(Y_{2}^{-})^{2}+i Q_{2} Y_{2}^{-}+g\pm\sqrt{\ko{g(Y_{2}^{-})^{2}+i Q_{2} Y_{2}^{-}+g}^{2}-4g^{2}(Y_{2}^{-})^{2}}}{2gY_{2}^{-}}\,.
\label{y' pm}
\end{equation}
Comparing them with $y_{\pm}^{(n)}$ in (\ref{y pm}), we see that as $n$ tends to $Q_{2}$\,, two branches $y_{\pm}^{(n)}$ approach $y'{}_{\pm}$ respectively.
Recall that in the previous section we found the only physical simple pole of $G(Q_{1}-Q_{2})$ in all three regions (the plane wave, giant magnon and Heisenberg regions) was $Y_{1}^{+}=Y_{2}^{+}$\,, and the physical double poles with $n=Q_{2}-1,\,Q_{2}-2,\dots$ should be close to it.
These observations lead us to conclude that, for $n$ close to $Q_{2}$\,, if $Y_{2}^{+}=y'{}_{+}$\,, the physical double pole is given by $Y_{1}^{+}=y_{+}^{(n)}$\,, whereas if $Y_{2}^{+}=y'{}_{-}$\,, it is given by $Y_{1}^{+}=y_{-}^{(n)}$\,, for all the three regions.

Which of $y_{\pm}^{(n)}$ should be singled out as the physical branch of $Y_{1}^{+}$ around $n\sim Q_{2}$ depends on the magnitude of $Q_{2}$ compared to $g$\,.
When $Q_{2}\ll g$\,, the two branches $y'{}_{+}$ and $y'{}_{-}$ approach $Y_{2}^{-}$ and $1/Y_{2}^{-}$\,, respectively.
Therefore, for the plane wave region where $Y_{2}^{+}\simeq Y_{2}^{-}$\,, we should single out $Y_{2}^{+}=y'{}_{+}$ so that the physical double pole corresponds to $Y_{1}^{+}=y_{+}^{(n)}$\,.
On the other hand, in the giant magnon region where $Y_{2}^{+}\simeq 1/Y_{2}^{-}$\,, we should single out $Y_{2}^{+}=y'{}_{-}$ so that the physical double pole corresponds to $Y_{1}^{+}=y_{-}^{(n)}$\,.
As for the Heisenberg spin-chain limit, there is no such limit we can take for the current $Q_{2}\ll g$ case.
The results for $Q_{2}\ll g$ case is summarised in Table \ref{tab:physical double poles 1}.
In the giant magnon region, the physical double pole remains to be $Y_{1}^{+}=y_{-}^{(n)}$ for all $n=1,\dots,Q_{2}-1$\,, while in the plane wave region, the physical double pole switches from $Y_{1}^{+}=y_{-}^{(n)}$ ($n\sim 1$) to $Y_{1}^{+}=y_{+}^{(n)}$ ($n\sim Q_{2}$) around some point.

\begin{table}[htbp]
\caption{\small The $Q_{2}\ll g$ case.}
\begin{center}
\begin{tabular}{|cl||c|c||c|c|}\hline
{} &  & \multicolumn{2}{c||}{$n\sim 1$} & \multicolumn{2}{c|}{$n\sim Q_{2}\, (\ll g)$}\\ \cline{3-6}
{} & {} & $Y_{1}^{+}=y_{+}^{(n)}$ & $Y_{1}^{+}=y_{-}^{(n)}$ & $Y_{1}^{+}=y_{+}^{(n)}$ & $Y_{1}^{+}=y_{-}^{(n)}$ \\ \hline
$(i)$ & Giant Magnon limit & \batsu & \check & \batsu & \check \\
$(ii)$ & Plane-Wave limit & \batsu & \check & \check & \batsu \\
$(iii)$ & Heisenberg spin-chain limit & - & - & - & - \\ \hline
\end{tabular}
\end{center}
\label{tab:physical double poles 1}
\end{table}

\paragraph{}
When $Q_{2}\gg g$\,, the two branches $y'{}_{+}$ and $y'{}_{-}$ reduce to $iQ_{2}/g$ and $0$\,, respectively.
Hence the physical double poles around $n\sim Q_{2}$ are singled out to be $Y_{1}^{+}=y_{+}^{(n)}$ corresponding to the $y'{}_{+}$ branch of $Y_{2}^{+}$\,, since the other root $y_{-}^{(n)}$ disappears just like the case with the Heisenberg spin-chain limit (\ref{Heisenberg limit}).
As a result, in all three regions, we conclude that $Y_{1}^{+}=y_{+}^{(n)}$ corresponds to the physical double pole when $n$ is close to $Q_{2}\, (\gg g)$\,.
The results are summarised in Table \ref{tab:physical double poles 2}.

\begin{table}[htbp]
\caption{\small The $Q_{2}\gg g$ case.}
\begin{center}
\begin{tabular}{|cl||c|c||c|c|}\hline
{} &  & \multicolumn{2}{c||}{$n\sim 1$} & \multicolumn{2}{c|}{$n\sim Q_{2}\, (\gg g)$}\\ \cline{3-6}
{} & {} & $Y_{1}^{+}=y_{+}^{(n)}$ & $Y_{1}^{+}=y_{-}^{(n)}$ & $Y_{1}^{+}=y_{+}^{(n)}$ & $Y_{1}^{+}=y_{-}^{(n)}$ \\ \hline
$(i)$ & Giant Magnon limit & \batsu & \check & \check & \batsu \\
$(ii)$ & Plane-Wave limit & \batsu & \check & \check & \batsu \\
$(iii)$ & Heisenberg spin-chain limit & \check & \batsu & \check & \batsu \\ \hline
\end{tabular}
\end{center}
\label{tab:physical double poles 2}
\end{table}

\paragraph{$\bullet$ The dressing part.}
Next let us turn to the double poles from the dressing phase, (\ref{u-u:2 x}).
Actually the analysis for this case is already basically done, since (\ref{u-u:2 x}) leads to the same equation (\ref{y pm}).
The only difference lies in the range of $n$\,, which runs $n=Q_{2}+1,\,Q_{2}+2,\dots$ in this case.
Therefore, in order to identify the physical double poles, we have only to refer to Tables \ref{tab:physical double poles 1} and \ref{tab:physical double poles 2}.
In the giant magnon region, $Y_{1}^{+}=y_{-}^{(n)}$ corresponds to the physical double pole when $n\ll g$\,, while when $n\gg g$\,, the other branch $Y_{1}^{+}=y_{+}^{(n)}$ plays the role.
In the plane wave region, $Y_{1}^{+}=y_{+}^{(n)}$ remains the physical double pole for all $n=Q_{2}+1,\,Q_{2}+2,\dots$\,.

\section{Decoding physical poles\label{sec:decode}}

For the conjectured boundstate S-matrix to be correct, there should
exist at least one physical process that 
accounts for each physical pole.
In other words, we should be able to draw at least one consistent
Landau diagram. 
Also, there should not be any Landau diagrams which lead to extra
poles in the physical region which are not seen in the S-matrix. 
In this section, we will draw Landau diagrams corresponding to the
physical poles we identified above and comment on the possible
occurence of other diagrams.

\begin{figure}[tb]
\begin{center}
\vspace{.5cm}
\hspace{-.0cm}\includegraphics[scale=0.7]{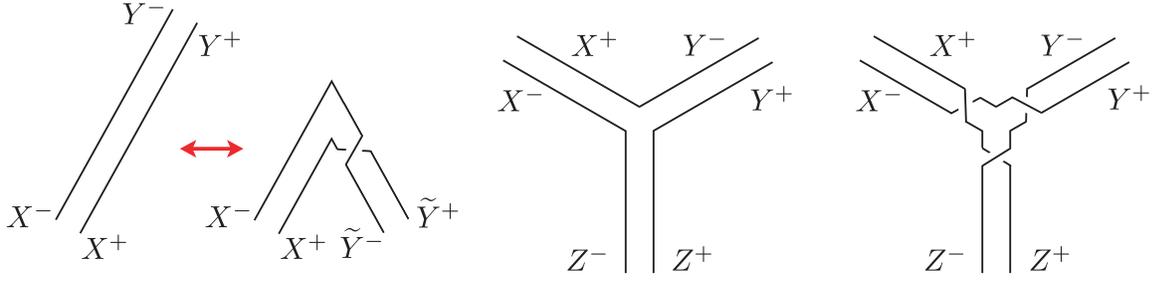}
\vspace{.0cm}
\caption{\small Building blocks of physical processes.
The ``double line'' notation of \cite{Dorey:2007xn} is employed, and time flows from bottom to top.
The dotted line indicates the corresponding particles carry negative charges.
}
\label{fig:Landau}
\end{center}
\end{figure}

The rules for constructing these diagrams are the same as given
in \cite{Dorey:2007xn}. Our current analysis generalises that of \cite{Dorey:2007xn} in that
we are analysing the situation where 
both the external (incoming/outgoing) particles carry generic (positive) charges, which we denote as $Q_{1}$ and $Q_{2}$ ($Q_{1}\geq Q_{2}$).
The building blocks of physical processes are the three particle
vertices shown in Figure
\ref{fig:Landau} which implement conservation of energy, momentum and
other quantum numbers. The left diagram shows the crossing transformation, $\widetilde Y^{\pm}=1/Y^{\pm}$\,. 
The other two diagrams describe two possible three-vertex diagrams.
The spectral parameters of the three particles
are related as 
$X^{+}=Y^{-}$\,, $X^{-}=Z^{-}$ and $Y^{+}=Z^{+}$ for the middle, and 
$X^{+}=Z^{+}$\,, $X^{-}=Y^{+}$ and $Y^{-}=Z^{-}$ for the right. All
lines in the diagram are on-shell. 

\subsection{Landau diagrams for simple poles}

As we saw in the previous section, there are three physical simple poles, and there must be at least one corresponding Landau diagram for each of them.
Let us first forget about the physicality condition and try to draw down the diagrams endowed with four simple pole conditions $Y_{1}^{+}=Y_{2}^{-}$ and $Y_{1}^{+}=1/Y_{2}^{-}$ (from $G(Q_{1}+Q_{2})$) and $Y_{1}^{+}=Y_{2}^{+}$ and $Y_{1}^{+}=1/Y_{2}^{+}$ (from $G(Q_{1}-Q_{2})$).
In any case, those simple poles describe formations of boundstate $Z^{\pm}$ either in the s- or t-channel.
Let us denote the multiplet number of the intermediate BPS particle $Q_{Z}$\,, which can be found out by the formula
\begin{equation}
Q_{Z}=\f{g}{i}\kko{\ko{Z^{+}+\f{1}{Z^{+}}}-\ko{Z^{-}+\f{1}{Z^{-}}}}\,.
\label{Qz}
\end{equation}
Since we assumed $Q_{1}>Q_{2}$\,, the multiplet number $Q_{Z}\, (>0)$ can only take values either $Q_{Z}=Q_{1}+Q_{2}$ or $Q_{Z}=Q_{1}-Q_{2}$\,, and the $U(1)$ charge carried by $Z^{\pm}$ is $Q_{1}+Q_{2}$ if the process is in the s-channel, and $Q_{1}-Q_{2}$ or $-(Q_{1}-Q_{2})$ if it is in the t-channel.
Notice in our convention the multiplet number $Q_{Z}$ must be positive while the $U(1)$ charge can take either positive or negative values, varying from $-Q_{Z}$ to $+Q_{Z}$\,.

\begin{figure}[tb]
\begin{center}
\vspace{.5cm}
\hspace{-.0cm}\includegraphics[scale=0.7]{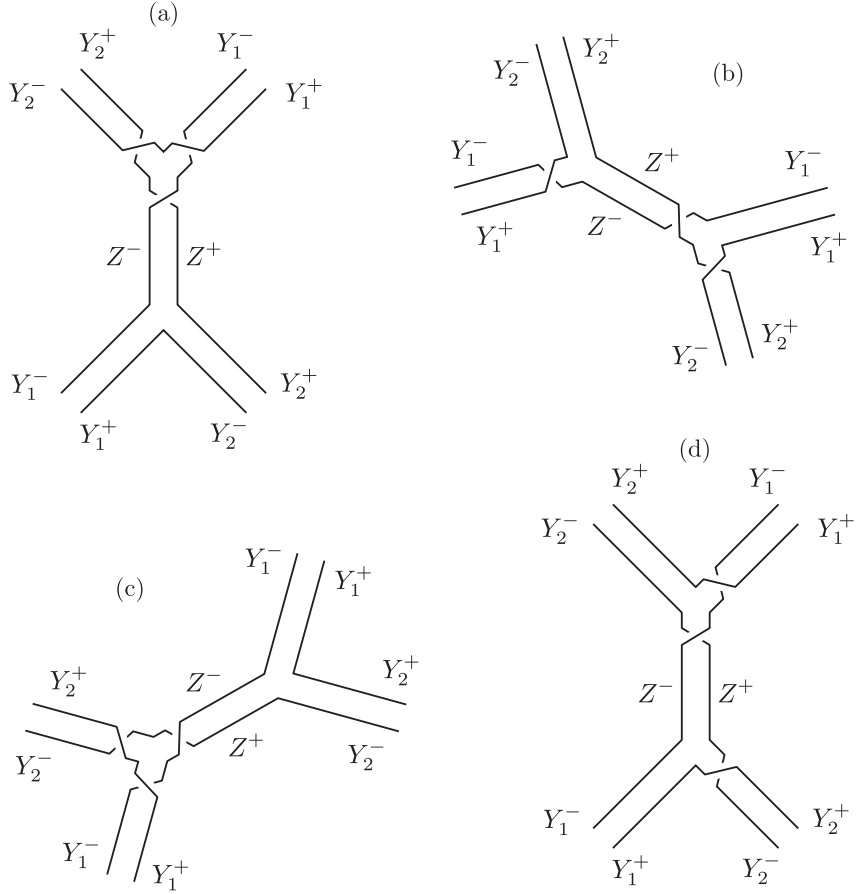}
\vspace{.0cm}
\caption{\small (Examples of) diagrams describing four simple poles $Y_{1}^{+}=Y_{2}^{-}$\,, $Y_{1}^{+}=1/Y_{2}^{-}$\,, $Y_{1}^{+}=Y_{2}^{+}$ and $Y_{1}^{+}=1/Y_{2}^{+}$\,.
They correspond to the diagrams (a)\,-\,(d) respectively.
The diagrams (a)\,-\,(c) describe physical processes, whereas (d) is not an allowed process.
}
\label{fig:simple}
\end{center}
\end{figure}

One can draw Landau diagrams corresponding to the poles $Y_{1}^{+}=Y_{2}^{-}$ and $Y_{1}^{+}=1/Y_{2}^{+}$ uniquely, which are shown in Figure \ref{fig:simple} (a) and (d), respectively.
In both cases the intermediate particle belongs to the multiplet $Q_{Z}=Q_{1}+Q_{2}$\,.
As for the rest two poles, for each $Y_{1}^{+}=1/Y_{2}^{-}$ and $Y_{1}^{+}=Y_{2}^{+}$\,, there are two diagrams possible\,;
one of them corresponds to the case where $U(1)$ charge carried by $Z^{\pm}$ is positive, while the other it is negative.
Still, in both cases the intermediate particle belongs to the multiplet $Q_{Z}=Q_{1}-Q_{2}>0$\,.
For each of these two simple poles, only one of the two possibilities is displayed in Figure \ref{fig:simple} (b) and (c), such that in (b) the $U(1)$ charge of $Z^{\pm}$ is negative, while in (c) it is positive.

The pole conditions, the multiplet number and the $U(1)$ charge of the intermediate particle $Z^{\pm}$ associated with the Landau diagrams (a)\,-\,(d) in Figure \ref{fig:simple} are summarised in Table \ref{tab:a-d}.
For example, for the boundstate formation process (a), by plugging the pole condition $Y_{1}^{+}=Y_{2}^{-}$ and the other constraints $Y_{1}^{-}=Z^{-}$\,, $Y_{2}^{+}=Z^{+}$ into (\ref{Qz}), one finds out $Q_{Z}=Q_{1}+Q_{2}$\,, and the $U(1)$ charge carried by $Z^{\pm}$ is $Q_{1}+Q_{2}$ since it is in the s-chanel.
The rest diagrams can be worked out in the same way.

\begin{table}[htbp]
\caption{\small Four simple poles and corresponding diagrams.
The first three (a)\,-\,(c) are physical process while the last (d) is not allowed.}
\begin{center}
\begin{tabular}{|c|c|c|c|c|c|}\hline
& simple pole & constraints & $Q_{Z}$ & charge of $Z^{\pm}$ & physcality \\ \hline
(a) & $Y_{1}^{+}=Y_{2}^{-}$ & $Y_{1}^{-}=Z^{-}\,,~ Y_{2}^{+}=Z^{+}$ & $Q_{1}+Q_{2}$ & $Q_{1}+Q_{2}$ & \check\\
(b) & $Y_{1}^{+}=1/Y_{2}^{-}$ & $Y_{1}^{-}=1/Z^{-}\,,~ Y_{2}^{+}=Z^{+}$ & $Q_{1}+Q_{2}$ & $Q_{1}-Q_{2}$ & \check\\
(c) & $Y_{1}^{+}=Y_{2}^{+}$ & $Y_{1}^{-}=Z^{-}\,,~ Y_{2}^{-}=Z^{+}$ & $Q_{1}-Q_{2}$ & $Q_{1}-Q_{2}$ & \check\\
(d) & $Y_{1}^{+}=1/Y_{2}^{+}$ & $Y_{1}^{-}=Z^{-}\,,~ Y_{2}^{-}=1/Z^{+}$ & $Q_{1}-Q_{2}$ & $Q_{1}+Q_{2}$ & \batsu \\ \hline
\end{tabular}
\end{center}
\label{tab:a-d}
\end{table}

We can now see that the case (d) is impossible since it corresponds to a process where the intermediate particle belongs to multiplet $Q_{1}-Q_{2}$ but has $U(1)$ charge $Q_{1}+Q_{2}$\,.
This fact indicates the simple pole $Y_{1}^{+}=Y_{2}^{+}$ is not a physical pole, which is consistent with what we found in the previous section.

\subsection{Landau diagrams for double poles}

The relevant diagrams are the ``box'' and ``bow-tie'' diagrams, 
which were studied in \cite{Dorey:2007xn} for the elementary magnon scattering case.

\subsubsection{``Box'' diagram}

There are two possibilities here concerning the charges of the intermediate states;
\begin{description}
\item[\rm Case (A)\,:]
Both intermediate particles carry positive charges.
\item[\rm Case (B)\,:]
One of them carries positive charge while the other negative.
\end{description}
We will examine both cases in turn, and see they give rise to double
poles in two 
complementary regions in the parameter space.

\paragraph{$\bullet$ Double poles in Case (A).}

\begin{figure}[tb]
\begin{center}
\vspace{.5cm}
\hspace{-.0cm}\includegraphics[scale=0.7]{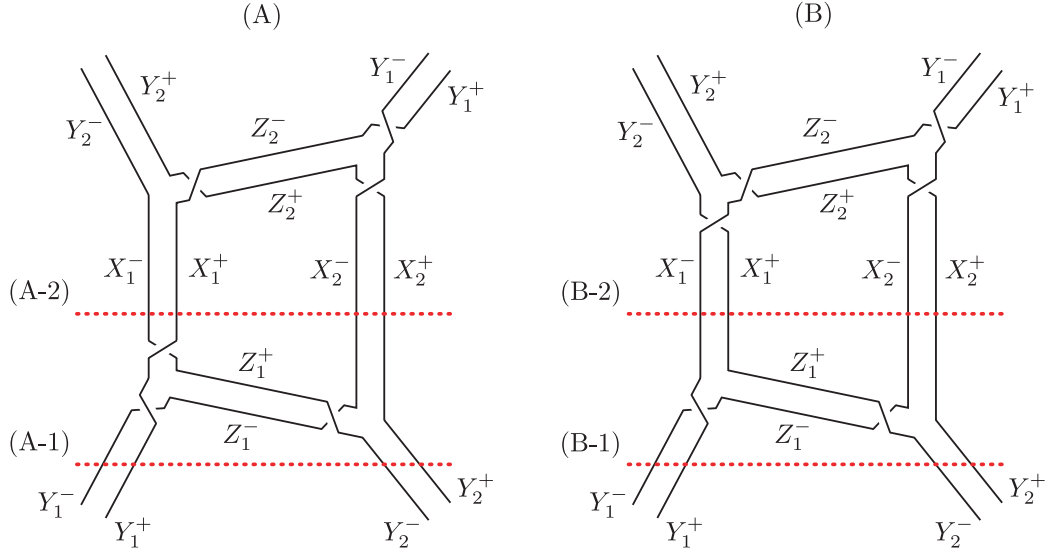}
\vspace{.0cm}
\caption{\small ``Box'' processes that give rise to double poles.
In (A), the absolute value of $X_{1}^{\pm}$ is greater than one, while in (B) it is smaller than one (see Figure \ref{fig:rap}).
In Section \ref{sec:Bethe String}, the two sets of time-slices, (A-1,2) and (B-1,2), will be interpreted as two different ways of viewing the same Bethe root configurations.
}
\label{fig:dp}
\end{center}
\end{figure}

The corresponding box diagram is shown in Figure \ref{fig:dp} (A).
We assigned spectral parameters $Y_{1}^{\pm}$ and $Y_{2}^{\pm}$ to the
two external particles, and 
$X_{1}^{\pm}$ and $X_{2}^{\pm}$ to the intermediate particles.
When the particle with $X_{1}^{\pm}$ carries positive charge $m$\,,
the two exchanged particles with spectral parameters $Z_{1}^{\pm}$ and
$Z_{2}^{\pm}$\,, carry negative charges $-(Q_{1}-m)$ and
$-(Q_{2}-m)$\,, respectively, 
in view of the charge conservation.
Here $m$ takes values $m=1,2,\dots, Q_{2}-1$ (we assumed $Q_{1}\geq Q_{2}$ as before).
Further by taking into account for the conservation of energy and
momentum at all vertices, 
one can show the spectral parameters must satisfy 
\begin{alignat}{3}
X_{2}^{+}&=Y_{2}^{+}=1/Z_{2}^{+}\,,&\qquad X_{2}^{-}&=Y_{1}^{-}=1/Z_{1}^{-}\,,\label{XYZ1}\\
X_{1}^{-}&=Y_{2}^{-}=1/Z_{1}^{+}\,,&\qquad X_{1}^{+}&=Y_{1}^{+}=1/Z_{2}^{-}\label{XYZ2}\,.
\end{alignat}
Using (\ref{XYZ1}), it is easy to verify that in this case (A) the double poles locate at
\begin{align}
U_{1}-U_{2}
&=\f{1}{Z^{+}}+Z^{+}-\f{1}{Z^{-}}-Z^{-}\cr
&=-\f{i}{2g}(Q_{1}+Q_{2}-2m)\,,\quad m=1,2,\dots, Q_{2}-1\,.
\label{u-u:1'}
\end{align}
where as before $U_{j}\eq u(Y_{j}^{\pm})$\,.
We see the number of double poles $Q_{2}-1$ is finite in this Case
(A), and the location exactly matches 
with the double poles in the BDS part of conjectured boundstate S-matrix, given in (\ref{u-u:3}).

\paragraph{$\bullet$ Double poles in region (B).}

The Figure \ref{fig:dp} (B) shows the box diagram process of Case (B),
where the particle with spextral parameter $X_{1}^{\pm}$ 
carries negative charge $-m<0$\,.
We assigned all the spectral parameters as the same as Case (A).
Then the energy and momentum conservation at all vertices imply, for one condition, the same as (\ref{XYZ1}), while on the other, 
\begin{alignat}{3}
1/X_{1}^{+}&=Y_{2}^{-}=1/Z_{1}^{+}\,,&\qquad 1/X_{1}^{-}&=Y_{1}^{+}=1/Z_{2}^{-}\label{XYZ22}
\end{alignat}
instead of (\ref{XYZ2}).
We see the spectral parameters $X_{1}^{\pm}$ in (\ref{XYZ2}) has replaced with $1/X_{1}^{\mp}$ in (\ref{XYZ22}), which is just the combination of the maps (\ref{interchange}) and (\ref{crossing}) that only flips the sign of the charge, unchanging energy and momentum.
Since now the two exchanged particles with $Z_{1}^{\pm}$ and $Z_{2}^{\pm}$ carry negative charges $-(Q_{1}+m)$ and $-(Q_{2}+m)$ respectively, the locations of the double poles become, in light of (\ref{XYZ1}), 
\begin{equation}
U_{1}-U_{2}
=-\f{i}{2g}(Q_{1}+Q_{2}+2m)\,,\quad m=1,2,\dots\,.
\label{u-u:1}
\end{equation}
This is an infinite series, and matches with the location of the double poles in the BES dressing part of conjectured 
boundstate S-matrix, given in (\ref{u-u:2}). 
The situation considered in \cite{Dorey:2007xn} corresponds to $Q_{1}=Q_{2}=1$ case.
Note also there is no double pole at
$U_{1}-U_{2}=-i(Q_{1}+Q_{2})/(2g)$\,; instead there is a single pole
there, corresponding a formation of boundstate with charge
$Q_{1}+Q_{2}$ in the s-channel process shown in Figure \ref{fig:simple} (a).

\paragraph{}
Let us summarise.
For the box diagram case, double poles are found in two separate regions for given $Q_{1}$ and $Q_{2}$ (with $Q_{1}\geq Q_{2}$), as
\begin{align}
&U_{1}-U_{2}=-\f{i}{2g}(Q_{1}+Q_{2}+2n)\,,\no\\
&\qquad \mbox{where}\quad 
n=
\left\{
\begin{array}{ll}
\ds -Q_{2}+1\,, \dots\,, -2\,, -1 & \mbox{for Case (A)}\,, \\
\ds 1\,, 2\,, \dots & \mbox{for Case (B)}\,.
\end{array}
\right.
\label{double poles}
\end{align}
The poles in (A) originate from the BDS part of the boundstate
S-matirix, whereas the ones in (B) comes from the dressing
factor. 
Each of these equations (\ref{double poles}) has two roots, and which of them corresponds to the physical double pole is summarised in Tables  \ref{tab:physical double poles 1} and \ref{tab:physical double poles 2}.

\subsubsection{``Bow-tie'' diagram}

\begin{figure}[htb]
\begin{center}
\vspace{.5cm}
\hspace{-.0cm}\includegraphics[scale=0.6]{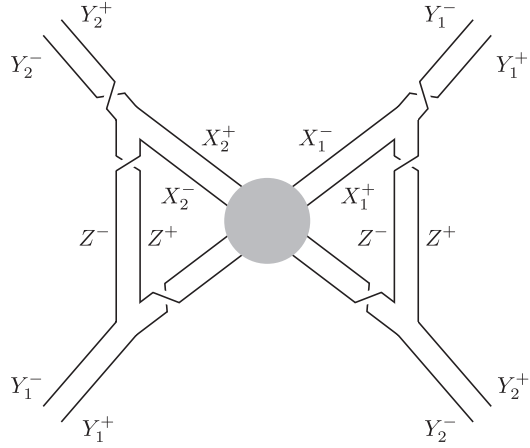}
\vspace{.0cm}
\caption{\small ``Bow-tie'' process that also rises to double poles \cite{Dorey:2007xn}.
There is a blob in the centre of the diagram, therefore special care has to be paid for the case both the intermediate plane wave magnons carry charge $Q_{1}+Q_{2}$\,.
}
\label{fig:wg}
\end{center}
\end{figure}

Generalisation of the other bow-tie shaped diagram to the boundstate scattering case is also straightforward.
Setting both the charges carried by the two intermediate giant magnons as $m$\,, the charges of plane wave magnons exchanged by the giants are $Q_{1}-m$ and $Q_{2}-m$\,, which are both negative. 
In the same manner as in the ``box'' diagram case, by using the relations
\begin{alignat}{3}
1/X_{1}^{+}&=Y_{2}^{+}=Z^{+}\,,&\qquad 1/X_{2}^{-}&=Y_{1}^{-}=Z^{-}\,,\\
1/X_{2}^{+}&=Y_{2}^{-}\,,&\qquad 1/X_{1}^{-}&=Y_{1}^{+}\,,
\end{alignat}
one can show that, setting $Q_{2}=\min\{Q_{1},Q_{2}\}$\,, the double poles apparently locate at
\begin{equation}
U_{1}-U_{2}= -\f{i}{2g}(Q_{1}+Q_{2}+2n)\,,\quad 
\mbox{with}\quad 
n\geq -Q_{2}+1\,.
\end{equation}
However, as in the case of \cite{Dorey:2007xn}, one has to take account of the effect of the blob at the centre of the diagram.
When $m=Q_{1}+Q_{2}$\,, the blob corresponds to the scattering of two anti-magnons with charges $-Q_{1}$ and $-Q_{2}$ with spectral parameters $X_{1}^{\pm}$ and $X_{2}^{\pm}$\,.
Then by considering the self-consistency condition of the diagram, the degree of the pole with $m=Q_{1}+Q_{2}$ turn out not two but one.
Therefore, again, there is a gap in the spectrum at $U_{1}-U_{2}=-i(Q_{1}+Q_{2})/(2g)$\,, and leads to the same spectrum as the ``box'' case (\ref{double poles}).

\subsubsection{Other diagrams}

We have so far been able to account for all the physical simple and double poles in the conjectured boundstate S-matrix, by finding (at least one) Landau diagrams for them.
Finally let us note that one can also 
draw lots of Landau diagrams leading to unphysical poles.
It is meaningful to note that they may or may not match the 
unphysical poles of the boundstate S-matrix.
It is also possible those diagrams which do not satisfy the physicality conditions lead to a set of double poles that coincides with the physical double poles (\ref{double poles}).
For example, one can draw ``sandglass'' shaped Landau diagrams which are obtained by rotating the bow-tie diagram by $90^{\circ}$\,.
There are many sandglass diagrams where all charges are conserved at
each vertex, but which do not satisfy the extra physicality
conditions. Some of these diagrams give rise to the same set of 
double poles as the physical ones.

\section{Bethe string interpretation\label{sec:Bethe String}}

In this section we are going to discuss how two time-slices (A-1,2) and (B-1,2) in Figure \ref{fig:dp} (the ``box'' diagram), which correspond to external and the internal on-shell states respectively, are interpreted as two different ways of viewing the same Bethe root configurations. 
In terms of the root configurations, the origins of double poles in both the BDS and the dressing pieces can be understood intuitively.

\subsubsection*{BDS part\,:~double poles from ``overlaps''}

\begin{figure}[tb]
\begin{center}
\vspace{.5cm}
\hspace{-.0cm}\includegraphics[scale=0.95]{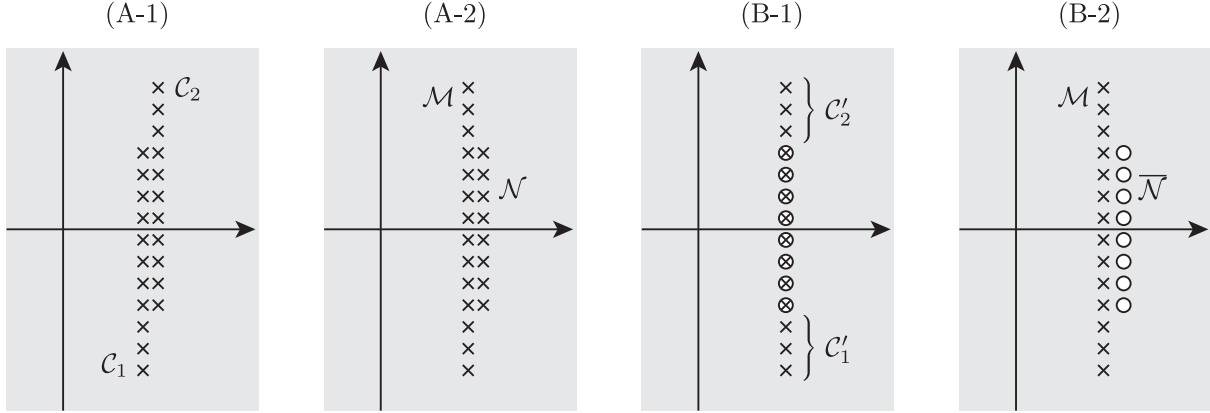}
\vspace{.0cm}
\caption{\small {[A-1]:} Bethe root configuration describing the two external particles in Figure \ref{fig:dp} (A).
\small {[A-2]:} The two intermediate BPS particles in Figure \ref{fig:dp} (A), both $\mathcal M$ and $\mathcal N$ carrying positive charges.
{[B-1]:} The two external particles in Figure \ref{fig:dp} (B).
\small {[B-2]:} The two intermediate BPS particles in Figure \ref{fig:dp} (B), where $\mathcal M$ carries positive charge while $\overline{\mathcal N}$ carries negative charge.}
\label{fig:rap}
\end{center}
\end{figure}

Let us first see the origin of the double poles (\ref{u-u:3}) in the rapidity plane.
Actually the same Bethe root configuration describing the double poles (\ref{u-u:3}) can be interpreted in two ways, each corresponding to two time-slices (A-1,2) of Figure \ref{fig:dp} (A).
The root configurations corresponding to these time-slices are shown in Figure \ref{fig:rap} (A-1,2), respectively.

\begin{figure}[tb]
\begin{center}
\vspace{.5cm}
\hspace{-.0cm}\includegraphics[scale=0.95]{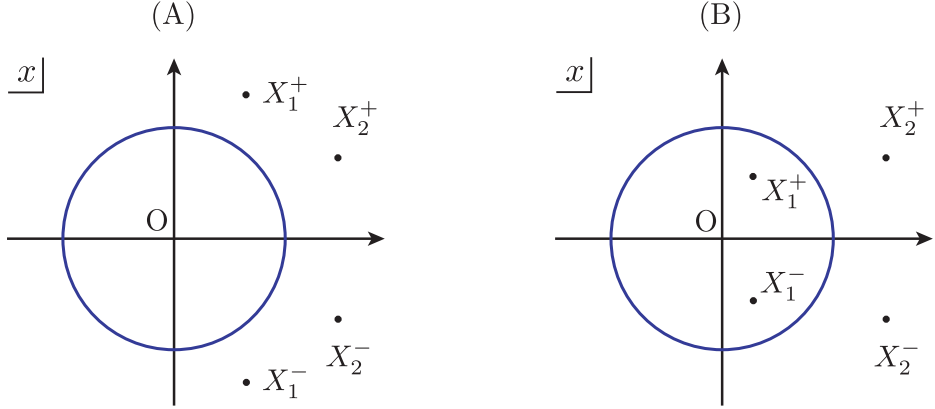}
\vspace{.0cm}
\caption{\small Locations of spectral parameters for Case (A) and (B).
When $X^{\pm}$ are outside the unit circle, the particle has positive charge, otherwise negative.}
\label{fig:spec}
\end{center}
\end{figure}

In Figure \ref{fig:rap} (A-1), the incoming particles are described by $\mathcal C_{1}(Y_{1}^{\pm})\cup \mathcal C_{2}(Y_{2}^{\pm})$\,, where
\begin{align}
{\mathcal C_{1}}(Y_{1}^{\pm};Q_{1})&=\Big\{u_{\tilde\jmath_{1}}\, \Big|\, u_{\tilde\jmath_{1}}-u_{\tilde\jmath_{1}+1}=i/g\,,
\,\,\, \tilde\jmath_{1}=\komoji{Q_{2}-m+1\,, \dots\,, Q_{1}+Q_{2}-m-1}\Big\}\,, \\
{\mathcal C_{2}}(Y_{2}^{\pm};Q_{2})&=\Big\{u_{\tilde\jmath_{2}}\, \Big|\, u_{\tilde\jmath_{2}}-u_{\tilde\jmath_{2}+1}=i/g\,,
\,\,\, \tilde\jmath_{2}=\komoji{1\,, \dots\,, Q_{2}-1}\Big\}\,.
\end{align}
Each cross $(\times)$ represents a pole of the BDS S-matrix.
There is an overlap of length $m-1$ units (one unit is of length $i/g$) running from $u_{Q_{2}-m}$ to $u_{Q_{2}-1}$\,.
One can view this configuration as physically equivalent to $\mathcal M(X_{2}^{\pm})\cup \mathcal N(X_{1}^{\pm})$ of Figure \ref{fig:rap} (A-2), where
\begin{align}
&{\mathcal M}(X_{2}^{\pm};Q_{1}+Q_{2}-m)=\Big\{u_{j_{1}}\, \Big|\, u_{j_{1}}-u_{j_{1}+1}=i/g\,,
\,\,\, j_{1}=\komoji{1\,, \dots\,, Q_{1}+Q_{2}-m-1}\Big\}\,, \\
&\mathcal N(X_{1}^{\pm};m)=\Big\{u_{j_{2}}\, \Big|\, u_{j_{2}}-u_{j_{2}+1}=i/g\,,
\,\,\, j_{2}=\komoji{Q_{2}-m+1\,, \dots\,, Q_{2}-1}\Big\} \,.
\end{align}
They correspond to the intermediate BPS particles, both carrying positive charges.
The locations of the spectral parameters $X_{1,2}^{\pm}$ are shown in Figure \ref{fig:spec} (A).
%
%
They can be expressed by the momenta and charges as
\begin{alignat}{3}
&X_{1}^{+}(P_{1},Q_{1})\eq x_{Q_{2}-m}^{+}=R_{1}\,e^{iP_{1}/2}\,,&\qquad &{}
X_{1}^{-}(P_{1},Q_{1})\eq x_{Q_{2}}^{-}=R_{1}\,e^{-iP_{1}/2}\,;\label{X1}\\
&X_{2}^{+}(P_{2},Q_{2})\eq x_{1}^{+}=R_{2}\,e^{iP_{2}/2}\,,&\qquad &{}
X_{2}^{-}(P_{2},Q_{2})\eq x_{Q_{1}+Q_{2}-m}^{-}=R_{2}\,e^{-iP_{2}/2}\,,\label{X2}
\end{alignat}
where $x=x(u)$ as in (\ref{x(u)}), and $R_{j}=R(P_{j}, Q_{j})$ as in (\ref{radius}).
In terms of these parameters, the charge, momentum and energy of $\mathcal M$ are given by $Q_{1}=Q(X_{1}^{\pm})$\,, $P_{1}=P(X_{1}^{\pm})$ and $E_{1}=E(X_{1}^{\pm})$\,, and the similar for ${\mathcal N}$\,.
For $\mathcal C_{1}$ and $\mathcal C_{2}$\,, we assign spectral parameters $Y_{1}^{\pm}$ and $Y_{2}^{\pm}$ defined by
\begin{alignat}{3}
& Y_{1}^{+}(\tilde P_{1},\tilde Q_{1})\eq x_{Q_{2}-m}^{+}\,,&\qquad &{}
Y_{1}^{-}(\tilde P_{2},\tilde Q_{2})\eq x_{Q_{1}+Q_{2}-m}^{-}\,;\label{Y1}\\
& Y_{2}^{+}(\tilde P_{1},\tilde Q_{1})\eq x_{1}^{+}\,,&\qquad &{}
Y_{2}^{-}(\tilde P_{2},\tilde Q_{2})\eq x_{Q_{2}}^{-}\,.\label{Y2}
\end{alignat}
In terms of these parameters, the charge, momentum and energy of $\mathcal C_{j}$ are given by $\tilde Q_{j}=Q(Y_{j}^{\pm})$\,, $\tilde P_{j}=P(Y_{j}^{\pm})$ and $\tilde E_{j}=E(Y_{j}^{\pm})$\,.
From the definitions (\ref{X1}\,-\,\ref{Y2}), we see the parameters $X_{1,2}^{\pm}$ and $Y_{1,2}^{\pm}$ are related as
\begin{equation}
Y_{1}^{+}=X_{1}^{+}\,,\quad 
Y_{1}^{-}=X_{2}^{-}\,,\quad 
Y_{2}^{+}=X_{2}^{+}\,,\quad 
Y_{2}^{-}=X_{1}^{-}\,.
\label{X<->Y 2}
\end{equation}
Using the relation (\ref{X<->Y 2}), one can easily check
\begin{equation}
Q_{1}+Q_{2}=\tilde Q_{1}+\tilde Q_{2}\,,\qquad 
E_{1}+E_{2}=\tilde E_{1}+\tilde E_{2}\,,\qquad 
P_{1}+P_{2}=\tilde P_{1}+\tilde P_{2}\,.
\label{identity}
\end{equation}
Therefore the assignments of spectral parameters (\ref{X1}\,-\,\ref{Y2}) are consistent with the condition that $\mathcal M\cup\mathcal N$ and $\mathcal C_{1}\cup\mathcal C_{2}$ are physically the same.
This condition is, of course, the same as (\ref{XYZ1}, \ref{XYZ2}) obtained by the physical process analysis.

%
%

\subsubsection*{Dressing part\,:~double poles from ``gaps''}

Let us see how the double poles (\ref{u-u:2}) are seen in the rapidity/spectral plane.
Just as in Case (A), one can interpret the same configuration in two different ways.
One is shown in Fugure \ref{fig:rap} (B-1), $\mathcal C'_{1}(Y_{1}^{\pm})\cup \mathcal C'_{2}(Y_{2}^{\pm})$\,, where
\begin{align}
{\mathcal C'_{1}}(Y_{1}^{\pm};Q_{1})&=\Big\{u_{\tilde\jmath_{1}}\, \Big|\, u_{\tilde\jmath_{1}}-u_{\tilde\jmath_{1}+1}=i/g\,,
\,\,\, \tilde\jmath_{1}=\komoji{Q_{2}+m+1\,, \dots\,, Q_{1}+Q_{2}+m-1}\Big\}\,, \\
{\mathcal C'_{2}}(Y_{2}^{\pm};Q_{2})&=\Big\{u_{\tilde\jmath_{2}}\, \Big|\, u_{\tilde\jmath_{2}}-u_{\tilde\jmath_{2}+1}=i/g\,,
\,\,\, \tilde\jmath_{2}=\komoji{1\,, \dots\,, Q_{2}-1}\Big\}\,,
\end{align}
with gap of length $m-1$ units running from $u_{Q_{2}+1}$ to $u_{Q_{2}+m}$\,.
They correspond to external particles in Figure \ref{fig:dp} (B), see the time-slice (B-1) of Figure \ref{fig:dp}.

The other configuration is shown in Fugure \ref{fig:rap} (B-2), $\mathcal M(X_{2}^{\pm})\cup \overline{\mathcal N}(X_{1}^{\pm})$\,, where
\begin{align}
&{\mathcal M}(X_{2}^{\pm};Q_{1}+Q_{2}+m)=\Big\{u_{j_{1}}\, \Big|\, u_{j_{1}}-u_{j_{1}+1}=i/g\,,
\,\,\, j_{1}=\komoji{1\,, \dots\,, Q_{1}+Q_{2}+m-1}\Big\}\,, \\
&\overline{\mathcal N}(X_{1}^{\pm};-m)=\Big\{u_{j_{2}}\, \Big|\, u_{j_{2}}-u_{j_{2}+1}=i/g\,,
\,\,\, j_{2}=\komoji{Q_{2}-1\,, \dots\,, Q_{2}+m-1}\Big\} \,.
\end{align}
The BPS boundstate with positive charge is described by $\mathcal M(X_{2}^{\pm})$\,, and the one with negative charge is $\overline{\mathcal N}(X_{1}^{\pm})$\,.
In Figure \ref{fig:rap} (B-2), we depict constituent magnons of $\overline{\mathcal N}$ by a white circle $(\,\circ\,)$ to distinguish it from one with positive charge ($\times$).

In this Case (B), we define the spectral parameters for $\mathcal M(X_{2}^{\pm})$ and $\overline{\mathcal N}(X_{1}^{\pm})$ as
\begin{alignat}{3}
&1/X_{1}^{+}(P_{1},Q_{1})\eq x_{Q_{2}}^{-}=R_{1}\,e^{iP_{1}/2}\,,&\qquad &{}
1/X_{1}^{-}(P_{1},Q_{1})\eq x_{Q_{2}+m}^{+}=R_{1}\,e^{-iP_{1}/2}\,;\label{X1b}\\
&X_{2}^{+}(P_{2},Q_{2})\eq x_{1}^{+}=R_{2}\,e^{iP_{2}/2}\,,&\qquad &{}
X_{2}^{-}(P_{2},Q_{2})\eq x_{Q_{1}+Q_{2}+m}^{-}=R_{2}\,e^{-iP_{2}/2}\,.\label{X2b}
\end{alignat}
Their locations are shown in Figure \ref{fig:spec} (B).
Notice $X_{2}^{\pm}$ reside inside the unit circle, reflecting the associated particle carries negative charge.
The spectral parameters for $\mathcal C'_{1}(Y_{1}^{\pm})$ and $\mathcal C'_{2}(Y_{2}^{\pm})$ are defined as
\begin{alignat}{3}
& Y_{1}^{+}(\tilde P_{1},\tilde Q_{1})\eq x_{Q_{2}+m}^{+}\,,&\qquad &{}
Y_{1}^{-}(\tilde P_{2},\tilde Q_{2})\eq x_{Q_{1}+Q_{2}+m}^{-}\,;\label{Y1b}\\
& Y_{2}^{+}(\tilde P_{1},\tilde Q_{1})\eq x_{1}^{+}\,,&\qquad &{}
Y_{2}^{-}(\tilde P_{2},\tilde Q_{2})\eq x_{Q_{2}}^{-}\,.\label{Y2b}
\end{alignat}
With the assignments of spectral parameters (\ref{X1b}\,-\,\ref{Y2b}), we see the parameters $X_{1,2}^{\pm}$ and $Y_{1,2}^{\pm}$ are now related as, contrast to (\ref{X<->Y 2}) of Case (A),
\begin{equation}
Y_{1}^{+}=1/X_{1}^{-}\,,\quad 
Y_{1}^{-}=X_{2}^{-}\,,\quad 
Y_{2}^{+}=X_{2}^{+}\,,\quad 
Y_{2}^{-}=1/X_{1}^{+}\,.
\end{equation}
This is of course consistent with (\ref{XYZ1}, \ref{XYZ22}).
Using this relation, one can again verify the same conservation conditions as (\ref{identity}).
In summary, the infinitely many possible lengths of the gap between $\mathcal C'_{1}$ and $\mathcal C'_{2}$ (or in other words, the number of roots in $\overline{\mathcal N}$) correspond to infinitely many double poles (\ref{u-u:2}) in the BES phase.

\subsubsection*{Dispersion relation}

Let us see how the dispersion relations for the configurations $\mathcal M\cup\mathcal N$${}={}$$\mathcal C_{1}\cup \mathcal C_{2}$ (Case (A)) and $\mathcal M\cup\overline{\mathcal N}$${}={}$$\mathcal C'_{1}\cup \mathcal C'_{2}$ (Case (B)) look like, in terms of $Q_{j}$\,, $P_{j}$ and $R_{j}$\,.
In both cases, the the total charge and energy are given by the same expressions,
\begin{align}
Q_{1}+Q_{2}&=2g\kko{\Big(R_{1}-\f{1}{R_{1}}\Big)\sin\Big(\f{P_{1}}{2}\Big)+\Big(R_{2}-\f{1}{R_{2}}\Big)\sin\Big(\f{P_{2}}{2}\Big)}\,,\\
E	&=2g\kko{\Big(R_{1}+\f{1}{R_{1}}\Big)\sin\Big(\f{P_{1}}{2}\Big)+\Big(R_{2}+\f{1}{R_{2}}\Big)\sin\Big(\f{P_{2}}{2}\Big)}\,,
\end{align}
and the dispersion relation becomes
\begin{align}
E=
	\sqrt{\ko{Q_{1}+Q_{2}}^{2}+16g^{2}
	\ko{\sin\Big(\f{P_{1}}{2}\Big)+\rho\sin\Big(\f{P_{2}}{2}\Big)}
	\ko{\sin\Big(\f{P_{1}}{2}\Big)+\f{1}{\rho}\sin\Big(\f{P_{2}}{2}\Big)}}\,,
\label{dispersion}
\end{align}
where we defined $\rho\eq R_{1}/R_{2}$\,.
A special case $P_{1}=P_{2}$ reproduces the result obtained in \cite{Spradlin:2006wk} (as a ``boundstate'' of two dyonic giant magnons) after setting $\rho=e^{q}$\,.
Notice that (\ref{dispersion}) can be also expressed as a sum of two BPS particles with positive energies,
\begin{equation}
E=\sqrt{Q_{1}^{2}+16g^{2}\sin\Big(\f{P_{1}}{2}\Big)}+\sqrt{Q_{2}^{2}+16g^{2}\sin\Big(\f{P_{2}}{2}\Big)}\,.
\end{equation}

\paragraph{}
Concerning Case (B), the analysis made in this section gives a support to the observation made in \cite{Dorey:2007xn} from a `quantised' point of view in the following sense.
If we only work in $\mathbb R\times S^{2}$ sector of the theory, the string solution obtained from a breather solution of sine-Gordon equation might seem like a non-BPS boundstate (as was indeed the case when they first appeared in \cite{Hofman:2006xt}), which is absent in the BPS spectrum.
However, it was correctly understood in \cite{Dorey:2007xn} that the ``breathing'' solutions can be and should be interpreted as, once embedded into a larger subspace $\mathbb R\times S^{3}$\,, a superposition of two BPS boundstates with opposite signs for $J_{2}$\,-charge.
This is indeed the picture we have obtained for Case (B);
If we only work in $SU(2)$ sector, the configuration $\mathcal C'_{1}\cup \mathcal C'_{2}$ (Figure \ref{fig:rap} [B-1]) corresponds to a non-BPS state which we cannot find in the BPS spectrum (\ref{E}), but once we enlarge the sector from $SU(2)$ to $SU(2)\times \overline{SU(2)}$ (which is the same symmetry as the isometry of the $S^{3}$ of string theory), one can view it as a superposition of two BPS boundstates, {\em i.e.}, $\mathcal M$ and $\overline{\mathcal N}$\,, each of which living in a different $SU(2)$ sector (Figure \ref{fig:rap} [B-2]).

\begin{figure}[htb]
\begin{center}
\vspace{.5cm}
\hspace{-.0cm}\includegraphics[scale=0.85]{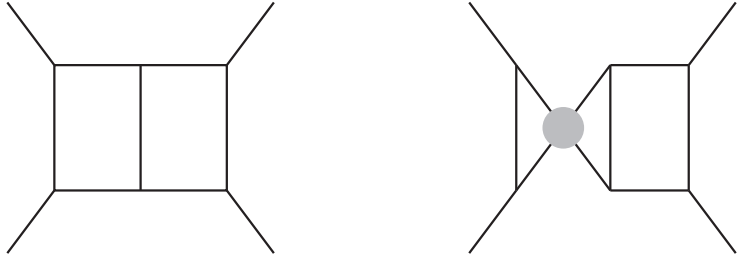}
\vspace{.0cm}
\caption{\small Examples of Landau diagrams that give rise to triple poles.}
\label{fig:tp}
\end{center}
\end{figure}

\paragraph{}
We have investigated the singularities of the
bound-state S-matrix which lie near the physical region of real,
postive energy and found a physical explanation for each of them in
terms of on-shell intermediate states. This
is further evidence in favour of the conjectured spectrum and S-matrix
of the ${\cal N}=4$ SYM spin-chain but is by no means a conclusive 
test. There are several ways in which our analysis could be made more
comprehensive. First it would be interesting to extend our analysis to 
triple or higher-order poles. The conjectured S-matrix does not appear
to have such singularities. However, one certainly can draw Landau
diagrams which seem to correspond to triple poles, 
and some of them are shown in Figure \ref{fig:tp}. For consistency 
each of these diagrams must represent an unphysical process for some
reason or cancel in some other way but this remains to be checked. 

Although, we have succeeded in determining the locations of 
physical poles, we have not determined their residues. As in a
relativistic theory, there may be additional constraints on these
residues for physical intermediate states. Finally, it would be nice
to confirm some of the singularity structure we have described here by
explicit calculations either in gauge theory or on the string
worldsheet.

\section[Appendix for Chapter \ref{chap:Singularities}]{Appendix for Chapter \ref{chap:Singularities}\,: Breathing magnons\label{sec:BM}}

As we saw in Chapter \ref{chap:DGM}, the string $O(4)$ sigma model with Virasoro constraints is classically equivalent to CsG model, and it is often useful to exploit this connection with CsG theory to construct classical strings.
Indeed, the Pohlmeyer reduction procedure has been efficiently utilised in the construction of various string solutions, see {\em e.g.}, \cite{Chen:2006ge, Okamura:2006zv, Hayashi:2007bq, Jevicki:2007aa}.
Our aim here is to realise the oscillating solutions of \cite{Hofman:2006xt, Spradlin:2006wk} from the standpoint of (C)sG solitons.
They correspond to Case (B) of the box diagram discussed in the main text.
In particular, we will concentrate on the $|Q_{1}-Q_{2}|=2$ case.

\subsection{``Minimal'' oscillating solutions from CsG solitons\,...}

The Complex sine-Gordon equation is given by
\begin{equation}
\pa_{+}\pa_{-}\psi +\psi^{*}\,\f{\pa_{+}\psi\pa_{-}\psi}{1-|\psi|^{2}}+\psi (1-|\psi|^{2})=0\,,
\end{equation}
where $\psi(t,x)$ is a complex field and $\pa_{\pm}=\pa_{t}\pm\pa_{x}$ with $(t,x)$ rescaled worldsheet variables $-\infty<t<\infty$ and $-\infty<x<\infty$\,.
It has kink soliton solutions of the form
\begin{equation}
\psi_{\rm K}(t,x)= \f{(\cos\al)\exp\kko{i(\sin\al)\ko{\cosh\Th\cdot t-\sinh\Th\cdot x}}}{\cosh\kko{(\cos\al)\ko{\cosh\Th\cdot x-\sinh\Th\cdot t}}}\,.
\end{equation}
They can be mapped to dyonic giant magnons via the Pohlmeyer reduction procedure, and they reproduce the dispersion relation for magnon boundstates under identifications (see (\ref{Eithetaalpha}) and (\ref{Qithetaalpha}))
\begin{align}
E=\f{4g\cos\al\cosh\Theta}{\cos^{2}\al+\sinh^{2}\Theta}\,,\qquad 
Q=\f{4g\cos\al\sin\al}{\cos^{2}\al+\sinh^{2}\Theta}\,.
\label{E,Q}
\end{align}
On string theory side, they represent the energies and the second spins of dyonic giant magnons, while on gauge theory side, they represent $\Delta-J_{1}$ and the number of constituent $SU(2)$ magnons in the boundstate, respectively.
The spectral parameters of the $Q$\,-magnon boundstates are expressed in terms of the CsG parameters as (see (\ref{X<->CsG})) \cite{Chen:2006gq}
\begin{equation}
X_{j}^{\pm}=\coth\kko{\f{\Th_{j}}{2}\pm i\ko{\f{\al_{j}}{2}-\f{\pi}{4}}}
=\f{\sinh\Th_{j}\pm i\cos\al_{j}}{\cosh\Th_{j}-\sin\al_{j}}\,.
\label{Lambda}
\end{equation}
Recall the parametrisation $X_{j}^{\pm}=R_{j}\,e^{iP_{j}/2}$ introduces before, then the above dictionary tells
\begin{equation}
R_{j}=\sqrt{\f{\cosh\Th_{j}+\sin\al_{j}}{\cosh\Th_{j}-\sin\al_{j}}}\,,\qquad 
\cot\ko{\f{P_{j}}{2}}=\f{\sinh\Th_{j}}{\cos\al_{j}}\,.
\label{R,P}
\end{equation}
Since we are interested in (C)sG description of the oscillating solutions, which are made up of two magnon boundstates with opposite charges, in view of (\ref{E,Q}) we should start with two CsG kinks $j=1,2$ having opposite signs for rotational parameters $\al_{j}$\,.
For simplicity, we restrict our analysis to the ``minimal'' case, $|Q_{1}-Q_{2}|=2$\,, where the length of 
corresponding boundstates $\mathcal M$ and $\overline{\mathcal N}$ differ only by two units.
In this case, in the first approximation in large-$g$\,, the two rotation parameters add up to zero, $\al_{1}=-\al_{2}$\,, and this condition implies the relation between radii as $R_{1}=1/R_{2}$ due to (\ref{R,P}).
Combining this with the condition that the string-centres of $\mathcal M$ and $\overline{\mathcal N}$ coincide, which reads in terms of CsG parameters
\begin{equation}
\f{\sinh(2\Th_{1})}{\cosh(2\Th_{1})+\cos(2\al_{1})}
=\f{\sinh(2\Th_{2})}{\cosh(2\Th_{2})+\cos(2\al_{2})}\,,
\label{U=U' in CsG}
\end{equation}
it follows that $\tanh\Th_{1}=\tanh\Th_{2}$ and $P_{1}=-P_{2}$\,.
In the CsG context, this means two kinks are moving in the same direction with the same velocity.
For notational simplicity, for this ``minimal'' case, we will here set as $R\eq R_{1}=1/R_{2}$\,, $P\eq P_{1}=-P_{2}$\,, $\al\eq \al_{1}=-\al_{2}$ and $\Th_{0}\eq \Th_{1}=\Th_{2}$\,.
The dispersion relation for this solution is then given by
\begin{equation}
E=4g\ko{R+\f{1}{R}}\sin\ko{\f{P}{2}}
\quad \mbox{with}\quad 
R=\sqrt{\f{\cosh\Th_{0}+\sin\al}{\cosh\Th_{0}-\sin\al}}\,,\quad 
\cot\ko{\f{P}{2}}=\f{\sinh\Th_{0}}{\cos\al}\,.
\label{special BM}
\end{equation}
This is the dispersion relation for the minimal oscillating string.
If we set $R=e^{q/2}$\,, it agrees with the dispersion relation obtained in \cite{Spradlin:2006wk} by the dressing method. 

It will become clear in the next section that under proper identification of parameters, the sine-Gordon breathers can be identified with the $\al_{1}=-\al_{2}$ case of CsG kink-kink scattering solutions we have examined, thus also corresponding to the same oscillating string with dispersion relation (\ref{special BM}).

\subsection{...\,and from sG breathers}

\begin{figure}[tb]
\begin{center}
\vspace{.5cm}
\hspace{-.0cm}\includegraphics[scale=0.76]{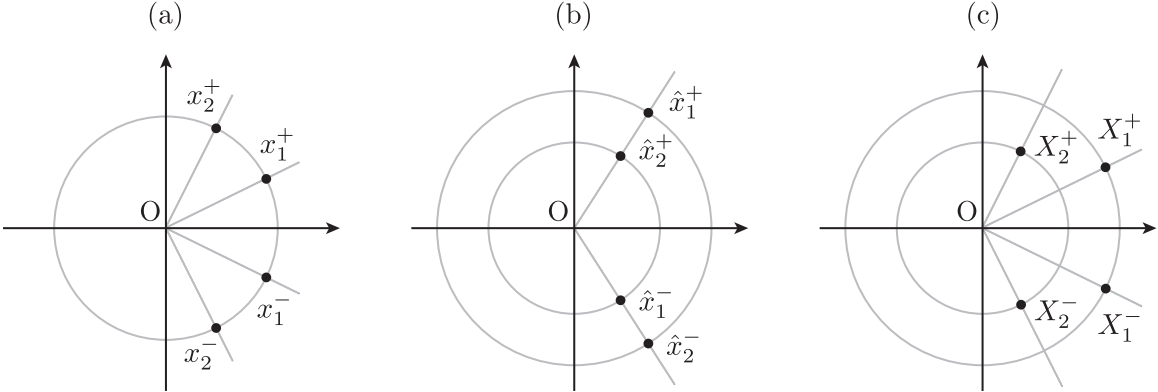}
\vspace{.0cm}
\caption{\small (a): A generic kink-antikink scattering solution of sine-Gordon equation.
(b): A breather solution.
Also describing a special case of CsG kink-kink scattering solution $\al_{1}=-\al_{2}$\,.
(c): A generic CsG kink-kink scattering solution.}
\label{fig:rapidity}
\end{center}
\end{figure}

Let us now turn to sine-Gordon (not ``Complex'') theory to see how this special case of the minimal oscillating string emerges from the sine-Gordon point of view.
The classical sine-Gordon equation,
\begin{equation}
\pa_{+}\pa_{-}\phi-\sin\phi=0\,,
\label{sG}
\end{equation}
has two types of finite-energy solutions.
One is a soliton, which is time-independent and topologically non-trivial solution.
The other is a breather, which is time-dependent and topologically trivial solution, and it can be viewed as a boundstate of a kink and an antikink oscillating in and out, namely, breathing.

\paragraph{}
Let us begin with sG kink-antikink scattering solution.
It is given by
\begin{equation}
\phi_{\rm K\overline{K}}(t,x)=2\arctan\kko{
\f{1}{\tanh\theta}
\f{\sinh\ko{\sinh\theta\cosh\theta_{0}\ko{t-\tanh\theta_{0}\cdot x}}}{\cosh\ko{\cosh\theta\cosh\theta_{0}\ko{x-\tanh\theta_{0}\cdot t}}}
}\,,
\label{kink-antikink}
\end{equation}
where the kink has velocity $\tanh\ko{\theta_{0}+\theta}$ and the antikink has $\tanh\ko{\theta_{0}-\theta}$\,.
Here $\tanh\theta_{0}$ is the velocity of the centre-of-mass and $\tanh\theta$ is the relative velocity.
It is convenient to introduce complex spectral parameters
\begin{equation}
x_{j}^{+}=\f{e^{\th_{j}}+i}{e^{\th_{j}}-i}\,,\qquad
x_{j}^{-}=\f{e^{\th_{j}}-i}{e^{\th_{j}}+i}
\end{equation}
with $\th_{1}=\th_{0}+\th$ and $\th_{2}=\th_{0}-\th$\,.
The parameters $x_{j}^{\pm}$ are located on a unit circle in the complex plane and satisfy reality conditions $x_{j}^{+}=(x_{j}^{-})^{*}$\,, see Figure \ref{fig:rapidity} (a).
We also introduce another parametrisation $x_{j}^{\pm}=e^{\pm ip_{j}/2}$\,.
This way of parametrisation is useful in discussing corresponding string solution and also its counterpart in gage theory.
In view of classical-string/sG dictionary, in the limit $t\to \pm\infty$\,, the profile (\ref{kink-antikink}) corresponds to two giant magnons having angular differences $p_{1}$ and $p_{2}$ between their endpoints.
They in turn correspond to two isolated magnons in an asymptotic SYM spin-chain, each of which having quasi-momenta $p_{1}$ and $p_{2}$\,, respectively.
Note also the relation between $p_{j}$ and $\th_{j}$ are the same as that of $\al_{j}\to 0$ limit of CsG case, see (\ref{R,P}).

\paragraph{}
The sG breather solution can be obtained as an analytic continuation of a kink-antikink scattering solution (\ref{kink-antikink}).
By setting $\theta= i\hat\theta$ in (\ref{kink-antikink}), we obtain
\begin{equation}
\phi_{\rm B}(t,x)=2\arctan\kko{
\f{1}{\tan\hat\theta}
\f{\sin\ko{\sin\hat\theta\cosh\theta_{0}\ko{t-\tanh\theta_{0}\cdot x}}}{\cosh\ko{\cos\hat\theta\cosh\theta_{0}\ko{x-\tanh\theta_{0}\cdot t}}}}\,.
\label{breather}
\end{equation}
The solution (\ref{breather}) represents a breather that is moving with velocity $\tanh\th_{0}$ and oscillating with frequency $f=\sin\hat\th/(2\pi \cosh\th_{0})$\,.
The kink and antikink have complex conjugate velocities $\hat v_{1}=\tanh(\th_{0}+\th)$ and $\hat v_{2}=\tanh(\th_{0}-\th)$\,.
It is again convenient to introduce parametrisations $\hat v_{1}=\cos(\hat p_{1}/2)$ and $\hat v_{2}=\cos(\hat p_{2}/2)$ with $\hat p_{1}=p-iq$ and $\hat p_{2}=p+iq$\,.
These two ways of parametrising the velocities are related through the relations
\begin{equation}
\tan\ko{\f{p}{2}}=\f{\cos\hat\th}{\sinh\th_{0}}\,,\qquad 
\tanh\ko{\f{q}{2}}=\f{\sin\hat\th}{\cosh\th_{0}}\,.
\label{p,q}
\end{equation}
In the kink-antikink scattering case, the rapidities $x_{1}^{\pm}$ (for the kink) and $x_{2}^{\pm}$ (for the antikink) satisfied $x_{j}^{+}=(x_{j}^{-})^{*}$\,, which meant both the kink and antikink are physical particles.
In the current breather case, after the analytic continuation, the rapidities become
\begin{align}
\hat x_{1}^{\pm}\eq e^{\pm i\hat p_{1}/2}&=e^{\pm q/2} e^{\pm ip/2}=\f{e^{\th_{0}} e^{i\hat\th}\pm i}{e^{\th_{0}} e^{i\hat\th}\mp i}\,,\\[2mm]
\hat x_{2}^{\pm}\eq e^{\pm i\hat p_{2}/2}&=e^{\mp q/2} e^{\pm ip/2}=\f{e^{\th_{0}} e^{-i\hat\th}\pm i}{e^{\th_{0}} e^{-i\hat\th}\mp i}\,.
\end{align}
This is shown in Figure \ref{fig:rapidity} (b).
In this case we have $\hat x_{1}^{+}=(\hat x_{2}^{-})^{*}$ and $\hat x_{1}^{-}=(\hat x_{2}^{+})^{*}$\,, which means particle $1$ and $2$ are no longer physical particles but instead $\hat x_{1}^{+}\mbox{\,-\,}\hat x_{2}^{-}$ pair (which we call particle $1'$) and $\hat x_{1}^{-}\mbox{\,-\,}\hat x_{2}^{+}$ pair (particle $2'$) represent physical particles.
Actually, these particles $1'$ and $2'$ can be identified with the CsG kinks with $\al_{1}=-\al_{2}$\,, so that they form a minimal breathing solution on the string theory side.
Explicitly, all physical constraints turn out to be exactly equivalent under identification $\th_{0}\eq\Th_{0}$ and $\hat\th\eq\al$\,.
If we define $\hat r_{1'}=|\hat x_{1}^{+}|=|\hat x_{2}^{-}|=e^{q/2}$ and $\hat r_{2'}=|\hat x_{1}^{-}|=|\hat x_{2}^{+}|=e^{-q/2}$\,, then one can also check $R_{1}=\hat r_{1'}$ and $R_{2}=\hat r_{2'}$ in this case.

\begin{figure}[thb]
\begin{center}
\vspace{.3cm}
\includegraphics[scale=0.9]{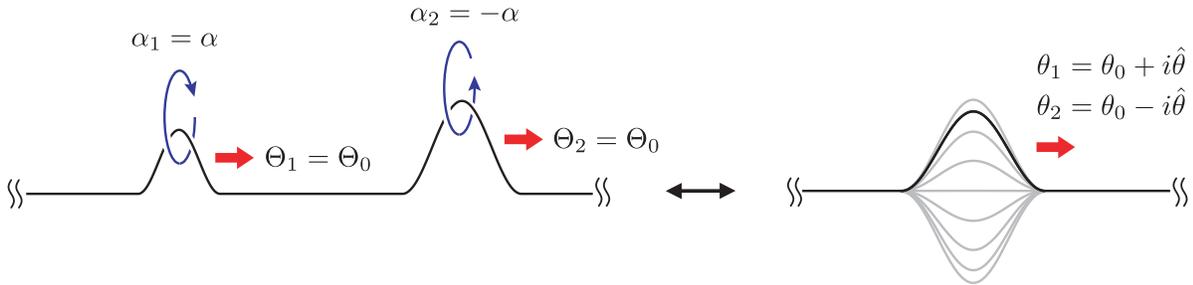}
\vspace{.3cm}
\caption{\small The ``minimal'' breathing magnon can be obtained either from a breather solution in sG theory or from a kink-kink soliton solution CsG theory, which are identical under the identification $\theta_{0}=\Theta_{0}$ and $\hat\theta=\al$\,.}
\label{fig:CsG_minimal.eps}
\end{center}
\end{figure}

\paragraph{}
In view of the second equality in (\ref{p,q}), the period of the oscillation is also expressed as $T=1/f=2\pi/\tanh(q/2)$\,.
This is different from the period of rotation by the factor $\tanh(q/2)$\,; these two kinds of periods agree only in the limit $q\to \infty$\,.
As a result, while the location of endpoints of a string on the equator are the same between $t=0$ and $t=T$\,, the shape of the string itself are not in general.

As was done in \cite{Hofman:2006xt}, one can relate the $q$ parameter which controls the period of breathing to the oscillation number $n$ of breather solution.
The energy of elementary magnon is given by the formula $\ep_{j}=4g\sin(\hat p_{j}/2)$\,, which is the large-$g$ limit of (\ref{Delta}) (or large-$g$ limit of (\ref{E}) with $Q=1$).
Each constituent magnon has complex energy, but since they are complex conjugate to each other, they sum up to give a real energy for the minimal breathing solution, 
$\ep_{\rm BM}=\ep_{1}+\ep_{2}$\,.
We can then define the oscillation number as the action variable associated with the breathing,
\begin{equation}
n=\int \f{T}{2\pi}\,d\ep_{\rm BM}
=8g \sin\ko{\f{p}{2}}\sinh\ko{\f{q}{2}}\,.
\end{equation}
Large $n$ thus means large value of $q$ for fixed $p$\,.
As noted before, in the limit $n\to \infty$ or $q\to \infty$\,, the two kinds of periods, the oscillation period and the rotation period become identical (both are equal to $2\pi$).
This means that when $p$ is near to $\pi$\,, the string looks like no more rotating but rather pulsating; staring from one point on the equator and sweeps whole the sphere and shrinks into its antipodal points, and again back to the original point by time reversal motion with only the orientation changed.

\paragraph{}
Let us consider the limit $q\to \infty$ ($n\to \infty$) with $p=\pi$\,.
In this case we can write the SYM dual for the minimal breathing magnon as, schematically, $\mathcal O \sim \cZ^{J_{1}/2}\, \cW\, \overline{\cZ}{}^{J_{1}/2}\, \overline{\cW}+\dots $ with $J_{1}\to \infty$ (note that $n$ essentially counts the number of $\overline{\cZ}$).
In this ``breathing'' operator, the infinite number of $\cZ$ fields correspond to the two coincident endpoints of the string which moves along the equator at the speed of light.
The infinite number of $\overline{\cZ}$ fields represent another intersecting point of the string with the equator that also moves along the equator at the speed of light, only in the opposite direction to the $\cZ$s.
The $\cW$ (resp.~$\overline{\cW}$) stands for the giant magnon on the upper (resp.~lower) hemisphere of the $S^{2}$\,.
Contrast to the scattering case, in this breathing case, periodic (trivial) scatterings between $\cW$ and $\overline{\cW}$ take place with period $2\pi$\,, like
\begin{equation}
\dots \cZ\,\, \cW\,\, \overline{\cZ}\dots \overline{\cZ}\,\, \overline{\cW}\,\, \cZ\dots
\qquad 
\longleftrightarrow
\qquad 
\dots \cZ\,\, \overline{\cW}\,\, \overline{\cZ}\dots \overline{\cZ}\,\, \cW\,\, \cZ\dots  ~~.
\end{equation}
Or more generally, one can argue that the breathing magnons can be thought of SYM operators of the form $\mathcal O \sim \cZ^{J_{1}/2}\, \cW^{Q_{1}}\, \overline{\cZ}{}^{J'_{1}/2}\, \overline{\cW}{}^{Q_{2}}+\dots$ with $J_{1}\to \infty$\,, where $Q_{1}$\,, $Q_{2}$ and $J'_{1}$ are not fixed number but vary because of the decay of $\cZ\,\overline{\cZ}$ into some singlet composites of $\cN=4$ SYM fields.

\begin{figure}[htbp]
\begin{center}
\vspace{-1.5cm}
\includegraphics{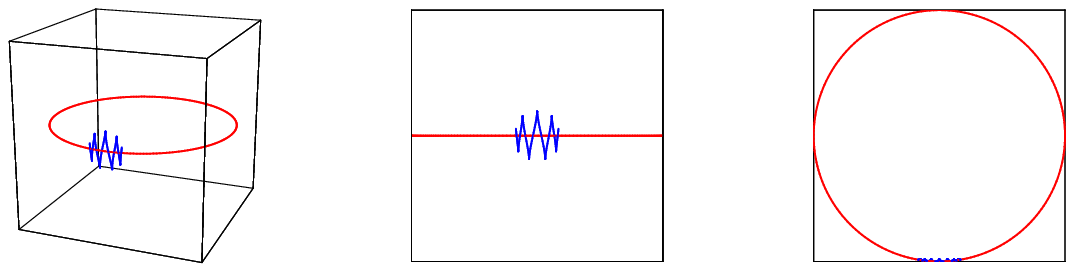}
\vspace{0.0cm}\hspace{0.3cm}\includegraphics{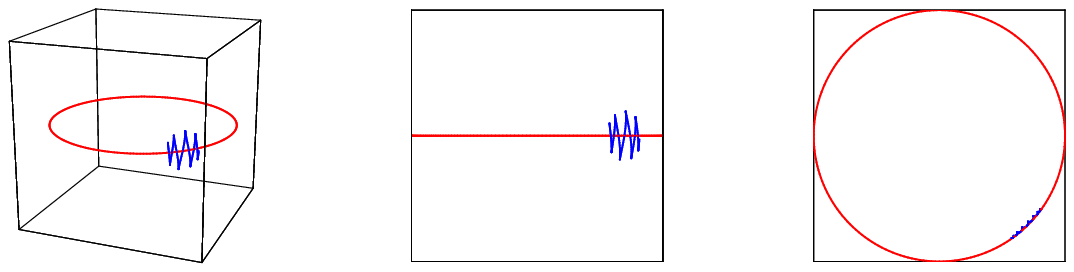}
\vspace{0.0cm}\hspace{0.3cm}\includegraphics{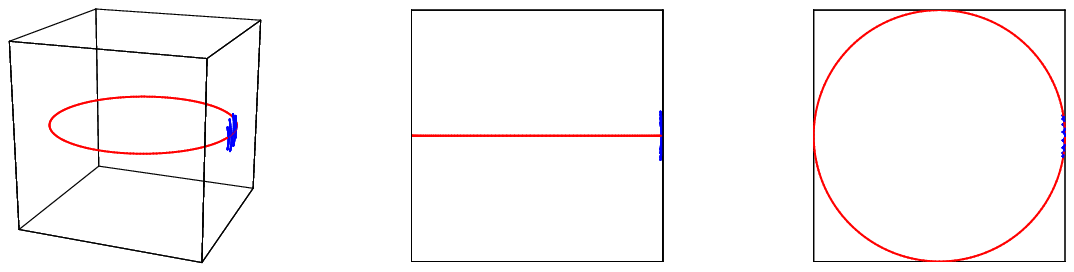}
\vspace{0.0cm}\hspace{0.3cm}\includegraphics{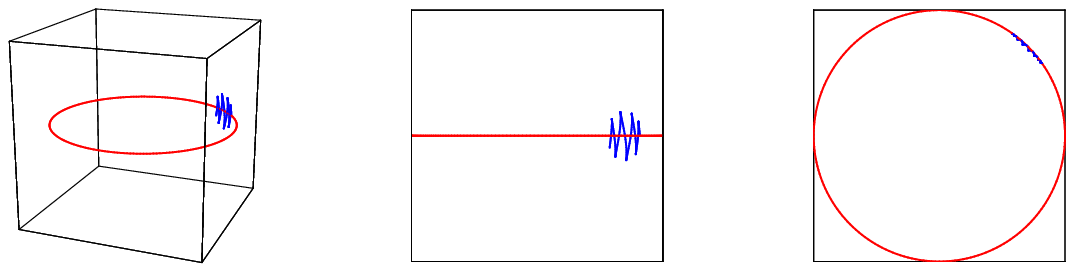}
\vspace{1.0cm}

\includegraphics{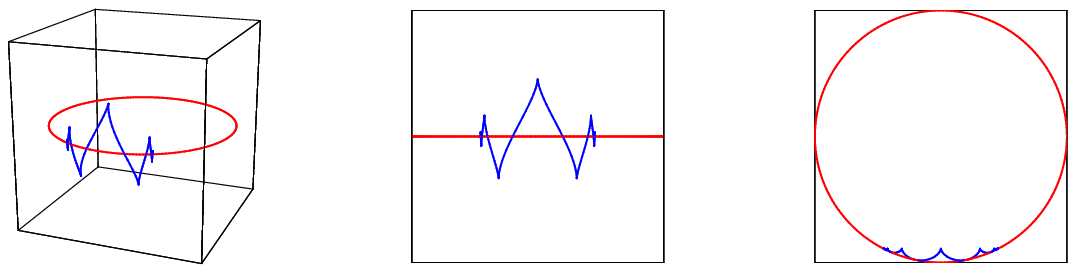}
\includegraphics{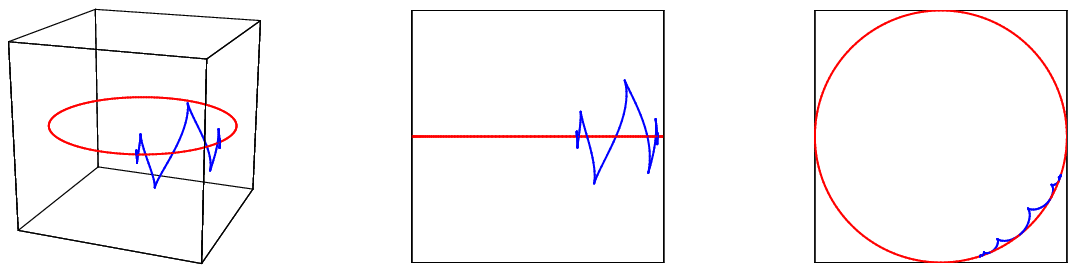}
\includegraphics{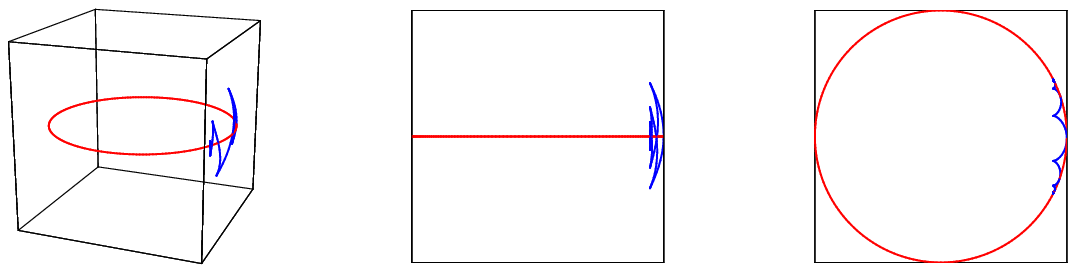}
\includegraphics{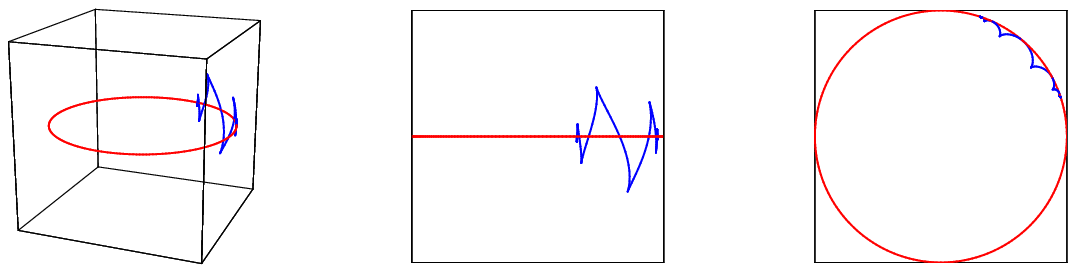}
\caption{\small Breathing magnon with $p=0.0625\,\pi$ (top four) and $p=0.15\,\pi$ (bottom four).}
\label{fig:middle}
\end{center}
\end{figure}

\begin{figure}[htbp]
\begin{center}
\vspace{-1.0cm}
\includegraphics{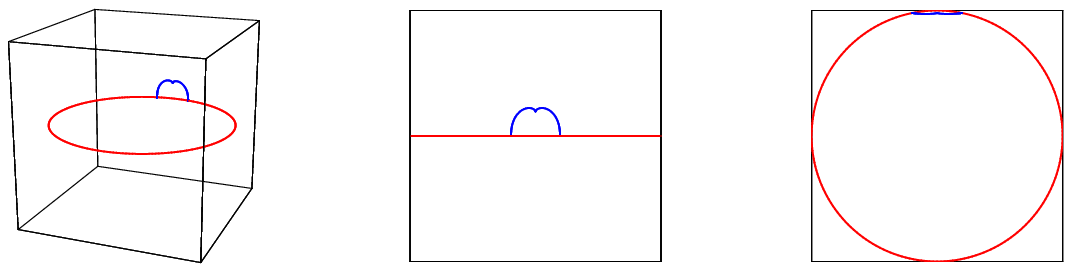}
\includegraphics{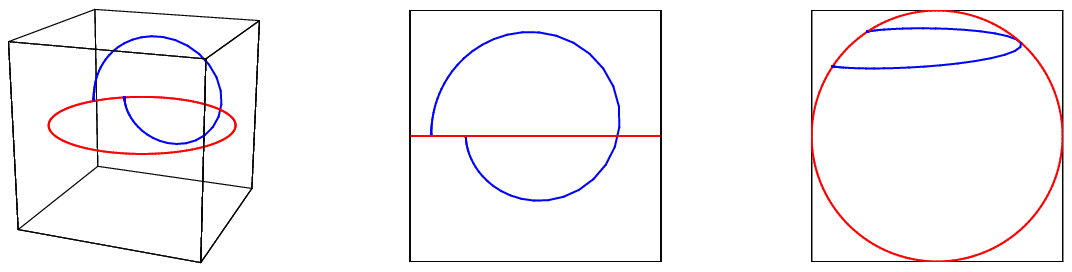}
\includegraphics{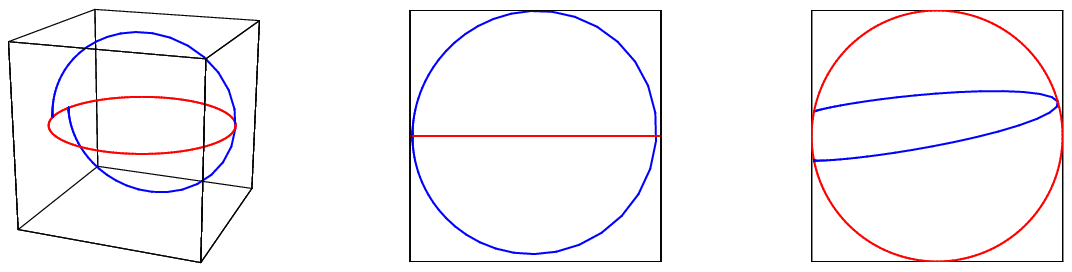}
\includegraphics{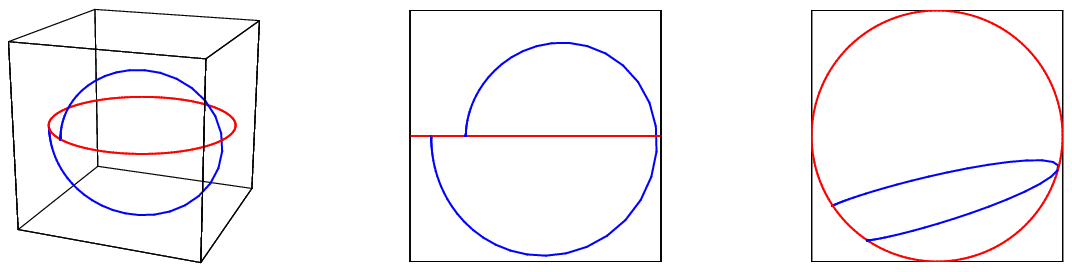}
\includegraphics{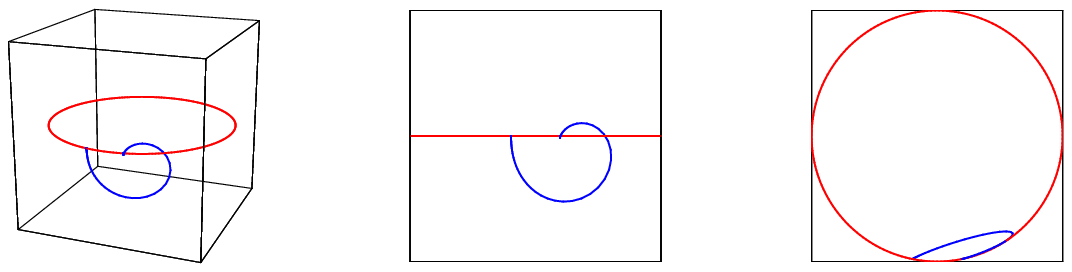}
\includegraphics{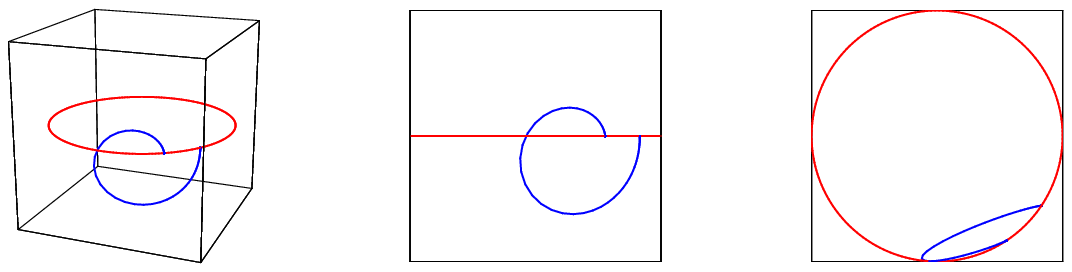}
\includegraphics{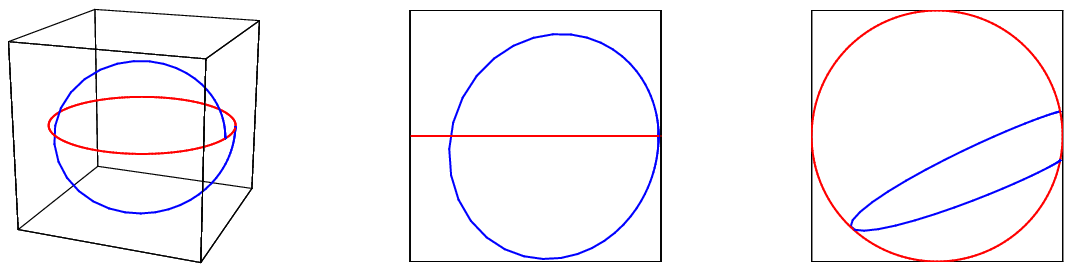}
\includegraphics{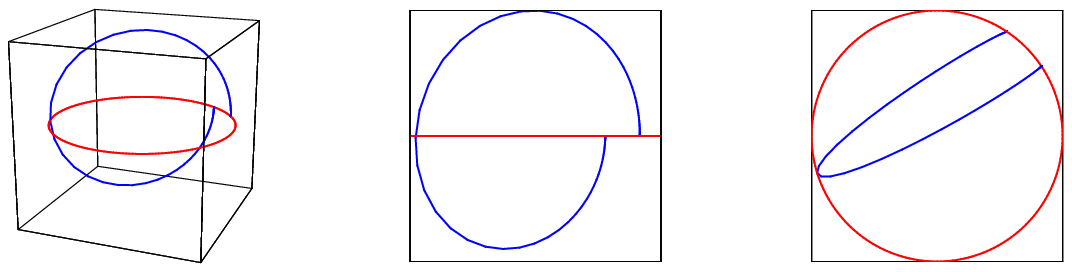}
\caption{\small Breathing magnon with $p=0.9375\,\pi$\,.}
\label{fig:nearly-pi}
\end{center}
\end{figure}

\begin{figure}[htbp]
\begin{center}
\vspace{-1.0cm}
\includegraphics{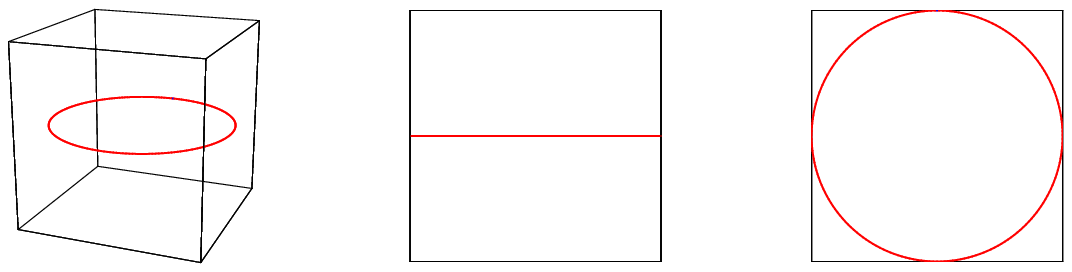}
\includegraphics{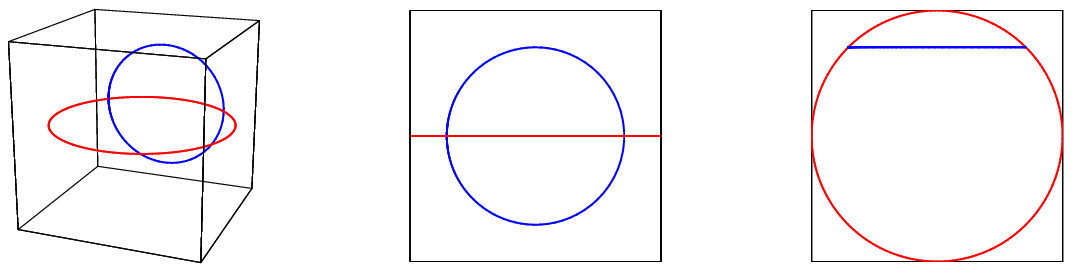}
\includegraphics{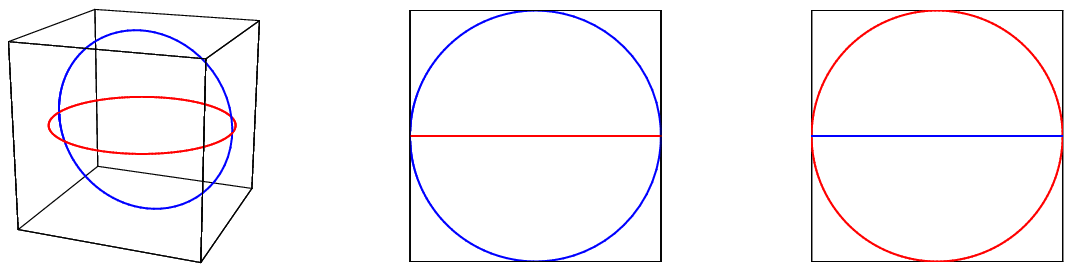}
\includegraphics{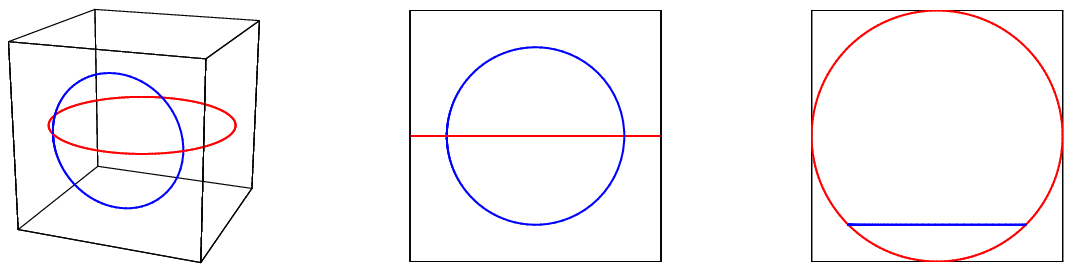}
\includegraphics{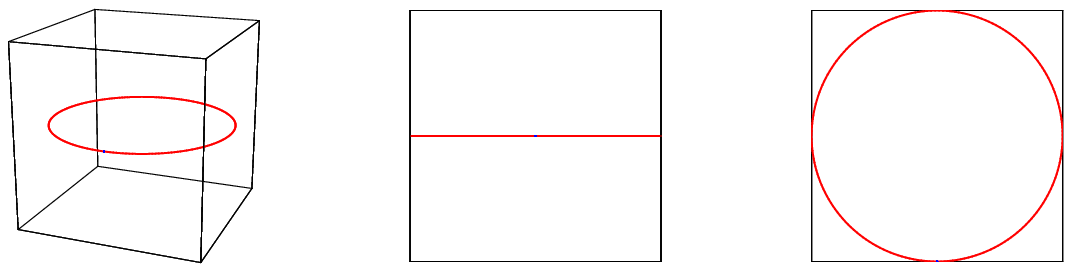}
\includegraphics{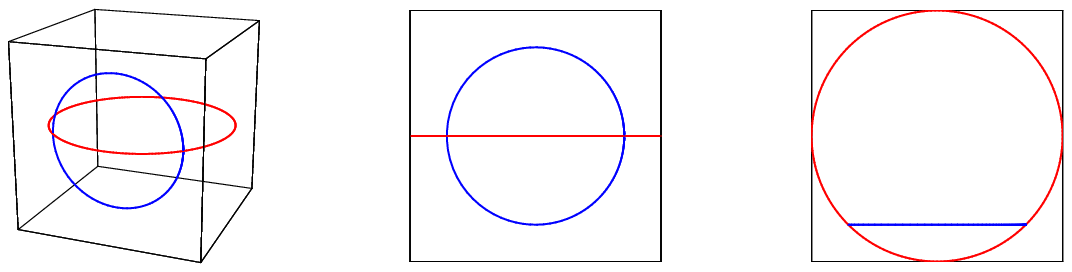}
\includegraphics{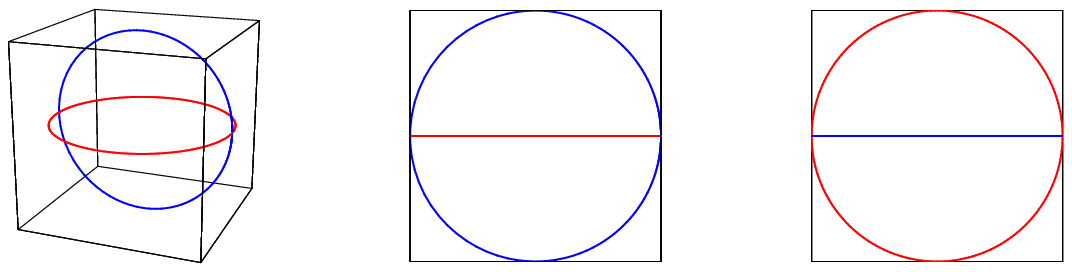}
\includegraphics{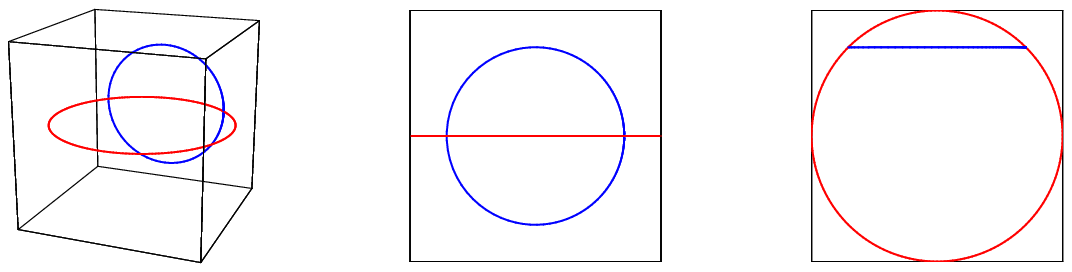}
\caption{\small Breathing magnon with $p=\pi$\,.
The value of $q$ is taken sufficiently large $(q=10.0)$\,. 
As a result the periods of the breathing and the rotation become almost the same.}
\label{fig:pi}
\end{center}
\end{figure}

\part[Conclusion]
	{Conclusion\label{part:6}}

\section*{Summary}

The emergence of integrability on each side of the AdS/CFT correspondence continues to provide remarkable improvements in our understanding of large-$N$ gauge theory and string theory.
Below we shall summarise what we have seen in this thesis.

\subsubsection*{\underline{Exploring near- and far-from-BPS sectors of AdS/CFT}}

\begin{itemize}
\item 
The AdS/CFT correspondence \cite{Maldacena:1997re,Gubser:1998bc,Witten:1998qj} states that type IIB string theory on $AdS_{5}\times S^{5}$ is a dual description of the four-dimensional, $\mathcal N=4$ super Yang-Mills (SYM)
theory.
One of the predictions of AdS/CFT is the exact matching of the spectra on both sides, namely the conformal dimensions of SYM operators with the energies of string states, (\ref{E=D}).
In the large\,-$N$ limit, these quantities are supposed to be connected
by some function of the 't Hooft coupling $\lambda$\,, but the
strong/weak nature of the AdS/CFT usually prevents us from a direct
comparison of the spectra. 

\item 
Berenstein, Maldacena and Nastase (BMN) \cite{Berenstein:2002jq}, proposed the plane-wave/SYM correspondence.
In the BMN limit, the string theory is reduced to a free, effectively massive two-dimensional model and thus can be quantised.
The resulting spectrum of string energies was shown to agree with the corresponding SYM conformal dimensions at the first few orders in the BMN coupling, see, {\em e.g.}, (\ref{BMN op ene}) and (\ref{BMN sqrt formula}) for the comparison at $\cO(\lam/J^{2})$\,.
Thus the BMN sector was expected to provide an overlapping perturbative regime where one can perturbatively access from both sides of the correspondence, despite the strong/weak nature of the duality.
(This expectation was, however, proved to be wrong as we already discussed.)

\item 
Part of the BMN results can be also obtained by the string sigma model approach of GKP \cite{Gubser:2002tv}.
In this picture, the BMN string corresponds to an almost collapsed, point-like solitonic closed string rotating on a great circle in $S^{5}$ with large spin $J$\,, which includes the first worldsheet quantum correction.
The BMN sector is ``almost BPS'', where the worldsheet momentum $p\sim 1/J$ is very small.
The large spin corresponds to large R-charge of a dual SYM operator, which is made up almost of a single flavour of complex scalar field, with few impurity fields (magnons).

\item 
The study of ``far-from-BPS'' sectors was further pursued by Frolov and Tseytlin \cite{Frolov:2003qc,Frolov:2003tu} and many applications followed.
The energies of semiclassical spinning/rotating string solutions were compared with the conformal dimensions of ``long'' composite operators of the SYM theory in the limit $\lam\to \infty$ with the effective coupling $\lam/J^{2}\ll 1$ kept fixed, now with macroscopic number of impurities.
Various types of string configurations were studied in this context \cite{Gubser:2002tv, Frolov:2002av, Frolov:2003qc, Frolov:2003tu, Frolov:2003xy, Minahan:2002rc, Arutyunov:2003uj}.
\end{itemize}

\subsubsection*{\underline{Bethe ansatz for integrable gauge/string spin-chains}}

\begin{itemize}

\item 
On the gauge theory side, by virtue of integrability, the computation of the anomalous dimensions can be achieved by solving a set of Bethe ansatz equations.
They can be formulated at higher loops as well.
This was first done for the scalar $SO(6)$ sector at the one-loop in the seminal work by Minahan and Zarembo \cite{Minahan:2002ve}.
The formulation was then extended to the full $PSU(2,2|4)$ model by Beisert and Staudacher \cite{Beisert:2003yb} at the one-loop level.
The integrability of gauge theory was established to the three-loop in \cite{Serban:2004jf}, by embeding the $SU(2)$ sector up to three-loops into the integrable Inozemtsev spin-chain model.
Assuming all-order perturbative integrability and also BMN scaling, the BDS model \cite{Beisert:2004hm} was proposed as the first candidate for the long-range integrable spin-chain for all-loop gauge theory, which, however, turned out not quite precise later.

\item 
A wealth of checks were performed for the spinning-string/spin-chains correspondence in the far-from-BPS sector of AdS/CFT.
Among them, we investigated the correspondence between the rational circular string and the ``one-cut'' configuration of SYM Bethe roots (Sections \ref{sec:Rational circular} and \ref{sec:rational 1-loop}) as well as the elliptic folded/circular strings and ``double contour''/``imaginary root'' configurations (Sections \ref{sec:elliptic} and \ref{sec:elliptic 3-loop}).
They provided important examples in testing the duality at the level of concrete solutions \cite{Beisert:2003xu,Beisert:2003ea}.
Perturbative expansion of their energies revealed a remarkable agreement with the SYM counterparts including and up to the two-loop level, in quite a non-trivial fashion.
At the three-loop level, however, the coefficients turn out to disagree, which is known as the ``three-loop discrepancy'' \cite{Serban:2004jf, Callan:2003xr, Callan:2004uv, Rej:2005qt}. 

\item 
In the algebro-geometric approach to the string equations of motion, classical string solutions are studied as finite-gap solutions.
In the KMMZ formalism \cite{Kazakov:2004qf}, every string solution is characterised by a spectral curve endowed with an Abelian integral called quasimomentum.
For the $SU(2)$ sector, by comparing the resulting classical Bethe ansatz equations (\ref{SBA rho}) with the gauge theory counterpart (\ref{BA sig}), that is the thermodynamic limit of SYM (discrete) Bathe equations, the agreement of all charges up to two-loops is manifest, while the discrepancy indeed starts at the three-loop level.

\item 
With the aim of obtaining the quantum string Bethe ansatz equations, which is the discretised version of the classical Bethe ansatz equations (\ref{SBA rho}), the so-called dressing factor was introduced by Arutyunov, Frolov and Staudacher (AFS) \cite{Arutyunov:2004vx} (see also \cite{Staudacher:2004tk}) that is to be multiplied to the BDS S-matrix.
Possible non-trivial interpolating feature of the dressing phase were expected to account for the three-loop discrepancy ``puzzle'' as well, which was later proved to be indeed the case. See the end of Section \ref{sec:weak-expansion} for the resolution.

\end{itemize}

\subsubsection*{\underline{Checking the conjectured AdS/CFT S-matrix}}

\begin{itemize}

\item 
Hofman and Maldacena \cite{Hofman:2006xt} considered a particular limit where the $\cN=4$ SYM spin-chain state becomes infinitely long.
Specifically, one considers a limit where the $U(1)_{\rm R}$-charge $J_{1}$ and the scaling dimension $\Delta$ of the operator go to infinity while their difference $\Delta-J_{1}$ and the 't Hooft coupling $\lambda$ held fixed. 
In this limit,  the worldsheet momentum $p$ is also kept fixed, and the spectrum corresponds to localised excitations which propagate almost freely on the infinite chain.
The remaining interactions between these excitations are governed by a factorisable S-matrix.

\item 
In the Hofman-Maldacena limit, both strings and dual spin-chains become infinitely long, and both sides of the correspondence are characterised by a centrally-extended SUSY algebra $({PSU(2|2)}\times {PSU(2|2)}) \ltimes {\mathbb R}^{3}$\,.
Asymptotic spin-chain spectrum for an elementary magnon case in this limit was derived by Beisert \cite{Beisert:2005tm}.
In \cite{Beisert:2005tm}, the structure of the asymptotic S-matrix was determined completely, up to an overall scalar phase factor.
The scalar phase turned out to be essentially the aforementioned dressing phase introduced by AFS \cite{Arutyunov:2004vx}.

\item 
The asymptotic spectrum was generalised to magnon boundstate case in \cite{Dorey:2006dq,Chen:2006gp}.
In \cite{Chen:2006gp}, we described the infinite tower of BPS boundstates appearing in the asymptotic spectrum of the ${\cal N}=4$ SYM spin-chain and identified the corresponding representation of supersymmetry in which they transform.
We also showed that the BPS condition of the extended SUSY algebra determines the dispersion relation for the magnon boundstate to be $\Delta-J_{1}=\sqrt{Q^{2}+f(\lam)\sin^{2}(P/2)}$\,, where $Q$ denotes the number of constituent magnons and $P$ is the momentum of the magnon boundstate along the spin-chain.
The function $f(\lambda)$ should be given by $f(\lam)=\lambda/\pi^2$ in view of the existing results of perturbative computations on the SYM (weak-coupling) side, resulting in (\ref{kdisprel}).

\item 
The dispersion relation for the SYM magnon boundstate was precisely reproduced from a classical string theory computation in \cite{Chen:2006gq}, where we considered a two-charge extension of giant magnon solution, which we named dyonic giant magnons.
We identified the BPS multiplet label $Q$ of order $\sqrt{\lam}$ with the second spin $J_2$\,, and the total momentum $P$ with the angular difference of two endpoints of the string.
In our construction of the dyonic giant magnon, we employed the Pohlmeyer reduction procedure \cite{Pohlmeyer:1975nb} as an efficient solution generating technique.
In static gauge, the string equations of motion are essentially those of a bosonic $O(4)$ sigma model supplemented by the Virasoro constraints, which is classically equivalent to Complex sine-Gordon (CsG) system. 
The dyonic giant magnon is so constructed that it corresponds to a kink soliton solution of CsG equation.
It was then shown that, in the large-$\lam$ limit, the conjectured AdS/CFT S-matrix for boundstates 
precisely agreed with the semiclassical S-matrix for scattering of dyonic giant magnons under an appropriate gauge choice \cite{Chen:2006gq}.

\item 
This idea of exploiting the equivalence between classical CsG system and classical $O(4)$ string sigma model was further utilised to construct more general classical string solutions on $\mathbb R\times S^{3}$\,, which are called type $(i)$ and $(ii)$ helical strings \cite{Okamura:2006zv}.
They are the most general genus-one classical string solutions that interpolate between two-spin folded/circular strings \cite{Frolov:2003xy} and dyonic giant magnons \cite{Chen:2006gq}.
They were also reformulated in terms of the finite-gap language \cite{Vicedo:2007rp}.

\item 
It was then noticed in \cite{Hayashi:2007bq} that, in the conformal gauge, starting from the type $(i)$ and $(ii)$ helical strings \cite{Okamura:2006zv} and by exchanging worldsheet variables $\tau$ and $\sig$ (``2D transformation'') on the $S^{5}$ side while keeping the temporal gauge $t=\kappa\tau$\,, one can reach another class of helical strings which actually together with the original ones complete the elliptic classical string solutions on $\mathbb R\times S^{3}$\,.
The new string solutions obtained via the 2D transformation, called type $(i)'$ and $(ii)'$ helical strings, were shown to interpolate pulsating strings and so-called single-spike strings \cite{Ishizeki:2007we}.
In \cite{Hayashi:2007bq}, we also interpreted type $(i)'$ and $(ii)'$ helical solutions as finite-gap solutions.
The effect of the 2D transformation can be interpreted as swapping the roles of quasi-momentum and quasi-energy endowed with the elliptic curve.

\item 
In \cite{Dorey:2007an}, we examined the singular structures of the refined conjectured S-matrix \cite{Beisert:2005fw,Beisert:2005tm,Janik:2006dc,Eden:2006rx,Arutyunov:2006iu,Beisert:2006ib,Beisert:2006ez,Dorey:2007xn} in order to perform further analyticity tests for it.
The conjectured S-matrix for magnon boundstates exhibits both simple and double poles at complex momenta.
Some of these poles lie parametrically close to the real axis in momentum space on the branch where particle energies are positive.
We showed that all such poles are precisely accounted for by physical processes involving one or more on-shell intermediate particles belonging to the BPS spectrum derived in \cite{Dorey:2006dq,Chen:2006gp}.

\end{itemize}

\section*{Outlook}

As we have stressed above, the new approach based on integrability allows a deeper exploration of the AdS/CFT duality and has led to many far-reaching developments in recent years.
We already discussed individual outlooks for each topic in the main text.
Here we will make one further general remark.

Our goal is to uncover the true nature of integrability in gauge and string theory, building bridges between the integrable structures found in gauge and string theory.
It was not clear at all at first sight how the two sides\, \----\, the integrable (spin-chain) model describing the $\cN=4$ SYM and the integrable sigma model describing string theory on $AdS_{5}\times S^{5}$\, \----\, are related to each other.
However, now that we have reached a considerably refined form of the conjecture for the AdS/CFT S-matrix, we have fertile setting and playground in order to test the conjecture.
All this could finally give a clue for quantising string theory on $AdS_{5}\times S^{5}$\,, which still remains a challenge.
Thus the study of integrability will surely continue to give us deep insights into the fundamental nature of string and gauge theory, as well as shed more light on the fine structure of AdS/CFT.

In conclusion, let us end this thesis with the following philosophical question posed by M.~Staudacher \cite{Staudacher:YITP:05}\,:

\bigskip
\quad {\sl ``Is there a physical reason for the integrability observed in gauge and string theory?''}

\bigskip
\noindent
It would be nice if we could, through unrelenting effort, nail down the message behind the discovered AdS/CFT integrability, giving an answer to this question still up in the air.

\appendix
\part*{Appendices}
\addcontentsline{toc}{part}{Appendices}

\chapter[Helical Strings on $AdS_{3}\times S^{1}$]
	{Helical Strings on $\boldsymbol{AdS_{3}\times S^{1}}$\label{app:AdS helicals}}

In this appendix we will discuss helical string solutions in the
$SL(2)$ sector. The construction almost parallels that in
\cite{Okamura:2006zv,Hayashi:2007bq}, however, non-compactness of the AdS space
leads to new non-trivial features compared to the sphere case.

\section[Classical strings on ${AdS_{3}\times S^{1}}$ and Complex sinh-Gordon model]
	{Classical strings on \bmt{AdS_{3}\times S^{1}} and Complex sinh-Gordon model}

A string theory on $AdS_{3}\times S^{1}\subset AdS_{5}\times
S^{5}$ spacetime is described by an $O(2,2)\times O(2)$ sigma
model. Let us denote the coordinates of the embedding space as
$\eta_0$\,, $\eta_1$ (for $AdS_{3}$) and $\xi_{1}$ (for $S^{1}$)
and set the radii of $AdS_{3}$ and $S^{1}$ both to unity,
\begin{equation}
\vec \eta \, {}^* \cdot \vec \eta \equiv
- \abs{\eta_0}^2 + \abs{\eta_1}^2 = - 1\,, \qquad
\abs{\xi_{1}}^2 = 1\,.
\label{norms}
\end{equation}
In the standard polar coordinates, the embedding coordinates are expressed as
\begin{alignat}{3}
\eta_0 &= \cosh \rho \, e^{i \ssp t} \,, &\quad
\eta_1 &= \sinh \rho \,  e^{i \ssp \phi_1}\,, &\quad
\xi_{1} &= e^{i \ssp \varphi_1} \,,
\end{alignat}
and all the charges of the string states are defined as Noether
charges associated with shifts of the angular variables. The
bosonic Polyakov action for the string on $AdS_{3}\times S^{1}$ is
given by
\begin{equation}
S = - \frac{\sqrt{\lambda}}{4\pi} \int d \sigma d \tau \cpare{
\gamma^{a b} \pare{ \partial_a \vec \eta \, {}^* \cdot \partial_b \vec \eta + \partial_a  \xi^* \cdot \partial_b \xi \,}
+ \widetilde\Lambda \big( \vec \eta \, {}^* \cdot \vec \eta + 1 \big)
+ \Lambda \big( \xi_{1}^* \cdot  \xi_{1} - 1 \big) }\,,
\label{action-AdS}
\end{equation}
and we take the same conformal gauge as in the \RS{3} case.
From the action (\ref{action-AdS}) we get the equations of motion
\begin{equation}
\partial_{a} \partial^{\ssp a} \vec \eta - (\partial_{a} \vec\eta \, {}^* \cdot \partial^{\ssp a} \vec \eta ) \, \vec \eta = 0\,, \qquad
\partial_{a} \partial^{\ssp a} \xi_{1} +(\partial_{a}  \xi_{1}^{*} \cdot \partial^{\ssp a}  \xi_{1}) \, \xi_{1} = 0 \,,
\label{str_eom}
\end{equation}
and Virasoro constraints
\begin{align}
0 &= {\cal T}_{\sigma \sigma} = {\cal T}_{\tau\tau} = \frac{\delta^{ab}}{2} \pare{ \partial_{a} \vec \eta \, {}^* \cdot \partial_{b} \vec \eta +
\partial_{a} \xi_{1}^* \cdot \partial_{b}  \xi_{1} }
\label{str_Vir1} \,, \\[2mm]
0 &= {\cal T}_{\tau\sigma} = {\cal T}_{\sigma\tau} = \Re \pare{ \partial_{\tau} \vec \eta \, {}^* \cdot \partial_{\sigma} \vec \eta +
\partial_{\tau}\xi_{1} \cdot \partial_{\sigma}\xi_{1}^{*} } \,.
\label{str_Vir2}
\end{align}

The Pohlmeyer reduction procedure, which we made use of in obtaining the
$O(4)$ sigma model solutions from Complex sine-Gordon solution in Chapter \ref{chap:DGM},
also works for the current case in much the same way. The $O(2,2)$
sigma model in conformal gauge is now related to what we call
Complex sinh-Gordon (CshG) model, which is defined by the
Lagrangian
\begin{equation}
{\cal L}_{\rm CshG} = \frac{\partial^{\ssp a} \psi^* \partial_a \psi}{1 + |\psi|^{2}} + |\psi|^{2} \,,
\label{CshG Lag}
\end{equation}
with $\psi = \psi (\tau, \sigma)$ being a complex field. It can be
viewed as a natural generalization of the well-known sinh-Gordon
model in the sense we describe below. By defining two real fields
$\alpha$ and $\beta$ of the CshG model through $\psi \equiv \sinh
\pare{\alpha/2} \exp (i \beta/2)$\,, the Lagrangian \eqref{CshG
Lag} is rewritten as
\begin{equation}
{\cal L}_{\rm CshG} = \frac14 \pare{\partial_a \alpha}^2 + \frac{\tanh^2 (\alpha/2)}{4} \pare{\partial_a \beta}^2 + \sinh^2 (\alpha/2)\,.
\label{CshG Lag2}
\end{equation}
The equations of motion that follow from the Lagrangian are
\begin{align}
&\partial^{\ssp a} \partial_a \psi - \psi^* \frac{\partial^{\ssp a} \psi \, \partial_a \psi}{1 + |\psi|^{2}} - \psi \pare{1 + |\psi|^{2}}=0\,,\label{CshG eom}\\[3mm]
&\quad \mbox{\em i.e.},\quad
\left\{
\begin{array}{l}
\ds \partial^{\ssp a} \partial_a \alpha - \frac{\sinh (\alpha/2)}{2 \cosh^3 (\alpha/2)} \pare{\partial_a \beta}^2 - \sinh \alpha = 0\,, \\[6mm]
\ds  \partial^{\ssp a} \partial_a \beta + \frac{2 \, \partial_a \alpha \, \partial^{\ssp a} \beta}{\sinh \alpha} = 0\,.
\end{array}
\right.\label{CshG eq}
\end{align}
We refer to the coupled equations \eqref{CshG eq} as Complex
sinh-Gordon (CshG) equations. If $\beta$ is a constant field, the
first equation in (\ref{CshG eq}) reduces to
\begin{equation}
\partial_a \partial^{\ssp a} \alpha - \sinh \alpha = 0\,.
\end{equation}
which is the ordinary sinh-Gordon equation. As readers familiar
with the Pohlmeyer reduction can easily imagine, it is this field
$\alpha$ that gets into a self-consistent potential in the
Schr\"{o}dinger equation this time. Namely, we can write the
string equations of motion given in \eqref{str_eom} as
\begin{equation}
\partial_{a} \partial^{\ssp a} \vec \eta - (\cosh \alpha) \, \vec \eta = 0\,, \qquad
\cosh \alpha \equiv \partial_{a} \vec\eta \, {}^* \cdot \partial^{\ssp a} \vec \eta\,,
\label{reduced_eom-AdS}
\end{equation}
with the same field $\alpha$ we introduced as the real part of the
CshG field $\psi$\,. What this means is that if $\{\vec \eta\,,
\xi\}$ is a consistent string solution which satisfies the Virasoro
conditions \eqref{str_Vir1} and \eqref{str_Vir2}, then $\psi =
\sinh \pare{\alpha/2} \exp (i \beta/2)$ defined via
\eqref{reduced_eom-AdS} and \eqref{AdS-PLR beta} solves the CshG
equations.

The derivation of this fact parallels the usual Pohlmeyer reduction
procedure. Let us define worldsheet light-cone coordinates as
$\sigma^{\pm}=\tau\pm\sigma$\,, and the embedding coordinates as
$\eta_0 = Y_0 + i Y_5$ and $\eta_1 = Y_1 + i Y_2$\,. Then consider
the equations of motion of the $O(2,2)$ nonlinear sigma model
through the constraints
\begin{equation}
\vec Y \cdot \vec Y = -1\,, \quad (\partial_+ \vec Y )^2 = -1\,,\quad
(\partial_- \vec Y )^2 = -1\,,\quad \partial_+ \vec Y \cdot \partial_- \vec Y \equiv - \cosh \alpha\,,
\end{equation}
where $\vec Y \cdot \vec Y \equiv (\vec Y)^2 \equiv - (Y_0)^2 + (Y_1)^2 + (Y_2)^2 - (Y_5)^2$\,.
A basis of $O(2,2)$-covariant vectors can be given by $Y_i$\,, $\partial_+ Y_i $\,, $\partial_- Y_i$ and $K_i \equiv \epsilon_{ijkl} Y^j \partial_+ Y^k \partial_- Y^l$\,.
By defining a pair of scalar functions $u$ and $v$ as
\begin{equation}
u \equiv \frac{\vec K \cdot \partial_+ ^{\ssp 2} \vec Y}{\sinh
\alpha} \,, \qquad v \equiv \frac{\vec K \cdot \partial_- ^{\ssp
2} \vec Y}{\sinh \alpha} \,,
\end{equation}
the equations of motion of the $O(2,2)$ sigma model are recast in
the form
\begin{equation}
\partial_- \partial_+ \alpha + \sinh \alpha + \frac{uv}{\sinh \alpha} = 0\,, \qquad
\partial_{-} u = \frac{v \, \partial_{+} \alpha}{\sinh \alpha} \,, \qquad
\partial_{+} v = \frac{u \, \partial_{-} \alpha}{\sinh \alpha} \,.
\label{Pohlmeyer_eom1}
\end{equation}
One can easily confirm that this set of equations is equivalent to
the pair of equations (\ref{CshG eq}) of CshG theory, under the
identifications
\begin{equation}
u = (\partial_+ \beta) \, \tanh \frac{\alpha}{2} \,,\qquad
v = - (\partial_- \beta) \, \tanh \frac{\alpha}{2} \,.
\label{AdS-PLR beta}
\end{equation}

\paragraph{}
Thus there is a classical equivalence between the $O(2,2)$ sigma
model $\leftrightarrow$ CshG as in the $O(4)\leftrightarrow
\mbox{CsG}$ case. Making use of the equivalence, one can construct
classical string solutions on $AdS_{3}\times S^{1}$ by the
following recipe\,:
\begin{enumerate}
\item Find a solution $\psi$ of CshG equation (\ref{CshG eom}).
\item Identify $\cosh\alpha\eq \partial_{a} \vec\eta \, {}^* \cdot
\partial^{\ssp a} \vec \eta$\,, where $\alpha$ appears in the real
part of the solution $\psi$\,, and $\eta$ are the embedding
coordinates of the corresponding string solution in $AdS_{3}$\,.
\item Solve the ``Schr\"{o}dinger equation'' (\ref{reduced_eom-AdS})
together with the Virasoro constraints (\ref{str_Vir1}) and
(\ref{str_Vir2}), under appropriate boundary conditions. \item
Resulting set of $\vec\eta$ (``wavefunction'') and $\xi_{1}$ gives
the corresponding string profile in $AdS_{3}\times S^{1}$\,.
\end{enumerate}

Let us start with Step 1. From the similarities between the CshG
equation and the CsG equation, it is easy to find helical-wave
solutions of the CshG equation. Here we give two such solutions
that will be important later. The first one is given by
\begin{equation}
\psi_{\rm cd}=kc\frac{\cn (c x_v)}{\dn (c x_v)}\exp \Big( i \sqrt{(1+c^2)(1+k^2 c^2)} \; t_v \Big)\,,
\label{CshG helical cd}
\end{equation}
and the second one is
\begin{equation}
\psi_{\rm ds}=c\frac{\dn (c x_v)}{\sn (c x_v)}\exp  \Big( i \sqrt{(1-k^2c^2)(1+c^2-k^2c^2)} \; t_v \Big)\,.
\label{CshG helical ds}
\end{equation}
By substituting the solution (\ref{CshG helical ds}) into the string equations of motion \eqref{reduced_eom-AdS}, we obtain
\begin{equation}
\cpare{- \partial_T ^2 + \partial_X ^2 - k^2 \pare{\frac{2}{k^2 \sn^2 (X|k)} - 1} } \, \vec \eta = U \vec \eta \,,
\label{GAL eom}
\end{equation}
under the identification of $(\mu \ssp \tau, \mu \ssp \sigma)
\equiv (c \ssp t, c \ssp x)$\,. The ``eigenenergy'' $U$ can be
treated as a free parameter as was the case in
\cite{Okamura:2006zv}. Different choices of helical-waves of CshG
equation simply correspond to taking different ranges of $U$\,.

\paragraph{}
We are now at the stage of constructing the corresponding string
solution by following Steps 2\,\--\,4 described above. However,
we do not need to do this literally. Since the metrics of \AdSS{3}
\begin{equation}
ds_{(AdS_{3}\times S^{1})}^2 = - \cosh^2 \! \tilde \rho \, d {\tilde t\,}{}^2 + d{\tilde \rho}^2 + \sinh^2 \! \tilde \rho \, d {\tilde \phi_1}^2 + d {\tilde \varphi_1}^2 \,,
\end{equation}
and of \RS{3}
\begin{equation}
ds_{(\mathbb R\times S^{3})}^2 = - d t^2 + d\gamma^2 + \cos^2 \! \gamma \, d \varphi_1^2 + \sin^2 \! \gamma \, d \varphi_2^2 \,,
\end{equation}
are related by the map
\begin{equation}
\tilde\rho  \leftrightarrow i\gamma ,\quad \tilde t \leftrightarrow \varphi _1 ,\quad \tilde \phi _1  \leftrightarrow \varphi _2 ,\quad \tilde \varphi _1  \leftrightarrow t \quad \Rightarrow \quad ds_{AdS_{3}\times S^{1}}^2 \leftrightarrow - ds_{\mathbb R\times S^{3}}^2 \,,
\label{an_cont}
\end{equation}
string solutions on both manifolds are related by a sort of
analytic continuation of global coordinates. Therefore, the
simplest way to obtain helical string solutions on \AdSS{3} is to
perform analytic continuation of helical string solutions on
\RS{3}, as will be done in the following sections. Large parts of
the calculation parallel the $\mathbb R \times S^{3}$ case, and the
most significant difference lies in the constraints imposed on the
solution of the equations of motion, such as the periodicity
conditions.

\section[Helical strings on ${AdS_{3}\times S^{1}}$]
	{Helical strings on \bmt{AdS_{3}\times S^{1}}}

In this section, we consider the analytic continuation of helical
strings on \RS{3} to those on \AdSS{3}. Among various possible
solutions, we will concentrate on two particular examples that
have clear connections with known string solutions of interest to
us. The first example, called type $(iii)$ helical string, is a
helical generalization of the folded string solution on \AdSS{3}
\cite{Frolov:2002av}. The second one, called type $(iv)$\,,
reproduces the $SL(2)$ ``giant magnon" solution
\cite{Minahan:2006bd, Ryang:2006yq} in the infinite-spin limit.

\subsection{Type $\bmt{(iii)}$ helical strings}

In \cite{Beisert:2003ea}, it was pointed out that $(S, J)$ folded
strings can be obtained from $(J_1, J_2)$ folded strings by
analytic continuation of the elliptic modulus squared, from $k^2
\ge 0$ to $k^2 \le 0$\,. Here we apply the same analytic
continuation to type $(i)$ helical strings to obtain solutions on
\AdSS{3}, which we call type $(iii)$ strings. For notational
simplicity, it is useful to introduce a new moduli parameter $q$
through the relation
\begin{equation}
k \equiv \frac{i q}{q'} \eq \frac{i q}{\sqrt{1 - q^2}} \,.
\label{moduli kq}
\end{equation}
If $k$ is located on the upper half of the imaginary axis, {\em
i.e.}, $k=i\kappa$ with $0\leq \kappa$\,, then $q$ is a real
parameter in the interval $[0,1]$\,.

\begin{figure}[htbp]
\begin{center}
\vspace{0.5cm}
\includegraphics[scale=0.9]{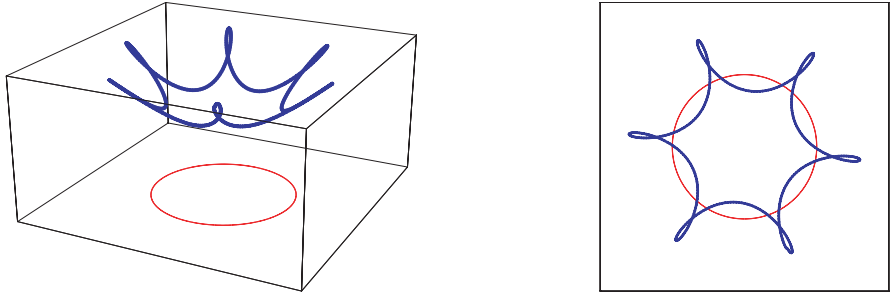}
\vspace{0.5cm}
\caption{\small Type $(iii)$ helical string ($q=0.700$\,, $U=12.0$\,, $\tilde \om_{0}=-0.505$\,, $\tilde\om_{1}=0.776$\,, $n=6$), projected onto $AdS_{2}$ spanned by $(\Re\!\eta_{1}, \Im\!\eta_{1}, |\eta_{0}|)$\,.
The circle represents a unit circle $|\eta_{1}|=1$ at $\eta_{0}=0$\,.}
\label{fig:III-gen}
\end{center}
\end{figure}

As shown in Appendix \ref{app:formula}, the transformation
\eqref{moduli kq} can be regarded as an $SL(2,\mathbb Z)$ $\mathsf T$-transformation
of the modulus $\tau$\,. Hence, by performing a $\mathsf
T$-transformation on the profile of type $(i)$ helical strings
(\ref{zf0}\,-\,\ref{zf2}), we obtain type $(iii)$ string solutions:
\begin{align}
\eta_0 &= \frac{C}{\sqrt{q q'}} \, \frac{\bTh_3 (0|q) \, \bTh_0 (\tilde X - \tomm{0}|q)}{\bTh_2 (\tomm{0}|q) \, \bTh_3 (\tilde X|q)} \,
\exp \Big( \eZ_2 (\tomm{0}|q) \tilde X + i \tilde u_0 \ssp \tilde T \Big) \,,  \label{e0_two} \\[2mm]
\eta_1 &= \frac{C}{\sqrt{q q'}} \, \frac{\bTh_3 (0|q) \, \bTh_1 (\tilde X - \tomm{1}|q)}{\bTh_3 (\tomm{1}|q) \, \bTh_3 (\tilde X|q)} \,
\exp \Big( \eZ_3 (\tomm{1}|q) \tilde X + i \tilde u_1 \ssp \tilde T \Big) \,,  \label{e1_two} \\[2mm]
\xi_1 &= \exp \pare{ i \ssp \tilde a \ssp \tilde T + i \ssp \tilde b \ssp \tilde X } \,,  \label{x1_two}
\end{align}
where we rescaled various parameters as
\begin{equation}
\tilde X = X/q' \,,\quad \tilde T = T/q' \,,\quad \tilde \omega_j = \omega_j / q' \,, \quad \tilde a = a \ssp q'\,, \quad \tilde b = b \ssp q' \,, \quad \tilde u_j = u_j \ssp q' \,.   \label{var tildes}
\end{equation}
We choose the constant $C$ so that they satisfy $\left| \eta_0
\right|^2 - \left| \eta_1 \right|^2 = 1$\,. One such possibility
is to choose\footnote{In contrast to the \RS{3} case, the RHS
of (\ref{cd_two}) is not always real for arbitrary real values of
$\tilde \omega_0$ and $\tilde \omega_1$\,. If $C^2 < 0$\,, we have
to interchange $\eta_0$ and $\eta_1$ to obtain a solution properly
normalised on \AdS{3}\,. }
\begin{equation}
C = \pare{\frac{1}{q^2 \cn^2 (\tomm{0})}  + \frac{\sn^2 (\tomm{1})}{\dn^2 (\tomm{1})} }^{-1/2} \,.  \label{cd_two}
\end{equation}
With the help of various formulae on elliptic functions, one can check that $\vec \eta$ in (\ref{e0_two}), (\ref{e1_two}) certainly solves the string equations of motion as
\begin{equation}
\cpare{- \partial_{\tilde T} ^2 + \partial_{\tilde X} ^2 + q^2 \pare{2 (1-q^2) \, \frac{\sn^2}{\dn^2} (\tilde X|q) - 1 }} \, \vec \eta = \tilde U \vec \eta \,,
\label{str eom_two}
\end{equation}
if the parameters are related as
\begin{equation}
\tilde u_0^2 = \tilde U - (1-q^2) \frac{\sn^2 (\tomm{0})}{\cn^2 (\tomm{0})} \,,\qquad
\tilde u_1^2 = \tilde U + \frac{1 - q^2}{\dn^2 (\tomm{1})} \,.
\end{equation}
As is clear from \eqref{str eom_two}, the type $(iii)$ solution is
related to the helical-wave solution of the CshG equation given in
\eqref{CshG helical cd}. The Virasoro constraints (\ref{str_Vir1}, \ref{str_Vir2}) impose constraints on $\tilde a$ and $\tilde
b$ in (\ref{x1_two})\,:\footnote{Note that the Virasoro
constraints require neither $a \ge b$ nor $a \le b$\,. This means
that both $\xi_1 = \exp \big(i \tilde a_0 \tilde T + i \tilde b_0
\tilde X\big)$ and $\exp\big(i \tilde b_0 \tilde T + i \tilde a_0
\tilde X\big)$ can be consistent string solutions. It can be viewed
as the $\ts$ transformation applied only to the $S^{1}\subset
S^{5}$ part while leaving the \AdS{3} part intact.}
\begin{alignat}{2}
&\tilde a^2 + \tilde b^2 & &= - q^2 - \tilde U - \frac{2 \ssp (1 - q^2)}{\cn^2 (\iomm{0})} + 2 \ssp \tilde u_1^2 \,,
\label{ab1_two}  \\
&\quad \tilde a \ssp \tilde b & &= i \, C ^2 \pare{\frac{\tilde u_0}{q^2} \, \frac{\sn (\iomm{0}) \dn (\iomm{0})}{\cn^3 (\iomm{0})} + \tilde u_1 \, \frac{\sn (\iomm{1}) \cn (\iomm{1})}{\dn^3 (\iomm{1})}}
\label{ab2_two}\,.
\end{alignat}
The reality of $\tilde a$ and $\tilde b$ must also hold.

Since we are interested in closed string solutions, we should
impose periodic boundary conditions. Let us define the period in
the $\sigma$ direction by
\begin{equation}
\Delta \sigma = \frac{2 \ssp \eK (k) \sqrt{1 - v^2}}{\mu}
= \frac{2 \ssp q' \, \eK (q) \sqrt{1 - v^2}}{\mu} \equiv 2 l \equiv \frac{2 \pi}{n}\,,
\label{one-hop def}
\end{equation}
which is equivalent to $\Delta \tilde X = 2 \ssp \eK (q)$ and $\Delta \tilde T = - 2 \ssp v \ssp \eK (q)$\,.
The closedness conditions for the AdS variables are written as
\begin{alignat}{2}
\Delta t &= 2 \eK (q) \bpare{ -i \eZ_2 (\tomm{0}) - v \ssp \tilde u_0 } + 2 \ssp n'_{\rm time} \ssp \pi \equiv \frac{2 \pi N_t}{n}\,,
\label{Dt_ads_dy} \\[2mm]
\Delta \phi_1 &= 2 \eK (q) \bpare{ -i \eZ_3 (\tomm{1}) - v \ssp \tilde u_1 } + \pare{2 \ssp n'_1 + 1} \pi \equiv \frac{2 \pi N_{\phi_{1}}}{n} \,.
\label{Dphi_ads_dy}
\end{alignat}
And from the periodicity in $\varphi_1$ direction, we have
\begin{equation}
N_{\varphi_{1}} = \mu \, \frac{\tilde b - v \ssp \tilde a}{\sqrt{1 - v^2}} \in \bb{Z} \,.
\label{Dvarphi_ads_dy}
\end{equation}

We must further require the timelike winding $N_t$ to be zero. Just as in the \RS{3} case, one can adjust the value of $v$ to fulfill this requirement.\footnote{Note that in $\mathbb R\times S^{3}$ case, the zero-$N_t$ condition was trivially solved by $v=b/a$\,.} 
The integer $n'_{\rm time}$ is evaluated as
\begin{equation}
2 \ssp n'_{\rm time} \ssp \pi = \frac{1}{2 \ssp i} \int_{-\eK}^{\eK} d\tilde X \; \frac{\partial}{\partial \tilde X} \cpare{ \ln \pare{\frac{\bTh_0 (\tilde X - \tomm{0})}{\bTh_0 (\tilde X + \tomm{0})}} } \,.
\end{equation}
Then, by solving the equation $N_t=0$\,, one finds an appropriate
value of $v=v_t$\,. The absolute value of the worldsheet boost
parameter $v_t$ may possibly exceed one (the speed of light). In
such cases, we have to perform the 2D transformation $\tau
\leftrightarrow \sigma$ on the AdS space to get $v_t \mapsto
-1/v_t$\,.

\bigskip
As usual, conserved charges for $n$ periods are defined by
\begin{alignat}{3}
E &\equiv \frac{\sqrt{\lambda}}{\pi} \, {\cal E} & &= \frac{n \sqrt{\lambda}}{2 \pi} \int_{-l}^{\ssp l} d \sigma \,
\Im\!\pare{\eta_0^{*} \, \partial_{\tau} \eta_0} \,,  \label{charge_def E}\\[2mm]
S &\equiv \frac{\sqrt{\lambda}}{\pi} \, {\cal S} & &= \frac{n \sqrt{\lambda}}{2 \pi} \int_{-l}^{\ssp l} d \sigma
\Im\!\pare{\eta_{1}^{*} \, \partial_{\tau} \eta_{1}} \,,  \label{charge_def S}\\[2mm]
J &\equiv \frac{\sqrt{\lambda}}{\pi} \, {\cal J} & &= \frac{n \sqrt{\lambda}}{2 \pi} \int_{-l}^{\ssp l} d \sigma
\Im\!\pare{\xi_{1}^{*} \, \partial_{\tau} \xi_{1}} \,.  \label{charge_def J}
\end{alignat}
which are evaluated as, for the current type $(iii)$ case,
\begin{align}
{\cal E} &= \frac{n \ssp C^2 \, \tilde u_0}{q^2 (1-q^2)} \cpare{ \eE
+ (1-q^2) \bpare{ \frac{\sn^2 (\tomm{0})}{\cn^2 (\tomm{0})} - \frac{iv}{\tilde u_0} \, \frac{\sn (\tomm{0}) \dn (\tomm{0})}{\cn^3 (\tomm{0})} } \eK } \,,  \label{chE two}  \\[2mm]
{\cal S} &= \frac{n \ssp C^2 \, \tilde u_1}{q^2 (1-q^2)} \cpare{ \eE
- (1-q^2) \bpare{ \frac{1}{\dn^2 (\tomm{1})} - \frac{iv \ssp q^2}{\tilde u_1} \, \frac{\sn (\tomm{1}) \cn (\tomm{1})}{\dn^3 (\tomm{1})} } \eK } \,,  \label{chS two}  \\[2mm]
{\cal J} &= n \pare{\tilde a - v \ssp \tilde b} \eK \,.
\label{chJ two}
\end{align}

\paragraph{}
It is interesting to see some of the limiting behaviors of this
type $(iii)$ helical string in detail.\footnote{It seems the
original ``spiky string'' solution of \cite{Kruczenski:2004wg} is
also contained in the type $(iii)$ class, although we have not been able to reproduce it analytically. 
}

\subsubsection*{$\bullet$ $\bmt{\tilde \om_{1,2}\to 0}$ limit\,:~Folded strings on $\bmt{AdS_{3}\times S^{1}}$}

In the $\tilde \om_{1,2}\to 0$ the timelike winding condition \eqref{Dt_ads_dy} requires $v = 0$\,, so the boosted worldsheet coordinates $(\tilde T, \tilde X)$ become
\begin{equation}
(\tilde T, \tilde X) \to \pare{{\mu \tau}/{q'}\,, {\mu \sigma}/{q'}} \equiv
(\tilde \mu \tau, \tilde \mu \sigma) \equiv (\tilde \tau, \tilde \sigma) \,.
\label{tilde ts}
\end{equation}
The periodicity condition \eqref{one-hop def} allows $\tilde \mu$
to take only a discrete set of values.

\begin{figure}[htbp]
\begin{center}
\vspace{0.5cm}
\includegraphics[scale=0.9]{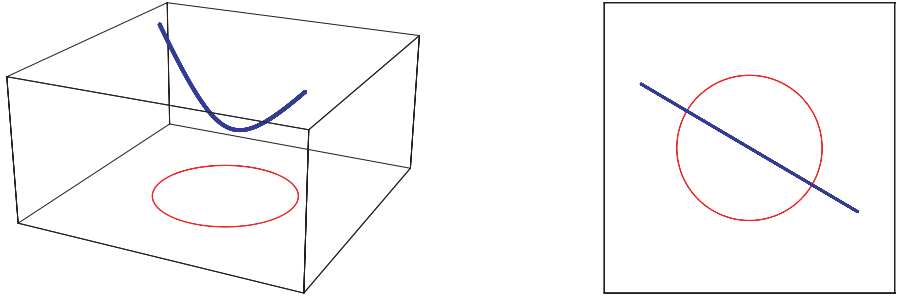}
\vspace{0.5cm}
\caption{\small $\tilde \om_{1,2}\to 0$ limit of type $(iii)$ helical string becomes a folded string studied in \cite{Frolov:2002av}.}
\label{fig:III-fold}
\end{center}
\end{figure}

The profile of type $(iii)$ strings now reduces to
\begin{equation}
\eta_0 = \frac{1}{\dn(\tilde \sigma | q)} \; e^{i \tilde u_0 \tilde \tau} \,, \qquad
\eta_1 = \frac{q \sn(\tilde \sigma | q)}{\dn(\tilde \sigma|q)} \; e^{i \tilde u_1 \tilde \tau} \,, \qquad
\xi_1 = \exp \pare{ i \sqrt{\tilde U - q^2} \; \tilde \tau}\,,
\label{stat_i}
\end{equation}
where $\tilde u_0^2 = \tilde U$ and $\tilde u_1^2= \tilde U + 1 - q^2$\,.
This solution is equivalent to $\mathsf T$-transformation
of $(J_1, J_2)$ folded strings of \cite{Frolov:2003xy}, namely, $(S, J)$ folded strings.\footnote{Note that the set, $\eta_{0,1}=\mbox{the same as (\ref{stat_i})}$ and $\xi_{1}=\exp\ko{ i \sqrt{\tilde U - q^2} \; \tilde \sigma}$\,, also gives a solution.}
The conserved charges of \eqref{stat_i} are computed as
\begin{equation}
{\cal E} = \frac{n \tilde u_0}{1-q^2} \, \eE (q) \,, \quad
{\cal S} = \frac{n \tilde u_1}{1-q^2} \Big( \eE (q) - (1-q^2) \eK (q) \Big) \,,\quad
{\cal J} = n \sqrt{\tilde U - q^2} \; \eK (q) \,.
\end{equation}
Rewriting these expressions in terms of the original imaginary
modulus $k$\,, we find the following relations among conserved
charges\,:
\begin{equation}
\left( {\frac{\cal J}{{\eK (k)}}} \right)^2  - \left( {\frac{\cal E}{{\eE (k)}}} \right)^2  = n^2 k^2 \,,\qquad
\left( {\frac{\cal S}{{\eK (k) - \eE (k)}}} \right)^2  - \left( {\frac{\cal J}{{\eK (k)}}} \right)^2  = n^2 (1 - k^2) \,,
\label{FT relation}
\end{equation}
as obtained in \cite{Beisert:2003ea}.

\subsubsection*{$\bullet$ $\bmt{q\to 1}$ limit\,:~ Logarithmic behavior}

Another interesting limit is to send the elliptic modulus $q$ to unity.
In this limit, the spikes of the type $(iii)$ string attach to the AdS boundary, and the energy $E$ and AdS spin $S$ become divergent.
Again, the condition of vanishing timelike winding is fulfilled by $v = 0$\,, and the periodicity condition \eqref{one-hop def} implies that $\tilde \mu$ given in \eqref{tilde ts} goes to infinity.
The profile becomes
\begin{equation}
\eta_0 = C \cosh(\tilde \sigma - \tomm{0}) \; e^{i \tilde u_0 \tilde \tau} \,, \quad
\eta_1 = C \sinh(\tilde \sigma - \tomm{1}) \; e^{i \tilde u_1 \tilde \tau} \,, \quad
\xi_1 = \exp \pare{ i \ssp \tilde a \ssp \tilde \tau + i \ssp \tilde b \ssp \tilde \sigma } \,,
\end{equation}
where $C = \pare{\cos^2 \tilde \omega_1 - \sin^2 \tilde \omega_0}^{-1/2}$ and $\tilde u_0^2 = \tilde u_1^2 = \tilde U$\,.
The constants $\tilde a$ and $\tilde b$ satisfy the constraints
\begin{equation}
\tilde a^2 + \tilde b^2 = - 1 + \tilde U \qquad
\mbox{and}
\qquad \tilde a \ssp \tilde b = C^2 \pare{ \tilde u_0 \ssp \sin \tilde\omega_0 \cos \tilde\omega_0 + \tilde u_1 \ssp \sin \tilde\omega_1 \cos \tilde\omega_1 } \,.
\end{equation}
The conserved charges are computed as
\begin{equation}
{\cal E} = n \ssp C^2 \ssp \tilde u_0 \Big( \Lambda - \sin^2 \tilde \omega_0 \, \eK (1) \Big) \,, \quad
{\cal S} = n \ssp C^2 \ssp \tilde u_1 \Big( \Lambda - \cos^2 \tilde \omega_1 \, \eK (1) \Big) \,, \quad
{\cal J} = n \ssp \tilde a \, \eK (1)\,,
\end{equation}
where we defined a cut-off $\Lambda \eq 1/(1 - q^2)$\,.

\paragraph{}
Let us pay special attention to the $\tilde u_0 = \tilde u_1 =
\sqrt{\tilde U}$ case. For this case the energy-spin relation
reads
\begin{equation}
{\cal E} - {\cal S} = n \sqrt{\tilde U} \; \eK (1)\,.
\label{inf iii E-S}
\end{equation}
Obviously the RHS is divergent, and careful examination reveals it
is logarithmic in ${\cal S}$\,. This can be seen by first
noticing, on one hand, that the complete elliptic integral of the
first kind $\eK (q) \equiv \eK (e^{-r})$ has asymptotic behavior
\begin{equation}
\eK (e^{-r}) = - \frac12 \ln \pare{\frac{r}{8}} + {\cal O} (r \ln r) \,,
\end{equation}
while on the other, the degree of divergence for the cutoff is $\Lambda = (1-q^2)^{-1} =(1 - e^{-2r})^{-1} \sim (2r)^{-1}$ as $r\to 0$\,.
Since the most divergent part of ${\cal S}$ is governed by
$\Lambda$ rather than $\eK(1)$\,, it follows that
\begin{equation}
\eK (e^{-r}) \sim \eK (1 - r) \sim - \frac12 \, \ln \pare{\frac{n \ssp C^2 \ssp \tilde u_1}{16 \ssp {\cal S}}}
\label{eK lnS}
\end{equation}
as $r\to 0$\,.
Then it follows that
\begin{equation}
{\cal E} - {\cal S} \sim - \frac{n \sqrt{\tilde U}}{2} \, \ln \pare{\frac{16 \cal S}{n \ssp C^2 \ssp \tilde u_1}}  \qquad (r \to 0)
\label{e-s-lns}
\end{equation}
as promised.

Let us consider the particular case $\tilde U = 1$\,, which is
equivalent to $\tilde{a}=\tilde{b}=0$ and $\tilde{\omega}_0 =
-\tilde{\omega}_1$\,. The above dispersion relation
\eqref{e-s-lns} now reduces to
\begin{equation}
E - S \sim \frac{n \sqrt{\lambda}}{2\pi} \; \ln S \,,
\label{log div}
\end{equation}
omitting the finite part. This result was first obtained in
\cite{Gubser:2002tv} for the $n=2$ case, and generalised to
generic $n$ case in \cite{Kruczenski:2004wg}.

\paragraph{}
One can also reproduce the double logarithm behavior of \cite{Frolov:2002av} (see also \cite{Beisert:2003ea, Belitsky:2006en, Frolov:2006qe, Casteill:2007ct}).
To see this, let us set $\tilde b=0$ and $\tilde a=\sqrt{\tilde U-1}$\,, and rewrite the relation \eqref{inf iii E-S} as
\begin{equation}
{\cal E} - {\cal S} = \sqrt{ \mathstrut {\cal J}^2 + n^2 \; \eK (1)^2} \sim \kko{\mathstrut {\cal J}^2 + \frac{n^2}{4} \ln^2 \pare{\frac{2 {\cal S}}{n \ssp C^2 \ssp \sqrt{\tilde U} }}}^{1/2} \,.
\label{inf iii E-S 2}
\end{equation}
There are two limits of special interest. The ``slow long string"
limit of \cite{Frolov:2006qe}, is reached by $\sqrt{U} \ll
\lambda$\,, so that in the strong coupling regime $\lambda \gg 1$
the RHS of (\ref{inf iii E-S 2}) becomes
\begin{equation}
{\cal E} - {\cal S} \sim \sqrt{\mathstrut {\cal J}^2 + \frac{n^2}{4} \ln^2 {\cal S} } \,.
\label{inf iii slowlong}
\end{equation}
Similarly, the ``fast long string" of \cite{Frolov:2006qe} is obtained by taking $\sqrt{U} \sim \lambda \gg 1$\,, resulting in
\begin{equation}
{\cal E} - {\cal S} \sim \kko{\mathstrut {\cal J}^2 + \frac{n^2}{4} \ko{\ln \pare{\frac{\cal S}{\cal J}} + \ln \pare{\ln r}}^2 }^{1/2} \sim \sqrt{\mathstrut {\cal J}^2 + \frac{n^2}{4} \ln^2 \pare{\frac{\cal S}{\cal J}} } \,,
\label{inf iii fastlong}
\end{equation}
where we neglected a term $\ln \pare{\ln r}$ which is relatively less divergent in the limit $r \to 0$\,.

\subsection{Type $\bmt{(iv)}$ helical strings}

Let us finally present another AdS helical solution which
incorporates the $SL(2)$ ``(dyonic) giant magnon'' of \cite{Minahan:2006bd,
Ryang:2006yq}. This solution, which we call the type $(iv)$
string, is obtained by applying a shift $X \to X + i \eK' (k)$ to
the type $(i)$ helical string. Its profile is given by
\begin{align}
\eta_0 &= \frac{C}{\sqrt{k}} \, \frac{\bTh _0 (0|k) \, \bTh _0 (X  - \iomm{0}|k)}{\bTh_0 (\iomm{0}|k) \, \bTh _1 (X|k)} \,
\exp \Big( \eZ_0 (\iomm{0}|k) X + i u_0 T \Big) \,,
\label{e0d_two} \\[2mm]
\eta_1 &= \frac{C}{\sqrt{k}} \, \frac{\bTh _0 (0|k) \, \bTh _3 (X  - \iomm{1}|k)}{\bTh_2 (\iomm{1}|k) \, \bTh _1 (X|k)} \,
\exp \Big( \eZ_3 (\iomm{1}|k) X + i u_1 T \Big) \,,
\label{e1d_two} \\[2mm]
\xi_1 &= \exp \pare{ i a T + i b X } \,.
\label{xi1d_two}
\end{align}
We omit displaying all the constraints among the parameters (they
can be obtained in a similar manner as in the type $(i)$ case).
The type $(iv)$ solution corresponds to the helical-wave
given in \eqref{CshG helical ds}, and satisfy the string equations
of motion of the form \eqref{GAL eom}. \footnote{This can be
easily checked by using a relation $1/k^2 \sn^2 (x|k) = \sn^2 \ko{
x +i\eK'(k)|k }$\,. }

\subsubsection*{$\bullet$ $\bmt{k\to 1}$ limit\,:~ $\bmt{SL(2)}$ ``dyonic giant magnon"}

The $SL(2)$ ``dyonic giant magnon'' is reproduced in the limit $k\to 1$\,, as
\begin{equation}
\eta_{0}=\f{\cosh(X-\iomm{0})}{\sinh X}\; e^{i(\tan\om_{0})X+i u_{0}T}\,,\quad
\eta_{1}=\f{\cos\om_{0}}{\sinh X}\; e^{i u_{1}T}\,,\quad
\xi_{1}=e^{i\hat a T + i\hat b X}\,,
\label{dual-GM}
\end{equation}
where
\begin{equation}
u_0^{2} = u_1^{2} + \frac{1}{\cos^{2} \omega_{0}} \,, \qquad
(\hat a, \hat b)=(u_{1},\tan\om_{0}) \ \ {\rm or}\ \ (\tan\om_{0}, u_{1})\,.
\end{equation}
Due to the non-compactness of AdS space, the conserved charges are
divergent. This is a UV divergence, and we regularise it by the following prescription.
First change the integration range for the charges (see (\ref{charge_def
E}\,-\,\ref{charge_def J})) from $\int_{0}^{2l}d\sig$ to
$\int_{\ep}^{2l-\ep}d\sig$\,, with $\epsilon>0$\,, to obtain
\begin{align}
\mathcal E &= u_{0}\cos^{2}\om_{0} \pare{\ep^{-1}-1} +\eK(1)(u_{0}-v\tan\om_{0})\,,\\
\mathcal S &= u_{1}\cos^{2}\om_{0} \pare{\ep^{-1}-1}\,,\\
{\mathcal J}&=\eK(1)(u_{0}-v\tan\om_{0})\,,
\end{align}
then drop the terms proportional to $\ep^{-1}$ by hand. This prescription yields
a regularised energy and an $S^{5}$ spin which are still IR divergent
due to the non-compactness of the worldsheet. However, their
difference becomes finite, leading to the energy-spin relation
\begin{equation}
(\mathcal E-\mathcal J)_{\rm reg}=-\sqrt{(\mathcal S)_{\rm reg}^{2}+\cos^{2}\om_{0}}\,.
\end{equation}
Note that in view of the AdS/CFT correspondence, ${\cal E} - {\cal J}$ must be positive, which in turn implies $({\cal E} - {\cal J})_{\rm reg}$ is negative.

\begin{figure}[htbp]
\begin{center}
\vspace{0.5cm}
\includegraphics[scale=0.9]{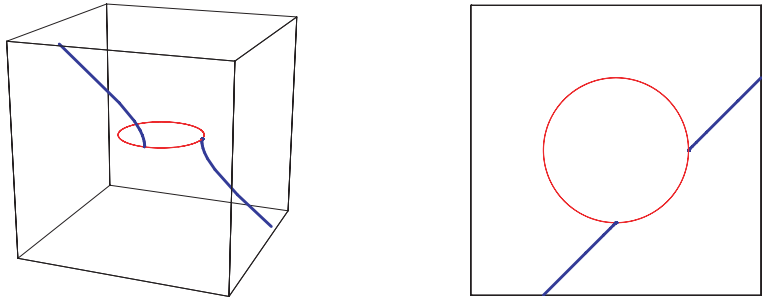}
\vspace{0.5cm}
\caption{\small $k\to 1$ limit of type $(iv)$ helical string ($\om_{0}=0.785$\,, $u_{0}=1.41$\,, $u_{1}=0$) becomes a ``giant magnon'' solution in AdS space.}
\label{fig:dual-GM}
\end{center}
\end{figure}

Let us take $v = \tan \omega_0/u_0$ in (\ref{dual-GM}), and
consider a rotating frame ${\eta}_{0}^{\rm new} = e^{- i \tilde
\tau} \eta_0 \equiv \tilde Y_0 + i \tilde Y_5$\,. We then find
$\tilde Y_5=-i\sin\om_{0}$ is independent of $\tilde \tau$ and
$\tilde\sigma$\,, showing that the ``shadow'' of the $SL(2)$
``dyonic giant magnon'' projected onto the $\tilde Y_0$-$\tilde Y_5$ plane is just
given by two semi-infinite straight lines on the same line.
Namely, the shadow is obtained by removing a finite segment from
an infinitely long line, where the two endpoints of the segment
are on the unit circle $|\eta_{0}|=1$ with angular difference
$\Delta t = \pi - 2\omega_0$\,. Figure \ref{fig:dual-GM} shows the
snapshot of the $SL(2)$ ``dyonic giant magnon'', projected onto the plane
spanned by $(\Re\!\eta_{0}, \Im\!\eta_{0}, |\eta_{1}|)$\,.

It is interesting to compare this situation with the usual giant magnon on $\mathbb R\times S^3$\,.
In the sphere case, the ``shadow'' of the giant magnon is just a straight line segment connecting two endpoints on the equatorial circle $|\xi_{1}|=1$\,.
So the ``shadows'' of $SU(2)$ and $SL(2)$ giant magnons are just complementary.
Using this picture of ``shadows on the Lin-Lunin-Maldacena (LLM) plane \cite{Lin:2004nb}'', one can further discuss the ``scattering'' of two $SL(2)$ ``(dyonic) giant magnons'' in the similar manner as in the $SU(2)$ case \cite{Hofman:2006xt}.\footnote{Scattering $SL(2)$ (dyonic) giant magnon solutions can be constructed from the scattering $SU(2)$ (dyonic) giant magnon solutions $\xi_{i}(u_{1},u_{2};v_{1},v_{2})$ \cite{Spradlin:2006wk} by performing $(u_{1},u_{2})\mapsto (u_{1}+i\pi/2,u_{2}+i\pi/2)$\,.}

Interestingly, these ``shadow'' pictures remind us of the corresponding
finite-gap representations of both solutions, resulting from the
$SU(2)$ and $SL(2)$ spin-chain analyses. While in the $SU(2)$
case, a condensate cut, or a Bethe string, has finite length in
the imaginary direction of the complex spectral parameter plane,
for the $SL(2)$ case, they are given by two semi-infinite lines in
the same imaginary direction \cite{Minahan:2006bd}. This
complementary feature reflects the structural symmetry between the
BDS parts of S-matrices, $S_{SU(2)}=S_{SL(2)}^{-1}$\,, see (\ref{all-loop rank-1}).

These ``shadow'' pictures also show up in matrix model context
\cite{Berenstein:2005jq,Vazquez:2006hd,Hatsuda:2006ty,Berenstein:2007zf}. In a
reduced matrix quantum mechanics setup obtained from $\mathcal N =
4$ SYM on $\mathbb R\times S^{3}$\,, a ``string-bit'' connecting
eigenvalues of background matrices forming $\mbox{\large $\f{1}{2}$}$-BPS circular
droplet can be viewed as the shadow of the corresponding string.
For the $SU(2)$ sector, it is true even for the boundstate (bound
``string-bits'') case \cite{Hatsuda:2006ty}. It would be
interesting to investigate the $SL(2)$ case along similar lines of
thoughts.

\chapter[Giant Spinons]
	{Giant Spinons\label{sec:gauge theory}}

In \cite{Hayashi:2007bq}, we claimed that the gauge theory dual operators of type $(i)'$ and $(ii)'$ of Section \ref{sec:helical large-winding}, which are oscillating strings with large winding numbers, are only found in a non-holomorphic sector.
Such a non-holomorphic sector has been much less explored than the holomorphic, large-spin sectors, because of its intractability mainly related with the non-closedness, or difficulty of perturbative computations. Nevertheless, since our results \cite{Okamura:2006zv,Hayashi:2007bq} complete the whole catalog of classical, elliptic strings on $\mathbb R\times S^{3}$\,, we hope they could shed more light not only on holomorphic but also non-holomorphic sectors of the string/spin-chain duality, for a deeper understanding of AdS/CFT.
As a first step, in this appendix, we give a possible arguments about the identification of the gauge theory duals of the 2D transformed strings, namely the oscillating strings with large windings.

\subsubsection*{Gauge theory duals of the 2D transformed strings}
As for Cases ${\rm I}$ and ${\rm II}$ we summarised in the end of Section \ref{sec:helical large-winding}, the gauge theory duals are
well-known. They are all of the form
\begin{equation}
\mathcal O\sim \tr\ko{\cZ^{L-M}\cW^{M}}+\dots\,,
\end{equation}
 with $L$ very large.
For example, for the type $(i)$ case, a BPS string ($k\to 0$) of
course corresponds to $M=0$\,, and a BMN string corresponds to $M$
very small. A dyonic giant magnon corresponds to an $M$-magnon
boundstate in the asymptotic SYM spin-chain ($L\to \infty$), which
is described by a straight Bethe string in rapidity plane
\cite{Dorey:2006dq, Chen:2006gq}. In the Bethe string, all $M$
roots are equally spaced in the imaginary direction, reflecting
the pole condition of the asymptotic S-matrix. As to the elliptic
folded/circular strings, they correspond to, respectively, the
so-called double-contour/imaginary-root distributions of Bethe
roots \cite{Beisert:2003xu}.

\paragraph{}
Now let us turn to the present oscillating case.
First we discuss the two-spin single-spike string case. 
In contrast to the dyonic giant magnon, it has finite spins $J_{i}$ $(i=1,2)$ and infinite
energy. This fact allows us to argue that the relevant dual SYM
operators should look like\footnote{Or one may just write (\ref{op}) as $\mathcal O_{\rm spi}=\tr\big(\cZ^{K''}\,\cW^{M}\,{\mathcal S}^{(L-K''-M)/2}\big)+\dots$ with $L\to \infty$ and finite $K''$ and $M$\,.
We put it as (\ref{op}) simply because we found it more convenient to do so for our purpose.}
\begin{equation}
\mathcal O_{\rm spi}= \tr\ko{\cZ^{K}\,\overline{\cZ}{}^{K'}\,\cW^{M}\,{\mathcal S}^{(L-K-K'-M)/2}}+\dots\,,\quad
L\,, K\,, K'\to \infty\,,\quad K-K'\,, M : \mbox{finite}\,.
\label{op}
\end{equation}
In (\ref{op}), the factor
${\mathcal S}$ appearing in (\ref{op}) is an $SO(6)$\,-singlet composite including $\cZ\overline{\cZ}$ {\rm etc}.\footnote{The $SO(6)$ sector is not closed beyond
one-loop level in $\lam$\,, and operator mixing occurs in the full
$PSU(2,2|4)$ sector due to the higher-loop effects. 
However, we can still expect that such mixing
into $PSU(2,2|4)$ is suppressed in our classical ($L\to \infty$)
setup as in \cite{Minahan:2004ds}. We would like to thank J.~Minahan for discussing this
point.}
One can easily understand that the pairs like $\cZ\overline{\cZ}$ give rise to oscillating motion in the sting side, since if we associate $\cZ$ to a particle rotating along a great circle of $S^{5}$ say clockwise, the other particle associated with $\overline{\cZ}$ rotates counterclockwise, thus making the string connecting these two points non-rigid and oscillating.
The dots in (\ref{op}) denotes terms that mix under
renormalisation. An important assumption is that $M$ $\cW$s form a
boundstate. Indeed loop-effects mix $\cZ\overline{\cZ}$ with other
neutral combinations $\cW\overline{\cW}$ and $\cY\overline{\cY}$\,, but it is assumed the boundstate condition still holds. 
Let $X^{\pm}$ be the spectral parameters assigned to the
boundstate. We write them as
\begin{alignat}{3}
X^{\pm}&=R\,e^{\pm iP/2}
&\quad &\mbox{with}\quad R=\f{M+\sqrt{M^{2}+16g^{2}\sin^{2}\ko{P/2}}}{4g\sin\ko{P/2}} ~ (>1)\,,
\end{alignat}
where $P$ is the momentum carried by the boundstate. Recall that
we took $\tr\ko{\cZ\cZ\dots}$ as the vacuum state, therefore $\cW$ is an
excitation above the vacuum with
$\Delta_{0}-J_{1}=1$\,, whereas $\overline{\cZ}$ is an
excitation with $\Delta_{0}-J_{1}=2$\,.\footnote{$\overline{\cZ}$ is not a fundamental excitation. It is an excitation corresponding to a two-magnon state, see the comment below (\ref{SO(6) fields}).} The composite ${\mathcal
S}$ also contributes to the spin-chain energy in some way, and we
must take all the contributions into account when evaluating the
total energy $\Delta_{\mathcal O_{\rm spi}}-J_{1}$ of (\ref{op}).
We assume that the contribution of $M$ $\cW$s results in two parts;
one is the boundstate energy that contributes in the same way as
in the case of an $SU(2)$ boundstate $\mathcal O_{\rm mag}\sim
\tr\ko{\cZ^{K}\cW^{M}}+\dots$ $(K\to \infty)$\,, and the other is its
interactions with other fields. One can then write down the total
energy as
\begin{align}
\Delta_{\mathcal O_{\rm spi}}-(K-K')=
\f{g}{i}\kko{\ko{X^{+}-\f{1}{X^{+}}}-\ko{X^{-}-\f{1}{X^{-}}}}
+\chi\,.
\label{dr1}
\end{align}
The first term in RHS comes from the boundstate $\cW^{M}$\,, while the last $\chi$ accounts for contributions concerning ${\mathcal S}$\,, $\overline{\cZ}$ and all their interactions with other fields, including $\cW$s\,. Currently we have no
knowledge of how the actual form of $\chi$ looks like, and so we leave it as some function of the coupling and boundstate momentum here (however, we will
later discuss its form in the strong coupling, infinite-winding
limit). One can also express the $J_{2}$\,-charge carried by the
boundstate in terms of the spectral parameters as
\begin{align}
M=\f{g}{i}\kko{\ko{X^{+}+\f{1}{X^{+}}}-\ko{X^{-}+\f{1}{X^{-}}}}\,.
\label{M1}
\end{align}

Now perform a change of basis for the spin-chain, and take $\tr\ko{\overline{\cZ}\,\overline{\cZ}\dots}$ as the vacuum state, instead of $\tr\ko{\cZ\cZ\dots}$\,.
This particular transformation of SUSY multiplet, namely the charge conjugation, maps the original $\cW^{M}$ to $\overline{\cW}{}^{M}$ with new spectral parameters
\begin{equation}
{\widetilde X}^{\pm}=1/X^{\pm}\,.
\end{equation}
This is actually a crossing transformation that maps a usual
particle to its conjugate particle (antiparticle)
\cite{Beisert:2005tm}.\footnote{This crossing transformation could be related to the $\ts$ flip considered in Section \ref{sec:helical large-winding}.
We thank M.~Staudacher for this remark.}
In the new basis, $\overline{\cW}$s, $\cZ$s and
$\overline {\mathcal S}={\mathcal S}$ play the role of excitations
above the new vacuum. The contribution of $\mathcal S$ to the new vacuum should be the same as in the old case since it is an $SO(6)$ singlet, and we assume the total contributions from all excitations to be the same as in the old case.
Then one obtains a relation similar to
(\ref{dr1}),
\begin{align}
\Delta_{\mathcal O_{\rm spi}}-(K'-K)=\f{g}{i}\kko{\ko{{\widetilde X}^{+}-\f{1}{{\widetilde X}^{+}}}-\ko{{\widetilde X}^{-}-\f{1}{{\widetilde X}^{-}}}}+\chi\,,
\label{dr2}
\end{align}
and similarly for the second charge. From (\ref{dr1})-(\ref{dr2}),
it follows that
\begin{align}
\Delta_{\mathcal O_{\rm spi}}=\chi
\qquad\mbox{and}\qquad
K'-K=\sqrt{M^{2}+16 g^{2}\sin^{2}\ko{\f{P}{2}}}\,.\label{Delta and K-K'}
\end{align}
Then if we identify naturally
\begin{equation}
K-K'\eq J_{1}\,,\quad
M\eq J_{2}\quad
\mbox{and}\quad
P\eq 2\pi m \pm 2\bar \theta\quad (m\in \mathbb Z\,; ~ 0\leq \bar{\theta}\leq \pi/2)\,,
\label{identification1}
\end{equation}
the second relation in (\ref{Delta and K-K'}) precisely reproduces
the dispersion relation (\ref{spin spin}) for single-spike strings, after
substituting $g^{2}=\lam/16\pi^{2}$\,. Here we included an integer
degree of freedom $m$ that plays the role of the winding number on
the string theory side. One can also deduce that
\begin{equation}
\f{J_{2}}{J_{1}}=\f{R^{2}-1}{R^{2}+1}\,,
\end{equation}
which corresponds to $\sin\gamma$ in the notation used in
\cite{Ishizeki:2007we}. In (\ref{identification1}), one may choose
either the plus/minus signs in $P$\,; they correspond to the
momenta of a particle/antiparticle.

Notice also the above argument, resulting in
\begin{align}
-J_{1}&=\f{g}{i}\kko{\ko{X^{+}-\f{1}{X^{+}}}-\ko{X^{-}-\f{1}{X^{-}}}}\,,\\
J_{2}&=\f{g}{i}\kko{\ko{X^{+}+\f{1}{X^{+}}}-\ko{X^{-}+\f{1}{X^{-}}}}\,,
\end{align}
is consistent with what we found in Section \ref{sec:FG}, equations (\ref{J_1 res-2}, \ref{J_2 res-2}), if we, as usual, identify the string theory spectral parameters $x_{1}$ and $\bar{x}_{1}$ (in finite-gap language) with the ones for gauge theory $X^{+}$ and $X^{-}$ (for the boundstate).

\paragraph{}
To proceed in the reasoning, suppose the asymptotic behavior of
$\chi$ in the strong coupling and infinite-``winding'' limit
becomes
\begin{equation}
\chi\sim 2gP=m\sqrt{\lam}\pm \f{\bar \theta}{\pi}\,,\qquad (m\to \infty)\,.
\label{c}
\end{equation}
We kept here $\pm\bar\theta/\pi$ term to ensure that $\chi$ is not
just given by $\mbox{(integer)}\times \sqrt{\lam}$ but contains
some continuous shift away from that. We will give more
explanations concerning this conjecture soon. The relation
(\ref{c}) then implies that
\begin{equation}
\Delta_{\mathcal O_{\rm spi}} - \f{\sqrt{\lam}}{2\pi}\cdot 2\pi m
=\pm\f{\sqrt{\lam}}{\pi}\,\bar\theta\,,
\label{ene-wind-2}
\end{equation}
where we used the identifications we made before. This can be
compared to the string theory result for the single-spike,
(\ref{ene-wind}). The integer $m$ here corresponds to the winding
number $N_{1}$ there (recall that for single spike case, we had
$\Delta \varphi_{1}=2\pi N_{1}$ due to the periodicity condition).
When there are $n$ boundstates
in the spin-chain all with the same momentum $P$\,, RHS of
(\ref{ene-wind-2}) is just multiplied by $n$ and modified to
$n(\sqrt{\lam}/\pi)\,\bar \theta$\,, which corresponds to an array
of $n$ single-spikes.

\paragraph{}
Let us explain the conjecture (\ref{c}) in greater detail. Of
course one of the motivations is that it reproduces the relation
(\ref{ene-wind-2}) of the string theory side, as we have just seen. Further
evidence can be found by considering particular sets of operators
contained in (\ref{op}) and checking for consistency.
For example, let us consider the limit $K-K'\to 0$ and $M\to 0$\,.
This takes the operator (\ref{op}) to the form $\tr\ko{(\cZ\overline{\cZ})^{K}{\mathcal
S}^{L/2-K}}+\dots$\,, which must sum up to the singlet operator $\tr{\mathcal
S}^{L/2}$ for it to be a solution of the Bethe ansatz equation.
In this limit, the ``angle'' $\bar \theta$ should vanish in view of the second
equation in (\ref{Delta and K-K'}) and (\ref{identification1}).
Therefore the relation (\ref{c}) together with the first equation
in (\ref{Delta and K-K'}) imply that the conformal dimension of
the singlet operator is just given by
\begin{equation}
\Delta_{\tr{\mathcal S^{L/2}}}\Big|_{L\to \infty}=m\sqrt{\lam}\qquad (m\to \infty)\,,
\label{k->1 pulsating}
\end{equation}
which agrees with the energy expression (\ref{winding energy}) of the $\ts$ transformed point-like BPS string (in the limit $\mu\sqrt{U}\to \infty$), under the identification $N_{2}=m$\,.

\paragraph{}
As we have seen, in contrast to the dyonic giant magnon vs.\
magnon boundstate $\mathcal O_{\rm mag}\sim
\tr\ko{\cZ^{\infty}\cW^{M}}+\dots$ case, the correspondence between two-spin
single-spike vs.\ $\mathcal O_{\rm spi}$ given in (\ref{op}) is
slightly more involved. In the former correspondence in the
infinite spin sector, the magnon boundstate is an excitation above
the BPS vacuum $\mathcal O_{\rm F}\sim \tr\ko{\cZ^{\infty}}$\,, and
one can think of the boundstate $\cW^{M}$ as the counterpart of the
corresponding dyonic giant magnon. For the latter case in the
infinite winding sector, however, it is not the boundstate $\cW^{M}$
alone but the ``$\cZ^{K}\,\overline{\cZ}{}^{K'}\,\cW^{M}+\dots$'' (or ``$\cZ^{K''}\,\cW^{M}+\dots$'' with $K''=K-K'$) part of
$\mathcal O_{\rm spi}$ that encodes the two-spin single-spike. It can be
viewed as an excitation above the $SO(6)$ singlet operator
$\mathcal O_{\rm AF}\sim \tr \mathcal S^{L/2}$\,. Actually this is the
``antiferromagnetic'' state of the $SO(6)$ spin-chain, which is
``the farthest from BPS'' (notice that a solution of the Bethe ansatz equation with
$J_{1}=J_{2}=J_{3}=0$ is nothing but the $SO(6)$ singlet state).
It is dual to the rational circular static string (\ref{rational static
prof}) obtained by performing a $\tau\leftrightarrow\sig$
transformation on the point-like BPS string.

\paragraph{}
To summarise, the gauge theory duals of the
$\ts$ transformed strings (derivatives of type $(i)'$ and $(ii)'$
helical strings) were identified with operators of the form
\begin{equation}
\mathcal O\sim \tr\ko{\cZ^{K}\,\overline{\cZ}{}^{K'}\,\cW^{M}\,{\mathcal S}^{(L-K-K'-M)/2}}+\dots
\label{op:gen}
\end{equation}
with $\mathcal S$ being an $SO(6)$ singlet composite.
The single-spike limit $k\to 1$ was identified with the
$K\,, K'\to \infty$ limit while keeping $K-K'$ and $M$ finite (see
(\ref{op})). In this limit, the ``$\cZ^{K}\,\overline{\cZ}{}^{K'}\,\cW^{M}+\dots$''
part in the operator, of which $\cW^{M}$ is assumed to form a boundstate, was claimed to be responsible for the transverse
excitation (spikes) of the string state winding infinitely many
times around a great circle of $S^{5}$\,. In other words, the
spikes are dual to excitations above the ``antiferromagnetic''
state $\tr \mathcal S^{L/2}$\,.
The ``antiferromagnetic'' state is the singlet
state of the $SO(6)$ spin-chain, and located at ``the farthest from BPS'' in the spin-chain spectrum.
One might be then tempted to call these ``spiky'' objects ``giant spinons''.
These features can be
compared to that of magnons in the large spin sector (impurity
above BPS vacuum) corresponding to the transverse excitations of the
point-like string orbiting around a great circle of $S^{5}$\,.

\paragraph{}

We have so far given a very naive discussion on the gauge theory dual operator for the 2D transformed helical strings. 
It would be interesting to check the prediction (\ref{c}) directly
by using the conjectured AdS/CFT Bethe ansatz equation. In the
$SU(2)$ sector where the number of operators is finite, the nature
of the antiferromagnetic state is better understood
\cite{Rej:2005qt}, and the upper bound on the energy is known
\cite{Zarembo:2005ur} (see also \cite{Roiban:2006jt}). It is
proportional to $\sqrt{\lam}$\,, which is the same behavior as our
conjecture (\ref{c}). Recall that we argued the $SO(6)$ singlet
state was dual to a large winding string state with
zero-spins, (\ref{rational static prof}). If the prediction
(\ref{c}) is correct, then we should be able to reproduce it by
the $SO(6)$ Bethe ansatz equation approach. An approach similar to
\cite{Zarembo:2005ur} would be useful. In this case, the ``(two-spin) giant spinon'' part ``$\cZ^{K}\,\overline{\cZ}{}^{K'}\,\cW^{M}+\dots$'' could be
understood as (macroscopic number of) ``holes'' made in the
continuous mode numbers associated with an $SO(6)$ singlet Bethe
root configuration.\footnote{See also the remarks in the last paragraph of this appendix.}
In the weak coupling regime, the
$SO(6)$ singlet Bethe root configuration and excitations above it
were studied in \cite{Minahan:2002ve, Engquist:2003rn,
Minahan:2004ds}.
An $SO(6)$ singlet state was also studied in \cite{Rej:2007vm}, where an integral equation for the Bethe root density was derived.
It would be interesting to study generic $SO(6)$ singlet states at strong coupling, and compare them with our results.\footnote{We thank M.~Staudacher for pointing this out to us.}

Since the $\ts$ transformed string solutions discussed in this
thesis are periodic classical solutions, one can define
corresponding action variables, namely the oscillation numbers. By
imposing the Bohr-Sommerfeld quantization condition, one obtains
integer valued action variables, which from lesson of the large
spin sector \cite{Chen:2006ge} we can again expect to correspond
to filling fractions defined for the $SO(6)$ spin-chain. It would
be interesting to understand this correspondence from the
finite-gap perspective along the lines of \cite{Dorey:2006zj,
Dorey:2006mx}.

It would be also interesting to compare the spectra of AdS/CFT
near the $SO(6)$ ``antiferromagnetic'' vacuum by an effective
sigma model approach (without any apparent use of integrability)
\cite{Kruczenski:2003gt}.\footnote{We gave a brief introduction of this approach in Appendix \ref{sec:Kruczenski}.}
In the $SU(2)$ case, a similar approach
was taken in \cite{Roiban:2006jt}, where a continuum limit of the
half-filled Hubbard chain was compared to an effective action for
``slow-moving'' strings with $J_{1}=J_{2}$\,. In our case, some
Hubbard-like model with $SO(6)$ symmetry would give clues.

\subsubsection*{``Dressing/nesting'' and ``magnons/spinons''}

Finally, we comment on some possible implications of a more recent paper \cite{Ishizeki:2007kh}, in which a scattering state of two single-spikes were constructed by the dressing method \cite{Spradlin:2006wk,Kalousios:2006xy}.
Using the scattering solution, the scattering phase-shift for two single-spikes were determined.
Remarkably, the phase-shift $\Th_{\rm spi}$ agreed with the one $\Th_{\rm mag}$ for dyonic giant magnon \cite{Chen:2006gq}, up to non-logarithmic terms.
We believe \cite{KO} that this feature is a manifestation of a close relation between the ``nesting'' and ``dressing'' \cite{Rej:2007vm} for the reason briefly explained below.

First recall from our argument that the single-spike string should be obtained as a solution of the nested Bethe ansatz equation in the strong coupling limit, presumably for the $SO(6)$ sector, while the giant magnon is a solution of a simple, unnested Bethe ansatz equation for the $SU(2)$ sector.
Recall also, as we saw in Chapter \ref{chap:S-matrices in HM}, that the scattering phase-shift for the two (dyonic) giant magnons is reproduced from the strong coupling limit of the conjectured $SU(2)$ S-matrix including the dressing phase.
Suppose if we could at all find out an ``antiferromagnetic'' ($SO(6)$ singlet) vacuum above which fundamental excitations scatter without acquiring the dressing phase.\footnote{See \cite{Sakai:2007rk, Sakai:2007ie} and also Appendix D of \cite{Rej:2007vm} for a possibly related point of view.}
The single-spike string, which can be viewed as an excited state above the purely winding string state (\ref{rational static prof}) \cite{Hayashi:2007bq, Ishizeki:2007kh}, would then correspond to a spinon excitation with particular fillings $\{ K_{i} \}$\,.
It is tempting to argue that the scattering phase-shift $\Th_{\rm mag}$ for the giant magnons and the one $\Th_{\rm spi}$ for the single-spikes (``giant spinons'') agree due to that the ``nesting kernel'' \cite{Rej:2007vm} coming from the nested Bethe ansatz equations
agrees with the ``dressing kernel'' \cite{Beisert:2006ez} for the dressing phase.
This scenario, if at all works, could thus give a possible resolution of the curious observation in \cite{Ishizeki:2007kh}.

Actually, in the strong coupling and the thermodynamic limit, a particular configuration of infinite number of Bethe roots, which is an array of two-strings (and holes) along the real axis of the rapidity plane, reproduces the known dispersion relation for the giant spinons \cite{KO}.
The infinite number of Bethe string serves as the ``Dirac sea''.
The angle $\bar \theta$ in (\ref{ene-wind}) can be identified with a specific geometrical angle in the spectral parameter plane mapped from the rapidity plane, and can be interpreted as a momentum just as in (\ref{identification1}).
This feature can be compared to what is observed in the giant magnon case\cite{Hofman:2006xt}, in which case the projection of the string profile onto the ``equatorial plane'' could be directly identified with a straight stick in the LLM plane \cite{Lin:2004nb} whose endpoints being located on the ``equatorial circle''.
It can be further identified with the finite-gap description in string theory, or the Bethe string configuration in gauge theory.
It would be interesting to understand more about the new plane on which the giant spinon profile is projected.

\chapter[Elliptic Functions and Elliptic Integrals]
	{Elliptic Functions and Elliptic Integrals\label{app:Elliptic}}

\section{Definitions and identities for elliptic functions\label{app:Elliptic Functions}}

Our conventions for the elliptic functions, elliptic integrals are presented below.

\paragraph{Elliptic theta functions.}
Let $Q=\prod\limits_{n = 1}^\infty  {\left( {1 - e^{2\pi in\tau } } \right)}$\,. 
We define elliptic theta functions by
\begin{align}
\vartheta _0 \left( {z|\tau } \right) &\defeq Q\prod\limits_{n = 1}^\infty  {\left( {1 - 2 \, e^{\pi i (2n - 1) \tau } \cos (2\pi z) + e^{2\pi i (2n - 1) \tau } } \right)}\,,  \\
\vartheta _1 \left( {z|\tau } \right) &\defeq 2 \, Q \, e^{i \pi \tau/4} \sin (2\pi z)\prod\limits_{n = 1}^\infty  {\left( {1 - 2 \, e^{2\pi in\tau } \cos (2\pi z) + e^{4\pi in\tau } } \right)}\,,  \\
\vartheta _2 \left( {z|\tau } \right) &\defeq 2 \, Q \, e^{i \pi \tau/4} \cos (2 \pi z) \prod\limits_{n = 1}^\infty  {\left( {1 + 2 \, e^{2\pi in\tau } \cos (2\pi z) + e^{4\pi in\tau } } \right)}\,,  \\
\vartheta _3 \left( {z|\tau } \right) &\defeq Q \prod\limits_{n = 1}^\infty  {\left( {1 + 2 \, e^{\pi i (2n - 1) \tau } \cos (2\pi z) + e^{2\pi i (2n - 1) \tau } } \right)}\,.
\end{align}
We also use an abbreviation $\vartheta _\nu ^0  \equiv \vartheta _\nu  (0|k)$\,.
The following functions are known as Jacobian theta and zeta functions, respectively:
\begin{equation}
\bTh _\nu  \left( {z|k} \right) \equiv \vartheta _\nu  \left( \left.\hf{z}{2\eK}\right| \tau  = \hf{{i\eK'}}{\eK} \right),\qquad
\eZ_\nu  \left( {z|k} \right) \equiv \frac{{\partial _z \bTh _\nu  \left( {z|k} \right)}}{{\bTh _\nu  \left( {z|k} \right)}}\,,\qquad 
\nu=0,1,2,3\,.
\end{equation}

\paragraph{Complete elliptic integrals.}
Complete elliptic integral of the first kind and its complement are defined as, respectively,
\begin{equation}
\eK(k) \defeq \int_0^1 \frac{dz}{\sqrt{(1-z^2)(1-k^2 z^2)}} \,,\qquad
\eK'(k) \defeq \eK (\sqrt{1 - k^2}) \,.
\end{equation}
We often write $\eK(k)$ as $\eK$\,. 
Likewise, we omit the moduli parameter $k$ of other elliptic functions or elliptic integrals as well.
There are alternative expressions for $\eK$ and $\eK'$ in terms of elliptic theta functions\,:
\begin{equation}
\eK(k) = \frac{\pi (\vartheta_3 ^0)^2}{2}\,, \qquad
\eK'(k) = - i \eK (k)\tau = \frac{\pi i \tau (\vartheta_3 ^0)^2}{2}\,.
\end{equation}
Complete elliptic integral of the second kind is defined as 
\begin{equation}
\eE(k) \defeq \int_0^1 dz\sqrt{\frac{1 - k^2 z^2}{1 - z^2}}
= \int_0^{\eK(k)} du\dn^2 u\,.
\end{equation}
In angle variable, they are also written as
\begin{equation}
\eK(k) =\int_{0}^{\pi/2} \frac{d \varphi}{\sqrt{1-k^{2}\sin^{2}\varphi}}\,,\qquad 
\eE(k) =\int_{0}^{\pi/2} d \varphi\,\sqrt{1-k^{2}\sin^{2}\varphi}\,.
\end{equation}
These are related to the hypergeometric functions as
\begin{equation}
 {}_2{\mathrm F}_1\ko{\hf{1}{2},\hf{1}{2};1;k}=\f{2}{\pi}\,\eK(k)\, ,\qquad 
 {}_2{\mathrm F}_1\ko{-\hf{1}{2},\hf{1}{2};1;k}=\f{2}{\pi}\,\eE(k)\, .
\end{equation}
\paragraph{Identities.}
The following modular transformations, 
\begin{alignat}{3}
&\eK\komoji{(1/k)}&{}&=k\left(\eK(k)-i \eK(1-k)\right)\,, \label{mod.trans-1}\\
&\eE\komoji{(1/k)}&{}&=k\left[ {\eE(k)+i \eE(1-k)-(1-k^{2})\eK(k)-i k^{2}\eK(1-k)}\right]\,,\label{mod.trans-2}\\
&\eK\komoji{(1-1/k)}&{}&=k\,\eK(1-k)\,,\label{mod.trans-3}\\
&\eE\komoji{(1-1/k)}&{}&=\f{1}{k}\,{\eE(1-k)}\,,\label{mod.trans-4}
\end{alignat}
and the Legendre relation,
\begin{equation}
\eK(k)\eE(1-k)-\eK(k)\eK(1-k)+\eE(k)\eK(1-k)=\f{\pi}{2}\,,\label{Legendre Relation}
\end{equation} 
are useful for the computation of two-cut finite gap problems.

\paragraph{Jacobian elliptic functions.}
Jacobian sn, dn and cn functions are defined as
\begin{gather}
\sn(z) \defeq \frac{{\vartheta _3^0 }}{{\vartheta _2^0 }} \, \frac{{\vartheta _1 (w)}}{{\vartheta _0 (w)}} \,,\quad
\dn(z) \defeq \frac{{\vartheta _0^0 }}{{\vartheta _3^0 }} \, \frac{{\vartheta _3 (w)}}{{\vartheta _0 (w)}} \,,\quad
\cn(z) \defeq \frac{{\vartheta _0^0 }}{{\vartheta _2^0 }} \, \frac{{\vartheta _2 (w)}}{{\vartheta _0 (w)}} \,, \label{sncndn_def}
\end{gather}
where $z = \pi \pare{\vartheta_3^0}^2 w = 2 \, \eK w$\,. In terms of Jacobian theta functions, they can be written as
\begin{gather}
\sn(z) = \frac{{\bTh _3 (0) }}{{\bTh _2 (0) }} \, \frac{{\bTh _1 (z)}}{{\bTh _0 (z)}} \,,\quad
\dn(z) = \frac{{\bTh _0 (0) }}{{\bTh _3 (0) }} \, \frac{{\bTh _3 (z)}}{{\bTh _0 (z)}} \,,\quad
\cn(z) = \frac{{\bTh _0 (0) }}{{\bTh _2 (0) }} \, \frac{{\bTh _2 (z)}}{{\bTh _0 (z)}} \,. \label{sncndn_def2}
\end{gather}
The moduli $k$ and $k'\equiv \sqrt{1-k^{2}}$ are related to the elliptic theta functions by 
\begin{equation}
k \equiv \left( {\frac{{\vartheta _2^0 }}{{\vartheta _3^0 }}} \right)^2 ,\qquad k' \equiv \left( {\frac{{\vartheta _0^0 }}{{\vartheta _3^0 }}} \right)^2\,.  \label{kck_def}
\end{equation}
The Jacobian elliptic functions satisfy the following relations\,:
\begin{equation}
\sn^2 (z|k) + \cn ^2 (z|k) = 1, \qquad k^2 \sn^2 (z|k) + \dn ^2 (z|k) = 1\,.
\label{sn cn dn}
\end{equation}
The period of $\mathrm{sn}\ko{u|k}$ and $\mathrm{cn}\ko{u|k}$ is $4\,\eK(k)$\,, while the period of $\mathrm{dn}\ko{u|k}$ is $2\,\eK(k)$\,:
\begin{alignat}{5}
\mathrm{sn}\ko{u+2\,\eK(k)|k}&=-\mathrm{sn}\ko{u|k}\,,\label{periodicity-sn}\\
\mathrm{cn}\ko{u+2\,\eK(k)|k}&=-\mathrm{cn}\ko{u|k}\,,\label{periodicity-cn}\\
\mathrm{dn}\ko{u+2\,\eK(k)|k}&=\mathrm{dn}\ko{u|k}\,.\label{periodicity-dn}
\end{alignat}

\section{Some details of calculations\label{app:Details of Calculations}}

Below we will collect some key formulae that are useful in performing the calculation involving the function of the form
\begin{align}
\Xi (X,T,w) &= \frac{\bTh _1 (X - X_0 - w + w_0)}{\bTh _0 (X - X_0) \ssp \bTh_0 (w - w_0)} \, 
\exp \Big( \ssp \eZ_0 (w - w_0) (X - X_0) + i \ssp u (T - T_0) \Big)\,.
\label{BAF}
\end{align}
Here $u^2 = U + \dn^2 (w - w_0)$ and $X$\,, $X_0$\,, $T$ and $T_0$ are all real, and $w$ and $w_0$ assumed to be purely imaginary.
The degrees of freedom of $(T_{0}, X_{0})$ correspond to the initial values for the phase of (\ref{BAF}), and in what follows, we will set them to zero.
We will also set $w_{0}=0$\,.

\paragraph{}
As a preliminary, we shall write down several useful formulae concerning elliptic functions.

\medskip \noindent $\bullet$\ \ 
One can express $\eZ_0 (z|k)$ in terms of Jacobian dn function and complete elliptic integrals of the first and second kind as
\begin{equation}
\eZ_0(z|k) = \int_0^z du\dn^2 (u|k)  - z \, \frac{{\eE}(k)}{{\eK}(k)}\,,\quad 
\mbox{\em i.e.},\quad \frac{\partial \eZ_0 (u|k)}{\partial u} = \dn^2 (u|k) - \frac{\eE (k)}{\eK (k)}\,.
\label{zeta0_1}
\end{equation}

\medskip \noindent $\bullet$\ \ 
By using an addition theorem
\begin{equation}
\eZ_0 (u + v) = \eZ_0 (u) + \eZ_0 (v) - k^2 \sn(u)\sn(v)\sn(u + v) \,,
\label{ZZZ}
\end{equation}
one can verify the following identities\,:
\begin{align}
\frac{1}{2} \Big( {\eZ_1 (x + y) + \eZ_1 (x - y)} \Big) &= \eZ_0 (x) + \frac{{\sn (x) \cn (x) \dn (x)}}{{\sn^2 (x) - \sn^2 (y)}} \,, \\[2mm]
\frac{1}{2} \Big( {\eZ_1 (x + y) - \eZ_1 (x - y)} \Big) &= \eZ_0 (y) - \frac{{\sn (y) \cn (y) \dn (y)}}{{\sn^2 (x) - \sn^2 (y)}} \,.
\end{align}

\medskip \noindent $\bullet$\ \ 
Concerning the absolute value of $\Xi (X,T,w)$\,, one can show that
\begin{equation}
\frac{\bTh _1 (z - w) \, \bTh _1 (z + w)}{\bTh _0 ^2 (z) \, \bTh _0 ^2 (w)} = \frac{k}{{\bTh _0 ^2 (0)}} \, \Big( \sn^2 (z) - \sn^2 (w) \Big) \,.
\label{norm_id1}
\end{equation}

\noindent 
With the help of those formulae, we can easily deduce the following relations\,:
\begin{gather}
\left| {\frac{{\partial_X \Xi}}{\Xi}} \right|^2  = \frac{\sn^2(X) \cn^2(X) \dn^2 (X) - \sn^2(w) \cn^2(w) \dn^2 (w)}{\left( \sn^2(X) - \sn^2(w) \right)^2 } \,, \\[2mm]
\Re \!\pare{ \frac{\partial_T \Xi^*}{\overline \Xi} \, \frac{\partial_X \Xi}{\Xi} } = - i u \, \frac{ \sn (w) \cn (w) \dn (w) }{\sn^2(X) - \sn^2 (w)}\,,
\\[2mm]
\Im \!\pare{\frac{\partial_X \Xi}{\Xi}} = \frac{1}{i} \frac{ \sn(w)  \cn(w)  \dn (w) }{\sn^2(X) - \sn^2 (w)}\,.
\end{gather}
These relations are useful in evaluating the consistency condition, Virasoro conditions and conserved charges of helical strings.

We can now discuss a generalization of the ansatz \eqref{ansatz xi}. 
In order for $\Xi (X,T,w)$ to be normalisable for all range of $X$\,, $\eZ_0 (w|k)$ must be purely imaginary.
When $k$ is real, this can be achieved if and only if $w = m \eK(k) + \iom$ with $m \in \bb{Z}$ and $\omega \in \bb{R}$\,. 
Therefore, under the ansatz (\ref{ansatz xi}), general solutions of \eqref{reduced_eom-AdS} are given by
\begin{alignat}{2}
\Xi^0 &=   \frac{{\bTh _1 (X  - i\omega )}}{{\bTh _0 (X ) \ssp \bTh_0 (\iom)}} \, 
\exp \Big( {\eZ_0 (i\omega )X + i u T} \Big) , &\hspace{4mm}
u^2 &= U + \dn^2 (\iom)\,,
\label{Xi-0} \\[2mm]
\Xi^1 &=   \frac{{\bTh _0 (X  - i\omega )}}{{\bTh _0 (X ) \ssp \bTh_1 (\iom)}} \,
\exp \Big( {\eZ_1 (i\omega )X + i u T} \Big) , &\hspace{4mm}
u^2 &= U - \frac{\cn^2 (\iom)}{\sn^2 (\iom)}\,,
\label{Xi-1} \\[2mm]
\Xi^2 &=   \frac{{\bTh _3 (X  - i\omega )}}{{\bTh _0 (X ) \ssp \bTh_2 (\iom)}} \,
\exp \Big( {\eZ_2 (i\omega )X + iuT} \Big) , &\hspace{4mm}
u^2 &= U - \frac{(1 - k^2) \sn^2 (\iom)}{\cn^2 (\iom)}\,,
\label{Xi-2}  \\[2mm]
\Xi^3 &=   \frac{{\bTh _2 (X  - i\omega )}}{{\bTh _0 (X ) \ssp \bTh_3 (\iom)}}\exp \Big( \eZ_3 (i\omega ) X+ i uT \Big) , &\hspace{4mm}
u^2 &= U + \frac{1-k^2}{\dn^2 (\iom)}\,.
\label{Xi-3}
\end{alignat}
These four functions are mutually related by a shift of $w$ as
\begin{alignat}{2}
\Xi^0(X,T;w) &= \Xi(X,T;\iom ) \,,&\quad 
\Xi^1(X,T;w) &= - \Xi(X,T;\iom - i\eK' ) \,,\no\\
\Xi^2(X,T;w) &= \Xi(X,T;\iom - \eK-i\eK' ) \,,&\quad 
\Xi^3(X,T;w) &= \Xi(X,T;\iom -\eK ) \,. \label{Xi} 
\end{alignat}
Notice that $\Xi^i$ are doubly periodic with respect to $w$\,:
\begin{equation}
\Xi^i  \to  - \Xi^i \quad \left( {w \to w + 2\eK} \right),\qquad
\Xi^i  \to \Xi^i \quad \left( {w \to w + 2i\eK'} \right),
\end{equation}
and quasi-periodic with respect to $X $\,:
\begin{alignat}{2}
\Xi^0 (X  + 2\eK) &=  - e^{2\eZ_0 (w)\eK } \, \Xi^0 (X) \,, &\qquad
\Xi^1 (X  + 2\eK) &= e^{2\eZ_1 (w)\eK } \, \Xi^1 (X) \,, \cr
\Xi^2 (X  + 2\eK) &= e^{2\eZ_2 (w)\eK } \, \Xi^2 (X) \,, &\qquad
\Xi^3 (X  + 2\eK) &= - e^{2\eZ_3 (w)\eK } \, \Xi^3 (X) \,.
\label{Xi_per_sig}
\end{alignat}
Note also that in $\omega \to 0$ limit, the functions $\Xi^{0}$\,, $\Xi^{2}$ and $\Xi^{3}$ reduce to $\sn(X)$\,, $\dn(X)$ and $\cn(X)$
with the angular velocity satisfying $u^2 = U + 1$\,, $U$ and $U + 1-k^{2}$\,, respectively.

\section{Useful formulae\label{app:formula}}

This appendix provides some more formulae that are useful for computation involving Jacobi elliptic functions and elliptic integrals in Chapter \ref{chap:OS} and Appendix \ref{app:AdS helicals}.

\subsubsection*{Elliptic functions and elliptic integrals near \bmt{k = 1}}

The behavior of Jacobi elliptic functions around $k = 1$ is
discussed below.\footnote{We make the elliptic moduli explicit
in this section, and use the same conventions as
\cite{Okamura:2006zv}. } We follow the method of
\cite{Lin:1962:EJE}, where they computed asymptotics around $k =
0$\,.

\paragraph{$\bullet$ Jacobi sn, cn and dn functions.}

The Jacobi sn function obeys an equation
\begin{equation}
u = \int_0^{\sn(u|k)} \frac{dt}{\sqrt{1-t^2} \sqrt{1-k^2t^2}} \,.
\end{equation}
Differentiating both sides with respect to $k$\,, one finds
\begin{equation}
\frac{\partial \sn(u|k)}{\partial k} = - \cn (u|k) \dn (u|k) \int_{0}^{\sn(u|k)} \frac{k \ssp t^2 \, dt}{\sqrt{1-t^2} \pare{1-k^2t^2}^{3/2}} \,.
\end{equation}
Taking the limit $k \to 1$ and substituting $u=i\omega$\,, we obtain
\begin{equation}
\left. \frac{\partial \sn(u|k)}{\partial k} \right|_{k \to 1} = \frac{i \pare{\omega - \sin \omega \cos \omega}}{2 \cos^2 \omega} \,,
\end{equation}
which is the first term in the expansion of the Jacobi sn function
around $k=1$\,.
The asymptotics of the Jacobi cn and dn functions can be
determined by (\ref{sn cn dn}).

\paragraph{$\bullet$ Jacobi zeta function.}

The Jacobi zeta function behaves around $k=1$ as
\begin{equation}
\eZ_0 (u|k=e^{-r}) = \tanh u + \frac{z_2 (u)}{\ln r} + r \ssp z_1 (u) + \ldots.
\end{equation}
The third term containing $z_1 (u)$ can be evaluated by using the formula \cite{Byrd:1971:HEI}:
\begin{equation}
\mathop {\rm lim}_{k \to 1} \ \eK(k) \pare{\eZ_0 (u|k) - \tanh u} = - u\,,
\label{leading Zeta}
\end{equation}
while in the second term, $z_2 (u)$ can be determined by (\ref{zeta0_1}) and (\ref{ZZZ}).

\paragraph{$\bullet$ Complete elliptic integrals.}

For actual use of the relations \eqref{leading Zeta} and (\ref{zeta0_1}), we need to know the asymptotics of complete ellitpic integrals.
They are given by
\begin{alignat}{2}
\eK (e^{-r}) &= - \hf{1}{2} \ssp \ln r + \hf{3}{2} \ssp \ln 2 - \hf{1}{4} \ssp r \ln r + \cO(r \ln^m r)\,, \ \ & & \\
\eE (e^{-r}) &= 1 - \hf{1}{2} \ssp r \ln r + \cO(r \ln^m r)\,, & &
\end{alignat}
with $m>1$\,. Changing the elliptic modulus from $k$ to
$e^{-r}$\,, the asymptotic behaviors of elliptic functions around
$r=0$ are given by
\begin{alignat}{1}
\sn (\iom| e^{-r}) &= i \tan \omega - ir \, \frac{\omega - \sin \omega \cos \omega}{2 \cos^2 \omega} + \cO (r^2)\,,   \label{expand sn} \\[1mm]
\cn (\iom| e^{-r}) &= \frac{1}{\cos \omega} - r\, \frac{\omega \sin \omega - \sin^2 \omega \cos \omega}{2 \cos^2 \omega} + \cO (r^2)\,,  \label{expand cn} \\[1mm]
\dn (\iom| e^{-r}) &= \frac{1}{\cos \omega} - r \, \frac{\omega \sin \omega + \sin^2 \omega \cos \omega}{2 \cos^2 \omega} + \cO (r^2)\,,  \label{expand dn} \\[1mm]
\eZ_0 (\iom| e^{-r}) &= i \tan \omega - ir \, \frac{\omega + \sin \omega \cos \omega}{2 \cos^2 \omega} + \frac{2 \ssp i \ssp \omega}{\ln r} + \cO (r^2)\,.
\label{expand Zeta}
\end{alignat}

\subsubsection*{Moduli transformations}\label{app:Ttransf}

We collect some formulae for $SL(2,\bb{Z})$ transformations acting
on elliptic functions.

Elliptic theta functions transform under the ${\sf T}$-transformation as
\begin{alignat}{2}
\vartheta _0 (z|\tau  + 1) &= \vartheta _3 (z|\tau )\,,&\qquad
\vartheta _1 (z|\tau  + 1) &= e^{\pi i/4} \, \vartheta _1 (z|\tau )\,, \\[1mm]
\vartheta _2 (z|\tau  + 1) &= e^{\pi i/4} \, \vartheta _2 (z|\tau )\,,&\qquad
\vartheta _3 (z|\tau  + 1) &= \vartheta _0 (z|\tau )\,,
\end{alignat}
and complete elliptic integrals with $q \ge 0$ transform as
\begin{equation}
\eK (q ) = k' \eK (k)\,,\qquad \eK '(q ) = k' \left( {\eK'(k) - i \eK (k)} \right)\,,\qquad \eE (q ) = \eE (k)/k'\,.
\end{equation}
Jacobian theta functions $\bTh _\nu  (z|k)$
transform as
\begin{alignat}{2}
\bTh _0 (z|\tau  + 1) &= \bTh _3 ( z/k' |\tau )\,,&\qquad
\bTh _1 (z|\tau  + 1) &= e^{\pi i/4} \, \bTh _1 ( z/k' |\tau )\,, \\[1mm]
\bTh _2 (z|\tau  + 1) &= e^{\pi i/4} \, \bTh _2 ( z/k' |\tau )\,,&\qquad
\bTh _3 (z|\tau  + 1) &= \bTh _0 ( z/k' |\tau )\,,
\end{alignat}
and Jacobian zeta functions $\eZ_\nu (z|k) \equiv \partial _z \ln \bTh _\nu (z|k)$ transform as
\begin{alignat}{2}
\eZ_0 (z|\tau  + 1) &= \eZ_3 (z/k' |\tau) / k' \,,&\qquad
\eZ_1 (z|\tau  + 1) &= \eZ_1 (z/k' |\tau) / k' \,, \\[1mm]
\eZ_2 (z|\tau  + 1) &= \eZ_2 (z/k' |\tau) / k' \,,&\qquad
\eZ_3 (z|\tau  + 1) &= \eZ_0 (z/k' |\tau) / k' \,.
\end{alignat}
Therefore, the ${\sf T}$-transformation acts on the elliptic
modulus $k$ as
\begin{alignat}{2}
q &\equiv \left( {\frac{{\bTh _2 (0|\tau  + 1)}}{{\bTh _3 (0|\tau  + 1)}}} \right)^2 = i \left( {\frac{{\bTh _2 (0|\tau)}}{{\bTh _0 (0|\tau)}}} \right)^2 & &= \frac{{ik}}{{k'}} \,, \\[2mm]
q ^\prime &\equiv \left( {\frac{{\bTh _0 (0|\tau  + 1)}}{{\bTh _3 (0|\tau  + 1)}}} \right)^2 = \left( {\frac{{\bTh _3 (0|\tau)}}{{\bTh _0 (0|\tau)}}} \right)^2 & &= \frac{1}{{k'}} \,.
\end{alignat}
In terms of the modulus $q$ defined in (\ref{moduli kq}),
Jacobian sn, cn and dn functions are written as
\begin{equation}
\sn(z|q ) = k' \frac{\sn (z/k'|k)}{\dn (z/k'|k)} \,, \quad
\cn(z|q ) = \frac{\cn (z/k'|k)}{\dn (z/k'|k)} \,, \quad
\dn(z|q ) = \frac{1}{\dn (z/k'|k)} \,.
\end{equation}



\providecommand{\href}[2]{#2}\begingroup\raggedright\endgroup

\end{document}